\documentclass[12pt]{article}
\usepackage{jheppub}

\pdfoutput=1

\usepackage{amsmath,bbm,array,amsfonts,graphicx,wrapfig,lscape,float,mathtools,multirow,longtable}
\usepackage[dvipsnames]{xcolor}
\usepackage{array}

\usepackage{tikz-cd}

\newcommand{\be}{\begin{equation}}
\newcommand{\ee}{\end{equation}}
\newcommand{\beq}{\begin{equation}}
\newcommand{\beql}[1]{\begin{equation}\label{#1}}
\newcommand{\eeq}{\end{equation}}
\newcommand{\ba}{\begin{array}}
\newcommand{\ea}{\end{array}}
\newcommand{\bea}{\begin{eqnarray}}
\newcommand{\beal}[1]{\begin{eqnarray}\label{#1}}
\newcommand{\eea}{\end{eqnarray}}
\newcommand{\ben}{\begin{enumerate}}
\newcommand{\een}{\end{enumerate}}
\newcommand{\bean}{\begin{eqnarray*}}
\newcommand{\eean}{\end{eqnarray*}}
\newcommand{\eref}[1]{(\ref{#1})}
\newcommand{\sref}[1]{\S\ref{#1}}
\newcommand{\tref}[1]{Table~\ref{#1}}
\newcommand{\nn}{\nonumber}

\newcommand{\fref}[1]{Figure \ref{#1}}
\newcommand{\btab}[1]{\begin{tabular}{#1}}
\newcommand{\etab}{\end{tabular}}

\newcommand{\comment}[1]{}

\newcommand{\qed}{\nobreak \ifvmode \relax \else
      \ifdim\lastskip<1.5em \hskip-\lastskip
      \hskip1.5em plus0em minus0.5em \fi \nobreak
      \vrule height0.75em width0.5em depth0.25em\fi}

\definecolor{darkspringgreen}{rgb}{0.09, 0.45, 0.27}
\definecolor{forestgreen}{rgb}{0.13, 0.55, 0.13}

\usepackage{array}
\usepackage{physics}
\usepackage{subcaption}
\usepackage{tikz,tikz-3dplot}
\usepackage[colorlinks=true]{hyperref}
\newcolumntype{C}[1]{>{\centering\let\newline\\\arraybackslash\hspace{0pt}}m{#1}}

\definecolor{yellow2}{rgb}{0.98, 0.80, 0.20}

\title{Abelian Orbifolds for Brane Brick Models} 
\author[a]{Juno Kwon}

\author[a,b]{and Rak-Kyeong Seong}

\affiliation[a]{
Department of Mathematical Sciences, and 
${}^{b}$Department of Physics,\\ 
Ulsan National Institute of Science and Technology,\\
50 UNIST-gil, Ulsan 44919, South Korea
}

\emailAdd{junokwon@unist.ac.kr}
\emailAdd{seong@unist.ac.kr}

\preprint{
\begin{flushright}
UNIST-MTH-26-RS-04 \\
\end{flushright}
}

\abstract{
We present a systematic procedure for constructing brane brick models corresponding to abelian orbifolds of toric Calabi-Yau 4-folds, 
extending the orbifold construction beyond the well-studied case of abelian orbifolds of $\mathbb{C}^4$. 
Given a parent brane brick model corresponding to a toric Calabi-Yau 4-fold $\mathcal{M}$, 
we show that the action of an abelian orbifold group $\Gamma$ on the generators of $\mathcal{M}$ induces an action on the chiral and Fermi fields 
as well as the $J$- and $E$-terms of the associated $2d$ $(0,2)$ supersymmetric gauge theory. 
The requirement that the orbifolded brane brick model remains consistent with the closed paths associated with the $J$- and $E$-terms, 
together with the chiral cycles formed by their products, 
precisely reproduces the Calabi-Yau condition on the orbifold action. 
This procedure yields an explicit formula for the $J$- and $E$-terms of the orbifolded brane brick model in terms of those of the parent theory. 
We apply our construction to the brane brick models corresponding to $Q^{1,1,1}$ and $D_3$, 
and present the resulting families of $2d$ $(0,2)$ quiver gauge theories. 
We also present explicit expressions for generating functions that count distinct abelian orbifolds of $Q^{1,1,1}$ and $D_3$. 
}

\begin{document}

\maketitle

%=================================================================
\section{Introduction \label{sec01}}

The discovery of a large family of $2d$ $(0,2)$ supersymmetric gauge theories known as brane brick models \cite{Franco:2015tna, Franco:2015tya} 
has led to a series of advancements in our understanding of the interplay between $2d$ $(0,2)$ theories, toric Calabi-Yau geometry and brane engineering in string theory \cite{Franco:2016nwv,Franco:2016qxh, Franco:2016fxm, Franco:2016tcm, He:2017gam, Franco:2017cjj, Franco:2018qsc, Closset:2017xsc, Franco:2019bmx, Franco:2019pum, Kho:2023dcm, Franco:2023tyf, Ghim:2024asj, Bao:2024nyu, Ghim:2025zhs, Carcamo:2025shw}. 
Brane brick models are worldvolume theories of D1-branes probing toric Calabi-Yau 4-folds, 
and they have been constructed for a variety of Calabi-Yau geometries, 
including cones over toric Fano 3-folds \cite{batyrev1993dual, borisov1993towards, Davey:2011mz, Franco:2022gvl}, 
cones over families of Sasaki-Einstein 7-manifolds such as $Y^{p,k}(\mathbb{CP}^1 \times \mathbb{CP}^1)$ and $Y^{p,k}(\mathbb{CP}^2)$ \cite{Martelli:2008rt, Sorokin:1985ap,Nilsson:1984bj,  Gauntlett:2004hh, Franco:2022isw}, 
as well as the cones over $Q^{1,1,1}$ and $D_3$ \cite{Nilsson:1984bj, Franco:2008um, Franco:2016nwv,Franco:2016qxh}.

Many of these brane brick models have been obtained via a process known as \textit{orbifold reduction} \cite{Franco:2016fxm} 
and its generalization known as \textit{$3d$ printing} \cite{Franco:2018qsc}. 
These methods start from a $4d$ $\mathcal{N}=1$ supersymmetric gauge theory living 
on a D3-brane probing a toric Calabi-Yau 3-fold \cite{fulton1993introduction, Leung:1997tw, Beasley:1999uz, Feng:2000mi}, also referred to as a brane tiling \cite{Franco:2005rj, Hanany:2005ve, Franco:2005sm, Hanany:2005ss, kenyon2003introduction, Kennaway:2007tq, Yamazaki:2008bt,  Hanany:2012hi},
and dimensionally reduce it to $2d$ while breaking the supersymmetry to $(0,2)$. 
While powerful, these methods effectively lift the vertices of a toric diagram in $\mathbb{Z}^2$, 
corresponding to a toric Calabi-Yau 3-fold, 
along the orthogonal direction of the $\mathbb{Z}^2$ plane embedded in $\mathbb{Z}^3$, 
in order to target toric diagrams in $\mathbb{Z}^3$ corresponding to toric Calabi-Yau 4-folds.

%------------------------------------------------------------------------------------------
\begin{figure}[H]
\centering
\includegraphics[width=0.9\linewidth]{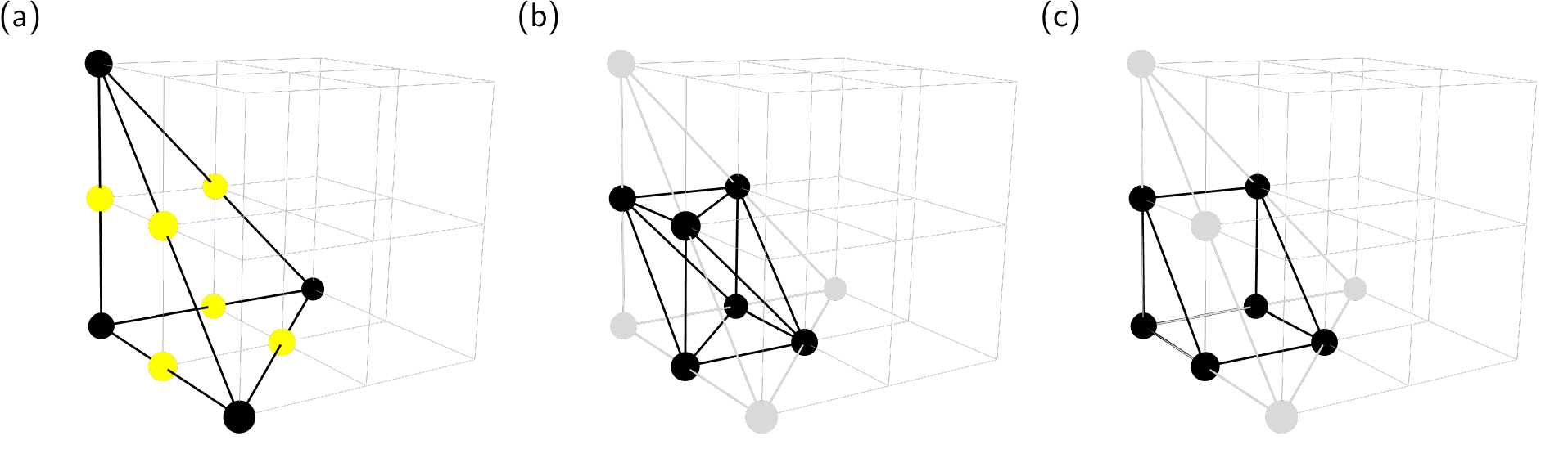}
\caption{
The toric diagram for (a) the abelian orbifold
of the form $\mathbb{C}^4/\mathbb{Z}_2 \times \mathbb{Z}_2 \times \mathbb{Z}_2$,
containing the toric diagram for (b) $Q^{1,1,1}$ and (c) $D_3$.
}
\label{fig_00}
\end{figure}
%------------------------------------------------------------------------------------------

Another method involves the use of worldvolume theories of D-branes at orbifold singularities \cite{Douglas:1996sw, Douglas:1997de, Klebanov:1998hh}. 
The method involves brane brick models corresponding to abelian orbifolds of $\mathbb{C}^4$, 
identified in \cite{Franco:2015tna}, 
and makes use of the Higgs mechanism. 
While abelian orbifolds of $\mathbb{C}^4$ have $3$-simplices in $\mathbb{Z}^3$ as their toric diagrams, 
giving a non-zero vacuum expectation value to a bifundamental chiral field and integrating out the resulting massive $J$- and $E$-terms in the corresponding 
$2d$ $(0,2)$ theory translates into a partial resolution of the orbifold singularity. 
This amounts to removing selected GLSM fields, or equivalently deleting appropriate vertices from the toric diagram,
allowing one in principle to reach any toric Calabi-Yau 4-fold whose toric diagram is contained within the toric diagram of the parent theory prior to the resolution. 
This method was used in the original work on brane brick models \cite{Franco:2015tna, Franco:2015tya} to construct the $2d$ $(0,2)$ theories corresponding to the cones over $Q^{1,1,1}$ and $D_3$, 
whose toric diagrams fit within the toric diagram of the abelian orbifold of the form $\mathbb{C}^4/\mathbb{Z}_2 \times \mathbb{Z}_2 \times \mathbb{Z}_2$, 
as illustrated in \fref{fig_00}. 
Abelian orbifolds of $\mathbb{C}^4$ therefore play a fundamental role in obtaining brane brick models for arbitrary toric Calabi-Yau 4-folds.

In fact, even before moving to general toric Calabi-Yau 4-folds, 
one needs to appreciate the richness of abelian orbifolds of $\mathbb{C}^d$ in general.
The classification and counting of abelian orbifolds of $\mathbb{C}^d$ 
has itself been the subject of a series of systematic studies \cite{Davey:2010px, Hanany:2010ne, Hanany:2010cx, Davey:2011dd, Hanany:2011iw}. 
Using various methods for distinguishing inequivalent abelian orbifolds of $\mathbb{C}^d$,
these works have enumerated distinct abelian orbifolds of up to $\mathbb{C}^6$, 
and generating functions for the counting of these distinct abelian orbifolds have been presented \cite{Davey:2010px, Hanany:2010ne, Hanany:2010cx, Davey:2011dd}.

%-------------------
\begin{table}[http!!]
\centering
\begin{tabular} {|c|cc|cccccc|cc|}
\hline
-& 0 & 1 & 2 & 3 & 4 & 5 & 6 & 7 & 8 & 9
\\
\hline
D4 & $\times$ & $\times$ & $\times$ & $\cdot$ & $\times$ & $\cdot$ & $\times$ & $\cdot$ & $\cdot$ & $\cdot$
\\
NS5 & $\times$ & $\times$  & \multicolumn{6}{c|}{\rule[0.5ex]{1.5cm}{0.4pt} $\Sigma$ \rule[0.5ex]{1.5cm}{0.4pt}} & $\cdot$ & $\cdot$
\\
\hline
\end{tabular}
\caption{The Type IIA brane configuration known as a brane brick model \cite{Franco:2015tna, Franco:2015tya}. \label{tab_02a01}}
\end{table}
%-------------------

The underlying Type IIA brane configuration of a brane brick model consists of D4-branes suspended between an NS5-brane that wraps a holomorphic surface $\Sigma$ \cite{Hori:2000ck, Cachazo:2001sg, Feng:2005gw, Franco:2016qxh,  Franco:2016tcm}, 
defined by the vanishing locus of the Newton polynomial $P(x,y,z)$ in $x,y,z \in \mathbb{C}^*$ associated to the toric diagram of the toric Calabi-Yau 4-fold. 
This brane configuration is T-dual to the original D1-brane probe configuration when one T-dualizes along the $(246)$-directions originally occupied by the Calabi-Yau cone, 
as summarized in \tref{tab_02a01}. 
It was shown in \cite{Franco:2015tna, Franco:2015tya} that for abelian orbifolds of $\mathbb{C}^4$, 
the NS5-brane wrapping $\Sigma$ tessellates the $3$-torus formed by the $(246)$-directions in terms of truncated octahedra, 
where the interior of each truncated octahedron corresponds to a brane brick associated with a $U(N)$ gauge group of the $2d$ $(0,2)$ theory. 
Bifundamental chiral and Fermi fields are then respectively encoded by the hexagonal and square faces shared between adjacent truncated octahedra, 
which provides a geometric realization of the Lagrangian of the corresponding $2d$ $(0,2)$ theory.

While the construction of brane brick models for abelian orbifolds of $\mathbb{C}^4$ is well established \cite{Franco:2015tna, Franco:2015tya},
an analogous systematic construction for brane brick models corresponding to abelian orbifolds of more general toric Calabi-Yau 4-folds has so far been missing. 
The aim of this work is to fill this gap by extending the orbifold construction to brane brick models corresponding to non-trivial toric Calabi-Yau 4-folds.
Beyond providing new families of $2d$ $(0,2)$ gauge theories, 
these abelian orbifolds enlarge the set of parent theories from which, 
via partial resolution and the Higgs mechanism, 
brane brick models for further toric Calabi-Yau 4-folds can be reached.

In this paper, we develop a systematic procedure for computing the brane brick model 
with its quiver and $J$- and $E$-terms
for an arbitrary abelian orbifold of the form $\mathcal{M}/\Gamma$, 
where $\mathcal{M}$ is a toric Calabi-Yau 4-fold, 
which corresponds to the mesonic moduli space of the parent abelian brane brick model prior to orbifolding with $U(1)$ gauge groups. 
The orbifold action of $\Gamma$ leads to an action on the generators of the mesonic moduli space $\mathcal{M}$, 
which in turn induces an action on the chiral and Fermi fields as well as the $J$- and $E$-terms of the brane brick model. 
We show that the requirement that the orbifolded brane brick model remains consistent with the closed paths associated to the $J$- and $E$-terms, 
as well as with the chiral cycles formed by their products, precisely reproduces the Calabi-Yau condition on the orbifold action. 
This procedure produces an explicit formula for the $J$- and $E$-terms of the orbifolded brane brick model in terms of the orbifold action. 
We apply this orbifolding procedure to the brane brick models corresponding to $Q^{1,1,1}$ and $D_3$ \cite{Franco:2016nwv,Franco:2016qxh} and present the resulting $2d$ $(0,2)$ quiver gauge theories.

The work is organized as follows. 
In section \sref{sec02}, we begin with a review of $2d$ $(0,2)$ quiver gauge theories and brane brick models, 
the structure of their mesonic moduli spaces, and the associated Hilbert series and global symmetries. 
Sections \sref{sec022} and \sref{sec023} review the brane brick models and gauge theories corresponding to $Q^{1,1,1}$ and $D_3$, respectively. 
Section \sref{sec03} is devoted to the general construction of abelian orbifolds of brane brick models. 
After reviewing in section \sref{sec031} the algebraic structure of abelian orbifolds of toric Calabi-Yau 4-folds 
and the constraints imposed on the orbifold action, 
we present in section \sref{sec0320} a systematic procedure for computing the $J$- and $E$-terms of the orbifolded brane brick model from a parent brane brick model. 
Sections \sref{sec032} and \sref{sec033} then apply this construction to obtain the general form of 
the abelian orbifolds of the brane brick models corresponding to $Q^{1,1,1}$ and $D_3$, respectively. 
In section \sref{sec035}, we present explicit expressions for generating functions that count distinct abelian orbifolds of the form $Q^{1,1,1}/\Gamma$ and $D_3/\Gamma$ for a given orbifold order $|\Gamma|$. 
In section \sref{sec04}, we illustrate and verify the general orbifold construction by working out a series of explicit examples of brane brick models and their corresponding $2d$ $(0,2)$ quiver gauge theories for abelian orbifolds of $Q^{1,1,1}$ and $D_3$. 
\\

%=================================================================
\section{Background \label{sec02}}

%=================================================================
\subsection{$2d$ $(0,2)$ Quiver Gauge Theory and Brane Brick Models \label{sec021}}

%=========================
\paragraph{Brane Brick Models.}
Worldvolume theories on D1-branes probing toric Calabi-Yau 4-folds form a large family of $2d$ $(0,2)$ quiver gauge theories \cite{Mohri:1997ef, Garcia-Compean:1998sla, Franco:2015tna, Franco:2015tya}.
By applying T-duality, the D1-branes become D4-branes suspended between an NS5-brane, 
which itself comes from the probed toric Calabi-Yau 4-fold under T-duality.
The NS5-brane wraps a holomorphic surface $\Sigma$ defined by the vanishing locus of the Newton polynomial $P(x,y,z)$
of the toric diagram that characterizes the probed toric Calabi-Yau 4-fold. 
The resulting brane configuration is summarized in \tref{tab_02a01} and is known as a \textit{brane brick model} \cite{Franco:2015tna, Franco:2015tya}.

Brane brick models encode the Lagrangian of the corresponding $2d$ $(0,2)$ quiver gauge theory
as well as the geometry of the associated toric Calabi-Yau 4-fold.
A brane brick model is subdivided into
brane bricks, brick faces and edges. The boundary of a brane brick is made of two kinds of faces, corresponding to bifundamental chiral fields and Fermi fields. 
Each brane brick corresponds to a $U(N)$ gauge group
and since each brick face is adjacent to two brane bricks, they can be associated to bifundamental chiral or Fermi fields.
Chiral fields are associated to oriented brick faces that
transform under the bifundamental representation of the two $U(N)$ gauge groups associated to the two adjacent brane bricks.
Similarly, certain brick faces remain unoriented along their boundary edges and are associated to Fermi fields.
\\

%=========================
\paragraph{$2d$ $(0,2)$ Gauge Theories and Quivers.}
The brane brick model fully encodes the Lagrangian of the corresponding $2d$ $(0,2)$ gauge theory.
The \textit{quiver} of a $2d$ $(0,2)$ gauge theory
consists of nodes corresponding to $U(N)$ gauge groups, and edges corresponding to either a chiral field or Fermi field. 
When the edge is oriented, 
then the orientation identifies the bifundamental representation of the associated chiral field,
whereas when the edge is unoriented, then it corresponds to a Fermi field.

Next to the quiver,
we also have for each Fermi field $\Lambda$ and its conjugate $\overline{\Lambda}$ a corresponding \textit{$J$- and $E$-term}.
In the brane brick model, every $J$- and $E$-term is given by binomial relations of the following form,
\beal{es02a01}
&\Lambda_{ij}& \ : \ J^{+}_{ji}-J^{-}_{ji}~,~
\nn\\
&\overline{\Lambda}_{ij}& \ : \ E^{+}_{ij}-E^{-}_{ij} ~,~
\eea
where $J^{\pm}_{ji}$ and $E^{\pm}_{ij}$ are monomials in terms of chiral fields, which combined with corresponding Fermi fields
form closed paths in the quiver of the brane brick model, 
\beal{es02a01b}
\Lambda_{ij} \cdot J^{+}_{ji} ~,~
\Lambda_{ij} \cdot J^{-}_{ji} ~,~
\overline{\Lambda}_{ij} \cdot E^{+}_{ij} ~,~
\overline{\Lambda}_{ij} \cdot E^{-}_{ij} ~,~
\eea
known as \textit{plaquettes}.
In the quiver, $J^{\pm}_{ji}$ corresponding to $\Lambda_{ij}$ can be considered as a directed path from node $j$ to node $i$, whereas $E^{\pm}_{ij}$ corresponding to $\overline{\Lambda}_{ij}$ can be considered as a directed path from node $i$ to node $j$.

\tref{tab_02a02} summarizes the dictionary between objects in the brane brick model, the corresponding quiver and the $2d$ $(0,2)$ theory.
\\

%-------------------
\begin{table}[H]
\centering
\begin{tabular} {|c|c|c|}
\hline
Brane brick model & Quiver & Gauge theory
\\
\hline
\hline
brane brick & node & $U(N)_i$ gauge group
\\
\hline
oriented brick face & directed edge & chiral field $X_{ij}$
\\
\hline
unoriented brick face & undirected edge & Fermi field $\Lambda_{ij}$ ($\overline{\Lambda}_{ij}$)
\\
\hline
edge & directed closed path & plaquettes $\Lambda_{ij} \cdot J^{\pm}_{ji}$ or $\overline{\Lambda}_{ij} \cdot E^{\pm}_{ij}$
\\
\hline
\end{tabular}
\caption{Dictionary between objects in the brane brick model, the corresponding quiver and the $2d$ $(0,2)$ theory. \label{tab_02a02}}
\end{table}
%-------------------

%=========================
\paragraph{Mesonic Moduli Space.}
For abelian $2d$ $(0,2)$ gauge theories with $U(1)$ gauge groups, 
the \textit{mesonic moduli space} \cite{Feng:2000mi, Forcella:2008bb, Kho:2023dcm} is given by an algebraic variety of the following form,
\beal{es02a02}
\mathcal{M}^{mes} = \mathrm{Spec}(\mathbb{C}[X_{ij}]/ \langle J, E\rangle)//U(1)^{G-1}~,~
\eea
where $X_{ij}$ are the chiral fields,
$\langle J, E \rangle$ is the binomial ideal given by the $J$- and $E$-terms,
and $G$ is the number of $U(1)$ gauge groups.
Since the $J$- and $E$-terms are binomial as defined in \eqref{es02a01}, the mesonic moduli space $\mathcal{M}^{mes}$ is toric.
In fact, the mesonic moduli space for the abelian brane brick model is precisely the probed toric Calabi-Yau 4-fold. 

By applying the \textit{forward algorithm} introduced in \cite{Franco:2015tna}, we can express the mesonic moduli space $\mathcal{M}^{mes}$ in terms of \textit{GLSM fields} $p_a$ \cite{Witten:1993yc}.
The $J$-, $E$- and $D$-terms of the brane brick models
can be expressed as $U(1)$ charges on the GLSM fields.
This allows us to rewrite the expression for the mesonic moduli space as the following symplectic quotient, 
\beal{es02a03}
\mathcal{M}^{mes} = (\mathrm{Spec}(\mathbb{C}[p_{a}])//Q_{JE})//Q_D~,~
\eea
where $Q_{JE}$ is the charge matrix from the $J$- and $E$-terms, $Q_D$ is the charge matrix from the $D$-terms, and $p_a$ are the GLSM fields.
\\

%=========================
\paragraph{Toric Diagram.}
A toric Calabi-Yau 4-fold and its associated toric variety are
characterized by a convex lattice polytope in $\mathbb{Z}^3$, also known as a \textit{toric diagram} \cite{fulton1993introduction, Leung:1997tw}.
Given that the mesonic moduli space of the abelian brane brick model is a toric Calabi-Yau 4-fold, 
its toric diagram can be obtained using the forward algorithm \cite{Franco:2015tna}.
The coordinates of the vertices of the toric diagram
correspond to columns in the $G_t$-matrix, which is given by, 
\beal{es02a05}
G_t = \text{ker} (Q_t) ~,~
\eea
where $Q_t = (Q_{JE}, Q_D)$ is the total charge matrix.
We note here that 
GLSM fields $p_a$ are associated to vertices in the toric diagram.
\\

%=========================
\paragraph{Hilbert Series.}
The \textit{Hilbert series} \cite{Benvenuti:2006qr, Hanany:2006uc, Butti:2007jv, Feng:2007ur, Kho:2024wcw} allows us to count mesonic gauge invariant operators of $\mathcal{M}^{mes}$.
By expressing the mesonic gauge invariant operators in terms of GLSM fields $p_a$,
whose associated fugacities in the Hilbert series are given by $t_a$, 
we define the Hilbert series using the Molien integral as follows,
\beal{es02a04}
g(t_\alpha; \mathcal{M}^{mes})=\prod_{i=1}^{c-4} \oint_{|z_i|=1} \frac{dz_i}{2 \pi i z_i} \prod_{\alpha=1}^{c} \frac{1}{1-t_\alpha \prod_{j=1}^{c-4} z^{(Q_t)_{j \alpha}}_j} ~,
\eea
where $c$ is the number of GLSM fields.
The generators and defining relations formed by the generators of the mesonic moduli space $\mathcal{M}^{mes}$
can be identified using the plethystic logarithm of the Hilbert series \cite{Benvenuti:2006qr, Hanany:2006uc, Butti:2007jv, Feng:2007ur}, which is given by,
\beal{es02a04b}
PL[g(t_\alpha; \mathcal{M}^{mes})]
= 
\sum_{k=1}^{\infty}
\frac{\mu(k)}{k}
\log\left[
g(t_\alpha^k; \mathcal{M}^{mes})
\right]
~,~
\eea
where $\mu(k)$ is the M\"obius function.
In the expansion of the plethystic logarithm ordered by the overall degree of the fugacities $t_a$, 
the first positive terms correspond to the generators of $\mathcal{M}^{mes}$, 
whereas the following negative terms correspond to the defining relations formed amongst them.
If the expansion of the plethystic logarithm is finite, the corresponding moduli space is a complete intersection.
\\

%=========================
\paragraph{Global Symmetries.}
The mesonic moduli space $\mathcal{M}^{mes}$ of a brane brick model as a toric Calabi-Yau 4-fold
has an isometry that is reflected as the mesonic flavor symmetry and the $U(1)_R$ symmetry of the corresponding $2d$ $(0,2)$ supersymmetric gauge theory. 
The possible choices of the mesonic flavor symmetry with the $U(1)_R$ symmetry are as follows, 
\begin{itemize}
\item $U(1)_{f_1} \times U(1)_{f_2} \times U(1)_{f_3} \times U(1)_R$
\item $SU(2)_{x} \times U(1)_{f_1} \times U(1)_{f_2} \times U(1)_R$
\item $SU(2)_{x_1} \times SU(2)_{x_2} \times U(1)_{f} \times U(1)_R$
\item $SU(2)_{x_1} \times SU(2)_{x_2} \times SU(2)_{x_3} \times U(1)_R$
\item $SU(3)_{x_1,x_2} \times U(1)_{f} \times U(1)_R$
\item $SU(3)_{x_1,x_2} \times SU(2)_{x_3} \times U(1)_R$
\item $SU(4)_{x_1,x_2, x_3} \times U(1)_R$
\end{itemize}
where the lower case indices on the mesonic flavor symmetries indicate fugacities in the Hilbert series
for charges under the symmetry, and the $U(1)_R$ symmetry has a fugacity given by $\bar{t}$.
In the examples below, we use $t$ as the degree-counting fugacity for extremal GLSM fields.
Here, we note that the enhancement of a $U(1)$ mesonic flavor symmetry to a $SU(2)$ and $SU(3)$ factor is indicated by repeated columns in the total charge matrix $Q_t= (Q_{JE}, Q_D)$.
\\

%=================================================================
\subsection{The $2d$ $(0,2)$ Quiver Gauge Theory corresponding to $Q^{1,1,1}$ \label{sec022}}

%------------------------------------------------------------------------------------------
\begin{figure}[H]
\centering
\includegraphics[width=0.55\linewidth]{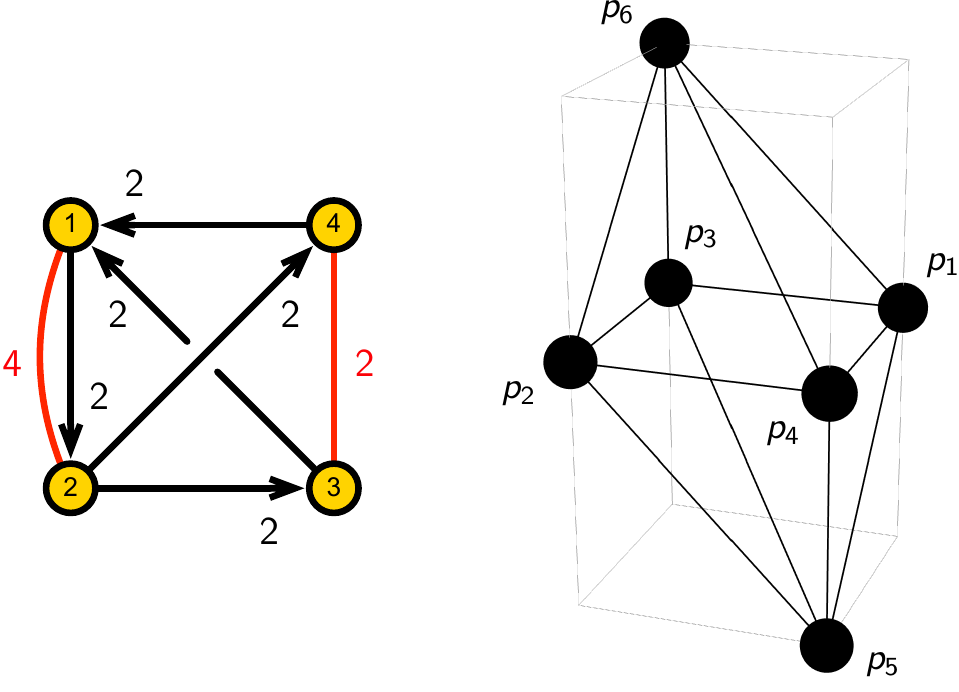}
\caption{
The quiver and toric diagram for the $Q^{1,1,1}$ brane brick model. 
}
\label{fig_01}
\end{figure}
%------------------------------------------------------------------------------------------

The quiver of the brane brick model corresponding to $Q^{1,1,1}$ \cite{Franco:2016nwv,Franco:2016qxh}
and the toric diagram are shown in \fref{fig_01}.
The  $J$- and $E$-terms for the $2d$ $(0,2)$ gauge theory are given by, 
\beal{es02b01}
\begin{array}{rrclcrcl}
& & J & & & & E &  \\
\Lambda^{1}_{21}: & 
D_{12} \cdot Z_{24} \cdot Y_{41} \cdot X_{12}  &-&  X_{12} \cdot Z_{23} \cdot D_{31} \cdot D_{12} 
& ~~ &
X_{23}\cdot Y_{31} &-& X_{24}\cdot D_{41} 
\\
\Lambda^{2}_{21}: &
X_{12} \cdot X_{24} \cdot Y_{41} \cdot D_{12} &-& D_{12} \cdot Z_{23} \cdot Y_{31} \cdot X_{12} 
& ~~ &
X_{23} \cdot D_{31} &-& Z_{24} \cdot D_{41} 
\\
\Lambda^{3}_{21}: &
D_{12} \cdot Z_{24} \cdot D_{41} \cdot X_{12} &-& X_{12} \cdot X_{23} \cdot D_{31} \cdot D_{12} 
& ~~ &
X_{24} \cdot Y_{41} &-& Z_{23} \cdot Y_{31}
 \\
\Lambda^{4}_{21}: &
D_{12} \cdot X_{23} \cdot Y_{31} \cdot X_{12} &-& X_{12} \cdot X_{24} \cdot D_{41} \cdot D_{12} 
& ~~ &
Z_{23} \cdot D_{31} &-& Z_{24} \cdot Y_{41} 
\\
\Lambda_{43}: &
D_{31} \cdot X_{12} \cdot X_{24} &-&Y_{31} \cdot X_{12} \cdot Z_{24} 
& ~~ &
D_{41} \cdot D_{12} \cdot Z_{23} &-& Y_{41} \cdot D_{12} \cdot X_{23} 
\\
\Lambda_{34}: &
Y_{41} \cdot X_{12} \cdot X_{23} &-& D_{41} \cdot X_{12} \cdot Z_{23} 
& ~~ &
D_{31} \cdot D_{12} \cdot X_{24} &-& Y_{31} \cdot D_{12} \cdot Z_{24}
\\
\end{array}
\nn\\
\eea

The forward algorithm allows us to calculate the GLSM field matrix as follows, 
\beal{es02b01m1}
P=
\left(
\begin{array}{c|cccccc|cc}
&p_1 & p_2 &p_3 & p_4 &p_5 & p_6 & o_1 & o_2 \\
\hline
 D_{12} & 1 & 0 & 0 & 0 & 0 & 0 & 0 & 0 \\
 X_{12} & 0 & 1 & 0 & 0 & 0 & 0 & 0 & 0 \\
 Y_{31} & 0 & 0 & 1 & 0 & 0 & 0 & 0 & 1 \\
 D_{31} & 0 & 0 & 0 & 1 & 0 & 0 & 0 & 1 \\
 D_{41} & 0 & 0 & 0 & 0 & 1 & 0 & 0 & 1 \\
 Y_{41} & 0 & 0 & 0 & 0 & 0 & 1 & 0 & 1 \\
 X_{24} & 0 & 0 & 1 & 0 & 0 & 0 & 1 & 0 \\
 Z_{24} & 0 & 0 & 0 & 1 & 0 & 0 & 1 & 0 \\
 X_{23} & 0 & 0 & 0 & 0 & 1 & 0 & 1 & 0 \\
 Z_{23} & 0 & 0 & 0 & 0 & 0 & 1 & 1 & 0 \\
\end{array}
\right)
~,~
\eea
where $p_a$ correspond to GLSM fields associated to extremal vertices in the toric diagram shown in \fref{fig_01}.
From the $P$-matrix, we are able to express the $J$- and $E$-terms as well as the $D$-terms 
of the $Q^{1,1,1}$ model in terms of $U(1)$ charges on the GLSM fields,
as summarized by the following total charge matrix,
\beal{es02b01m2}
Q_t = 
\left(
\begin{array}{cccccc|cc}
p_1 & p_2 &p_3 & p_4 &p_5 & p_6 & o_1 & o_2 \\
\hline
 0 & 0 & -1 & -1 & -1 & -1 & 1 & 1 \\
 -1 & -1 & 1 & 1 & 1 & 1 & -1 & 0 \\
 1 & 1 & 0 & 0 & 0 & 0 & -1 & 0 \\
 0 & 0 & -1 & -1 & 0 & 0 & 1 & 0 \\
\end{array}
\right)
~.~
\eea
From the $Q_t=(Q_{JE}, Q_D)$ charge matrix, 
we identify that the global symmetry of the mesonic moduli space $\mathcal{M}^{mes}$ of the $Q^{1,1,1}$ model is enhanced to 
\beal{es02b01m2b}
SU(2)_{x} \times SU(2)_{y} \times SU(2)_z \times U(1)_R~.~
\eea
The charges under the enhanced mesonic flavor symmetry on the extremal GLSM field $p_a$ of the $Q^{1,1,1}$ model are summarized in \tref{tab_02b01}.

%-------------------
\begin{table}[htt!!]
\centering
\begin{tabular}{|c|c|c|c|l|}
\hline
& $SU(2)_x$ & $SU(2)_y$ & $SU(2)_z$ & fugacity
\\
\hline
$p_1$ & 1 & 0 & 0 & $t_1=xt$ \\
$p_2$ & -1 & 0 & 0 & $t_2=x^{-1}t$\\
$p_3$ & 0 & 1 & 0 & $t_3=yt$\\
$p_4$ & 0 & -1 & 0 &$t_4=y^{-1}t$\\
$p_5$ & 0 & 0 & 1 &$t_5=zt$\\
$p_6$ & 0 & 0 & -1 &$t_6=z^{-1}t$\\
\hline
\end{tabular}
\caption{The charges under the enhanced mesonic flavor symmetry $SU(2)_{x} \times SU(2)_{y} \times SU(2)_z$
on the extremal GLSM fields $p_a$ for the $Q^{1,1,1}$ model. Here, $t$ is the fugacity counting the degree in $p_a$.
\label{tab_02b01}
}
\end{table}
%-------------------

The toric diagram for the $Q^{1,1,1}$ model is given by, 
\beal{es02b01m3}
G_t = 
\left(
\begin{array}{cccccc|cc}
p_1 & p_2 &p_3 & p_4 &p_5 & p_6 & o_1 & o_2 \\
\hline
 1 & 0 & 1 & 0 & 0 & 1 & 1 & 1 \\
 0 & 1 & 1 & 0 & 0 & 1 & 1 & 1 \\
 0 & 0 & 0 & 0 & -1 & 1 & 0 & 0 \\
 \hline
 1 & 1 & 1 & 1 & 1 & 1 & 2 & 2 \\
\end{array}
\right)
~.~
\eea
Here, we note that the extra GLSM fields $o_1, o_2$ do not contribute to the toric diagram of the $Q^{1,1,1}$ model \cite{Franco:2016nwv,Franco:2016qxh}.
The toric diagram of the $Q^{1,1,1}$ is illustrated in \fref{fig_01}.

The Hilbert series for $\mathcal{M}^{mes}$ can be written in terms of characters of irreducible representations of the enhanced mesonic flavor symmetry as follows, 
\beal{es02b01a}
g(x,y,z,t; \mathcal{M}^{mes}) 
=
\sum_{n=0}^{\infty}
[n; n; n] t^{3n} ~,~
\eea
where $t$ is here the fugacity counting degrees of extremal GLSM fields $p_a$,
and $[m; n; k] = [m]_x [n]_y [k]_z$ are characters of irreducible representations of the enhanced mesonic flavor symmetry $SU(2)_{x} \times SU(2)_{y} \times SU(2)_z$ with highest weight $(m)(n)(k)$. 
The plethystic logarithm of the Hilbert series takes the following form, 
\beal{es02b01b}
PL[g(x,y,z,t; \mathcal{M}^{mes})]
=
[1]_x [1]_y [1]_z t^3 - ([2]_x + [2]_y + [2]_z) t^6 + \dots ~,~
\eea
where the infinite expansion indicates that $\mathcal{M}^{mes}$ is not a complete intersection.

%-------------------
\begin{table}[htt!!]
\centering
\begin{tabular}{|c|c|ccc|}
\hline
PL term & generator & $SU(2)_x$ & $SU(2)_y$ & $SU(2)_z$  
\\
\hline 
\multirow{8}{*}{$+[1]_x [1]_y [1]_z t^3$} 
& $A_{111}=p_1 p_3 p_5 $ & $1$ & $1$ & $1$   \\
& $A_{211}=p_2 p_3 p_5 $ & $-1$ & $1$ & $1$   \\
& $A_{121}=p_1 p_4 p_5 $ & $1$ & $-1$ & $1$   \\
& $A_{221}=p_2 p_4 p_5 $ & $-1$ & $-1$ & $1$   \\
& $A_{112}=p_1 p_3 p_6 $ & $1$ & $1$ & $-1$   \\
& $A_{212}=p_2 p_3 p_6 $ & $-1$ & $1$ & $-1$   \\
& $A_{122}=p_1 p_4 p_6 $ & $1$ & $-1$ & $-1$   \\
& $A_{222}=p_2 p_4 p_6 $ & $-1$ & $-1$ & $-1$   \\

\hline
\end{tabular}
\caption{
Generators of the $Q^{1,1,1}$ model in terms of extremal GLSM fields $p_a$ and their corresponding mesonic flavor symmetry charges. The fugacity $t$ counts the degree in $p_a$. 
\label{tab_02b02}
}
\end{table}
%-------------------

From the plethystic logarithm, we identify 8 generators that form 9 defining relations. 
We are therefore able to write the mesonic moduli space in the following form,
\beal{es02b02}
\mathcal{M}^{mes}=\mathrm{Spec}(\mathbb{C}[A_{ijk}]/\mathcal{I})
\eea
where $i,j,k=1,2$ are the $SU(2)_{x} \times SU(2)_{y} \times SU(2)_z$ mesonic flavor indices.
The binomial ideal formed by the relations is given by,
\beal{es02b03}
\mathcal{I}=
\langle 
\epsilon^{ab} A_{ajk} A_{blm}
~,~
\epsilon^{ab} A_{iak} A_{jbm}
~,~
\epsilon^{ab} A_{ija} A_{lmb}
\rangle
~,~
\eea
where $\epsilon^{ab}$ is the standard anti-symmetric invariant tensor of $SU(2)$.
By relabelling the chiral fields using $SU(2)$ indices $i,j,k=1,2$, 
such that
\beal{es02b04b}
&&
(D_{12}, D_{31}, D_{41}, X_{12}, X_{23}, X_{24}, Y_{31}, Y_{41}, Z_{23}, Z_{24})
\nn\\
&&\leftrightarrow
(X^1_{12}, X^2_{31}, X^1_{41}, X^2_{12},X^1_{23}, X^1_{24}, X^1_{31}, X^2_{41}, X^2_{23}, X^2_{24})
~,~
\eea
we can succinctly write the generators of the mesonic moduli space $\mathcal{M}^{mes}$
in terms of chiral fields as follows,
\beal{es02b04}
A_{ijk} = X_{12}^i X_{24}^j X_{41}^k = X_{12}^i X_{23}^k X_{31}^j
~.~
\eea
\tref{tab_02b02} summarizes the generators 
and expresses them also in terms of GLSM fields with their corresponding charges under the mesonic flavor symmetry.
\\

%=================================================================
\subsection{The $2d$ $(0,2)$ Quiver Gauge Theory corresponding to $D_{3}$ \label{sec023}}

%------------------------------------------------------------------------------------------
\begin{figure}[H]
\centering
\includegraphics[width=0.55\linewidth]{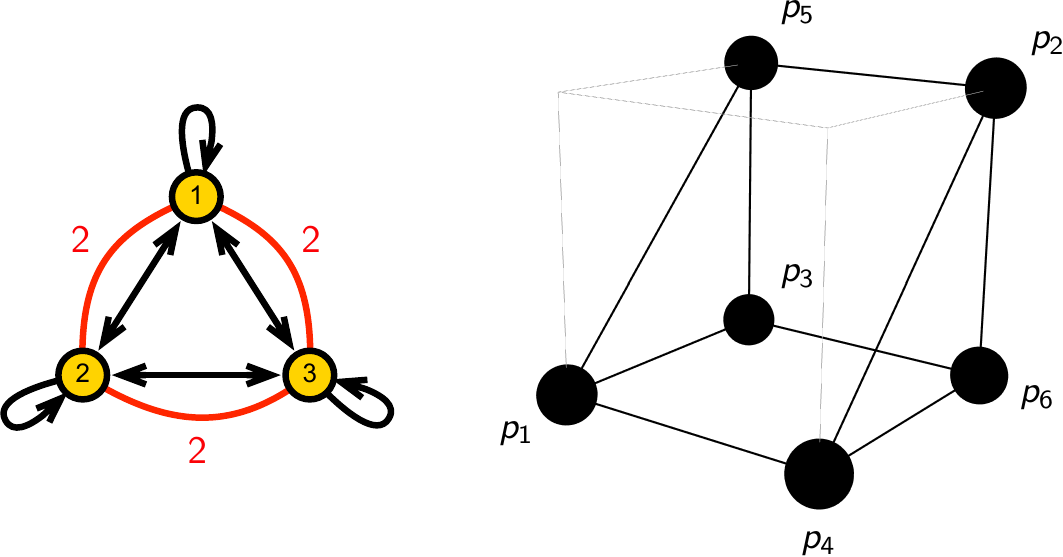}
\caption{
The quiver and toric diagram for the $D_3$ brane brick model. 
}
\label{fig_02}
\end{figure}
%------------------------------------------------------------------------------------------

\fref{fig_02} illustrates the quiver and the toric diagram for the $D_3$ brane brick model \cite{Franco:2016nwv,Franco:2016qxh}.
The $J$- and $E$-terms for the $2d$ $(0,2)$ theory are given by,
\beal{es02c01}
\begin{array}{rrclcrcl}
& & J & & & & E &  \\
\Lambda_{21}: & 
X_{13}\cdot X_{31} \cdot Y_{12} &-& Y_{12}\cdot X_{22} 
& ~~ & 
D_{23}\cdot Z_{32} \cdot Z_{21} &-& Z_{21}\cdot D_{11}
\\
\Lambda_{12}: &
Z_{21}\cdot X_{13} \cdot X_{31} &-& X_{22}\cdot Z_{21} 
& ~~ &
D_{11} \cdot Y_{12} &-& Y_{12}\cdot D_{23} \cdot Z_{32}
\\
\Lambda_{31}: &
X_{13}\cdot Y_{33} &-& Y_{12}\cdot Z_{21}\cdot X_{13} 
& ~~ &
X_{31}\cdot D_{11} &-& Z_{32}\cdot D_{23}\cdot X_{31} 
\\
\Lambda_{13}: &
X_{31}\cdot Y_{12} \cdot Z_{21} &-& Y_{33}\cdot X_{31} 
& ~~ &
D_{11}\cdot X_{13} &-& X_{13}\cdot Z_{32}\cdot D_{23} 
\\
\Lambda^{1}_{23}: &
Y_{33}\cdot Z_{32} &-& Z_{32}\cdot Z_{21} \cdot Y_{12} 
& ~~ &
D_{23}\cdot X_{31}\cdot X_{13} &-& X_{22}\cdot D_{23} 
\\
\Lambda^{2}_{23}: &
Z_{32}\cdot X_{22} &-& X_{31}\cdot X_{13}\cdot Z_{32} 
& ~~ &
D_{23}\cdot Y_{33} &-& Z_{21}\cdot Y_{12}\cdot D_{23}
\end{array}
\eea

Using the forward algorithm, we obtain the following $P$-matrix 
that summarizes the GLSM fields in terms of the chiral fields of the $D_3$ model, 
\beal{es02m01}
P=\left(
\begin{array}{c|cccccc}
&p_1 & p_2 &p_3 & p_4 &p_5 & p_6 \\
\hline
 D_{11} & 1 & 0 & 0 & 1 & 0 & 0 \\
 X_{22} & 0 & 1 & 0 & 0 & 1 & 0 \\
 Y_{33} & 0 & 0 & 1 & 0 & 0 & 1 \\
 D_{23} & 1 & 0 & 0 & 0 & 0 & 0 \\
 Z_{32} & 0 & 0 & 0 & 1 & 0 & 0 \\
 X_{13} & 0 & 1 & 0 & 0 & 0 & 0 \\
 X_{31} & 0 & 0 & 0 & 0 & 1 & 0 \\
 Y_{12} & 0 & 0 & 1 & 0 & 0 & 0 \\
 Z_{21} & 0 & 0 & 0 & 0 & 0 & 1 \\
\end{array}
\right)
~,~
\eea
where $p_a$ correspond to extremal GLSM fields. 
In terms of the GLSM fields, we are able to express the $J$- and $E$-terms, as well as the $D$-terms as $U(1)$ charges as follows, 
\beal{es02m02}
Q_t = 
\left(
\begin{array}{cccccc}
p_1 & p_2 & p_3 & p_4 & p_5 & p_6 \\
\hline
 -1 & 0 & 1 & 1 & 0 & -1 \\
 0 & -1 & -1 & 0 & 1 & 1 \\
\end{array}
\right)
~.~
\eea
Based on the $Q_t$ matrix, 
we note that the global symmetry of the mesonic moduli space 
is 
\beal{es02m02bb}
U(1)_{f_1} \times U(1)_{f_2} \times U(1)_{f_3} \times U(1)_R~.~
\eea
\tref{tab_02c01} summarizes the charges under the mesonic flavor symmetry $U(1)_{f_1} \times U(1)_{f_2} \times U(1)_{f_3}$ on the extremal GLSM fields $p_a$.

The toric diagram of the $D_3$ model is given by, 
\beal{es02m02b}
G_t = 
\left(
\begin{array}{cccccc}
p_1 & p_2 & p_3 & p_4 & p_5 & p_6 \\
\hline
 1 & 0 & 1 & 0 & 1 & 0 \\
 1 & 0 & 0 & 1 & 0 & 0 \\
 0 & 1 & 0 & 0 & 1 & 0 \\
  \hline
 1 & 1 & 1 & 1 & 1 & 1 \\
\end{array}
\right)
~,~
\eea
where columns of the $G_t$-matrix correspond to coordinates of vertices of the toric diagram as well as the associated GLSM fields.
\fref{fig_02} illustrates the toric diagram for the $D_3$ model.

%-------------------
\begin{table}[htt!!]
\centering
\begin{tabular}{|c|c|c|c|l|}
\hline
& $U(1)_{f_1}$ & $U(1)_{f_2}$ & $U(1)_{f_3}$ & fugacity
\\
\hline
$p_1$ & 1 & 0 & 0 & $t_1=f_1t$ \\
$p_2$ & -1 & 0 & 0 & $t_2={f_1}^{-1}t$\\
$p_3$ & 0 & 1 & 0 & $t_3=f_2t$\\
$p_4$ & 0 & -1 & 0 &$t_4={f_2}^{-1}t$\\
$p_5$ & 0 & 0 & 1 &$t_5=f_3t$\\
$p_6$ & 0 & 0 & -1 &$t_6={f_3}^{-1}t$\\
\hline
\end{tabular}
\caption{
Mesonic flavor symmetry of the $D_3$ model and charges on extremal GLSM fields $p_a$.
Here, the fugacity $t$ counts the degree in $p_a$. 
 \label{tab_02c01}
 }
\end{table}
%-------------------

The Hilbert series for the mesonic moduli space 
$\mathcal{M}^{mes}$ takes the following form,
\beal{es02m02c}
&&
g(t_a;\mathcal{M}^{mes})
=
\frac{
1-  t_1 t_2 t_3 t_4 t_5 t_6
}{
(1 -  t_1 t_4) 
(1 -  t_3 t_6) 
(1 -  t_2 t_5)  
(1 -  t_1 t_3 t_5) 
(1 -  t_2 t_4 t_6) 
}~,~
\eea
where the fugacities $t_a$ correspond to GLSM fields $p_a$.
When refined under the mesonic flavor symmetry summarized in \tref{tab_02c01}, 
the Hilbert series takes the following form,
\beal{es02m03}
&&
g(f_1,f_2,f_3,t;\mathcal{M}^{mes})
=
\nn\\
&&
\hspace{1cm}
\frac{
1-t^6
}{
(1 -  f_1 f_2^{-1} t^2) 
(1 -  f_2 f_3^{-1} t^2) 
(1 -  f_1^{-1} f_3 t^2)  
(1 -  f_1 f_2 f_3 t^3) 
(1 -  f_1^{-1} f_2^{-1} f_3^{-1} t^3) 
}
~,~
\nn\\
\eea
where $t$ counts degrees in extremal GLSM fields $p_a$ and $f_1, f_2, f_3$ are fugacities corresponding to the charges under the mesonic flavor symmetry $U(1)_{f_1} \times U(1)_{f_2} \times U(1)_{f_3}$.
The corresponding plethystic logarithm takes the form, 
\beal{es02m04}
PL[g(f_1,f_2,f_3,t;\mathcal{M}^{mes})]
&=&
(f_1 f_2^{-1} + f_1^{-1} f_3 + f_2 f_3^{-1}) t^2 
\nn\\
&&
+ (f_1 f_2 f_3 + f_1^{-1} f_2^{-1} f_3^{-1}) t^3
- t^6
~,~
\eea
which indicates that $\mathcal{M}^{mes}$ is a complete intersection. 

In fact, the plethystic logarithm shows that the mesonic moduli space $\mathcal{M}^{mes}$ has 5 generators satisfying a single defining relation, which allows us to express the mesonic moduli space as follows, 
\beal{es02c02}
\mathcal{M}^{mes}=\mathrm{Spec}(\mathbb{C}[A_{1},A_{2},A_{3},B_{1},B_{2}]/\langle A_{1} A_{2} A_{3}-B_{1} B_{2}\rangle)
~,~
\eea
where
\beal{es02c03}
\begin{array}{rclccclcc}
A_{1}&=&D_{11}=D_{23} Z_{32}\\
A_{2}&=&X_{22}=X_{13} X_{31}\\
A_{3}&=&Y_{33}=Y_{12} Z_{21}\\
B_{1}&=&D_{23} X_{31} Y_{12}\\
B_{2}&=&Z_{21} X_{13} Z_{32}
\end{array}
~.~
\eea
\tref{tab_02c02} summarizes the generators also in terms of the extremal GLSM fields with the corresponding charges under the mesonic flavor symmetry. 
\\

%-------------------
\begin{table}[htt!!]
\centering
\begin{tabular}{|c|c|ccc|}
\hline
PL term & generator &  $U(1)_{f_1}$ & $U(1)_{f_2}$ & $U(1)_{f_3}$  
\\
\hline 
$f_1 f_2^{-1} t^2$ & $A_1 = p_1 p_4$ & 1 & -1 & 0 \\
\hline
$f_1^{-1} f_3 t^2$ & $A_2 = p_2 p_5$ & -1 & 0 & 1 \\
\hline
$f_2 f_3^{-1} t^2$ & $A_3 = p_3 p_6$ & 0 & 1 & -1 \\
\hline
$f_1 f_2 f_3 t^3$ & $B_1= p_1 p_3 p_5$ & 1 & 1 & 1 \\
\hline
$f_1^{-1} f_2^{-1} f_3^{-1} t^3$ & $B_2= p_2 p_4 p_6$ & -1 & -1 & -1 \\
\hline
\end{tabular}
\caption{
Generators of the $D_3$ model in terms of extremal GLSM fields $p_a$ and their corresponding mesonic flavor symmetry charges. 
Here, the fugacity $t$ counts the degree in $p_a$. 
\label{tab_02c02}}
\end{table}
%-------------------

%=================================================================
\section{Abelian Orbifolds for Brane Brick Models \label{sec03}}

%=================================================================
\subsection{Toric Calabi-Yau 4-folds \label{sec031}}

%-------------------------------------------------------
\paragraph{Toric Calabi-Yau 4-fold.}
A non-compact toric Calabi-Yau 4-fold is given by,
\beal{es02d0001}
\mathcal{M} = \text{Spec}(\mathbb{C}[z_1, \dots, z_d] / \mathcal{I}) ~,~
\eea
where $z_1, \dots, z_d$ are the generators of the coordinate ring $\mathbb{C}[z_1, \dots, z_d]$.
In other words, we note that $\mathcal{M}$ is embedded into $\mathbb{C}^d$
as an affine variety of dimension 4.
The quotient ideal $\mathcal{I}$ takes the form,
\beal{es02d0002}
\mathcal{I} = \langle
R_1, \dots, R_{M}
\rangle
~.~
\eea
Here, $\mathcal{I}$
is a binomial ideal, where 
$R_l$ form binomial relations in $z_1, \dots, z_d$.
These take the following form,
\beal{es02d12}
R_l = \left( \prod_{u \in I_{l}^{+} } z_{u} \right) - \left( \prod_{v \in I_{l}^{-} } z_{v}\right)
~,~
\eea
where $I_l^\pm$ are the set of indices for the coordinates $z_1, \dots, z_d$
constituting, respectively, the positive and negative terms of the binomial relation $R_l$
with $l = 1, \dots, M$.
We note here that for abelian brane brick models, the mesonic moduli space
defined in \eref{es02a02} in section \sref{sec021}
precisely takes the form of a toric Calabi-Yau 4-fold given in \eref{es02d0001},
where $z_1, \dots, z_d$ become the generators
and the relations $R_l$ are the defining binomial relations amongst the generators of the mesonic moduli space.
\\

%-------------------------------------------------------
\paragraph{Orbifold Group.}

In this work, we discuss 
abelian orbifolds of the form $\mathcal{M}/\Gamma$, 
with $\mathcal{M}$ being the toric Calabi-Yau 4-fold
and $\Gamma$ being a finite abelian subgroup of $SU(4)$. 
We refer to $\Gamma$
as the \textit{orbifold group} or \textit{quotienting group}, which takes the following general form,
\beal{es02d000}
\Gamma = \mathbb{Z}_{n_{1}}\times\mathbb{Z}_{n_{2}}\times \mathbb{Z}_{n_{3}} ~,~
\eea
where the order of the orbifold is given by $n=|\Gamma| = \prod_{j=1}^3 n_j$.
Here, we denote $(s_1, s_2 , s_3) \in \Gamma$, 
where $s_j \in \mathbb{Z}_{n_j}$ and $s_j \sim s_j + n_j$.
\\

%-------------------------------------------------------
\paragraph{Orbifold Action.}

Given an abelian orbifold of the form $\mathcal{M}/\Gamma$, 
we need to identify how $\Gamma$ acts on the coordinates $z_1, \dots, z_d$.
This is specified by the corresponding 
\textit{orbifold action}, which takes the following general form,
\beal{es02d001}
a_i = (a_{i1}, a_{i2}, a_{i3}) \in \Gamma~,~ 
\eea
where $a_{ij} \in \mathbb{Z}_{n_j}$ with $a_{ij}  \sim a_{ij} + n_j$,
and $i=1,\dots, d$.
We note that
$a_i$ labels here a character of 
$\Gamma$.
For an element $(s_1, s_2, s_3) \in \Gamma$,
the orbifold action $(a_1, a_2, \dots, a_d)$ acts on the coordinates $z_i$ of $\mathbb{C}^d$ 
as follows, 
\beal{es02d01}
z_{i}\mapsto 
\exp\left[2\pi i \left(\frac{a_{i1}}{n_{1}} s_{1}+\frac{a_{i2}}{n_{2}} s_{2}+\frac{a_{i3}}{n_{3}} s_{3}\right)\right] z_{i}
~.~
\eea

Since only the effective action is physically relevant, 
we assume from now on that the action of $\Gamma$ on the coordinates $z_1, \dots, z_d$ is faithful.
In summary, 
we can specify an abelian orbifold of $\mathcal{M}$
as follows,
\beal{es02d02}
\mathcal{M} / \Gamma ~ (a_1, \dots, a_d) ~,~
\eea
where $\mathcal{M}$ is the toric Calabi-Yau 4-fold with coordinates $z_i$ of $\mathbb{C}^d$, 
the quotienting group $\Gamma$, 
and the orbifold action $(a_1, \dots, a_d)$.
\\

%-------------------------------------------------------
\paragraph{Constraints on the Orbifold Action.}
Due to the algebro-geometric properties of the toric Calabi-Yau 4-fold $\mathcal{M}$ that are required to be preserved under orbifolding, 
the way the quotienting group $\Gamma$ acts on $\mathcal{M}$ determined by the orbifold action 
undergoes certain fundamental constraints.

First, 
we have $\Gamma$ 
as a finite abelian subgroup of $SU(4)$.
We also have the necessary condition for $\Gamma$ to preserve the canonical holomorphic top form $\Omega_{\mathcal{M}}$,
which guarantees 
that the quotient $\mathcal{M}/\Gamma$ remains \textit{Calabi-Yau}. 
The invariant top form condition then leads to an overall constraint on the orbifold action of the following form,
\beal{es02d04}
\sum_{i=1}^{d} a_{ij} = 0 ~~~(\bmod{~n_j})
~,~
\eea
for all $j=1, 2, 3$.

%------------------------------------------------------------------------------------------
\begin{figure}[ht!!]
\centering
\includegraphics[width=0.65\linewidth]{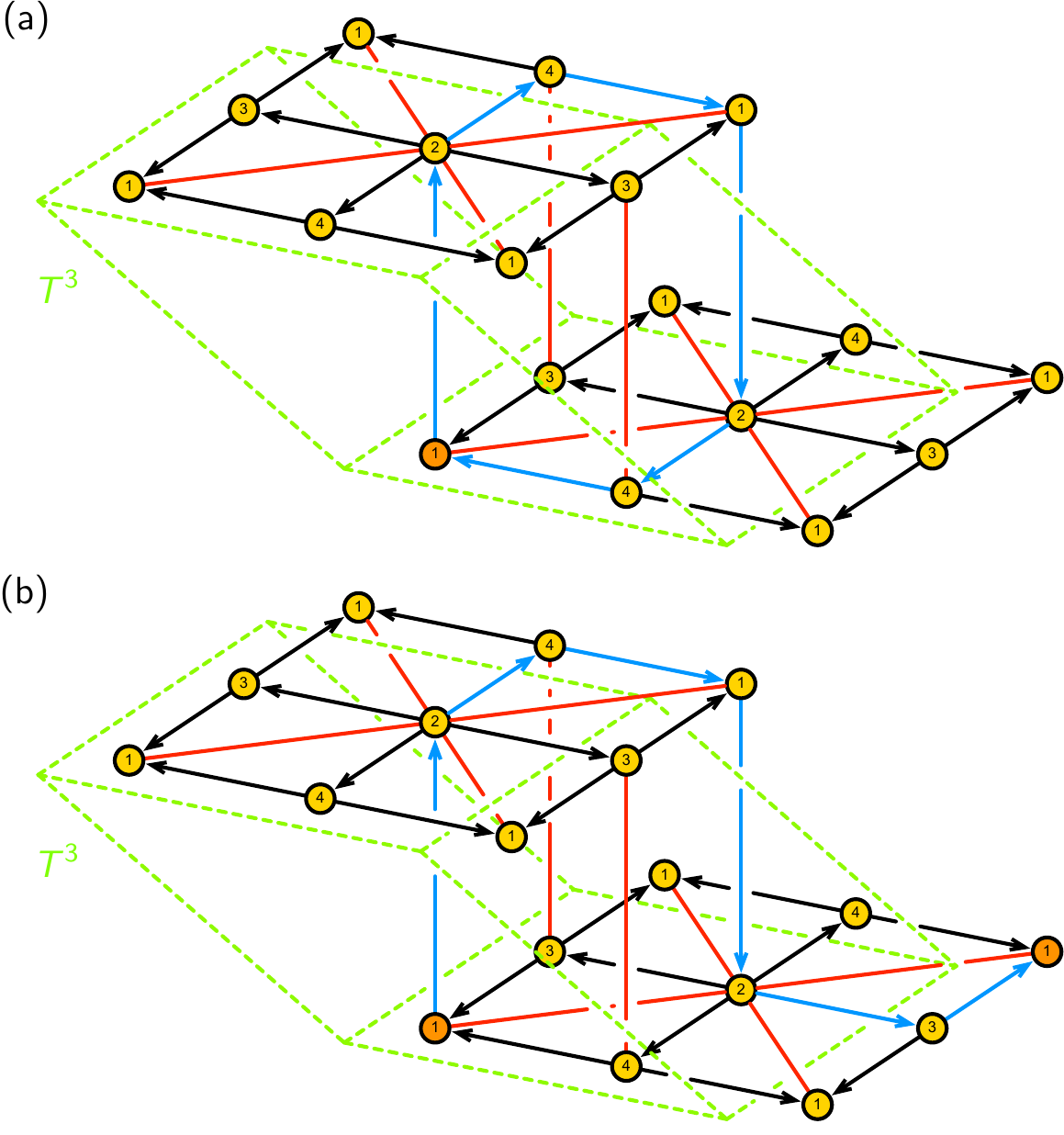}
\caption{
(a) A closed path in the quiver that is closed in the periodic quiver, even after orbifolding, 
and (b) a closed path in the quiver that is not closed in the periodic quiver
\cite{Franco:2016nwv,Franco:2016qxh}.
}
\label{fig_loops}
\end{figure}
%------------------------------------------------------------------------------------------

Additionally, the orbifold action of $\Gamma$ needs to preserve the \textit{toric ideal} $\mathcal{I}$ for $\mathcal{M}$. 
This means that the orbifold action is required to preserve the defining relations in \eref{es02d12} such that,
\beal{es02d13}
\sum_{u\in I_l^{+}} a_{uj} 
=
\sum_{v\in I_l^{-}} a_{vj}
~~~(\bmod{~n_j}) ~,~
\eea
for each $R_l$ with $l = 1, \dots, M$. 
Here, we note that 
the requirement for the holomorphic top form $\Omega_{\mathcal{M}}$ to remain invariant under $\Gamma$
further requires that certain constraints in \eref{es02d13}
vanish $\bmod~n_j$.
These 
correspond to products of mesonic gauge-invariant operators
that are generators of $\mathcal{M}$.
These products form a closed path in the periodic quiver 
and remain a closed path in the periodic quiver even after orbifolding. 
Such closed paths are illustrated in \fref{fig_loops}.

Finally, the action of $\Gamma$ must be faithful.  Equivalently,
for every non-identity element $s=(s_1,s_2,s_3)\in\Gamma$, there must
exist at least one coordinate $z_i$ such that
\beal{es02d05}
\sum_{j=1}^3 \frac{a_{ij}}{n_j}s_j \notin \mathbb{Z}~.~
\eea
In other words,
\beal{es02d05b}
\left\{
s\in\Gamma \ \middle|\
\sum_{j=1}^3 \frac{a_{ij}}{n_j}s_j\in\mathbb{Z}
\text{ for all } i
\right\}
=
\{0\}~.~
\eea
A necessary, but not sufficient, consequence is
\beal{es02d06}
\gcd(a_{1j},a_{2j},\ldots,a_{dj},n_j)=1~,~
\eea
for each cyclic factor $j$.
\\

%=================================================================
\subsection{Brane Brick Models \label{sec0320}}

%-------------------------------------------------------
\paragraph{Brane Brick Model.}
A brane brick model contains a quiver
\beal{es02f01}
Q = (Q_0,\, Q_1^\chi,\, Q_1^{\Lambda})~,
\eea
where $Q_0$ is the set of nodes, $Q_1^\chi$ is the set of directed edges, and $Q_1^{\Lambda}$ is the set of undirected edges. 
A node $i \in Q_0$ corresponds to a gauge group $U(N)_i$, with $i = 1,\dots,G$ labeling the $G$ gauge groups of the $2d$ $(0,2)$ supersymmetric gauge theory. 
A directed edge corresponds to a bifundamental chiral field $X_{ij}$, where $i$ and $j$ identify the gauge groups under which the chiral field transforms in the anti-fundamental and fundamental representations, respectively. 
An undirected edge corresponds to a Fermi field $\Lambda_{ij}$ and its conjugate $\overline{\Lambda}_{ij}$.

For notational convenience, we label each directed edge by a variable $\mathsf{X}_m \in Q_1^\chi$ with $m=1,\dots,E$, where $E = |Q_1^\chi|$. 
Writing $\mathsf{t}_m = \text{tail}(\mathsf{X}_m)$ and $\mathsf{h}_m = \text{head}(\mathsf{X}_m)$,
\beal{es02d20}
\mathsf{X}_m ~:~ \mathsf{t}_{m} \rightarrow \mathsf{h}_m ~,~
\eea
with $\mathsf{t}_{m},\mathsf{h}_{m} \in Q_0$. 
When a chiral field $X_{ij}$ is identified with a directed edge $\mathsf{X}_m$, we identify the labels $i \equiv \mathsf{t}_m$ and $j \equiv \mathsf{h}_m$.
\\

%------------------------------------------------------------------------------------------
\begin{figure}[ht!!]
\centering
\includegraphics[width=0.95\linewidth]{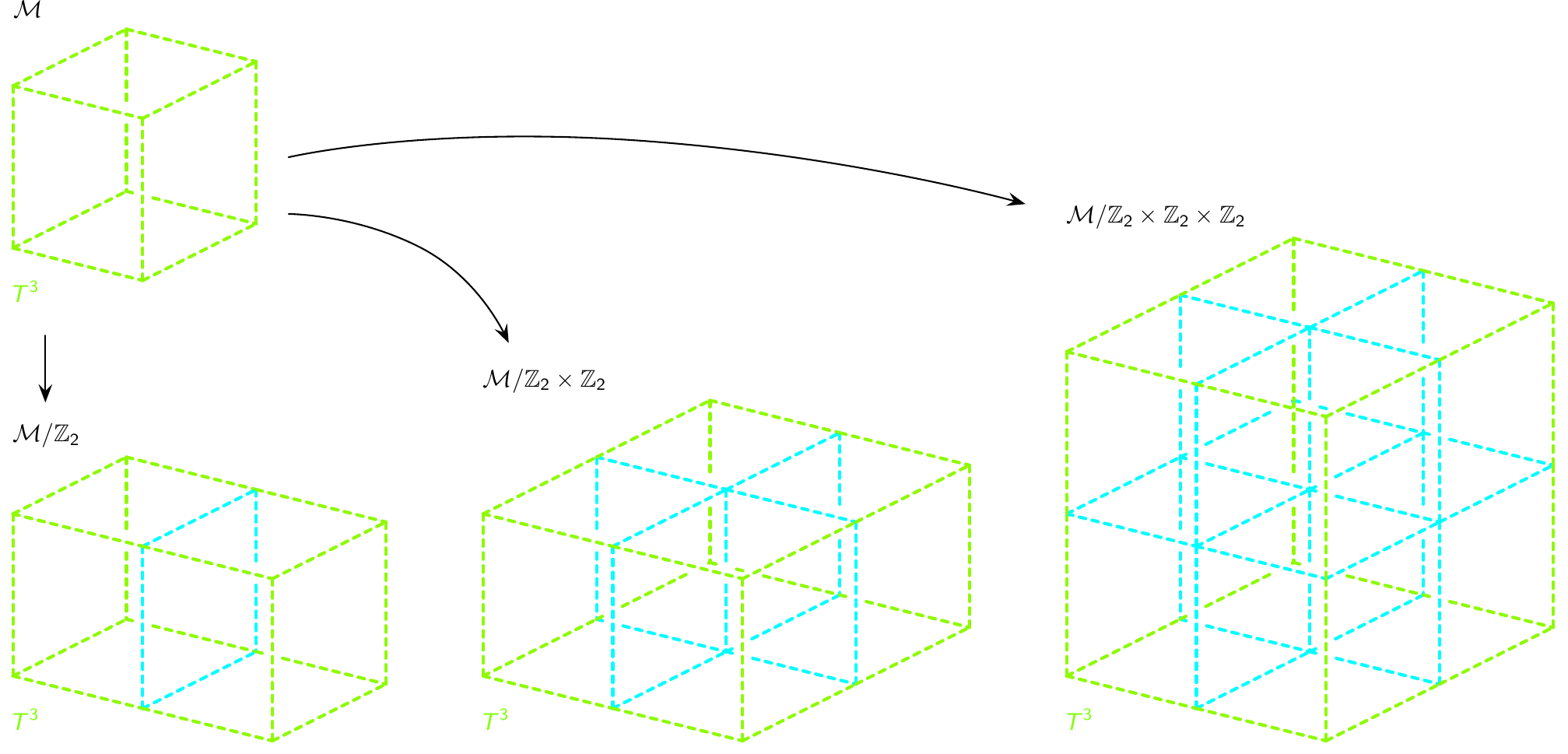}
\caption{
Unit cell enlargement under orbifolding for brane brick models corresponding to an abelian orbifold of a toric Calabi-Yau 4-fold \cite{Davey:2010px, Hanany:2010ne, Hanany:2010cx, Davey:2011dd, Hanany:2011iw}.
}
\label{fig_03}
\end{figure}
%------------------------------------------------------------------------------------------

%-------------------------------------------------------
\paragraph{Orbifolding via Unit Cell Enlargement.}
As illustrated in \fref{fig_03},
the brane brick model for an abelian orbifold $\mathcal{M}/\Gamma$ can be obtained from the brane brick model for $\mathcal{M}$ by enlarging the unit cell of the latter \cite{Davey:2010px, Hanany:2010ne, Hanany:2010cx, Davey:2011dd, Hanany:2011iw}. 
Under this enlargement, every closed path along directed edges in $Q$ in the periodic quiver of the brane brick model for $\mathcal{M}$ 
remains a closed path in the brane brick model for $\mathcal{M}/\Gamma$. 
This basic observation translates into constraints on the orbifold action of $\Gamma$ as reviewed in section \sref{sec031}
and discussed in the following section in more detail from the point of view of the brane brick model and its quiver and $J$- and $E$-terms.
Since the mesonic moduli space of an abelian brane brick model is a toric Calabi-Yau 4-fold, 
these constraints reproduce precisely the Calabi-Yau condition for $\mathcal{M}/\Gamma$.
In the following paragraphs, 
we focus first on the orbifold action on chiral fields and the resulting quiver of the orbifolded brane brick model. 
We illustrate how the orbifold action on Fermi fields and on the corresponding $J$- and $E$-terms is determined by the action on chiral fields.
\\

%-------------------------------------------------------
\paragraph{Orbifold Action on Gauge Groups and Chiral Fields.}

As reviewed in section \sref{sec031}, 
an abelian orbifold of the form $\mathcal{M}/\Gamma$ with orbifold action $(a_1,\dots,a_d)$ 
acts on the coordinates $z_i$ of $\mathcal{M}$. 
In this work, 
$\mathcal{M}$ is the mesonic moduli space of the abelian brane brick model defined in \eref{es02a03}, and the coordinates $z_i$ are its generators -- gauge invariant operators in the chiral fields of the $2d$ $(0,2)$ gauge theory. 
We have in terms of edge variables,
\beal{es02d21}
z_i = \prod_{q \in I_{h}(z_i)} \mathsf{X}_{q} ~,
\eea
where $I_h(z_i)$ is the set of indices of edge variables whose corresponding chiral fields form a gauge invariant product identified with $z_i$. 
Because the $J$- and $E$-terms of the brane brick model relate distinct products of chiral fields, 
a single generator $z_i$ may admit several such expressions, labeled here by $h = 1,\dots, n(i)$.

Equation \eqref{es02d21} implies that the orbifold action $(a_1,\dots,a_d)$ on the coordinates decomposes into an action $(b_1,\dots,b_E)$ on the edge variables. 
We have for each fixed $h$, 
\beal{es02e04}
a_i = \sum_{q \in I_{h}(z_i)} b_q ~,
\eea
where $b_m = (b_{m1},b_{m2},b_{m3}) \in \Gamma$ with $b_{mj} \in \mathbb{Z}_{n_j}$ and $b_{mj} \sim b_{mj} + n_j$. 
We can think of $(b_1,\dots,b_E) \in \Gamma^E$ as the orbifold action on the edge variables induced by $(a_1,\dots,a_d)$. 

In order to track the orbifold action on both nodes and directed edges of the quiver $Q$, 
we label nodes of the orbifolded quiver $Q^\prime$ by pairs that include
the original node label from $Q$ together with an element of $\Gamma$ recording the orbifold action. 
For an element $(s_1,s_2,s_3) \in \Gamma$ with $s_j \in \mathbb{Z}_{n_j}$ and $s_j \sim s_j + n_j$, 
the directed edge variable $\mathsf{X}_m^s$ in $Q^\prime$ takes the following general form,
\beal{es02e05}
\mathsf{X}_{m}^s ~:~ [\mathsf{t}_{m}, s] \rightarrow [\mathsf{h}_m, s + b_m]~,
\eea
where $b_m \in \Gamma$ is the orbifold action on the original directed edge variable $\mathsf{X}_m \in Q_1^\chi$.

In summary, after orbifolding by $\Gamma$ with orbifold action $(a_1,\dots,a_d)$, 
the orbifolded quiver $Q^\prime$ contains,
\beal{es02e06}
Q_{0}^\prime 
= 
\{ [i,s] ~|~ i \in Q_0,\; s \in \Gamma \}
~,~
Q_{1}^{\chi\prime} 
=
\{ \mathsf{X}_{m}^{s} ~|~ \mathsf{X}_{m} \in Q_1^\chi,\; s \in \Gamma \}
~,~
\eea
with $G \cdot |\Gamma|$ nodes and $E \cdot |\Gamma|$ directed edges. 
The nodes $[i,s] \in Q_0^\prime$ correspond to gauge groups $U(N)_{[i,s]}$, 
and the directed edge variables $\mathsf{X}_m^s \in Q_1^{\chi\prime}$ 
correspond to bifundamental chiral fields of the form $X_{[\mathsf{t}_m,s] [\mathsf{h}_m,s+b_m]}$ in the orbifolded brane brick model.
\\

%-------------------------------------------------------
\paragraph{Orbifold Action on $J$- and $E$-Terms and Fermi Fields.}
We now turn to closed paths in the periodic quiver of $Q$. 
Because $Q^\prime$ is obtained by enlarging the unit cell of $Q$ on the 3-torus, 
every closed path in the periodic quiver of $Q$ remains closed in $Q^\prime$. 
This requirement is what reproduces the Calabi-Yau condition discussed in section \sref{sec031} for an abelian orbifold of the form $\mathcal{M}/\Gamma$.

\begin{figure}[ht!!]
\centering
\includegraphics[width=1\linewidth]{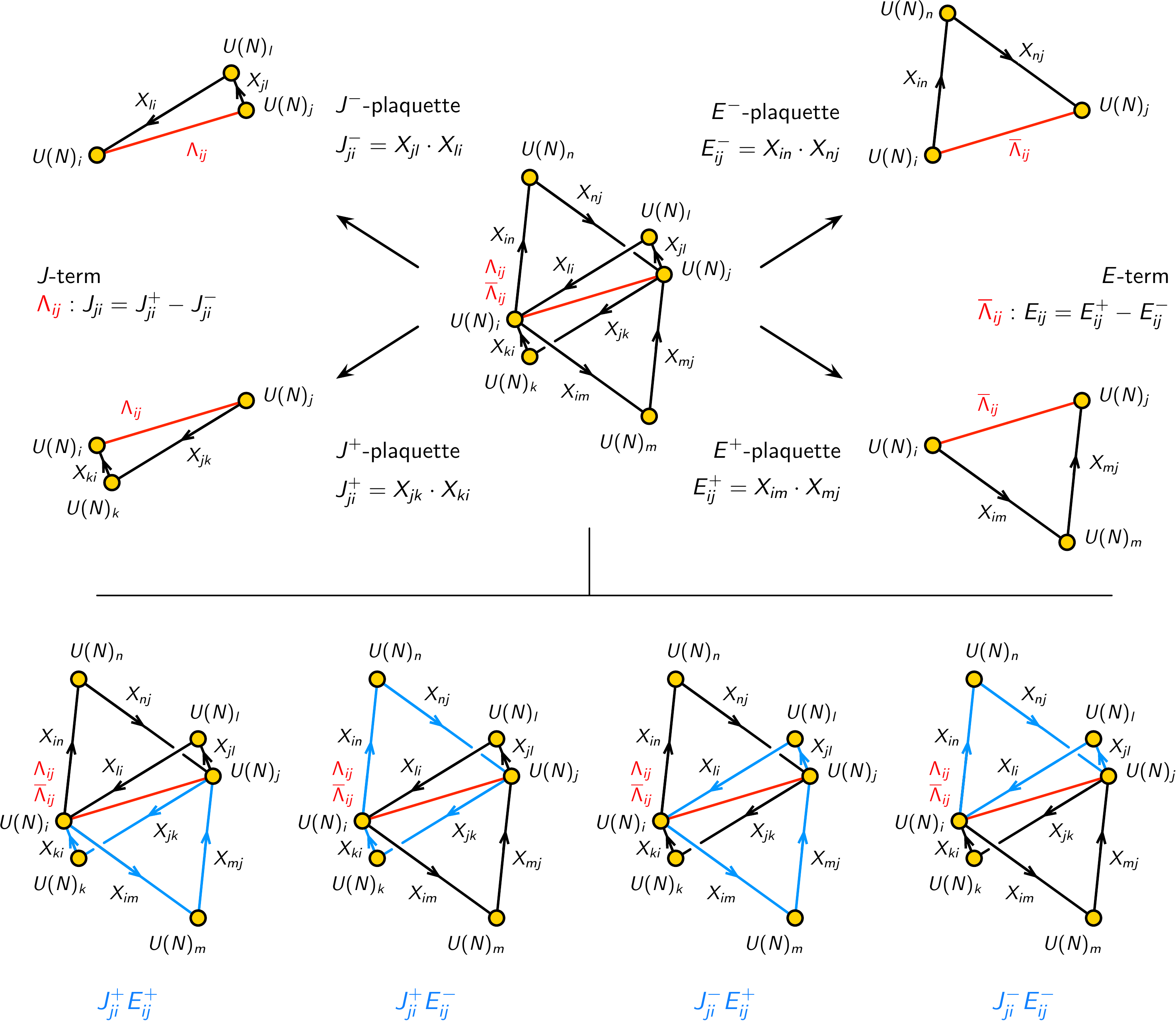}
\caption{
Chiral cycles $J^{\pm}_{ji} E^{\pm}_{ij}$ (blue) as closed paths in the periodic quiver
with plaquettes from the $J$- and $E$-terms.
}
\label{fig_plaquette}
\end{figure}

A Fermi field $\Lambda_{ij}$ corresponds to an undirected edge of $Q$ and comes with a pair of $J$- and $E$-terms in a brane brick model.
These take the following form,
\beal{es02e07}
J_{ji} = 
J_{ji}^{+} - J_{ji}^{-}~,~
E_{ij} = 
E_{ij}^{+} - E_{ij}^{-}~,~
\eea
where each monomial 
$J_{ji}^\pm$ is a directed path along directed edges from nodes $j$ to $i$ in $Q$, 
and each monomial $E_{ij}^\pm$ is a directed path along directed edges from nodes $i$ to $j$. 
Together with the undirected edge associated with $\Lambda_{ij}$ and its conjugate $\overline{\Lambda}_{ij}$, 
the products $\Lambda_{ij} J_{ji}^\pm$ and $\overline{\Lambda}_{ij} E_{ij}^\pm$, also referred to as \textit{plaquettes} \cite{Franco:2015tna, Franco:2015tya} of the brane brick model,
form four distinct closed paths along edges in the periodic quiver of the brane brick model on the 3-torus. 
Additionally, each term in the expansion
\beal{es02e08}
J_{ji} E_{ij} = J_{ji}^{+} E_{ij}^{+} - J_{ji}^{+} E_{ij}^{-} - J_{ji}^{-} E_{ij}^{+} + J_{ji}^{-} E_{ij}^{-}~
\eea
is a closed directed path along directed edges in the periodic quiver. 
We refer to such closed directed paths as \emph{chiral cycles} \cite{Franco:2023tly}.
\fref{fig_plaquette} illustrates the closed paths from plaquettes as well as the chiral cycles given by \eref{es02e08}.

%------------------------------------------------------------------------------------------
%------------------------------------------------------------------------------------------

We have in terms of directed edge variables, 
\beal{es02e09}
J^\pm_{ji} = \prod_{q \in I(J^\pm_{ji})} \mathsf{X}_{q}~,~ \qquad
E^\pm_{ij} = \prod_{q \in I(E^\pm_{ij})} \mathsf{X}_{q}~,~
\eea
where $I(J^\pm_{ji})$ and $I(E^\pm_{ij})$ are the index sets of the corresponding edge products. 
By construction, we have the following correspondences between gauge node indices of the quiver and endpoints of the directed edge paths $J_{ji}^\pm$ and $E_{ij}^\pm$, 
\beal{es02e10}
j &=& \text{tail}(J^+_{ji}) = \text{tail}(J^-_{ji}) = \text{head}(E^+_{ij}) = \text{head}(E^-_{ij})~,~
\nn\\
i &=& \text{head}(J^+_{ji}) = \text{head}(J^-_{ji}) = \text{tail}(E^+_{ij}) = \text{tail}(E^-_{ij})~.~
\eea

Under the orbifold action on the edge variables given by $(b_1,\dots,b_E)$, 
the $E$-term paths $E_{ij}^\pm$ transform as follows, 
\beal{es02e11}
(E^+_{ij})^s &:& 
[i,s] \rightarrow \Big[j, s + \sum_{q \in I(E^+_{ij})} b_q\Big]
~,~
\nn\\
(E^-_{ij})^s &:& 
[i,s] \rightarrow \Big[j, s + \sum_{q \in I(E^-_{ij})} b_q\Big]
~.~
\eea
Because $(E^+_{ij})^s$ and $(E^-_{ij})^s$ must end at the same node in the orbifolded quiver $Q^\prime$,
we have the following constraint on the orbifold action $(b_1,\dots,b_E)$, 
\beal{es02e12}
\sum_{q \in  I(E^+_{ij})} b_q ~=~ \sum_{q \in I(E^-_{ij})} b_q~.
\eea
This also dictates how Fermi fields themselves transform under the orbifold action given by $(b_1,\dots,b_E)$.
 $\Lambda_{ij} \in Q_1^\Lambda$ becomes under orbifolding
 $\Lambda_{ij}^s \in Q_1^{\Lambda\prime}$, which is an undirected edge connecting node $[i,s]$ to $[j, s + \sum_{q \in I(E^+_{ij})} b_q]$ in the orbifolded quiver $Q^\prime$.

The $J$-terms transform analogously, but their starting node is the head of the corresponding $E$-term. 
Therefore, we have the following under an orbifold action $(b_1,\dots,b_E)$, 
\beal{es02e13}
(J^+_{ji})^s &:& 
\Big[j, s + \sum_{q \in I(E^+_{ij})} b_q\Big] 
\rightarrow \Big[i, s + \sum_{q \in I(E^+_{ij})} b_q + \sum_{q \in I(J^+_{ji})} b_q
\Big]~,~
\nn\\
(J^-_{ji})^s &:& 
\Big[j, s + \sum_{q \in I(E^-_{ij})} b_q\Big] 
\rightarrow \Big[i, s + \sum_{q \in I(E^-_{ij})} b_q + \sum_{q \in I(J^-_{ji})} b_q\Big]
~.~
\eea
Since $(J^+_{ji})^s$ and $(J^-_{ji})^s$ must end at the same node in the orbifolded quiver $Q^\prime$, we have the following additional constraint on the orbifold action $(b_1,\dots,b_E)$, 
\beal{es02e14}
\sum_{q \in I(J^+_{ji})} b_q ~=~ \sum_{q \in I(J^-_{ji})} b_q
~.~
\eea

Besides the constraints coming from each of the plaquettes of the brane brick model, 
we have an additional constraint on the orbifold action $(b_1,\dots,b_E)$
due to the fact that the
chiral cycles from the terms in \eref{es02e08}
are closed paths in the periodic quiver of the brane brick model on the 3-torus.
These constraints are as follows,
\beal{es02e15}
\sum_{q \in I(E^+_{ij})} b_q
~+~
\sum_{q \in I(J^+_{ji})} b_q
~=~
0
\pmod{n_j}
~.~
\eea

%-------------------------------------------------------
\paragraph{Calabi-Yau Condition on the Orbifolded Brane Brick Model.}

Combining \eqref{es02e04}, \eqref{es02e12}, \eqref{es02e14} and \eqref{es02e15} produces 
the precise Calabi-Yau condition on the orbifold action for an abelian orbifold of the form $\mathcal{M}/\Gamma$ as discussed in section \sref{sec031}. 
In particular, \eqref{es02e12} and \eqref{es02e14} ensure that the decomposition \eqref{es02e04} of $(a_1,\dots,a_d)$
in terms of directed edge variables is consistent across all $h = 1,\dots,n(i)$, so that the orbifold action $(a_1,\dots,a_d)$ 
on the mesonic generators $z_1, \dots, z_d$ is well defined.

This procedure not only yields the Calabi-Yau condition itself, but also gives a systematic recipe for computing 
the $J$- and $E$-terms of the orbifolded brane brick model from those of the parent theory corresponding to $\mathcal{M}$, 
for any orbifold action $(a_1,\dots,a_d)$ for an abelian orbifold of the form $\mathcal{M}/\Gamma$.
Following this recipe, we present in the following sections the brane brick models
with their $J$- and $E$-terms for abelian orbifolds of $Q^{1,1,1}$ and $D_3$. 
\\

%=================================================================
\subsection{Abelian Orbifolds of $Q^{1,1,1}$ \label{sec032}}

We now consider the brane brick model for $Q^{1,1,1}$, which is reviewed in section \sref{sec022}.
We recall that the brane brick model for $Q^{1,1,1}$
has in total 10 bifundamental chiral fields $X_{ij}$ that form 
8 gauge invariant generators $A_{ijk}$ for the mesonic moduli space of the abelian brane brick model. 
\\

%-------------------------------------------------------
\paragraph{Orbifold Action.}
Let us consider an abelian orbifold of the form $Q^{1,1,1} / \mathbb{Z}_{n_1} \times \mathbb{Z}_{n_2} \times \mathbb{Z}_{n_3}$ with orbifold action $(a_1,\dots, a_8)$
acting on the 8 generators $A_{ijk}$ of the $Q^{1,1,1}$ model.
We note here that the generators $A_{ijk}$ are associated to the following coordinates,
\beal{es03b00b0}
&
z_1 = A_{111}~,~
z_2 = A_{211}~,~
z_3 = A_{121}~,~
z_4 = A_{221}~,~
&
\nn\\
&
z_5 = A_{112}~,~
z_6 = A_{212}~,~
z_7 = A_{122}~,~
z_8 = A_{222}~,~
&
\eea
which satisfy the following binomial relations
for $Q^{1,1,1}$ based on \eref{es02b03},
\beal{es03b01b0}
\begin{array}{rcl@{\qquad}rcl@{\qquad}rcl}
z_3 z_5 &=& z_1 z_7 ~,~ &
z_4 z_5 &=& z_1 z_8 ~,~ &
z_4 z_6 &=& z_2 z_8 ~,~
\\
z_2 z_5 &=& z_1 z_6 ~,~ &
z_2 z_7 &=& z_1 z_8 ~,~ &
z_4 z_7 &=& z_3 z_8 ~,~
\\
z_2 z_3 &=& z_1 z_4 ~,~ &
z_3 z_6 &=& z_1 z_8 ~,~ &
z_6 z_7 &=& z_5 z_8 ~.~
\end{array}
\eea
Based on the general formula in \eref{es02d01},
we note here that for an abelian orbifold of the form $Q^{1,1,1} / \mathbb{Z}_{n_1} \times \mathbb{Z}_{n_2} \times \mathbb{Z}_{n_3}$
the action $(a_1,\dots, a_8)$ 
on the coordinates in \eref{es03b00b0} takes the form 
$
z_i \mapsto \exp\left[
2\pi i~ \left(
\frac{a_{i1}}{n_1} s_{1}
+ \frac{a_{i2}}{n_2} s_{2}
+ \frac{a_{i3}}{n_3} s_{3}
\right)
\right]
z_i
$,
where $(s_1,s_2,s_3)\in  \mathbb{Z}_{n_1} \times \mathbb{Z}_{n_2} \times \mathbb{Z}_{n_3}$.

As discussed in \eref{es02e04},
we can decompose the orbifold action $(a_1,\dots, a_8)$
as an action $(b_1, \dots, b_{10})$ on the 
10 bifundamental chiral fields of the $Q^{1,1,1}$ model.
The generators of the $Q^{1,1,1}$ mesonic moduli space can be expressed in terms of the chiral fields
as shown in \eref{es02b04}
and as a result the orbifold action $(a_1,\dots, a_8)$ gives,
\beal{es03b12}
\begin{array}{rclcl@{\qquad}rclcl}
a_{1} &=& b_{1} + b_{6}    + b_{3} &=& b_{1} + b_{5} + b_{7} ~,~&
a_{2} &=& b_{4} + b_{6}    + b_{3} &=& b_{4} + b_{5} + b_{7} ~,~\\
a_{3} &=& b_{1} + b_{10} + b_{3} &=& b_{1} + b_{5} + b_{2} ~,~&
a_{4} &=& b_{4} + b_{10} + b_{3} &=& b_{4} + b_{5} + b_{2} ~,~\\
a_{5} &=& b_{1} + b_{6}    + b_{8} &=& b_{1} + b_{9} + b_{7} ~,~&
a_{6} &=& b_{4} + b_{6}    + b_{8} &=& b_{4} + b_{9} + b_{7} ~,~\\
a_{7} &=& b_{1} + b_{10} + b_{8} &=& b_{1} + b_{9} + b_{2} ~,~&
a_{8} &=& b_{4} + b_{10} + b_{8} &=& b_{4} + b_{9} + b_{2} ~,~
\end{array}
\eea
which hold 
under $(\bmod{~n_j})$ with $j=1,2,3$,  $a_{i} = (a_{i1}, a_{i2}, a_{i3})$ and $b_{k} = (b_{k1}, b_{k2}, b_{k3})$.
The resulting effective action on the indices of the bifundamental chiral fields
is as follows, 
\beal{es03b12b}
&
D_{[1,s][2,s+b_1]}~,~ 
D_{[3,s][1,s+b_2]}~,~
D_{[4,s][1,s+b_3]}~,~ 
X_{[1,s][2,s+b_4]}~,~ 
X_{[2,s][3,s+b_5]}~,~ 
X_{[2,s][4,s+b_6]}~,~ 
&
\nn\\
&
Y_{[3,s][1,s+b_7]}~,~
Y_{[4,s][1,s+b_8]}~,~ 
Z_{[2,s][3,s+b_9]}~,~ 
Z_{[2,s][4,s+b_{10}]} ~,~
&
\eea
where  $s=(s_1, s_2, s_3)$,
$b_{k} = (b_{k1}, b_{k2}, b_{k3})$, 
and the labels are under $(\bmod{~n_j})$ with $j=1,2,3$.
\\

%-------------------------------------------------------
\paragraph{Constraints on the Orbifold Action.}
The orbifold action $(a_1, \dots, a_8)$ satisfies several constraints that have been summarized in section \sref{sec031}.
First, we have the constraint from \eref{es02d04}, which is given by,
\beal{es03b00b1}
\sum_{i=1}^{8} a_{ij} = 0 ~~~(\bmod{~n_j})
~,~
\eea
for $j=1,2,3$.
Additionally, 
the binomial relations amongst the generators of the $Q^{1,1,1}$ model 
in \eref{es03b01b0}
set the following constraints on the orbifold action,
\beal{es03b04}
\begin{array}{rcl@{\qquad}rcl@{\qquad}rcl}
a_3+a_5 &=& a_1+a_7 ~,~ &
a_4+a_5 &=& a_1+a_8 ~,~ &
a_4+a_6 &=& a_2+a_8 ~,~
\\
a_2+a_5 &=& a_1+a_6 ~,~ &
a_2+a_7 &=& a_1+a_8 ~,~ &
a_4+a_7 &=& a_3+a_8 ~,~
\\
a_2+a_3 &=& a_1+a_4 ~,~ &
a_3+a_6 &=& a_1+a_8 ~,~ &
a_6+a_7 &=& a_5+a_8 ~,~
\end{array}
\eea
where $a_{i} = (a_{i1}, a_{i2}, a_{i3})$ and 
the constraints on the orbifold action $(a_1, \dots, a_8)$ hold 
under $(\bmod{~n_j})$ with $j=1,2,3$.

%------------------------------------------------------------------------------------------
\begin{figure}[ht!!]
\centering
\includegraphics[width=0.9\linewidth]{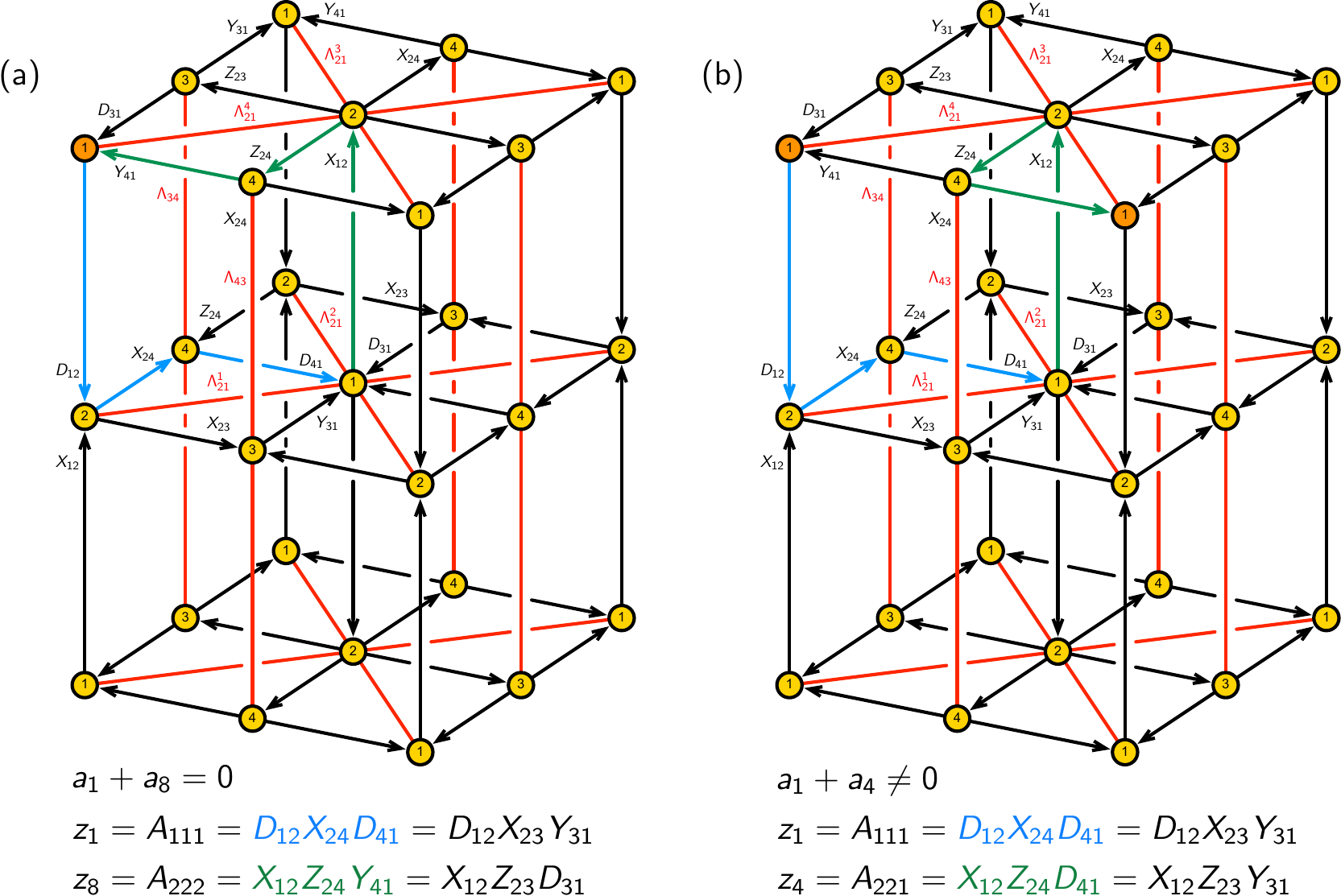}
\caption{
(a) The vanishing constraint $a_1+a_8 = 0$
on the orbifold action for abelian orbifolds of $Q^{1,1,1}$
is associated with the gauge invariant $z_1 z_8 = A_{111} A_{222}$
that forms a closed path in the periodic quiver of the $Q^{1,1,1}$ brane brick model.
(b) In contrast, 
the combination $a_1+a_4 \neq 0$, which is not constrained to vanish in general,
is associated with the gauge invariant $z_1 z_4 = A_{111} A_{221}$,
which does not form a closed path in the periodic quiver. 
}
\label{fig_q111_loops}
\end{figure}
%------------------------------------------------------------------------------------------

Moreover, 
in order to ensure that the orbifold itself remains Calabi-Yau, we
have to have a look at the 
holomorphic top form $\Omega_\mathcal{M}$
for $Q^{1,1,1}$.
In the patch when $z_1 \neq 0$, 
we can write the top form $\Omega_\mathcal{M}$
in terms of local independent coordinates, 
\beal{es03b01b3}
z_1 ~,~ \frac{z_2}{z_1} ~,~ \frac{z_3}{z_1} ~,~ \frac{z_5}{z_1} ~,~
\eea
which give,
\beal{es03b01b4}
\Omega_{\mathcal{M}} = 
z_1 d z_1 \wedge 
d\left(\frac{z_2}{z_1}\right) \wedge
d\left(\frac{z_3}{z_1}\right) \wedge
d\left(\frac{z_5}{z_1}\right) ~.~
\eea
This simplifies to the following holomorphic top form,
\beal{es03b01b1}
\Omega_{\mathcal{M}} = 
\frac{
d z_1 \wedge d z_2 \wedge d z_3 \wedge d z_5
}{
z_1^2
}
~,~
\eea
which needs to remain invariant under the orbifold. 
This leads to an additional constraint on the orbifold action, 
\beal{es03b01b2b}
a_1 + a_2 + a_3 + a_5 - 2 a_1 = 0
~,~
\eea
which combined with the constraints in \eref{es03b04}
simplifies to,
\beal{es03b01b2}
&
a_{2} - a_{1} = a_{4} - a_{3} = a_{6} - a_{5} = a_{8} - a_{7} ~,~
&
\nn\\
&
a_{3} - a_{1} = a_{4} - a_{2} = a_{7} - a_{5} = a_{8} - a_{6} ~,~
&
\nn\\
&
a_{1} + a_{8} = a_{2} + a_{7} = a_{3} + a_{6} = a_{4} + a_{5} = 0 ~.~
&
\eea
where $a_{i} = (a_{i1}, a_{i2}, a_{i3})$ and the above relations hold
under $(\bmod{~n_j})$ with $j=1,2,3$.
We note here that the vanishing constraints in \eref{es03b01b2}
on the orbifold action correspond to products of coordinates $z_i$
associated with products of gauge invariant generators
of the mesonic moduli space whose chiral field content form closed paths in the periodic quiver of the brane brick model as illustrated in \fref{fig_q111_loops}.
The vanishing constraint in \eref{es03b01b2} implies that these closed paths remain closed paths in the periodic quiver even after orbifolding.

Lastly, as stated in \eref{es02d06}, 
we need the orbifold action on the coordinates $z_1, \dots, z_8$ in \eref{es03b00b0}
to be faithful, leading to the following necessary constraint on the orbifold action, 
\beal{es03b00b2}
\gcd (a_{1j}, \dots, a_{8j}, n_j) = 1 ~,~
\eea
where $j=1,2,3$.
\\

%-------------------------------------------------------
\paragraph{$J$- and $E$-Terms of the Abelian Orbifold.}
We are now in a position to derive the general form of the
$J$- and $E$-terms for the brane brick model corresponding to an abelian orbifold of the form 
$Q^{1,1,1}/\Gamma$ with orbifold action $(a_{1},a_{2},a_{3},a_{4},a_{5},a_{6},a_{7},a_{8})$.
The $J$- and $E$-terms take the following general form,
\beal{es03b01}
\begin{array}{rcl}
J^{+}[\Lambda^{1}_{[2,a_{8}+s][1,s]}]
&=&
D_{[1,s][2,s]}  Z_{[2,s][4,a_{3}+s]}  Y_{[4,a_{3}+s][1,a_{7}+s]}  X_{[1,a_{7}+s][2,a_{8}+s]}
\\
J^{-}[\Lambda^{1}_{[2,a_{8}+s][1,s]}]
&=&
X_{[1,s][2,a_{2}-a_{1}+s]}  Z_{[2,a_{2}-a_{1}+s][3,a_{8}+s]}  D_{[3,a_{8}+s][1,a_{8}+s]}  D_{[1,a_{8}+s][2,a_{8}+s]} 
\\
E^{+}[\Lambda^{1}_{[2,a_{8}+s][1,s]}]
&=&
X_{[2,a_{8}+s][3,a_{8}+a_{3}+s]}  Y_{[3,a_{8}+a_{3}+s][1,s]}
\\
E^{-}[\Lambda^{1}_{[2,a_{8}+s][1,s]}]
&=&
X_{[2,a_{8}+s][4,s]}  D_{[4,s][1,s]} 
\\
\\
J^{+}[\Lambda^{2}_{[2,a_{6}+s][1,s]}]
&=&
X_{[1,s][2,a_{2}-a_{1}+s]}  X_{[2,a_{2}-a_{1}+s][4,a_{2}+s]}  Y_{[4,a_{2}+s][1,a_{6}+s]}  D_{[1,a_{6}+s][2,a_{6}+s]}
\\
J^{-}[\Lambda^{2}_{[2,a_{6}+s][1,s]}]
&=&
D_{[1,s][2,s]}  Z_{[2,s][3,a_{7}+s]}  Y_{[3,a_{7}+s][1,a_{5}+s]}  X_{[1,a_{5}+s][2,a_{6}+s]} 
\\
E^{+}[\Lambda^{2}_{[2,a_{6}+s][1,s]}]
&=&
X_{[2,a_{6}+s][3,s]}  D_{[3,s][1,s]}
\\
E^{-}[\Lambda^{2}_{[2,a_{6}+s][1,s]}]
&=&
Z_{[2,a_{6}+s][4,s]}  D_{[4,s][1,s]} 
\\
\\
J^{+}[\Lambda^{3}_{[2,a_{4}+s][1,s]}]
&=&
D_{[1,s][2,s]}  Z_{[2,s][4,a_{3}+s]}  D_{[4,a_{3}+s][1,a_{3}+s]}  X_{[1,a_{3}+s][2,a_{4}+s]}
\\
J^{-}[\Lambda^{3}_{[2,a_{4}+s][1,s]}]
&=&
X_{[1,s][2,a_{2}-a_{1}+s]}  X_{[2,a_{2}-a_{1}+s][3,a_{4}+s]}  D_{[3,a_{4}+s][1,a_{4}+s]}  D_{[1,a_{4}+s][2,a_{4}+s]} 
\\
E^{+}[\Lambda^{3}_{[2,a_{4}+s][1,s]}]
&=&
X_{[2,a_{4}+s][4,a_{4}+a_{1}+s]}  Y_{[4,a_{4}+a_{1}+s][1,s]}
\\
E^{-}[\Lambda^{3}_{[2,a_{4}+s][1,s]}]
&=&
Z_{[2,a_{4}+s][3,a_{4}+a_{7}+s]}  Y_{[3,a_{4}+a_{7}+s][1,s]} 
\\
\\
J^{+}[\Lambda^{4}_{[2,a_{2}+s][1,s]}]
&=&
D_{[1,s][2,s]}  X_{[2,s][3,a_{3}+s]}  Y_{[3,a_{3}+s][1,a_{1}+s]}  X_{[1,a_{1}+s][2,a_{2}+s]}
\\
J^{-}[\Lambda^{4}_{[2,a_{2}+s][1,s]}]
&=&
X_{[1,s][2,a_{2}-a_{1}+s]}  X_{[2,a_{2}-a_{1}+s][4,a_{2}+s]}  D_{[4,a_{2}+s][1,a_{2}+s]}  D_{[1,a_{2}+s][2,a_{2}+s]} 
\\
E^{+}[\Lambda^{4}_{[2,a_{2}+s][1,s]}]
&=&
Z_{[2,a_{2}+s][3,s]}  D_{[3,s][1,s]}
\\
E^{-}[\Lambda^{4}_{[2,a_{2}+s][1,s]}]
&=&
Z_{[2,a_{2}+s][4,a_{2}+a_{3}+s]}  Y_{[4,a_{2}+a_{3}+s][1,s]} 
\\
\\
J^{+}[\Lambda^{5}_{[4,a_{2}+s][3,s]}]
&=&
D_{[3,s][1,s]}  X_{[1,s][2,a_{2}-a_{1}+s]}   X_{[2,a_{2}-a_{1}+s][4,a_{2}+s]}
\\
J^{-}[\Lambda^{5}_{[4,a_{2}+s][3,s]}]
&=&
Y_{[3,s][1,a_{1}-a_{3}+s]}  X_{[1,a_{1}-a_{3}+s][2,a_{2}-a_{3}+s]}  Z_{[2,a_{2}-a_{3}+s][4,a_{2}+s]} 
\\
E^{+}[\Lambda^{5}_{[4,a_{2}+s][3,s]}]
&=&
D_{[4,a_{2}+s][1,a_{2}+s]}  D_{[1,a_{2}+s][2,a_{2}+s]}  Z_{[2,a_{2}+s][3,s]}
\\
E^{-}[\Lambda^{5}_{[4,a_{2}+s][3,s]}]
&=&
Y_{[4,a_{2}+s][1,a_{6}+s]}  D_{[1,a_{6}+s][2,a_{6}+s]}  X_{[2,a_{6}+s][3,s]} 
\\
\\
J^{+}[\Lambda^{6}_{[3,a_{8}+s][4,s]}]
&=&
Y_{[4,s][1,a_{5}-a_{1}+s]}  X_{[1,a_{5}-a_{1}+s][2,a_{6}-a_{1}+s]}  X_{[2,a_{6}-a_{1}+s][3,a_{8}+s]}
\\
J^{-}[\Lambda^{6}_{[3,a_{8}+s][4,s]}]
&=&
D_{[4,s][1,s]}  X_{[1,s][2,a_{2}-a_{1}+s]}   Z_{[2,a_{2}-a_{1}+s][3,a_{8}+s]} 
\\
E^{+}[\Lambda^{6}_{[3,a_{8}+s][4,s]}]
&=&
D_{[3,a_{8}+s][1,a_{8}+s]}  D_{[1,a_{8}+s][2,a_{8}+s]}  X_{[2,a_{8}+s][4,s]}
\\
E^{-}[\Lambda^{6}_{[3,a_{8}+s][4,s]}]
&=&
Y_{[3,a_{8}+s][1,a_{6}+s]}  D_{[1,a_{6}+s][2,a_{6}+s]}  Z_{[2,a_{6}+s][4,s]} 
\end{array}
\eea
where $s$ is running over every element in $\Gamma$,
$a_{i} = (a_{i1}, a_{i2}, a_{i3})$
and the labels are all under $(\bmod{~n_j})$ with $j=1,2,3$.
The number of Fermi fields is $6\cdot |\Gamma|$ and the number of chiral fields is $10\cdot |\Gamma|$.
\\

%=================================================================
\subsection{Abelian Orbifolds of $D_3$ \label{sec033}}

In this section, we revisit the brane brick model for
$D_3$, which is reviewed in section \sref{sec023}.
We recall that the brane brick model has
9 bifundamental chiral fields $X_{ij}$,
which form 5 gauge invariant generators $A_{1}, A_{2}, A_{3}, B_{1}, B_{2}$
for the mesonic moduli space of the abelian brane brick model of $D_3$.
\\

%-------------------------------------------------------
\paragraph{Orbifold Action.}
For an abelian orbifold of the form $D_3 / \mathbb{Z}_{n_1} \times \mathbb{Z}_{n_2} \times \mathbb{Z}_{n_3}$
with orbifold action $(a_1, \dots, a_5)$, 
we first note that $(a_1, \dots, a_5)$ acts on the generators, which are identified with the following coordinates,
\beal{es03c01c01}
z_1 = A_1 ~,~
z_2 = A_2 ~,~
z_3 = A_3 ~,~
z_4 = B_1 ~,~
z_5 = B_2 ~.~
\eea
Based on \eref{es02c02},
these generators satisfy the following binomial relation for the $D_3$ model, 
\beal{es03c01c03}
z_1 z_2 z_3 = z_4 z_5 ~.~
\eea
Following \eref{es02d01},
we note here that the orbifold action $(a_1, \dots, a_5)$
acts on the above coordinates as 
$
z_i \mapsto \exp\left[
2\pi i~ \left(
\frac{a_{i1}}{n_1} s_{1}
+ \frac{a_{i2}}{n_2} s_{2}
+ \frac{a_{i3}}{n_3} s_{3}
\right)
\right]
z_i
$,
where $(s_1,s_2,s_3)\in  \mathbb{Z}_{n_1} \times \mathbb{Z}_{n_2} \times \mathbb{Z}_{n_3}$.

We can decompose the orbifold action $(a_1, \dots, a_5)$
acting on the coordinates $z_1, \dots, z_5$
as an orbifold action acting on the 9 bifundamental chiral fields of the $D_3$ model.
In order to do so, we first note that the generators of the mesonic moduli space of the abelian $D_3$ model
can be expressed in terms of gauge invariant products of chiral fields as shown in \eref{es02c03}, 
leading to the following identifications, 
\beal{es03c03}
\begin{array}{rcl@{\qquad}rcl@{\qquad}rcl}
a_1 &=& b_1 = b_2+b_9 ~,~ &
a_2 &=& b_4 = b_3+b_5 ~,~ &
a_3 &=& b_7 = b_6+b_8 ~,~
\\
a_4 &=& b_2+b_5+b_6 ~,~ &
a_5 &=& b_8+b_3+b_9 ~,~ &
&&
\end{array}
\eea
where $(b_1, \dots, b_9)$ forms the orbifold action on the chiral fields.
Here, $a_{i} = (a_{i1}, a_{i2}, a_{i3})$ and $b_{k} = (b_{k1}, b_{k2}, b_{k3})$, 
and the relations hold under $(\bmod{~n_j})$ with $j=1,2,3$.
The resulting effective 
orbifold action on the indices of the bifundamental chiral fields is as follows,
\beal{es03c_fields}
&
D_{[1,s] [1,s+b_1]}~,~
D_{[2,s] [3,s+b_2]}~,~
X_{[1,s] [3,s+b_3]}~,~
X_{[2,s] [2,s+b_4]}~,~
X_{[3,s] [1,s+b_5]}~,~
&
\nn\\
&
Y_{[1,s] [2,s+b_6]}~,~
Y_{[3,s] [3,s+b_7]}~,~
Z_{[2,s] [1,s+b_8]}~,~
Z_{[3,s] [2,s+b_9]}~,~
&
\eea
where  $s=(s_1, s_2, s_3)$,
$b_{k} = (b_{k1}, b_{k2}, b_{k3})$, 
and the labels are under $(\bmod{~n_j})$ with $j=1,2,3$.
\\

%-------------------------------------------------------
\paragraph{Constraints on the Orbifold Action.}
The orbifold action $(a_1, \dots, a_5)$
for the abelian orbifold of the form $D_3 / \mathbb{Z}_{n_1} \times \mathbb{Z}_{n_2} \times \mathbb{Z}_{n_3}$
undergoes several constraints
as reviewed in section \sref{sec031}.
The first constraint comes from \eref{es02d04}
and takes the following form, 
\beal{es03c01c02}
\sum_{i=1}^{5} a_{ij} = 0 ~~~(\bmod{~n_j})
~,~
\eea
for $j=1,2,3$.
Additionally, 
the orbifold action is required to preserve the defining relation for $D_3$
as given in \eref{es03c01c03}.
This leads to the following constraint, 
\beal{es03c06}
a_1 + a_2 + a_3 = a_4 + a_5 ~,~
\eea
where $a_{i} = (a_{i1}, a_{i2}, a_{i3})$ and 
the constraint on the orbifold action $(a_1, \dots, a_5)$ holds
under $(\bmod{~n_j})$ with $j=1,2,3$.

%------------------------------------------------------------------------------------------
\begin{figure}[ht!!]
\centering
\includegraphics[width=0.75\linewidth]{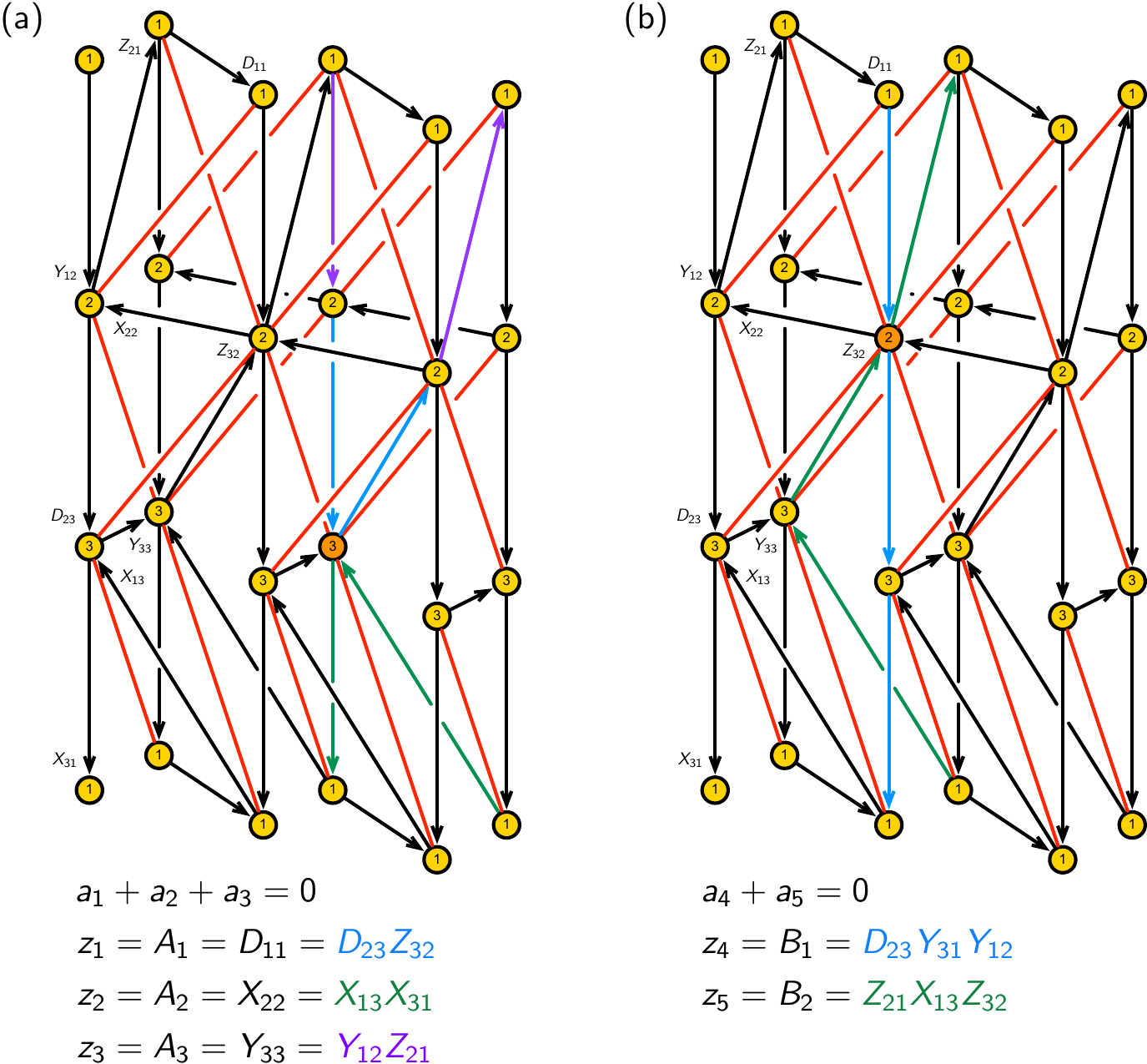}
\caption{
The vanishing constraints (a) $a_1 + a_2 + a_3 = 0$
and (b) $a_4 + a_5 = 0$
on the orbifold action for abelian orbifolds of $D_3$
are associated with gauge invariants $z_1 z_2 z_3 = A_1 A_2 A_3$
and $z_4 z_5 = B_1 B_2$, respectively, which each form closed paths in the periodic quiver of the $D_3$ brane brick model.
}
\label{fig_d3_loops}
\end{figure}
%------------------------------------------------------------------------------------------

We also have to ensure that the orbifold of $D_3$ remains Calabi-Yau.
For this to be the case, the holomorphic top form $\Omega_\mathcal{M}$ for $D_3$ 
needs to stay invariant under the orbifold action.
In the patch when $z_5 \neq 0$, we have $z_4 = z_1 z_2 z_3 z_5^{-1}$ from \eref{es03c01c03} allowing us to write the holomorphic top form $\Omega_\mathcal{M}$ in terms of local independent coordinates, 
\beal{es03c06b01}
z_1 ~,~ 
z_2 ~,~
z_3 ~,~
z_5 ~.~
\eea
Accordingly, we have,
\beal{es03c06b02}
\Omega_\mathcal{M}
=
\frac{
d z_1 \wedge
d z_2 \wedge
d z_3 \wedge
d z_5 
}{
z_5
}
~,~
\eea
which needs to remain invariant under the orbifold.
This leads to an additional constraint on the orbifold action given by,
\beal{es03c06b03b}
a_1 + a_2 + a_3 + a_5 - a_5 = 0
~,~
\eea
which combined with the constraints in \eref{es03c06} simplifies to, 
\beal{es03c06b03}
a_1 + a_2 + a_3 = a_4 + a_5  = 0
~,~
\eea
where $a_i = (a_{i1}, a_{i2}, a_{i3})$ and the above relations hold under $(\bmod ~n_j)$ with $j=1,2,3$.
We note here that the vanishing constraints in \eref{es03c06b03}
on the orbifold action
correspond to products of coordinates $z_i$ associated with products of gauge invariant generators of the mesonic moduli space
whose chiral field content forms closed paths in the periodic quiver of the brane brick model as illustrated in \fref{fig_d3_loops}.
The vanishing constraints in \eref{es03c06b03} imply that these closed paths remain closed paths in the periodic quiver even after orbifolding. 

As stated in \eref{es02d06}, 
we also need the orbifold action $(a_1, \dots, a_5)$ on the coordinates $z_1, \dots, z_5$
to be faithful, 
leading to the following necessary constraint, 
\beal{es03c01c02b}
\gcd (a_{1j}, \dots, a_{5j}, n_j) = 1 ~,~
\eea
for $j=1,2,3$.
\\

Following the above discussion on the form and constraints of the orbifold actions, 
we can now derive the general form of the $J$- and $E$-terms of the brane brick model corresponding to an 
abelian orbifold of the form
$D_3 / \mathbb{Z}_{n_1} \times \mathbb{Z}_{n_2} \times \mathbb{Z}_{n_3}$
with orbifold action $(a_1, \dots, a_5)$.
The $J$- and $E$-terms take the following form,
\beal{es03c07}
\begin{array}{rcl}
J^+[\Lambda^{1}_{[2,a_{4}+s][1,s]}]
&=&
X_{[1,s][3,s]}  X_{[3,s][1,a_{2}+s]}  Y_{[1,a_{2}+s][2,a_{4}+s]}
\\
J^-[\Lambda^{1}_{[2,a_{4}+s][1,s]}]
&=&
Y_{[1,s][2,a_{4}-a_{2}+s]}   X_{[2,a_{4}-a_{2}+s][2,a_{4}+s]}
\\
E^+[\Lambda^{1}_{[2,a_{4}+s][1,s]}]
&=&
D_{[2,a_{4}+s][3,a_{4}+s]}  Z_{[3,a_{4}+s][2,a_{4}+a_{1}+s]}   Z_{[2,a_{4}+a_{1}+s][1,s]}
\\
E^-[\Lambda^{1}_{[2,a_{4}+s][1,s]}]
&=&
Z_{[2,a_{4}+s][1,-a_{1}+s]}  D_{[1,-a_{1}+s][1,s]}
\\
\\
J^{+}[\Lambda^{2}_{[1,a_{5}-a_{1}+a_{2}+s][2,s]}]
&=&
Z_{[2,s][1,a_{5}-a_{1}+s]}  X_{[1,a_{5}-a_{1}+s][3,a_{5}-a_{1}+s]}   X_{[3,a_{5}-a_{1}+s][1,a_{5}-a_{1}+a_{2}+s]}
\\
J^{-}[\Lambda^{2}_{[1,a_{5}-a_{1}+a_{2}+s][2,s]}]
&=&
X_{[2,s][2,a_{2}+s]}  Z_{[2,a_{2}+s][1,a_{5}-a_{1}+a_{2}+s]}
\\
E^{+}[\Lambda^{2}_{[1,a_{5}-a_{1}+a_{2}+s][2,s]}]
&=&
D_{[1,a_{5}-a_{1}+a_{2}+s][1,a_{5}+a_{2}+s]}  Y_{[1,a_{5}+a_{2}+s][2,s]}
\\
E^{-}[\Lambda^{2}_{[1,a_{5}-a_{1}+a_{2}+s][2,s]}]
&=&
Y_{[1,a_{5}-a_{1}+a_{2}+s][2,-a_{1}+s]}  D_{[2,-a_{1}+s][3,-a_{1}+s]}  Z_{[3,-a_{1}+s][2,s]}
\\
\\
J^{+}[\Lambda^{3}_{[3,a_{3}+s][1,s]}]
&=&
X_{[1,s][3,s]}   Y_{[3,s][3,a_{3}+s]}
\\
J^{-}[\Lambda^{3}_{[3,a_{3}+s][1,s]}]
&=&
Y_{[1,s][2,a_{4}-a_{2}+s]}   Z_{[2,a_{4}-a_{2}+s][1,a_{3}+s]}   X_{[1,a_{3}+s][3,a_{3}+s]}
\\
E^{+}[\Lambda^{3}_{[3,a_{3}+s][1,s]}]
&=&
X_{[3,a_{3}+s][1,-a_{1}+s]}   D_{[1,-a_{1}+s][1,s]}
\\
E^{-}[\Lambda^{3}_{[3,a_{3}+s][1,s]}]
&=&
Z_{[3,a_{3}+s][2,-a_{2}+s]}   D_{[2,-a_{2}+s][3,-a_{2}+s]}   X_{[3,-a_{2}+s][1,s]}
\\
\\
J^{+}[\Lambda^{4}_{[1,-a_{1}+s][3,s]}]
&=&
X_{[3,s][1,a_{2}+s]}   Y_{[1,a_{2}+s][2,a_{4}+s]}   Z_{[2,a_{4}+s][1,-a_{1}+s]}
\\
J^{-}[\Lambda^{4}_{[1,-a_{1}+s][3,s]}]
&=&
Y_{[3,s][3,a_{3}+s]}   X_{[3,a_{3}+s][1,-a_{1}+s]}
\\
E^{+}[\Lambda^{4}_{[1,-a_{1}+s][3,s]}]
&=&
D_{[1,-a_{1}+s][1,s]}   X_{[1,s][3,s]}
\\
E^{-}[\Lambda^{4}_{[1,-a_{1}+s][3,s]}]
&=&
X_{[1,-a_{1}+s][3,-a_{1}+s]}   Z_{[3,-a_{1}+s][2,s]}   D_{[2,s][3,s]}
\\
\\
J^{+}[\Lambda^{5}_{[2,-a_{2}+s][3,s]}]
&=&
Y_{[3,s][3,a_{3}+s]}   Z_{[3,a_{3}+s][2,-a_{2}+s]}
\\
J^{-}[\Lambda^{5}_{[2,-a_{2}+s][3,s]}]
&=&
Z_{[3,s][2,a_{1}+s]}   Z_{[2,a_{1}+s][1,a_{5}+s]}   Y_{[1,a_{5}+s][2,-a_{2}+s]}
\\
E^{+}[\Lambda^{5}_{[2,-a_{2}+s][3,s]}]
&=&
D_{[2,-a_{2}+s][3,-a_{2}+s]}   X_{[3,-a_{2}+s][1,s]}   X_{[1,s][3,s]}
\\
E^{-}[\Lambda^{5}_{[2,-a_{2}+s][3,s]}]
&=&
X_{[2,-a_{2}+s][2,s]}   D_{[2,s][3,s]}
\\
\\
J^{+}[\Lambda^{6}_{[2,-a_{3}+s][3,s]}]
&=&
Z_{[3,s][2,a_{1}+s]}   X_{[2,a_{1}+s][2,-a_{3}+s]}
\\
J^{-}[\Lambda^{6}_{[2,-a_{3}+s][3,s]}]
&=&
X_{[3,s][1,a_{2}+s]}   X_{[1,a_{2}+s][3,a_{2}+s]}   Z_{[3,a_{2}+s][2,-a_{3}+s]}
\\
E^{+}[\Lambda^{6}_{[2,-a_{3}+s][3,s]}]
&=&
D_{[2,-a_{3}+s][3,-a_{3}+s]}   Y_{[3,-a_{3}+s][3,s]}
\\
E^{-}[\Lambda^{6}_{[2,-a_{3}+s][3,s]}]
&=&
Z_{[2,-a_{3}+s][1,a_{5}+a_{2}+s]}   Y_{[1,a_{5}+a_{2}+s][2,s]}   D_{[2,s][3,s]}
\end{array}
\eea
where $s$ is running over every element in $\Gamma$,
$a_{i} = (a_{i1}, a_{i2}, a_{i3})$
and the labels are all under $(\bmod{~n_j})$ with $j=1,2,3$.
The number of Fermi fields is $6\cdot |\Gamma|$ and the number of chiral fields is $9\cdot |\Gamma|$.
\\

%=================================================================
\subsection{On Counting Abelian Orbifolds of $Q^{1,1,1}$ and $D_3$ \label{sec035}}

We are able to count distinct abelian orbifolds of $Q^{1,1,1}$
and $D_3$ using the constraints and conditions on the orbifold actions discussed in sections \sref{sec032} and \sref{sec033}.
In order to count distinct orbifolds, 
we make use of the techniques introduced in \cite{Davey:2010px, Hanany:2010ne, Hanany:2010cx, Davey:2011dd, Hanany:2011iw}, which we review in part in this section.
\\

%------------------------------------------------------------
\paragraph{Sequences and Series Convolutions.}
We first note that a sequence of the following form, 
\beal{es50a01}
\mathsf{g} = \{ \mathsf{g}(1), \mathsf{g}(2), \mathsf{g}(3), \dots \} ~,~
\eea
can be expressed in terms of its corresponding generating function given by, 
\beal{es50a02}
g(t) = \sum_{n=1}^{\infty} \mathsf{g}(n) t^n ~,~
\eea
where $t$ is the fugacity corresponding to the index $n$ of the sequence $\mathsf{g}$.

Given two sequences $\mathsf{r}$ and $\mathsf{s}$, 
we can define the convolution $\mathsf{q} = \mathsf{r} \ast \mathsf{s}$ whose corresponding generating function is given by, 
\beal{es50a05}
q(t) = \sum_{m,k=1}^{\infty} \mathsf{r}(m) \mathsf{s}(k) t^{mk} ~.~
\eea
Here, the inversion is given by,
\beal{es50a06}
r(t) = \sum_{k=1}^{\infty}  \mu(k)  \mathsf{s}(k)  q(t^k)~,~
\eea
where $\mu(k)$ is the M\"obius function that we also introduced for the plethystic logarithm in section \sref{sec021}.

As discussed in \cite{Davey:2010px, Hanany:2010ne, Hanany:2010cx, Davey:2011dd, Hanany:2011iw},
we can make use of convolutions of basic multiplicative sequences 
in order to obtain more complicated sequences.
Such multiplicative sequences include the following basic examples:
\begin{itemize}

\item Unit sequence:
\beal{es50a10}
\mathsf{u} = \{1,1,1, \dots \} 
~\leftrightarrow~
u(t) = \sum_{n=1}^\infty t^n
~,~
\eea

\item Natural number sequence:
\beal{es50a11}
\mathsf{N} = \{1,2,3, \dots \} 
~\leftrightarrow~
N(t) = \sum_{n=1}^\infty n t^n
~,~
\eea

\item Powers of natural number sequences:
\beal{es50a12}
\mathsf{N}^d = \{1^d,2^d,3^d, \dots \} 
~\leftrightarrow~
N^d(t) = \sum_{n=1}^\infty n^d t^n
~,~
\eea

\item Dirichlet character $\chi_{k,m}$ modulo $k$ and index $m$, which is defined under the following conditions,
\beal{es50a15}
\chi_{k,m}(1) &=& 1
\nn\\
\chi_{k,m}(a) &=& \chi_{k,m}(a+k)
\nn\\
\chi_{k,m}(a) \chi_{k,m}(b) &=& \chi_{k,m}(ab)
\nn\\
\chi_{k,m}(a) &=&0~~\text{if $\gcd(k,a) \neq 1$}
~.~
\eea
Some of the simple Dirichlet characters used in this work are as follows,
\beal{es50a16}
\chi_{1,1} &=& \mathsf{u} \nn\\
\chi_{2,1} &=& \{1,0,\dots \} \nn\\
\chi_{3,1} &=& \{1,1,0,\dots \} \nn\\
\chi_{3,2} &=& \{1,-1,0,\dots \} \nn\\
\chi_{4,1} &=& \{1,0,1,0,\dots \} \nn\\
\chi_{4,2} &=& \{1,0,-1,0,\dots \} ~,~
\eea
where each displayed block is extended periodically.

\end{itemize}

%------------------------------------------------------------
\paragraph{Counting Sublattices and Hermite Normal Forms.}
The enumeration of abelian orbifolds of a toric Calabi-Yau 4-fold $\mathcal{M}$
corresponds to the enumeration of sublattices in $\mathbb{Z}^3$ of index $n$ 
as originally observed in \cite{Davey:2010px, Hanany:2010ne, Hanany:2010cx, Davey:2011dd, Hanany:2011iw}.
Every such sublattice in 
$\mathbb{Z}^3$ of index $n$
is represented by a unique Hermite
normal form matrix $H$ with $\det H = n$. 
These matrices take the following general form,
\beal{es50a20}
H = \left(
\ba{ccc}
a & d & e \\
0 & b & f \\
0 & 0  & c
\ea
\right)
~,~ a,b,c \geq 1 ~,~ abc = n ~,~
\eea
where the remaining components satisfy, 
\beal{es50a21}
0 \leq d < b~,~
0 \leq e < c~,~
0 \leq f < c~.~
\eea
We refer here to $\mathcal{H}_n$ as the finite set
of Hermite normal form matrices at order $n$.
The rows of the individual matrices in $\mathcal{H}_n$
form a basis generating sublattices of $\mathbb{Z}^3$ of index $n$.

The set $\mathcal{H}_n$ enumerates more sublattices of $\mathbb{Z}^3$ 
than there are distinct abelian orbifolds of the toric Calabi-Yau 4-fold $\mathcal{M}$
at orbifold order $n=|\Gamma|$.
Because the toric diagram of $\mathcal{M}$
carries a discrete symmetry
which corresponds to lattice automorphisms that map the toric
diagram to itself,
sublattices corresponding to two matrices in $\mathcal{H}_n$
related by such an automorphism correspond to the same abelian orbifold of $\mathcal{M}$.
These automorphisms form a finite group $G\subset GL(3,\mathbb{Z})$.
Two sublattices related by an element of $G$
describe the same abelian orbifold of $\mathcal{M}$, 
and hence the number of inequivalent abelian orbifolds of $\mathcal{M}$ at order
$n$ is the number of orbits of $G$ acting on the set $\mathcal{H}_n$.

%------------------------------------------------------------------------------------------
\begin{table}[htt!!]
\centering
\begin{tabular}{|r|c|l|l|}
\hline
$[g]$ & $|[g]|$ & cycle label & sequence\\
\hline\hline
$E$              & 1 & $x_{1}^{8}$ & $\mathsf{g}_{x_1^8}$
\\
\hline
$\sigma_{d}$    & 6 & $x_{1}^{4}x_{2}^{2}$ & $\mathsf{g}_{x_1^4 x_2^2}$
\\
\hline
$C_{3}$         & 8 & $x_{1}^{2}x_{3}^{2}$ & $\mathsf{g}_{x_1^2 x_3^2}$
\\
\hline
Inv              & 1 & $x_{2}^{4}~ (a)$ & $\mathsf{g}_{x_2^4}^{(a)}$
\\
$\sigma_{h}$    & 3 & $x_{2}^{4}~ (b)$ & $\mathsf{g}_{x_2^4}^{(b)}$
\\
$C_{2}$         & 3 & $x_{2}^{4}~ (c)$ & $\mathsf{g}_{x_2^4}^{(c)}$
\\
$C_{2}'$        & 6 & $x_{2}^{4}~ (d)$ & $\mathsf{g}_{x_2^4}^{(d)}$
\\
\hline
$S_{4}$         & 6 & $x_{4}^{2}~ (a)$ & $\mathsf{g}_{x_4^2}^{(a)}$
\\
$C_{4}$         & 6 & $x_{4}^{2}~ (b)$ & $\mathsf{g}_{x_4^2}^{(b)}$
\\
\hline
$S_{6}$         & 8 & $x_{2}x_{6}$ & $\mathsf{g}_{x_2 x_6}$
\\
\hline
\end{tabular}
\caption{
The conjugacy classes $[g]$ and cycle labels for the octahedral group $O_h$ for abelian orbifolds of $Q^{1,1,1}$.}
\label{tab_q111_conj}
\end{table}
%------------------------------------------------------------------------------------------

%------------------------------------------------------------------------------------------
\begin{table}[htt!!]
\centering
\begin{tabular}{|r|c|l|}
\hline
sequence & $|[g]|$ & convolution \\
\hline \hline
$\mathsf{g}_{x_1^{8}}$        & $1$ & $\mathsf{u}\ast\mathsf{N}\ast\mathsf{N}^{2}$ \\
\hline
$\mathsf{g}_{x_1^{4}x_2^{2}}$ & $6$ & $\mathsf{u}\ast\mathsf{u}\ast\mathsf{N}\ast(t-t^{2}+4t^{4})$ \\
\hline
$\mathsf{g}_{x_1^{2}x_3^{2}}$ & $8$ & $\mathsf{u}\ast\mathsf{u}\ast\chi_{3,2}\ast(t-t^{3}+3t^{9})$ \\
\hline
$\mathsf{g}_{x_2^{4}}^{(a)}$  & $1$ & $\mathsf{u}\ast\mathsf{N}\ast\mathsf{N}^{2}$ \\
$\mathsf{g}_{x_2^{4}}^{(b)}$  & $3$ & $\mathsf{u}\ast\mathsf{u}\ast\mathsf{N}\ast(t-t^{2}+4t^{4})$ \\
$\mathsf{g}_{x_2^{4}}^{(c)}$  & $3$ & $\mathsf{u}\ast\mathsf{u}\ast\mathsf{N}\ast(t-t^{2}+4t^{4})$ \\
$\mathsf{g}_{x_2^{4}}^{(d)}$  & $6$ & $\mathsf{u}\ast\mathsf{u}\ast\mathsf{N}\ast(t-t^{2}+4t^{4})$ \\
\hline
$\mathsf{g}_{x_4^{2}}^{(a)}$     & $6$ & $\mathsf{u}\ast\mathsf{u}\ast\chi_{4,2}\ast(t-t^{2}+2t^{4})$ \\
$\mathsf{g}_{x_4^{2}}^{(b)}$     & $6$ & $\mathsf{u}\ast\mathsf{u}\ast\chi_{4,2}\ast(t-t^{2}+2t^{4})$ \\
\hline
$\mathsf{g}_{x_2 x_6}$        & $8$ & $\mathsf{u}\ast\mathsf{u}\ast\chi_{3,2}\ast(t-t^{3}+3t^{9})$ \\
\hline
\end{tabular}
\caption{Sequence convolutions 
for each of the subsequences 
contributing to the total number of distinct abelian orbifolds of $Q^{1,1,1}$.}
\label{tab_q111_convolutions}
\end{table}
%------------------------------------------------------------------------------------------

The action is well defined since every $g \in G$ permutes the set of sublattices corresponding to $\mathcal{H}_n$. By Burnside's lemma, the number of orbits and hence the number of distinct abelian orbifolds of the form 
$\mathcal{M}/\Gamma$ at order $n=|\Gamma|$
equals the average number of
fixed points of the elements of $G$,
\beal{es50a25}
\mathsf{g}_\text{orbits}(n) =
\frac{1}{|G|}
\sum_{g \in G} \bigl| F_n(g) \bigr|
~,~
\eea
where the set of sublattices fixed by $g$ is given by,
\beal{es50a26}
F_n(g)=
\left\{
H \in \mathcal{H}_n ~\middle|~
H\, g\, H^{-1} \in GL(3,\mathbb{Z})
\right\} ~.~
\eea
Equivalently, $H \in F_n(g)$ whenever $H g$ and $H$ correspond to equivalent
sublattices. 

The average in \eqref{es50a25}
can be written in terms of the conjugacy classes $[g]$ of $G$, weighted by their
sizes $|[g]|$.
We have, 
\beal{es50a27}
\mathsf{g}_\text{orbits}(n) =
\frac{1}{|G|}
\sum_{[g] \subset G} \bigl| [g] \bigr| \,
\bigl| F_n(g) \bigr|
~,~
\eea
where the sum runs over the conjugacy classes $[g]$ of $G$ 
with $|[g]|$
denoting the number of elements of the conjugacy class $[g]$
and
$g$ being a chosen representative of $[g]$. 
\\

%------------------------------------------------------------
\paragraph{Counting Abelian Orbifolds of $Q^{1,1,1}$.}
The toric diagram of $Q^{1,1,1}$
illustrated in \fref{fig_01} and given by \eref{es02b01m3},
has a discrete symmetry given by the octahedral group $O_h$
of order $|O_h| = 48$.
When we think of each of the elements of $O_h$
as permutations of the 8 boundary triangles of the toric diagram, the conjugacy classes and the corresponding cycle structures of $O_h$
can be summarized as shown in \tref{tab_q111_conj}.
Here, we label the conjugacy classes in terms of their cycle structure.
When multiple conjugacy classes have the same cycle structure, we refine the label with an additional numbering as shown in \tref{tab_q111_conj}.

%------------------------------------------------------------------------------------------
\begin{table}[ht!]
\centering
\renewcommand{\arraystretch}{1.3}
\setlength{\tabcolsep}{4.5pt}
\begin{tabular}{r|p{1cm}p{1cm}p{1cm}p{1cm}p{1cm}p{1cm}p{1cm}p{1cm}p{1cm}p{1cm}}
\hline
$n$  & $1$ & $2$ & $3$ & $4$ & $5$ & $6$ & $7$ & $8$ & $9$ & $10$ \\
\hline\hline
$\mathsf{g}_{x_1^{8}}$         & $1$ & $7$ & $13$ & $35$ & $31$ & $91$ & $57$ & $155$ & $130$ & $217$ \\
\hline
$\mathsf{g}_{x_1^{4}x_2^{2}}$ & $1$ & $3$ & $5$  & $11$ & $7$  & $15$ & $9$  & $31$  & $18$  & $21$ \\
\hline
$\mathsf{g}_{x_1^{2}x_3^{2}}$  & $1$ & $1$ & $1$  & $2$  & $1$  & $1$  & $3$  & $2$   & $4$   & $1$ \\
\hline
$\mathsf{g}_{x_2^{4}}^{(a)}$ & $1$ & $7$ & $13$ & $35$ & $31$ & $91$ & $57$ & $155$ & $130$ & $217$ \\
$\mathsf{g}_{x_2^{4}}^{(b)}$  & $1$ & $3$ & $5$  & $11$ & $7$  & $15$ & $9$  & $31$  & $18$  & $21$ \\
$\mathsf{g}_{x_2^{4}}^{(c)}$  & $1$ & $3$ & $5$  & $11$ & $7$  & $15$ & $9$  & $31$  & $18$  & $21$ \\
$\mathsf{g}_{x_2^{4}}^{(d)}$  & $1$ & $3$ & $5$  & $11$ & $7$  & $15$ & $9$  & $31$  & $18$  & $21$ \\
\hline
$\mathsf{g}_{x_4^2}^{(a)}$       & $1$ & $1$ & $1$  & $3$  & $3$  & $1$  & $1$  & $5$   & $2$   & $3$ \\
$\mathsf{g}_{x_4^2}^{(b)}$       & $1$ & $1$ & $1$  & $3$  & $3$  & $1$  & $1$  & $5$   & $2$   & $3$ \\
\hline
$\mathsf{g}_{x_2 x_6}$       & $1$ & $1$ & $1$  & $2$  & $1$  & $1$  & $3$  & $2$   & $4$   & $1$ \\
\hline \hline
$\mathsf{g}_{Q^{1,1,1}}$       & $1$ & $2$ & $3$  & $7$  & $5$  & $10$  & $7$  & $20$   & $14$   & $18$ \\
\hline
\end{tabular}
\caption{Sequences for $n=1, \dots, 10$, where $n=|\Gamma|$,
contributing to the total number given by $\mathsf{g}_{Q^{1,1,1}}$ of distinct abelian orbifolds of the form $Q^{1,1,1}/\Gamma$.}
\label{tab_q111_counts}
\end{table}
%------------------------------------------------------------------------------------------

Using Burnside's lemma and the cycle label for conjugacy classes in \tref{tab_q111_conj}, 
the sequence of the number of distinct abelian orbifolds of the form $Q^{1,1,1}/\Gamma$
at order $n=|\Gamma|$
is given by,
\beal{es50a30}
\mathsf{g}_{Q^{1,1,1}}
&
=
&
\frac{1}{48}
\Big(
\mathsf{g}_{x_1^{8}}
+ 6 \mathsf{g}_{x_1^{4} x_2^{2}}
+ 8 \mathsf{g}_{x_1^{2} x_3^{2}}
+ \mathsf{g}_{x_2^{4}}^{(a)}
+ 3 \mathsf{g}_{x_2^{4}}^{(b)}
+ 3 \mathsf{g}_{x_2^{4}}^{(c)}
+ 6 \mathsf{g}_{x_2^{4}}^{(d)}
\nn\\
&&
\hspace{1cm}
+ 6 \mathsf{g}_{x_4^{2}}^{(a)}
+ 6 \mathsf{g}_{x_4^{2}}^{(b)}
+ 8 \mathsf{g}_{x_2 x_6}
\Big)
~,~
\eea
where $\mathsf{g}_{[g]}(n) = |F_n(g)|$.
The subsequences $\mathsf{g}_{[g]}$
corresponding to the conjugacy classes $[g]$
expressed in terms of convolutions of standard multiplicative sequences
are summarized in \tref{tab_q111_convolutions}.
The actual sequences with the counting of distinct abelian orbifolds of $Q^{1,1,1}$
are presented in \tref{tab_q111_counts}
for orders $n=1, \dots, 10$.
We note that the sequence enumerating distinct abelian orbifolds of $Q^{1,1,1}$ 
also appears in a different context in \cite{OEIS_A159842}
and matches exactly the
number of distinct abelian orbifolds of $\mathbb{C}^4$ \cite{Davey:2010px, Hanany:2010ne, Hanany:2010cx, Davey:2011dd}.

%------------------------------------------------------------------------------------------
\begin{table}[htt!!]
\centering
\begin{tabular}{|>{\centering\arraybackslash}m{1cm}
                |>{\centering\arraybackslash}m{2cm}
                |>{\centering\arraybackslash}m{4cm}
                |>{\centering\arraybackslash}m{4cm}|}
\hline
$n$ & Orbifold & Orbifold Action & Toric Diagram\\
\hline\hline
1 & $Q^{1,1,1}$ &
$(0,0,0,0,0,0,0,0)$ &
\includegraphics[width=0.2\linewidth]{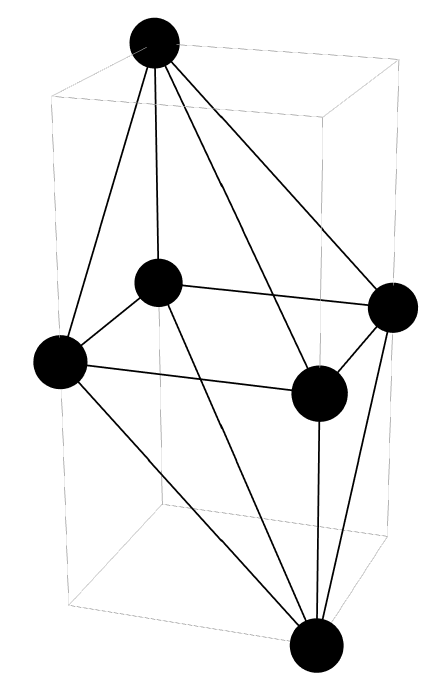}
\\
\hline
2 & $Q^{1,1,1}/\mathbb Z_2$ &
$(1,1,1,1,1,1,1,1)$ &
\includegraphics[width=0.35\linewidth]{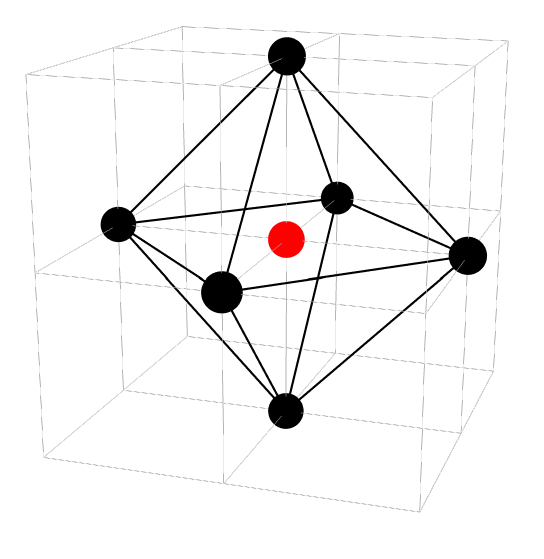}
\\
\hline
2 & $Q^{1,1,1}/\mathbb Z_2$ &
$(0,0,1,1,1,1,0,0)$ &
\includegraphics[width=0.4\linewidth]{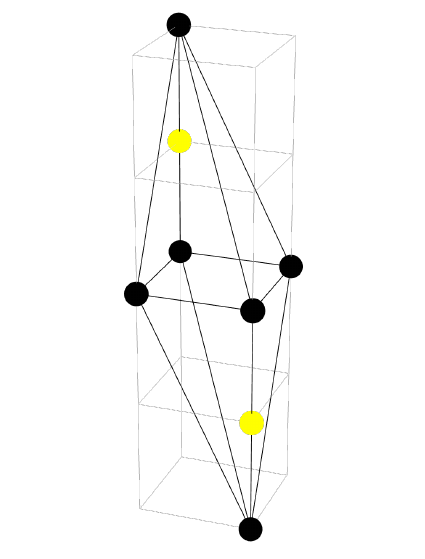}
\\
\hline
3 & $Q^{1,1,1}/\mathbb Z_3$ &
$(1,1,1,1,2,2,2,2)$ &
\includegraphics[width=0.45\linewidth]{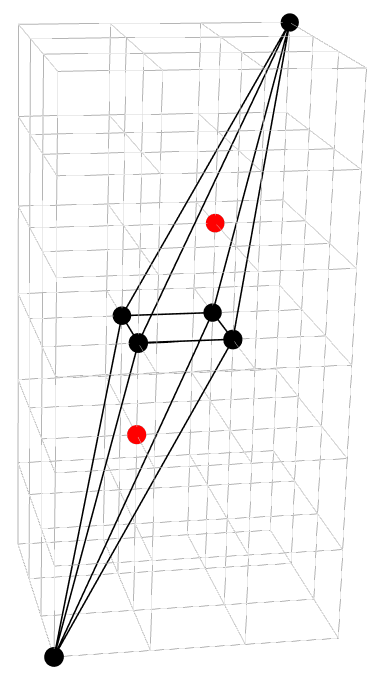}
\\
\hline
3 & $Q^{1,1,1}/\mathbb Z_3$ &
$(0,0,1,1,2,2,0,0)$ &
\includegraphics[width=0.2\linewidth]{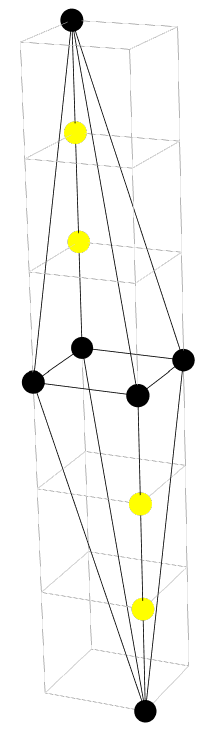}
\\
\hline
3 & $Q^{1,1,1}/\mathbb Z_3$ &
$(0,1,1,2,1,2,2,0)$ &
\includegraphics[width=0.35\linewidth]{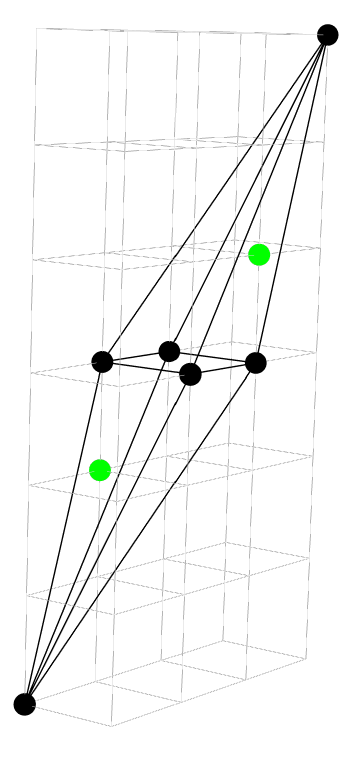}
\\
\hline
\end{tabular}
\caption{
Distinct abelian orbifolds
of $Q^{1,1,1}$
for $n=1,2,3$
with the corresponding orbifold actions and toric diagrams.
}
\label{tab_q111_orbifolds}
\end{table}
%------------------------------------------------------------------------------------------

%------------------------------------------------------------------------------------------
\begin{table}[htt!!]
\centering
\begin{tabular}{|r|c|l|l|}
\hline
$[g]$ & $|[g]|$ & cycle label & sequence\\
\hline\hline
$E$              & 1 & $x_{1}^{5}$ & $\mathsf{g}_{x_1^5}$
\\
\hline
$C_{3}$          & 2 & $x_{1}^{2}x_{3}$ & $\mathsf{g}_{x_1^2 x_3}$
\\
\hline
$C_{2}$          & 3 & $x_{1}x_{2}^{2}$ & $\mathsf{g}_{x_1 x_2^2}$
\\
\hline
$\sigma_{h}$     & 1 & $x_{1}^{3}x_{2}~ (a)$ & $\mathsf{g}_{x_1^3 x_2}^{(a)}$
\\
$\sigma_{v}$     & 3 & $x_{1}^{3}x_{2}~ (b)$ & $\mathsf{g}_{x_1^3 x_2}^{(b)}$
\\
\hline
$S_{3}$          & 2 & $x_{2}x_{3}$ & $\mathsf{g}_{x_2 x_3}$
\\
\hline
\end{tabular}
\caption{
The conjugacy classes and cycle structures for the prismatic group $D_{3h}$ for abelian orbifolds of $D_3$.}
\label{tab_d3_conj}
\end{table}
%------------------------------------------------------------------------------------------

Remarkably, the counting of distinct abelian orbifolds of $Q^{1,1,1}$
from sublattice counting 
matches the identification of distinct and consistent orbifold actions for abelian orbifolds of $Q^{1,1,1}$
following the consistency conditions of orbifold actions discussed in section \sref{sec032}.
\tref{tab_q111_orbifolds} summarizes the first few distinct abelian orbifolds of the form $Q^{1,1,1}/\Gamma$
with their corresponding inequivalent orbifold actions. 
\\

%------------------------------------------------------------
\paragraph{Counting Abelian Orbifolds of $D_3$.}
The toric diagram of $D_3$,
illustrated in \fref{fig_02} and given by \eref{es02m02b},
has the shape of a triangular prism whose boundary consists of
$3$ rectangular and $2$ triangular faces.
Its discrete symmetry is the prismatic group $D_{3h}$
of order $|D_{3h}| = 12$.
Viewing each element of $D_{3h}$ as a permutation of the $5$ boundary
faces of the toric diagram, the conjugacy classes and the associated
cycle structures of $D_{3h}$ are summarized in \tref{tab_d3_conj}.
As in the $Q^{1,1,1}$ case, we label the conjugacy classes by their cycle structure,
and whenever two conjugacy classes carry the same cycle structure we
distinguish them with an additional numbering as shown in \tref{tab_d3_conj}.

%------------------------------------------------------------------------------------------
\begin{table}[htt!!]
\centering
\begin{tabular}{|r|c|l|}
\hline
sequence & $|[g]|$ & convolution \\
\hline \hline
$\mathsf{g}_{x_1^{5}}$         & $1$ & $\mathsf{u}\ast\mathsf{N}\ast\mathsf{N}^{2}$ \\
\hline
$\mathsf{g}_{x_1^{2}x_3}$      & $2$ & $\mathsf{u}\ast\mathsf{u}\ast\chi_{3,2}\ast(t+2t^{3})$ \\
\hline
$\mathsf{g}_{x_1 x_2^{2}}$     & $3$ & $\mathsf{u}\ast\mathsf{u}\ast\mathsf{N}\ast(t-t^{2}+4t^{4})$ \\
\hline
$\mathsf{g}_{x_1^{3}x_2}^{(a)}$ & $1$ & $\mathsf{u}\ast\mathsf{u}\ast\mathsf{N}\ast(t+3t^{2})$ \\
$\mathsf{g}_{x_1^{3}x_2}^{(b)}$ & $3$ & $\mathsf{u}\ast\mathsf{u}\ast\mathsf{N}\ast(t-t^{2}+4t^{4})$ \\
\hline
$\mathsf{g}_{x_2 x_3}$         & $2$ & $\mathsf{u}\ast\mathsf{u}\ast\chi_{3,2}$ \\
\hline
\end{tabular}
\caption{Sequence convolutions
for each of the subsequences
contributing to the total number of distinct abelian orbifolds of $D_3$.}
\label{tab_d3_convolutions}
\end{table}
%------------------------------------------------------------------------------------------

%------------------------------------------------------------------------------------------
\begin{table}[ht!]
\centering
\renewcommand{\arraystretch}{1.3}
\setlength{\tabcolsep}{4.5pt}
\begin{tabular}{r|p{1cm}p{1cm}p{1cm}p{1cm}p{1cm}p{1cm}p{1cm}p{1cm}p{1cm}p{1cm}}
\hline
$n$  & $1$ & $2$ & $3$ & $4$ & $5$ & $6$ & $7$ & $8$ & $9$ & $10$ \\
\hline\hline
$\mathsf{g}_{x_1^{5}}$          & $1$ & $7$ & $13$ & $35$ & $31$ & $91$ & $57$ & $155$ & $130$ & $217$ \\
\hline
$\mathsf{g}_{x_1^{2}x_3}$       & $1$ & $1$ & $4$  & $2$  & $1$  & $4$  & $3$  & $2$   & $7$   & $1$ \\
\hline
$\mathsf{g}_{x_1 x_2^{2}}$      & $1$ & $3$ & $5$  & $11$ & $7$  & $15$ & $9$  & $31$  & $18$  & $21$ \\
\hline
$\mathsf{g}_{x_1^{3}x_2}^{(a)}$ & $1$ & $7$ & $5$  & $23$ & $7$  & $35$ & $9$  & $59$  & $18$  & $49$ \\
$\mathsf{g}_{x_1^{3}x_2}^{(b)}$ & $1$ & $3$ & $5$  & $11$ & $7$  & $15$ & $9$  & $31$  & $18$  & $21$ \\
\hline
$\mathsf{g}_{x_2 x_3}$          & $1$ & $1$ & $2$  & $2$  & $1$  & $2$  & $3$  & $2$   & $3$   & $1$ \\
\hline \hline
$\mathsf{g}_{D_3}$              & $1$ & $3$ & $5$  & $11$ & $7$  & $19$ & $11$ & $34$  & $23$  & $33$ \\
\hline
\end{tabular}
\caption{Sequences for $n=1, \dots, 10$, where $n=|\Gamma|$,
contributing to the total number given by $\mathsf{g}_{D_3}$ of distinct abelian orbifolds of the form $D_3/\Gamma$.}
\label{tab_d3_counts}
\end{table}
%------------------------------------------------------------------------------------------

%------------------------------------------------------------------------------------------
\begin{table}[htt!!]
\centering
\begin{tabular}{|>{\centering\arraybackslash}m{1cm}
                |>{\centering\arraybackslash}m{2cm}
                |>{\centering\arraybackslash}m{4cm}
                |>{\centering\arraybackslash}m{4cm}|}
\hline
$n$ & Orbifold & Orbifold Action & Toric Diagram\\
\hline\hline
1 & $D_3$ &
$(0,0,0,0,0)$ &
\includegraphics[width=0.3\linewidth]{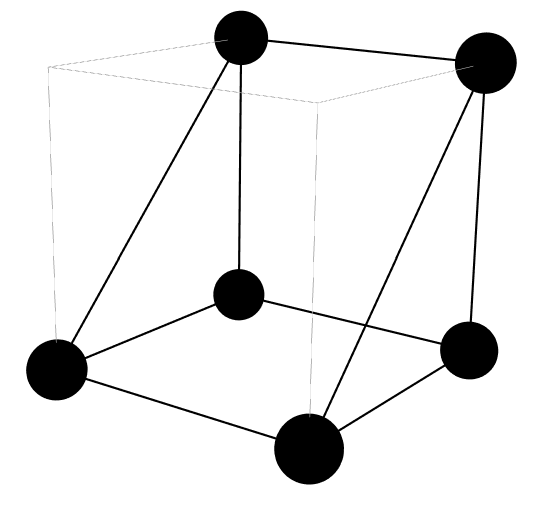}
\\
\hline
2 & $D_3/\mathbb Z_2$ &
$(0,0,0,1,1)$ &
\includegraphics[width=0.4\linewidth]{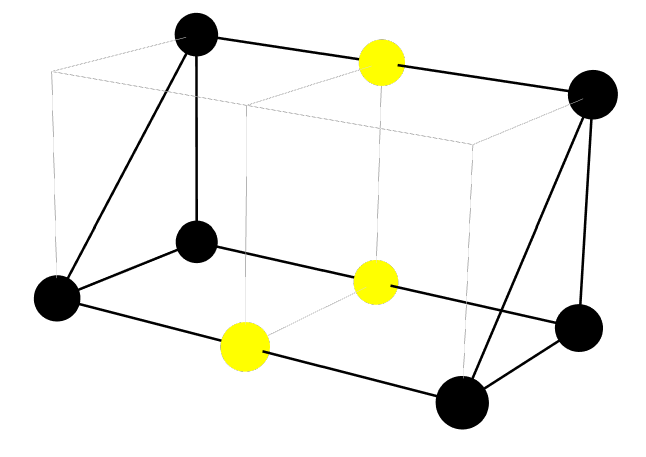}
\\
\hline
2 & $D_3/\mathbb Z_2$ &
$(0,1,1,0,0)$ &
\includegraphics[width=0.28\linewidth]{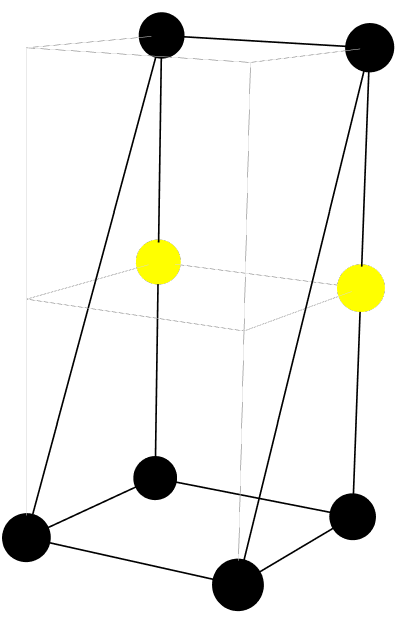}
\\
\hline
2 & $D_3/\mathbb Z_2$ &
$(0,1,1,1,1)$ &
\includegraphics[width=0.45\linewidth]{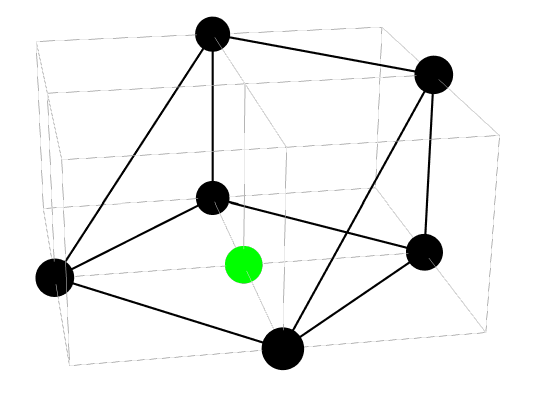}
\\
\hline
3 & $D_3/\mathbb Z_3$ &
$(0,0,0,1,2)$ &
\includegraphics[width=0.55\linewidth]{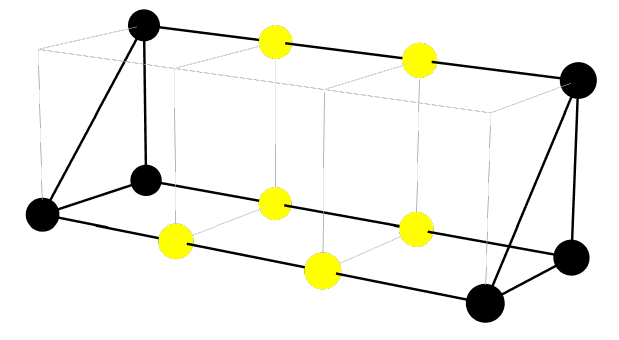}
\\
\hline
3 & $D_3/\mathbb Z_3$ &
$(0,1,2,0,0)$ &
\includegraphics[width=0.28\linewidth]{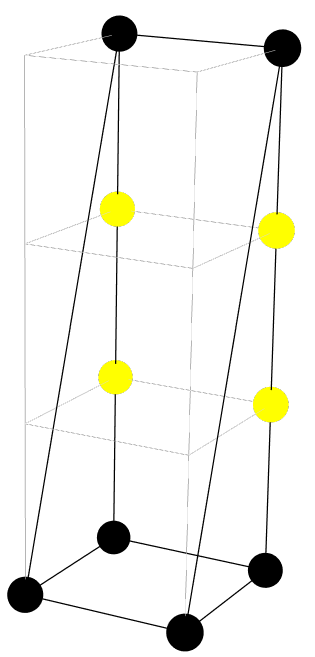}
\\
\hline
3 & $D_3/\mathbb Z_3$ &
$(0,1,2,1,2)$ &
\includegraphics[width=0.43\linewidth]{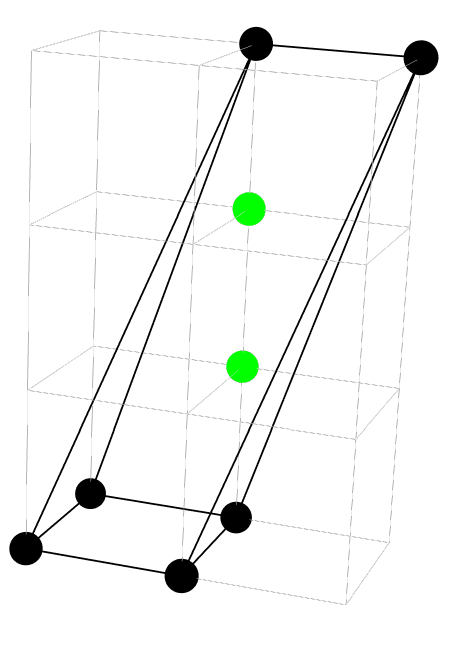}
\\
\hline
3 & $D_3/\mathbb Z_3$ &
$(1,1,1,0,0)$ &
\includegraphics[width=0.38\linewidth]{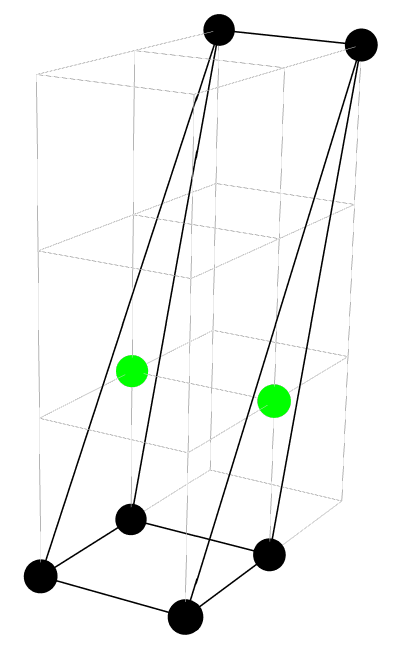}
\\
\hline
3 & $D_3/\mathbb Z_3$ &
$(1,1,1,1,2)$ &
\includegraphics[width=0.55\linewidth]{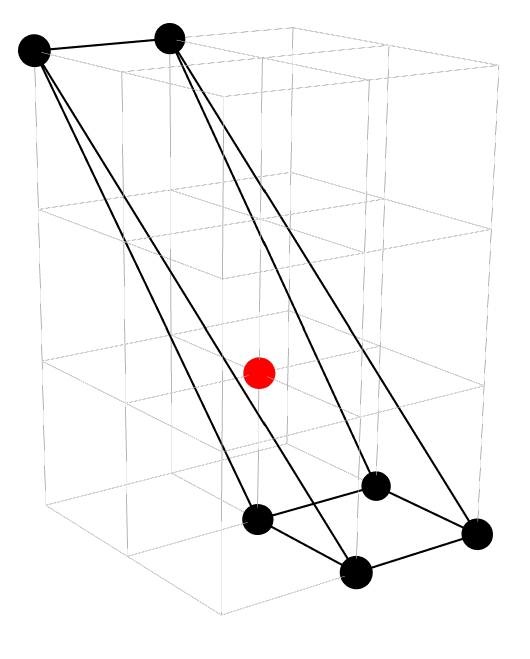}
\\
\hline
\end{tabular}
\caption{
Distinct abelian orbifolds
of $D_3$
for $n=1,2,3$
with the corresponding orbifold actions and toric diagrams.
}
\label{tab_d3_orbifolds}
\end{table}
%------------------------------------------------------------------------------------------

Using Burnside's lemma and the conjugacy classes in \tref{tab_d3_conj},
the sequence counting the number of distinct abelian orbifolds of the form $D_3/\Gamma$
at order $n=|\Gamma|$
takes the following form,
\beal{es50a40}
\mathsf{g}_{D_3}
&
=
&
\frac{1}{12}
\Big(
\mathsf{g}_{x_1^{5}}
+ 2 \mathsf{g}_{x_1^{2} x_3}
+ 3 \mathsf{g}_{x_1 x_2^{2}}
+ \mathsf{g}_{x_1^{3} x_2}^{(a)}
+ 3 \mathsf{g}_{x_1^{3} x_2}^{(b)}
+ 2 \mathsf{g}_{x_2 x_3}
\Big)
~,~
\eea
where $\mathsf{g}_{[g]}(n) = |F_n(g)|$.
The subsequences $\mathsf{g}_{[g]}$ corresponding to the conjugacy classes [g]
expressed in terms of convolutions of standard multiplicative sequences are summarized in
\tref{tab_d3_convolutions}.
The sequence
enumerating distinct abelian orbifolds of $D_3$ is shown in \tref{tab_d3_counts} and 
also appears in a different context in \cite{OEIS_A300783}.

The counting of distinct abelian orbifolds of $D_3$
from sublattice counting 
matches the identification of distinct and consistent orbifold actions for abelian orbifolds of $D_3$
following the consistency conditions of orbifold actions discussed in section \sref{sec033}.
\tref{tab_d3_orbifolds} summarizes the first few distinct abelian orbifolds of the form $D_3/\Gamma$
with their corresponding inequivalent orbifold actions. 
\\

%=================================================================
\section{Examples \label{sec04}}

%=================================================================
\subsection{Abelian Orbifolds of $Q^{1,1,1}$ \label{sec04sub}}

%=================================================================
\subsubsection{$Q^{1,1,1}/ \mathbb{Z}_2 \ (1,1,1,1,1,1,1,1)$ \label{sec041}}

%-------------------
\begin{figure}[H]
    \centering
    \includegraphics[width=0.3\textwidth]{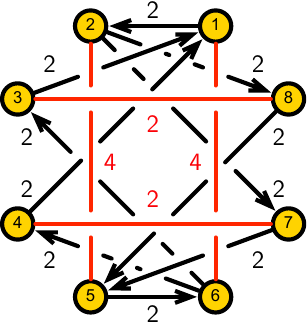}
    \caption{
    The quiver for the $Q^{1,1,1}/\mathbb{Z}_2 \ (1,1,1,1,1,1,1,1)$ model.
    \label{fig_411_quiver}
    }
\end{figure}
%-------------------

The $J$- and $E$-terms for the brane brick model corresponding to the abelian orbifold of the form
$Q^{1,1,1}/ \mathbb{Z}_2 \ (1,1,1,1,1,1,1,1)$ are as follows, 
\beal{es04a01}
\resizebox{0.95\textwidth}{!}{$
\begin{array}{rclccclcc}
&& &J& && &E& \\
\Lambda^{(1)}_{61}&:&~~ &D_{12} \cdot Z_{28} \cdot Y_{85} \cdot X_{56} - X_{12} \cdot Z_{27} \cdot D_{75} \cdot D_{56}& ~~&&~~ &X_{63} \cdot Y_{31} - X_{64} \cdot D_{41}& \\
\Lambda^{(1)}_{25}&:&~~ &D_{56} \cdot Z_{64} \cdot Y_{41} \cdot X_{12} - X_{56} \cdot Z_{63} \cdot D_{31} \cdot D_{12}& ~~&&~~ &X_{27} \cdot Y_{75} - X_{28} \cdot D_{85}& \\
\Lambda^{(2)}_{61}&:&~~ &X_{12} \cdot X_{28} \cdot Y_{85} \cdot D_{56} - D_{12} \cdot Z_{27} \cdot Y_{75} \cdot X_{56}& ~~&&~~ &X_{63} \cdot D_{31} - Z_{64} \cdot D_{41}& \\
\Lambda^{(2)}_{25}&:&~~ &X_{56} \cdot X_{64} \cdot Y_{41} \cdot D_{12} - D_{56} \cdot Z_{63} \cdot Y_{31} \cdot X_{12}& ~~&&~~ &X_{27} \cdot D_{75} - Z_{28} \cdot D_{85}& \\
\Lambda^{(3)}_{61}&:&~~ &D_{12} \cdot Z_{28} \cdot D_{85} \cdot X_{56} - X_{12} \cdot X_{27} \cdot D_{75} \cdot D_{56}& ~~&&~~ &X_{64} \cdot Y_{41} - Z_{63} \cdot Y_{31}& \\
\Lambda^{(3)}_{25}&:&~~ &D_{56} \cdot Z_{64} \cdot D_{41} \cdot X_{12} - X_{56} \cdot X_{63} \cdot D_{31} \cdot D_{12}& ~~&&~~ &X_{28} \cdot Y_{85} - Z_{27} \cdot Y_{75}& \\
\Lambda^{(4)}_{61}&:&~~ &D_{12} \cdot X_{27} \cdot Y_{75} \cdot X_{56} - X_{12} \cdot X_{28} \cdot D_{85} \cdot D_{56}& ~~&&~~ &Z_{63} \cdot D_{31} - Z_{64} \cdot Y_{41}& \\
\Lambda^{(4)}_{25}&:&~~ &D_{56} \cdot X_{63} \cdot Y_{31} \cdot X_{12} - X_{56} \cdot X_{64} \cdot D_{41} \cdot D_{12}& ~~&&~~ &Z_{27} \cdot D_{75} - Z_{28} \cdot Y_{85}& \\
\Lambda^{(5)}_{83}&:&~~ &D_{31} \cdot X_{12} \cdot X_{28} - Y_{31} \cdot X_{12} \cdot Z_{28}& ~~&&~~ &D_{85} \cdot D_{56} \cdot Z_{63} - Y_{85} \cdot D_{56} \cdot X_{63}& \\
\Lambda^{(5)}_{47}&:&~~ &D_{75} \cdot X_{56} \cdot X_{64} - Y_{75} \cdot X_{56} \cdot Z_{64}& ~~&&~~ &D_{41} \cdot D_{12} \cdot Z_{27} - Y_{41} \cdot D_{12} \cdot X_{27}& \\
\Lambda^{(6)}_{74}&:&~~ &Y_{41} \cdot X_{12} \cdot X_{27} - D_{41} \cdot X_{12} \cdot Z_{27}& ~~&&~~ &D_{75} \cdot D_{56} \cdot X_{64} - Y_{75} \cdot D_{56} \cdot Z_{64}& \\
\Lambda^{(6)}_{38}&:&~~ &Y_{85} \cdot X_{56} \cdot X_{63} - D_{85} \cdot X_{56} \cdot Z_{63}& ~~&&~~ &D_{31} \cdot D_{12} \cdot X_{28} - Y_{31} \cdot D_{12} \cdot Z_{28}&
\end{array}
$}
~.~
\nn\\
\eea
The corresponding quiver diagram is shown in \fref{fig_411_quiver}.
The $J$- and $E$-terms
come from \eqref{es03b01} 
using the following relabelling of indices, 
\beal{es04a02}
&
[1,0] \rightarrow 1~,~
[2,0] \rightarrow 2~,~  
[3,0] \rightarrow 3~,~ 
[4,0] \rightarrow 4~,~
&
\nn\\
&
[1,1] \rightarrow 5~,~ 
[2,1] \rightarrow 6~,~ 
[3,1] \rightarrow 7~,~ 
[4,1] \rightarrow 8~.~
&
\eea

\begin{figure}[httt!]
    \centering
    \includegraphics[width=0.35\textwidth]{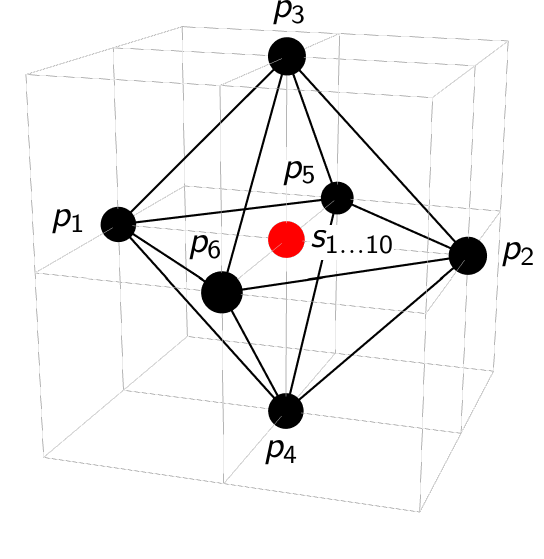}
    \caption{
    The toric diagram for the $Q^{1,1,1}/\mathbb{Z}_2 \ (1,1,1,1,1,1,1,1)$ model.
    \label{fig_411_toric}
    }
\end{figure}

Using the forward algorithm for brane brick models, 
we obtain the
$P$-matrix, which takes the form,
\beal{es04a03}
P=
\left(
\ba{c|cccccc|cccccccccc}
&p_{1} & p_{2} & p_{3} & p_{4} & p_{5} & p_{6} & s_{1} & s_{2} & s_{3} 
& s_{4} & s_{5} & s_{6} & s_{7} & s_{8} & s_{9} & s_{10} \\
\hline
 D_{12} & 0 & 1 & 0 & 0 & 0 & 0 & 0 & 0 & 0 & 1 & 0 & 0 & 0 & 0 & 0 &
0 \\
 D_{31} & 0 & 0 & 0 & 0 & 1 & 0 & 0 & 0 & 0 & 0 & 0 & 0 & 0 & 1 & 0 &
1 \\
 D_{41} & 0 & 0 & 0 & 1 & 0 & 0 & 0 & 0 & 0 & 0 & 0 & 0 & 1 & 1 & 0 &
0 \\
 D_{56} & 0 & 1 & 0 & 0 & 0 & 0 & 1 & 0 & 0 & 0 & 0 & 0 & 0 & 0 & 0 &
0 \\
 D_{75} & 0 & 0 & 0 & 0 & 1 & 0 & 0 & 0 & 1 & 0 & 0 & 1 & 0 & 0 & 0 &
0 \\
 D_{85} & 0 & 0 & 0 & 1 & 0 & 0 & 0 & 1 & 1 & 0 & 0 & 0 & 0 & 0 & 0 &
0 \\
 X_{12} & 1 & 0 & 0 & 0 & 0 & 0 & 0 & 0 & 0 & 1 & 0 & 0 & 0 & 0 & 0 &
0 \\
 X_{27} & 0 & 0 & 0 & 1 & 0 & 0 & 0 & 1 & 0 & 0 & 1 & 0 & 0 & 0 & 0 &
0 \\
 X_{28} & 0 & 0 & 0 & 0 & 0 & 1 & 0 & 0 & 0 & 0 & 1 & 1 & 0 & 0 & 0 &
0 \\
 X_{56} & 1 & 0 & 0 & 0 & 0 & 0 & 1 & 0 & 0 & 0 & 0 & 0 & 0 & 0 & 0 &
0 \\
 X_{63} & 0 & 0 & 0 & 1 & 0 & 0 & 0 & 0 & 0 & 0 & 0 & 0 & 1 & 0 & 1 &
0 \\
 X_{64} & 0 & 0 & 0 & 0 & 0 & 1 & 0 & 0 & 0 & 0 & 0 & 0 & 0 & 0 & 1 &
1 \\
 Y_{31} & 0 & 0 & 0 & 0 & 0 & 1 & 0 & 0 & 0 & 0 & 0 & 0 & 0 & 1 & 0 &
1 \\
 Y_{41} & 0 & 0 & 1 & 0 & 0 & 0 & 0 & 0 & 0 & 0 & 0 & 0 & 1 & 1 & 0 &
0 \\
 Y_{75} & 0 & 0 & 0 & 0 & 0 & 1 & 0 & 0 & 1 & 0 & 0 & 1 & 0 & 0 & 0 &
0 \\
 Y_{85} & 0 & 0 & 1 & 0 & 0 & 0 & 0 & 1 & 1 & 0 & 0 & 0 & 0 & 0 & 0 &
0 \\
 Z_{27} & 0 & 0 & 1 & 0 & 0 & 0 & 0 & 1 & 0 & 0 & 1 & 0 & 0 & 0 & 0 &
0 \\
 Z_{28} & 0 & 0 & 0 & 0 & 1 & 0 & 0 & 0 & 0 & 0 & 1 & 1 & 0 & 0 & 0 &
0 \\
 Z_{63} & 0 & 0 & 1 & 0 & 0 & 0 & 0 & 0 & 0 & 0 & 0 & 0 & 1 & 0 & 1 &
0 \\
 Z_{64} & 0 & 0 & 0 & 0 & 1 & 0 & 0 & 0 & 0 & 0 & 0 & 0 & 0 & 0 & 1 &
1 \\
\end{array}
\right)~.~
\eea
The $U(1)$ charges on the GLSM fields
corresponding to the $J$- and $E$-terms
are given by the following charge matrix, 
\beal{es04a04}
Q_{JE}=
\left(
\ba{cccccc|cccccccccc}
p_{1} & p_{2} & p_{3} & p_{4} & p_{5} & p_{6} & s_{1} & s_{2} & s_{3} 
& s_{4} & s_{5} & s_{6} & s_{7} & s_{8} & s_{9} & s_{10} \\
\hline
 0 & 0 & 0 & 0 & -1 & -1 & 0 & -1 & 1 & 0 & 1 & 0 & 0 & 0 & 0 & 1 \\
 0 & 0 & -1 & -1 & -1 & -1 & 0 & 0 & 1 & 0 & 1 & 0 & 0 & 1 & 1 & 0 \\
 0 & 0 & -1 & -1 & 0 & 0 & 0 & 1 & 0 & 0 & 0 & 0 & 1 & 0 & 0 & 0 \\
 0 & 0 & 0 & 0 & 0 & 0 & 0 & 1 & -1 & 0 & -1 & 1 & 0 & 0 & 0 & 0 \\
 -1 & -1 & 0 & 0 & 0 & 0 & 1 & 0 & 0 & 1 & 0 & 0 & 0 & 0 & 0 & 0 \\
\ea
\right) ~.~
\eea
The $D$-term charge matrix takes the following form, 
\beal{es04a05}
Q_{D}=
\left(
\begin{array}{cccccc|cccccccccc}
p_{1} & p_{2} & p_{3} & p_{4} & p_{5} & p_{6} & s_{1} & s_{2} & s_{3} 
& s_{4} & s_{5} & s_{6} & s_{7} & s_{8} & s_{9} & s_{10} \\
\hline
 -1 & -1 & 0 & 0 & 0 & 0 & 1 & 0 & 0 & 0 & 0 & 0 & 0 & 1 & 0 & 0 \\
 1 & 1 & 0 & 0 & 0 & 0 & -1 & 0 & 0 & 0 & -1 & 0 & 0 & 0 & 0 & 0 \\
 0 & 0 & 1 & 1 & 0 & 0 & 0 & -1 & 0 & 0 & 0 & 0 & 0 & -1 & 0 & 0 \\
 0 & 0 & 0 & 0 & 1 & 1 & 0 & 1 & -1 & 0 & -1 & 0 & 0 & -1 & 0 & 0 \\
 0 & 0 & 0 & 0 & 0 & 0 & -1 & 0 & 1 & 0 & 0 & 0 & 0 & 0 & 0 & 0 \\
 0 & 0 & -1 & -1 & -1 & -1 & 1 & 0 & 1 & 0 & 1 & 0 & 0 & 1 & 0 & 0 \\
 0 & 0 & 0 & 0 & 0 & 0 & 0 & 1 & -1 & 0 & 0 & 0 & 0 & 0 & 0 & 0 \\
\end{array}
\right)~.~
\eea
The toric diagram of 
the abelian orbifold of the form $Q^{1,1,1}/\mathbb{Z}_2 \ (1,1,1,1,1,1,1,1)$
is shown in \fref{fig_411_toric} and is given by, 
\beal{es04a06}
G_{t}=
\left(
\begin{array}{cccccc|cccccccccc}
p_{1} & p_{2} & p_{3} & p_{4} & p_{5} & p_{6} & s_{1} & s_{2} & s_{3} 
& s_{4} & s_{5} & s_{6} & s_{7} & s_{8} & s_{9} & s_{10} \\
\hline
 -1 & 1 & 0 & 0 & 0 & 0 & 0 & 0 & 0 & 0 & 0 & 0 & 0 & 0 & 0 & 0 \\
 0 & 0 & 0 & 0 & 1 & -1 & 0 & 0 & 0 & 0 & 0 & 0 & 0 & 0 & 0 & 0 \\
 0 & 0 & 1 & -1 & 0 & 0 & 0 & 0 & 0 & 0 & 0 & 0 & 0 & 0 & 0 & 0 \\
 \hline
 1 & 1 & 1 & 1 & 1 & 1 & 1 & 1 & 1 & 1 & 1 & 1 & 1 & 1 & 1 & 1 \\
\end{array}
\right)~.~
\eea
In the literature, the abelian orbifold of the form $Q^{1,1,1}/\mathbb{Z}_2~(1,1,1,1,1,1,1,1)$
is also simply referred to as $Q^{1,1,1}/\mathbb{Z}_2$
and its toric diagram is one of the reflexive polytopes in $\mathbb{Z}^3$ \cite{batyrev1993dual, borisov1993towards, Davey:2011mz, Franco:2022gvl}.

%-------------------
%-------------------

From the $Q_{JE}$ and $Q_D$ charge matrices, 
we see that the global symmetry of the brane brick model is enhanced to the following form, 
\beal{es04a07}
SU(2)_x \times SU(2)_y \times SU(2)_z \times U(1)_R 
~.~
\eea
Here, the factors $SU(2)_x \times SU(2)_y \times SU(2)_z$
form the mesonic flavor symmetry,
and \tref{tab_04a01} summarizes the charges under this symmetry on the extremal GLSM fields $p_a$.

%-------------------
\begin{table}[H]
\centering
\begin{tabular}{|c|c|c|c|l|}
\hline
\; & $SU(2)_x$ & $SU(2)_y$ & $SU(2)_z$ & fugacity \\
\hline
$p_1$ & $+1$ & $0$ & $0$ & $t_1=x t$ \\
$p_2$ & $-1$ & $0$ & $0$  & $t_2=x^{-1} t$\\
$p_3$ & $0$ & $+1$ & $0$  & $t_3=y t$\\
$p_4$ & $0$ & $-1$ & $0$  & $t_4=y^{-1} t$ \\
$p_5$ & $0$ & $0$ & $+1$  & $t_5=z t$ \\
$p_6$ & $0$ & $0$ & $-1$ & $t_6=z^{-1} t$ \\
\hline
\end{tabular}
\caption{
Mesonic flavor symmetry of the $Q^{1,1,1}/\mathbb{Z}_2 \ (1,1,1,1,1,1,1,1)$ model
and charges on the extremal GLSM fields $p_a$. 
Here, the fugacity $t$ counts the degree in extremal GLSM fields $p_a$. 
\label{tab_04a01}
}
\end{table}
%-------------------

The Hilbert series of the mesonic moduli space of the brane brick model for $Q^{1,1,1}/\mathbb{Z}_2 \ (1,1,1,1,1,1,1,1)$
takes the following form,
\beal{es04a08}
&&
g(t_a; \mathcal{M}^{mes})=
\frac{
P(t_a; \mathcal{M}^{mes}) \left(1 + t_1 t_2 t_3 t_4 t_5 t_6\right)
}{
\left(1-t_1^2 t_3^2 t_5^2\right)
\left(1-t_2^2 t_3^2 t_5^2\right) 
\left(1-t_1^2 t_4^2 t_5^2\right)  \left(1-t_2^2 t_4^2 t_5^2\right)}
\nn\\
&&
\hspace{1cm}
\times
\frac{1}{ \left(1-t_1^2 t_3^2 t_6^2\right) 
\left(1 - t_2^2 t_3^2 t_6^2\right)  \left(1-t_1^2 t_4^2 t_6^2\right) \left(1-t_2^2 t_4^2 t_6^2\right)}
~,~
\eea
where $t_a$ is the fugacity associated to the extremal GLSM field $p_a$.
The remaining numerator factor $P(t_a ; \mathcal{M}^{mes})$ in \eref{es04a08} is presented in appendix \sref{appCa}. 
When unrefined by setting $t_a=t$, the Hilbert series takes the following form,
\beal{es04a09}
g(t; \mathcal{M}^{mes})=\frac{1+23t^6 +23t^{12} + t^{18}}{(1-t^6)^4}
~,~
\eea
where the palindromic numerator indicates that the mesonic moduli space is Calabi-Yau. 

The Hilbert series can be expressed in terms of characters of irreducible representations 
of the mesonic flavor symmetry $SU(2)_x \times SU(2)_y \times SU(2)_z$
using the fugacity assignment summarized in \tref{tab_04a01}.
The corresponding highest weight generating function is given by,
\beal{es04a10}
h(\mu,\nu,\lambda , t; \mathcal{M}^{mes})=\frac{1}{1-\mu^2 \nu^2 \lambda^2 t^6}
~,~
\eea
where $\mu^m \nu^n \lambda^l$ counts the character of the form $[m]_x [n]_y [l]_z$
corresponding to irreducible representations of $SU(2)_x \times SU(2)_y \times SU(2)_z$ with highest weight $(m)_x (n)_y (l)_z$.

%-------------------
\begin{table}[htt!!!]
\centering
\begin{tabular}{|c|c|ccc|}
\hline
PL term & generator & $SU(2)_x$ & $SU(2)_y$ & $SU(2)_z$  
\\
\hline 
\multirow{27}{*}{$+[2]_x [2]_y [2]_z t^6$} 
& $p_2^2 p_4^2 p_6^2 ~s$ & $-2$ & $-2$ & $-2$   \\

& $p_2^2 p_4^2 p_5 p_6 ~s$ & $-2$ & $-2$ & $0$   \\

& $p_2^2 p_4^2 p_5^2 ~s$ & $-2$ & $-2$ & $2$   \\

& $p_2^2 p_3 p_4 p_6^2 ~s$ & $-2$ & $0$ & $-2$   \\

& $p_2^2 p_3 p_4 p_5 p_6 ~s$ & $-2$ & $0$ & $0$  \\

& $p_2^2 p_3 p_4 p_5^2 ~s$ & $-2$ & $0$ & $2$   \\

& $p_2^2 p_3^2 p_6^2 ~s$ & $-2$ & $2$ & $-2$   \\

& $p_2^2 p_3^2 p_5 p_6 ~s$ & $-2$ & $2$ & $0$   \\

& $p_2^2 p_3^2 p_5^2 ~s$ & $-2$ & $2$ & $2$  \\

& $p_1 p_2 p_4^2 p_6^2 ~s$ & $0$ & $-2$ & $-2$   \\

& $p_1 p_2 p_4^2 p_5 p_6 ~s$ & $0$ & $-2$ & $0$   \\

& $p_1 p_2 p_4^2 p_5^2 ~s$ & $0$ & $-2$ & $2$   \\

& $p_1 p_2 p_3 p_4 p_6^2 ~s$ & $0$ & $0$ & $-2$  \\

& $p_1 p_2 p_3 p_4 p_5 p_6 ~s$ & $0$ & $0$ & $0$ \\

& $p_1 p_2 p_3 p_4 p_5^2 ~s$ & $0$ & $0$ & $2$  \\

& $p_1 p_2 p_3^2 p_6^2 ~s$ & $0$ & $2$ & $-2$   \\

& $p_1 p_2 p_3^2 p_5 p_6 ~s$ & $0$ & $2$ & $0$  \\

& $p_1 p_2 p_3^2 p_5^2 ~s$ & $0$ & $2$ & $2$   \\

& $p_1^2 p_4^2 p_6^2 ~s$ & $2$ & $-2$ & $-2$  \\

& $p_1^2 p_4^2 p_5 p_6 ~s$ & $2$ & $-2$ & $0$   \\

& $p_1^2 p_4^2 p_5^2 ~s$ & $2$ & $-2$ & $2$   \\

& $p_1^2 p_3 p_4 p_6^2 ~s$ & $2$ & $0$ & $-2$   \\

& $p_1^2 p_3 p_4 p_5 p_6 ~s$ & $2$ & $0$ & $0$   \\

& $p_1^2 p_3 p_4 p_5^2 ~s$ & $2$ & $0$ & $2$   \\

& $p_1^2 p_3^2 p_6^2 ~s$ & $2$ & $2$ & $-2$   \\

& $p_1^2 p_3^2 p_5 p_6 ~s$ & $2$ & $2$ & $0$   \\

& $p_1^2 p_3^2 p_5^2 ~s$ & $2$ & $2$ & $2$   \\
\hline
\end{tabular}
\caption{
Generators of the $Q^{1,1,1}/\mathbb{Z}_2 \ (1,1,1,1,1,1,1,1)$ model in terms of GLSM fields and their corresponding mesonic flavor charges. 
Here, we denote $s=\prod_{i=1}^{10}s_i$. 
\label{tab_04a02}
}
\end{table}
%-------------------

The plethystic logarithm of the refined Hilbert series with the corresponding highest weight generating function given in \eref{es04a10}
takes the following form,
\beal{es04a12}
&&
\text{PL}[g(x,y,z,t; \mathcal{M}^{mes})]=
[2]_x [2]_y [2]_z t^6
-( 1+ [4]_x  + [4]_y +   [4]_z 
\nn\\
&&
\hspace{1cm}
+ [2]_x [2]_y  + [2]_x  [2]_z +  [2]_y [2]_z + [4]_x [2]_y [2]_z + [2]_x [4]_y [2]_z + [2]_x [2]_y [4]_z 
\nn\\
&&
\hspace{1cm}
+ [4]_x [4]_y  + [4]_x  [4]_z +  [4]_y [4]_z) t^{12}
+\dots
~.~
\eea
We can see from the infinite expansion of the plethystic logarithm that the mesonic moduli space of the
$Q^{1,1,1}/\mathbb{Z}_2 \ (1,1,1,1,1,1,1,1)$ model is not a complete intersection.
The first positive terms of the expansion correspond to the generators of the mesonic moduli space,
which are summarized with their mesonic flavor charges in \tref{tab_04a02}. 

We note that the abelian orbifold $Q^{1,1,1}/\mathbb{Z}_2 \ (1,1,1,1,1,1,1,1)$
is part of a family of abelian orbifolds of $Q^{1,1,1}$
of the form $Q^{1,1,1}/\mathbb{Z}_n \ (1,1,1,1,-1,-1,-1,-1)$, whose 
toric diagram has extremal vertices with coordinates given by, 
\beal{es04a20}
&
(-1,0,0)~,~
(0,-1,0)~,~
(0,0,1)~,~
&
\nn\\
&
(n-2,n-1,0)~,~
(n-1,n-2,0)~,~
(n-2,n-2,-1)~.~
&
\eea
In section \sref{sec043}, 
we summarize the brane brick model for the abelian orbifold of the form $Q^{1,1,1}/\mathbb{Z}_3$ with orbifold action $(1,1,1,1,2,2,2,2)$,
which is also part of this family of abelian orbifolds of $Q^{1,1,1}$.
\\

%=================================================================
\subsubsection{$Q^{1,1,1}/ \mathbb{Z}_2 \ (0,0,1,1,1,1,0,0)$ \label{sec042}}

%-------------------
\begin{figure}[H]
    \centering
    \includegraphics[width=0.3\textwidth]{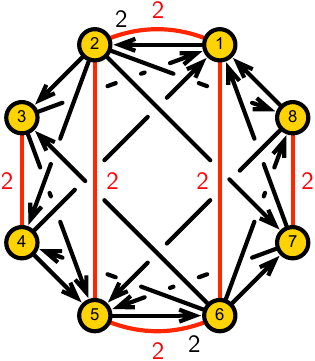}
    \caption{
    The quiver for the $Q^{1,1,1}/ \mathbb{Z}_2 \ (0,0,1,1,1,1,0,0)$ model.
    \label{fig_412_quiver}
    }
\end{figure}
%-------------------

The $J$- and $E$-terms for the abelian orbifold of the form 
$Q^{1,1,1}/ \mathbb{Z}_2 \ (0,0,1,1,1,1,0,0)$ are as follows, 
\beal{es04b01}
\resizebox{0.95\textwidth}{!}{$
\begin{array}{rclccclcccc}
&& && &J& && &E& \\
&&\Lambda^{(1)}_{2 1} &:&~~ & D_{1 2} \cdot Z_{2 8} \cdot Y_{8 1} \cdot X_{1 2} - X_{1 2} \cdot Z_{2 3} \cdot D_{3 1} \cdot D_{1 2}& ~~&&~~ & X_{2 7} \cdot Y_{7 1} - X_{2 4} \cdot D_{4 1}& \\ 
 &&\Lambda^{(1)}_{6 5} &:&~~ & D_{5 6} \cdot Z_{6 4} \cdot Y_{4 5} \cdot X_{5 6} - X_{5 6} \cdot Z_{6 7} \cdot D_{7 5} \cdot D_{5 6}& ~~&&~~ & X_{6 3} \cdot Y_{3 5} - X_{6 8} \cdot D_{8 5}& \\ 
 &&\Lambda^{(2)}_{6 1} &:&~~ & X_{1 2} \cdot X_{2 4} \cdot Y_{4 5} \cdot D_{5 6} - D_{1 2} \cdot Z_{2 3} \cdot Y_{3 5} \cdot X_{5 6}& ~~&&~~ & X_{6 3} \cdot D_{3 1} - Z_{6 4} \cdot D_{4 1}& \\ 
 &&\Lambda^{(2)}_{2 5} &:&~~ & X_{5 6} \cdot X_{6 8} \cdot Y_{8 1} \cdot D_{1 2} - D_{5 6} \cdot Z_{6 7} \cdot Y_{7 1} \cdot X_{1 2}& ~~&&~~ & X_{2 7} \cdot D_{7 5} - Z_{2 8} \cdot D_{8 5}& \\ 
 &&\Lambda^{(3)}_{6 1} &:&~~ & D_{1 2} \cdot Z_{2 8} \cdot D_{8 5} \cdot X_{5 6} - X_{1 2} \cdot X_{2 7} \cdot D_{7 5} \cdot D_{5 6}& ~~&&~~ & X_{6 8} \cdot Y_{8 1} - Z_{6 7} \cdot Y_{7 1}& \\ 
 &&\Lambda^{(3)}_{2 5} &:&~~ & D_{5 6} \cdot Z_{6 4} \cdot D_{4 1} \cdot X_{1 2} - X_{5 6} \cdot X_{6 3} \cdot D_{3 1} \cdot D_{1 2}& ~~&&~~ & X_{2 4} \cdot Y_{4 5} - Z_{2 3} \cdot Y_{3 5}& \\ 
 &&\Lambda^{(4)}_{2 1} &:&~~ & D_{1 2} \cdot X_{2 7} \cdot Y_{7 1} \cdot X_{1 2} - X_{1 2} \cdot X_{2 4} \cdot D_{4 1} \cdot D_{1 2}& ~~&&~~ & Z_{2 3} \cdot D_{3 1} - Z_{2 8} \cdot Y_{8 1}& \\ 
 &&\Lambda^{(4)}_{6 5} &:&~~ & D_{5 6} \cdot X_{6 3} \cdot Y_{3 5} \cdot X_{5 6} - X_{5 6} \cdot X_{6 8} \cdot D_{8 5} \cdot D_{5 6}& ~~&&~~ & Z_{6 7} \cdot D_{7 5} - Z_{6 4} \cdot Y_{4 5}& \\ 
 &&\Lambda^{(5)}_{4 3} &:&~~ & D_{3 1} \cdot X_{1 2} \cdot X_{2 4} - Y_{3 5} \cdot X_{5 6} \cdot Z_{6 4}& ~~&&~~ & D_{4 1} \cdot D_{1 2} \cdot Z_{2 3} - Y_{4 5} \cdot D_{5 6} \cdot X_{6 3}& \\ 
 &&\Lambda^{(5)}_{8 7} &:&~~ & D_{7 5} \cdot X_{5 6} \cdot X_{6 8} - Y_{7 1} \cdot X_{1 2} \cdot Z_{2 8}& ~~&&~~ & D_{8 5} \cdot D_{5 6} \cdot Z_{6 7} - Y_{8 1} \cdot D_{1 2} \cdot X_{2 7}& \\ 
 &&\Lambda^{(6)}_{3 4} &:&~~ & Y_{4 5} \cdot X_{5 6} \cdot X_{6 3} - D_{4 1} \cdot X_{1 2} \cdot Z_{2 3}& ~~&&~~ & D_{3 1} \cdot D_{1 2} \cdot X_{2 4} - Y_{3 5} \cdot D_{5 6} \cdot Z_{6 4}& \\ 
 &&\Lambda^{(6)}_{7 8} &:&~~ & Y_{8 1} \cdot X_{1 2} \cdot X_{2 7} - D_{8 5} \cdot X_{5 6} \cdot Z_{6 7}& ~~&&~~ & D_{7 5} \cdot D_{5 6} \cdot X_{6 8} - Y_{7 1} \cdot D_{1 2} \cdot Z_{2 8}&
\end{array}
$}
~.~
\nn\\
\eea
The corresponding quiver is shown in \fref{fig_412_quiver}. 
The $J$- and $E$-terms
are obtained from the general formula in \eqref{es03b01} 
with the additional relabelling of indices as follows, 
\beal{es04b02}
&
[1,0] \rightarrow 1~,~
[2,0] \rightarrow 2~,~  
[3,0] \rightarrow 3~,~ 
[4,0] \rightarrow 4~,~
&
\nn\\
&
[1,1] \rightarrow 5~,~ 
[2,1] \rightarrow 6~,~ 
[3,1] \rightarrow 7~,~ 
[4,1] \rightarrow 8~.~
&
\eea

\begin{figure}[H]
\centering
\includegraphics[width=0.3\textwidth]{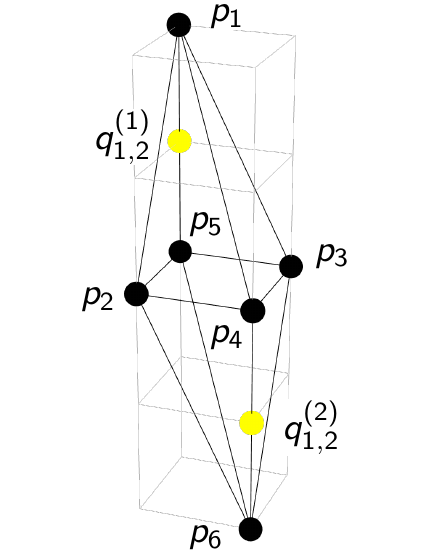}
\caption{
The toric diagram for the $Q^{1,1,1}/\mathbb{Z}_2 \ (0,0,1,1,1,1,0,0)$ model.
\label{fig_412_toric}
}
\end{figure}

The $P$-matrix is obtained using the forward algorithm.
It takes the following form,
\beal{es04b03}
&&
P=
\resizebox{0.75\textwidth}{!}{$
\left(
\begin{array}{c|cccccc|cc|cc|ccc|ccc|ccc}
& p_{1} & p_{2} & p_{3} & p_{4} & p_{5} & p_{6}
& q^{(1)}_{1} & q^{(1)}_{2} & q^{(2)}_{1} & q^{(2)}_{2}
& o^{(1)}_{1} & \cdots & o^{(1)}_{6}
& o^{(2)}_{1} & \cdots & o^{(2)}_{6}
& o^{(3)}_{1} & \cdots & o^{(3)}_{16} \\
\hline
 D_{12} & 0 & 0 & 1 & 0 & 0 & 0 & 0 & 0 & 0 & 0 & 0 & \cdots & 1 & 0 & \cdots & 1 & 0 & \cdots & 1 \\
 D_{31} & 0 & 0 & 0 & 1 & 0 & 0 & 0 & 0 & 0 & 1 & 0 & \cdots & 0 & 0 & \cdots & 0 & 0 & \cdots & 0 \\
 D_{41} & 1 & 0 & 0 & 0 & 0 & 0 & 0 & 1 & 0 & 0 & 0 & \cdots & 0 & 0 & \cdots & 0 & 0 & \cdots & 0 \\
 D_{56} & 0 & 0 & 1 & 0 & 0 & 0 & 0 & 0 & 0 & 0 & 0 & \cdots & 0 & 0 & \cdots & 0 & 0 & \cdots & 0 \\
 D_{75} & 0 & 0 & 0 & 1 & 0 & 0 & 0 & 0 & 1 & 0 & 0 & \cdots & 0 & 1 & \cdots & 1 & 0 & \cdots & 1 \\
 D_{85} & 1 & 0 & 0 & 0 & 0 & 0 & 1 & 0 & 0 & 0 & 0 & \cdots & 0 & 1 & \cdots & 1 & 0 & \cdots & 1 \\
 X_{12} & 0 & 1 & 0 & 0 & 0 & 0 & 0 & 0 & 0 & 0 & 0 & \cdots & 1 & 0 & \cdots & 1 & 0 & \cdots & 1 \\
 X_{24} & 0 & 0 & 0 & 0 & 1 & 0 & 1 & 0 & 0 & 0 & 1 & \cdots & 0 & 1 & \cdots & 0 & 1 & \cdots & 0 \\
 X_{27} & 1 & 0 & 0 & 0 & 0 & 0 & 1 & 0 & 0 & 0 & 0 & \cdots & 0 & 1 & \cdots & 0 & 1 & \cdots & 0 \\
 X_{56} & 0 & 1 & 0 & 0 & 0 & 0 & 0 & 0 & 0 & 0 & 0 & \cdots & 0 & 0 & \cdots & 0 & 0 & \cdots & 0 \\
 X_{63} & 1 & 0 & 0 & 0 & 0 & 0 & 0 & 1 & 0 & 0 & 1 & \cdots & 0 & 1 & \cdots & 1 & 1 & \cdots & 0 \\
 X_{68} & 0 & 0 & 0 & 0 & 1 & 0 & 0 & 1 & 0 & 0 & 1 & \cdots & 1 & 0 & \cdots & 0 & 1 & \cdots & 0 \\
 Y_{35} & 0 & 0 & 0 & 0 & 1 & 0 & 1 & 0 & 0 & 0 & 0 & \cdots & 1 & 0 & \cdots & 0 & 0 & \cdots & 1 \\
 Y_{45} & 0 & 0 & 0 & 0 & 0 & 1 & 0 & 0 & 1 & 0 & 0 & \cdots & 1 & 0 & \cdots & 0 & 0 & \cdots & 1 \\
 Y_{71} & 0 & 0 & 0 & 0 & 1 & 0 & 0 & 1 & 0 & 0 & 1 & \cdots & 0 & 0 & \cdots & 0 & 0 & \cdots & 0 \\
 Y_{81} & 0 & 0 & 0 & 0 & 0 & 1 & 0 & 0 & 0 & 1 & 1 & \cdots & 0 & 0 & \cdots & 0 & 0 & \cdots & 0 \\
 Z_{23} & 0 & 0 & 0 & 0 & 0 & 1 & 0 & 0 & 1 & 0 & 1 & \cdots & 0 & 1 & \cdots & 0 & 1 & \cdots & 0 \\
 Z_{28} & 0 & 0 & 0 & 1 & 0 & 0 & 0 & 0 & 1 & 0 & 0 & \cdots & 0 & 1 & \cdots & 0 & 1 & \cdots & 0 \\
 Z_{64} & 0 & 0 & 0 & 1 & 0 & 0 & 0 & 0 & 0 & 1 & 1 & \cdots & 0 & 1 & \cdots & 1 & 1 & \cdots & 0 \\
 Z_{67} & 0 & 0 & 0 & 0 & 0 & 1 & 0 & 0 & 0 & 1 & 1 & \cdots & 1 & 0 & \cdots & 0 & 1 & \cdots & 0 \\
\end{array}
\right)
$}
~,~
\eea
where $o_k^{(1)}$, $o_l^{(2)}$ and $o_m^{(3)}$ are extra GLSM fields \cite{Witten:1993yc}.
The $J$- and $E$-term charge matrix is
as follows,
\beal{es04b04}
&&
Q_{JE}=
\resizebox{0.7\textwidth}{!}{$
\left(
\begin{array}{cccccc|cc|cc|ccc|ccc|ccc}
p_{1} & p_{2} & p_{3} & p_{4} & p_{5} & p_{6} & q^{(1)}_{1} & q^{(1)}_{2} & q^{(2)}_{1} & q^{(2)}_{2} & o^{(1)}_{1} & \cdots & o^{(1)}_{6} & o^{(2)}_{1} & \cdots & o^{(2)}_{6} & o^{(3)}_{1} & \cdots & o^{(3)}_{16} \\
\hline
 0 & 1 & 1 & 0 & 0 & 0 & 0 & 0 & 0 & 0 & 0 & \cdots & 0 & 0 & \cdots & 0 & 1 & \cdots & 0 \\
 0 & 0 & 0 & 0 & 0 & 1 & 0 & 0 & -1 & 0 & 0 & \cdots & 0 & 0 & \cdots & 0 & 2 & \cdots & 1 \\
 0 & 0 & 0 & 0 & 0 & 1 & 0 & 0 & -1 & 0 & 0 & \cdots & 0 & 0 & \cdots & 1 & 1 & \cdots & 0 \\
 0 & 0 & 0 & 0 & 0 & 0 & 0 & 0 & 0 & 0 & 0 & \cdots & 1 & 0 & \cdots & 0 & 1 & \cdots & 0 \\
 0 & 0 & 0 & 0 & 0 & 1 & 0 & 0 & 0 & -1 & -1 & \cdots & 0 & 0 & \cdots & 0 & 2 & \cdots & 0 \\
 0 & 0 & 0 & 0 & 0 & 1 & 0 & 0 & 0 & -1 & 0 & \cdots & 0 & 0 & \cdots & 0 & 1 & \cdots & 0 \\
 0 & 0 & 0 & 0 & 0 & 2 & 0 & 0 & -1 & -1 & -1 & \cdots & 0 & 0 & \cdots & 0 & 3 & \cdots & 0 \\
 0 & 0 & 0 & 0 & 0 & 2 & 0 & 0 & -1 & -1 & 0 & \cdots & 0 & 0 & \cdots & 0 & 2 & \cdots & 0 \\
 0 & 0 & 0 & 0 & 0 & 2 & 0 & 0 & -1 & -1 & -1 & \cdots & 0 & 0 & \cdots & 0 & 2 & \cdots & 0 \\
 0 & 0 & 0 & 0 & 0 & 2 & 0 & 0 & -1 & -1 & -1 & \cdots & 0 & 0 & \cdots & 0 & 2 & \cdots & 0 \\
 0 & 0 & 0 & 0 & 0 & 1 & 0 & 0 & 0 & -1 & 0 & \cdots & 0 & 0 & \cdots & 0 & 2 & \cdots & 0 \\
 0 & 0 & 0 & 0 & 0 & 1 & 0 & 0 & 0 & -1 & -1 & \cdots & 0 & 0 & \cdots & 0 & 2 & \cdots & 0 \\
 0 & 0 & 0 & 0 & 0 & 1 & 0 & 0 & 0 & -1 & 0 & \cdots & 0 & 0 & \cdots & 0 & 1 & \cdots & 0 \\
 0 & 0 & 0 & 0 & 0 & 1 & 0 & 0 & 0 & -1 & 0 & \cdots & 0 & 0 & \cdots & 0 & 0 & \cdots & 0 \\
 0 & 0 & 0 & 0 & 0 & 2 & 0 & 0 & -1 & -1 & -1 & \cdots & 0 & 0 & \cdots & 0 & 2 & \cdots & 0 \\
 0 & 0 & 0 & 1 & 0 & 1 & 0 & 0 & -1 & -1 & 0 & \cdots & 0 & 0 & \cdots & 0 & 0 & \cdots & 0 \\
 0 & 0 & 0 & 0 & 0 & 1 & 0 & 0 & 0 & -1 & 0 & \cdots & 0 & 0 & \cdots & 0 & 0 & \cdots & 0 \\
 0 & 0 & 0 & 0 & 0 & 1 & 0 & 0 & -1 & 0 & -1 & \cdots & 0 & 0 & \cdots & 0 & 2 & \cdots & 0 \\
 1 & 0 & 0 & 0 & 0 & 1 & 0 & 0 & 0 & 0 & 0 & \cdots & 0 & 0 & \cdots & 0 & 1 & \cdots & 0 \\
 0 & 0 & 0 & 0 & 0 & 1 & 0 & 1 & 0 & 0 & -1 & \cdots & 0 & 0 & \cdots & 0 & 1 & \cdots & 0 \\
 0 & 0 & 0 & 0 & 0 & 0 & 0 & 0 & 0 & 0 & -1 & \cdots & 0 & 0 & \cdots & 0 & 1 & \cdots & 0 \\
 0 & 0 & 0 & 0 & 0 & 1 & 0 & 0 & -1 & 0 & -1 & \cdots & 0 & 0 & \cdots & 0 & 2 & \cdots & 0 \\
 0 & 0 & 0 & 0 & 0 & 1 & 0 & 0 & -1 & 0 & -1 & \cdots & 0 & 0 & \cdots & 0 & 1 & \cdots & 0 \\
 0 & 0 & 0 & 0 & 0 & 1 & 0 & 0 & -1 & 0 & 0 & \cdots & 0 & 1 & \cdots & 0 & 0 & \cdots & 0 \\
 0 & 0 & 0 & 0 & 0 & 1 & 0 & 0 & -1 & 0 & -1 & \cdots & 0 & 0 & \cdots & 0 & 0 & \cdots & 0 \\
 0 & 0 & 0 & 0 & 0 & 1 & 1 & 0 & 0 & 0 & 0 & \cdots & 0 & 0 & \cdots & 0 & 1 & \cdots & 0 \\
 0 & 0 & 0 & 0 & 1 & 1 & 0 & 0 & 0 & 0 & -1 & \cdots & 0 & 0 & \cdots & 0 & 1 & \cdots & 0 \\
\end{array}
\right)
$}
~,~
\eea
and
the $D$-term charge matrix takes the form,
\beal{es04b05}
&&
Q_{D}=
\resizebox{0.7\textwidth}{!}{$
\left(
\begin{array}{cccccc|cc|cc|ccc|ccc|ccc}
p_{1} & p_{2} & p_{3} & p_{4} & p_{5} & p_{6} & q^{(1)}_{1} & q^{(1)}_{2} & q^{(2)}_{1} & q^{(2)}_{2} & o^{(1)}_{1} & \cdots & o^{(1)}_{6} & o^{(2)}_{1} & \cdots & o^{(2)}_{6} & o^{(3)}_{1} & \cdots & o^{(3)}_{16} \\
\hline
 0 & 0 & 0 & 0 & 0 & -1 & 0 & 0 & 0 & 1 & 1  & \cdots & 0 & 0  & \cdots & 0 & -1 & \cdots  & 0 \\
 0 & 0 & 0 & 0 & 0 & 0 & 0 & 0 & 0 & 0 & 0  & \cdots & 0 & 0  & \cdots & 0 & -1  & \cdots & 0 \\
 0 & 0 & 0 & 0 & 0 & 1 & 0 & 0 & 0 & -1 & 0  & \cdots & 0 & 0  & \cdots & 0 & 1  & \cdots & 0 \\
 0 & 0 & 0 & 0 & 0 & 0 & 0 & 0 & 0 & 0 & 0  & \cdots & 0 & 0  & \cdots & 0 & 1  & \cdots & 0 \\
 0 & 0 & 0 & 0 & 0 & -1 & 0 & 0 & 1 & 0 & 0  & \cdots & 0 & 0  & \cdots & 0 & -1  & \cdots & 0 \\
 0 & 0 & 0 & 0 & 0 & 0 & 0 & 0 & 0 & 0 & 0  & \cdots & 0 & 0  & \cdots & 0 & -1  & \cdots & 0 \\
 0 & 0 & 0 & 0 & 0 & 1 & 0 & 0 & -1 & 0 & -1  & \cdots & 0 & 0  & \cdots & 0 & 1  & \cdots & 0 \\
\end{array}
\right)
$}
~.~
\eea
The toric diagram for the $Q^{1,1,1}/ \mathbb{Z}_2 \ (0,0,1,1,1,1,0,0)$ model shown in \fref{fig_412_toric} is 
given by, 
\beal{es04b06}
&&
G_{t}=
\resizebox{0.7\textwidth}{!}{$
\left(
\begin{array}{cccccc|cc|cc|ccc|ccc|ccc}
p_{1} & p_{2} & p_{3} & p_{4} & p_{5} & p_{6}
& q^{(1)}_{1} & q^{(1)}_{2} & q^{(2)}_{1} & q^{(2)}_{2}
& o^{(1)}_{1} & \cdots & o^{(1)}_{6}
& o^{(2)}_{1} & \cdots & o^{(2)}_{6}
& o^{(3)}_{1} & \cdots & o^{(3)}_{16} \\
\hline
1 & 0 & 1 & 0 & 1 & 0 & 1 & 1 & 0 & 0 & 1 & \cdots & 1 & 1 & \cdots & 1 & 1 & \cdots & 1 \\
0 & 0 & 1 & 1 & 0 & 1 & 0 & 0 & 1 & 1 & 1 & \cdots & 1 & 1 & \cdots & 1 & 1 & \cdots & 1 \\
2 & 0 & 0 & 0 & 0 & -2 & 1 & 1 & -1 & -1 & -1 & \cdots & -1 & 1 & \cdots & 1 & 0 & \cdots & 0 \\
\hline
1 & 1 & 1 & 1 & 1 & 1 & 1 & 1 & 1 & 1 & 2 & \cdots & 2 & 2 & \cdots & 2 & 2 & \cdots & 2 \\
\end{array}
\right)
$}
~.~
\eea

%-------------------
%-------------------

The global symmetry of the 
$Q^{1,1,1}/\mathbb{Z}_2 \ (0,0,1,1,1,1,0,0)$ model
takes the following enhanced form,
\beal{es04b07}
SU(2)_x \times U(1)_{f_1} \times U(1)_{f_2} \times U(1)_R
~.~
\eea
The enhancement can be identified from the $Q_{JE}$ and $Q_{D}$ charge matrices.
\tref{tab_04b01} summarizes the charges under the mesonic flavor symmetry $SU(2)_x \times U(1)_{f_1} \times U(1)_{f_2}$
on the extremal GLSM fields $p_a$ of the brane brick model.

%-------------------
\begin{table}[httt!!]
\centering
\begin{tabular}{|c|c|c|c|l|}
\hline
\; & $SU(2)_x$ & $U(1)_{f_1}$ & $U(1)_{f_2}$ & fugacity \\
\hline
$p_1$ & $0$ & $0$ & $-1$  & $t_1=f_2^{-1} t$ \\
$p_2$ & $-1$ & $0$ & $0$  & $t_2=x^{-1} t$\\
$p_3$ & $+1$ & $0$ & $0$  & $t_3=x t$\\
$p_4$ & $0$ & $-1$ & $0$  & $t_4=f_1^{-1} t$ \\
$p_5$ & $0$ & $+1$ & $0$  & $t_5=f_1 t$ \\
$p_6$ & $0$ & $0$ & $+1$  & $t_6=f_2 t$ \\
\hline
\end{tabular}
\caption{
Mesonic flavor symmetry of the $Q^{1,1,1}/\mathbb{Z}_2 \ (0,0,1,1,1,1,0,0)$ model 
and charges on the extremal GLSM fields $p_a$. 
Here, the fugacity $t$ counts the degree in extremal GLSM fields $p_a$. 
\label{tab_04b01}
}
\end{table}
%-------------------

The Hilbert series of the mesonic moduli space
of the brane brick model corresponding to $Q^{1,1,1}/\mathbb{Z}_2 \ (0,0,1,1,1,1,0,0)$ 
takes the following form,
\beal{es04b08a}
&&
g(t_a; \mathcal{M}^{mes})=
\frac{
P(t_a ; \mathcal{M}^{mes})
}{
\left(1- t_1^2 t_2^2 t_4^2\right)  \left(1-t_1^2 t_3^2 t_4^2\right)
\left(1 - t_1 t_2 t_5\right) \left(1 - t_1 t_3 t_5\right)
\left(1 - t_2 t_4 t_6\right) }
\nn\\
&&
\hspace{1cm}
\times \frac{1}{\left(1 - t_3 t_4 t_6\right) \left(1 - t_2^2 t_5^2 t_6^2\right)
\left(1-t_3^2 t_5^2 t_6^2\right) }
~,~
\eea
where $t_a$ is the fugacity corresponding to the extremal GLSM field $p_a$.
The numerator given by 
$P(t_a ; \mathcal{M}^{mes})$
is fully presented in appendix \sref{appCa}.
When unrefined by setting $t_a=t$, 
the Hilbert series takes the form,
\beal{es04b08}
g(t; \mathcal{M}^{mes})=\frac{1+2t^3 +6t^6 + 2t^9+t^{12}}{(1-t^3)^2 (1-t^6)^2}
~,~
\eea
where the palindromic numerator indicates that the mesonic moduli space is Calabi-Yau.

The Hilbert series of the mesonic moduli space can be refined using the fugacities corresponding to the mesonic flavor fugacities
summarized in \tref{tab_04b01}.
The corresponding highest weight generating function is given by, 
\beal{es04b09}
&&
h(\mu, f_1, f_2 , t; \mathcal{M}^{mes})=
\nn\\
&&
\hspace{1cm}
\frac{1- \mu^4 t^{12}}{(1- \mu f_1 f_2^{-1} t^3)(1-\mu f_1^{-1} f_2 t^3)(1- \mu^2 f_1^2 f_2^2 t^6)(1-\mu^2 f_1^{-2} f_2^{-2} t^6)}
~,~
\eea
where $\mu^m$ is fugacity for counting the $SU(2)_x$ character of the form $[m]_x$,
and $f_1, f_2$ are the fugacities associated to $U(1)_{f_1}$ and $U(1)_{f_2}$, respectively.

%------------------ -
 \begin {table}[httt!]
\centering
\begin {tabular} {|c|c|ccc|}
\hline
PL term & generator & $SU(2)_x$ & $U(1)_{f_1}$ & $U(1)_{f_2}$ 
\\
\hline 
\multirow{2}{*}{$+[1]_x f_1 f_2^{-1} t^3$}
& $p_1 p_3 p_5 ~q_{(1)} $ & $1$ & $1$ & $-1$  \\
& $p_1 p_2 p_5 ~q_{(1)} $ & $-1$ & $1$ & $-1$  \\
\hline
\multirow{2}{*}{$+[1]_x f_1^{-1} f_2 t^3$}
& $p_3 p_4 p_6 ~q_{(2)} $ & $1$ & $-1$ & $1$  \\
& $p_2 p_4 p_6 ~q_{(2)} $ & $-1$ & $-1$ & $1$  \\
\hline
\multirow{3}{*}{$+[2]_x f_1^{2} f_2^{2} t^6$}
& $p_3^2 p_5^2 p_6^2 ~q_{(1)} q_{(2)} $ & $2$ & $2$ & $2$  \\
& $p_2 p_3 p_5^2 p_6^2 ~q_{(1)} q_{(2)} $ & $0$ & $2$ & $2$  \\
& $p_2^2 p_5^2 p_6^2 ~q_{(1)} q_{(2)} $ & $-2$ & $2$ & $2$  \\
\hline
\multirow{3}{*}{$+[2]_x f_1^{-2} f_2^{-2} t^6$}
& $p_1^2 p_3^2 p_4^2 ~q_{(1)} q_{(2)} $ & $2$ & $-2$ & $-2$  \\
& $p_1^2 p_2 p_3 p_4^2 ~q_{(1)} q_{(2)} $ & $0$ & $-2$ & $-2$  \\
& $p_1^2 p_2^2 p_4^2 ~q_{(1)} q_{(2)} $ & $-2$ & $-2$ & $-2$  \\
\hline
\end{tabular}
\caption{
Generators of the $Q^{1,1,1}/\mathbb{Z}_2 \ (0,0,1,1,1,1,0,0)$ model in terms of GLSM fields 
and their corresponding mesonic flavor charges. 
Here, we denote $q_{(1)}=q_1^{(1)}q_2^{(1)}$,  $q_{(2)}=q_1^{(2)} q_2^{(2)}$, 
and exclude contributions from extra GLSM fields. 
\label{tab_04b02}
}
\end{table}
%-------------------

The plethystic logarithm of the mesonic flavor symmetry refined Hilbert series 
takes the following form, 
\beal{es04b10}
&&
\text{PL}[g(x,f_1,f_2,t; \mathcal{M}^{mes})]=
([1]_x f_1 f_2^{-1} +[1]_x f_1^{-1} f_2  )t^3 + ([2]_x f_1^{2} f_2^{2}  +[2]_x f_1^{-2} f_2^{-2}  )t^6
\nn\\
&&
\hspace{1cm}
- t^6 
- ( [1]_x f_1^{-1}f_2^{-3} +[1]_x f_1^{3}f_2 + [1]_x f_1 f_2^{3} + [1]_x f_1^{-3}f_2^{-1}) t^9
- ( 1 + f_1^{-4} f_2^{-4} 
\nn\\
&&
\hspace{1cm}
+ f_1^{4} f_2^{4} + [2]_x + [4]_x ) t^{12}
+\dots
~,~
\eea
where $[m]_x$ is the character of the irreducible representation of $SU(2)_x$ with highest weight $(m)_x$.
The infinite expansion of the plethystic logarithm indicates that the mesonic moduli space of the $Q^{1,1,1}/\mathbb{Z}_2 \ (0,0,1,1,1,1,0,0)$  model is not a complete intersection.
The generators of the mesonic moduli space with their mesonic flavor charges are summarized in \tref{tab_04b02}.

We note that the abelian orbifold $Q^{1,1,1}/\mathbb{Z}_2 \ (0,0,1,1,1,1,0,0)$ is part of a family of abelian orbifolds of $Q^{1,1,1}$
of the form $Q^{1,1,1}/\mathbb{Z}_n \ (0,0,1,1,-1,-1,0,0)$, whose toric diagram has extremal vertices with coordinates given by,
\beal{es04b20}
&
(0,0,0)~,~
(1,0,0)~,~
(0,1,0)~,~
(1,1,0)~,~
(1,0,n)~,~
(0,1,-n)~.~
&
\eea
In section \sref{sec044},
we summarize the brane brick model for the abelian orbifold of the form 
$Q^{1,1,1}/\mathbb{Z}_3 \ (0,0,1,1,2,2,0,0)$, which is also part of this family of abelian orbifolds of $Q^{1,1,1}$.
\\

%=================================================================
\subsubsection{$Q^{1,1,1}/ \mathbb{Z}_3 \ (1,1,1,1,2,2,2,2)$ \label{sec043}}

%-------------------
\begin{figure}[H]
    \centering
    \includegraphics[width=0.4\textwidth]{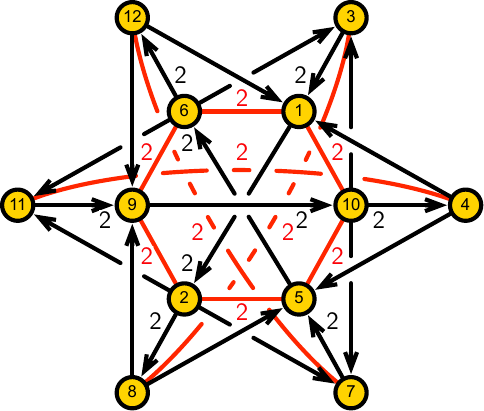}
    \caption{
    The quiver for the $Q^{1,1,1}/ \mathbb{Z}_3 \ (1,1,1,1,2,2,2,2)$ model.
    \label{fig_413_quiver}
    }
\end{figure}
%-------------------

\fref{fig_413_quiver} shows the quiver for the $Q^{1,1,1}/ \mathbb{Z}_3 \ (1,1,1,1,2,2,2,2)$ model.
The corresponding $J$- and $E$-terms are given as follows, 
\beal{es04c01}
\resizebox{0.95\textwidth}{!}{$
\begin{array}{rclccclcccc}
&& && &J& && &E& \\
&&\Lambda^{(1)}_{10, 1} &:& & D_{1, 2} \cdot Z_{2, 8} \cdot Y_{8, 9} \cdot X_{9, 10} - X_{1, 2} \cdot Z_{2, 11} \cdot D_{11, 9} \cdot D_{9, 10}& && & X_{10, 3} \cdot Y_{3, 1} - X_{10, 4} \cdot D_{4, 1}& \\ 
 &&\Lambda^{(1)}_{2, 5} &:& & D_{5, 6} \cdot Z_{6, 12} \cdot Y_{12, 1} \cdot X_{1, 2} - X_{5, 6} \cdot Z_{6, 3} \cdot D_{3, 1} \cdot D_{1, 2}& && & X_{2, 7} \cdot Y_{7, 5} - X_{2, 8} \cdot D_{8, 5}& \\ 
 &&\Lambda^{(1)}_{6, 9} &:& & D_{9, 10} \cdot Z_{10, 4} \cdot Y_{4, 5} \cdot X_{5, 6} - X_{9, 10} \cdot Z_{10, 7} \cdot D_{7, 5} \cdot D_{5, 6}& && & X_{6, 11} \cdot Y_{11, 9} - X_{6, 12} \cdot D_{12, 9}& \\ 
 &&\Lambda^{(2)}_{10, 1} &:& & X_{1, 2} \cdot X_{2, 8} \cdot Y_{8, 9} \cdot D_{9, 10} - D_{1, 2} \cdot Z_{2, 11} \cdot Y_{11, 9} \cdot X_{9, 10}& && & X_{10, 3} \cdot D_{3, 1} - Z_{10, 4} \cdot D_{4, 1}& \\ 
 &&\Lambda^{(2)}_{2, 5} &:& & X_{5, 6} \cdot X_{6, 12} \cdot Y_{12, 1} \cdot D_{1, 2} - D_{5, 6} \cdot Z_{6, 3} \cdot Y_{3, 1} \cdot X_{1, 2}& && & X_{2, 7} \cdot D_{7, 5} - Z_{2, 8} \cdot D_{8, 5}& \\ 
 &&\Lambda^{(2)}_{6, 9} &:& & X_{9, 10} \cdot X_{10, 4} \cdot Y_{4, 5} \cdot D_{5, 6} - D_{9, 10} \cdot Z_{10, 7} \cdot Y_{7, 5} \cdot X_{5, 6}& && & X_{6, 11} \cdot D_{11, 9} - Z_{6, 12} \cdot D_{12, 9}& \\ 
 &&\Lambda^{(3)}_{6, 1} &:& & D_{1, 2} \cdot Z_{2, 8} \cdot D_{8, 5} \cdot X_{5, 6} - X_{1, 2} \cdot X_{2, 7} \cdot D_{7, 5} \cdot D_{5, 6}& && & X_{6, 12} \cdot Y_{12, 1} - Z_{6, 3} \cdot Y_{3, 1}& \\ 
 &&\Lambda^{(3)}_{10, 5} &:& & D_{5, 6} \cdot Z_{6, 12} \cdot D_{12, 9} \cdot X_{9, 10} - X_{5, 6} \cdot X_{6, 11} \cdot D_{11, 9} \cdot D_{9, 10}& && & X_{10, 4} \cdot Y_{4, 5} - Z_{10, 7} \cdot Y_{7, 5}& \\ 
 &&\Lambda^{(3)}_{2, 9} &:& & D_{9, 10} \cdot Z_{10, 4} \cdot D_{4, 1} \cdot X_{1, 2} - X_{9, 10} \cdot X_{10, 3} \cdot D_{3, 1} \cdot D_{1, 2}& && & X_{2, 8} \cdot Y_{8, 9} - Z_{2, 11} \cdot Y_{11, 9}& \\ 
 &&\Lambda^{(4)}_{6, 1} &:& & D_{1, 2} \cdot X_{2, 7} \cdot Y_{7, 5} \cdot X_{5, 6} - X_{1, 2} \cdot X_{2, 8} \cdot D_{8, 5} \cdot D_{5, 6}& && & Z_{6, 3} \cdot D_{3, 1} - Z_{6, 12} \cdot Y_{12, 1}& \\ 
 &&\Lambda^{(4)}_{10, 5} &:& & D_{5, 6} \cdot X_{6, 11} \cdot Y_{11, 9} \cdot X_{9, 10} - X_{5, 6} \cdot X_{6, 12} \cdot D_{12, 9} \cdot D_{9, 10}& && & Z_{10, 7} \cdot D_{7, 5} - Z_{10, 4} \cdot Y_{4, 5}& \\ 
 &&\Lambda^{(4)}_{2, 9} &:& & D_{9, 10} \cdot X_{10, 3} \cdot Y_{3, 1} \cdot X_{1, 2} - X_{9, 10} \cdot X_{10, 4} \cdot D_{4, 1} \cdot D_{1, 2}& && & Z_{2, 11} \cdot D_{11, 9} - Z_{2, 8} \cdot Y_{8, 9}& \\ 
 &&\Lambda^{(5)}_{8, 3} &:& & D_{3, 1} \cdot X_{1, 2} \cdot X_{2, 8} - Y_{3, 1} \cdot X_{1, 2} \cdot Z_{2, 8}& && & D_{8, 5} \cdot D_{5, 6} \cdot Z_{6, 3} - Y_{8, 9} \cdot D_{9, 10} \cdot X_{10, 3}& \\ 
 &&\Lambda^{(5)}_{12, 7} &:& & D_{7, 5} \cdot X_{5, 6} \cdot X_{6, 12} - Y_{7, 5} \cdot X_{5, 6} \cdot Z_{6, 12}& && & D_{12, 9} \cdot D_{9, 10} \cdot Z_{10, 7} - Y_{12, 1} \cdot D_{1, 2} \cdot X_{2, 7}& \\ 
 &&\Lambda^{(5)}_{4, 11} &:& & D_{11, 9} \cdot X_{9, 10} \cdot X_{10, 4} - Y_{11, 9} \cdot X_{9, 10} \cdot Z_{10, 4}& && & D_{4, 1} \cdot D_{1, 2} \cdot Z_{2, 11} - Y_{4, 5} \cdot D_{5, 6} \cdot X_{6, 11}& \\ 
 &&\Lambda^{(6)}_{11, 4} &:& & Y_{4, 5} \cdot X_{5, 6} \cdot X_{6, 11} - D_{4, 1} \cdot X_{1, 2} \cdot Z_{2, 11}& && & D_{11, 9} \cdot D_{9, 10} \cdot X_{10, 4} - Y_{11, 9} \cdot D_{9, 10} \cdot Z_{10, 4}& \\ 
 &&\Lambda^{(6)}_{3, 8} &:& & Y_{8, 9} \cdot X_{9, 10} \cdot X_{10, 3} - D_{8, 5} \cdot X_{5, 6} \cdot Z_{6, 3}& && & D_{3, 1} \cdot D_{1, 2} \cdot X_{2, 8} - Y_{3, 1} \cdot D_{1, 2} \cdot Z_{2, 8}& \\ 
 &&\Lambda^{(6)}_{7, 12} &:& & Y_{12, 1} \cdot X_{1, 2} \cdot X_{2, 7} - D_{12, 9} \cdot X_{9, 10} \cdot Z_{10, 7}& && & D_{7, 5} \cdot D_{5, 6} \cdot X_{6, 12} - Y_{7, 5} \cdot D_{5, 6} \cdot Z_{6, 12}&
\end{array}
$}
~.~
\nn\\
\eea
Here, we made use of the general form of the $J$- and $E$-terms given in \eqref{es03b01} 
with a relabelling of indices as shown below, 
\beal{es04c02}
&
[1,0] \rightarrow 1~,~
[2,0] \rightarrow 2~,~  
[3,0] \rightarrow 3~,~ 
[4,0] \rightarrow 4~,~
[1,1] \rightarrow 5~,~ 
[2,1] \rightarrow 6~,~
&
\nn\\
& 
[3,1] \rightarrow 7~,~ 
[4,1] \rightarrow 8~,~
[1,2] \rightarrow 9~,~ 
[2,2] \rightarrow 10~,~ 
[3,2] \rightarrow 11~,~ 
[4,2] \rightarrow 12~.~
&
\eea

The corresponding $P$-matrix 
takes the following form,
\beal{es04c03}
&&
P=
\nn\\
&&
\resizebox{0.95\textwidth}{!}{$
\left(
\begin{array}{c|cccccc|ccccccccccccccc|ccccccccccccccc|ccc|ccc|ccc}
 & p_{1} & p_{2} & p_{3} & p_{4} & p_{5} & p_{6} & s^{(1)}_{1} & s^{(1)}_{2} & s^{(1)}_{3} & s^{(1)}_{4} & s^{(1)}_{5} & s^{(1)}_{6} & s^{(1)}_{7} & s^{(1)}_{8} & s^{(1)}_{9} & s^{(1)}_{10} & s^{(1)}_{11} & s^{(1)}_{12} & s^{(1)}_{13} & s^{(1)}_{14} & s^{(1)}_{15} & s^{(2)}_{1} & s^{(2)}_{2} & s^{(2)}_{3} & s^{(2)}_{4} & s^{(2)}_{5} & s^{(2)}_{6} & s^{(2)}_{7} & s^{(2)}_{8} & s^{(2)}_{9} & s^{(2)}_{10} & s^{(2)}_{11} & s^{(2)}_{12} & s^{(2)}_{13} & s^{(2)}_{14} & s^{(2)}_{15} & o^{(1)}_{1} & \cdots & o^{(1)}_{44} & o^{(2)}_{1} & \cdots & o^{(2)}_{3} & o^{(3)}_{1} & \cdots & o^{(3)}_{3} \\
\hline
D_{1,2} & 0 & 0 & 1 & 0 & 0 & 0 & 0 & 0 & 1 & 0 & 0 & 0 & 0 & 0 & 0 & 0 & 0 & 0 & 0
   & 0 & 0 & 0 & 0 & 0 & 0 & 0 & 0 & 0 & 0 & 1 & 0 & 0 & 0 & 0 & 0 & 0 & 1 & \cdots & 0 & 1 & \cdots & 1 &
   1 & \cdots & 1 \\
 D_{3,1} & 0 & 0 & 0 & 1 & 0 & 0 & 0 & 0 & 0 & 0 & 0 & 0 & 0 & 0 & 0 & 0 & 0 & 1 & 1
   & 0 & 0 & 0 & 0 & 0 & 0 & 0 & 0 & 0 & 0 & 0 & 0 & 0 & 0 & 0 & 1 & 1 & 0 & \cdots & 1 & 0 & \cdots & 1 &
   0 & \cdots & 1 \\
 D_{4,1} & 1 & 0 & 0 & 0 & 0 & 0 & 0 & 0 & 0 & 0 & 0 & 0 & 1 & 0 & 1 & 1 & 1 & 1 & 1
   & 0 & 1 & 0 & 0 & 0 & 0 & 0 & 0 & 0 & 0 & 0 & 0 & 0 & 0 & 1 & 0 & 1 & 0 & \cdots & 0 & 0 & \cdots & 0 &
   0 & \cdots & 0 \\
 D_{5,6} & 0 & 0 & 1 & 0 & 0 & 0 & 0 & 0 & 0 & 0 & 0 & 0 & 1 & 0 & 0 & 0 & 0 & 0 & 0
   & 0 & 0 & 1 & 0 & 0 & 0 & 0 & 0 & 0 & 0 & 0 & 0 & 0 & 0 & 0 & 0 & 0 & 1 & \cdots & 0 & 0 & \cdots & 1 &
   1 & \cdots & 0 \\
 D_{7,5} & 0 & 0 & 0 & 1 & 0 & 0 & 0 & 0 & 0 & 0 & 0 & 0 & 0 & 1 & 1 & 0 & 0 & 0 & 0
   & 0 & 0 & 0 & 1 & 1 & 0 & 0 & 0 & 0 & 0 & 0 & 0 & 0 & 0 & 0 & 0 & 0 & 0 & \cdots & 0 & 1 & \cdots & 0 &
   1 & \cdots & 1 \\
 D_{8,5} & 1 & 0 & 0 & 0 & 0 & 0 & 0 & 1 & 0 & 0 & 1 & 1 & 0 & 1 & 1 & 0 & 0 & 0 & 0
   & 1 & 1 & 0 & 0 & 1 & 0 & 0 & 0 & 0 & 0 & 0 & 0 & 1 & 0 & 0 & 0 & 0 & 0 & \cdots & 1 & 0 & \cdots & 0 &
   0 & \cdots & 1 \\
 D_{9,10} & 0 & 0 & 1 & 0 & 0 & 0 & 0 & 0 & 0 & 0 & 0 & 1 & 0 & 0 & 0 & 0 & 0 & 0 &
   0 & 0 & 0 & 0 & 0 & 0 & 0 & 0 & 0 & 0 & 1 & 0 & 0 & 0 & 0 & 0 & 0 & 0 & 0 & \cdots & 0 & 1 & \cdots & 0
   & 0 & \cdots & 1 \\
 D_{11,9} & 0 & 0 & 0 & 1 & 0 & 0 & 1 & 1 & 0 & 0 & 0 & 0 & 0 & 0 & 0 & 0 & 0 & 0 &
   0 & 0 & 0 & 0 & 0 & 0 & 1 & 1 & 0 & 0 & 0 & 0 & 0 & 0 & 0 & 0 & 0 & 0 & 1 & \cdots & 0 & 1 & \cdots & 1
   & 1 & \cdots & 0 \\
 D_{12,9} & 1 & 0 & 0 & 0 & 0 & 0 & 1 & 1 & 1 & 1 & 1 & 0 & 0 & 0 & 0 & 0 & 1 & 0 &
   1 & 0 & 0 & 0 & 0 & 0 & 0 & 1 & 0 & 1 & 0 & 0 & 0 & 0 & 0 & 0 & 0 & 0 & 1 & \cdots & 0 & 0 & \cdots & 0
   & 1 & \cdots & 0 \\
 X_{1,2} & 0 & 1 & 0 & 0 & 0 & 0 & 0 & 0 & 1 & 0 & 0 & 0 & 0 & 0 & 0 & 0 & 0 & 0 & 0
   & 0 & 0 & 0 & 0 & 0 & 0 & 0 & 0 & 0 & 0 & 1 & 0 & 0 & 0 & 0 & 0 & 0 & 1 & \cdots & 0 & 1 & \cdots & 1 &
   1 & \cdots & 1 \\
 X_{2,7} & 1 & 0 & 0 & 0 & 0 & 0 & 1 & 1 & 0 & 1 & 1 & 1 & 0 & 0 & 0 & 0 & 0 & 0 & 0
   & 1 & 1 & 0 & 0 & 0 & 0 & 0 & 0 & 0 & 0 & 0 & 1 & 1 & 0 & 0 & 0 & 0 & 0 & \cdots & 1 & 0 & \cdots & 0 &
   0 & \cdots & 0 \\
 X_{2,8} & 0 & 0 & 0 & 0 & 1 & 0 & 1 & 0 & 0 & 1 & 0 & 0 & 0 & 0 & 0 & 0 & 0 & 0 & 0
   & 0 & 0 & 0 & 1 & 0 & 0 & 0 & 0 & 0 & 0 & 0 & 1 & 0 & 0 & 0 & 0 & 0 & 0 & \cdots & 0 & 1 & \cdots & 0 &
   1 & \cdots & 0 \\
 X_{5,6} & 0 & 1 & 0 & 0 & 0 & 0 & 0 & 0 & 0 & 0 & 0 & 0 & 1 & 0 & 0 & 0 & 0 & 0 & 0
   & 0 & 0 & 1 & 0 & 0 & 0 & 0 & 0 & 0 & 0 & 0 & 0 & 0 & 0 & 0 & 0 & 0 & 1 & \cdots & 0 & 0 & \cdots & 1 &
   1 & \cdots & 0 \\
 X_{6,11} & 1 & 0 & 0 & 0 & 0 & 0 & 0 & 0 & 1 & 1 & 1 & 0 & 0 & 0 & 0 & 1 & 1 & 1 &
   1 & 0 & 0 & 0 & 0 & 0 & 0 & 0 & 1 & 1 & 0 & 0 & 0 & 0 & 0 & 0 & 0 & 0 & 0 & \cdots & 1 & 0 & \cdots & 0
   & 0 & \cdots & 1 \\
 X_{6,12} & 0 & 0 & 0 & 0 & 1 & 0 & 0 & 0 & 0 & 0 & 0 & 0 & 0 & 0 & 0 & 1 & 0 & 1 &
   0 & 0 & 0 & 0 & 0 & 0 & 1 & 0 & 1 & 0 & 0 & 0 & 0 & 0 & 0 & 0 & 0 & 0 & 0 & \cdots & 1 & 1 & \cdots & 1
   & 0 & \cdots & 1 \\
 X_{9,10} & 0 & 1 & 0 & 0 & 0 & 0 & 0 & 0 & 0 & 0 & 0 & 1 & 0 & 0 & 0 & 0 & 0 & 0 &
   0 & 0 & 0 & 0 & 0 & 0 & 0 & 0 & 0 & 0 & 1 & 0 & 0 & 0 & 0 & 0 & 0 & 0 & 0 & \cdots & 0 & 1 & \cdots & 0
   & 0 & \cdots & 1 \\
 X_{10,3} & 1 & 0 & 0 & 0 & 0 & 0 & 0 & 0 & 0 & 0 & 0 & 0 & 1 & 1 & 1 & 1 & 1 & 0 &
   0 & 1 & 1 & 0 & 0 & 0 & 0 & 0 & 0 & 0 & 0 & 0 & 0 & 0 & 1 & 1 & 0 & 0 & 0 & \cdots & 0 & 0 & \cdots & 0
   & 1 & \cdots & 0 \\
 X_{10,4} & 0 & 0 & 0 & 0 & 1 & 0 & 0 & 0 & 0 & 0 & 0 & 0 & 0 & 1 & 0 & 0 & 0 & 0 &
   0 & 1 & 0 & 0 & 0 & 0 & 0 & 0 & 0 & 0 & 0 & 0 & 0 & 0 & 1 & 0 & 1 & 0 & 0 & \cdots & 1 & 0 & \cdots & 1
   & 1 & \cdots & 1 \\
 Y_{3,1} & 0 & 0 & 0 & 0 & 1 & 0 & 0 & 0 & 0 & 0 & 0 & 0 & 0 & 0 & 0 & 0 & 0 & 1 & 1
   & 0 & 0 & 0 & 0 & 0 & 0 & 0 & 0 & 0 & 0 & 0 & 0 & 0 & 0 & 0 & 1 & 1 & 0 & \cdots & 1 & 0 & \cdots & 1 &
   0 & \cdots & 1 \\
 Y_{4,5} & 0 & 0 & 0 & 0 & 0 & 1 & 0 & 0 & 0 & 0 & 0 & 0 & 0 & 0 & 1 & 0 & 0 & 0 & 0
   & 0 & 1 & 0 & 1 & 1 & 0 & 0 & 0 & 0 & 0 & 1 & 1 & 1 & 0 & 1 & 0 & 1 & 0 & \cdots & 0 & 1 & \cdots & 0 &
   0 & \cdots & 0 \\
 Y_{7,5} & 0 & 0 & 0 & 0 & 1 & 0 & 0 & 0 & 0 & 0 & 0 & 0 & 0 & 1 & 1 & 0 & 0 & 0 & 0
   & 0 & 0 & 0 & 1 & 1 & 0 & 0 & 0 & 0 & 0 & 0 & 0 & 0 & 0 & 0 & 0 & 0 & 0 & \cdots & 0 & 1 & \cdots & 0 &
   1 & \cdots & 1 \\
 Y_{8,9} & 0 & 0 & 0 & 0 & 0 & 1 & 0 & 1 & 0 & 0 & 1 & 0 & 0 & 0 & 0 & 0 & 0 & 0 & 0
   & 0 & 0 & 1 & 0 & 1 & 1 & 1 & 1 & 1 & 0 & 0 & 0 & 1 & 0 & 0 & 0 & 0 & 1 & \cdots & 1 & 0 & \cdots & 1 &
   0 & \cdots & 0 \\
 Y_{11,9} & 0 & 0 & 0 & 0 & 1 & 0 & 1 & 1 & 0 & 0 & 0 & 0 & 0 & 0 & 0 & 0 & 0 & 0 &
   0 & 0 & 0 & 0 & 0 & 0 & 1 & 1 & 0 & 0 & 0 & 0 & 0 & 0 & 0 & 0 & 0 & 0 & 1 & \cdots & 0 & 1 & \cdots & 1
   & 1 & \cdots & 0 \\
 Y_{12,1} & 0 & 0 & 0 & 0 & 0 & 1 & 0 & 0 & 0 & 0 & 0 & 0 & 0 & 0 & 0 & 0 & 1 & 0 &
   1 & 0 & 0 & 0 & 0 & 0 & 0 & 1 & 0 & 1 & 1 & 0 & 0 & 0 & 1 & 1 & 1 & 1 & 0 & \cdots & 0 & 0 & \cdots & 0
   & 0 & \cdots & 0 \\
 Z_{2,8} & 0 & 0 & 0 & 1 & 0 & 0 & 1 & 0 & 0 & 1 & 0 & 0 & 0 & 0 & 0 & 0 & 0 & 0 & 0
   & 0 & 0 & 0 & 1 & 0 & 0 & 0 & 0 & 0 & 0 & 0 & 1 & 0 & 0 & 0 & 0 & 0 & 0 & \cdots & 0 & 1 & \cdots & 0 &
   1 &  \cdots &0 \\
 Z_{2,11} & 0 & 0 & 0 & 0 & 0 & 1 & 0 & 0 & 0 & 1 & 1 & 0 & 0 & 0 & 0 & 0 & 0 & 0 &
   0 & 0 & 0 & 1 & 1 & 1 & 0 & 0 & 1 & 1 & 0 & 0 & 1 & 1 & 0 & 0 & 0 & 0 & 0 & \cdots & 1 & 0 & \cdots & 0
   & 0 & \cdots & 0 \\
 Z_{6,3} & 0 & 0 & 0 & 0 & 0 & 1 & 0 & 0 & 0 & 0 & 0 & 0 & 0 & 0 & 0 & 1 & 1 & 0 & 0
   & 0 & 0 & 0 & 0 & 0 & 1 & 1 & 1 & 1 & 1 & 0 & 0 & 0 & 1 & 1 & 0 & 0 & 0 & \cdots & 0 & 1 & \cdots & 0 &
   0 & \cdots & 0 \\
 Z_{6,12} & 0 & 0 & 0 & 1 & 0 & 0 & 0 & 0 & 0 & 0 & 0 & 0 & 0 & 0 & 0 & 1 & 0 & 1 &
   0 & 0 & 0 & 0 & 0 & 0 & 1 & 0 & 1 & 0 & 0 & 0 & 0 & 0 & 0 & 0 & 0 & 0 & 0 & \cdots & 1 & 1 & \cdots & 1
   & 0 & \cdots & 1 \\
 Z_{10,4} & 0 & 0 & 0 & 1 & 0 & 0 & 0 & 0 & 0 & 0 & 0 & 0 & 0 & 1 & 0 & 0 & 0 & 0 &
   0 & 1 & 0 & 0 & 0 & 0 & 0 & 0 & 0 & 0 & 0 & 0 & 0 & 0 & 1 & 0 & 1 & 0 & 0 & \cdots & 1 & 0 & \cdots & 1
   & 1 & \cdots & 1 \\
 Z_{10,7} & 0 & 0 & 0 & 0 & 0 & 1 & 0 & 0 & 0 & 0 & 0 & 0 & 0 & 0 & 0 & 0 & 0 & 0 &
   0 & 1 & 1 & 0 & 0 & 0 & 0 & 0 & 0 & 0 & 0 & 1 & 1 & 1 & 1 & 1 & 1 & 1 & 0 & \cdots & 1 & 0 & \cdots & 1
   & 0 & \cdots & 0 \\
\end{array}
\right)
$}
~,~
\nn\\
\eea
where $o_k^{(1)}$, $o_l^{(2)}$ and $o_m^{(3)}$ are extra GLSM fields \cite{Witten:1993yc}.
The $J$- and $E$-term charge matrix takes the following form,
\beal{es04c04}
&&
Q_{JE}=
\nn\\
&&
\resizebox{0.95\textwidth}{!}{$
\left(
\begin{array}{cccccc|ccccccccccccccc|ccccccccccccccc|ccc|ccc|ccc}
p_{1} & p_{2} & p_{3} & p_{4} & p_{5} & p_{6} & s^{(1)}_{1} & s^{(1)}_{2} & s^{(1)}_{3} & s^{(1)}_{4} & s^{(1)}_{5} & s^{(1)}_{6} & s^{(1)}_{7} & s^{(1)}_{8} & s^{(1)}_{9} & s^{(1)}_{10} & s^{(1)}_{11} & s^{(1)}_{12} & s^{(1)}_{13} & s^{(1)}_{14} & s^{(1)}_{15} & s^{(2)}_{1} & s^{(2)}_{2} & s^{(2)}_{3} & s^{(2)}_{4} & s^{(2)}_{5} & s^{(2)}_{6} & s^{(2)}_{7} & s^{(2)}_{8} & s^{(2)}_{9} & s^{(2)}_{10} & s^{(2)}_{11} & s^{(2)}_{12} & s^{(2)}_{13} & s^{(2)}_{14} & s^{(2)}_{15} & o^{(1)}_{1} & \cdots & o^{(1)}_{44} & o^{(2)}_{1} & \cdots & o^{(2)}_{3} & o^{(3)}_{1} & \cdots & o^{(3)}_{3} \\
\hline
0 & 0 & 0 & -1 & -1 & -1 & -1 & 0 & 0 & 0 & 0 & 0 & 0 & 0 & 0 & 0 & 0 & 0 & 0 & 0 & 0 &
   0 & 2 & -1 & 1 & 1 & -1 & 0 & 0 & 0 & 0 & 0 & 0 & 0 & 0 & 0 & 0 & \cdots & 1 & 0 & \cdots & 0 & 0 & \cdots & 0
   \\
 0 & 0 & 0 & -1 & -1 & -1 & -1 & 0 & 0 & 0 & 0 & 0 & 0 & 0 & 0 & 0 & 0 & 0 & 0 & 0 & 0 &
   0 & 1 & 0 & 1 & 1 & -1 & 0 & 0 & 0 & 0 & 0 & 0 & 0 & 0 & 0 & 0 & \cdots & 0 & 0 & \cdots & 0 & 0 & \cdots & 0 \\
 0 & 0 & 0 & -1 & -1 & -1 & 0 & 0 & 0 & 0 & 0 & 0 & 0 & 0 & 0 & 0 & 0 & 0 & 0 & 0 & 0 &
   1 & 1 & 0 & 1 & 1 & -1 & 0 & 0 & 0 & 0 & 0 & 0 & 0 & 0 & 0 & -1 & \cdots & 0 & 0 & \cdots & 0 & 0 & \cdots & 0
   \\
 0 & 0 & 0 & -1 & -1 & -1 & -1 & 0 & 0 & 0 & 0 & 0 & -1 & 0 & 0 & 0 & 0 & 0 & 0 & 0 & 0
   & 0 & 1 & -1 & 0 & 1 & 0 & 0 & 0 & 0 & 0 & 0 & 0 & 0 & 0 & 0 & 0 & \cdots & 0 & 0 & \cdots & 0 & 0 & \cdots & 0
   \\
 0 & 0 & 0 & -1 & -1 & -1 & -1 & 0 & 0 & 0 & 0 & 0 & 0 & 0 & 0 & 0 & 0 & 0 & 0 & 0 & 0 &
   0 & 2 & -1 & 0 & 1 & 0 & 0 & 0 & 0 & 0 & 0 & 0 & 0 & 0 & 0 & 0 & \cdots & 0 & 0 & \cdots & 0 & 0 & \cdots & 0 \\
 0 & 0 & 0 & -1 & -1 & -1 & -1 & 0 & 0 & 0 & 0 & 0 & -1 & 0 & 0 & 0 & 0 & 0 & 0 & 0 & 0
   & 0 & 0 & 0 & 0 & 1 & 0 & 0 & 0 & 0 & 0 & 0 & 0 & 0 & 0 & 0 & 0 & \cdots & 0 & 0 & \cdots & 0 & 0 & \cdots & 0
   \\
 0 & 0 & 0 & -1 & -1 & -1 & -1 & 0 & 0 & 0 & 0 & 0 & 0 & 0 & 0 & 0 & 0 & 0 & 0 & 0 & 0 &
   0 & 1 & 0 & 0 & 1 & 0 & 0 & 0 & 0 & 0 & 0 & 0 & 0 & 0 & 0 & 0 & \cdots & 0 & 0 & \cdots & 0 & 0 & \cdots & 0 \\
 0 & 0 & 0 & -1 & -1 & -1 & 0 & 0 & 0 & 0 & 0 & 0 & 0 & 0 & 0 & 0 & 0 & 0 & 0 & 0 & 0 &
   0 & 1 & 0 & 0 & 1 & 0 & 0 & 0 & 0 & 0 & 0 & 0 & 0 & 0 & 0 & -1 & \cdots & 0 & 0 & \cdots & 1 & 0 & \cdots & 0 \\
 0 & 0 & 0 & -1 & -1 & -1 & 0 & 0 & 0 & 0 & 0 & 0 & -1 & 0 & 0 & 0 & 0 & 0 & 0 & 0 & 0 &
   0 & 0 & 0 & 1 & 0 & 0 & 0 & 0 & 0 & 0 & 0 & 0 & 0 & 0 & 1 & 0 & \cdots & 0 & 0 & \cdots & 0 & 0 & \cdots & 0 \\
 0 & 0 & 0 & -1 & -1 & -1 & 0 & 0 & 0 & 0 & 0 & 0 & 0 & 0 & 0 & 0 & 0 & 0 & 0 & 0 & 0 &
   0 & 1 & 0 & 1 & 0 & 0 & 0 & 0 & 0 & 0 & 0 & 0 & 0 & 1 & 0 & 0 & \cdots & 0 & 0 & \cdots & 0 & 0 & \cdots & 0 \\
 0 & 0 & 0 & 0 & 0 & -1 & -1 & 0 & 0 & 0 & 0 & 0 & 0 & 0 & 0 & 0 & 0 & 0 & 0 & 0 & 0 & 1
   & 1 & 0 & 0 & 1 & -1 & 0 & 0 & 0 & 0 & 0 & 0 & 0 & 0 & 0 & 0 & \cdots & 0 & 0 & \cdots & 0 & 0 & \cdots & 0 \\
 0 & 0 & 0 & 0 & 0 & -1 & 0 & 0 & 0 & 0 & 0 & 0 & 0 & 0 & 0 & 0 & 0 & 0 & 0 & 0 & 0 & 2
   & 1 & 0 & 0 & 1 & -1 & 0 & 0 & 0 & 0 & 0 & 0 & 0 & 0 & 0 & -1 & \cdots & 0 & 0 & \cdots & 0 & 0 & \cdots & 0 \\
 0 & 0 & 0 & 0 & 0 & -1 & -1 & 0 & 0 & 0 & 0 & 0 & -1 & 0 & 0 & 0 & 0 & 0 & 0 & 0 & 0 &
   1 & 0 & 0 & -1 & 1 & 0 & 0 & 0 & 0 & 0 & 0 & 0 & 0 & 0 & 0 & 0 & \cdots & 0 & 0 & \cdots & 0 & 0 & \cdots & 0 \\
 0 & 0 & 0 & 0 & 0 & -1 & -1 & 0 & 0 & 0 & 0 & 0 & 0 & 0 & 0 & 0 & 0 & 0 & 0 & 0 & 0 & 1
   & 1 & 0 & -1 & 1 & 0 & 0 & 0 & 0 & 0 & 0 & 0 & 0 & 0 & 0 & 0 & \cdots & 0 & 0 & \cdots & 0 & 0 & \cdots & 0 \\
 0 & 0 & 0 & 0 & 0 & -1 & -1 & 0 & 0 & 0 & 0 & 0 & -1 & 0 & 0 & 0 & 0 & 0 & 0 & 0 & 1 &
   1 & 1 & -1 & 0 & 1 & 0 & 0 & 0 & 0 & 0 & 0 & 0 & 0 & 0 & 0 & 0 & \cdots & 0 & 0 & \cdots & 0 & 0 & \cdots & 0 \\
 0 & 0 & 0 & 0 & 0 & -1 & -1 & 0 & 0 & 0 & 0 & 0 & 0 & 0 & 0 & 0 & 0 & 0 & 0 & 1 & 0 & 1
   & 2 & -1 & 0 & 1 & 0 & 0 & 0 & 0 & 0 & 0 & 0 & 0 & 0 & 0 & 0 & \cdots & 0 & 0 & \cdots & 0 & 0 & \cdots & 0 \\
 0 & 0 & 0 & 0 & 0 & -1 & -1 & 0 & 0 & 0 & 0 & 0 & 0 & 0 & 0 & 0 & 0 & 0 & 0 & 0 & 0 & 0
   & 1 & 0 & 0 & 1 & 0 & 0 & 0 & 0 & 0 & 0 & 0 & 0 & 0 & 0 & 0 & \cdots & 0 & 0 & \cdots & 0 & 0 & \cdots & 0 \\
 0 & 0 & 0 & 0 & 0 & -1 & 0 & 0 & 0 & 0 & 0 & 0 & 0 & 0 & 0 & 0 & 0 & 0 & 0 & 0 & 0 & 1
   & 1 & 0 & 0 & 1 & 0 & 0 & 0 & 0 & 0 & 0 & 0 & 0 & 0 & 0 & -1 & \cdots & 0 & 0 & \cdots & 0 & 0 & \cdots & 0 \\
 0 & 0 & 0 & 0 & 0 & -1 & 0 & 0 & 0 & 0 & 0 & 0 & -1 & 0 & 0 & 0 & 0 & 0 & 0 & 0 & 0 & 1
   & 0 & 0 & 0 & 0 & 0 & 0 & 0 & 0 & 0 & 0 & 0 & 1 & 0 & 0 & 0 & \cdots & 0 & 0 & \cdots & 0 & 0 & \cdots & 0 \\
 0 & 0 & 0 & 0 & 0 & -1 & 0 & 0 & 0 & 0 & 0 & 0 & 0 & 0 & 0 & 0 & 0 & 0 & 0 & 0 & 0 & 1
   & 1 & 0 & 0 & 0 & 0 & 0 & 0 & 0 & 0 & 0 & 1 & 0 & 0 & 0 & 0 & \cdots & 0 & 0 & \cdots & 0 & 0 & \cdots & 0 \\
 0 & 0 & 0 & 0 & 0 & -1 & -1 & 0 & 0 & 0 & 0 & 0 & 0 & 0 & 0 & 0 & 0 & 0 & 0 & 0 & 0 & 0
   & 1 & -1 & 0 & 1 & 0 & 0 & 0 & 0 & 0 & 1 & 0 & 0 & 0 & 0 & 0 & \cdots & 0 & 0 & \cdots & 0 & 0 & \cdots & 0 \\
 0 & 0 & 0 & 0 & 0 & -1 & -1 & 0 & 0 & 0 & 0 & 0 & 0 & 0 & 0 & 0 & 0 & 0 & 0 & 0 & 0 & 0
   & 0 & 0 & 0 & 1 & 0 & 0 & 0 & 0 & 1 & 0 & 0 & 0 & 0 & 0 & 0 & \cdots & 0 & 0 & \cdots & 0 & 0 & \cdots & 0 \\
 0 & 0 & 0 & 0 & 0 & -1 & 0 & 0 & 0 & 0 & 0 & 0 & 0 & 0 & 0 & 0 & 0 & 0 & 0 & 0 & 0 & 1
   & 0 & 0 & 0 & 1 & 0 & 0 & 0 & 1 & 0 & 0 & 0 & 0 & 0 & 0 & -1 & \cdots & 0 & 0 & \cdots & 0 & 0 & \cdots & 0 \\
 0 & -1 & -1 & -1 & -1 & 0 & 0 & 0 & 0 & 0 & 0 & 0 & 0 & 0 & 0 & 0 & 0 & 0 & 0 & 0 & 0 &
   1 & 1 & -1 & 1 & 0 & -1 & 0 & 0 & 0 & 0 & 0 & 0 & 0 & 0 & 0 & 0 & \cdots & 0 & 0 & \cdots & 0 & 0 & \cdots & 1
   \\
 0 & -1 & -1 & -1 & -1 & 0 & -1 & 0 & 0 & 0 & 0 & 0 & -1 & 0 & 0 & 0 & 0 & 0 & 0 & 0 & 0
   & -1 & 0 & 0 & 0 & 0 & 0 & 0 & 0 & 0 & 0 & 0 & 0 & 0 & 0 & 0 & 1 & \cdots & 0 & 0 & \cdots & 0 & 0 & \cdots & 0
   \\
 0 & -1 & -1 & -1 & -1 & 0 & 0 & 0 & 0 & 0 & 0 & 0 & -1 & 0 & 0 & 0 & 0 & 0 & 0 & 0 & 0
   & 0 & 0 & -1 & 1 & -1 & 0 & 0 & 0 & 0 & 0 & 0 & 0 & 0 & 0 & 0 & 1 & \cdots & 0 & 0 & \cdots & 0 & 0 & \cdots & 0
   \\
 0 & -1 & -1 & -1 & -1 & 0 & 0 & 0 & 0 & 0 & 0 & 0 & 0 & 0 & 0 & 0 & 0 & 0 & 0 & 0 & 0 &
   0 & 1 & -1 & 1 & -1 & 0 & 0 & 0 & 0 & 0 & 0 & 0 & 0 & 0 & 0 & 1 & \cdots & 0 & 0 & \cdots & 0 & 0 & \cdots & 0
   \\
 0 & -1 & -1 & -1 & -1 & 0 & 0 & 0 & 0 & 0 & 0 & 0 & 0 & 0 & 0 & 0 & 0 & 0 & 0 & 0 & 0 &
   -1 & 0 & 0 & 1 & -1 & 0 & 0 & 0 & 0 & 0 & 0 & 0 & 0 & 0 & 0 & 1 & \cdots & 0 & 0 & \cdots & 0 & 0 & \cdots & 0
   \\
 0 & -1 & -1 & -1 & -1 & 0 & 0 & 0 & 0 & 0 & 0 & 0 & -1 & 0 & 0 & 0 & 0 & 0 & 0 & 0 & 0
   & -1 & 0 & 0 & 1 & -1 & 0 & 0 & 0 & 0 & 0 & 0 & 0 & 0 & 0 & 0 & 1 & \cdots & 0 & 0 & \cdots & 0 & 0 & \cdots & 0
   \\
 0 & 0 & 0 & -1 & -1 & 0 & 0 & 0 & 0 & 0 & 0 & 0 & 0 & 0 & 0 & 0 & 0 & 0 & 0 & 0 & 0 & 0
   & 1 & -1 & 2 & -1 & -1 & 0 & 0 & 0 & 0 & 0 & 0 & 0 & 0 & 0 & 0 & \cdots & 0 & 0 & \cdots & 0 & 0 & \cdots & 0 \\
 0 & 0 & 0 & -1 & -1 & 0 & 0 & 0 & 0 & 0 & 0 & 0 & 0 & 0 & 0 & 0 & 0 & 0 & 0 & 0 & 0 & 0
   & 0 & 0 & 2 & -1 & -1 & 0 & 0 & 0 & 0 & 0 & 0 & 0 & 0 & 0 & 0 & \cdots & 0 & 0 & \cdots & 0 & 0 & \cdots & 0 \\
 0 & 0 & 0 & -1 & -1 & 0 & 0 & 0 & 0 & 0 & 0 & 0 & -1 & 0 & 0 & 0 & 0 & 0 & 1 & 0 & 0 &
   0 & 0 & 0 & 2 & -1 & -1 & 0 & 0 & 0 & 0 & 0 & 0 & 0 & 0 & 0 & 0 & \cdots & 0 & 0 & \cdots & 0 & 0 & \cdots & 0
   \\
 0 & 0 & 0 & -1 & -1 & 0 & 0 & 0 & 0 & 0 & 0 & 0 & 0 & 0 & 0 & 0 & 0 & 0 & 0 & 0 & 0 & 0
   & 1 & -1 & 1 & 0 & -1 & 0 & 0 & 0 & 0 & 0 & 0 & 0 & 0 & 0 & 0 & \cdots & 0 & 0 & \cdots & 0 & 0 & \cdots & 0 \\
 0 & 0 & 0 & -1 & -1 & 0 & 0 & 0 & 0 & 0 & 0 & 0 & 0 & 0 & 0 & 0 & 0 & 0 & 0 & 0 & 0 & 0
   & 0 & 0 & 1 & 0 & -1 & 0 & 0 & 0 & 0 & 0 & 0 & 0 & 0 & 0 & 0 & \cdots & 0 & 0 & \cdots & 0 & 0 & \cdots & 0 \\
 0 & 0 & 0 & -1 & -1 & 0 & 0 & 0 & 0 & 0 & 0 & 0 & -1 & 0 & 0 & 0 & 0 & 1 & 0 & 0 & 0 &
   0 & 0 & 0 & 1 & 0 & -1 & 0 & 0 & 0 & 0 & 0 & 0 & 0 & 0 & 0 & 0 & \cdots & 0 & 0 & \cdots & 0 & 0 & \cdots & 0 \\
 0 & 0 & 0 & -1 & -1 & 0 & 0 & 0 & 0 & 0 & 0 & 0 & -1 & 0 & 0 & 0 & 0 & 0 & 0 & 0 & 0 &
   0 & 0 & -1 & 1 & -1 & 0 & 0 & 0 & 0 & 0 & 0 & 0 & 0 & 0 & 0 & 0 & \cdots & 0 & 0 & \cdots & 0 & 0 & \cdots & 0
   \\
 0 & 0 & 0 & -1 & -1 & 0 & 0 & 0 & 0 & 0 & 0 & 0 & 0 & 0 & 0 & 0 & 0 & 0 & 0 & 0 & 0 & 0
   & 1 & -1 & 1 & -1 & 0 & 0 & 0 & 0 & 0 & 0 & 0 & 0 & 0 & 0 & 0 & \cdots & 0 & 0 & \cdots & 0 & 0 & \cdots & 0 \\
 0 & 0 & 0 & -1 & -1 & 0 & 0 & 0 & 0 & 0 & 0 & 0 & -1 & 0 & 0 & 0 & 0 & 0 & 0 & 0 & 0 &
   0 & -1 & 0 & 1 & -1 & 0 & 0 & 0 & 0 & 0 & 0 & 0 & 0 & 0 & 0 & 0 & \cdots & 0 & 0 & \cdots & 0 & 0 & \cdots & 0
   \\
 0 & 0 & 0 & -1 & -1 & 0 & 0 & 0 & 0 & 0 & 0 & 0 & 0 & 0 & 0 & 0 & 0 & 0 & 0 & 0 & 0 & 0
   & 0 & 0 & 1 & -1 & 0 & 0 & 0 & 0 & 0 & 0 & 0 & 0 & 0 & 0 & 0 & \cdots & 0 & 0 & \cdots & 0 & 0 & \cdots & 0 \\
 0 & 0 & 0 & -1 & -1 & 0 & 0 & 0 & 0 & 0 & 0 & 0 & -1 & 0 & 0 & 0 & 0 & 0 & 0 & 0 & 0 &
   -1 & 0 & 0 & 1 & -1 & 0 & 0 & 0 & 0 & 0 & 0 & 0 & 0 & 0 & 0 & 0 & \cdots & 0 & 0 & \cdots & 0 & 0 & \cdots & 0
   \\
 0 & 0 & 0 & -1 & -1 & 0 & 0 & 0 & 0 & 0 & 0 & 0 & -1 & 0 & 0 & 0 & 0 & 0 & 0 & 0 & 0 &
   0 & 0 & -1 & 0 & 0 & 0 & 0 & 0 & 0 & 0 & 0 & 0 & 0 & 0 & 0 & 0 & \cdots & 0 & 0 & \cdots & 0 & 0 & \cdots & 0 \\
 0 & 0 & 0 & -1 & -1 & 0 & 0 & 0 & 0 & 0 & 0 & 0 & 0 & 0 & 0 & 0 & 0 & 0 & 0 & 0 & 0 & 0
   & 1 & -1 & 0 & 0 & 0 & 0 & 0 & 0 & 0 & 0 & 0 & 0 & 0 & 0 & 0 & \cdots & 0 & 0 & \cdots & 0 & 0 & \cdots & 0 \\
 0 & 0 & 0 & -1 & -1 & 0 & 0 & 0 & 0 & 0 & 0 & 0 & -1 & 0 & 0 & 0 & 0 & 0 & 0 & 0 & 0 &
   0 & -1 & 0 & 0 & 0 & 0 & 0 & 0 & 0 & 0 & 0 & 0 & 0 & 0 & 0 & 0 & \cdots & 0 & 0 & \cdots & 0 & 0 & \cdots & 0 \\
 0 & 0 & 0 & -1 & -1 & 0 & 0 & 0 & 0 & 0 & 0 & 0 & -1 & 0 & 0 & 0 & 0 & 0 & 0 & 0 & 0 &
   -1 & 0 & 0 & 0 & 0 & 0 & 0 & 0 & 0 & 0 & 0 & 0 & 0 & 0 & 0 & 0 & \cdots & 0 & 0 & \cdots & 0 & 0 & \cdots & 0 \\
 1 & 0 & 0 & 0 & 0 & 0 & -1 & 0 & 0 & 0 & 0 & 0 & -1 & 0 & 0 & 0 & 0 & 0 & 0 & 0 & 0 & 1
   & 1 & -1 & 1 & 0 & -1 & 0 & 0 & 0 & 0 & 0 & 0 & 0 & 0 & 0 & 0 & \cdots & 0 & 0 & \cdots & 0 & 0 & \cdots & 0 \\
 0 & 0 & 0 & 0 & 0 & 0 & 0 & 0 & 0 & 0 & 0 & 0 & 0 & 0 & 0 & 0 & 0 & 0 & 0 & 0 & 0 & 1 &
   0 & 0 & 1 & -1 & -1 & 0 & 0 & 0 & 0 & 0 & 0 & 0 & 0 & 0 & 0 & \cdots & 0 & 0 & \cdots & 0 & 0 & \cdots & 0 \\
 0 & 0 & 0 & 0 & 0 & 0 & 0 & 0 & 0 & 0 & 0 & 0 & -1 & 0 & 0 & 0 & 1 & 0 & 0 & 0 & 0 & 1
   & 0 & 0 & 1 & -1 & -1 & 0 & 0 & 0 & 0 & 0 & 0 & 0 & 0 & 0 & 0 & \cdots & 0 & 0 & \cdots & 0 & 0 & \cdots & 0 \\
 0 & 0 & 0 & 0 & 0 & 0 & 0 & 0 & 0 & 0 & 0 & 0 & 0 & 0 & 0 & 0 & 0 & 0 & 0 & 0 & 0 & 1 &
   0 & 0 & 0 & 0 & -1 & 0 & 0 & 0 & 0 & 0 & 0 & 0 & 0 & 0 & 0 & \cdots & 0 & 0 & \cdots & 0 & 0 & \cdots & 0 \\
 0 & 0 & 0 & 0 & 0 & 0 & 0 & 0 & 0 & 0 & 0 & 0 & -1 & 0 & 0 & 1 & 0 & 0 & 0 & 0 & 0 & 1
   & 0 & 0 & 0 & 0 & -1 & 0 & 0 & 0 & 0 & 0 & 0 & 0 & 0 & 0 & 0 & \cdots & 0 & 0 & \cdots & 0 & 0 & \cdots & 0 \\
 0 & 0 & 0 & 0 & 0 & 0 & 0 & 0 & 0 & 0 & 0 & 0 & 0 & 0 & 0 & 0 & 0 & 0 & 0 & 0 & 0 & 1 &
   0 & 0 & 0 & 0 & 0 & 0 & 0 & 0 & 0 & 0 & 0 & 0 & 0 & 0 & -1 & \cdots & 0 & 0 & \cdots & 0 & 1 & \cdots & 0 \\
 0 & 0 & 0 & 0 & 0 & 0 & 0 & 0 & 0 & 0 & 0 & 0 & -1 & 0 & 0 & 0 & 0 & 0 & 0 & 0 & 0 & 1
   & -1 & 0 & 0 & -1 & 0 & 0 & 0 & 0 & 0 & 0 & 0 & 0 & 0 & 0 & 0 & \cdots & 0 & 0 & \cdots & 0 & 0 & \cdots & 0 \\
 0 & 0 & 0 & 0 & 0 & 0 & 0 & 0 & 0 & 0 & 0 & 0 & 0 & 0 & 0 & 0 & 0 & 0 & 0 & 0 & 0 & 1 &
   0 & 0 & 0 & -1 & 0 & 0 & 0 & 0 & 0 & 0 & 0 & 0 & 0 & 0 & 0 & \cdots & 0 & 0 & \cdots & 0 & 0 & \cdots & 0 \\
 0 & 0 & 0 & 0 & 0 & 0 & 0 & 0 & 0 & 0 & 0 & 0 & -1 & 0 & 0 & 0 & 0 & 0 & 0 & 0 & 0 & 1
   & -1 & 0 & -1 & 0 & 0 & 0 & 0 & 0 & 0 & 0 & 0 & 0 & 0 & 0 & 0 & \cdots & 0 & 0 & \cdots & 0 & 0 & \cdots & 0 \\
 0 & 0 & 0 & 0 & 0 & 0 & 0 & 0 & 0 & 0 & 0 & 0 & 0 & 0 & 0 & 0 & 0 & 0 & 0 & 0 & 0 & 1 &
   0 & 0 & -1 & 0 & 0 & 0 & 0 & 0 & 0 & 0 & 0 & 0 & 0 & 0 & 0 & \cdots & 0 & 0 & \cdots & 0 & 0 & \cdots & 0 \\
 0 & 0 & 0 & 0 & 0 & 0 & 0 & 0 & 0 & 0 & 0 & 0 & -1 & 0 & 1 & 0 & 0 & 0 & 0 & 0 & 0 & 1
   & 0 & -1 & 0 & 0 & 0 & 0 & 0 & 0 & 0 & 0 & 0 & 0 & 0 & 0 & 0 & \cdots & 0 & 0 & \cdots & 0 & 0 & \cdots & 0 \\
 0 & 0 & 0 & 0 & 0 & 0 & 0 & 0 & 0 & 0 & 0 & 0 & 0 & 1 & 0 & 0 & 0 & 0 & 0 & 0 & 0 & 1 &
   1 & -1 & 0 & 0 & 0 & 0 & 0 & 0 & 0 & 0 & 0 & 0 & 0 & 0 & 0 & \cdots & 0 & 0 & \cdots & 0 & 0 & \cdots & 0 \\
 0 & -1 & -1 & 0 & 0 & 0 & -1 & 0 & 0 & 0 & 0 & 0 & 0 & 0 & 0 & 0 & 0 & 0 & 0 & 0 & 0 &
   0 & 0 & 0 & 0 & 0 & -1 & 0 & 0 & 0 & 0 & 0 & 0 & 0 & 0 & 0 & 1 & \cdots & 0 & 0 & \cdots & 0 & 0 & \cdots & 0 \\
 0 & -1 & -1 & 0 & 0 & 0 & 0 & 0 & 0 & 0 & 0 & 0 & 0 & 0 & 0 & 0 & 0 & 0 & 0 & 0 & 0 & 1
   & 0 & 0 & 0 & 0 & -1 & 0 & 0 & 0 & 0 & 0 & 0 & 0 & 0 & 0 & 0 & \cdots & 0 & 0 & \cdots & 0 & 0 & \cdots & 0 \\
 0 & -1 & -1 & 0 & 0 & 0 & -1 & 0 & 0 & 0 & 0 & 0 & 0 & 0 & 0 & 0 & 0 & 0 & 0 & 0 & 0 &
   0 & 0 & 0 & -1 & 0 & 0 & 0 & 0 & 0 & 0 & 0 & 0 & 0 & 0 & 0 & 1 & \cdots & 0 & 0 & \cdots & 0 & 0 & \cdots & 0 \\
 0 & -1 & -1 & 0 & 0 & 0 & -1 & 0 & 0 & 0 & 0 & 1 & 0 & 0 & 0 & 0 & 0 & 0 & 0 & 0 & 0 &
   0 & 1 & -1 & 0 & 0 & 0 & 0 & 0 & 0 & 0 & 0 & 0 & 0 & 0 & 0 & 1 & \cdots & 0 & 0 & \cdots & 0 & 0 & \cdots & 0 \\
 0 & -1 & -1 & 0 & 0 & 0 & -1 & 0 & 0 & 0 & 0 & 0 & 0 & 0 & 0 & 0 & 0 & 0 & 0 & 0 & 0 &
   -1 & 0 & 0 & 0 & 0 & 0 & 0 & 0 & 0 & 0 & 0 & 0 & 0 & 0 & 0 & 1 & \cdots & 0 & 0 & \cdots & 0 & 0 & \cdots & 0 \\
 0 & -1 & -1 & 0 & 0 & 0 & 0 & 0 & 0 & 0 & 0 & 0 & 0 & 0 & 0 & 0 & 0 & 0 & 0 & 0 & 0 & 1
   & -1 & 0 & -1 & 0 & 0 & 0 & 0 & 0 & 0 & 0 & 0 & 0 & 0 & 0 & 0 & \cdots & 0 & 1 & \cdots & 0 & 0 & \cdots & 0 \\
 0 & -1 & -1 & 0 & 0 & 0 & 0 & 0 & 0 & 0 & 0 & 0 & 0 & 0 & 0 & 0 & 0 & 0 & 0 & 0 & 0 & 1
   & 0 & -1 & 0 & 0 & 0 & 0 & 0 & 0 & 0 & 0 & 0 & 0 & 0 & 0 & 0 & \cdots & 0 & 0 & \cdots & 0 & 0 & \cdots & 0 \\
 0 & -1 & -1 & 0 & 0 & 0 & 0 & 0 & 0 & 0 & 0 & 0 & 0 & 0 & 0 & 0 & 0 & 0 & 0 & 0 & 0 & 0
   & 0 & 0 & 0 & -1 & 0 & 0 & 1 & 0 & 0 & 0 & 0 & 0 & 0 & 0 & 1 & \cdots & 0 & 0 & \cdots & 0 & 0 & \cdots & 0 \\
 0 & 0 & 0 & 0 & 0 & 0 & -1 & 0 & 0 & 0 & 1 & 0 & 0 & 0 & 0 & 0 & 0 & 0 & 0 & 0 & 0 & 0
   & 1 & -1 & 1 & 0 & -1 & 0 & 0 & 0 & 0 & 0 & 0 & 0 & 0 & 0 & 0 & \cdots & 0 & 0 & \cdots & 0 & 0 & \cdots & 0 \\
 0 & 0 & 0 & 0 & 0 & 0 & -1 & 0 & 0 & 1 & 0 & 0 & 0 & 0 & 0 & 0 & 0 & 0 & 0 & 0 & 0 & 0
   & 0 & 0 & 1 & 0 & -1 & 0 & 0 & 0 & 0 & 0 & 0 & 0 & 0 & 0 & 0 & \cdots & 0 & 0 & \cdots & 0 & 0 & \cdots & 0 \\
 0 & 0 & 0 & 0 & 0 & 0 & 0 & 0 & 1 & 0 & 0 & 0 & 0 & 0 & 0 & 0 & 0 & 0 & 0 & 0 & 0 & 1 &
   0 & 0 & 1 & 0 & -1 & 0 & 0 & 0 & 0 & 0 & 0 & 0 & 0 & 0 & -1 & \cdots & 0 & 0 & \cdots & 0 & 0 & \cdots & 0 \\
 0 & 0 & 0 & 0 & 0 & 0 & 0 & 0 & 0 & 0 & 0 & 0 & 0 & 0 & 0 & 0 & 0 & 0 & 0 & 0 & 0 & 0 &
   0 & 0 & 1 & -1 & -1 & 1 & 0 & 0 & 0 & 0 & 0 & 0 & 0 & 0 & 0 & \cdots & 0 & 0 & \cdots & 0 & 0 & \cdots & 0 \\
 0 & 0 & 0 & 0 & 0 & 0 & -1 & 1 & 0 & 0 & 0 & 0 & 0 & 0 & 0 & 0 & 0 & 0 & 0 & 0 & 0 & 0
   & 1 & -1 & 0 & 0 & 0 & 0 & 0 & 0 & 0 & 0 & 0 & 0 & 0 & 0 & 0 & \cdots & 0 & 0 & \cdots & 0 & 0 & \cdots & 0 \\
 0 & 0 & 0 & 0 & 0 & 0 & 0 & 0 & 0 & 0 & 0 & 0 & 0 & 0 & 0 & 0 & 0 & 0 & 0 & 0 & 0 & 1 &
   0 & -1 & 0 & 0 & 0 & 0 & 0 & 0 & 0 & 0 & 0 & 0 & 0 & 0 & -1 & \cdots & 0 & 0 & \cdots & 0 & 0 & \cdots & 0 \\
 0 & 0 & 0 & 0 & 0 & 0 & 0 & 0 & 0 & 0 & 0 & 0 & 0 & 0 & 0 & 0 & 0 & 0 & 0 & 0 & 0 & 1 &
   -1 & 0 & 0 & 0 & 0 & 0 & 0 & 0 & 0 & 0 & 0 & 0 & 0 & 0 & -1 & \cdots & 0 & 0 & \cdots & 0 & 0 & \cdots & 0 \\
\end{array}
\right)
$}
~,~
\nn\\
\eea
and
the $D$-term charge matrix takes the form,
\beal{es04c05}
&&
Q_{D}=
\nn\\
&&
\resizebox{1\textwidth}{!}{$
\left(
\begin{array}{cccccc|ccccccccccccccc|ccccccccccccccc|ccc|ccc|ccc}
p_{1} & p_{2} & p_{3} & p_{4} & p_{5} & p_{6} & s^{(1)}_{1} & s^{(1)}_{2} & s^{(1)}_{3} & s^{(1)}_{4} & s^{(1)}_{5} & s^{(1)}_{6} & s^{(1)}_{7} & s^{(1)}_{8} & s^{(1)}_{9} & s^{(1)}_{10} & s^{(1)}_{11} & s^{(1)}_{12} & s^{(1)}_{13} & s^{(1)}_{14} & s^{(1)}_{15} & s^{(2)}_{1} & s^{(2)}_{2} & s^{(2)}_{3} & s^{(2)}_{4} & s^{(2)}_{5} & s^{(2)}_{6} & s^{(2)}_{7} & s^{(2)}_{8} & s^{(2)}_{9} & s^{(2)}_{10} & s^{(2)}_{11} & s^{(2)}_{12} & s^{(2)}_{13} & s^{(2)}_{14} & s^{(2)}_{15} & o^{(1)}_{1} & \cdots & o^{(1)}_{44} & o^{(2)}_{1} & \cdots & o^{(2)}_{3} & o^{(3)}_{1} & \cdots & o^{(3)}_{3} \\
\hline
 0 & 0 & 0 & 1 & 1 & 0 & 0 & 0 & 0 & 0 & 0 & 0 & 1 & 0 & 0 & 0 & 0 & 0 & 0 & 0 & 0 & 1 &
   0 & 0 & -1 & 1 & 0 & 0 & 0 & 0 & 0 & 0 & 0 & 0 & 0 & 0 & -1 & \cdots & 0 & 0 & \cdots & 0 & 0 & \cdots & 0 \\
 0 & 0 & 0 & 0 & 0 & 0 & -1 & 0 & 0 & 0 & 0 & 0 & 0 & 0 & 0 & 0 & 0 & 0 & 0 & 0 & 0 & -1
   & 0 & 0 & 0 & 0 & 0 & 0 & 0 & 0 & 0 & 0 & 0 & 0 & 0 & 0 & 1 & \cdots & 0 & 0 & \cdots & 0 & 0 & \cdots & 0 \\
 0 & 0 & 0 & -1 & -1 & 0 & 0 & 0 & 0 & 0 & 0 & 0 & 0 & 0 & 0 & 0 & 0 & 0 & 0 & 0 & 0 &
   -1 & 0 & 0 & 1 & 0 & 0 & 0 & 0 & 0 & 0 & 0 & 0 & 0 & 0 & 0 & 0 & \cdots & 0 & 0 & \cdots & 0 & 0 & \cdots & 0 \\
 0 & 0 & 0 & 0 & 0 & 0 & 0 & 0 & 0 & 0 & 0 & 0 & -1 & 0 & 0 & 0 & 0 & 0 & 0 & 0 & 0 & 0
   & -1 & 0 & 0 & 0 & 0 & 0 & 0 & 0 & 0 & 0 & 0 & 0 & 0 & 0 & 0 & \cdots & 0 & 0 & \cdots & 0 & 0 & \cdots & 0 \\
 0 & 0 & 0 & 0 & 0 & 0 & 0 & 0 & 0 & 0 & 0 & 0 & 0 & 0 & 0 & 0 & 0 & 0 & 0 & 0 & 0 & -1
   & 0 & 1 & 0 & 0 & 0 & 0 & 0 & 0 & 0 & 0 & 0 & 0 & 0 & 0 & 0 & \cdots & 0 & 0 & \cdots & 0 & 0 & \cdots & 0 \\
 0 & 0 & 0 & 0 & 0 & 0 & 0 & 0 & 0 & 0 & 0 & 0 & 0 & 0 & 0 & 0 & 0 & 0 & 0 & 0 & 0 & 1 &
   0 & 0 & 0 & 0 & -1 & 0 & 0 & 0 & 0 & 0 & 0 & 0 & 0 & 0 & 0 & \cdots & 0 & 0 & \cdots & 0 & 0 & \cdots & 0 \\
 0 & 0 & 0 & 0 & 0 & 1 & 1 & 0 & 0 & 0 & 0 & 0 & 0 & 0 & 0 & 0 & 0 & 0 & 0 & 0 & 0 & 0 &
   -1 & 0 & 0 & -1 & 0 & 0 & 0 & 0 & 0 & 0 & 0 & 0 & 0 & 0 & 0 & \cdots & 0 & 0 & \cdots & 0 & 0 & \cdots & 0 \\
 0 & 0 & 0 & 0 & 0 & 0 & 0 & 0 & 0 & 0 & 0 & 0 & 0 & 0 & 0 & 0 & 0 & 0 & 0 & 0 & 0 & 0 &
   1 & -1 & 0 & 0 & 0 & 0 & 0 & 0 & 0 & 0 & 0 & 0 & 0 & 0 & 0 & \cdots & 0 & 0 & \cdots & 0 & 0 & \cdots & 0 \\
 0 & -1 & -1 & 0 & 0 & 0 & 0 & 0 & 0 & 0 & 0 & 0 & 0 & 0 & 0 & 0 & 0 & 0 & 0 & 0 & 0 & 0
   & 0 & 0 & 0 & 0 & 0 & 0 & 0 & 0 & 0 & 0 & 0 & 0 & 0 & 0 & 1 & \cdots & 0 & 0 & \cdots & 0 & 0 & \cdots & 0 \\
 0 & 1 & 1 & 0 & 0 & -1 & 0 & 0 & 0 & 0 & 0 & 0 & 0 & 0 & 0 & 0 & 0 & 0 & 0 & 0 & 0 & 1
   & 1 & 0 & 0 & 1 & 0 & 0 & 0 & 0 & 0 & 0 & 0 & 0 & 0 & 0 & -1 & \cdots & 0 & 0 & \cdots & 0 & 0 & \cdots & 0 \\
 0 & 0 & 0 & 0 & 0 & 0 & 0 & 0 & 0 & 0 & 0 & 0 & 0 & 0 & 0 & 0 & 0 & 0 & 0 & 0 & 0 & 0 &
   0 & 0 & -1 & 0 & 1 & 0 & 0 & 0 & 0 & 0 & 0 & 0 & 0 & 0 & 0 & \cdots & 0 & 0 & \cdots & 0 & 0 & \cdots & 0 \\
\end{array}
\right)
$}
~,~
\nn\\
\eea

\begin{figure}[H]
\centering
\includegraphics[width=0.3\textwidth]{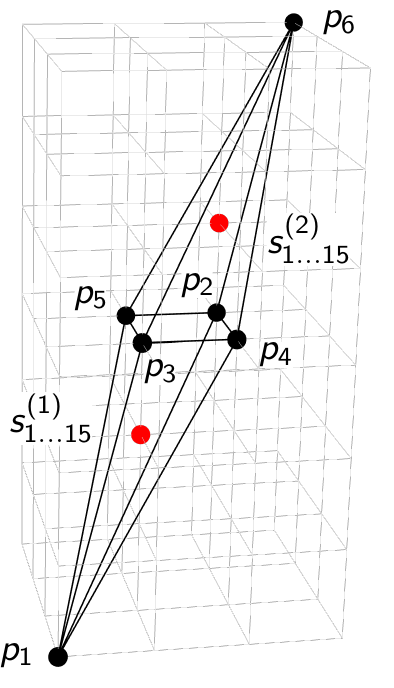}
\caption{
The toric diagram for the $Q^{1,1,1}/\mathbb{Z}_3 \ (1,1,1,1,2,2,2,2)$ model.
\label{fig_413_toric}
}
\end{figure}

The toric diagram of the $Q^{1,1,1}/\mathbb{Z}_3 \ (1,1,1,1,2,2,2,2)$ model 
is shown in \fref{fig_413_toric} and is given by, 
\beal{es04c06}
&&
G_{t}=
\resizebox{0.8\textwidth}{!}{$
\left(
\begin{array}{cccccc|ccc|ccc|ccc|ccc|ccc}
p_{1} & p_{2} & p_{3} & p_{4} & p_{5} & p_{6}
 & s^{(1)}_{1} & \cdots & s^{(1)}_{15}
 & s^{(2)}_{1} & \cdots & s^{(2)}_{15}
 & o^{(1)}_{1} & \cdots & o^{(1)}_{44}
 & o^{(2)}_{1} & \cdots & o^{(2)}_{3}
 & o^{(3)}_{1} & \cdots & o^{(3)}_{3} \\
\hline
-3 & 0 & 0 & 0 & 0 &  3 & -1 & \cdots & -1 & 1 & \cdots & 1 & 0 & \cdots & 0 &  1 & \cdots &  1 & -1 & \cdots & -1 \\
 2 & 0 & 1 & 0 & 1 & -1 &  1 & \cdots &  1 & 0 & \cdots & 0 & 1 & \cdots & 1 &  1 & \cdots &  1 &  2 & \cdots &  2 \\
 2 & 0 & 1 & 1 & 0 & -1 &  1 & \cdots &  1 & 0 & \cdots & 0 & 1 & \cdots & 1 &  1 & \cdots &  1 &  2 & \cdots &  2 \\
\hline
 1 & 1 & 1 & 1 & 1 &  1 &  1 & \cdots &  1 & 1 & \cdots & 1 & 2 & \cdots & 2 &  3 & \cdots &  3 &  3 & \cdots &  3 \\
\end{array}
\right)
$}
~.~
\nn\\
\eea

%-------------------
%-------------------

The global symmetry of the mesonic moduli space of the $Q^{1,1,1}/\mathbb{Z}_3 \ (1,1,1,1,2,2,2,2)$ model
takes the following enhanced form,
\beal{es04c07}
SU(2)_x \times SU(2)_y \times U(1)_f \times U(1)_R
~,~
\eea
where $SU(2)_x \times SU(2)_y \times U(1)_f $ is the mesonic flavor symmetry.
\tref{tab_04c01} summarizes the charges under the mesonic flavor symmetry
on the extremal GLSM fields $p_a$ of the brane brick model.

%-------------------
\begin{table}[httt!!]
\centering
\begin{tabular}{|c|c|c|c|l|}
\hline
\; & $SU(2)_x$ & $SU(2)_{y}$ & $U(1)_{f}$  & fugacity \\
\hline
$p_1$ & $0$ & $0$ & $+1$  & $t_1=f t$ \\
$p_2$ & $-1$ & $0$ & $0$  & $t_2=x^{-1} t$\\
$p_3$ & $+1$ & $0$ & $0$  & $t_3=x t$\\
$p_4$ & $0$ & $-1$ & $0$  & $t_4=y^{-1} t$ \\
$p_5$ & $0$ & $+1$ & $0$ & $t_5=y t$ \\
$p_6$ & $0$ & $0$ & $-1$ & $t_6=f^{-1} t$ \\
\hline
\end{tabular}
\caption{
Mesonic flavor symmetry of the $Q^{1,1,1}/\mathbb{Z}_3 \ (1,1,1,1,2,2,2,2)$ model 
and charges on the extremal GLSM fields $p_a$. 
Here, the fugacity $t$ counts the degree in extremal GLSM fields $p_a$. 
\label{tab_04c01}}
\end{table}
%-------------------

The unrefined Hilbert series of the moduli space of the $Q^{1,1,1}/\mathbb{Z}_3 \ (1,1,1,1,2,2,2,2)$ model is given by,
\beal{es04c08}
g(t; \mathcal{M}^{mes})=\frac{1-t^3 + 9t^6 +20 t^9 -4 t^{12} +20 t^{15} + 9t^{18} -t^{21} +t^{24}}{(1-t^3) (1-t^9)^3}
~,~
\eea
where the palindromic numerator indicates that the mesonic moduli space is Calabi-Yau.
Here, the fugacity $t$ counts the degree in extremal GLSM fields $p_a$. 

The Hilbert series of the mesonic moduli space can be refined under the fugacity assignment 
based on the mesonic flavor symmetry 
as summarized in \tref{tab_04c01}.
The corresponding highest weight generating function is given by,
\beal{es04c09}
h(\mu, \nu, f , t; \mathcal{M}^{mes})=\frac{1-\mu^6 \nu^6 t^{18}}{(1-\mu^2 \nu^2 t^6)(1- \mu^3 \nu^3 f^3 t^9)(1- \mu^3 \nu^3 f^{-3} t^9)}
~,~
\eea
where $\mu^m \nu^n$ is fugacity for counting the $SU(2)_x \times SU(2)_y$ character of the form $[m]_x [n]_y$, 
and $f$ is the fugacity associated to $U(1)_f$.

%-------------------
\begin {table}[hhtt!!]
\centering
\resizebox{0.55\textwidth}{!}{
\begin{tabular}{|c|c|ccc|}
\hline
PL term & generator & $SU(2)_x$ & $SU(2)_y$ & $U(1)_{f}$ 
\\
\hline
\multirow{9}{*}{$+[2]_x [2]_y t^6$}
& $p_1 p_3^2 p_5^2 p_6 ~s_{{(1)}} s_{{(2)}} $ & $2$ & $2$ & $0$  \\
& $p_1 p_3^2 p_4 p_5 p_6 ~s_{{(1)}} s_{{(2)}} $ & $2$ & $0$ & $0$ \\
& $p_1 p_3^2 p_4^2 p_6 ~s_{{(1)}} s_{{(2)}} $ & $2$ & $-2$ & $0$  \\
& $p_1 p_2 p_3 p_5^2 p_6 ~s_{{(1)}} s_{{(2)}} $ & $0$ & $2$ & $0$  \\
& $p_1 p_2 p_3 p_4 p_5 p_6 ~s_{{(1)}} s_{{(2)}} $ & $0$ & $0$ & $0$  \\
& $p_1 p_2 p_3 p_4^2 p_6 ~s_{{(1)}} s_{{(2)}} $ & $0$ & $-2$ & $0$  \\
& $p_1 p_2^2 p_5^2 p_6 ~s_{{(1)}} s_{{(2)}} $ & $-2$ & $2$ & $0$  \\
& $p_1 p_2^2 p_4 p_5 p_6 ~s_{{(1)}} s_{{(2)}} $ & $-2$ & $0$ & $0$  \\
& $p_1 p_2^2 p_4^2 p_6 ~s_{{(1)}} s_{{(2)}} $ & $-2$ & $-2$ & $0$  \\
\hline
\multirow{16}{*}{$+[3]_x [3]_y f^3 t^9$}
& $p_1^3 p_3^3 p_5^3 ~s_{{(1)}}^2 s_{{(2)}} $ & $3$ & $3$ & $3$  \\

& $p_1^3 p_3^3 p_4 p_5^2 ~s_{{(1)}}^2 s_{{(2)}} $ & $3$ & $1$ & $3$  \\

& $p_1^3 p_3^3 p_4^2 p_5 ~s_{{(1)}}^2 s_{{(2)}} $ & $3$ & $-1$ & $3$  \\

& $p_1^3 p_3^3 p_4^3 ~s_{{(1)}}^2 s_{{(2)}} $ & $3$ & $-3$ & $3$  \\

& $p_1^3 p_2 p_3^2 p_5^3 ~s_{{(1)}}^2 s_{{(2)}} $ & $1$ & $3$ & $3$  \\

& $p_1^3 p_2 p_3^2 p_4 p_5^2 ~s_{{(1)}}^2 s_{{(2)}} $ & $1$ & $1$ & $3$  \\

& $p_1^3 p_2 p_3^2 p_4^2 p_5 ~s_{{(1)}}^2 s_{{(2)}} $ & $1$ & $-1$ & $3$  \\

& $p_1^3 p_2 p_3^2 p_4^3 ~s_{{(1)}}^2 s_{{(2)}} $ & $1$ & $-3$ & $3$  \\

& $p_1^3 p_2^2 p_3 p_5^3 ~s_{{(1)}}^2 s_{{(2)}} $ & $-1$ & $3$ & $3$  \\

& $p_1^3 p_2^2 p_3 p_4 p_5^2 ~s_{{(1)}}^2 s_{{(2)}} $ & $-1$ & $1$ & $3$  \\

& $p_1^3 p_2^2 p_3 p_4^2 p_5 ~s_{{(1)}}^2 s_{{(2)}} $ & $-1$ & $-1$ & $3$  \\

& $p_1^3 p_2^2 p_3 p_4^3 ~s_{{(1)}}^2 s_{{(2)}} $ & $-1$ & $-3$ & $3$  \\

& $p_1^3 p_2^3 p_5^3 ~s_{{(1)}}^2 s_{{(2)}} $ & $-3$ & $3$ & $3$  \\

& $p_1^3 p_2^3 p_4 p_5^2 ~s_{{(1)}}^2 s_{{(2)}} $ & $-3$ & $1$ & $3$  \\

& $p_1^3 p_2^3 p_4^2 p_5 ~s_{{(1)}}^2 s_{{(2)}} $ & $-3$ & $-1$ & $3$  \\

& $p_1^3 p_2^3 p_4^3 ~s_{{(1)}}^2 s_{{(2)}} $ & $-3$ & $-3$ & $3$  \\
\hline
\multirow{16}{*}{$+[3]_x [3]_y f^{-3} t^9$}
& $p_3^3 p_5^3 p_6^3 ~s_{{(1)}} s_{{(2)}}^2 $ & $3$ & $3$ & $-3$  \\

& $p_3^3 p_4 p_5^2 p_6^3 ~s_{{(1)}} s_{{(2)}}^2 $ & $3$ & $1$ & $-3$  \\

& $p_3^3 p_4^2 p_5 p_6^3 ~s_{{(1)}} s_{{(2)}}^2 $ & $3$ & $-1$ & $-3$  \\

& $p_3^3 p_4^3 p_6^3 ~s_{{(1)}} s_{{(2)}}^2 $ & $3$ & $-3$ & $-3$  \\

& $p_2 p_3^2 p_5^3 p_6^3 ~s_{{(1)}} s_{{(2)}}^2 $ & $1$ & $3$ & $-3$  \\

& $p_2 p_3^2 p_4 p_5^2 p_6^3 ~s_{{(1)}} s_{{(2)}}^2 $ & $1$ & $1$ & $-3$  \\

& $p_2 p_3^2 p_4^2 p_5 p_6^3 ~s_{{(1)}} s_{{(2)}}^2 $ & $1$ & $-1$ & $-3$  \\

& $p_2 p_3^2 p_4^3 p_6^3 ~s_{{(1)}} s_{{(2)}}^2 $ & $1$ & $-3$ & $-3$  \\

& $p_2^2 p_3 p_5^3 p_6^3 ~s_{{(1)}} s_{{(2)}}^2 $ & $-1$ & $3$ & $-3$  \\

& $p_2^2 p_3 p_4 p_5^2 p_6^3 ~s_{{(1)}} s_{{(2)}}^2 $ & $-1$ & $1$ & $-3$  \\

& $p_2^2 p_3 p_4^2 p_5 p_6^3 ~s_{{(1)}} s_{{(2)}}^2 $ & $-1$ & $-1$ & $-3$  \\

& $p_2^2 p_3 p_4^3 p_6^3 ~s_{{(1)}} s_{{(2)}}^2 $ & $-1$ & $-3$ & $-3$  \\

& $p_2^3 p_5^3 p_6^3 ~s_{{(1)}} s_{{(2)}}^2 $ & $-3$ & $3$ & $-3$  \\

& $p_2^3 p_4 p_5^2 p_6^3 ~s_{{(1)}} s_{{(2)}}^2 $ & $-3$ & $1$ & $-3$  \\

& $p_2^3 p_4^2 p_5 p_6^3 ~s_{{(1)}} s_{{(2)}}^2 $ & $-3$ & $-1$ & $-3$  \\

& $p_2^3 p_4^3 p_6^3 ~s_{{(1)}} s_{{(2)}}^2 $ & $-3$ & $-3$ & $-3$  \\
\hline
\end{tabular}
}
\caption{
Generators of the $Q^{1,1,1}/\mathbb{Z}_3 \ (1,1,1,1,2,2,2,2)$ model in terms of GLSM fields and their corresponding mesonic flavor charges. Here, we denote $s_{(1)}=\prod_{i=1}^{15}s_i^{(1)}$,  $s_{(2)}=\prod_{i=1}^{15} s_i^{(2)}$, and set extra GLSM fields to 1. 
\label{tab_04c02}
}
\end{table}
%-------------------

The plethystic logarithm of the mesonic flavor symmetry refined Hilbert series 
takes the following form, 
\beal{es04c10}
&&
\text{PL}[g(x,y,f,t; \mathcal{M}^{mes})]=
[2]_x [2]_y t^6 + ([3]_x [3]_y f^3 + [3]_x [3]_y f^{-3}) t^9
-( [2]_x [2]_y +  [4]_y
\nn\\
&&
\hspace{1cm}
+ [4]_x  +  1 ) t^{12}
-( [3]_x [5]_y f^3 + [1]_x [5]_y f^3
+ [3]_x [5]_y f^{-3} + [1]_x [5]_y f^{-3}
+ [5]_x [3]_y f^3
\nn\\
&&
\hspace{1cm}
 + [5]_x [3]_y f^{-3}
+ [3]_x [3]_y f^3 + [3]_x [3]_y f^{-3}
+ [1]_x [3]_y f^3 + [1]_x [3]_y f^{-3}
+ [5]_x [1]_y f^3
\nn\\
&&
\hspace{1cm}
 + [5]_x [1]_y f^{-3}
+ [3]_x [1]_y f^3 + [3]_x [1]_y f^{-3}
+ [1]_x [1]_y f^3 + [1]_x [1]_y f^{-3} ) t^{15}
\nn\\
&&
\hspace{1cm}
- ( [6]_x [6]_y + [6]_x [4]_y + [4]_x [6]_y
+[2]_x [6]_y + [6]_x [2]_y + [4]_x [4]_y
+ [2]_x [6]_y f^6 
\nn\\
&&
\hspace{1cm}
+ [2]_x [6]_y f^{-6} +  [6]_y
+  [4]_y +  [2]_y + [4]_x [4]_y f^6
+ [2]_x [4]_y + [4]_x [2]_y + [4]_x [4]_y f^{-6}
\nn\\
&&
\hspace{1cm}
+ [6]_x [2]_y f^{-6} + [2]_x [2]_y + [6]_x [2]_y f^6
+ [6]_x  + [4]_x  + [2]_x  +  [4]_y f^{-6}
+  [4]_y f^6 
\nn\\
&&
\hspace{1cm}
+ [4]_x  f^6 
+ [4]_x  f^{-6} + [2]_x [2]_y f^6
+   f^6 +   f^{-6} + [2]_x [2]_y f^{-6} ) t^{18}
+ \dots
~,~
\eea
where $[m]_x [n]_y$ is the character of the irreducible representation of $SU(2)_x \times SU(2)_y$ with highest weight $(m)_x (n)_y$.
The infinite expansion of the plethystic logarithm indicates that the mesonic moduli space of the $Q^{1,1,1}/\mathbb{Z}_3 \ (1,1,1,1,2,2,2,2)$  model is not a complete intersection.
The generators of the mesonic moduli space with their mesonic flavor charges are summarized in \tref{tab_04c02}.

We note here that the abelian orbifold $Q^{1,1,1}/\mathbb{Z}_3 \ (1,1,1,1,2,2,2,2)$ is part of a family of abelian orbifolds of the form 
$Q^{1,1,1}/\mathbb{Z}_n \ (1,1,1,1,-1,-1,-1,-1)$ as discussed in section \sref{sec041}. 
\\

%=================================================================
\subsubsection{$Q^{1,1,1}/ \mathbb{Z}_3 \ (0,0,1,1,2,2,0,0)$ \label{sec044}}

%-------------------
\begin{figure}[H]
    \centering
    \includegraphics[width=0.4\textwidth]{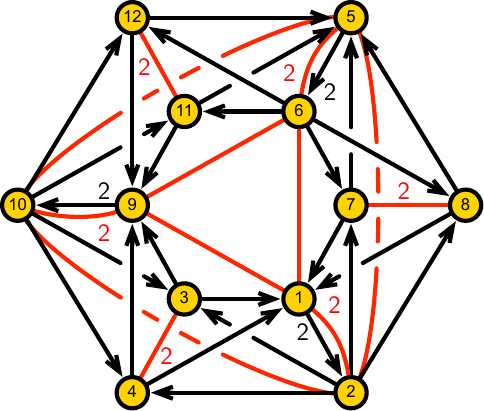}
    \caption{
    The quiver for the $Q^{1,1,1}/ \mathbb{Z}_3 \ (0,0,1,1,2,2,0,0)$ model.
    \label{fig_414_quiver}
    }
\end{figure}
%-------------------

The $J$- and $E$-terms for the brane brick model corresponding to the abelian orbifold of the form
$Q^{1,1,1}/ \mathbb{Z}_3 \ (0,0,1,1,2,2,0,0)$ are as follows, 
\beal{es04d01}
\resizebox{1\textwidth}{!}{$
\begin{array}{rclccclcccc}
&& && &J& && &E& \\
&&\Lambda^{(1)}_{2, 1} &:& & D_{1, 2} \cdot Z_{2, 8} \cdot Y_{8, 1} \cdot X_{1, 2} - X_{1, 2} \cdot Z_{2, 3} \cdot D_{3, 1} \cdot D_{1, 2}& && & X_{2, 7} \cdot Y_{7, 1} - X_{2, 4} \cdot D_{4, 1}& \\ 
 &&\Lambda^{(1)}_{6, 5} &:& & D_{5, 6} \cdot Z_{6, 12} \cdot Y_{12, 5} \cdot X_{5, 6} - X_{5, 6} \cdot Z_{6, 7} \cdot D_{7, 5} \cdot D_{5, 6}& && & X_{6, 11} \cdot Y_{11, 5} - X_{6, 8} \cdot D_{8, 5}& \\ 
 &&\Lambda^{(1)}_{10, 9} &:& & D_{9, 10} \cdot Z_{10, 4} \cdot Y_{4, 9} \cdot X_{9, 10} - X_{9, 10} \cdot Z_{10, 11} \cdot D_{11, 9} \cdot D_{9, 10}& && & X_{10, 3} \cdot Y_{3, 9} - X_{10, 12} \cdot D_{12, 9}& \\ 
 &&\Lambda^{(2)}_{10, 1} &:& & X_{1, 2} \cdot X_{2, 4} \cdot Y_{4, 9} \cdot D_{9, 10} - D_{1, 2} \cdot Z_{2, 3} \cdot Y_{3, 9} \cdot X_{9, 10}& && & X_{10, 3} \cdot D_{3, 1} - Z_{10, 4} \cdot D_{4, 1}& \\ 
 &&\Lambda^{(2)}_{2, 5} &:& & X_{5, 6} \cdot X_{6, 8} \cdot Y_{8, 1} \cdot D_{1, 2} - D_{5, 6} \cdot Z_{6, 7} \cdot Y_{7, 1} \cdot X_{1, 2}& && & X_{2, 7} \cdot D_{7, 5} - Z_{2, 8} \cdot D_{8, 5}& \\ 
 &&\Lambda^{(2)}_{6, 9} &:& & X_{9, 10} \cdot X_{10, 12} \cdot Y_{12, 5} \cdot D_{5, 6} - D_{9, 10} \cdot Z_{10, 11} \cdot Y_{11, 5} \cdot X_{5, 6}& && & X_{6, 11} \cdot D_{11, 9} - Z_{6, 12} \cdot D_{12, 9}& \\ 
 &&\Lambda^{(3)}_{6, 1} &:& & D_{1, 2} \cdot Z_{2, 8} \cdot D_{8, 5} \cdot X_{5, 6} - X_{1, 2} \cdot X_{2, 7} \cdot D_{7, 5} \cdot D_{5, 6}& && & X_{6, 8} \cdot Y_{8, 1} - Z_{6, 7} \cdot Y_{7, 1}& \\ 
 &&\Lambda^{(3)}_{10, 5} &:& & D_{5, 6} \cdot Z_{6, 12} \cdot D_{12, 9} \cdot X_{9, 10} - X_{5, 6} \cdot X_{6, 11} \cdot D_{11, 9} \cdot D_{9, 10}& && & X_{10, 12} \cdot Y_{12, 5} - Z_{10, 11} \cdot Y_{11, 5}& \\ 
 &&\Lambda^{(3)}_{2, 9} &:& & D_{9, 10} \cdot Z_{10, 4} \cdot D_{4, 1} \cdot X_{1, 2} - X_{9, 10} \cdot X_{10, 3} \cdot D_{3, 1} \cdot D_{1, 2}& && & X_{2, 4} \cdot Y_{4, 9} - Z_{2, 3} \cdot Y_{3, 9}& \\ 
 &&\Lambda^{(4)}_{2, 1} &:& & D_{1, 2} \cdot X_{2, 7} \cdot Y_{7, 1} \cdot X_{1, 2} - X_{1, 2} \cdot X_{2, 4} \cdot D_{4, 1} \cdot D_{1, 2}& && & Z_{2, 3} \cdot D_{3, 1} - Z_{2, 8} \cdot Y_{8, 1}& \\ 
 &&\Lambda^{(4)}_{6, 5} &:& & D_{5, 6} \cdot X_{6, 11} \cdot Y_{11, 5} \cdot X_{5, 6} - X_{5, 6} \cdot X_{6, 8} \cdot D_{8, 5} \cdot D_{5, 6}& && & Z_{6, 7} \cdot D_{7, 5} - Z_{6, 12} \cdot Y_{12, 5}& \\ 
 &&\Lambda^{(4)}_{10, 9} &:& & D_{9, 10} \cdot X_{10, 3} \cdot Y_{3, 9} \cdot X_{9, 10} - X_{9, 10} \cdot X_{10, 12} \cdot D_{12, 9} \cdot D_{9, 10}& && & Z_{10, 11} \cdot D_{11, 9} - Z_{10, 4} \cdot Y_{4, 9}& \\ 
 &&\Lambda^{(5)}_{4, 3} &:& & D_{3, 1} \cdot X_{1, 2} \cdot X_{2, 4} - Y_{3, 9} \cdot X_{9, 10} \cdot Z_{10, 4}& && & D_{4, 1} \cdot D_{1, 2} \cdot Z_{2, 3} - Y_{4, 9} \cdot D_{9, 10} \cdot X_{10, 3}& \\ 
 &&\Lambda^{(5)}_{8, 7} &:& & D_{7, 5} \cdot X_{5, 6} \cdot X_{6, 8} - Y_{7, 1} \cdot X_{1, 2} \cdot Z_{2, 8}& && & D_{8, 5} \cdot D_{5, 6} \cdot Z_{6, 7} - Y_{8, 1} \cdot D_{1, 2} \cdot X_{2, 7}& \\ 
 &&\Lambda^{(5)}_{12, 11} &:& & D_{11, 9} \cdot X_{9, 10} \cdot X_{10, 12} - Y_{11, 5} \cdot X_{5, 6} \cdot Z_{6, 12}& && & D_{12, 9} \cdot D_{9, 10} \cdot Z_{10, 11} - Y_{12, 5} \cdot D_{5, 6} \cdot X_{6, 11}& \\ 
 &&\Lambda^{(6)}_{3, 4} &:& & Y_{4, 9} \cdot X_{9, 10} \cdot X_{10, 3} - D_{4, 1} \cdot X_{1, 2} \cdot Z_{2, 3}& && & D_{3, 1} \cdot D_{1, 2} \cdot X_{2, 4} - Y_{3, 9} \cdot D_{9, 10} \cdot Z_{10, 4}& \\ 
 &&\Lambda^{(6)}_{7, 8} &:& & Y_{8, 1} \cdot X_{1, 2} \cdot X_{2, 7} - D_{8, 5} \cdot X_{5, 6} \cdot Z_{6, 7}& && & D_{7, 5} \cdot D_{5, 6} \cdot X_{6, 8} - Y_{7, 1} \cdot D_{1, 2} \cdot Z_{2, 8}& \\ 
 &&\Lambda^{(6)}_{11, 12} &:& & Y_{12, 5} \cdot X_{5, 6} \cdot X_{6, 11} - D_{12, 9} \cdot X_{9, 10} \cdot Z_{10, 11}& && & D_{11, 9} \cdot D_{9, 10} \cdot X_{10, 12} - Y_{11, 5} \cdot D_{5, 6} \cdot Z_{6, 12}&
\end{array}
$}
\nn\\
\eea
The corresponding quiver diagram is shown in \fref{fig_414_quiver}.
The $J$- and $E$-terms
come from \eqref{es03b01} 
with the following additional relabelling of indices, 
\beal{es04d02}
&
[1,0] \rightarrow 1 ~,~
[2,0] \rightarrow 2 ~,~
[3,0] \rightarrow 3 ~,~
[4,0] \rightarrow 4 ~,~
&
\nn\\
&
[1,1] \rightarrow 5 ~,~
[2,1] \rightarrow 6 ~,~
[3,1] \rightarrow 7 ~,~
[4,1] \rightarrow 8 ~,~
&
\nn\\
&
[1,2] \rightarrow 9 ~,~
[2,2] \rightarrow 10 ~,~
[3,2] \rightarrow 11 ~,~
[4,2] \rightarrow 12 ~.~
&
\eea

The $P$-matrix of the $Q^{1,1,1}/ \mathbb{Z}_3 \ (0,0,1,1,2,2,0,0)$ model takes the following form,
\beal{es04d03}
&&
P=
\nn\\
&&
\resizebox{0.95\textwidth}{!}{$
\left(
\begin{array}{c|cccccc|ccc|ccc|ccc|ccc|ccc|ccc|ccc|ccc|ccc}
 & p_{1} & p_{2} & p_{3} & p_{4} & p_{5} & p_{6} & q^{(1)}_{1} & q^{(1)}_{2} & q^{(1)}_{3} & q^{(2)}_{1} & q^{(2)}_{2} & q^{(2)}_{3} & q^{(3)}_{1} & q^{(3)}_{2} & q^{(3)}_{3} & q^{(4)}_{1} & q^{(4)}_{2} & q^{(4)}_{3} & o^{(1)}_{1} & \cdots & o^{(1)}_{9} & o^{(2)}_{1} & \cdots & o^{(2)}_{9} & o^{(3)}_{1} & \cdots & o^{(3)}_{57} & o^{(4)}_{1} & \cdots & o^{(4)}_{57} & o^{(5)}_{1} & \cdots & o^{(5)}_{104} \\
\hline
D_{1,2} & 0 & 1 & 0 & 0 & 0 & 0 & 0 & 0 & 0 & 0 & 0 & 0 & 0 & 0 & 0 & 0 & 0 & 0 & 0
   & \cdots & 0 & 0 & \cdots & 0 & 0 & \cdots & 0 & 0 & \cdots & 0 & 0 & \cdots & 0 \\
 D_{3,1} & 0 & 0 & 0 & 0 & 1 & 0 & 0 & 0 & 0 & 0 & 0 & 0 & 1 & 0 & 1 & 0 & 1 & 0 & 1
   & \cdots & 1 & 0 & \cdots & 0 & 0 & \cdots & 1 & 1 & \cdots & 1 & 1 & \cdots & 1 \\
 D_{4,1} & 1 & 0 & 0 & 0 & 0 & 0 & 1 & 0 & 1 & 0 & 1 & 0 & 0 & 0 & 0 & 0 & 0 & 0 & 1
   & \cdots & 1 & 0 & \cdots & 0 & 0 & \cdots & 1 & 1 & \cdots & 1 & 1 & \cdots & 1 \\
 D_{5,6} & 0 & 1 & 0 & 0 & 0 & 0 & 0 & 0 & 0 & 0 & 0 & 0 & 0 & 0 & 0 & 0 & 0 & 0 & 0
   & \cdots & 0 & 0 & \cdots & 0 & 1 & \cdots & 0 & 0 & \cdots & 0 & 1 & \cdots & 0 \\
 D_{7,5} & 0 & 0 & 0 & 0 & 1 & 0 & 0 & 0 & 0 & 0 & 0 & 0 & 0 & 1 & 1 & 0 & 0 & 1 & 1
   & \cdots & 1 & 0 & \cdots & 0 & 0 & \cdots & 1 & 0 & \cdots & 1 & 0 & \cdots & 1 \\
 D_{8,5} & 1 & 0 & 0 & 0 & 0 & 0 & 0 & 1 & 1 & 0 & 0 & 1 & 0 & 0 & 0 & 0 & 0 & 0 & 1
   & \cdots & 1 & 0 & \cdots & 0 & 0 & \cdots & 0 & 1 & \cdots & 1 & 0 & \cdots & 1 \\
 D_{9,10} & 0 & 1 & 0 & 0 & 0 & 0 & 0 & 0 & 0 & 0 & 0 & 0 & 0 & 0 & 0 & 0 & 0 & 0 &
   0 & \cdots & 0 & 1 & \cdots & 0 & 1 & \cdots & 0 & 1 & \cdots & 0 & 1 & \cdots & 0 \\
 D_{11,9} & 0 & 0 & 0 & 0 & 1 & 0 & 0 & 0 & 0 & 0 & 0 & 0 & 1 & 1 & 0 & 1 & 0 & 0 &
   1 & \cdots & 0 & 0 & \cdots & 0 & 0 & \cdots & 0 & 0 & \cdots & 0 & 0 & \cdots & 0 \\
 D_{12,9} & 1 & 0 & 0 & 0 & 0 & 0 & 1 & 1 & 0 & 1 & 0 & 0 & 0 & 0 & 0 & 0 & 0 & 0 &
   1 & \cdots & 0 & 0 & \cdots & 0 & 0 & \cdots & 0 & 0 & \cdots & 1 & 0 & \cdots & 0 \\
 X_{1,2} & 0 & 0 & 1 & 0 & 0 & 0 & 0 & 0 & 0 & 0 & 0 & 0 & 0 & 0 & 0 & 0 & 0 & 0 & 0
   & \cdots & 0 & 0 & \cdots & 0 & 0 & \cdots & 0 & 0 & \cdots & 0 & 0 & \cdots & 0 \\
 X_{2,4} & 0 & 0 & 0 & 1 & 0 & 0 & 0 & 1 & 0 & 1 & 0 & 1 & 0 & 0 & 0 & 0 & 0 & 0 & 0
   & \cdots & 0 & 1 & \cdots & 1 & 1 & \cdots & 0 & 0 & \cdots & 0 & 0 & \cdots & 0 \\
 X_{2,7} & 1 & 0 & 0 & 0 & 0 & 0 & 0 & 1 & 1 & 0 & 0 & 1 & 0 & 0 & 0 & 0 & 0 & 0 & 0
   & \cdots & 1 & 0 & \cdots & 1 & 0 & \cdots & 0 & 1 & \cdots & 1 & 0 & \cdots & 1 \\
 X_{5,6} & 0 & 0 & 1 & 0 & 0 & 0 & 0 & 0 & 0 & 0 & 0 & 0 & 0 & 0 & 0 & 0 & 0 & 0 & 0
   & \cdots & 0 & 0 & \cdots & 0 & 1 & \cdots & 0 & 0 & \cdots & 0 & 1 & \cdots & 0 \\
 X_{6,8} & 0 & 0 & 0 & 1 & 0 & 0 & 1 & 0 & 0 & 1 & 1 & 0 & 0 & 0 & 0 & 0 & 0 & 0 & 0
   & \cdots & 0 & 1 & \cdots & 1 & 0 & \cdots & 1 & 0 & \cdots & 0 & 0 & \cdots & 0 \\
 X_{6,11} & 1 & 0 & 0 & 0 & 0 & 0 & 1 & 1 & 0 & 1 & 0 & 0 & 0 & 0 & 0 & 0 & 0 & 0 &
   1 & \cdots & 1 & 0 & \cdots & 0 & 0 & \cdots & 0 & 1 & \cdots & 1 & 0 & \cdots & 0 \\
 X_{9,10} & 0 & 0 & 1 & 0 & 0 & 0 & 0 & 0 & 0 & 0 & 0 & 0 & 0 & 0 & 0 & 0 & 0 & 0 &
   0 & \cdots & 0 & 1 & \cdots & 0 & 1 & \cdots & 0 & 1 & \cdots & 0 & 1 & \cdots & 0 \\
 X_{10,3} & 1 & 0 & 0 & 0 & 0 & 0 & 1 & 0 & 1 & 0 & 1 & 0 & 0 & 0 & 0 & 0 & 0 & 0 &
   1 & \cdots & 1 & 0 & \cdots & 0 & 0 & \cdots & 0 & 0 & \cdots & 0 & 0 & \cdots & 0 \\
 X_{10,12} & 0 & 0 & 0 & 1 & 0 & 0 & 0 & 0 & 1 & 0 & 1 & 1 & 0 & 0 & 0 & 0 & 0 & 0 &
   0 & \cdots & 1 & 0 & \cdots & 1 & 0 & \cdots & 1 & 0 & \cdots & 0 & 0 & \cdots & 1 \\
 Y_{3,9} & 0 & 0 & 0 & 1 & 0 & 0 & 0 & 1 & 0 & 1 & 0 & 1 & 0 & 0 & 0 & 0 & 0 & 0 & 0
   & \cdots & 0 & 0 & \cdots & 1 & 0 & \cdots & 1 & 0 & \cdots & 1 & 0 & \cdots & 1 \\
 Y_{4,9} & 0 & 0 & 0 & 0 & 0 & 1 & 0 & 0 & 0 & 0 & 0 & 0 & 0 & 1 & 0 & 1 & 0 & 1 & 0
   & \cdots & 0 & 0 & \cdots & 1 & 0 & \cdots & 1 & 0 & \cdots & 1 & 0 & \cdots & 1 \\
 Y_{7,1} & 0 & 0 & 0 & 1 & 0 & 0 & 1 & 0 & 0 & 1 & 1 & 0 & 0 & 0 & 0 & 0 & 0 & 0 & 1
   & \cdots & 0 & 1 & \cdots & 0 & 1 & \cdots & 1 & 0 & \cdots & 0 & 1 & \cdots & 0 \\
 Y_{8,1} & 0 & 0 & 0 & 0 & 0 & 1 & 0 & 0 & 0 & 0 & 0 & 0 & 1 & 0 & 0 & 1 & 1 & 0 & 1
   & \cdots & 0 & 1 & \cdots & 0 & 1 & \cdots & 0 & 1 & \cdots & 0 & 1 & \cdots & 0 \\
 Y_{11,5} & 0 & 0 & 0 & 1 & 0 & 0 & 0 & 0 & 1 & 0 & 1 & 1 & 0 & 0 & 0 & 0 & 0 & 0 &
   0 & \cdots & 0 & 1 & \cdots & 1 & 0 & \cdots & 1 & 0 & \cdots & 0 & 0 & \cdots & 1 \\
 Y_{12,5} & 0 & 0 & 0 & 0 & 0 & 1 & 0 & 0 & 0 & 0 & 0 & 0 & 0 & 0 & 1 & 0 & 1 & 1 &
   0 & \cdots & 0 & 1 & \cdots & 1 & 0 & \cdots & 1 & 0 & \cdots & 1 & 0 & \cdots & 1 \\
 Z_{2,3} & 0 & 0 & 0 & 0 & 0 & 1 & 0 & 0 & 0 & 0 & 0 & 0 & 0 & 1 & 0 & 1 & 0 & 1 & 0
   & \cdots & 0 & 1 & \cdots & 1 & 1 & \cdots & 0 & 0 & \cdots & 0 & 0 & \cdots & 0 \\
 Z_{2,8} & 0 & 0 & 0 & 0 & 1 & 0 & 0 & 0 & 0 & 0 & 0 & 0 & 0 & 1 & 1 & 0 & 0 & 1 & 0
   & \cdots & 1 & 0 & \cdots & 1 & 0 & \cdots & 1 & 0 & \cdots & 1 & 0 & \cdots & 1 \\
 Z_{6,7} & 0 & 0 & 0 & 0 & 0 & 1 & 0 & 0 & 0 & 0 & 0 & 0 & 1 & 0 & 0 & 1 & 1 & 0 & 0
   & \cdots & 0 & 1 & \cdots & 1 & 0 & \cdots & 0 & 1 & \cdots & 0 & 0 & \cdots & 0 \\
 Z_{6,12} & 0 & 0 & 0 & 0 & 1 & 0 & 0 & 0 & 0 & 0 & 0 & 0 & 1 & 1 & 0 & 1 & 0 & 0 &
   1 & \cdots & 1 & 0 & \cdots & 0 & 0 & \cdots & 0 & 1 & \cdots & 0 & 0 & \cdots & 0 \\
 Z_{10,4} & 0 & 0 & 0 & 0 & 1 & 0 & 0 & 0 & 0 & 0 & 0 & 0 & 1 & 0 & 1 & 0 & 1 & 0 &
   1 & \cdots & 1 & 0 & \cdots & 0 & 0 & \cdots & 0 & 0 & \cdots & 0 & 0 & \cdots & 0 \\
 Z_{10,11} & 0 & 0 & 0 & 0 & 0 & 1 & 0 & 0 & 0 & 0 & 0 & 0 & 0 & 0 & 1 & 0 & 1 & 1 &
   0 & \cdots & 1 & 0 & \cdots & 1 & 0 & \cdots & 1 & 0 & \cdots & 1 & 0 & \cdots & 1 \\
\end{array}
\right)
$}
~,~
\nn\\
\eea
where $o^{(1)}_k, \dots, o^{(5)}_m$ are extra GLSM fields \cite{Witten:1993yc}.
The kernel of the $P$-matrix
gives the $J$- and $E$-term charge matrix $Q_{JE}$
under the forward algorithm for brane brick models.
For the $Q^{1,1,1}/ \mathbb{Z}_3 \ (0,0,1,1,2,2,0,0)$ model, the $Q_{JE}$-matrix is a $239\times 254$ matrix, which we choose not to present in this work.
In addition to the $Q_{JE}$-matrix, 
we also have the $D$-term charge matrix, which takes the form,
\beal{es04d05}
&&
Q_{D}=
\nn\\
&&
\resizebox{0.9\textwidth}{!}{$
\left(
\begin{array}{cccccc|ccc|ccc|ccc|ccc|ccc|ccc|ccc|ccc|ccc}
p_{1} & p_{2} & p_{3} & p_{4} & p_{5} & p_{6} & q^{(1)}_{1} & q^{(1)}_{2} & q^{(1)}_{3} & q^{(2)}_{1} & q^{(2)}_{2} & q^{(2)}_{3} & q^{(3)}_{1} & q^{(3)}_{2} & q^{(3)}_{3} & q^{(4)}_{1} & q^{(4)}_{2} & q^{(4)}_{3} & o^{(1)}_{1} & \cdots & o^{(1)}_{9} & o^{(2)}_{1} & \cdots & o^{(2)}_{9} & o^{(3)}_{1} & \cdots & o^{(3)}_{57} & o^{(4)}_{1} & \cdots & o^{(4)}_{57} & o^{(5)}_{1} & \cdots & o^{(5)}_{104} \\
\hline
 0 & -1 & -1 & 0 & 0 & 0 & 0 & 0 & 0 & 0 & 0 & 0 & 0 & 0 & 0 & 0 & 0 & 0 & 0 & \cdots & 0 & 0 & \cdots & 0
   & 0 & \cdots & 0 & 0 & \cdots & 0 & 1 & \cdots & 0 \\
 0 & 1 & 1 & 0 & 0 & 0 & 0 & 0 & 0 & 0 & 0 & 0 & 0 & 0 & 0 & 0 & 0 & 0 & 0 & \cdots & 0 & 0 & \cdots & 0 &
   0 & \cdots & 0 & 0 & \cdots & 0 & 0 & \cdots & 0 \\
 1 & 0 & 0 & 0 & 0 & 0 & 0 & -1 & 0 & 0 & 0 & 0 & 0 & 0 & 0 & 0 & 0 & 0 & 0 & \cdots & 0 & 0 & \cdots & 0
   & 1 & \cdots & 0 & 0 & \cdots & 0 & -1 & \cdots & 0 \\
 0 & 0 & 0 & 0 & 0 & 0 & 0 & 0 & 0 & 0 & 0 & 0 & 1 & 0 & 0 & -1 & 0 & 0 & 0 & \cdots & 0 & 0 & \cdots & 0
   & 1 & \cdots & 0 & 0 & \cdots & 0 & -1 & \cdots & 0 \\
 0 & 0 & 0 & 0 & 0 & 0 & 0 & 0 & 0 & 0 & 0 & 0 & 0 & 0 & 0 & 0 & 0 & 0 & 0 & \cdots & 0 & 0 & \cdots & 0 &
   -1 & \cdots & 0 & 0 & \cdots & 0 & 0 & \cdots & 0 \\
 0 & 0 & 0 & 0 & 0 & 0 & 0 & 0 & 0 & 0 & 0 & 0 & 0 & 0 & 0 & 0 & 0 & 0 & 0 & \cdots & 0 & 0 & \cdots & 0 &
   1 & \cdots & 0 & 0 & \cdots & 0 & 0 & \cdots & 0 \\
 0 & 0 & 0 & 0 & 0 & 0 & 0 & 0 & 0 & 0 & 0 & 0 & 0 & 0 & 0 & 0 & 0 & 0 & 0 & \cdots & 0 & 0 & \cdots & 0 &
   0 & \cdots & 0 & 0 & \cdots & 0 & 0 & \cdots & 0 \\
 0 & 0 & 0 & 0 & 0 & 0 & 0 & 0 & 0 & 0 & 0 & 0 & 0 & 0 & 0 & 0 & 0 & 0 & 0 & \cdots & 0 & 0 & \cdots & 0 &
   -1 & \cdots & 0 & 0 & \cdots & 0 & 0 & \cdots & 0 \\
 0 & 0 & 0 & 0 & 0 & 0 & 0 & 1 & 0 & 0 & 0 & 0 & 0 & 0 & 0 & 1 & 0 & 0 & 0 & \cdots & 0 & 0 & \cdots & 0 &
   0 & \cdots & 0 & 0 & \cdots & 0 & 0 & \cdots & 0 \\
 0 & 0 & 0 & 0 & 0 & -1 & 0 & 0 & -1 & 0 & 0 & 0 & -1 & 0 & 0 & 1 & 0 & 0 & 0 & \cdots & 0 & 0 & \cdots &
   0 & -1 & \cdots & 0 & 0 & \cdots & 0 & 1 & \cdots & 0 \\
 0 & 0 & 0 & 0 & 0 & 1 & 0 & 0 & 0 & 0 & 0 & 0 & 0 & 0 & 0 & -1 & 0 & 0 & 0 & \cdots & 0 & 0 & \cdots & 0
   & 0 & \cdots & 0 & 0 & \cdots & 0 & 0 & \cdots & 0 \\
\end{array}
\right)
$}
~.~
\nn\\
\eea

\begin{figure}[htt!!]
    \centering
    \includegraphics[width=0.2\textwidth]{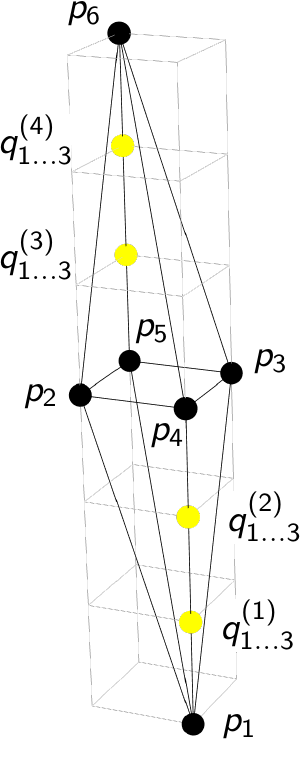}
    \caption{The toric diagram for the $Q^{1,1,1}/\mathbb{Z}_3 \ (0,0,1,1,2,2,0,0)$ model.
    \label{fig_414_toric}}
\end{figure}

The toric diagram of 
the abelian orbifold of the form $Q^{1,1,1}/\mathbb{Z}_3 \ (0,0,1,1,2,2,0,0)$
is shown in \fref{fig_414_toric} and is given by, 
\beal{es04d03}
&&
G_{t}=
\nn\\
&&
\resizebox{0.9\textwidth}{!}{$
\left(
\begin{array}{cccccc|ccc|ccc|ccc|ccc|ccc|ccc|ccc|ccc|ccc}
p_1&p_2&p_3&p_4&p_5&p_6&q^{(1)}_1 & q^{(1)}_2 & q^{(1)}_3&q^{(2)}_1 & q^{(2)}_2 & q^{(2)}_3&q^{(3)}_1 &q^{(3)}_2 & q^{(3)}_3&q^{(4)}_1 & q^{(4)}_2 & q^{(4)}_3 & o^{(1)}_1 & \dots & o^{(1)}_9&o^{(2)}_1 & \dots & o^{(2)}_9&o^{(3)}_1 & \dots & o^{(3)}_{57}&o^{(4)}_1 & \dots & o^{(4)}_{57}&o^{(5)}_1 & \dots & o^{(5)}_{104} \\
\hline
 0 & 0 & 1 & 0 & 1 & 1 & 0 & 0 & 0 & 0 & 0 & 0 & 1 & 1 & 1 & 1 & 1 & 1 & 1 & \dots & 1 & 1 & \dots & 1 & 1 & \dots & 1 & 1 & \dots & 1 & 1 & \dots & 1 \\
 1 & 0 & 1 & 1 & 0 & 0 & 1 & 1 & 1 & 1 & 1 & 1 & 0 & 0 & 0 & 0 & 0 & 0 & 1 & \dots & 1 & 1 & \dots & 1 & 1 & \dots & 1 & 1 & \dots & 1 & 1 & \dots & 1 \\
  -3 & 0 & 0 & 0 & 0 & 3 & -2 & -2 & -2 & -1 & -1 & -1 & 1 & 1 & 1 & 2 & 2 & 2 & -2 & \dots & -2 & 2 & \dots & 2 & 1  & \dots & 1 & -1 & \dots & -1 & 0 & \dots & 0 \\
 \hline
 1 & 1 & 1 & 1 & 1 & 1 & 1 & 1 & 1 & 1 & 1 & 1 & 1 & 1 & 1 & 1 & 1 & 1 & 2 & \dots & 2 & 2 & \dots & 2 & 2 & \dots & 2 & 2 & \dots & 2 & 2 & \dots & 2 \\
\end{array}
\right)
$}
~.~
\nn\\
\eea

%------------------
%------------------

The global symmetry of the 
$Q^{1,1,1}/\mathbb{Z}_3 \ (0,0,1,1,2,2,0,0)$ model
takes the following enhanced form,
\beal{es04d04}
SU(2)_x \times U(1)_{f_1} \times U(1)_{f_2} \times U(1)_R ~,~
\eea
where the factor $SU(2)_x \times U(1)_{f_1} \times U(1)_{f_2}$ is the mesonic flavor symmetry.
The charges under the mesonic flavor symmetry on the extremal GLSM fields $p_a$ are summarized in \tref{tab_04d01}.

%-------------------
\begin{table}[htt!!]
\centering
\begin{tabular}{|c|c|c|c|l|}
\hline
\; & $SU(2)_x$ & $U(1)_{f_1}$ & $U(1)_{f_2}$ & fugacity \\
\hline
$p_1$ & $0$ & $0$ & $+1$  & $t_1=f_2 t$ \\
$p_2$ & $+1$ & $0$ & $0$  & $t_2=x t$\\
$p_3$ & $-1$ & $0$ & $0$  & $t_3=x^{-1} t$\\
$p_4$ & $0$ & $+1$ & $0$  & $t_4=f_1 t$ \\
$p_5$ & $0$ & $-1$ & $0$  & $t_5=f_1^{-1} t$ \\
$p_6$ & $0$ & $0$ & $-1$  & $t_6=f_2^{-1} t$ \\
\hline
\end{tabular}
\caption{
Mesonic flavor symmetry of the $Q^{1,1,1}/\mathbb{Z}_3 \ (0,0,1,1,2,2,0,0)$ model 
and charges on the extremal GLSM fields $p_a$.
Here, the fugacity $t$ counts the degree in extremal GLSM fields $p_a$. 
\label{tab_04d01}}
\end{table}
%-------------------
The Hilbert series of the mesonic moduli space
of the brane brick model corresponding to $Q^{1,1,1}/\mathbb{Z}_3 \ (0,0,1,1,2,2,0,0)$ 
takes the following form,
\beal{es04d05b}
&&
g(t_a; \mathcal{M}^{mes})=
\frac{
P(t_a ;\mathcal{M}^{mes})
}{
\left( 1 - t_1 t_2 t_4\right) \left( 1 - t_1 t_3 t_4\right) \left(1 - t_1^3 t_2^3 t_5^3\right) \left(1 - t_1^3 t_3^3 t_5^3\right) 
}
\nn\\
&&
\hspace{1cm}
\times
\frac{1}{\left(1 - t_2 t_5 t_6\right) \left(1 - t_3 t_5 t_6\right) \left(1 - t_2^3 t_4^3 t_6^3\right) \left(1 - t_3^3 t_4^3 t_6^3\right)}
~,~
\eea
where $t_a$ is the fugacity corresponding to the extremal GLSM field $p_a$.
The numerator given by 
$P(t_a ; \mathcal{M}^{mes})$
is fully presented in appendix \sref{appCa}.
When unrefined by setting $t_a=t$, 
the Hilbert series takes the form,
\beal{es04d06}
g(t; \mathcal{M}^{mes})=\frac{1+2t^3 + 2t^6 +8t^9 +2t^{12} +2t^{15} + t^{18}}{(1-t^3)^2 (1-t^9)^2}~,~
\eea
where the palindromic numerator indicates that the mesonic moduli space is Calabi-Yau.

The Hilbert series of the mesonic moduli space can be refined under the mesonic flavor symmetry fugacity assignment
as summarized in \tref{tab_04d01}.
The corresponding highest weight generating function is given by,
\beal{es04d07}
&&
h(\mu, f_1, f_2, t; \mathcal{M}^{mes})= 
\nn\\
&&
\hspace{1cm}
\frac{1-\mu^6  t^{18}}{(1-\mu f_1 f_2 t^3)(1-\mu f_1^{-1} f_2^{-1} t^3) (1- \mu^3 f_1^3 f_2^{-3} t^9)(1- \mu^3 f_1^{-3} f_2^3 t^9)}
~,~
\eea
where $\mu^m$ is fugacity for counting the $SU(2)_x$ character of the form $[m]_x$,
and $f_1, f_2$ are the fugacities associated to $U(1)_{f_1}$ and $U(1)_{f_2}$, respectively.

%-------------------
 \begin {table}[htt!!]
\centering
\begin {tabular} {|c|c|ccc|}
\hline
PL term & generator & $SU(2)_x$ & $U(1)_{f_1}$ & $U(1)_{f_2}$ 
\\
\hline 
\multirow{2}{*}{$+[1]_x f_1 f_2  t^3$}
& $p_1 p_2 p_4 ~q_{(1)} q_{(2)} $ & $1$ & $1$ & $1$ \\
& $p_1 p_3 p_4 ~q_{(1)} q_{(2)} $ & $-1$ & $1$ & $1$  \\
\hline
\multirow{2}{*}{$+[1]_x f_1^{-1} f_2^{-1} t^3$}
& $p_2 p_5 p_6 ~q_{{(3)}} q_{{(4)}} $ & $1$ & $-1$ & $-1$  \\
& $p_3 p_5 p_6 ~q_{{(3)}} q_{{(4)}} $ & $-1$ & $-1$ & $-1$  \\
\hline
\multirow{4}{*}{$+[3]_x f_1^{3} f_2^{-3} t^9$}
& $p_3^3 p_4^3 p_6^3 ~q_{(1)} q_{(2)}^2 q_{{(3)}} q_{{(4)}}^2 $ & $-3$ & $3$ & $-3$  \\

& $p_2 p_3^2 p_4^3 p_6^3 ~q_{(1)} q_{(2)}^2 q_{{(3)}} q_{{(4)}}^2 $ & $-1$ & $3$ & $-3$ \\

& $p_2^2 p_3 p_4^3 p_6^3 ~q_{(1)} q_{(2)}^2 q_{{(3)}} q_{{(4)}}^2 $ & $1$ & $3$ & $-3$  \\

& $p_2^3 p_4^3 p_6^3 ~q_{(1)} q_{(2)}^2 q_{{(3)}} q_{{(4)}}^2 $ & $3$ & $3$ & $-3$  \\
\hline
\multirow{4}{*}{$+[3]_x f_1^{-3} f_2^{3} t^9$}
& $p_1^3 p_3^3 p_5^3 ~q_{(1)}^2 q_{(2)} q_{{(3)}}^2 q_{{(4)}} $ & $-3$ & $-3$ & $3$  \\

& $p_1^3 p_2 p_3^2 p_5^3 ~q_{(1)}^2 q_{(2)} q_{{(3)}}^2 q_{{(4)}} $ & $-1$ & $-3$ & $3$  \\

& $p_1^3 p_2^2 p_3 p_5^3 ~q_{(1)}^2 q_{(2)} q_{{(3)}}^2 q_{{(4)}} $ & $1$ & $-3$ & $3$  \\

& $p_1^3 p_2^3 p_5^3 ~q_{(1)}^2 q_{(2)} q_{{(3)}}^2 q_{{(4)}} $ & $3$ & $-3$ & $3$  \\
\hline
\end{tabular}
\caption{Generators of the $Q^{1,1,1}/\mathbb{Z}_3 \ (0,0,1,1,2,2,0,0)$ model in terms of GLSM fields and their corresponding mesonic flavor charges. Here, we denote $q_{(1)}=\prod_{i=1}^{3}q_i^{(1)}$, $q_{(2)}=\prod_{i=1}^{3}q_i^{(2)}$, $q_{(3)}=\prod_{i=1}^{3}q_i^{(3)}$, $q_{(4)}=\prod_{i=1}^{3}q_i^{(4)}$, and set extra GLSM fields to 1.
\label{tab_04d02}}
\end{table}
%-------------------

The plethystic logarithm of the mesonic flavor symmetry refined Hilbert series with the corresponding highest weight generating function given in \eref{es04d07}
takes the following form,
\beal{es04d08}
&&
\text{PL}[g(x,f_1,f_2,t; \mathcal{M}^{mes})]=
( [1]_x f_1 f_2 + [1]_x f_1^{-1} f_2^{-1} )t^3
+ ( [3]_x f_1^3 f_2^{-3} + [3]_x f_1^{-3} f_2^{3} )t^9
\nn\\
&&
\hspace{1cm}
-  t^6
- ( [2]_x f_1^{-2} f_2^4 + [2]_x f_1^{-4} f_2^2
+ [2]_x f_1^4 f_2^{-2} + [2]_x f_1^2 f_2^{-4}) t^{12}
\nn\\
&&
\hspace{1cm}
- ( [2]_x f_1^{-6} f_2^6 + [2]_x f_1^6 f_2^{-6}
+ [6]_x + [4]_x + [2]_x + 1 ) t^{18}
+ \dots
~.~
\eea
We can see from the infinite expansion of the plethystic logarithm that the mesonic moduli space of the
$Q^{1,1,1}/\mathbb{Z}_3 \ (0,0,1,1,2,2,0,0)$ model is not a complete intersection.
The first positive terms of the expansion correspond to the generators of the mesonic moduli space,
which are summarized with their mesonic flavor charges in \tref{tab_04d02}. 

We note here that the abelian orbifold $Q^{1,1,1}/\mathbb{Z}_3 \ (0,0,1,1,2,2,0,0)$
is part of a family of abelian orbifolds of the form $Q^{1,1,1}/\mathbb{Z}_n \ (0,0,1,1,-1,-1,0,0)$
as discussed in section \sref{sec042}. 
\\

%=================================================================
\subsubsection{$Q^{1,1,1}/ \mathbb{Z}_3 \ (0,1,1,2,1,2,2,0)$ \label{sec045}}

%-------------------
\begin{figure}[H]
    \centering
    \includegraphics[width=0.4\textwidth]{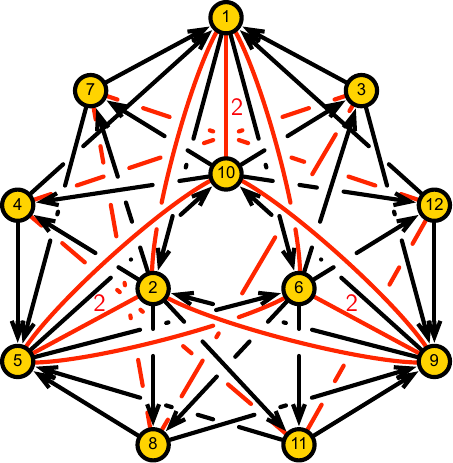}
    \caption{
    The quiver for the $Q^{1,1,1}/ \mathbb{Z}_3 \ (0,1,1,2,1,2,2,0)$ model.
    \label{fig_415_quiver}
    }
\end{figure}
%-------------------
The $J$- and $E$-terms for the brane brick model corresponding to the abelian orbifold of the form
$Q^{1,1,1}/ \mathbb{Z}_3 \ (0,1,1,2,1,2,2,0)$ are as follows, 
\beal{es04e01}
\resizebox{0.95\textwidth}{!}{$
\begin{array}{rclccclcccc}
&& && &J& && &E& \\
&&\Lambda^{(1)}_{2, 1} &:& & D_{1, 2} \cdot Z_{2, 8} \cdot Y_{8, 9} \cdot X_{9, 2} - X_{1, 6} \cdot Z_{6, 3} \cdot D_{3, 1} \cdot D_{1, 2}& && & X_{2, 7} \cdot Y_{7, 1} - X_{2, 4} \cdot D_{4, 1}& \\ 
 &&\Lambda^{(1)}_{6, 5} &:& & D_{5, 6} \cdot Z_{6, 12} \cdot Y_{12, 1} \cdot X_{1, 6} - X_{5, 10} \cdot Z_{10, 7} \cdot D_{7, 5} \cdot D_{5, 6}& && & X_{6, 11} \cdot Y_{11, 5} - X_{6, 8} \cdot D_{8, 5}& \\ 
 &&\Lambda^{(1)}_{10, 9} &:& & D_{9, 10} \cdot Z_{10, 4} \cdot Y_{4, 5} \cdot X_{5, 10} - X_{9, 2} \cdot Z_{2, 11} \cdot D_{11, 9} \cdot D_{9, 10}& && & X_{10, 3} \cdot Y_{3, 9} - X_{10, 12} \cdot D_{12, 9}& \\ 
 &&\Lambda^{(2)}_{10, 1} &:& & X_{1, 6} \cdot X_{6, 8} \cdot Y_{8, 9} \cdot D_{9, 10} - D_{1, 2} \cdot Z_{2, 11} \cdot Y_{11, 5} \cdot X_{5, 10}& && & X_{10, 3} \cdot D_{3, 1} - Z_{10, 4} \cdot D_{4, 1}& \\ 
 &&\Lambda^{(2)}_{2, 5} &:& & X_{5, 10} \cdot X_{10, 12} \cdot Y_{12, 1} \cdot D_{1, 2} - D_{5, 6} \cdot Z_{6, 3} \cdot Y_{3, 9} \cdot X_{9, 2}& && & X_{2, 7} \cdot D_{7, 5} - Z_{2, 8} \cdot D_{8, 5}& \\ 
 &&\Lambda^{(2)}_{6, 9} &:& & X_{9, 2} \cdot X_{2, 4} \cdot Y_{4, 5} \cdot D_{5, 6} - D_{9, 10} \cdot Z_{10, 7} \cdot Y_{7, 1} \cdot X_{1, 6}& && & X_{6, 11} \cdot D_{11, 9} - Z_{6, 12} \cdot D_{12, 9}& \\ 
 &&\Lambda^{(3)}_{10, 1} &:& & D_{1, 2} \cdot Z_{2, 8} \cdot D_{8, 5} \cdot X_{5, 10} - X_{1, 6} \cdot X_{6, 11} \cdot D_{11, 9} \cdot D_{9, 10}& && & X_{10, 12} \cdot Y_{12, 1} - Z_{10, 7} \cdot Y_{7, 1}& \\ 
 &&\Lambda^{(3)}_{2, 5} &:& & D_{5, 6} \cdot Z_{6, 12} \cdot D_{12, 9} \cdot X_{9, 2} - X_{5, 10} \cdot X_{10, 3} \cdot D_{3, 1} \cdot D_{1, 2}& && & X_{2, 4} \cdot Y_{4, 5} - Z_{2, 11} \cdot Y_{11, 5}& \\ 
 &&\Lambda^{(3)}_{6, 9} &:& & D_{9, 10} \cdot Z_{10, 4} \cdot D_{4, 1} \cdot X_{1, 6} - X_{9, 2} \cdot X_{2, 7} \cdot D_{7, 5} \cdot D_{5, 6}& && & X_{6, 8} \cdot Y_{8, 9} - Z_{6, 3} \cdot Y_{3, 9}& \\ 
 &&\Lambda^{(4)}_{6, 1} &:& & D_{1, 2} \cdot X_{2, 7} \cdot Y_{7, 1} \cdot X_{1, 6} - X_{1, 6} \cdot X_{6, 8} \cdot D_{8, 5} \cdot D_{5, 6}& && & Z_{6, 3} \cdot D_{3, 1} - Z_{6, 12} \cdot Y_{12, 1}& \\ 
 &&\Lambda^{(4)}_{10, 5} &:& & D_{5, 6} \cdot X_{6, 11} \cdot Y_{11, 5} \cdot X_{5, 10} - X_{5, 10} \cdot X_{10, 12} \cdot D_{12, 9} \cdot D_{9, 10}& && & Z_{10, 7} \cdot D_{7, 5} - Z_{10, 4} \cdot Y_{4, 5}& \\ 
 &&\Lambda^{(4)}_{2, 9} &:& & D_{9, 10} \cdot X_{10, 3} \cdot Y_{3, 9} \cdot X_{9, 2} - X_{9, 2} \cdot X_{2, 4} \cdot D_{4, 1} \cdot D_{1, 2}& && & Z_{2, 11} \cdot D_{11, 9} - Z_{2, 8} \cdot Y_{8, 9}& \\ 
 &&\Lambda^{(5)}_{8, 3} &:& & D_{3, 1} \cdot X_{1, 6} \cdot X_{6, 8} - Y_{3, 9} \cdot X_{9, 2} \cdot Z_{2, 8}& && & D_{8, 5} \cdot D_{5, 6} \cdot Z_{6, 3} - Y_{8, 9} \cdot D_{9, 10} \cdot X_{10, 3}& \\ 
 &&\Lambda^{(5)}_{12, 7} &:& & D_{7, 5} \cdot X_{5, 10} \cdot X_{10, 12} - Y_{7, 1} \cdot X_{1, 6} \cdot Z_{6, 12}& && & D_{12, 9} \cdot D_{9, 10} \cdot Z_{10, 7} - Y_{12, 1} \cdot D_{1, 2} \cdot X_{2, 7}& \\ 
 &&\Lambda^{(5)}_{4, 11} &:& & D_{11, 9} \cdot X_{9, 2} \cdot X_{2, 4} - Y_{11, 5} \cdot X_{5, 10} \cdot Z_{10, 4}& && & D_{4, 1} \cdot D_{1, 2} \cdot Z_{2, 11} - Y_{4, 5} \cdot D_{5, 6} \cdot X_{6, 11}& \\ 
 &&\Lambda^{(6)}_{3, 4} &:& & Y_{4, 5} \cdot X_{5, 10} \cdot X_{10, 3} - D_{4, 1} \cdot X_{1, 6} \cdot Z_{6, 3}& && & D_{3, 1} \cdot D_{1, 2} \cdot X_{2, 4} - Y_{3, 9} \cdot D_{9, 10} \cdot Z_{10, 4}& \\ 
 &&\Lambda^{(6)}_{7, 8} &:& & Y_{8, 9} \cdot X_{9, 2} \cdot X_{2, 7} - D_{8, 5} \cdot X_{5, 10} \cdot Z_{10, 7}& && & D_{7, 5} \cdot D_{5, 6} \cdot X_{6, 8} - Y_{7, 1} \cdot D_{1, 2} \cdot Z_{2, 8}& \\ 
 &&\Lambda^{(6)}_{11, 12} &:& & Y_{12, 1} \cdot X_{1, 6} \cdot X_{6, 11} - D_{12, 9} \cdot X_{9, 2} \cdot Z_{2, 11}& && & D_{11, 9} \cdot D_{9, 10} \cdot X_{10, 12} - Y_{11, 5} \cdot D_{5, 6} \cdot Z_{6, 12}&
\end{array}
$}
~.~
\nn\\
\eea
The corresponding quiver is shown in \fref{fig_415_quiver}. 
The $J$- and $E$-terms
are obtained from the general formula in \eqref{es03b01} 
with the additional relabelling of indices as follows, 
\beal{es04e02}
&
[1,0] \rightarrow 1 ~,~ 
[2,0] \rightarrow 2 ~,~
[3,0] \rightarrow 3 ~,~
[4,0] \rightarrow 4 ~,~
&
\nn\\
&
[1,1] \rightarrow 5 ~,~
[2,1] \rightarrow 6 ~,~
[3,1] \rightarrow 7 ~,~
[4,1] \rightarrow 8 ~,~
&
\nn\\
&
[1,2] \rightarrow 9 ~,~
[2,2] \rightarrow 10 ~,~
[3,2] \rightarrow 11 ~,~ 
[4,2] \rightarrow 12 ~.~
&
\eea

Using the forward algorithm for brane brick models, 
we obtain the $P$-matrix for the $Q^{1,1,1}/ \mathbb{Z}_3 \ (0,1,1,2,1,2,2,0)$ model.
It takes the following form, 
\beal{es04e03}
&&
P=
\resizebox{0.55\textwidth}{!}{$
\left(
\begin{array}{c|cccccc|ccc|ccc|ccc}
 & p_{1} & p_{2} & p_{3} & p_{4} & p_{5} & p_{6} & q^{(1)}_{1} & q^{(1)}_{2} & q^{(1)}_{3} & q^{(2)}_{1} & q^{(2)}_{2} & q^{(2)}_{3} & o^{(1)}_{1} & \cdots & o^{(21)}_{27} \\
\hline
 D_{1,2} & 1 & 0 & 0 & 0 & 0 & 0 & 0 & 0 & 1 & 0 & 0 & 0 & 1 & \cdots & 0 \\
 D_{3,1} & 0 & 1 & 0 & 0 & 0 & 0 & 0 & 0 & 0 & 0 & 0 & 1 & 0 & \cdots & 1 \\
 D_{4,1} & 0 & 0 & 0 & 1 & 0 & 0 & 1 & 0 & 0 & 0 & 0 & 0 & 0 &\cdots & 2 \\
 D_{5,6} & 1 & 0 & 0 & 0 & 0 & 0 & 1 & 0 & 0 & 0 & 0 & 0 & 0 & \cdots & 0 \\
 D_{7,5} & 0 & 1 & 0 & 0 & 0 & 0 & 0 & 0 & 0 & 1 & 0 & 0 & 1 & \cdots & 0 \\
 D_{8,5} & 0 & 0 & 0 & 1 & 0 & 0 & 0 & 1 & 0 & 0 & 0 & 0 & 0 & \cdots & 1 \\
 D_{9,10} & 1 & 0 & 0 & 0 & 0 & 0 & 0 & 1 & 0 & 0 & 0 & 0 & 1 & \cdots & 0 \\
 D_{11,9} & 0 & 1 & 0 & 0 & 0 & 0 & 0 & 0 & 0 & 0 & 1 & 0 & 0 & \cdots &0 \\
 D_{12,9} & 0 & 0 & 0 & 1 & 0 & 0 & 0 & 0 & 1 & 0 & 0 & 0 & 0 &\cdots & 0 \\
 X_{1,6} & 0 & 0 & 0 & 0 & 0 & 1 & 0 & 0 & 0 & 1 & 0 & 0 & 0 & \cdots & 0 \\
 X_{2,4} & 0 & 0 & 0 & 0 & 1 & 0 & 0 & 1 & 0 & 0 & 0 & 0 & 0 & \cdots & 0 \\
 X_{2,7} & 0 & 0 & 0 & 1 & 0 & 0 & 0 & 1 & 0 & 0 & 0 & 0 & 0 &\cdots & 1 \\
 X_{5,10} & 0 & 0 & 0 & 0 & 0 & 1 & 0 & 0 & 0 & 0 & 1 & 0 & 0 & \cdots & 0 \\
 X_{6,8} & 0 & 0 & 0 & 0 & 1 & 0 & 0 & 0 & 1 & 0 & 0 & 0 & 1 & \cdots &1 \\
 X_{6,11} & 0 & 0 & 0 & 1 & 0 & 0 & 0 & 0 & 1 & 0 & 0 & 0 & 1 & \cdots & 1 \\
 X_{9,2} & 0 & 0 & 0 & 0 & 0 & 1 & 0 & 0 & 0 & 0 & 0 & 1 & 0 &\cdots & 1 \\
 X_{10,3} & 0 & 0 & 0 & 1 & 0 & 0 & 1 & 0 & 0 & 0 & 0 & 0 & 0 &\cdots & 1 \\
 X_{10,12} & 0 & 0 & 0 & 0 & 1 & 0 & 1 & 0 & 0 & 0 & 0 & 0 & 0 & \cdots & 2 \\
 Y_{3,9} & 0 & 0 & 0 & 0 & 1 & 0 & 0 & 0 & 1 & 0 & 0 & 0 & 0 & \cdots &1 \\
 Y_{4,5} & 0 & 0 & 1 & 0 & 0 & 0 & 0 & 0 & 0 & 1 & 0 & 0 & 1 & \cdots &1 \\
 Y_{7,1} & 0 & 0 & 0 & 0 & 1 & 0 & 1 & 0 & 0 & 0 & 0 & 0 & 0 & \cdots &1 \\
 Y_{8,9} & 0 & 0 & 1 & 0 & 0 & 0 & 0 & 0 & 0 & 0 & 1 & 0 & 0 & \cdots & 0 \\
 Y_{11,5} & 0 & 0 & 0 & 0 & 1 & 0 & 0 & 1 & 0 & 0 & 0 & 0 & 0 & \cdots & 1 \\
 Y_{12,1} & 0 & 0 & 1 & 0 & 0 & 0 & 0 & 0 & 0 & 0 & 0 & 1 & 0 & \cdots & 0 \\
 Z_{2,8} & 0 & 1 & 0 & 0 & 0 & 0 & 0 & 0 & 0 & 1 & 0 & 0 & 1 & \cdots & 0 \\
 Z_{2,11} & 0 & 0 & 1 & 0 & 0 & 0 & 0 & 0 & 0 & 1 & 0 & 0 & 1 & \cdots & 0 \\
 Z_{6,3} & 0 & 0 & 1 & 0 & 0 & 0 & 0 & 0 & 0 & 0 & 1 & 0 & 1 & \cdots & 0 \\
 Z_{6,12} & 0 & 1 & 0 & 0 & 0 & 0 & 0 & 0 & 0 & 0 & 1 & 0 & 1 & \cdots & 1 \\
 Z_{10,4} & 0 & 1 & 0 & 0 & 0 & 0 & 0 & 0 & 0 & 0 & 0 & 1 & 0 & \cdots & 0 \\
 Z_{10,7} & 0 & 0 & 1 & 0 & 0 & 0 & 0 & 0 & 0 & 0 & 0 & 1 & 0 & \cdots & 1 \\
\end{array}
\right)
$}
~,~
\nn\\
\eea
where there are in total 386 extra GLSM fields
partitioned into 21 sets identified by $o_k^{(1)}, \dots , o_l^{(21)}$.
The $J$- and $E$-term charge matrix 
$Q_{JE}$ can be obtained from the kernel of the $P$-matrix above.
For the $Q^{1,1,1}/ \mathbb{Z}_3 \ (0,1,1,2,1,2,2,0)$ model, 
the $Q_{JE}$-matrix is a $383 \times 398$ dimensional matrix, which we choose not to present here.
Additionally, we have the $D$-term charge matrix, which takes the following form,
\beal{es04e05b}
&&
Q_{D}=
\resizebox{0.5\textwidth}{!}{$
\left(
\begin{array}{cccccc|ccc|ccc|ccc}
p_{1} & p_{2} & p_{3} & p_{4} & p_{5} & p_{6} & q^{(1)}_{1} & q^{(1)}_{2} & q^{(1)}_{3} & q^{(2)}_{1} & q^{(2)}_{2} & q^{(2)}_{3} & o^{(1)}_{1} & \cdots & o^{(21)}_{27} \\
\hline
 -1 & 0 & 0 & 0 & 0 & 0 & 0 & 0 & 0 & -1 & 0 & 0 & 0 & \cdots & 0 \\
 1 & 0 & 0 & 0 & 0 & 0 & -1 & 0 & 0 & 0 & 0 & 0 & 0 & \cdots & 0 \\
 0 & 0 & 0 & 0 & 0 & 0 & 0 & 0 & 0 & 0 & 0 & 0 & 0 & \cdots & 0 \\
 0 & 0 & 0 & 0 & 0 & 0 & 0 & 0 & 0 & 0 & 0 & 0 & 0 & \cdots & 0 \\
 0 & 0 & 0 & 0 & 0 & 0 & -1 & 0 & 0 & 0 & -1 & 0 & 0 & \cdots & 0 \\
 0 & 0 & 0 & 0 & 0 & 0 & 1 & 0 & 0 & 1 & 0 & 0 & 0 & \cdots & 0 \\
 0 & 0 & 1 & 0 & 0 & 0 & 1 & 0 & 0 & 0 & 0 & 0 & 0 & \cdots & 0 \\
 0 & 0 & 0 & 0 & 0 & 0 & 0 & 0 & 0 & 0 & 0 & 0 & 0 & \cdots & 0 \\
 0 & 1 & 0 & 0 & 0 & 0 & 1 & 0 & 0 & 0 & 0 & 0 & 0 & \cdots & 0 \\
 0 & -1 & -1 & 0 & 0 & 0 & -1 & 0 & 0 & 0 & 1 & 0 & 0 & \cdots & 0 \\
 0 & 0 & 0 & 0 & 0 & 0 & 0 & 0 & 0 & 0 & 0 & 0 & 0 & \cdots & 0 \\
\end{array}
\right)
$}
~.~
\eea

\begin{figure}[htt!!]
    \centering
    \includegraphics[width=0.3\textwidth]{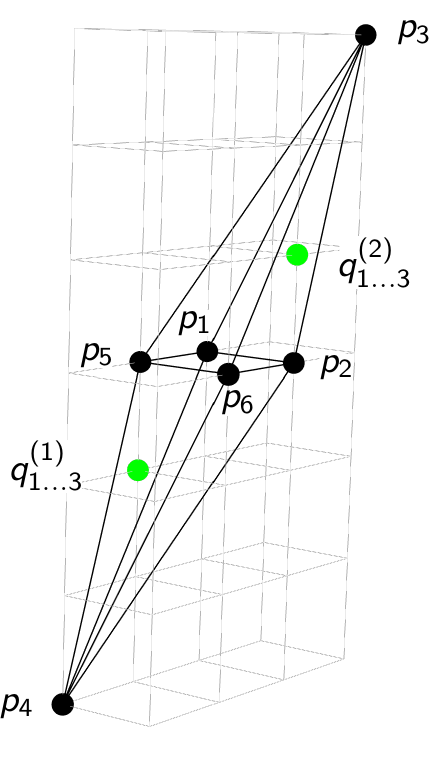}
    \caption{The toric diagram for the $Q^{1,1,1}/\mathbb{Z}_3 \ (0,1,1,2,1,2,2,0)$ model.
    \label{fig_415_toric}}
\end{figure}

The toric diagram of the $Q^{1,1,1}/\mathbb{Z}_3 \ (0,1,1,2,1,2,2,0)$ model 
is shown in \fref{fig_415_toric} and is given by, 
\beal{es04e03b}
&&
G_{t}=
\resizebox{0.5\textwidth}{!}{$
\left(
\begin{array}{cccccc|ccc|ccc|ccc}
p_1&p_2&p_3&p_4&p_5&p_6&
q^{(1)}_1&q^{(1)}_2&q^{(1)}_3&
q^{(2)}_1&q^{(2)}_2&q^{(2)}_3&
o^{(1)}_1& \cdots & o^{(21)}_{27} \\
\hline
0 & 0 & 3 & -3 & 0 & 0 &
-1 & -1 & -1 &
1 & 1 & 1 &
1 & \cdots  & -2 \\
1 & 1 & 2 & -1 & 0 & 0 &
0 & 0 & 0 &
1 & 1 & 1 &
2 & \cdots  & 0 \\
0 & 1 & 1 & 0 & 0 & 1 &
0 & 0 & 0 &
1 & 1 & 1 &
1 & \cdots  & 1 \\
\hline
1 & 1 & 1 & 1 & 1 & 1 &
1 & 1 & 1 &
1 & 1 & 1 &
2 & \cdots & 3
\end{array}
\right)
$}
~.~
\eea

%-------------------
%-------------------

The global symmetry of the brane brick model is not enhanced and takes the following form,
\beal{es04e04}
U(1)_{f_1} \times U(1)_{f_2} \times U(1)_{f_3} \times U(1)_R
~.~
\eea
\tref{tab_04e01} summarizes the charges on the extremal GLSM fields $p_a$
under the mesonic flavor symmetry factor $U(1)_{f_1} \times U(1)_{f_2} \times U(1)_{f_3}$.

%-------------------
\begin{table}[htt!!]
\centering
\begin{tabular}{|c|c|c|c|l|}
\hline
\; & $U(1)_{f_1}$ & $U(1)_{f_2}$ & $U(1)_{f_3}$  & fugacity \\
\hline
$p_1$ & $+1$ & $0$ & $0$  & $t_1=f_1 t$ \\
$p_2$ & $0$ & $+1$ & $0$  & $t_2=f_2 t$\\
$p_3$ & $0$ & $0$ & $+1$  & $t_3=f_3 t$\\
$p_4$ & $0$ & $0$ & $-1$  & $t_4=f_3^{-1} t$ \\
$p_5$ & $0$ & $-1$ & $0$  & $t_5=f_2^{-1} t$ \\
$p_6$ & $-1$ & $0$ & $0$  & $t_6=f_1^{-1} t$ \\
\hline
\end{tabular}
\caption{Mesonic flavor symmetry of the $Q^{1,1,1}/\mathbb{Z}_3 \ (0,1,1,2,1,2,2,0)$ model 
and charges on the extremal GLSM fields $p_a$.
Here, the fugacity $t$ counts the degree in extremal GLSM fields $p_a$. 
 \label{tab_04e01}}
\end{table}
%-------------------

The Hilbert series of the mesonic moduli space for the $Q^{1,1,1}/\mathbb{Z}_3 \ (0,1,1,2,1,2,2,0)$ model
takes the following form,
\beal{es04e05}
&&
g(t_a ; \mathcal{M}^{mes})=
\frac{
P(t_a; \mathcal{M}^{mes} )
}{
\left(1 - t_1^3 t_2^3 t_3^3\right) \left(1 - t_1^3 t_2^3 t_4^3\right) \left(1 - t_1 t_4 t_5\right) \left(1 - t_1^3 t_3^3 t_5^3\right) }
\nn\\
&&
\hspace{1cm}
\times
\frac{1}{
\left(1 - t_2 t_3 t_6\right) \left(1 - t_2^3 t_4^3 t_6^3\right) \left(1 - t_3^3 t_5^3 t_6^3\right) \left(1 - t_4^3 t_5^3 t_6^3\right)
}
~,~
\eea
where $t_a$ is the fugacity associated to the extremal GLSM field $p_a$.
The numerator factor $P(t_a ; \mathcal{M}^{mes})$ in \eref{es04e05} is presented fully in appendix \sref{appCa}. 
When unrefined by setting $t_a=t$, the Hilbert series takes the following form,
\beal{es04e06}
g(t; \mathcal{M}^{mes})=\frac{1-t^3+6t^6-t^9 +t^{12}}{(1-t^3)^3 (1-t^9)}
~,~
\eea
where the palindromic numerator indicates that the mesonic moduli space is Calabi-Yau. 

The Hilbert series refined in terms of the mesonic flavor symmetry fugacities summarized in \tref{tab_04e01}
is given as follows,
\beal{es04e07}
&&
g(f_1,f_2,f_3,t; \mathcal{M}^{mes})
=
\frac{
P(f_1,f_2,f_3,t; \mathcal{M}^{mes})
}{
(1-f_1 f_2^{-1} f_3^{-1} t^3) (1-f_1^{-1} f_2 f_3 t^3) (1-f_1^{-3} f_2^{-3} f_3^{-3}t^9) 
}
\nn\\
&&
\hspace{1cm}
\times \frac{
1
}{
(1-f_1^{-3} f_2^3 f_3^{-3} t^9)(1-f_1^3 f_2^3 f_3^{-3} t^9) (1-f_1^{-3} f_2^{-3} f_3^3 t^9) 
}
\nn\\
&&
\hspace{1cm}
\times \frac{
1
}{
(1-f_1^3 f_2^{-3} f_3^3 t^9) (1-f_1^3 f_2^3 f_3^3 t^9)
}
~,~
\eea
where the
numerator factor $P(f_1,f_2,f_3,t;\mathcal{M}^{mes})$ is presented in appendix \sref{appCa}. 

%-------------------
 \begin {table}[httt!!]
\centering
\begin {tabular} {|c|c|ccc|}
\hline
PL term & generator & $U(1)_{f_1}$ & $U(1)_{f_2}$ & $U(1)_{f_3}$ 
\\
\hline 
\multirow{1}{*}{$+ f_1^{-3} f_2^{-3} f_3^{-3} t^9$}
& $p_4^3 p_5^3 p_6^3 ~q_{(1)}^2 q_{(2)} $ & $-3$ & $-3$ & $-3$  \\
\hline
\multirow{1}{*}{$+ f_1^{-2} f_2^{-2}  t^{6}$}
& $p_3 p_4 p_5^2 p_6^2 ~q_{(1)} q_{(2)} $ & $-2$ & $-2$ & $0$  \\
\hline
\multirow{1}{*}{$+f_1^{-3} f_2^{-3} f_3^{3} t^{9} $}
& $p_3^3 p_5^3 p_6^3 ~q_{(1)} q_{(2)}^2 $ & $-3$ & $-3$ & $3$  \\
\hline
\multirow{1}{*}{$+  f_1^{-2}  f_3^{-2} t^{6}$}
& $p_2 p_4^2 p_5 p_6^2 ~q_{(1)} q_{(2)} $ & $-2$ & $0$ & $-2$  \\
\hline
\multirow{1}{*}{$+ f_1^{-1} f_2 f_3 t^3$}
& $p_2 p_3 p_6 ~q_{(2)} $ & $-1$ & $1$ & $1$  \\
\hline
\multirow{1}{*}{$+ f_1^{-3} f_2^{3} f_3^{-3} t^{9}$}
& $p_2^3 p_4^3 p_6^3 ~q_{(1)} q_{(2)}^2 $ & $-3$ & $3$ & $-3$  \\
\hline
\multirow{1}{*}{$+ f_1 f_2^{-1} f_3^{-1} t^3$}
& $p_1 p_4 p_5 ~q_{(1)} $ & $1$ & $-1$ & $-1$  \\
\hline
\multirow{1}{*}{$+f_2^{-2} f_3^{2} t^{6} $}
& $p_1 p_3^2 p_5^2 p_6 ~q_{(1)} q_{(2)} $ & $0$ & $-2$ & $2$ \\
\hline
\multirow{1}{*}{$+  f_2^{2} f_3^{-2} t^{6}$}
& $p_1 p_2^2 p_4^2 p_6 ~q_{(1)} q_{(2)} $ & $0$ & $2$ & $-2$  \\
\hline
\multirow{1}{*}{$+f_1^{2}  f_3^{2} t^{6} $}
& $p_1^2 p_2 p_3^2 p_5 ~q_{(1)} q_{(2)} $ & $2$ & $0$ & $2$  \\
\hline
\multirow{1}{*}{$+f_1^{2} f_2^{2}  t^{6} $}
& $p_1^2 p_2^2 p_3 p_4 ~q_{(1)} q_{(2)} $ & $2$ & $2$ & $0$  \\
\hline
\multirow{1}{*}{$+ f_1^{3} f_2^{-3} f_3^{3} t^{9}$}
& $p_1^3 p_3^3 p_5^3 ~q_{(1)}^2 q_{(2)} $ & $3$ & $-3$ & $3$  \\
\hline
\multirow{1}{*}{$+f_1^{3} f_2^{3} f_3^{-3} t^{9} $}
& $p_1^3 p_2^3 p_4^3 ~q_{(1)}^2 q_{(2)} $ & $3$ & $3$ & $-3$  \\
\hline
\multirow{1}{*}{$+f_1^{3} f_2^{3} f_3^{3} t^{9} $}
& $p_1^3 p_2^3 p_3^3 ~q_{(1)} q_{(2)}^2 $ & $3$ & $3$ & $3$  \\
\hline
\end{tabular}
\caption{Generators of the $Q^{1,1,1}/\mathbb{Z}_3 \ (0,1,1,2,1,2,2,0)$ model in terms of GLSM fields and their corresponding mesonic flavor charges. Here, we denote $q_{(1)}=\prod_{i=1}^{3}q_i^{(1)}$,  $q_{(2)}=\prod_{i=1}^{3} q_i^{(2)}$, and set the extra GLSM fields to 1. \label{tab_04e02}}
\end{table}
%-------------------

The plethystic logarithm of the refined Hilbert series in \eref{es04e07}
takes the following form,
\beal{es04e08}
&&
\text{PL}[g(f_1,f_2,f_3,t; \mathcal{M}^{mes})]=
( f_1 f_2^{-1} f_3^{-1} + f_1^{-1} f_2 f_3 ) t^3
+ ( f_1^{-2} f_2^{-2}+f_1^{2} f_2^{2}+f_1^{-2} f_3^{-2}
\nn\\
&&
\hspace{1cm}
+f_2^{2}
f_3^{-2}+f_1^{2} f_3^{2}+f_2^{-2} f_3^{2} ) t^6
+( f_1^{-3} f_2^{-3} f_3^{-3}+f_1^{-3} f_2^{3} f_3^{-3}+f_1^{3} f_2^{3} f_3^{-3}
\nn\\
&&
\hspace{1cm}
+f_1^{-3} f_2^{-3} f_3^{3}+f_1^{3} f_2^{-3} f_3^{3}+f_1^{3} f_2^{3} f_3^{3}) t^9
-( 3+f_1^{-2} f_2^{-2}+f_1^{2} f_2^{2}
+f_1^{-2} f_2^{2} f_3^{-4}
\nn\\
&&
\hspace{1cm}
+f_1^{-2} f_3^{-2}
+f_1^{-4} f_2^{-2} f_3^{-2}+f_2^{2} f_3^{-2}+f_1^{2} f_2^{4} f_3^{-2}+f_1^{2} f_3^{2}
+f_1^{-2} f_2^{-4} f_3^{2}+f_2^{-2} f_3^{2}
\nn\\
&&
\hspace{1cm}
+f_1^{4} f_2^{2} f_3^{2}+f_1^{2} f_2^{-2} f_3^{4})t^{12}
+\dots
~,~
\eea
where we can see from the infinite expansion of the plethystic logarithm that the mesonic moduli space of the
$Q^{1,1,1}/\mathbb{Z}_3 \ (0,1,1,2,1,2,2,0)$ model is not a complete intersection.
The first positive terms of the expansion correspond to the generators of the mesonic moduli space,
which are summarized with their mesonic flavor symmetry charges in \tref{tab_04e02}. 
\\

%=================================================================
\subsubsection{$Q^{1,1,1}/ \mathbb{Z}_2 \times \mathbb{Z}_2 \ \left(\begin{array}{@{}c@{,\;}c@{,\;}c@{,\;}c@{,\;}c@{,\;}c@{,\;}c@{,\;}c@{}} 1 & 1 & 0 & 0 & 0 & 0 & 1 & 1 \\ 0 & 0 & 1 & 1 & 1 & 1 & 0 & 0 \end{array}\right)$ 
\label{sec046}}

%-------------------
\begin{figure}[H]
    \centering
    \includegraphics[width=0.45\textwidth]{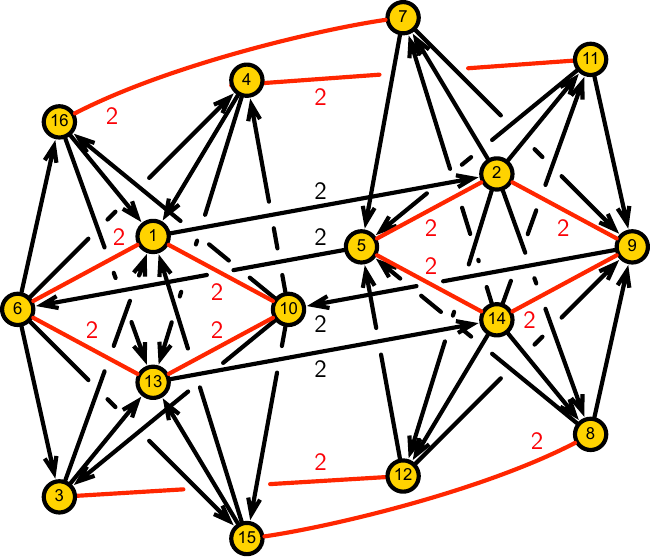}
    \caption{
    The quiver for the $Q^{1,1,1}/ \mathbb{Z}_2 \times \mathbb{Z}_2 \ \left(\begin{array}{@{}c@{,\;}c@{,\;}c@{,\;}c@{,\;}c@{,\;}c@{,\;}c@{,\;}c@{}} 1 & 1 & 0 & 0 & 0 & 0 & 1 & 1 \\ 0 & 0 & 1 & 1 & 1 & 1 & 0 & 0 \end{array}\right)$ model.
    \label{fig_416_quiver}
    }
\end{figure}
%-------------------

The $J$- and $E$-terms for the brane brick model corresponding to the abelian orbifold of the form
$Q^{1,1,1}/ \mathbb{Z}_2 \times \mathbb{Z}_2 \ \left(\begin{array}{@{}c@{,\;}c@{,\;}c@{,\;}c@{,\;}c@{,\;}c@{,\;}c@{,\;}c@{}} 1 & 1 & 0 & 0 & 0 & 0 & 1 & 1 \\ 0 & 0 & 1 & 1 & 1 & 1 & 0 & 0 \end{array}\right)$ are as follows, 
\beal{es04f01}
\resizebox{0.95\textwidth}{!}{$
\begin{array}{rclccclcccc}
&& && &J& && &E& \\
&&\Lambda^{(1)}_{10, 1} &:& & D_{1, 2} \cdot Z_{2, 8} \cdot Y_{8, 9} \cdot X_{9, 10} - X_{1, 2} \cdot Z_{2, 11} \cdot D_{11, 9} \cdot D_{9, 10}& && & X_{10, 15} \cdot Y_{15, 1} - X_{10, 4} \cdot D_{4, 1}& \\ 
 &&\Lambda^{(1)}_{14, 5} &:& & D_{5, 6} \cdot Z_{6, 4} \cdot Y_{4, 13} \cdot X_{13, 14} - X_{5, 6} \cdot Z_{6, 15} \cdot D_{15, 13} \cdot D_{13, 14}& && & X_{14, 11} \cdot Y_{11, 5} - X_{14, 8} \cdot D_{8, 5}& \\ 
 &&\Lambda^{(1)}_{2, 9} &:& & D_{9, 10} \cdot Z_{10, 16} \cdot Y_{16, 1} \cdot X_{1, 2} - X_{9, 10} \cdot Z_{10, 3} \cdot D_{3, 1} \cdot D_{1, 2}& && & X_{2, 7} \cdot Y_{7, 9} - X_{2, 12} \cdot D_{12, 9}& \\ 
 &&\Lambda^{(1)}_{6, 13} &:& & D_{13, 14} \cdot Z_{14, 12} \cdot Y_{12, 5} \cdot X_{5, 6} - X_{13, 14} \cdot Z_{14, 7} \cdot D_{7, 5} \cdot D_{5, 6}& && & X_{6, 3} \cdot Y_{3, 13} - X_{6, 16} \cdot D_{16, 13}& \\ 
 &&\Lambda^{(2)}_{6, 1} &:& & X_{1, 2} \cdot X_{2, 12} \cdot Y_{12, 5} \cdot D_{5, 6} - D_{1, 2} \cdot Z_{2, 11} \cdot Y_{11, 5} \cdot X_{5, 6}& && & X_{6, 3} \cdot D_{3, 1} - Z_{6, 4} \cdot D_{4, 1}& \\ 
 &&\Lambda^{(2)}_{2, 5} &:& & X_{5, 6} \cdot X_{6, 16} \cdot Y_{16, 1} \cdot D_{1, 2} - D_{5, 6} \cdot Z_{6, 15} \cdot Y_{15, 1} \cdot X_{1, 2}& && & X_{2, 7} \cdot D_{7, 5} - Z_{2, 8} \cdot D_{8, 5}& \\ 
 &&\Lambda^{(2)}_{14, 9} &:& & X_{9, 10} \cdot X_{10, 4} \cdot Y_{4, 13} \cdot D_{13, 14} - D_{9, 10} \cdot Z_{10, 3} \cdot Y_{3, 13} \cdot X_{13, 14}& && & X_{14, 11} \cdot D_{11, 9} - Z_{14, 12} \cdot D_{12, 9}& \\ 
 &&\Lambda^{(2)}_{10, 13} &:& & X_{13, 14} \cdot X_{14, 8} \cdot Y_{8, 9} \cdot D_{9, 10} - D_{13, 14} \cdot Z_{14, 7} \cdot Y_{7, 9} \cdot X_{9, 10}& && & X_{10, 15} \cdot D_{15, 13} - Z_{10, 16} \cdot D_{16, 13}& \\ 
 &&\Lambda^{(3)}_{6, 1} &:& & D_{1, 2} \cdot Z_{2, 8} \cdot D_{8, 5} \cdot X_{5, 6} - X_{1, 2} \cdot X_{2, 7} \cdot D_{7, 5} \cdot D_{5, 6}& && & X_{6, 16} \cdot Y_{16, 1} - Z_{6, 15} \cdot Y_{15, 1}& \\ 
 &&\Lambda^{(3)}_{2, 5} &:& & D_{5, 6} \cdot Z_{6, 4} \cdot D_{4, 1} \cdot X_{1, 2} - X_{5, 6} \cdot X_{6, 3} \cdot D_{3, 1} \cdot D_{1, 2}& && & X_{2, 12} \cdot Y_{12, 5} - Z_{2, 11} \cdot Y_{11, 5}& \\ 
 &&\Lambda^{(3)}_{14, 9} &:& & D_{9, 10} \cdot Z_{10, 16} \cdot D_{16, 13} \cdot X_{13, 14} - X_{9, 10} \cdot X_{10, 15} \cdot D_{15, 13} \cdot D_{13, 14}& && & X_{14, 8} \cdot Y_{8, 9} - Z_{14, 7} \cdot Y_{7, 9}& \\ 
 &&\Lambda^{(3)}_{10, 13} &:& & D_{13, 14} \cdot Z_{14, 12} \cdot D_{12, 9} \cdot X_{9, 10} - X_{13, 14} \cdot X_{14, 11} \cdot D_{11, 9} \cdot D_{9, 10}& && & X_{10, 4} \cdot Y_{4, 13} - Z_{10, 3} \cdot Y_{3, 13}& \\ 
 &&\Lambda^{(4)}_{10, 1} &:& & D_{1, 2} \cdot X_{2, 7} \cdot Y_{7, 9} \cdot X_{9, 10} - X_{1, 2} \cdot X_{2, 12} \cdot D_{12, 9} \cdot D_{9, 10}& && & Z_{10, 3} \cdot D_{3, 1} - Z_{10, 16} \cdot Y_{16, 1}& \\ 
 &&\Lambda^{(4)}_{14, 5} &:& & D_{5, 6} \cdot X_{6, 3} \cdot Y_{3, 13} \cdot X_{13, 14} - X_{5, 6} \cdot X_{6, 16} \cdot D_{16, 13} \cdot D_{13, 14}& && & Z_{14, 7} \cdot D_{7, 5} - Z_{14, 12} \cdot Y_{12, 5}& \\ 
 &&\Lambda^{(4)}_{2, 9} &:& & D_{9, 10} \cdot X_{10, 15} \cdot Y_{15, 1} \cdot X_{1, 2} - X_{9, 10} \cdot X_{10, 4} \cdot D_{4, 1} \cdot D_{1, 2}& && & Z_{2, 11} \cdot D_{11, 9} - Z_{2, 8} \cdot Y_{8, 9}& \\ 
 &&\Lambda^{(4)}_{6, 13} &:& & D_{13, 14} \cdot X_{14, 11} \cdot Y_{11, 5} \cdot X_{5, 6} - X_{13, 14} \cdot X_{14, 8} \cdot D_{8, 5} \cdot D_{5, 6}& && & Z_{6, 15} \cdot D_{15, 13} - Z_{6, 4} \cdot Y_{4, 13}& \\ 
 &&\Lambda^{(5)}_{12, 3} &:& & D_{3, 1} \cdot X_{1, 2} \cdot X_{2, 12} - Y_{3, 13} \cdot X_{13, 14} \cdot Z_{14, 12}& && & D_{12, 9} \cdot D_{9, 10} \cdot Z_{10, 3} - Y_{12, 5} \cdot D_{5, 6} \cdot X_{6, 3}& \\ 
 &&\Lambda^{(5)}_{16, 7} &:& & D_{7, 5} \cdot X_{5, 6} \cdot X_{6, 16} - Y_{7, 9} \cdot X_{9, 10} \cdot Z_{10, 16}& && & D_{16, 13} \cdot D_{13, 14} \cdot Z_{14, 7} - Y_{16, 1} \cdot D_{1, 2} \cdot X_{2, 7}& \\ 
 &&\Lambda^{(5)}_{4, 11} &:& & D_{11, 9} \cdot X_{9, 10} \cdot X_{10, 4} - Y_{11, 5} \cdot X_{5, 6} \cdot Z_{6, 4}& && & D_{4, 1} \cdot D_{1, 2} \cdot Z_{2, 11} - Y_{4, 13} \cdot D_{13, 14} \cdot X_{14, 11}& \\ 
 &&\Lambda^{(5)}_{8, 15} &:& & D_{15, 13} \cdot X_{13, 14} \cdot X_{14, 8} - Y_{15, 1} \cdot X_{1, 2} \cdot Z_{2, 8}& && & D_{8, 5} \cdot D_{5, 6} \cdot Z_{6, 15} - Y_{8, 9} \cdot D_{9, 10} \cdot X_{10, 15}& \\ 
 &&\Lambda^{(6)}_{11, 4} &:& & Y_{4, 13} \cdot X_{13, 14} \cdot X_{14, 11} - D_{4, 1} \cdot X_{1, 2} \cdot Z_{2, 11}& && & D_{11, 9} \cdot D_{9, 10} \cdot X_{10, 4} - Y_{11, 5} \cdot D_{5, 6} \cdot Z_{6, 4}& \\ 
 &&\Lambda^{(6)}_{15, 8} &:& & Y_{8, 9} \cdot X_{9, 10} \cdot X_{10, 15} - D_{8, 5} \cdot X_{5, 6} \cdot Z_{6, 15}& && & D_{15, 13} \cdot D_{13, 14} \cdot X_{14, 8} - Y_{15, 1} \cdot D_{1, 2} \cdot Z_{2, 8}& \\ 
 &&\Lambda^{(6)}_{3, 12} &:& & Y_{12, 5} \cdot X_{5, 6} \cdot X_{6, 3} - D_{12, 9} \cdot X_{9, 10} \cdot Z_{10, 3}& && & D_{3, 1} \cdot D_{1, 2} \cdot X_{2, 12} - Y_{3, 13} \cdot D_{13, 14} \cdot Z_{14, 12}& \\ 
 &&\Lambda^{(6)}_{7, 16} &:& & Y_{16, 1} \cdot X_{1, 2} \cdot X_{2, 7} - D_{16, 13} \cdot X_{13, 14} \cdot Z_{14, 7}& && & D_{7, 5} \cdot D_{5, 6} \cdot X_{6, 16} - Y_{7, 9} \cdot D_{9, 10} \cdot Z_{10, 16}&
\end{array}
$}
~.~
\nn\\
\eea
The corresponding quiver diagram is shown in \fref{fig_416_quiver}.
The $J$- and $E$-terms
come from the general formula in \eqref{es03b01} 
with the following additional relabelling of indices,
\beal{es04f02}
&
\left[1,\ba{c}0 \\ 0\ea\right] \rightarrow 1 ~,~
\left[2,\ba{c}0 \\ 0\ea\right] \rightarrow 2 ~,~
\left[3,\ba{c}0 \\ 0\ea\right] \rightarrow 3 ~,~
\left[4,\ba{c}0 \\ 0\ea\right] \rightarrow 4 ~,~
&
\nn\\
&
\left[1,\ba{c}0 \\ 1\ea\right] \rightarrow 5 ~,~
\left[2,\ba{c}0 \\ 1\ea\right] \rightarrow 6 ~,~
\left[3,\ba{c}0 \\ 1\ea\right] \rightarrow 7 ~,~
\left[4,\ba{c}0 \\ 1\ea\right] \rightarrow 8 ~,~
&
\nn\\
&
\left[1,\ba{c}1 \\ 0\ea\right] \rightarrow 9 ~,~
\left[2,\ba{c}1 \\ 0\ea\right] \rightarrow 10 ~,~
\left[3,\ba{c}1 \\ 0\ea\right] \rightarrow 11 ~,~
\left[4,\ba{c}1 \\ 0\ea\right] \rightarrow 12 ~,~
&
\nn\\
&
\left[1,\ba{c}1 \\ 1\ea\right] \rightarrow 13 ~,~
\left[2,\ba{c}1 \\ 1\ea\right] \rightarrow 14 ~,~
\left[3,\ba{c}1 \\ 1\ea\right] \rightarrow 15 ~,~
\left[4,\ba{c}1 \\ 1\ea\right] \rightarrow 16 ~.~
\eea

Using the forward algorithm for brane brick models, 
we obtain the $P$-matrix as follows,
\beal{es04f03}
&&
P=
\nn\\
&&
\resizebox{0.95\textwidth}{!}{$
\left(
\begin{array}{c|cccccc|cc|cc|cc|cc|*{46}{c}|ccc}
& p_1 & p_2 & p_3 & p_4 & p_5 & p_6 & q^{(1)}_{1} & q^{(1)}_{2} & q^{(2)}_{1} & q^{(2)}_{2} & q^{(3)}_{1} & q^{(3)}_{2} & q^{(4)}_{1} & q^{(4)}_{2} & s_{1} & s_{2} & s_{3} & s_{4} & s_{5} & s_{6} & s_{7} & s_{8} & s_{9} & s_{10} & s_{11} & s_{12} & s_{13} & s_{14} & s_{15} & s_{16} & s_{17} & s_{18} & s_{19} & s_{20} & s_{21} & s_{22} & s_{23} & s_{24} & s_{25} & s_{26} & s_{27} & s_{28} & s_{29} & s_{30} & s_{31} & s_{32} & s_{33} & s_{34} & s_{35} & s_{36} & s_{37} & s_{38} & s_{39} & s_{40} & s_{41} & s_{42} & s_{43} & s_{44} & s_{45} & s_{46} & o^{(1)}_{1} & \cdots & o^{(12)}_{28} \\
\hline
D_{11,9} & 1 & 0 & 0 & 0 & 0 & 0 & 0 & 1 & 0 & 0 & 0 & 0 & 0 & 1 & 0 & 1 & 0 & 0 & 0 & 0 & 0 & 0 & 0 & 0 & 1 & 0 & 1 & 1 & 1 & 0 & 0 & 0 & 0 & 0 & 1 & 1 & 1 & 1 & 0 & 0 & 0 & 0 & 0 & 0 & 0 & 0 & 0 & 0 & 0 & 0 & 1 & 0 & 0 & 0 & 0 & 0 & 0 & 0 & 0 & 0 & 1 & \cdots & 0 \\
D_{12,9} & 0 & 0 & 0 & 1 & 0 & 0 & 0 & 1 & 0 & 0 & 0 & 1 & 0 & 0 & 0 & 1 & 0 & 0 & 0 & 0 & 0 & 0 & 0 & 0 & 0 & 0 & 0 & 0 & 0 & 0 & 0 & 0 & 0 & 0 & 1 & 1 & 1 & 1 & 0 & 0 & 1 & 1 & 1 & 1 & 0 & 0 & 0 & 0 & 0 & 0 & 1 & 0 & 0 & 0 & 0 & 0 & 0 & 0 & 0 & 0 & 0 & \cdots & 2 \\
D_{13,14} & 0 & 0 & 1 & 0 & 0 & 0 & 0 & 0 & 0 & 0 & 0 & 0 & 0 & 0 & 0 & 0 & 0 & 0 & 0 & 0 & 0 & 0 & 1 & 0 & 0 & 0 & 0 & 0 & 0 & 1 & 0 & 0 & 0 & 0 & 0 & 0 & 0 & 0 & 0 & 0 & 0 & 0 & 0 & 0 & 0 & 0 & 0 & 0 & 1 & 0 & 0 & 0 & 0 & 0 & 0 & 0 & 0 & 0 & 0 & 0 & 0 & \cdots & 0 \\
D_{15,13} & 1 & 0 & 0 & 0 & 0 & 0 & 0 & 1 & 0 & 0 & 0 & 0 & 1 & 0 & 1 & 0 & 0 & 0 & 0 & 0 & 0 & 0 & 0 & 0 & 0 & 1 & 0 & 0 & 0 & 0 & 0 & 0 & 0 & 0 & 0 & 0 & 0 & 0 & 0 & 0 & 0 & 0 & 0 & 0 & 1 & 1 & 1 & 1 & 0 & 0 & 0 & 0 & 0 & 0 & 0 & 0 & 1 & 1 & 1 & 1 & 1 & \cdots & 0 \\
D_{16,13} & 0 & 0 & 0 & 1 & 0 & 0 & 0 & 1 & 0 & 0 & 1 & 0 & 0 & 0 & 1 & 0 & 0 & 0 & 1 & 0 & 1 & 0 & 0 & 0 & 0 & 1 & 0 & 0 & 0 & 0 & 0 & 0 & 0 & 0 & 0 & 0 & 0 & 0 & 0 & 0 & 0 & 0 & 0 & 0 & 0 & 1 & 0 & 1 & 0 & 0 & 0 & 0 & 0 & 1 & 0 & 1 & 0 & 1 & 0 & 1 & 0 & \cdots & 0 \\
D_{1,2} & 0 & 0 & 1 & 0 & 0 & 0 & 0 & 0 & 0 & 0 & 0 & 0 & 0 & 0 & 1 & 0 & 0 & 0 & 0 & 0 & 0 & 0 & 1 & 1 & 0 & 0 & 0 & 0 & 0 & 0 & 0 & 0 & 0 & 0 & 0 & 0 & 0 & 0 & 0 & 0 & 0 & 0 & 0 & 0 & 0 & 0 & 0 & 0 & 0 & 0 & 0 & 0 & 0 & 0 & 0 & 0 & 0 & 0 & 0 & 0 & 0 & \cdots & 1 \\
D_{3,1} & 1 & 0 & 0 & 0 & 0 & 0 & 1 & 0 & 0 & 0 & 0 & 0 & 0 & 1 & 0 & 0 & 0 & 0 & 0 & 1 & 1 & 0 & 0 & 0 & 0 & 0 & 0 & 0 & 0 & 0 & 0 & 0 & 0 & 0 & 0 & 0 & 0 & 0 & 0 & 0 & 0 & 0 & 0 & 0 & 0 & 0 & 1 & 1 & 1 & 1 & 0 & 0 & 0 & 0 & 1 & 1 & 0 & 0 & 1 & 1 & 1 & \cdots & 0 \\
D_{4,1} & 0 & 0 & 0 & 1 & 0 & 0 & 1 & 0 & 0 & 0 & 0 & 1 & 0 & 0 & 0 & 0 & 0 & 1 & 1 & 1 & 1 & 0 & 0 & 0 & 0 & 0 & 0 & 0 & 0 & 0 & 0 & 0 & 0 & 0 & 0 & 0 & 0 & 0 & 0 & 0 & 0 & 0 & 0 & 0 & 1 & 1 & 1 & 1 & 1 & 1 & 0 & 0 & 0 & 0 & 0 & 0 & 0 & 0 & 0 & 0 & 1 & \cdots & 0 \\
D_{5,6} & 0 & 0 & 1 & 0 & 0 & 0 & 0 & 0 & 0 & 0 & 0 & 0 & 0 & 0 & 0 & 1 & 1 & 0 & 0 & 0 & 0 & 1 & 0 & 0 & 0 & 0 & 0 & 0 & 0 & 0 & 0 & 0 & 0 & 0 & 0 & 0 & 0 & 0 & 0 & 0 & 0 & 0 & 0 & 0 & 0 & 0 & 0 & 0 & 0 & 0 & 0 & 0 & 0 & 0 & 0 & 0 & 0 & 0 & 0 & 0 & 1 & \cdots & 0 \\
D_{7,5} & 1 & 0 & 0 & 0 & 0 & 0 & 1 & 0 & 0 & 0 & 0 & 0 & 1 & 0 & 0 & 0 & 0 & 0 & 0 & 0 & 0 & 0 & 0 & 0 & 0 & 0 & 1 & 0 & 1 & 0 & 0 & 1 & 0 & 1 & 0 & 1 & 0 & 1 & 1 & 1 & 0 & 1 & 0 & 1 & 0 & 0 & 0 & 0 & 0 & 0 & 0 & 0 & 0 & 0 & 0 & 0 & 0 & 0 & 0 & 0 & 1 & \cdots & 0 \\
D_{8,5} & 0 & 0 & 0 & 1 & 0 & 0 & 1 & 0 & 0 & 0 & 1 & 0 & 0 & 0 & 0 & 0 & 0 & 0 & 0 & 0 & 0 & 0 & 0 & 0 & 0 & 0 & 0 & 1 & 1 & 0 & 0 & 0 & 1 & 1 & 0 & 0 & 1 & 1 & 1 & 1 & 0 & 0 & 1 & 1 & 0 & 0 & 0 & 0 & 0 & 0 & 0 & 0 & 0 & 0 & 0 & 0 & 0 & 0 & 0 & 0 & 0 & \cdots & 0 \\
D_{9,10} & 0 & 0 & 1 & 0 & 0 & 0 & 0 & 0 & 0 & 0 & 0 & 0 & 0 & 0 & 0 & 0 & 1 & 0 & 0 & 0 & 0 & 0 & 0 & 0 & 0 & 0 & 0 & 0 & 0 & 0 & 0 & 0 & 0 & 0 & 0 & 0 & 0 & 0 & 1 & 0 & 0 & 0 & 0 & 0 & 0 & 0 & 0 & 0 & 0 & 0 & 0 & 1 & 0 & 0 & 0 & 0 & 0 & 0 & 0 & 0 & 0 & \cdots & 0 \\
X_{10,15} & 0 & 0 & 0 & 1 & 0 & 0 & 1 & 0 & 0 & 0 & 1 & 0 & 0 & 0 & 0 & 0 & 0 & 1 & 1 & 1 & 1 & 1 & 0 & 0 & 0 & 0 & 0 & 0 & 0 & 0 & 0 & 0 & 0 & 0 & 0 & 0 & 0 & 0 & 0 & 1 & 0 & 0 & 0 & 0 & 0 & 0 & 0 & 0 & 0 & 0 & 0 & 0 & 1 & 1 & 1 & 1 & 0 & 0 & 0 & 0 & 1 & \cdots & 0 \\
X_{10,4} & 0 & 0 & 0 & 0 & 0 & 1 & 0 & 0 & 0 & 1 & 1 & 0 & 0 & 0 & 0 & 0 & 0 & 0 & 0 & 0 & 0 & 1 & 0 & 0 & 0 & 0 & 0 & 0 & 0 & 0 & 0 & 0 & 0 & 0 & 0 & 0 & 0 & 0 & 0 & 1 & 0 & 0 & 0 & 0 & 0 & 0 & 0 & 0 & 0 & 0 & 0 & 0 & 1 & 1 & 1 & 1 & 1 & 1 & 1 & 1 & 0 & \cdots & 1 \\
X_{13,14} & 0 & 1 & 0 & 0 & 0 & 0 & 0 & 0 & 0 & 0 & 0 & 0 & 0 & 0 & 0 & 0 & 0 & 0 & 0 & 0 & 0 & 0 & 1 & 0 & 0 & 0 & 0 & 0 & 0 & 1 & 0 & 0 & 0 & 0 & 0 & 0 & 0 & 0 & 0 & 0 & 0 & 0 & 0 & 0 & 0 & 0 & 0 & 0 & 1 & 0 & 0 & 0 & 0 & 0 & 0 & 0 & 0 & 0 & 0 & 0 & 0 & \cdots & 0 \\
X_{14,11} & 0 & 0 & 0 & 1 & 0 & 0 & 1 & 0 & 0 & 0 & 0 & 1 & 0 & 0 & 0 & 0 & 0 & 0 & 0 & 0 & 0 & 0 & 0 & 1 & 0 & 0 & 0 & 0 & 0 & 0 & 1 & 1 & 1 & 1 & 0 & 0 & 0 & 0 & 0 & 0 & 1 & 1 & 1 & 1 & 0 & 0 & 0 & 0 & 0 & 1 & 0 & 0 & 0 & 0 & 0 & 0 & 0 & 0 & 0 & 0 & 0 & \cdots & 2 \\
X_{14,8} & 0 & 0 & 0 & 0 & 0 & 1 & 0 & 0 & 0 & 1 & 0 & 1 & 0 & 0 & 0 & 0 & 0 & 0 & 0 & 0 & 0 & 0 & 0 & 1 & 1 & 0 & 1 & 0 & 0 & 0 & 1 & 1 & 0 & 0 & 1 & 1 & 0 & 0 & 0 & 0 & 1 & 1 & 0 & 0 & 0 & 0 & 0 & 0 & 0 & 1 & 0 & 0 & 0 & 0 & 0 & 0 & 0 & 0 & 0 & 0 & 0 & \cdots & 2 \\
X_{1,2} & 0 & 1 & 0 & 0 & 0 & 0 & 0 & 0 & 0 & 0 & 0 & 0 & 0 & 0 & 1 & 0 & 0 & 0 & 0 & 0 & 0 & 0 & 1 & 1 & 0 & 0 & 0 & 0 & 0 & 0 & 0 & 0 & 0 & 0 & 0 & 0 & 0 & 0 & 0 & 0 & 0 & 0 & 0 & 0 & 0 & 0 & 0 & 0 & 0 & 0 & 0 & 0 & 0 & 0 & 0 & 0 & 0 & 0 & 0 & 0 & 0 & \cdots & 1 \\
X_{2,12} & 0 & 0 & 0 & 0 & 0 & 1 & 0 & 0 & 1 & 0 & 1 & 0 & 0 & 0 & 0 & 0 & 0 & 0 & 0 & 0 & 0 & 0 & 0 & 0 & 1 & 1 & 1 & 1 & 1 & 1 & 1 & 1 & 1 & 1 & 0 & 0 & 0 & 0 & 0 & 0 & 0 & 0 & 0 & 0 & 0 & 0 & 0 & 0 & 0 & 0 & 0 & 0 & 0 & 0 & 0 & 0 & 0 & 0 & 0 & 0 & 0 & \cdots & 0 \\
X_{2,7} & 0 & 0 & 0 & 1 & 0 & 0 & 0 & 1 & 0 & 0 & 1 & 0 & 0 & 0 & 0 & 0 & 0 & 0 & 0 & 0 & 0 & 0 & 0 & 0 & 1 & 1 & 0 & 1 & 0 & 1 & 1 & 0 & 1 & 0 & 1 & 0 & 1 & 0 & 0 & 0 & 1 & 0 & 1 & 0 & 0 & 0 & 0 & 0 & 0 & 0 & 0 & 0 & 0 & 0 & 0 & 0 & 0 & 0 & 0 & 0 & 0 & \cdots & 0 \\
X_{5,6} & 0 & 1 & 0 & 0 & 0 & 0 & 0 & 0 & 0 & 0 & 0 & 0 & 0 & 0 & 0 & 1 & 1 & 0 & 0 & 0 & 0 & 1 & 0 & 0 & 0 & 0 & 0 & 0 & 0 & 0 & 0 & 0 & 0 & 0 & 0 & 0 & 0 & 0 & 0 & 0 & 0 & 0 & 0 & 0 & 0 & 0 & 0 & 0 & 0 & 0 & 0 & 0 & 0 & 0 & 0 & 0 & 0 & 0 & 0 & 0 & 1 & \cdots & 0 \\
X_{6,16} & 0 & 0 & 0 & 0 & 0 & 1 & 0 & 0 & 1 & 0 & 0 & 1 & 0 & 0 & 0 & 0 & 0 & 1 & 0 & 1 & 0 & 0 & 0 & 0 & 0 & 0 & 0 & 0 & 0 & 0 & 0 & 0 & 0 & 0 & 0 & 0 & 0 & 0 & 0 & 0 & 0 & 0 & 0 & 0 & 1 & 0 & 1 & 0 & 0 & 0 & 1 & 1 & 1 & 0 & 1 & 0 & 1 & 0 & 1 & 0 & 0 & \cdots & 2 \\
X_{6,3} & 0 & 0 & 0 & 1 & 0 & 0 & 0 & 1 & 0 & 0 & 0 & 1 & 0 & 0 & 0 & 0 & 0 & 1 & 1 & 0 & 0 & 0 & 0 & 0 & 0 & 0 & 0 & 0 & 0 & 0 & 0 & 0 & 0 & 0 & 0 & 0 & 0 & 0 & 0 & 0 & 0 & 0 & 0 & 0 & 1 & 1 & 0 & 0 & 0 & 0 & 1 & 1 & 1 & 1 & 0 & 0 & 1 & 1 & 0 & 0 & 0 & \cdots & 1 \\
X_{9,10} & 0 & 1 & 0 & 0 & 0 & 0 & 0 & 0 & 0 & 0 & 0 & 0 & 0 & 0 & 0 & 0 & 1 & 0 & 0 & 0 & 0 & 0 & 0 & 0 & 0 & 0 & 0 & 0 & 0 & 0 & 0 & 0 & 0 & 0 & 0 & 0 & 0 & 0 & 1 & 0 & 0 & 0 & 0 & 0 & 0 & 0 & 0 & 0 & 0 & 0 & 0 & 1 & 0 & 0 & 0 & 0 & 0 & 0 & 0 & 0 & 0 & \cdots & 0 \\
Y_{11,5} & 0 & 0 & 0 & 0 & 0 & 1 & 0 & 0 & 0 & 1 & 1 & 0 & 0 & 0 & 0 & 0 & 0 & 0 & 0 & 0 & 0 & 0 & 0 & 0 & 1 & 0 & 1 & 1 & 1 & 0 & 0 & 0 & 0 & 0 & 1 & 1 & 1 & 1 & 1 & 1 & 0 & 0 & 0 & 0 & 0 & 0 & 0 & 0 & 0 & 0 & 0 & 0 & 0 & 0 & 0 & 0 & 0 & 0 & 0 & 0 & 0 & \cdots & 0 \\
Y_{12,5} & 0 & 0 & 0 & 0 & 1 & 0 & 0 & 0 & 0 & 1 & 0 & 0 & 1 & 0 & 0 & 0 & 0 & 0 & 0 & 0 & 0 & 0 & 0 & 0 & 0 & 0 & 0 & 0 & 0 & 0 & 0 & 0 & 0 & 0 & 1 & 1 & 1 & 1 & 1 & 1 & 1 & 1 & 1 & 1 & 0 & 0 & 0 & 0 & 0 & 0 & 0 & 0 & 0 & 0 & 0 & 0 & 0 & 0 & 0 & 0 & 0 & \cdots & 1 \\
Y_{15,1} & 0 & 0 & 0 & 0 & 0 & 1 & 0 & 0 & 0 & 1 & 0 & 1 & 0 & 0 & 0 & 0 & 0 & 0 & 0 & 0 & 0 & 0 & 0 & 0 & 0 & 0 & 0 & 0 & 0 & 0 & 0 & 0 & 0 & 0 & 0 & 0 & 0 & 0 & 0 & 0 & 0 & 0 & 0 & 0 & 1 & 1 & 1 & 1 & 1 & 1 & 0 & 0 & 0 & 0 & 0 & 0 & 1 & 1 & 1 & 1 & 0 & \cdots & 1 \\
Y_{16,1} & 0 & 0 & 0 & 0 & 1 & 0 & 0 & 0 & 0 & 1 & 0 & 0 & 0 & 1 & 0 & 0 & 0 & 0 & 1 & 0 & 1 & 0 & 0 & 0 & 0 & 0 & 0 & 0 & 0 & 0 & 0 & 0 & 0 & 0 & 0 & 0 & 0 & 0 & 0 & 0 & 0 & 0 & 0 & 0 & 0 & 1 & 0 & 1 & 1 & 1 & 0 & 0 & 0 & 1 & 0 & 1 & 0 & 1 & 0 & 1 & 0 & \cdots & 0 \\
Y_{3,13} & 0 & 0 & 0 & 0 & 0 & 1 & 0 & 0 & 1 & 0 & 1 & 0 & 0 & 0 & 1 & 0 & 0 & 0 & 0 & 1 & 1 & 0 & 0 & 0 & 0 & 1 & 0 & 0 & 0 & 0 & 0 & 0 & 0 & 0 & 0 & 0 & 0 & 0 & 0 & 0 & 0 & 0 & 0 & 0 & 0 & 0 & 1 & 1 & 0 & 0 & 0 & 0 & 0 & 0 & 1 & 1 & 0 & 0 & 1 & 1 & 0 & \cdots & 1 \\
Y_{4,13} & 0 & 0 & 0 & 0 & 1 & 0 & 0 & 0 & 1 & 0 & 0 & 0 & 1 & 0 & 1 & 0 & 0 & 1 & 1 & 1 & 1 & 0 & 0 & 0 & 0 & 1 & 0 & 0 & 0 & 0 & 0 & 0 & 0 & 0 & 0 & 0 & 0 & 0 & 0 & 0 & 0 & 0 & 0 & 0 & 1 & 1 & 1 & 1 & 0 & 0 & 0 & 0 & 0 & 0 & 0 & 0 & 0 & 0 & 0 & 0 & 1 & \cdots & 0 \\
Y_{7,9} & 0 & 0 & 0 & 0 & 0 & 1 & 0 & 0 & 1 & 0 & 0 & 1 & 0 & 0 & 0 & 1 & 0 & 0 & 0 & 0 & 0 & 0 & 0 & 0 & 0 & 0 & 1 & 0 & 1 & 0 & 0 & 1 & 0 & 1 & 0 & 1 & 0 & 1 & 0 & 0 & 0 & 1 & 0 & 1 & 0 & 0 & 0 & 0 & 0 & 0 & 1 & 0 & 0 & 0 & 0 & 0 & 0 & 0 & 0 & 0 & 0 & \cdots & 2 \\
Y_{8,9} & 0 & 0 & 0 & 0 & 1 & 0 & 0 & 0 & 1 & 0 & 0 & 0 & 0 & 1 & 0 & 1 & 0 & 0 & 0 & 0 & 0 & 0 & 0 & 0 & 0 & 0 & 0 & 1 & 1 & 0 & 0 & 0 & 1 & 1 & 0 & 0 & 1 & 1 & 0 & 0 & 0 & 0 & 1 & 1 & 0 & 0 & 0 & 0 & 0 & 0 & 1 & 0 & 0 & 0 & 0 & 0 & 0 & 0 & 0 & 0 & 0 & \cdots & 1 \\
Z_{10,16} & 1 & 0 & 0 & 0 & 0 & 0 & 1 & 0 & 0 & 0 & 0 & 0 & 1 & 0 & 0 & 0 & 0 & 1 & 0 & 1 & 0 & 1 & 0 & 0 & 0 & 0 & 0 & 0 & 0 & 0 & 0 & 0 & 0 & 0 & 0 & 0 & 0 & 0 & 0 & 1 & 0 & 0 & 0 & 0 & 1 & 0 & 1 & 0 & 0 & 0 & 0 & 0 & 1 & 0 & 1 & 0 & 1 & 0 & 1 & 0 & 2 & \cdots & 0 \\
Z_{10,3} & 0 & 0 & 0 & 0 & 1 & 0 & 0 & 0 & 0 & 1 & 0 & 0 & 1 & 0 & 0 & 0 & 0 & 1 & 1 & 0 & 0 & 1 & 0 & 0 & 0 & 0 & 0 & 0 & 0 & 0 & 0 & 0 & 0 & 0 & 0 & 0 & 0 & 0 & 0 & 1 & 0 & 0 & 0 & 0 & 1 & 1 & 0 & 0 & 0 & 0 & 0 & 0 & 1 & 1 & 0 & 0 & 1 & 1 & 0 & 0 & 1 & \cdots & 0 \\
Z_{14,12} & 1 & 0 & 0 & 0 & 0 & 0 & 1 & 0 & 0 & 0 & 0 & 0 & 0 & 1 & 0 & 0 & 0 & 0 & 0 & 0 & 0 & 0 & 0 & 1 & 1 & 0 & 1 & 1 & 1 & 0 & 1 & 1 & 1 & 1 & 0 & 0 & 0 & 0 & 0 & 0 & 0 & 0 & 0 & 0 & 0 & 0 & 0 & 0 & 0 & 1 & 0 & 0 & 0 & 0 & 0 & 0 & 0 & 0 & 0 & 0 & 1 & \cdots & 0 \\
Z_{14,7} & 0 & 0 & 0 & 0 & 1 & 0 & 0 & 0 & 0 & 1 & 0 & 0 & 0 & 1 & 0 & 0 & 0 & 0 & 0 & 0 & 0 & 0 & 0 & 1 & 1 & 0 & 0 & 1 & 0 & 0 & 1 & 0 & 1 & 0 & 1 & 0 & 1 & 0 & 0 & 0 & 1 & 0 & 1 & 0 & 0 & 0 & 0 & 0 & 0 & 1 & 0 & 0 & 0 & 0 & 0 & 0 & 0 & 0 & 0 & 0 & 0 & \cdots & 1 \\
Z_{2,11} & 0 & 0 & 0 & 0 & 1 & 0 & 0 & 0 & 1 & 0 & 0 & 0 & 1 & 0 & 0 & 0 & 0 & 0 & 0 & 0 & 0 & 0 & 0 & 0 & 0 & 1 & 0 & 0 & 0 & 1 & 1 & 1 & 1 & 1 & 0 & 0 & 0 & 0 & 0 & 0 & 1 & 1 & 1 & 1 & 0 & 0 & 0 & 0 & 0 & 0 & 0 & 0 & 0 & 0 & 0 & 0 & 0 & 0 & 0 & 0 & 0 & \cdots & 1 \\
Z_{2,8} & 1 & 0 & 0 & 0 & 0 & 0 & 0 & 1 & 0 & 0 & 0 & 0 & 1 & 0 & 0 & 0 & 0 & 0 & 0 & 0 & 0 & 0 & 0 & 0 & 1 & 1 & 1 & 0 & 0 & 1 & 1 & 1 & 0 & 0 & 1 & 1 & 0 & 0 & 0 & 0 & 1 & 1 & 0 & 0 & 0 & 0 & 0 & 0 & 0 & 0 & 0 & 0 & 0 & 0 & 0 & 0 & 0 & 0 & 0 & 0 & 1 & \cdots & 0 \\
Z_{6,15} & 0 & 0 & 0 & 0 & 1 & 0 & 0 & 0 & 1 & 0 & 0 & 0 & 0 & 1 & 0 & 0 & 0 & 1 & 1 & 1 & 1 & 0 & 0 & 0 & 0 & 0 & 0 & 0 & 0 & 0 & 0 & 0 & 0 & 0 & 0 & 0 & 0 & 0 & 0 & 0 & 0 & 0 & 0 & 0 & 0 & 0 & 0 & 0 & 0 & 0 & 1 & 1 & 1 & 1 & 1 & 1 & 0 & 0 & 0 & 0 & 0 & \cdots & 1 \\
Z_{6,4} & 1 & 0 & 0 & 0 & 0 & 0 & 0 & 1 & 0 & 0 & 0 & 0 & 0 & 1 & 0 & 0 & 0 & 0 & 0 & 0 & 0 & 0 & 0 & 0 & 0 & 0 & 0 & 0 & 0 & 0 & 0 & 0 & 0 & 0 & 0 & 0 & 0 & 0 & 0 & 0 & 0 & 0 & 0 & 0 & 0 & 0 & 0 & 0 & 0 & 0 & 1 & 1 & 1 & 1 & 1 & 1 & 1 & 1 & 1 & 1 & 0 & \cdots & 1 \\
\end{array}
\right)
$}
~.~
\nn\\
\eea
where there are in total $376$ extra GLSM fields
partitioned into 28 sets identified by $o_k^{(1)}, \dots, o_l^{(28)}$.
The $J$- and $E$-term charge matrix $Q_{JE}$
can be obtained from the kernel of the $P$-matrix above.
Here, the $Q_{JE}$-matrix is a 
$417 \times 436$ dimensional matrix, which we choose not to present here.
In addition, we have the $D$-term charge matrix, which takes the following form,
\beal{es04f05}
&&
Q_{D}=
\nn\\
&&
\resizebox{0.9\textwidth}{!}{$
\left(
\begin{array}{cccccc|cc|cc|cc|cc|cccccccccccccccccccccccccccccccccccccccccccccc|ccc}
p_1 & p_2 & p_3 & p_4 & p_5 & p_6 & q^{(1)}_{1} & q^{(1)}_{2} & q^{(2)}_{1} & q^{(2)}_{2} & q^{(3)}_{1} & q^{(3)}_{2} & q^{(4)}_{1} & q^{(4)}_{2} & s_{1} & s_{2} & s_{3} & s_{4} & s_{5} & s_{6} & s_{7} & s_{8} & s_{9} & s_{10} & s_{11} & s_{12} & s_{13} & s_{14} & s_{15} & s_{16} & s_{17} & s_{18} & s_{19} & s_{20} & s_{21} & s_{22} & s_{23} & s_{24} & s_{25} & s_{26} & s_{27} & s_{28} & s_{29} & s_{30} & s_{31} & s_{32} & s_{33} & s_{34} & s_{35} & s_{36} & s_{37} & s_{38} & s_{39} & s_{40} & s_{41} & s_{42} & s_{43} & s_{44} & s_{45} & s_{46} & o^{(1)}_{1} & \cdots & o^{(12)}_{28} \\
\hline
0 & 0 & 0 & 0 & 0 & 0 & 0 & 0 & 0 & 0 & 0 & 0 & 0 & 0 & -1 & 0 & 0 & -2 & 1 & 1 & 0 & 0 & 0 & 0 & 0 & 0 & 0 & 0 & 0 & 0 & 0 & 0 & 0 & 0 & 0 & 0 & 0 & 0 & 0 & 0 & 0 & 0 & 0 & 0 & 1 & 0 & 0 & 0 & 0 & 0 & 0 & 0 & 0 & 0 & 0 & 0 & 0 & 0 & 0 & 0 & 0 & \cdots & 0 \\
0 & 0 & 0 & 0 & 0 & 0 & 0 & 0 & -1 & 0 & 0 & 0 & 0 & 0 & 1 & 1 & 0 & 1 & 0 & 0 & 0 & 0 & 0 & 0 & 0 & 0 & 0 & 0 & 0 & 0 & 0 & 0 & 0 & 0 & 0 & 0 & 0 & 0 & 0 & 0 & 0 & 0 & 0 & 0 & 0 & 0 & 0 & 0 & 0 & 0 & 0 & 0 & 0 & 0 & 0 & 0 & 0 & 0 & 0 & 0 & 0 & \cdots & 0 \\
0 & 0 & 0 & 0 & 0 & 0 & 0 & 0 & 0 & 0 & 0 & 0 & 0 & 0 & 0 & 0 & 0 & 1 & 0 & -1 & 0 & 0 & 0 & 0 & 0 & 0 & 0 & 0 & 0 & 0 & 0 & 0 & 0 & 0 & 0 & 0 & 0 & 0 & 0 & 0 & 0 & 0 & 0 & 0 & 0 & 0 & 0 & 0 & 0 & 0 & 0 & 0 & 0 & 0 & 0 & 0 & 0 & 0 & 0 & 0 & 0 & \cdots & 0 \\
0 & 0 & 0 & 0 & 0 & 0 & 0 & 0 & 0 & 0 & 0 & 0 & 0 & 0 & 0 & -1 & 0 & -1 & 0 & 0 & 0 & 1 & 0 & 0 & 0 & 0 & 0 & 0 & 0 & 0 & 0 & 0 & 0 & 0 & 0 & 0 & 0 & 0 & 0 & 0 & 0 & 0 & 0 & 0 & 0 & 0 & 0 & 0 & 0 & 0 & 1 & 0 & 0 & 0 & 0 & 0 & 0 & 0 & 0 & 0 & 0 & \cdots & 0 \\
0 & 0 & 0 & 1 & 1 & 0 & 0 & 0 & -1 & 0 & 0 & 0 & 0 & 0 & 1 & 1 & 0 & 1 & -1 & 0 & 0 & 0 & 0 & -1 & 1 & 0 & 0 & 0 & 0 & 0 & 0 & 0 & 0 & 0 & 0 & 0 & 0 & 0 & 0 & 0 & 0 & 0 & 0 & 0 & 0 & 0 & 0 & 0 & 0 & 0 & 0 & 0 & 0 & 0 & 0 & 0 & 0 & 0 & 0 & 0 & 0 & \cdots & 0 \\
0 & 0 & 0 & 0 & 0 & 0 & 0 & 0 & 0 & 0 & 0 & 0 & 0 & 0 & 0 & 1 & 0 & 0 & 0 & 0 & 0 & 0 & 0 & 0 & 0 & 0 & 0 & 0 & 0 & 0 & 0 & 0 & 0 & 0 & 0 & 0 & 0 & 0 & 0 & 0 & 0 & 0 & 0 & 0 & 0 & 0 & 0 & 0 & 0 & 0 & -1 & 0 & 0 & 0 & 0 & 0 & 0 & 0 & 0 & 0 & 0 & \cdots & 0 \\
0 & 0 & 0 & 0 & 0 & 0 & 0 & 0 & 0 & 0 & 0 & 0 & 0 & 0 & 0 & -1 & 0 & 0 & 0 & 0 & 0 & 0 & 0 & 0 & 0 & 0 & 0 & 0 & 0 & 0 & 0 & 0 & 0 & 0 & 0 & 0 & 0 & 0 & 0 & 0 & 0 & 0 & 0 & 0 & 0 & 0 & 0 & 0 & 0 & 0 & 0 & 0 & 0 & 0 & 0 & 0 & 0 & 0 & 0 & 0 & 0 & \cdots & 0 \\
0 & 0 & 0 & -1 & 0 & 0 & 0 & 0 & 0 & 0 & 0 & 0 & 0 & 0 & 0 & 0 & 0 & 0 & 0 & 0 & 0 & 0 & 0 & 1 & 0 & 0 & 0 & 0 & 0 & 0 & 0 & 0 & 0 & 0 & 0 & 0 & 0 & 0 & 0 & 0 & 0 & 0 & 0 & 0 & 0 & 0 & 0 & 0 & 0 & 0 & 0 & 0 & 0 & 0 & 0 & 0 & 0 & 0 & 0 & 0 & 0 & \cdots & 0 \\
0 & 0 & 0 & 0 & 0 & 0 & 0 & 0 & 0 & 0 & 0 & 0 & 0 & 0 & 0 & 1 & -1 & 0 & 0 & 0 & 0 & 0 & 0 & 0 & 0 & 0 & 0 & 0 & 0 & 0 & 0 & 0 & 0 & 0 & 0 & 0 & 0 & 0 & 0 & 0 & 0 & 0 & 0 & 0 & 0 & 0 & 0 & 0 & 0 & 0 & 0 & 0 & 0 & 0 & 0 & 0 & 0 & 0 & 0 & 0 & 0 & \cdots & 0 \\
0 & 0 & 0 & 0 & 0 & 0 & 0 & 0 & 0 & 0 & 0 & 0 & 0 & 0 & 0 & 0 & 1 & 0 & 0 & 0 & 0 & -1 & 0 & 0 & 0 & 0 & 0 & 0 & 0 & 0 & 0 & 0 & 0 & 0 & 0 & 0 & 0 & 0 & 0 & 0 & 0 & 0 & 0 & 0 & 0 & 0 & 0 & 0 & 0 & 0 & 0 & 0 & 0 & 0 & 0 & 0 & 0 & 0 & 0 & 0 & 0 & \cdots & 0 \\
0 & 0 & 0 & 0 & 0 & 0 & 0 & 0 & 1 & 0 & 0 & 0 & 0 & 0 & -1 & -1 & 0 & -1 & 0 & 0 & 0 & 0 & 0 & 1 & -1 & 0 & 0 & 0 & 0 & 0 & 0 & 0 & 0 & 0 & 0 & 0 & 0 & 0 & 0 & 0 & 0 & 0 & 0 & 0 & 0 & 0 & 0 & 0 & 0 & 0 & 0 & 0 & 0 & 0 & 0 & 0 & 0 & 0 & 0 & 0 & 0 & \cdots & 0 \\
0 & 0 & 0 & 0 & -1 & 0 & 0 & 0 & 1 & 0 & 0 & 0 & 0 & 0 & -1 & -1 & 0 & -1 & 1 & 0 & 0 & 0 & 0 & 0 & 0 & 0 & 0 & 0 & 0 & 0 & 0 & 0 & 0 & 0 & 0 & 0 & 0 & 0 & 0 & 0 & 0 & 0 & 0 & 0 & 0 & 0 & 0 & 0 & 0 & 0 & 0 & 0 & 0 & 0 & 0 & 0 & 0 & 0 & 0 & 0 & 0 & \cdots & 0 \\
0 & -1 & -1 & 0 & 0 & 0 & 0 & 0 & 0 & 0 & 0 & 0 & 0 & 0 & 1 & 0 & 1 & 0 & 0 & 0 & 0 & 0 & 0 & 0 & 0 & 0 & 0 & 0 & 0 & 0 & 0 & 0 & 0 & 0 & 0 & 0 & 0 & 0 & 0 & 0 & 0 & 0 & 0 & 0 & 0 & 0 & 0 & 0 & 0 & 0 & 0 & 0 & 0 & 0 & 0 & 0 & 0 & 0 & 0 & 0 & 0 & \cdots & 0 \\
0 & 1 & 1 & 0 & 0 & 0 & 0 & 0 & 0 & 0 & 0 & 0 & 0 & 0 & 0 & 0 & -1 & 0 & 0 & 0 & 0 & 0 & 0 & -1 & 0 & 0 & 0 & 0 & 0 & 0 & 0 & 0 & 0 & 0 & 0 & 0 & 0 & 0 & 0 & 0 & 0 & 0 & 0 & 0 & 0 & 0 & 0 & 0 & 0 & 0 & 0 & 0 & 0 & 0 & 0 & 0 & 0 & 0 & 0 & 0 & 0 & \cdots & 0 \\
0 & 0 & 0 & 0 & 0 & 0 & 0 & 0 & 0 & 0 & 0 & 0 & 0 & 0 & 0 & 0 & 0 & 1 & 0 & 0 & 0 & 0 & 0 & 0 & 0 & 0 & 0 & 0 & 0 & 0 & 0 & 0 & 0 & 0 & 0 & 0 & 0 & 0 & 0 & 0 & 0 & 0 & 0 & 0 & -1 & 0 & 0 & 0 & 0 & 0 & 0 & 0 & 0 & 0 & 0 & 0 & 0 & 0 & 0 & 0 & 0 & \cdots & 0 \\
\end{array}
\right)
$}
~,~
\nn\\
\eea

\begin{figure}[htt!!]
    \centering
    \includegraphics[width=0.4\textwidth]{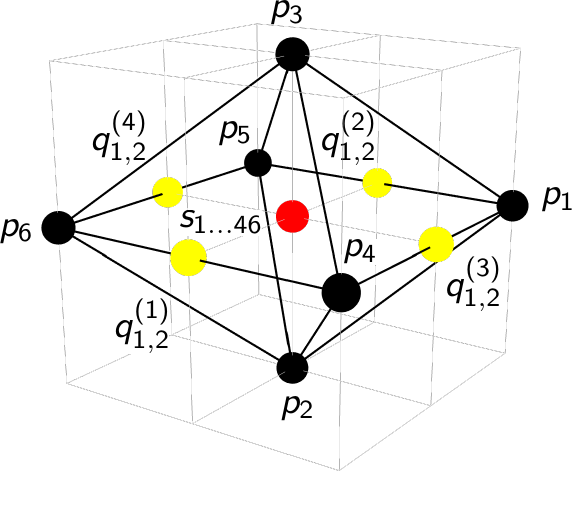}
    \caption{The toric diagram for the $Q^{1,1,1}/ \mathbb{Z}_2 \times \mathbb{Z}_2 \ \left(\begin{array}{@{}c@{,\;}c@{,\;}c@{,\;}c@{,\;}c@{,\;}c@{,\;}c@{,\;}c@{}} 1 & 1 & 0 & 0 & 0 & 0 & 1 & 1 \\ 0 & 0 & 1 & 1 & 1 & 1 & 0 & 0 \end{array}\right)$ model.
\label{fig_416_toric}}
\end{figure}

The toric diagram for the $Q^{1,1,1}/ \mathbb{Z}_2 \times \mathbb{Z}_2 \ \left(\begin{array}{@{}c@{,\;}c@{,\;}c@{,\;}c@{,\;}c@{,\;}c@{,\;}c@{,\;}c@{}} 1 & 1 & 0 & 0 & 0 & 0 & 1 & 1 \\ 0 & 0 & 1 & 1 & 1 & 1 & 0 & 0 \end{array}\right)$ model shown in \fref{fig_416_toric} is 
given by, 
\beal{es04f06b}
&&
G_{t}=
\resizebox{0.5\textwidth}{!}{$
\left(
\begin{array}{cccccc|cc|cc|cc|cc|ccc|ccc}
p_{1} & p_{2} & p_{3} & p_{4} & p_{5} & p_{6} &
q^{(1)}_{1} & q^{(1)}_2 &
q^{(2)}_{1} & q^{(2)}_2 &
q^{(3)}_{1} & q^{(3)}_2 &
q^{(4)}_1 & q^{(4)}_2 &
s_{1} & \cdots & s_{46} &
o^{(1)}_{1} & \cdots & o^{(12)}_{28}
\\
\hline
-1 & 0 & 0 & -1 & 1 & 1 &
-1 & -1 &
1 & 1 &
0 & 0 &
0 & 0 &
0 & \cdots & 0 &
-1 & \cdots & 1
\\
1 & 0 & 0 & -1 & 1 & -1 &
0 & 0 &
0 & 0 &
-1 & -1 &
1 & 1 &
0 & \cdots & 0 &
1 & \cdots & -1
\\
0 & -1 & 1 & 0 & 0 & 0 &
0 & 0 &
0 & 0 &
0 & 0 &
0 & 0 &
0 & \cdots & 0 &
0 & \cdots & 0
\\
\hline
1 & 1 & 1 & 1 & 1 & 1 &
1 & 1 &
1 & 1 &
1 & 1 &
1 & 1 &
1 & \cdots & 1 &
2 & \cdots & 3
\end{array}
\right)
$}
~.~
\eea

%-------------------
%-------------------

The global symmetry of the 
$Q^{1,1,1}/ \mathbb{Z}_2 \times \mathbb{Z}_2 \ \left(\begin{array}{@{}c@{,\;}c@{,\;}c@{,\;}c@{,\;}c@{,\;}c@{,\;}c@{,\;}c@{}} 1 & 1 & 0 & 0 & 0 & 0 & 1 & 1 \\ 0 & 0 & 1 & 1 & 1 & 1 & 0 & 0 \end{array}\right)$ model
takes the following enhanced form,
\beal{es04f07}
SU(2)_x \times U(1)_{f_1} \times U(1)_{f_2} \times U(1)_R
~.~
\eea
The charges under the mesonic flavor symmetry on the extremal GLSM fields $p_a$
as well as the corresponding fugacity assignment are summarized in \tref{tab_04f01}.

%-------------------
\begin{table}[htt!!]
\centering
\begin{tabular}{|c|c|c|c|l|}
\hline
\; & $SU(2)_x$ & $U(1)_{f_1}$ & $U(1)_{f_2}$  & fugacity \\
\hline
$p_1$ & $0$ & $+1$ & $0$  & $t_1=f_1 t$ \\
$p_2$ & $-1$ & $0$ & $0$  & $t_2=x^{-1} t$\\
$p_3$ & $+1$ & $0$ & $0$ & $t_3=x t$\\
$p_4$ & $0$ & $0$ & $+1$  & $t_4=f_2 t$ \\
$p_5$ & $0$ & $0$ & $-1$  & $t_5=f_2^{-1} t$ \\
$p_6$ & $0$ & $-1$ & $0$  & $t_6=f_1^{-1} t$ \\
\hline
\end{tabular}
\caption{
Mesonic flavor symmetry of the
 $Q^{1,1,1}/ \mathbb{Z}_2 \times \mathbb{Z}_2 \ \left(\begin{array}{@{}c@{,\;}c@{,\;}c@{,\;}c@{,\;}c@{,\;}c@{,\;}c@{,\;}c@{}} 1 & 1 & 0 & 0 & 0 & 0 & 1 & 1 \\ 0 & 0 & 1 & 1 & 1 & 1 & 0 & 0 \end{array}\right)$ model and charges on the extremal GLSM fields $p_a$. 
 Here, the fugacity $t$ counts the degree in extremal GLSM fields $p_a$. 
 \label{tab_04f01}
 }
\end{table}
%-------------------

The Hilbert series of the mesonic moduli space of the brane brick model for $Q^{1,1,1}/ \mathbb{Z}_2 \times \mathbb{Z}_2 \ \left(\begin{array}{@{}c@{,\;}c@{,\;}c@{,\;}c@{,\;}c@{,\;}c@{,\;}c@{,\;}c@{}} 1 & 1 & 0 & 0 & 0 & 0 & 1 & 1 \\ 0 & 0 & 1 & 1 & 1 & 1 & 0 & 0 \end{array}\right)$
takes the following form,
\beal{es04f10}
&&
g(t_a ; \mathcal{M}^{mes})=
\frac{
P(t_a ; \mathcal{M}^{mes}) \left(1 + t_1 t_2 t_3 t_4 t_5 t_6\right)
}{
\left(1 - t_1^2 t_2^2 t_3^2\right) \left(1 - t_1^2 t_2^2 t_4^2\right) \left(1 - t_1^2 t_3^2 t_5^2\right) \left(1 - t_1^2 t_4^2 t_5^2\right)
}
\nn\\
&&
\hspace{1cm}
\times
\frac{1}{
\left(1 - t_2^2 t_3^2 t_6^2\right) \left(1 - t_2^2 t_4^2 t_6^2\right) \left(1 - t_3^2 t_5^2 t_6^2\right) \left(1 - t_4^2 t_5^2 t_6^2\right)
}
~,~
\eea
where $t_a$ is the fugacity associated to the extremal GLSM field $p_a$.
The numerator factor $P(t_a ; \mathcal{M}^{mes})$ in \eref{es04f10} is presented fully in appendix \sref{appCa}. 
When unrefined by setting $t_a=t$, the Hilbert series takes the following form,
\beal{es04f06}
g(t; \mathcal{M}^{mes})=\frac{1+11t^6+11t^{12}+t^{18}}{(1-t^6)^4}
~,~
\eea
where the palindromic numerator indicates that the mesonic moduli space is Calabi-Yau.

The Hilbert series of the mesonic moduli space can be refined under the mesonic flavor symmetry
as summarized in \tref{tab_04f01}.
The corresponding highest weight generating function is given by,
\beal{es04f11}
&&
h(\mu,f_1,f_2,t; \mathcal{M}^{mes})
=
\frac{
(1-\mu^4 t^{12})^2}{(1- \mu^2 t^6)(1-\mu^2 f_1^{-2} f_2^{-2} t^6)
}
\nn\\
&&
\hspace{1cm}
\times
\frac{1}{
(1-\mu^2 f_1^{-2} f_2^{2} t^6)(1-\mu^2 f_1^{2} f_2^{-2} t^6)(1-\mu^2 f_1^{2} f_2^{2} t^6)
}
~,~
\eea
where $\mu^m$ is fugacity for counting the $SU(2)_x$ character of the form $[m]_x$,
and $f_1, f_2$ are the fugacities associated to $U(1)_{f_1}$ and $U(1)_{f_2}$, respectively.

%-------------------
 \begin {table}[httt!!]
\centering
\begin {tabular} {|c|c|ccc|c|}
\hline
PL term & generator & $SU(2)_x$ & $U(1)_{f_1}$ & $U(1)_{f_2}$ 
\\
\hline 
\multirow{3}{*}{$+[2]_x t^6$}
& $ p_1 p_2^2 p_4 p_5 p_6 ~s~ q_{(1)} q_{(2)} q_{(3)} q_{(4)}    $ & $-2$ & $0$ & $0$  \\
& $ p_1 p_2 p_3 p_4 p_5 p_6 ~s~  q_{{(1)}} q_{{(2)}} q_{{(3)}} q_{{(4)}} $ & $0$ & $0$ & $0$  \\
& $ p_1 p_3^2 p_4 p_5 p_6  ~s~  q_{{(1)}} q_{{(2)}} q_{{(3)}} q_{{(4)}} $ & $2$ & $0$ & $0$  \\
\hline
\multirow{3}{*}{$+[2]_x f_1^2 f_2^2 t^6$}
& $p_1^2 p_2^2 p_4^2 ~s~  q_{{(1)}}^2 q_{{(3)}} q_{{(4)}} $ & $-2$ & $2$ & $2$  \\
& $ p_1^2 p_2 p_3 p_4^2 ~s~  q_{{(1)}}^2 q_{{(3)}} q_{{(4)}} $ & $0$ & $2$ & $2$  \\
& $p_1^2 p_3^2 p_4^2 ~s~  q_{{(1)}}^2 q_{{(3)}} q_{{(4)}} $ & $2$ & $2$ & $2$  \\
\hline
\multirow{3}{*}{$+[2]_x f_1^2 f_2^{-2} t^6$}
& $p_1^2 p_2^2 p_5^2 ~s~  q_{{(1)}} q_{{(2)}} q_{{(4)}}^2 $ & $-2$ & $2$ & $-2$  \\
& $p_1^2 p_2 p_3 p_5^2 ~s~  q_{{(1)}} q_{{(2)}} q_{{(4)}}^2 $ & $0$ & $2$ & $-2$  \\
& $ p_1^2 p_3^2 p_5^2 ~s~  q_{{(1)}} q_{{(2)}} q_{{(4)}}^2 $ & $2$ & $2$ & $-2$  \\
\hline
\multirow{3}{*}{$+[2]_x f_1^{-2} f_2^{2} t^6$}
& $p_2^2 p_4^2 p_6^2 ~s~  q_{{(1)}} q_{{(2)}} q_{{(3)}}^2 $ & $-2$ & $-2$ & $2$  \\
& $p_2 p_3 p_4^2 p_6^2 ~s~  q_{{(1)}} q_{{(2)}} q_{{(3)}}^2 $ & $0$ & $-2$ & $2$  \\
& $ p_3^2 p_4^2 p_6^2 ~s~  q_{{(1)}} q_{{(2)}} q_{{(3)}}^2 $ & $2$ & $-2$ & $2$  \\
\hline
\multirow{3}{*}{$+[2]_x f_1^{-2} f_2^{-2} t^6$}
& $p_2^2 p_5^2 p_6^2 ~s~  q_{{(2)}}^2 q_{{(3)}} q_{{(4)}} $ & $-2$ & $-2$ & $-2$  \\
& $p_2 p_3 p_5^2 p_6^2~s~  q_{{(2)}}^2 q_{{(3)}} q_{{(4)}} $ & $0$ & $-2$ & $-2$  \\
& $p_3^2 p_5^2 p_6^2 ~s~  q_{{(2)}}^2 q_{{(3)}} q_{{(4)}} $ & $2$ & $-2$ & $-2$  \\
\hline
\end{tabular}
\caption{Generators of the  $Q^{1,1,1}/ \mathbb{Z}_2 \times \mathbb{Z}_2 \ \left(\begin{array}{@{}c@{,\;}c@{,\;}c@{,\;}c@{,\;}c@{,\;}c@{,\;}c@{,\;}c@{}} 1 & 1 & 0 & 0 & 0 & 0 & 1 & 1 \\ 0 & 0 & 1 & 1 & 1 & 1 & 0 & 0 \end{array}\right)$ model in terms of GLSM fields and their corresponding mesonic flavor charges. Here, we denote $q_{(1)}=q_1^{(1)}q_2^{(1)}$,  $q_{(2)}=q_1^{(2)} q_2^{(2)}$, $q_{(3)}=q_1^{(3)}q_2^{(3)}$,  $q_{(4)}=q_1^{(4)} q_2^{(4)}$, $s=\prod_{i=1}^{46}s_i$, and set extra GLSM fields to 1. \label{tab_04f02}}
\end{table}
%-------------------

The plethystic logarithm of the mesonic flavor symmetry refined Hilbert series
takes the following form,
\beal{es04f12}
&&
\text{PL}[g(x,f_1,f_2,t; \mathcal{M}^{mes})]=
( [2]_x + [2]_x f_1^2 f_2^2 + [2]_x f_1^2 f_2^{-2} + [2]_x f_1^{-2} f_2^2 + [2]_x f_1^{-2} f_2^{-2} )t^6
\nn\\
&&
\hspace{1cm}
- ( [2]_x f_1^4 + [2]_x f_2^4 + [2]_x f_1^{-4} + [2]_x f_2^{-4}
+ [2]_x f_1^2 f_2^2 + [2]_x f_1^{-2} f_2^{-2} 
+ [2]_x f_1^2 f_2^{-2} 
\nn\\
&&
\hspace{1cm}
+ [2]_x f_1^{-2} f_2^2
+ 2[4]_x + 2 [2]_x + 3 
+  f_1^{-4} +  f_1^{4}
+  f_2^{-4} +  f_2^4 
+  f_1^{-4} f_2^{-4} +  f_1^{4} f_2^{-4} 
\nn\\
&&
\hspace{1cm}
+  f_1^{-2} f_2^{-2}
+  f_1^{2} f_2^{-2} +  f_1^{-2} f_2^{2} +  f_1^{2} f_2^{2} +  f_1^{-4} f_2^{4} +  f_1^{4} f_2^{4}) t^{12}
+\dots
~,~
\eea
where $[m]_x$ is the character of the irreducible representation of $SU(2)_x$ with highest weight $(m)_x$.
The infinite expansion of the plethystic logarithm indicates that the mesonic moduli space of the $Q^{1,1,1}/ \mathbb{Z}_2 \times \mathbb{Z}_2 \ \left(\begin{array}{@{}c@{,\;}c@{,\;}c@{,\;}c@{,\;}c@{,\;}c@{,\;}c@{,\;}c@{}} 1 & 1 & 0 & 0 & 0 & 0 & 1 & 1 \\ 0 & 0 & 1 & 1 & 1 & 1 & 0 & 0 \end{array}\right)$ model is not a complete intersection.
The generators of the mesonic moduli space with their mesonic flavor charges are summarized in \tref{tab_04f02}.
\\

%=================================================================
\subsection{Abelian Orbifolds of $D_3$ \label{sec04sub2}}

%=================================================================
\subsubsection{$D_3/ \mathbb{Z}_2 \ (0,0,0,1,1)$ \label{sec047}}

%-------------------
\begin{figure}[H]
    \centering
    \includegraphics[width=0.3\textwidth]{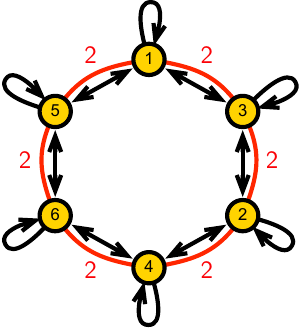}
    \caption{
    The quiver for the $D_3/ \mathbb{Z}_2 \ (0,0,0,1,1)$ model.
    \label{fig_421_quiver}
    }
\end{figure}
%-------------------

The $J$- and $E$-terms for the brane brick model corresponding to the abelian orbifold of the form
$D_3/ \mathbb{Z}_2 \ (0,0,0,1,1)$ are as follows, 
\beal{es04g01}
\begin{array}{rclccclcccc}
&& && &J& && &E& \\
&&\Lambda^{(1)}_{5 1} &:& & X_{1 3} \cdot X_{3 1} \cdot Y_{1 5} - Y_{1 5} \cdot X_{5 5}& && & D_{5 6} \cdot Z_{6 5} \cdot Z_{5 1} - Z_{5 1} \cdot D_{1 1}& \\ 
 &&\Lambda^{(1)}_{2 4} &:& & X_{4 6} \cdot X_{6 4} \cdot Y_{4 2} - Y_{4 2} \cdot X_{2 2}& && & D_{2 3} \cdot Z_{3 2} \cdot Z_{2 4} - Z_{2 4} \cdot D_{4 4}& \\ 
 &&\Lambda^{(2)}_{4 2} &:& & Z_{2 4} \cdot X_{4 6} \cdot X_{6 4} - X_{2 2} \cdot Z_{2 4}& && & D_{4 4} \cdot Y_{4 2} - Y_{4 2} \cdot D_{2 3} \cdot Z_{3 2}& \\ 
 &&\Lambda^{(2)}_{1 5} &:& & Z_{5 1} \cdot X_{1 3} \cdot X_{3 1} - X_{5 5} \cdot Z_{5 1}& && & D_{1 1} \cdot Y_{1 5} - Y_{1 5} \cdot D_{5 6} \cdot Z_{6 5}& \\ 
 &&\Lambda^{(3)}_{3 1} &:& & X_{1 3} \cdot Y_{3 3} - Y_{1 5} \cdot Z_{5 1} \cdot X_{1 3}& && & X_{3 1} \cdot D_{1 1} - Z_{3 2} \cdot D_{2 3} \cdot X_{3 1}& \\ 
 &&\Lambda^{(3)}_{6 4} &:& & X_{4 6} \cdot Y_{6 6} - Y_{4 2} \cdot Z_{2 4} \cdot X_{4 6}& && & X_{6 4} \cdot D_{4 4} - Z_{6 5} \cdot D_{5 6} \cdot X_{6 4}& \\ 
 &&\Lambda^{(4)}_{1 3} &:& & X_{3 1} \cdot Y_{1 5} \cdot Z_{5 1} - Y_{3 3} \cdot X_{3 1}& && & D_{1 1} \cdot X_{1 3} - X_{1 3} \cdot Z_{3 2} \cdot D_{2 3}& \\ 
 &&\Lambda^{(4)}_{4 6} &:& & X_{6 4} \cdot Y_{4 2} \cdot Z_{2 4} - Y_{6 6} \cdot X_{6 4}& && & D_{4 4} \cdot X_{4 6} - X_{4 6} \cdot Z_{6 5} \cdot D_{5 6}& \\ 
 &&\Lambda^{(5)}_{2 3} &:& & Y_{3 3} \cdot Z_{3 2} - Z_{3 2} \cdot Z_{2 4} \cdot Y_{4 2}& && & D_{2 3} \cdot X_{3 1} \cdot X_{1 3} - X_{2 2} \cdot D_{2 3}& \\ 
 &&\Lambda^{(5)}_{5 6} &:& & Y_{6 6} \cdot Z_{6 5} - Z_{6 5} \cdot Z_{5 1} \cdot Y_{1 5}& && & D_{5 6} \cdot X_{6 4} \cdot X_{4 6} - X_{5 5} \cdot D_{5 6}& \\ 
 &&\Lambda^{(6)}_{2 3} &:& & Z_{3 2} \cdot X_{2 2} - X_{3 1} \cdot X_{1 3} \cdot Z_{3 2}& && & D_{2 3} \cdot Y_{3 3} - Z_{2 4} \cdot Y_{4 2} \cdot D_{2 3}& \\ 
 &&\Lambda^{(6)}_{5 6} &:& & Z_{6 5} \cdot X_{5 5} - X_{6 4} \cdot X_{4 6} \cdot Z_{6 5}& && & D_{5 6} \cdot Y_{6 6} - Z_{5 1} \cdot Y_{1 5} \cdot D_{5 6}&
\end{array}
~.~
\eea
The corresponding quiver diagram is shown in \fref{fig_421_quiver}.
The $J$- and $E$-terms
come from the general formula in \eqref{es03c07} 
with the following relabelling of indices, 
\beal{es04g02}
&
[1,0] \rightarrow 1~,~ 
[2,0] \rightarrow 2 ~,~
[3,0] \rightarrow 3 ~,~ 
&
\nn\\
&
[1,1] \rightarrow 4~,~
[2,1] \rightarrow 5 ~,~
[3,1] \rightarrow 6 ~.~
&
\eea

\begin{figure}[htt!!]
    \centering
    \includegraphics[width=0.4\textwidth]{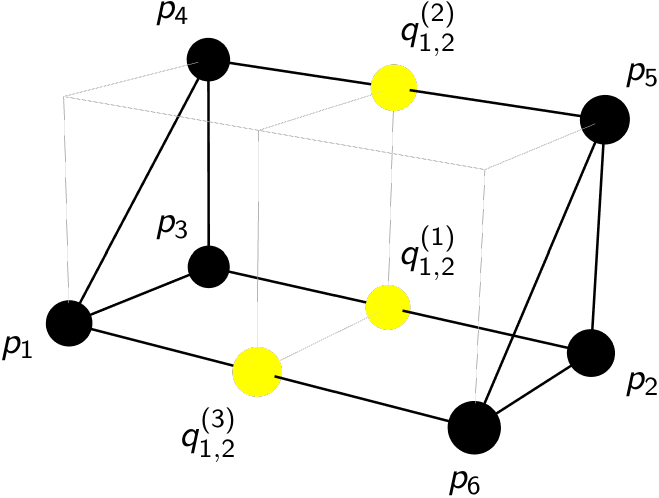}
    \caption{The toric diagram for the $D_3/\mathbb{Z}_2 \ (0,0,0,1,1)$ model.
    \label{fig_421_toric}}
\end{figure}

Using the forward algorithm for brane brick models, 
we obtain the
$P$-matrix, which takes the form,
\beal{es04g03}
P=
\left(
\begin{array}{c|cccccc|cc|cc|cc}
&p_{1} & p_{2} & p_{3} & p_{4} & p_{5} & p_{6} & q^{(1)}_{1} & q^{(1)}_{2} & q^{(2)}_{1} & q^{(2)}_{2} & q^{(3)}_{1} & q^{(3)}_{2} \\
\hline
 D_{11} & 1 & 0 & 0 & 0 & 0 & 1 & 0 & 0 & 0 & 0 & 1 & 1 \\
 D_{23} & 1 & 0 & 0 & 0 & 0 & 0 & 0 & 0 & 0 & 0 & 0 & 1 \\
 D_{44} & 1 & 0 & 0 & 0 & 0 & 1 & 0 & 0 & 0 & 0 & 1 & 1 \\
 D_{56} & 1 & 0 & 0 & 0 & 0 & 0 & 0 & 0 & 0 & 0 & 1 & 0 \\
 X_{13} & 0 & 1 & 0 & 0 & 0 & 0 & 0 & 1 & 0 & 0 & 0 & 0 \\
 X_{22} & 0 & 1 & 1 & 0 & 0 & 0 & 1 & 1 & 0 & 0 & 0 & 0 \\
 X_{31} & 0 & 0 & 1 & 0 & 0 & 0 & 1 & 0 & 0 & 0 & 0 & 0 \\
 X_{46} & 0 & 1 & 0 & 0 & 0 & 0 & 1 & 0 & 0 & 0 & 0 & 0 \\
 X_{55} & 0 & 1 & 1 & 0 & 0 & 0 & 1 & 1 & 0 & 0 & 0 & 0 \\
 X_{64} & 0 & 0 & 1 & 0 & 0 & 0 & 0 & 1 & 0 & 0 & 0 & 0 \\
 Y_{15} & 0 & 0 & 0 & 1 & 0 & 0 & 0 & 0 & 0 & 1 & 0 & 0 \\
 Y_{33} & 0 & 0 & 0 & 1 & 1 & 0 & 0 & 0 & 1 & 1 & 0 & 0 \\
 Y_{42} & 0 & 0 & 0 & 1 & 0 & 0 & 0 & 0 & 1 & 0 & 0 & 0 \\
 Y_{66} & 0 & 0 & 0 & 1 & 1 & 0 & 0 & 0 & 1 & 1 & 0 & 0 \\
 Z_{24} & 0 & 0 & 0 & 0 & 1 & 0 & 0 & 0 & 0 & 1 & 0 & 0 \\
 Z_{32} & 0 & 0 & 0 & 0 & 0 & 1 & 0 & 0 & 0 & 0 & 1 & 0 \\
 Z_{51} & 0 & 0 & 0 & 0 & 1 & 0 & 0 & 0 & 1 & 0 & 0 & 0 \\
 Z_{65} & 0 & 0 & 0 & 0 & 0 & 1 & 0 & 0 & 0 & 0 & 0 & 1 \\
\end{array}
\right)
~.~
\eea
The $U(1)$ charges on the GLSM fields
corresponding to the $J$- and $E$-terms
are given by the following charge matrix, 
\beal{es04g04}
Q_{JE}=
\left(
\begin{array}{cccccc|cc|cc|cc}
p_{1} & p_{2} & p_{3} & p_{4} & p_{5} & p_{6} & q^{(1)}_{1} & q^{(1)}_{2} & q^{(2)}_{1} & q^{(2)}_{2} & q^{(3)}_{1} & q^{(3)}_{2} \\
\hline
 -1 & 0 & 0 & 0 & 0 & -1 & 0 & 0 & 0 & 0 & 1 & 1 \\
 0 & 0 & 0 & -1 & -1 & 0 & 0 & 0 & 1 & 1 & 0 & 0 \\
 0 & 1 & 1 & 0 & 0 & 0 & -1 & -1 & 0 & 0 & 0 & 0 \\
\end{array}
\right)
~.~
\eea
The $D$-term charge matrix takes the following form, 
\beal{es04g05}
Q_{D}=
\left(
\begin{array}{cccccc|cc|cc|cc}
p_{1} & p_{2} & p_{3} & p_{4} & p_{5} & p_{6} & q^{(1)}_{1} & q^{(1)}_{2} & q^{(2)}_{1} & q^{(2)}_{2} & q^{(3)}_{1} & q^{(3)}_{2} \\
\hline
 0 & -1 & 0 & -1 & 0 & 0 & 1 & 0 & 1 & 0 & 0 & 0 \\
 -1 & 0 & 0 & 0 & -1 & 0 & 0 & 0 & 1 & 0 & 1 & 0 \\
 1 & 1 & 0 & 0 & 0 & 0 & -1 & 0 & 0 & 0 & -1 & 0 \\
 0 & -1 & 0 & 0 & 1 & 0 & 0 & 1 & -1 & 0 & 0 & 0 \\
 0 & 0 & 0 & 1 & 0 & 1 & 0 & 0 & -1 & 0 & -1 & 0 \\
\end{array}
\right)
~.~
\eea
The toric diagram of 
the abelian orbifold of the form $D_3/ \mathbb{Z}_2 \ (0,0,0,1,1)$
is shown in \fref{fig_421_toric} and is given by, 
\beal{es04g06}
G_{t}=
\left(
\begin{array}{cccccc|cc|cc|cc}
p_{1} & p_{2} & p_{3} & p_{4} & p_{5} & p_{6} & q^{(1)}_{1} & q^{(1)}_{2} & q^{(2)}_{1} & q^{(2)}_{2} & q^{(3)}_{1} & q^{(3)}_{2} \\
\hline
 1 & -1 & 1 & 1 & -1 & -1 & 0 & 0 & 0 & 0 & 0 & 0 \\
 1 & 0 & 0 & 0 & 0 & 1 & 0 & 0 & 0 & 0 & 1 & 1 \\
 0 & 0 & 0 & 1 & 1 & 0 & 0 & 0 & 1 & 1 & 0 & 0 \\
 \hline
 1 & 1 & 1 & 1 & 1 & 1 & 1 & 1 & 1 & 1 & 1 & 1 \\
\end{array}
\right)
~.~
\eea

%-------------------
%-------------------

%-------------------
\begin{table}[ht!!!]
\centering
\begin{tabular}{|c|c|c|c|l|}
\hline
\; & $U(1)_{f_1}$ & $U(1)_{f_2}$ & $U(1)_{f_3}$ & fugacity \\
\hline
$p_1$ & $+1$ & $0$ & $0$  & $t_1=f_1 t$ \\
$p_2$ & $-1$ & $0$ & $0$  & $t_2=f_1^{-1} t$\\
$p_3$ & $0$ & $+1$ & $0$  & $t_3=f_2 t$\\
$p_4$ & $0$ & $-1$ & $0$  & $t_4=f_2^{-1} t$ \\
$p_5$ & $0$ & $0$ & $+1$  & $t_5=f_3 t$ \\
$p_6$ & $0$ & $0$ & $-1$  & $t_6=f_3^{-1} t$ \\
\hline
\end{tabular}
\caption{
Mesonic flavor symmetry of the $D_3/\mathbb{Z}_2 \ (0,0,0,1,1)$ model
and charges on the extremal GLSM fields $p_a$. 
Here, the fugacity $t$ counts the degree in extremal GLSM fields $p_a$. 
\label{tab_04g01}}
\end{table}
%-------------------

The global symmetry of the brane brick model is not enhanced and accordingly has the following form,
\beal{es04g07}
U(1)_{f_1} \times U(1)_{f_2} \times U(1)_{f_3} \times U(1)_R
~.~
\eea
The Hilbert series of the mesonic moduli space for the $D_3/\mathbb{Z}_2 \ (0,0,0,1,1)$ model
takes the following form,
\beal{es04g08}
&&
g(t_a; \mathcal{M}^{mes})=
\frac{
1- t_1^2 t_2^2 t_3^2 t_4^2 t_5^2 t_6^2  
}{
\left(1 - t_2 t_3\right)  \left(1- t_4 t_5\right) \left(1 - t_1 t_6\right) 
\left(1 - t_1^2 t_3^2 t_4^2\right)
\left(1 - t_2^2 t_5^2 t_6^2\right)}
~,~
\eea
where $t_a$ is the fugacity associated to the extremal GLSM field $p_a$.
When unrefined by setting $t_a=t$, the Hilbert series takes the following form,
\beal{es04g09}
g(t; \mathcal{M}^{mes})=\frac{1+t^6}{(1-t^2)^3 (1-t^6)}
~,~
\eea
where the palindromic numerator indicates that the mesonic moduli space is Calabi-Yau. 

The Hilbert series refined using the mesonic flavor symmetry fugacities as summarized in \tref{tab_04g01}
takes the following form,
\beal{es04g10}
&&
g(f_1,f_2,f_3,t; \mathcal{M}^{mes})= 
\frac{
1-t^{12}
}{
(1-f_1f_3^{-1}t^2)(1-f_1^{-1}f_2 t^2)(1-f_2^{-1}f_3 t^2)
}
\nn\\
&&
\hspace{1cm}
\times
\frac{1}{
(1-f_1 ^{-2} t^6)(1-f_1^2 t^6)
}
~.~
\eea

%-------------------
 \begin {table}[htt!!]
\centering
\begin {tabular} {|c|c|ccc|}
\hline
PL term & generator & $U(1)_{f_1}$ & $U(1)_{f_2}$ & $U(1)_{f_3}$ 
\\
\hline
\multirow{1}{*}{$+f_1  f_3^{-1} t^2$}
& $A_1 = p_1 p_6 ~q_{{(3)}}$ & $1$ & $0$ & $-1$  \\
\hline
\multirow{1}{*}{$+f_1^{-1} f_2  t^2$}
& $A_2 = p_2 p_3 ~q_{(1)}$ & $-1$ & $1$ & $0$  \\
\hline
\multirow{1}{*}{$+f_2^{-1} f_3 t^2$}
& $A_3 = p_4 p_5 ~q_{(2)}$ & $0$ & $-1$ & $1$  \\
\hline
\multirow{1}{*}{$+f_1^{2}   t^{6}$}
& $B_1 = p_1^2 p_3^2 p_4^2 ~q_{(1)} q_{(2)} q_{{(3)}}$ & $2$ & $0$ & $0$  \\
\hline
\multirow{1}{*}{$+f_1^{-2}   t^{6}$}
& $B_2 = p_2^2 p_5^2 p_6^2 ~q_{(1)} q_{(2)} q_{{(3)}}$ & $-2$ & $0$ & $0$  \\
\hline
\end{tabular}
\caption{Generators of the $D_3/\mathbb{Z}_2 \ (0,0,0,1,1)$ model in terms of GLSM fields and their corresponding mesonic flavor charges. Here, we denote $q_{(1)}=q_1^{(1)}q_2^{(1)}$,  $q_{(2)}=q_1^{(2)} q_2^{(2)}$, $q_{(3)}=q_1^{(3)} q_2^{(3)}$. \label{tab_04g02}}
\end{table}
%-------------------

The plethystic logarithm of the refined Hilbert series in \eref{es04g10}
takes the following form,
\beal{es04g11}
\text{PL}[g(f_1,f_2,f_3,t; \mathcal{M}^{mes})]=(f_1^{-1}f_2 + f_2^{-1} f_3 + f_1 f_3^{-1})t^2 + (f_1^2 + f_1^{-2}) t^6 -t^{12}
~.~
\eea
We can see from the plethystic logarithm that the mesonic moduli space is a complete intersection.
The first positive terms of the expansion correspond to the generators of the mesonic moduli space,
which are summarized with their mesonic flavor charges in \tref{tab_04g02}.
These generators form a single unique defining relation for $D_3/\mathbb{Z}_2 \ (0,0,0,1,1)$, allowing us to write 
the mesonic moduli space as, 
\beal{es04g12}
\mathcal{M}^{mes} = 
\mathrm{Spec}~\mathbb{C}[A_1, A_2, A_3, B_1, B_2]/ \langle A_1^2 A_2^2 A_3^2 - B_1 B_2 \rangle~,~
\eea
where the generators $A_1,A_2,A_3,B_1,B_2$ are expressed in terms of GLSM fields in \tref{tab_04g02}.

We note here that the abelian orbifold $D_3/\mathbb{Z}_2 \ (0,0,0,1,1)$
is part of a family of abelian orbifolds of the form $D_3/\mathbb{Z}_n \ (0,0,0,1,-1)$, 
which are all complete intersections of the following form, 
\beal{es04g20}
\mathcal{M}^{mes} = 
\mathrm{Spec}~\mathbb{C}[A_1, A_2, A_3, B_1, B_2]/ \langle A_1^n A_2^n A_3^n - B_1 B_2 \rangle~.~
\eea
The corresponding toric diagram has extremal vertices with the following coordinates, 
\beal{es04g21}
&
(0,0,0)~,~
(1,0,0)~,~
(0,0,1)~,~
(0,n,0)~,~
(1,n,0)~,~
(0,n,1)~.~
&
\eea
In section \sref{sec0410}, 
we summarize the brane brick model for the abelian orbifold of the form $D_3/\mathbb{Z}_3 \ (0,0,0,1,2)$, 
which is also part of this family of abelian orbifolds of $D_3$.
\\

%=================================================================
\subsubsection{$D_3/ \mathbb{Z}_2 \ (0,1,1,0,0)$ \label{sec049}}

%-------------------
\begin{figure}[H]
    \centering
    \includegraphics[width=0.3\textwidth]{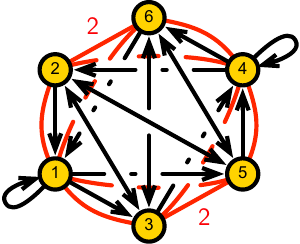}
    \caption{
    The quiver for the $D_3/ \mathbb{Z}_2 \ (0,1,1,0,0)$ model.
    \label{fig_423_quiver}
    }
\end{figure}
%-------------------

The quiver for the $D_3/ \mathbb{Z}_2 \ (0,1,1,0,0)$ model is shown in \fref{fig_423_quiver}.
The corresponding $J$- and $E$-terms are given below, 
\beal{es04i01}
\begin{array}{rclccclcccc}
&& && &J& && &E& \\
&&\Lambda^{(1)}_{2 1} &:& & X_{1 3} \cdot X_{3 4} \cdot Y_{4 2} - Y_{1 5} \cdot X_{5 2}& && & D_{2 3} \cdot Z_{3 2} \cdot Z_{2 1} - Z_{2 1} \cdot D_{1 1}& \\ 
 &&\Lambda^{(1)}_{5 4} &:& & X_{4 6} \cdot X_{6 1} \cdot Y_{1 5} - Y_{4 2} \cdot X_{2 5}& && & D_{5 6} \cdot Z_{6 5} \cdot Z_{5 4} - Z_{5 4} \cdot D_{4 4}& \\ 
 &&\Lambda^{(2)}_{4 2} &:& & Z_{2 1} \cdot X_{1 3} \cdot X_{3 4} - X_{2 5} \cdot Z_{5 4}& && & D_{4 4} \cdot Y_{4 2} - Y_{4 2} \cdot D_{2 3} \cdot Z_{3 2}& \\ 
 &&\Lambda^{(2)}_{1 5} &:& & Z_{5 4} \cdot X_{4 6} \cdot X_{6 1} - X_{5 2} \cdot Z_{2 1}& && & D_{1 1} \cdot Y_{1 5} - Y_{1 5} \cdot D_{5 6} \cdot Z_{6 5}& \\ 
 &&\Lambda^{(3)}_{6 1} &:& & X_{1 3} \cdot Y_{3 6} - Y_{1 5} \cdot Z_{5 4} \cdot X_{4 6}& && & X_{6 1} \cdot D_{1 1} - Z_{6 5} \cdot D_{5 6} \cdot X_{6 1}& \\ 
 &&\Lambda^{(3)}_{3 4} &:& & X_{4 6} \cdot Y_{6 3} - Y_{4 2} \cdot Z_{2 1} \cdot X_{1 3}& && & X_{3 4} \cdot D_{4 4} - Z_{3 2} \cdot D_{2 3} \cdot X_{3 4}& \\ 
 &&\Lambda^{(4)}_{1 3} &:& & X_{3 4} \cdot Y_{4 2} \cdot Z_{2 1} - Y_{3 6} \cdot X_{6 1}& && & D_{1 1} \cdot X_{1 3} - X_{1 3} \cdot Z_{3 2} \cdot D_{2 3}& \\ 
 &&\Lambda^{(4)}_{4 6} &:& & X_{6 1} \cdot Y_{1 5} \cdot Z_{5 4} - Y_{6 3} \cdot X_{3 4}& && & D_{4 4} \cdot X_{4 6} - X_{4 6} \cdot Z_{6 5} \cdot D_{5 6}& \\ 
 &&\Lambda^{(5)}_{5 3} &:& & Y_{3 6} \cdot Z_{6 5} - Z_{3 2} \cdot Z_{2 1} \cdot Y_{1 5}& && & D_{5 6} \cdot X_{6 1} \cdot X_{1 3} - X_{5 2} \cdot D_{2 3}& \\ 
 &&\Lambda^{(5)}_{2 6} &:& & Y_{6 3} \cdot Z_{3 2} - Z_{6 5} \cdot Z_{5 4} \cdot Y_{4 2}& && & D_{2 3} \cdot X_{3 4} \cdot X_{4 6} - X_{2 5} \cdot D_{5 6}& \\ 
 &&\Lambda^{(6)}_{5 3} &:& & Z_{3 2} \cdot X_{2 5} - X_{3 4} \cdot X_{4 6} \cdot Z_{6 5}& && & D_{5 6} \cdot Y_{6 3} - Z_{5 4} \cdot Y_{4 2} \cdot D_{2 3}& \\ 
 &&\Lambda^{(6)}_{2 6} &:& & Z_{6 5} \cdot X_{5 2} - X_{6 1} \cdot X_{1 3} \cdot Z_{3 2}& && & D_{2 3} \cdot Y_{3 6} - Z_{2 1} \cdot Y_{1 5} \cdot D_{5 6}&
\end{array}
~,~
\eea
which are obtained from the general formula in \eqref{es03c07}
with the additional relabelling of indices as follows, 
\beal{es04i02}
&
[1,0] \rightarrow 1~,~ [2,0] \rightarrow 2~,~ [3,0] \rightarrow 3~,~
&
\nn\\
&
[1,1] \rightarrow 4~,~[2,1] \rightarrow 5~,~ [3,1] \rightarrow 6~.~
&
\eea

The $P$-matrix is obtained using the forward algorithm.
It takes the following form,
\beal{es04i03}
P=
\left(
\begin{array}{c|cccccc|cc|cc|cccc}
&p_{1} & p_{2} & p_{3} & p_{4} & p_{5} & p_{6} & q^{(1)}_{1} & q^{(1)}_{2} & q^{(2)}_{1} & q^{(2)}_{2} & o_{1} & o_{2} & o_{3} & o_{4} \\
\hline
 D_{11} & 0 & 0 & 0 & 1 & 1 & 0 & 0 & 0 & 0 & 0 & 1 & 1 & 1 & 1 \\
 D_{23} & 0 & 0 & 0 & 0 & 1 & 0 & 0 & 0 & 0 & 0 & 0 & 0 & 1 & 1 \\
 D_{44} & 0 & 0 & 0 & 1 & 1 & 0 & 0 & 0 & 0 & 0 & 1 & 1 & 1 & 1 \\
 D_{56} & 0 & 0 & 0 & 0 & 1 & 0 & 0 & 0 & 0 & 0 & 1 & 1 & 0 & 0 \\
 X_{13} & 0 & 0 & 0 & 0 & 0 & 1 & 0 & 1 & 0 & 0 & 0 & 0 & 0 & 1 \\
 X_{25} & 1 & 0 & 0 & 0 & 0 & 1 & 1 & 0 & 0 & 1 & 0 & 0 & 1 & 1 \\
 X_{34} & 1 & 0 & 0 & 0 & 0 & 0 & 0 & 0 & 0 & 1 & 0 & 1 & 0 & 0 \\
 X_{46} & 0 & 0 & 0 & 0 & 0 & 1 & 1 & 0 & 0 & 0 & 1 & 0 & 0 & 0 \\
 X_{52} & 1 & 0 & 0 & 0 & 0 & 1 & 0 & 1 & 1 & 0 & 1 & 1 & 0 & 0 \\
 X_{61} & 1 & 0 & 0 & 0 & 0 & 0 & 0 & 0 & 1 & 0 & 0 & 0 & 1 & 0 \\
 Y_{15} & 0 & 0 & 1 & 0 & 0 & 0 & 0 & 0 & 0 & 1 & 0 & 0 & 0 & 1 \\
 Y_{36} & 0 & 1 & 1 & 0 & 0 & 0 & 1 & 0 & 0 & 1 & 1 & 1 & 0 & 0 \\
 Y_{42} & 0 & 0 & 1 & 0 & 0 & 0 & 0 & 0 & 1 & 0 & 1 & 0 & 0 & 0 \\
 Y_{63} & 0 & 1 & 1 & 0 & 0 & 0 & 0 & 1 & 1 & 0 & 0 & 0 & 1 & 1 \\
 Z_{21} & 0 & 1 & 0 & 0 & 0 & 0 & 1 & 0 & 0 & 0 & 0 & 0 & 1 & 0 \\
 Z_{32} & 0 & 0 & 0 & 1 & 0 & 0 & 0 & 0 & 0 & 0 & 1 & 1 & 0 & 0 \\
 Z_{54} & 0 & 1 & 0 & 0 & 0 & 0 & 0 & 1 & 0 & 0 & 0 & 1 & 0 & 0 \\
 Z_{65} & 0 & 0 & 0 & 1 & 0 & 0 & 0 & 0 & 0 & 0 & 0 & 0 & 1 & 1 \\
\end{array}
\right)
~,~
\eea
where $o_1, \dots, o_4$
are extra GLSM fields \cite{Witten:1993yc}.
The $J$- and $E$-term charge matrix is given by, 
\beal{es04i04}
Q_{JE}=
\left(
\begin{array}{cccccc|cc|cc|cccc}
p_{1} & p_{2} & p_{3} & p_{4} & p_{5} & p_{6} & q^{(1)}_{1} & q^{(1)}_{2} & q^{(2)}_{1} & q^{(2)}_{2} & o_{1} & o_{2} & o_{3} & o_{4} \\
\hline
 0 & 0 & 0 & 1 & 1 & 0 & 1 & 0 & 1 & 0 & -1 & 0 & -1 & 0 \\
 0 & 1 & -1 & 0 & 0 & 0 & 0 & -1 & 1 & 0 & 0 & 0 & -1 & 1 \\
 0 & -1 & 1 & 0 & 0 & 0 & 1 & 0 & 0 & -1 & -1 & 1 & 0 & 0 \\
 0 & 1 & 0 & 0 & 0 & 1 & -1 & -1 & 0 & 0 & 0 & 0 & 0 & 0 \\
 1 & 0 & 1 & 0 & 0 & 0 & 0 & 0 & -1 & -1 & 0 & 0 & 0 & 0 \\
\end{array}
\right)
~,~
\eea
and the corresponding $D$-term charge matrix is given by, 
\beal{es04i05}
Q_{D}=
\left(
\begin{array}{cccccc|cc|cc|cccc}
p_{1} & p_{2} & p_{3} & p_{4} & p_{5} & p_{6} & q^{(1)}_{1} & q^{(1)}_{2} & q^{(2)}_{1} & q^{(2)}_{2} & o_{1} & o_{2} & o_{3} & o_{4} \\
\hline
 0 & 1 & -1 & 0 & 0 & 0 & 0 & -1 & 1 & 0 & 0 & 0 & 0 & 0 \\
 0 & 0 & 0 & 1 & 0 & 0 & 0 & 0 & 1 & 0 & 0 & 0 & -1 & 0 \\
 0 & -1 & 1 & -1 & 0 & 0 & 0 & 1 & -1 & -1 & 0 & 0 & 1 & 0 \\
 0 & 1 & -1 & 0 & 0 & 0 & -1 & 0 & 0 & 1 & 0 & 0 & 0 & 0 \\
 0 & -1 & 1 & 1 & 0 & 0 & 1 & 0 & 0 & 0 & -1 & 0 & 0 & 0 \\
\end{array}
\right)~.~
\eea

\begin{figure}[htt!!]
\centering
\includegraphics[width=0.3\textwidth]{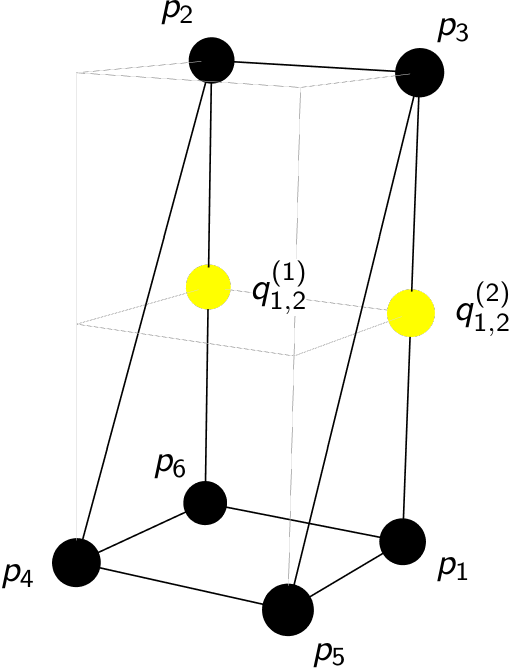}
\caption{
The toric diagram for the $D_3/\mathbb{Z}_2 \ (0,1,1,0,0)$ model.
\label{fig_423_toric}
}
\end{figure}

The toric diagram for $D_3/\mathbb{Z}_2 \ (0,1,1,0,0)$ model
is given by,
\beal{es04i06}
G_{t}=
\left(
\begin{array}{cccccc|cc|cc|cccc}
p_{1} & p_{2} & p_{3} & p_{4} & p_{5} & p_{6} & q^{(1)}_{1} & q^{(1)}_{2} & q^{(2)}_{1} & q^{(2)}_{2} & o_{1} & o_{2} & o_{3} & o_{4} \\
\hline
 1 & 1 & 1 & 0 & 0 & 1 & 1 & 1 & 1 & 1 & 1 & 1 & 1 & 1 \\
 1 & 0 & 1 & 0 & 1 & 0 & 0 & 0 & 1 & 1 & 1 & 1 & 1 & 1 \\
 0 & 2 & 2 & 0 & 0 & 0 & 1 & 1 & 1 & 1 & 1 & 1 & 1 & 1 \\
 \hline
 1 & 1 & 1 & 1 & 1 & 1 & 1 & 1 & 1 & 1 & 2 & 2 & 2 & 2 \\
\end{array}
\right)
~,~
\eea
and is illustrated in \fref{fig_423_toric}.

%-------------------
%-------------------

%-------------------
\begin{table}[htt!!!]
\centering
\begin{tabular}{|c|c|c|c|l|}
\hline
\; & $U(1)_{f_1}$ & $U(1)_{f_2}$ & $U(1)_{f_3}$ &  fugacity \\
\hline
$p_1$ & $+1$ & $0$ & $0$ &  $t_1=f_1 t$ \\
$p_2$ & $-1$ & $0$ & $0$ &  $t_2=f_1^{-1} t$\\
$p_3$ & $0$ & $+1$ & $0$ &  $t_3=f_2 t$\\
$p_4$ & $0$ & $-1$ & $0$ & $t_4=f_2^{-1} t$ \\
$p_5$ & $0$ & $0$ & $+1$ & $t_5=f_3 t$ \\
$p_6$ & $0$ & $0$ & $-1$ & $t_6=f_3^{-1} t$ \\
\hline
\end{tabular}
\caption{
Mesonic flavor symmetry of the $D_3/\mathbb{Z}_2 \ (0,1,1,0,0)$ model
and charges on the extremal GLSM fields $p_a$. 
Here, the fugacity $t$ counts the degree in extremal GLSM fields $p_a$. 
\label{tab_04i01}}
\end{table}
%-------------------

The global symmetry of 
the $D_3/\mathbb{Z}_2 \ (0,1,1,0,0)$ model
takes the following form,
\beal{es04i07}
U(1)_{f_1} \times U(1)_{f_2} \times U(1)_{f_3} \times U(1)_R
~.~
\eea
The Hilbert series of the mesonic moduli space
for the $D_3/\mathbb{Z}_2 \ (0,1,1,0,0)$ model
is given by, 
\beal{es04i08}
&&
g(t_a; \mathcal{M}^{mes})
=
\frac{
\left(1 + t_1 t_2 t_3 t_6\right) \left(1 - t_1 t_2 t_3 t_4 t_5 t_6\right)
}{
\left(1 - t_2^2 t_3^2\right)  \left(1 - t_1 t_3 t_5\right) \left(1 - t_4 t_5\right)  
\left(1 - t_1^2 t_6^2\right)  \left(1 - t_2 t_4 t_6\right)
}
~,~
\eea
where $t_a$ is the fugacity corresponding to the extremal GLSM field $p_a$.
When unrefined by setting $t_a=t$, the Hilbert series takes the following form,
\beal{es04i09}
g(t; \mathcal{M}^{mes})
= 
\frac{1-t +t^2 +t^4 -t^5 +t^6}{(1-t) (1-t^3) (1-t^4)^2}
~,~
\eea
where the palindromic numerator indicates that the mesonic moduli space is Calabi-Yau.

The Hilbert series refined in terms of the mesonic flavor symmetry fugacities summarized in \tref{tab_04i01}
takes the following form,
\beal{es04i10}
&&
g(f_1,f_2,f_3,t; \mathcal{M}^{mes})= 
\frac{
(1-t^6) (1-f_2^2 f_3^{-2} t^8)
}{
(1-f_1^2 f_3^{-2} t^4) (1-f_1^{-2} f_2^2 t^4) (1-f_2 f_3^{-1} t^4) 
}
\nn\\
&&
\hspace{1cm}
\times
\frac{1}{
(1-f_2^{-1} f_3 t^2) (1-f_1^{-1} f_2^{-1} f_3^{-1} t^3) (1-f_1f_2f_3 t^3)
}
~.~
\eea

%-------------------
 \begin {table}[htt!!!]
\centering
\begin {tabular} {|c|c|ccc|}
\hline
PL term & generator & $U(1)_{f_1}$ & $U(1)_{f_2}$ & $U(1)_{f_3}$ 
\\
\hline
\multirow{1}{*}{$+ f_2^{-1} f_3 t^{2}$}
& $A=  p_4 p_5~$ & $0$ & $-1$ & $1$  \\
\hline
\multirow{1}{*}{$+f_1 f_2 f_3 t^{3}$}
& $B_1 = p_1 p_3 p_5 ~q_{(2)} $ & $1$ & $1$ & $1$  \\
\hline
\multirow{1}{*}{$+f_1^{-1} f_2^{-1} f_3^{-1} t^{3}$}
& $B_2 = p_2 p_4 p_6 ~q_{(1)} $ & $-1$ & $-1$ & $-1$  \\
\hline
\multirow{1}{*}{$+f_1^{2}  f_3^{-2} t^{4}$}
& $C_1 = p_1^2 p_6^2 ~q_{(1)} q_{(2)}$ & $2$ & $0$ & $-2$  \\
\hline
\multirow{1}{*}{$+f_1^{-2} f_2^{2}  t^{4}$}
& $C_2 = p_2^2 p_3^2 ~q_{(1)} q_{(2)} $ & $-2$ & $2$ & $0$  \\
\hline
\multirow{1}{*}{$+ f_2 f_3^{-1} t^{4}$}
& $C_3 = p_1 p_2 p_3 p_6 ~q_{(1)} q_{(2)}$ & $0$ & $1$ & $-1$  \\
\hline
\end{tabular}
\caption{
Generators of the $D_3/\mathbb{Z}_2 \ (0,1,1,0,0)$ model in terms of GLSM fields and their corresponding mesonic flavor symmetry charges. Here, we denote $q_{(1)}=q_1^{(1)}q_2^{(1)}$,  $q_{(2)}=q_1^{(2)} q_2^{(2)}$, 
and set the extra GLSM fields to 1. 
\label{tab_04i02}}
\end{table}
%-------------------

The plethystic logarithm of the refined Hilbert series in \eref{es04i10}
is given by,
\beal{es04i12}
&&
\text{PL}[g(f_1,f_2,f_3,t)]=
f_2^{-1} f_3 t^2 +  (f_1^{-1} f_2^{-1} f_3^{-1}+f_1 f_2 f_3) t^{3}
\nn\\
&&
\hspace{1cm}
+ (f_1^{-2} f_2^{2}+f_1^{2} f_3^{-2}+f_2 f_3^{-1}) t^4
- t^6
- f_2^2 f_3^{-2} t^8
~,~
\eea
where the finite expansion indicates that the mesonic moduli space 
is a complete intersection.
The first positive terms correspond to the generators of the mesonic moduli space,
which are summarized with their mesonic flavor charges in \tref{tab_04i02}.
The generators form 2 defining relations for $D_3/\mathbb{Z}_2 \ (0,1,1,0,0)$.
In terms of these relations, we are able to summarize the mesonic moduli space as follows, 
\beal{es04i15}
\mathcal{M}^{mes}
=
\mathrm{Spec}~\mathbb{C}[A, B_1, B_2, C_1, C_2, C_3]/ \langle A C_3 - B_1 B_2 ~,~ C_1 C_2 - C_3^2 \rangle~,~
\eea
where the generators $A, B_1, B_2, C_1, C_2, C_3$
are expressed in terms of GLSM fields in \tref{tab_04i02}.

We note here that the abelian orbifold $D_3/\mathbb{Z}_2 \ (0,1,1,0,0)$
is part of a larger family of abelian orbifolds of the form $D_3/\mathbb{Z}_n \ (0,1,-1,0,0)$, 
with the mesonic moduli space taking the general form, 
\beal{es04i20}
\mathcal{M}^{mes}
=
\mathrm{Spec}~\mathbb{C}[A, B_1, B_2, C_1, C_2, C_3]/ \langle A C_3 - B_1 B_2 ~,~ C_1 C_2 - C_3^n \rangle~.~
\eea
The corresponding toric diagrams have extremal vertices whose coordinates take the following form, 
\beal{es04i21}
&
(0,0,0)~,~
(1,0,0)~,~
(0,1,0)~,~
(1,1,0)~,~
(0,1,n)~,~
(1,1,n)~.~
&
\eea
In section \sref{sec0411},
we summarize the brane brick model for $D_3/\mathbb{Z}_3 \ (0,1,2,0,0)$, 
which is also part of this family of abelian orbifolds of $D_3$.
\\

%=================================================================
\subsubsection{$D_3/ \mathbb{Z}_2 \ (0,1,1,1,1)$ \label{sec048}}

%-------------------
\begin{figure}[H]
    \centering
   \includegraphics[width=0.3\textwidth]{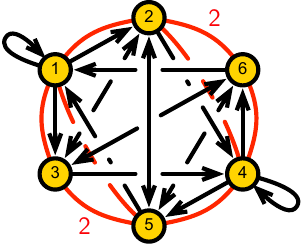}
    \caption{
    The quiver for the $D_3/ \mathbb{Z}_2 \ (0,1,1,1,1)$ model.
    \label{fig_422_quiver}
    }
\end{figure}
%-------------------

The $J$- and $E$-terms for the brane brick model corresponding to the abelian orbifold of the form
$D_3/ \mathbb{Z}_2 \ (0,1,1,1,1)$ are as follows, 
\beal{es04h01}
\begin{array}{rclccclcccc}
&& && &J& && &E& \\
&&\Lambda^{(1)}_{5 1} &:& & X_{1 3} \cdot X_{3 4} \cdot Y_{4 5} - Y_{1 2} \cdot X_{2 5}& && & D_{5 6} \cdot Z_{6 5} \cdot Z_{5 1} - Z_{5 1} \cdot D_{1 1}& \\ 
 &&\Lambda^{(1)}_{2 4} &:& & X_{4 6} \cdot X_{6 1} \cdot Y_{1 2} - Y_{4 5} \cdot X_{5 2}& && & D_{2 3} \cdot Z_{3 2} \cdot Z_{2 4} - Z_{2 4} \cdot D_{4 4}& \\ 
 &&\Lambda^{(2)}_{1 2} &:& & Z_{2 4} \cdot X_{4 6} \cdot X_{6 1} - X_{2 5} \cdot Z_{5 1}& && & D_{1 1} \cdot Y_{1 2} - Y_{1 2} \cdot D_{2 3} \cdot Z_{3 2}& \\ 
 &&\Lambda^{(2)}_{4 5} &:& & Z_{5 1} \cdot X_{1 3} \cdot X_{3 4} - X_{5 2} \cdot Z_{2 4}& && & D_{4 4} \cdot Y_{4 5} - Y_{4 5} \cdot D_{5 6} \cdot Z_{6 5}& \\ 
 &&\Lambda^{(3)}_{6 1} &:& & X_{1 3} \cdot Y_{3 6} - Y_{1 2} \cdot Z_{2 4} \cdot X_{4 6}& && & X_{6 1} \cdot D_{1 1} - Z_{6 5} \cdot D_{5 6} \cdot X_{6 1}& \\ 
 &&\Lambda^{(3)}_{3 4} &:& & X_{4 6} \cdot Y_{6 3} - Y_{4 5} \cdot Z_{5 1} \cdot X_{1 3}& && & X_{3 4} \cdot D_{4 4} - Z_{3 2} \cdot D_{2 3} \cdot X_{3 4}& \\ 
 &&\Lambda^{(4)}_{1 3} &:& & X_{3 4} \cdot Y_{4 5} \cdot Z_{5 1} - Y_{3 6} \cdot X_{6 1}& && & D_{1 1} \cdot X_{1 3} - X_{1 3} \cdot Z_{3 2} \cdot D_{2 3}& \\ 
 &&\Lambda^{(4)}_{4 6} &:& & X_{6 1} \cdot Y_{1 2} \cdot Z_{2 4} - Y_{6 3} \cdot X_{3 4}& && & D_{4 4} \cdot X_{4 6} - X_{4 6} \cdot Z_{6 5} \cdot D_{5 6}& \\ 
 &&\Lambda^{(5)}_{5 3} &:& & Y_{3 6} \cdot Z_{6 5} - Z_{3 2} \cdot Z_{2 4} \cdot Y_{4 5}& && & D_{5 6} \cdot X_{6 1} \cdot X_{1 3} - X_{5 2} \cdot D_{2 3}& \\ 
 &&\Lambda^{(5)}_{2 6} &:& & Y_{6 3} \cdot Z_{3 2} - Z_{6 5} \cdot Z_{5 1} \cdot Y_{1 2}& && & D_{2 3} \cdot X_{3 4} \cdot X_{4 6} - X_{2 5} \cdot D_{5 6}& \\ 
 &&\Lambda^{(6)}_{5 3} &:& & Z_{3 2} \cdot X_{2 5} - X_{3 4} \cdot X_{4 6} \cdot Z_{6 5}& && & D_{5 6} \cdot Y_{6 3} - Z_{5 1} \cdot Y_{1 2} \cdot D_{2 3}& \\ 
 &&\Lambda^{(6)}_{2 6} &:& & Z_{6 5} \cdot X_{5 2} - X_{6 1} \cdot X_{1 3} \cdot Z_{3 2}& && & D_{2 3} \cdot Y_{3 6} - Z_{2 4} \cdot Y_{4 5} \cdot D_{5 6}&
\end{array}
~.~
\eea
The corresponding quiver diagram is shown in \fref{fig_422_quiver}.
The $J$- and $E$-terms
are obtained from the general formula in \eqref{es03c07} 
with the following additional relabelling of indices, 
\beal{es04h02}
&
[1,0] \rightarrow 1 ~,~
[2,0] \rightarrow 2 ~,~ 
[3,0] \rightarrow 3 ~,~
&
\nn\\
&
[1,1] \rightarrow 4 ~,~
[2,1] \rightarrow 5 ~,~
[3,1] \rightarrow 6 ~.~
&
\eea

The $P$-matrix is obtained using the forward algorithm for brane brick models.
It takes the following form,
\beal{es04h03}
P=
\left(
\begin{array}{c|cccccc|cccc|cc|cc}
&p_{1} & p_{2} & p_{3} & p_{4} & p_{5} & p_{6} & q_{1} & q_{2} & q_{3} & q_{4} & o^{(1)}_{1} & o^{(1)}_{2} & o^{(2)}_{1} & o^{(2)}_{2} \\
\hline
 D_{11} & 0 & 1 & 1 & 0 & 0 & 0 & 0 & 0 & 0 & 0 & 1 & 1 & 1 & 1 \\
 D_{23} & 0 & 1 & 0 & 0 & 0 & 0 & 0 & 0 & 0 & 0 & 0 & 1 & 0 & 1 \\
 D_{44} & 0 & 1 & 1 & 0 & 0 & 0 & 0 & 0 & 0 & 0 & 1 & 1 & 1 & 1 \\
 D_{56} & 0 & 1 & 0 & 0 & 0 & 0 & 0 & 0 & 0 & 0 & 1 & 0 & 1 & 0 \\
 X_{13} & 0 & 0 & 0 & 0 & 1 & 0 & 0 & 0 & 0 & 1 & 0 & 0 & 0 & 1 \\
 X_{25} & 0 & 0 & 0 & 1 & 1 & 0 & 0 & 1 & 1 & 0 & 0 & 1 & 0 & 1 \\
 X_{34} & 0 & 0 & 0 & 1 & 0 & 0 & 0 & 0 & 1 & 0 & 1 & 0 & 0 & 0 \\
 X_{46} & 0 & 0 & 0 & 0 & 1 & 0 & 0 & 1 & 0 & 0 & 0 & 0 & 1 & 0 \\
 X_{52} & 0 & 0 & 0 & 1 & 1 & 0 & 1 & 0 & 0 & 1 & 1 & 0 & 1 & 0 \\
 X_{61} & 0 & 0 & 0 & 1 & 0 & 0 & 1 & 0 & 0 & 0 & 0 & 1 & 0 & 0 \\
 Y_{12} & 0 & 0 & 0 & 0 & 0 & 1 & 0 & 0 & 0 & 1 & 1 & 0 & 0 & 0 \\
 Y_{36} & 1 & 0 & 0 & 0 & 0 & 1 & 0 & 1 & 1 & 0 & 1 & 0 & 1 & 0 \\
 Y_{45} & 0 & 0 & 0 & 0 & 0 & 1 & 0 & 1 & 0 & 0 & 0 & 1 & 0 & 0 \\
 Y_{63} & 1 & 0 & 0 & 0 & 0 & 1 & 1 & 0 & 0 & 1 & 0 & 1 & 0 & 1 \\
 Z_{24} & 1 & 0 & 0 & 0 & 0 & 0 & 0 & 0 & 1 & 0 & 0 & 0 & 0 & 1 \\
 Z_{32} & 0 & 0 & 1 & 0 & 0 & 0 & 0 & 0 & 0 & 0 & 1 & 0 & 1 & 0 \\
 Z_{51} & 1 & 0 & 0 & 0 & 0 & 0 & 1 & 0 & 0 & 0 & 0 & 0 & 1 & 0 \\
 Z_{65} & 0 & 0 & 1 & 0 & 0 & 0 & 0 & 0 & 0 & 0 & 0 & 1 & 0 & 1 \\
\end{array}
\right)~,~
\eea
where $o_k^{(1)}$ and $o_l^{(2)}$ are extra GLSM fields \cite{Witten:1993yc}.
The $J$- and $E$-term charge matrix is as follows,
\beal{es04h04}
Q_{JE}=
\left(
\begin{array}{cccccc|cccc|cc|cc}
p_{1} & p_{2} & p_{3} & p_{4} & p_{5} & p_{6} & q_{1} & q_{2} & q_{3} & q_{4} & o^{(1)}_{1} & o^{(1)}_{2} & o^{(2)}_{1} & o^{(2)}_{2} \\
\hline
 0 & 1 & 1 & 0 & 0 & 0 & 1 & 1 & 0 & 0 & 0 & -1 & -1 & 0 \\
 -1 & 0 & 0 & 0 & 0 & 1 & 1 & 0 & 0 & -1 & 0 & -1 & 0 & 1 \\
 1 & 0 & 0 & 0 & 0 & -1 & 0 & 1 & -1 & 0 & 1 & 0 & -1 & 0 \\
 0 & 0 & 0 & 0 & 1 & 1 & 0 & -1 & 0 & -1 & 0 & 0 & 0 & 0 \\
 1 & 0 & 0 & 1 & 0 & 0 & -1 & 0 & -1 & 0 & 0 & 0 & 0 & 0 \\
\end{array}
\right)
~,~
\eea
and the $D$-term charge matrix takes the following form,
\beal{es04h05}
Q_{D}=
\left(
\begin{array}{cccccc|cccc|cc|cc}
p_{1} & p_{2} & p_{3} & p_{4} & p_{5} & p_{6} & q_{1} & q_{2} & q_{3} & q_{4} & o^{(1)}_{1} & o^{(1)}_{2} & o^{(2)}_{1} & o^{(2)}_{2} \\
\hline
 0 & 0 & 0 & 0 & 0 & 0 & 1 & 0 & 0 & -1 & 0 & 0 & 0 & 0 \\
 -1 & 0 & 1 & 0 & 0 & 1 & 1 & 0 & 0 & 0 & 0 & -1 & 0 & 0 \\
 1 & 0 & -1 & 0 & 0 & -1 & -1 & 0 & -1 & 1 & 0 & 1 & 0 & 0 \\
 0 & 0 & 0 & 0 & 0 & 0 & 0 & -1 & 1 & 0 & 0 & 0 & 0 & 0 \\
 0 & 0 & 1 & 0 & 0 & 0 & 0 & 1 & 0 & 0 & 0 & 0 & -1 & 0 \\
\end{array}
\right)~.~
\eea

\begin{figure}[htt!!]
\centering
\includegraphics[width=0.4\textwidth]{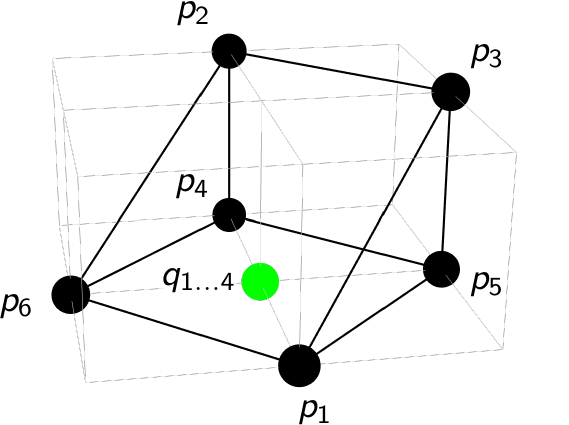}
\caption{
The toric diagram for the $D_3/\mathbb{Z}_2 \ (0,1,1,1,1)$ model.
\label{fig_422_toric}}
\end{figure}

The toric diagram is given by, 
\beal{es04h06}
G_{t}=
\left(
\begin{array}{cccccc|cccc|cc|cc}
p_{1} & p_{2} & p_{3} & p_{4} & p_{5} & p_{6} & q_{1} & q_{2} & q_{3} & q_{4} & o^{(1)}_{1} & o^{(1)}_{2} & o^{(2)}_{1} & o^{(2)}_{2} \\
\hline
 0 & 1 & 1 & 0 & 0 & 0 & 0 & 0 & 0 & 0 & 1 & 1 & 1 & 1 \\
 0 & 0 & 1 & 0 & 1 & -1 & 0 & 0 & 0 & 0 & 0 & 0 & 1 & 1 \\
 1 & -1 & 0 & -1 & 0 & 0 & 0 & 0 & 0 & 0 & -1 & -1 & 0 & 0 \\
 \hline
 1 & 1 & 1 & 1 & 1 & 1 & 1 & 1 & 1 & 1 & 2 & 2 & 2 & 2 \\
\end{array}
\right)
~.~
\eea
\fref{fig_422_toric} illustrates the toric diagram for the $D_3/\mathbb{Z}_2 \ (0,1,1,1,1)$ model.

%-------------------
%-------------------

%-------------------
\begin{table}[htt!!]
\centering
\begin{tabular}{|c|c|c|c|l|}
\hline
\; & $U(1)_{f_1}$ & $U(1)_{f_2}$ & $U(1)_{f_3}$ &  fugacity \\
\hline
$p_1$ & $+1$ & $0$ & $0$  & $t_1=f_1 t$ \\
$p_2$ & $-1$ & $0$ & $0$  & $t_2=f_1^{-1} t$\\
$p_3$ & $0$ & $+1$ & $0$  & $t_3=f_2 t$\\
$p_4$ & $0$ & $-1$ & $0$  & $t_4=f_2^{-1} t$ \\
$p_5$ & $0$ & $0$ & $+1$  & $t_5=f_3 t$ \\
$p_6$ & $0$ & $0$ & $-1$  & $t_6=f_3^{-1} t$ \\
\hline
\end{tabular}
\caption{
Mesonic flavor symmetry of the $D_3/\mathbb{Z}_2 \ (0,1,1,1,1)$ model 
and charges on the extremal GLSM fields $p_a$. 
Here, the fugacity $t$ counts the degree in extremal GLSM fields $p_a$. 
\label{tab_04h01}}
\end{table}
%-------------------

The global symmetry of the $D_3/\mathbb{Z}_2 \ (0,1,1,1,1)$ model
is not enhanced and takes the following form,
\beal{es04h07}
U(1)_{f_1} \times U(1)_{f_2} \times U(1)_{f_3} \times U(1)_R
~.~
\eea
The Hilbert series of the mesonic moduli space 
for the $D_3/\mathbb{Z}_2 \ (0,1,1,1,1)$ model takes the following form,
\beal{es04h08}
&&
g(t_a; \mathcal{M}^{mes})=
(1 - t_1 t_2 t_3 t_4 t_5 t_6)
(1+ t_1 t_3 t_4 t_5^2+ t_1^2 t_3 t_5 t_6+ t_1 t_4 t_5 t_6
\nn\\
&&
\hspace{1cm}
+ t_1 t_2 t_3 t_4 t_5 t_6+ t_2 t_4^2 t_5 t_6
+ t_1 t_2 t_4 t_6^2
+ t_1^2 t_2 t_3 t_4^2 t_5^2 t_6^2)
\nn\\
&&
\hspace{1cm}
\times
\frac{1}{
\left(1 - t_2 t_3\right) 
\left(1 - t_1^2 t_3^2 t_5^2\right)
\left(1 - t_4^2 t_5^2\right)
\left(1 - t_1^2 t_6^2\right)  
\left(1 - t_2^2 t_4^2 t_6^2\right) 
}
~,~
\eea
where $t_a$ is the fugacity associated to the extremal GLSM field $p_a$.
When unrefined by setting $t_a=t$, the Hilbert series takes the form, 
\beal{es04h09}
g(t; \mathcal{M}^{mes})
=
\frac{
1+t^4+4t^5+t^6+t^{10}
}{
(1-t^2)(1-t^4)^2(1-t^6)
}
~,~
\eea
where the palindromic numerator indicates that the mesonic moduli space is Calabi-Yau.

%-------------------
 \begin {table}[httt!!]
\centering
\begin {tabular} {|c|c|ccc|}
\hline
PL term & generator & $U(1)_{f_1}$ & $U(1)_{f_2}$ & $U(1)_{f_3}$ 
\\
\hline
\multirow{1}{*}{$+ f_2^{-2} f_3^{2} t^{4}$}
& $p_4^2 p_5^2 ~q $ & $0$ & $-2$ & $2$  \\
\hline
\multirow{1}{*}{$+f_1^{-1} f_2^{-2}  t^{5}$}
& $p_2 p_4^2 p_5 p_6 ~q $ & $-1$ & $-2$ & $0$  \\
\hline
\multirow{1}{*}{$+f_1^{-1} f_2  t^{2}$}
& $p_2 p_3 ~$ & $-1$ & $1$ & $0$ \\
\hline
\multirow{1}{*}{$+f_1^{-2} f_2^{-2} f_3^{-2} t^{6}$}
& $p_2^2 p_4^2 p_6^2 ~q $ & $-2$ & $-2$ & $-2$  \\
\hline
\multirow{1}{*}{$+f_1 f_2^{-1}  t^{4}$}
& $p_1 p_4 p_5 p_6 ~q $ & $1$ & $-1$ & $0$  \\
\hline
\multirow{1}{*}{$+f_1  f_3^{2} t^{5}$}
& $p_1 p_3 p_4 p_5^2 ~q ^2$ & $1$ & $0$ & $2$  \\
\hline
\multirow{1}{*}{$+ f_2^{-1} f_3^{-2} t^{5}$}
& $p_1 p_2 p_4 p_6^2 ~q $ & $0$ & $-1$ & $-2$  \\
\hline
\multirow{1}{*}{$+f_1^{2}  f_3^{-2} t^{4}$}
& $p_1^2 p_6^2 ~q $ & $2$ & $0$ & $-2$  \\
\hline
\multirow{1}{*}{$+f_1^{2} f_2  t^{5}$}
& $p_1^2 p_3 p_5 p_6 ~q ^2$ & $2$ & $1$ & $0$  \\
\hline
\multirow{1}{*}{$+f_1^{2} f_2^{2} f_3^{2} t^{6}$}
& $p_1^2 p_3^2 p_5^2 ~q ^3$ & $2$ & $2$ & $2$  \\
\hline
\end{tabular}
\caption{
Generators of the $D_3/\mathbb{Z}_2 \ (0,1,1,1,1)$ model in terms of GLSM fields and their corresponding mesonic flavor symmetry charges. Here, we denote $q=\prod_{i=1}^{4}q_i$, 
and set extra GLSM fields to 1.
\label{tab_04h02}}
\end{table}

%-------------------

The Hilbert series refined using mesonic flavor fugacities as summarized in \tref{tab_04h01}
is given by, 
\beal{es04h10}
&&
g(f_1,f_2,f_3,t; \mathcal{M}^{mes})
=
(1-t^6) (1+f_1f_2^{-1}t^4 + f_2^{-1} f_3^{-2}t^5 + f_1^{-1}f_2^{-2}t^5
 + f_1^2f_2t^5 
\nn\\
&&
\hspace{1cm}
+f_1 f_3^2 t^5 + t^6 + f_1f_2^{-1} t^{10})
\times
\frac{
1
}{
(1-f_1^{-1}f_2 t^2) (1-f_1^2 f_3^{-2} t^4) 
}
\nn\\
&&
\hspace{1cm}
\times
\frac{
1
}{
(1-f_2^{-2} f_3^2 t^4) (1-f_1^{-2} f_2^{-2} f_3^{-2} t^6) (1- f_1^2 f_2^2 f_3^2 t^6)
}
~.~
\eea
The corresponding plethystic logarithm takes the following form,
\beal{es04h11}
&&
\text{PL}[g(f_1,f_2,f_3,t; \mathcal{M}^{mes})] =
f_1^{-1} f_2 t^2
+ (f_1 f_2^{-1} 
+ f_1^{2} f_3^{-2} 
+ f_2^{-2} f_3^{2}) t^4
+ (f_1^{-1} f_2^{-2} 
\nn\\
&&
\hspace{1cm}
+ f_1^{2} f_2 
+ f_2^{-1} f_3^{-2} 
+ f_1 f_3^{2}) t^5
+ (f_1^{-2} f_2^{-2} f_3^{-2} 
+ f_1^{2} f_2^{2} f_3^{2}) t^6
- f_1^2 f_2^{-2} t^8
\nn\\
&&
\hspace{1cm}
- (f_1^3 + f_2^{-3} + f_1 f_2^{-2} f_3^{-2} + f_1^{2} f_2^{-1} f_3^{2} ) t^9
- (2 f_1 f_2^{-1} + f_1^{-2} f_2^{-4}+f_1^{4} f_2^{2}
\nn\\
&&
\hspace{1cm}
+f_2^{-2} f_3^{-4}+f_1^{2} f_3^{-2}+f_1^{-1} f_2^{-3} f_3^{-2}+f_2^{-2} f_3^{2}+f_1^{3} f_2 f_3^{2}+f_1^{2} f_3^{4}) t^{10}
\nn\\
&&
\hspace{1cm}
- (  f_1^{-1} f_2^{-2} 
+ f_1^{2} f_2+f_2^{-1} f_3^{-2}+f_1 f_3^{2})t^{11} - t^{12}
+ \dots
~,~
\eea
where the infinite expansion of the plethystic logarithm indicates that the mesonic moduli space is not a complete intersection.
The generators of the mesonic moduli space with their mesonic flavor charges are summarized in \tref{tab_04h02}. 

We note here that the abelian orbifold $D_3/\mathbb{Z}_2 \ (0,1,1,1,1)$ is part of a family of abelian orbifolds of the form 
$D_3/\mathbb{Z}_n \ (0,1,-1,1,-1)$, whose toric diagrams have extremal vertices with the following coordinates,
\beal{es04h20}
&
(-1,0,0)~,~
(0,-1,0)~,~
(-1,0,1)~,~
(0,-1,1)~,~
(0,n-1,0)~,~
(1,n-2,0)~.~
&
\eea
In section \sref{sec0412}, we review the brane brick model for the abelian orbifold of the form $D_3/\mathbb{Z}_3 \ (0,1,2,1,2)$, which is also part of this family of abelian orbifolds of $D_3$. 
\\

%=================================================================
\subsubsection{$D_3/ \mathbb{Z}_3 \ (0,0,0,1,2)$ \label{sec0410}}

%-------------------
\begin{figure}[H]
    \centering
    \includegraphics[width=0.4\textwidth]{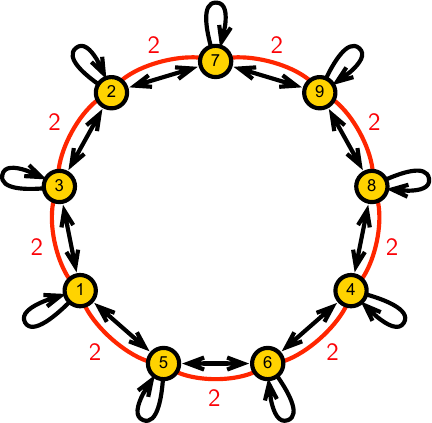}
    \caption{
    The quiver for the $D_3/ \mathbb{Z}_3 \ (0,0,0,1,2)$ model.
    \label{fig_424_quiver}
    }
\end{figure}
%-------------------

The $J$- and $E$-terms
for the abelian orbifold of the form $D_3/ \mathbb{Z}_3 \ (0,0,0,1,2)$ are as follows, 
\beal{es04j01}
\begin{array}{rclccclcccc}
&& && &J& && &E& \\
&&\Lambda^{(1)}_{5 1} &:& & X_{1 3} \cdot X_{3 1} \cdot Y_{1 5} - Y_{1 5} \cdot X_{5 5}& && & D_{5 6} \cdot Z_{6 5} \cdot Z_{5 1} - Z_{5 1} \cdot D_{1 1}& \\ 
 &&\Lambda^{(1)}_{8 4} &:& & X_{4 6} \cdot X_{6 4} \cdot Y_{4 8} - Y_{4 8} \cdot X_{8 8}& && & D_{8 9} \cdot Z_{9 8} \cdot Z_{8 4} - Z_{8 4} \cdot D_{4 4}& \\ 
 &&\Lambda^{(1)}_{2 7} &:& & X_{7 9} \cdot X_{9 7} \cdot Y_{7 2} - Y_{7 2} \cdot X_{2 2}& && & D_{2 3} \cdot Z_{3 2} \cdot Z_{2 7} - Z_{2 7} \cdot D_{7 7}& \\ 
 &&\Lambda^{(2)}_{7 2} &:& & Z_{2 7} \cdot X_{7 9} \cdot X_{9 7} - X_{2 2} \cdot Z_{2 7}& && & D_{7 7} \cdot Y_{7 2} - Y_{7 2} \cdot D_{2 3} \cdot Z_{3 2}& \\ 
 &&\Lambda^{(2)}_{1 5} &:& & Z_{5 1} \cdot X_{1 3} \cdot X_{3 1} - X_{5 5} \cdot Z_{5 1}& && & D_{1 1} \cdot Y_{1 5} - Y_{1 5} \cdot D_{5 6} \cdot Z_{6 5}& \\ 
 &&\Lambda^{(2)}_{4 8} &:& & Z_{8 4} \cdot X_{4 6} \cdot X_{6 4} - X_{8 8} \cdot Z_{8 4}& && & D_{4 4} \cdot Y_{4 8} - Y_{4 8} \cdot D_{8 9} \cdot Z_{9 8}& \\ 
 &&\Lambda^{(3)}_{3 1} &:& & X_{1 3} \cdot Y_{3 3} - Y_{1 5} \cdot Z_{5 1} \cdot X_{1 3}& && & X_{3 1} \cdot D_{1 1} - Z_{3 2} \cdot D_{2 3} \cdot X_{3 1}& \\ 
 &&\Lambda^{(3)}_{6 4} &:& & X_{4 6} \cdot Y_{6 6} - Y_{4 8} \cdot Z_{8 4} \cdot X_{4 6}& && & X_{6 4} \cdot D_{4 4} - Z_{6 5} \cdot D_{5 6} \cdot X_{6 4}& \\ 
 &&\Lambda^{(3)}_{9 7} &:& & X_{7 9} \cdot Y_{9 9} - Y_{7 2} \cdot Z_{2 7} \cdot X_{7 9}& && & X_{9 7} \cdot D_{7 7} - Z_{9 8} \cdot D_{8 9} \cdot X_{9 7}& \\ 
 &&\Lambda^{(4)}_{1 3} &:& & X_{3 1} \cdot Y_{1 5} \cdot Z_{5 1} - Y_{3 3} \cdot X_{3 1}& && & D_{1 1} \cdot X_{1 3} - X_{1 3} \cdot Z_{3 2} \cdot D_{2 3}& \\ 
 &&\Lambda^{(4)}_{4 6} &:& & X_{6 4} \cdot Y_{4 8} \cdot Z_{8 4} - Y_{6 6} \cdot X_{6 4}& && & D_{4 4} \cdot X_{4 6} - X_{4 6} \cdot Z_{6 5} \cdot D_{5 6}& \\ 
 &&\Lambda^{(4)}_{7 9} &:& & X_{9 7} \cdot Y_{7 2} \cdot Z_{2 7} - Y_{9 9} \cdot X_{9 7}& && & D_{7 7} \cdot X_{7 9} - X_{7 9} \cdot Z_{9 8} \cdot D_{8 9}& \\ 
 &&\Lambda^{(5)}_{2 3} &:& & Y_{3 3} \cdot Z_{3 2} - Z_{3 2} \cdot Z_{2 7} \cdot Y_{7 2}& && & D_{2 3} \cdot X_{3 1} \cdot X_{1 3} - X_{2 2} \cdot D_{2 3}& \\ 
 &&\Lambda^{(5)}_{5 6} &:& & Y_{6 6} \cdot Z_{6 5} - Z_{6 5} \cdot Z_{5 1} \cdot Y_{1 5}& && & D_{5 6} \cdot X_{6 4} \cdot X_{4 6} - X_{5 5} \cdot D_{5 6}& \\ 
 &&\Lambda^{(5)}_{8 9} &:& & Y_{9 9} \cdot Z_{9 8} - Z_{9 8} \cdot Z_{8 4} \cdot Y_{4 8}& && & D_{8 9} \cdot X_{9 7} \cdot X_{7 9} - X_{8 8} \cdot D_{8 9}& \\ 
 &&\Lambda^{(6)}_{2 3} &:& & Z_{3 2} \cdot X_{2 2} - X_{3 1} \cdot X_{1 3} \cdot Z_{3 2}& && & D_{2 3} \cdot Y_{3 3} - Z_{2 7} \cdot Y_{7 2} \cdot D_{2 3}& \\ 
 &&\Lambda^{(6)}_{5 6} &:& & Z_{6 5} \cdot X_{5 5} - X_{6 4} \cdot X_{4 6} \cdot Z_{6 5}& && & D_{5 6} \cdot Y_{6 6} - Z_{5 1} \cdot Y_{1 5} \cdot D_{5 6}& \\ 
 &&\Lambda^{(6)}_{8 9} &:& & Z_{9 8} \cdot X_{8 8} - X_{9 7} \cdot X_{7 9} \cdot Z_{9 8}& && & D_{8 9} \cdot Y_{9 9} - Z_{8 4} \cdot Y_{4 8} \cdot D_{8 9}&
\end{array}
~.~
\eea
The corresponding quiver is shown in \fref{fig_424_quiver}.
The $J$- and $E$-terms are obtained from the general formula in \eref{es03c07}
with the following additional relabelling of indices,
\beal{es04j02}
&
[1,0] \rightarrow 1~,~ [2,0] \rightarrow 2~,~ [3,0] \rightarrow 3~,~
&
\nn\\
&
[1,1] \rightarrow 4~,~ [2,1] \rightarrow 5~,~ [3,1] \rightarrow 6~,~
&
\nn\\
&
[1,2] \rightarrow 7~,~ [2,2] \rightarrow 8~,~ [3,2] \rightarrow 9~.~
&
\eea

The $P$-matrix is given by,
\beal{es04j03}
&&
P=
\nn\\
&&
\resizebox{0.95\textwidth}{!}{$
\left(
\begin{array}{c|cccccc|ccc|ccc|ccc|ccc|ccc|ccc}
& p_{1} & p_{2} & p_{3} & p_{4} & p_{5} & p_{6} & q^{(1)}_{1} & q^{(1)}_{2} & q^{(1)}_{3} & q^{(2)}_{1} & q^{(2)}_{2} & q^{(2)}_{3} & q^{(3)}_{1} & q^{(3)}_{2} & q^{(3)}_{3} & q^{(4)}_{1} & q^{(4)}_{2} & q^{(4)}_{3} & q^{(5)}_{1} & q^{(5)}_{2} & q^{(5)}_{3} & q^{(6)}_{1} & q^{(6)}_{2} & q^{(6)}_{3} \\
\hline
 D_{11} & 1 & 0 & 0 & 0 & 0 & 1 & 0 & 0 & 0 & 0 & 0 & 0 & 0 & 0 & 0 & \
0 & 0 & 0 & 1 & 1 & 1 & 1 & 1 & 1 \\
 D_{23} & 1 & 0 & 0 & 0 & 0 & 0 & 0 & 0 & 0 & 0 & 0 & 0 & 0 & 0 & 0 & \
0 & 0 & 0 & 0 & 1 & 0 & 1 & 0 & 1 \\
 D_{44} & 1 & 0 & 0 & 0 & 0 & 1 & 0 & 0 & 0 & 0 & 0 & 0 & 0 & 0 & 0 & \
0 & 0 & 0 & 1 & 1 & 1 & 1 & 1 & 1 \\
 D_{56} & 1 & 0 & 0 & 0 & 0 & 0 & 0 & 0 & 0 & 0 & 0 & 0 & 0 & 0 & 0 & \
0 & 0 & 0 & 0 & 0 & 1 & 0 & 1 & 1 \\
 D_{77} & 1 & 0 & 0 & 0 & 0 & 1 & 0 & 0 & 0 & 0 & 0 & 0 & 0 & 0 & 0 & \
0 & 0 & 0 & 1 & 1 & 1 & 1 & 1 & 1 \\
 D_{89} & 1 & 0 & 0 & 0 & 0 & 0 & 0 & 0 & 0 & 0 & 0 & 0 & 0 & 0 & 0 & \
0 & 0 & 0 & 1 & 0 & 0 & 1 & 1 & 0 \\
 X_{13} & 0 & 1 & 0 & 0 & 0 & 0 & 0 & 1 & 1 & 0 & 0 & 1 & 0 & 0 & 0 & \
0 & 0 & 0 & 0 & 0 & 0 & 0 & 0 & 0 \\
 X_{22} & 0 & 1 & 1 & 0 & 0 & 0 & 1 & 1 & 1 & 1 & 1 & 1 & 0 & 0 & 0 & \
0 & 0 & 0 & 0 & 0 & 0 & 0 & 0 & 0 \\
 X_{31} & 0 & 0 & 1 & 0 & 0 & 0 & 1 & 0 & 0 & 1 & 1 & 0 & 0 & 0 & 0 & \
0 & 0 & 0 & 0 & 0 & 0 & 0 & 0 & 0 \\
 X_{46} & 0 & 1 & 0 & 0 & 0 & 0 & 1 & 0 & 1 & 0 & 1 & 0 & 0 & 0 & 0 & \
0 & 0 & 0 & 0 & 0 & 0 & 0 & 0 & 0 \\
 X_{55} & 0 & 1 & 1 & 0 & 0 & 0 & 1 & 1 & 1 & 1 & 1 & 1 & 0 & 0 & 0 & \
0 & 0 & 0 & 0 & 0 & 0 & 0 & 0 & 0 \\
 X_{64} & 0 & 0 & 1 & 0 & 0 & 0 & 0 & 1 & 0 & 1 & 0 & 1 & 0 & 0 & 0 & \
0 & 0 & 0 & 0 & 0 & 0 & 0 & 0 & 0 \\
 X_{79} & 0 & 1 & 0 & 0 & 0 & 0 & 1 & 1 & 0 & 1 & 0 & 0 & 0 & 0 & 0 & \
0 & 0 & 0 & 0 & 0 & 0 & 0 & 0 & 0 \\
 X_{88} & 0 & 1 & 1 & 0 & 0 & 0 & 1 & 1 & 1 & 1 & 1 & 1 & 0 & 0 & 0 & \
0 & 0 & 0 & 0 & 0 & 0 & 0 & 0 & 0 \\
 X_{97} & 0 & 0 & 1 & 0 & 0 & 0 & 0 & 0 & 1 & 0 & 1 & 1 & 0 & 0 & 0 & \
0 & 0 & 0 & 0 & 0 & 0 & 0 & 0 & 0 \\
 Y_{15} & 0 & 0 & 0 & 1 & 0 & 0 & 0 & 0 & 0 & 0 & 0 & 0 & 0 & 1 & 0 & \
1 & 0 & 1 & 0 & 0 & 0 & 0 & 0 & 0 \\
 Y_{33} & 0 & 0 & 0 & 1 & 1 & 0 & 0 & 0 & 0 & 0 & 0 & 0 & 1 & 1 & 1 & \
1 & 1 & 1 & 0 & 0 & 0 & 0 & 0 & 0 \\
 Y_{48} & 0 & 0 & 0 & 1 & 0 & 0 & 0 & 0 & 0 & 0 & 0 & 0 & 0 & 0 & 1 & \
0 & 1 & 1 & 0 & 0 & 0 & 0 & 0 & 0 \\
 Y_{66} & 0 & 0 & 0 & 1 & 1 & 0 & 0 & 0 & 0 & 0 & 0 & 0 & 1 & 1 & 1 & \
1 & 1 & 1 & 0 & 0 & 0 & 0 & 0 & 0 \\
 Y_{72} & 0 & 0 & 0 & 1 & 0 & 0 & 0 & 0 & 0 & 0 & 0 & 0 & 1 & 0 & 0 & \
1 & 1 & 0 & 0 & 0 & 0 & 0 & 0 & 0 \\
 Y_{99} & 0 & 0 & 0 & 1 & 1 & 0 & 0 & 0 & 0 & 0 & 0 & 0 & 1 & 1 & 1 & \
1 & 1 & 1 & 0 & 0 & 0 & 0 & 0 & 0 \\
 Z_{27} & 0 & 0 & 0 & 0 & 1 & 0 & 0 & 0 & 0 & 0 & 0 & 0 & 0 & 1 & 1 & \
0 & 0 & 1 & 0 & 0 & 0 & 0 & 0 & 0 \\
 Z_{32} & 0 & 0 & 0 & 0 & 0 & 1 & 0 & 0 & 0 & 0 & 0 & 0 & 0 & 0 & 0 & \
0 & 0 & 0 & 1 & 0 & 1 & 0 & 1 & 0 \\
 Z_{51} & 0 & 0 & 0 & 0 & 1 & 0 & 0 & 0 & 0 & 0 & 0 & 0 & 1 & 0 & 1 & \
0 & 1 & 0 & 0 & 0 & 0 & 0 & 0 & 0 \\
 Z_{65} & 0 & 0 & 0 & 0 & 0 & 1 & 0 & 0 & 0 & 0 & 0 & 0 & 0 & 0 & 0 & \
0 & 0 & 0 & 1 & 1 & 0 & 1 & 0 & 0 \\
 Z_{84} & 0 & 0 & 0 & 0 & 1 & 0 & 0 & 0 & 0 & 0 & 0 & 0 & 1 & 1 & 0 & \
1 & 0 & 0 & 0 & 0 & 0 & 0 & 0 & 0 \\
 Z_{98} & 0 & 0 & 0 & 0 & 0 & 1 & 0 & 0 & 0 & 0 & 0 & 0 & 0 & 0 & 0 & \
0 & 0 & 0 & 0 & 1 & 1 & 0 & 0 & 1 \\
\end{array}
\right)
$}~.~
\nn\\
\eea
The $J$- and $E$-term charge matrix takes the following form,
\beal{es04j04}
&&
Q_{JE}=
\nn\\
&&
\resizebox{0.9\textwidth}{!}{$
\left(
\begin{array}{cccccc|ccc|ccc|ccc|ccc|ccc|ccc}
p_{1} & p_{2} & p_{3} & p_{4} & p_{5} & p_{6} & q^{(1)}_{1} & q^{(1)}_{2} & q^{(1)}_{3} & q^{(2)}_{1} & q^{(2)}_{2} & q^{(2)}_{3} & q^{(3)}_{1} & q^{(3)}_{2} & q^{(3)}_{3} & q^{(4)}_{1} & q^{(4)}_{2} & q^{(4)}_{3} & q^{(5)}_{1} & q^{(5)}_{2} & q^{(5)}_{3} & q^{(6)}_{1} & q^{(6)}_{2} & q^{(6)}_{3} \\
\hline
 0 & 0 & 0 & 0 & 0 & -1 & 0 & 0 & 0 & 0 & 0 & 0 & 0 & 0 & 0 & 0 & 0 & 0 & 2 & 0 & 0 & -1
   & -1 & 1 \\
 0 & 0 & 0 & 0 & 0 & -1 & 0 & 0 & 0 & 0 & 0 & 0 & 0 & 0 & 0 & 0 & 0 & 0 & 1 & 0 & 1 & 0 &
   -1 & 0 \\
 0 & 0 & 0 & 0 & 0 & -1 & 0 & 0 & 0 & 0 & 0 & 0 & 0 & 0 & 0 & 0 & 0 & 0 & 1 & 1 & 0 & -1
   & 0 & 0 \\
 0 & 0 & 0 & 0 & -1 & 0 & 0 & 0 & 0 & 0 & 0 & 0 & 2 & 0 & 0 & -1 & -1 & 1 & 0 & 0 & 0 & 0
   & 0 & 0 \\
 0 & 0 & 0 & 0 & -1 & 0 & 0 & 0 & 0 & 0 & 0 & 0 & 1 & 0 & 1 & 0 & -1 & 0 & 0 & 0 & 0 & 0
   & 0 & 0 \\
 0 & 0 & 0 & 0 & -1 & 0 & 0 & 0 & 0 & 0 & 0 & 0 & 1 & 1 & 0 & -1 & 0 & 0 & 0 & 0 & 0 & 0
   & 0 & 0 \\
 0 & 0 & 0 & 1 & 0 & 0 & 0 & 0 & 0 & 0 & 0 & 0 & 1 & 0 & 0 & -1 & -1 & 0 & 0 & 0 & 0 & 0
   & 0 & 0 \\
 0 & 2 & 1 & 0 & 0 & 0 & -1 & -1 & -1 & 0 & 0 & 0 & 0 & 0 & 0 & 0 & 0 & 0 & 0 & 0 & 0 & 0
   & 0 & 0 \\
 0 & 1 & 0 & 0 & 0 & 0 & 0 & -1 & -1 & 0 & 0 & 1 & 0 & 0 & 0 & 0 & 0 & 0 & 0 & 0 & 0 & 0
   & 0 & 0 \\
 0 & 1 & 0 & 0 & 0 & 0 & -1 & 0 & -1 & 0 & 1 & 0 & 0 & 0 & 0 & 0 & 0 & 0 & 0 & 0 & 0 & 0
   & 0 & 0 \\
 0 & 1 & 0 & 0 & 0 & 0 & -1 & -1 & 0 & 1 & 0 & 0 & 0 & 0 & 0 & 0 & 0 & 0 & 0 & 0 & 0 & 0
   & 0 & 0 \\
 1 & 0 & 0 & 0 & 0 & 0 & 0 & 0 & 0 & 0 & 0 & 0 & 0 & 0 & 0 & 0 & 0 & 0 & 1 & 0 & 0 & -1 &
   -1 & 0 \\
\end{array}
\right)
$}~,~
\nn\\
\eea
and the $D$-term charge matrix is given by,
\beal{es04j05}
&&
Q_{D}=
\nn\\
&&
\resizebox{0.9\textwidth}{!}{$
\left(
\begin{array}{cccccc|ccc|ccc|ccc|ccc|ccc|ccc}
p_{1} & p_{2} & p_{3} & p_{4} & p_{5} & p_{6} & q^{(1)}_{1} & q^{(1)}_{2} & q^{(1)}_{3} & q^{(2)}_{1} & q^{(2)}_{2} & q^{(2)}_{3} & q^{(3)}_{1} & q^{(3)}_{2} & q^{(3)}_{3} & q^{(4)}_{1} & q^{(4)}_{2} & q^{(4)}_{3} & q^{(5)}_{1} & q^{(5)}_{2} & q^{(5)}_{3} & q^{(6)}_{1} & q^{(6)}_{2} & q^{(6)}_{3} \\
\hline
 0 & -1 & 0 & 0 & 0 & 0 & 1 & 0 & 0 & 0 & 0 & 0 & 1 & 0 & 0 & -1 & 0 & 0 & 0 & 0 & 0 & 0
   & 0 & 0 \\
 0 & 0 & 0 & 0 & -1 & 0 & 0 & 0 & 0 & 0 & 0 & 0 & 1 & 0 & 0 & 0 & 0 & 0 & 1 & 0 & 0 & -1
   & 0 & 0 \\
 0 & 1 & 0 & 0 & 0 & 0 & -1 & 0 & 0 & 0 & 0 & 0 & 0 & 0 & 0 & 0 & 0 & 0 & -1 & 0 & 0 & 1
   & 0 & 0 \\
 0 & -1 & 0 & 0 & 0 & 0 & 0 & 1 & 0 & 0 & 0 & 0 & 1 & 0 & 0 & 0 & -1 & 0 & 0 & 0 & 0 & 0
   & 0 & 0 \\
 0 & 0 & 0 & 0 & 0 & 0 & 0 & 0 & 0 & 0 & 0 & 0 & -1 & 0 & 0 & 1 & 0 & 0 & 1 & 0 & 0 & 0 &
   -1 & 0 \\
 0 & 1 & 0 & 0 & 0 & 0 & 0 & -1 & 0 & 0 & 0 & 0 & 0 & 0 & 0 & 0 & 0 & 0 & -1 & 0 & 0 & 0
   & 1 & 0 \\
 0 & -1 & 0 & 0 & 1 & 0 & 0 & 0 & 1 & 0 & 0 & 0 & -1 & 0 & 0 & 0 & 0 & 0 & 0 & 0 & 0 & 0
   & 0 & 0 \\
 0 & 0 & 0 & 0 & 0 & 1 & 0 & 0 & 0 & 0 & 0 & 0 & -1 & 0 & 0 & 0 & 1 & 0 & -1 & 0 & 0 & 0
   & 0 & 0 \\
\end{array}
\right)
$}~,~
\nn\\
\eea

\begin{figure}[htt!!]
\centering
\includegraphics[width=0.4\textwidth]{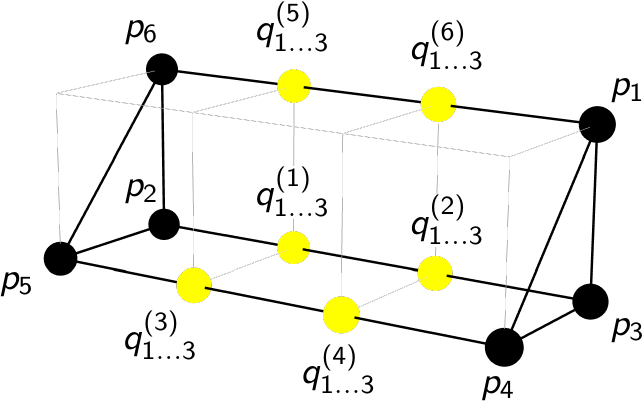}
\caption{
The toric diagram for the $D_3/\mathbb{Z}_3 \ (0,0,0,1,2)$ model.
\label{fig_424_toric}
}
\end{figure}

The toric diagram for the $D_3/ \mathbb{Z}_3 \ (0,0,0,1,2)$ model shown in \fref{fig_424_toric}
is given by, 
\beal{es04j06}
&&
G_{t}=
\nn\\
&&
\resizebox{0.9\textwidth}{!}{$
\left(
\begin{array}{cccccc|ccc|ccc|ccc|ccc|ccc|ccc}
p_{1} & p_{2} & p_{3} & p_{4} & p_{5} & p_{6} & q^{(1)}_{1} & q^{(1)}_{2} & q^{(1)}_{3} & q^{(2)}_{1} & q^{(2)}_{2} & q^{(2)}_{3} & q^{(3)}_{1} & q^{(3)}_{2} & q^{(3)}_{3} & q^{(4)}_{1} & q^{(4)}_{2} & q^{(4)}_{3} & q^{(5)}_{1} & q^{(5)}_{2} & q^{(5)}_{3} & q^{(6)}_{1} & q^{(6)}_{2} & q^{(6)}_{3} \\
\hline
 1 & 0 & 0 & 0 & 0 & 1 & 0 & 0 & 0 & 0 & 0 & 0 & 0 & 0 & 0 & 0 & 0 & 0 & 1 & 1 & 1 & 1 &
   1 & 1 \\
 0 & 3 & 0 & 0 & 3 & 3 & 2 & 2 & 2 & 1 & 1 & 1 & 2 & 2 & 2 & 1 & 1 & 1 & 2 & 2 & 2 & 1 &
   1 & 1 \\
 0 & 0 & 0 & 1 & 1 & 0 & 0 & 0 & 0 & 0 & 0 & 0 & 1 & 1 & 1 & 1 & 1 & 1 & 0 & 0 & 0 & 0 &
   0 & 0 \\
   \hline
 1 & 1 & 1 & 1 & 1 & 1 & 1 & 1 & 1 & 1 & 1 & 1 & 1 & 1 & 1 & 1 & 1 & 1 & 1 & 1 & 1 & 1 &
   1 & 1 \\
\end{array}
\right)
$}~.~
\nn\\
\eea

%-------------------
%-------------------

%-------------------
\begin{table}[htt!!]
\centering
\begin{tabular}{|c|c|c|c|l|}
\hline
\; & $U(1)_{f_1}$ & $U(1)_{f_2}$ & $U(1)_{f_3}$ &  fugacity \\
\hline
$p_1$ & $+1$ & $0$ & $0$ &  $t_1=f_1 t$ \\
$p_2$ & $-1$ & $0$ & $0$ &  $t_2=f_1^{-1} t$\\
$p_3$ & $0$ & $+1$ & $0$ &  $t_3=f_2 t$\\
$p_4$ & $0$ & $-1$ & $0$ & $t_4=f_2^{-1} t$ \\
$p_5$ & $0$ & $0$ & $+1$ & $t_5=f_3 t$ \\
$p_6$ & $0$ & $0$ & $-1$ & $t_6=f_3^{-1} t$ \\
\hline
\end{tabular}
\caption{
Mesonic flavor symmetry of the $D_3/\mathbb{Z}_3 \ (0,0,0,1,2)$ model 
and charges on the extremal GLSM fields $p_a$. 
Here, the fugacity $t$ counts the degree in extremal GLSM fields $p_a$. 
\label{tab_04j01}}
\end{table}
%-------------------

The global symmetry of the $D_3/\mathbb{Z}_3 \ (0,0,0,1,2)$ model takes the following form,
\beal{es04j07}
U(1)_{f_1} \times U(1)_{f_2} \times U(1)_{f_3} \times U(1)_R
~.~
\eea
The Hilbert series of the mesonic moduli space for the $D_3/\mathbb{Z}_3 \ (0,0,0,1,2)$ model 
is given by, 
\beal{es04j08}
&&
g(t_a; \mathcal{M}^{mes})
=
\frac{1 - t_1^3 t_2^3 t_3^3 t_4^3 t_5^3 t_6^3}{\left(1 - t_2 t_3\right) \left(1 - t_1^3 t_3^3 t_4^3\right) \left(1 - t_4 t_5\right) \left(1 - t_1 t_6\right) \left(1 - t_2^3 t_5^3 t_6^3\right)}
~,~
\eea
where $t_a$ is the fugacity corresponding to the extremal GLSM field $p_a$.
By setting $t_a=t$, 
the Hilbert series becomes unrefined and takes the following form, 
\beal{es04j09}
g(t; \mathcal{M}^{mes})
=
\frac{1-t^{18}}{(1-t^2)^3 (1-t^9)^2}
~.~
\eea

The Hilbert series refined by mesonic flavor fugacities summarized in \tref{tab_04j01}
is given by,
\beal{es04j10}
&&
g(f_1,f_2,f_3,t; \mathcal{M}^{mes})
=
\frac{1-t^{18}}{(1-f_1 f_3^{-1} t^2)(1-f_1^{-1} f_2 t^2)(1-f_2^{-1}f_3 t^2)}
\nn\\
&&
\hspace{1cm}
\times
\frac{1}{(1-f_1^{-3}t^9)(1-f_1^3 t^9)}
~.~
\eea

%-------------------
 \begin {table}[htt!!]
\centering
\begin {tabular} {|c|c|ccc|}
\hline
PL term & generator & $U(1)_{f_1}$ & $U(1)_{f_2}$ & $U(1)_{f_3}$ 
\\
\hline
\multirow{1}{*}{$+f_1  f_3^{-1} t^{2}$}
& $A_1 = p_1 p_6 ~q_{{(5)}} q_{{(6)}}$ & $1$ & $0$ & $-1$  \\
\hline
\multirow{1}{*}{$+f_1^{-1} f_2  t^{2}$}
& $A_2 = p_2 p_3 ~q_{(1)} q_{(2)}$ & $-1$ & $1$ & $0$  \\
\hline
\multirow{1}{*}{$+ f_2^{-1} f_3 t^{2}$}
& $A_3 = p_4 p_5 ~q_{{(3)}} q_{{(4)}}$ & $0$ & $-1$ & $1$  \\
\hline
\multirow{1}{*}{$+f_1^{3}   t^{9}$}
& $B_1 = p_1^3 p_3^3 p_4^3 ~q_{(1)} q_{(2)}^2 q_{{(3)}} q_{{(4)}}^2 q_{{(5)}} q_{{(6)}}^2$ & $3$ & $0$ & $0$ \\
\hline
\multirow{1}{*}{$+f_1^{-3}   t^{9}$}
& $B_2 = p_2^3 p_5^3 p_6^3 ~q_{(1)}^2 q_{(2)} q_{{(3)}}^2 q_{{(4)}} q_{{(5)}}^2 q_{{(6)}}$ & $-3$ & $0$ & $0$  \\
\hline
\end{tabular}
\caption{
Generators of the $D_3/\mathbb{Z}_3 \ (0,0,0,1,2)$ model in terms of GLSM fields and their corresponding mesonic flavor symmetry charges. Here, we denote $q_{(1)}=\prod_{i=1}^{3}q_i^{(1)}$, $q_{(2)}=\prod_{i=1}^{3}q_i^{(2)}$, $q_{(3)}=\prod_{i=1}^{3}q_i^{(3)}$, $q_{(4)}=\prod_{i=1}^{3}q_i^{(4)}$, $q_{(5)}=\prod_{i=1}^{3}q_i^{(5)}$, $q_{(6)}=\prod_{i=1}^{3}q_i^{(6)}$.
\label{tab_04j02}}
\end{table}
%-------------------

The plethystic logarithm of the Hilbert series refined under the mesonic flavor symmetry in \eref{es04j10}
takes the following form, 
\beal{es04j12}
\text{PL}[g(f_1,f_2,f_3,t; \mathcal{M}^{mes})]=
(f_1^{-1} f_2+f_1 f_3^{-1}+f_2^{-1} f_3) t^2
+ (f_1^3 + f_1^{-3}) t^9
- t^{18}
~,~
\nn\\
\eea
where the finite expansion of the plethystic logarithm indicates that the mesonic moduli space is a complete intersection.
The first positive terms in the expansion of the plethystic logarithm correspond to the generators of the mesonic moduli space, which are summarized with their mesonic flavor
charges in \tref{tab_04j02}.

These generators form a single defining relation for $D_3/\mathbb{Z}_3 \ (0,0,0,1,2)$, 
allowing us to write the mesonic moduli space as, 
\beal{es04j13}
\mathcal{M}^{mes} =
\mathrm{Spec}~\mathbb{C}[A_1,A_2,A_3,B_1,B_2]/ \langle A_1^3 A_2^3 A_3^3 - B_1 B_2 \rangle~.~
\eea
We note here that the abelian orbifold $D_3/\mathbb{Z}_3 \ (0,0,0,1,2)$
is part of a family of abelian orbifolds of the form $D_3/\mathbb{Z}_n \ (0,0,0,1,-1)$
as discussed in section \sref{sec047}.
\\

%=================================================================
\subsubsection{$D_3/ \mathbb{Z}_3 \ (0,1,2,0,0)$ \label{sec0411}}

%-------------------
\begin{figure}[H]
    \centering
    \includegraphics[width=0.5\textwidth]{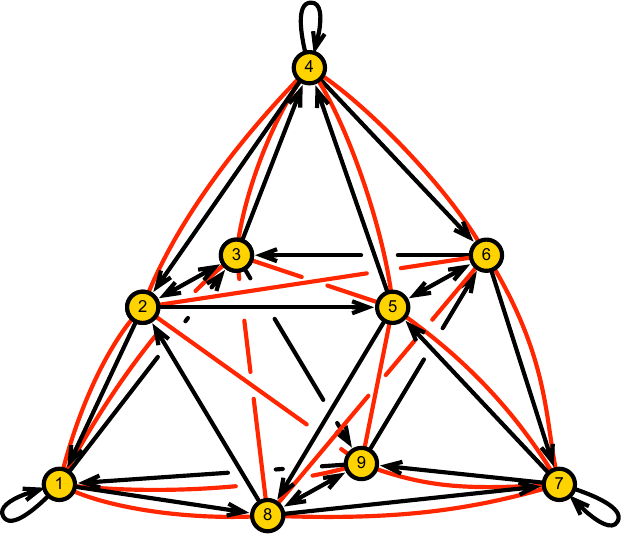}
    \caption{
    The quiver for the $D_3/ \mathbb{Z}_3 \ (0,1,2,0,0)$ model.
    \label{fig_425_quiver}
    }
\end{figure}
%-------------------

The $J$- and $E$-terms of the brane brick model for the abelian orbifold of the form $D_3/ \mathbb{Z}_3 \ (0,1,2,0,0)$
are given as follows, 
\beal{es04k01}
\begin{array}{rclccclcccc}
&& && &J& && &E& \\
&&\Lambda^{(1)}_{2 1} &:& & X_{1 3} \cdot X_{3 4} \cdot Y_{4 2} - Y_{1 8} \cdot X_{8 2}& && & D_{2 3} \cdot Z_{3 2} \cdot Z_{2 1} - Z_{2 1} \cdot D_{1 1}& \\ 
 &&\Lambda^{(1)}_{5 4} &:& & X_{4 6} \cdot X_{6 7} \cdot Y_{7 5} - Y_{4 2} \cdot X_{2 5}& && & D_{5 6} \cdot Z_{6 5} \cdot Z_{5 4} - Z_{5 4} \cdot D_{4 4}& \\ 
 &&\Lambda^{(1)}_{8 7} &:& & X_{7 9} \cdot X_{9 1} \cdot Y_{1 8} - Y_{7 5} \cdot X_{5 8}& && & D_{8 9} \cdot Z_{9 8} \cdot Z_{8 7} - Z_{8 7} \cdot D_{7 7}& \\ 
 &&\Lambda^{(2)}_{4 2} &:& & Z_{2 1} \cdot X_{1 3} \cdot X_{3 4} - X_{2 5} \cdot Z_{5 4}& && & D_{4 4} \cdot Y_{4 2} - Y_{4 2} \cdot D_{2 3} \cdot Z_{3 2}& \\ 
 &&\Lambda^{(2)}_{7 5} &:& & Z_{5 4} \cdot X_{4 6} \cdot X_{6 7} - X_{5 8} \cdot Z_{8 7}& && & D_{7 7} \cdot Y_{7 5} - Y_{7 5} \cdot D_{5 6} \cdot Z_{6 5}& \\ 
 &&\Lambda^{(2)}_{1 8} &:& & Z_{8 7} \cdot X_{7 9} \cdot X_{9 1} - X_{8 2} \cdot Z_{2 1}& && & D_{1 1} \cdot Y_{1 8} - Y_{1 8} \cdot D_{8 9} \cdot Z_{9 8}& \\ 
 &&\Lambda^{(3)}_{9 1} &:& & X_{1 3} \cdot Y_{3 9} - Y_{1 8} \cdot Z_{8 7} \cdot X_{7 9}& && & X_{9 1} \cdot D_{1 1} - Z_{9 8} \cdot D_{8 9} \cdot X_{9 1}& \\ 
 &&\Lambda^{(3)}_{3 4} &:& & X_{4 6} \cdot Y_{6 3} - Y_{4 2} \cdot Z_{2 1} \cdot X_{1 3}& && & X_{3 4} \cdot D_{4 4} - Z_{3 2} \cdot D_{2 3} \cdot X_{3 4}& \\ 
 &&\Lambda^{(3)}_{6 7} &:& & X_{7 9} \cdot Y_{9 6} - Y_{7 5} \cdot Z_{5 4} \cdot X_{4 6}& && & X_{6 7} \cdot D_{7 7} - Z_{6 5} \cdot D_{5 6} \cdot X_{6 7}& \\ 
 &&\Lambda^{(4)}_{1 3} &:& & X_{3 4} \cdot Y_{4 2} \cdot Z_{2 1} - Y_{3 9} \cdot X_{9 1}& && & D_{1 1} \cdot X_{1 3} - X_{1 3} \cdot Z_{3 2} \cdot D_{2 3}& \\ 
 &&\Lambda^{(4)}_{4 6} &:& & X_{6 7} \cdot Y_{7 5} \cdot Z_{5 4} - Y_{6 3} \cdot X_{3 4}& && & D_{4 4} \cdot X_{4 6} - X_{4 6} \cdot Z_{6 5} \cdot D_{5 6}& \\ 
 &&\Lambda^{(4)}_{7 9} &:& & X_{9 1} \cdot Y_{1 8} \cdot Z_{8 7} - Y_{9 6} \cdot X_{6 7}& && & D_{7 7} \cdot X_{7 9} - X_{7 9} \cdot Z_{9 8} \cdot D_{8 9}& \\ 
 &&\Lambda^{(5)}_{8 3} &:& & Y_{3 9} \cdot Z_{9 8} - Z_{3 2} \cdot Z_{2 1} \cdot Y_{1 8}& && & D_{8 9} \cdot X_{9 1} \cdot X_{1 3} - X_{8 2} \cdot D_{2 3}& \\ 
 &&\Lambda^{(5)}_{2 6} &:& & Y_{6 3} \cdot Z_{3 2} - Z_{6 5} \cdot Z_{5 4} \cdot Y_{4 2}& && & D_{2 3} \cdot X_{3 4} \cdot X_{4 6} - X_{2 5} \cdot D_{5 6}& \\ 
 &&\Lambda^{(5)}_{5 9} &:& & Y_{9 6} \cdot Z_{6 5} - Z_{9 8} \cdot Z_{8 7} \cdot Y_{7 5}& && & D_{5 6} \cdot X_{6 7} \cdot X_{7 9} - X_{5 8} \cdot D_{8 9}& \\ 
 &&\Lambda^{(6)}_{5 3} &:& & Z_{3 2} \cdot X_{2 5} - X_{3 4} \cdot X_{4 6} \cdot Z_{6 5}& && & D_{5 6} \cdot Y_{6 3} - Z_{5 4} \cdot Y_{4 2} \cdot D_{2 3}& \\ 
 &&\Lambda^{(6)}_{8 6} &:& & Z_{6 5} \cdot X_{5 8} - X_{6 7} \cdot X_{7 9} \cdot Z_{9 8}& && & D_{8 9} \cdot Y_{9 6} - Z_{8 7} \cdot Y_{7 5} \cdot D_{5 6}& \\ 
 &&\Lambda^{(6)}_{2 9} &:& & Z_{9 8} \cdot X_{8 2} - X_{9 1} \cdot X_{1 3} \cdot Z_{3 2}& && & D_{2 3} \cdot Y_{3 9} - Z_{2 1} \cdot Y_{1 8} \cdot D_{8 9}&
\end{array}
~,~
\eea
where the corresponding quiver is shown in \fref{fig_425_quiver}.
These $J$- and $E$-terms are obtained from the general formula in \eref{es03c07}
with additional relabelling of indices given as follows, 
\beal{es04k02}
&
[1,0] \rightarrow 1~,~ [2,0] \rightarrow 2~,~ [3,0] \rightarrow 3~,~
&
\nn\\
&
[1,1] \rightarrow 4~,~ [2,1] \rightarrow 5~,~ [3,1] \rightarrow 6~,~
&
\nn\\
&
[1,2] \rightarrow 7~,~ [2,2] \rightarrow 8~,~ [3,2] \rightarrow 9~.~
&
\eea

The $P$-matrix for the $D_3/ \mathbb{Z}_3 \ (0,1,2,0,0)$ model is given by, 
\beal{es04k03}
&&
P=
\nn\\
&&
\resizebox{0.95\textwidth}{!}{$
\left(
\begin{array}{c|cccccc|ccc|ccc|ccc|ccc|ccc|ccc}
 & p_{1} & p_{2} & p_{3} & p_{4} & p_{5} & p_{6} & q^{(1)}_{1} & q^{(1)}_{2} & q^{(1)}_{3} & q^{(2)}_{1} & q^{(2)}_{2} & q^{(2)}_{3} & q^{(3)}_{1} & q^{(3)}_{2} & q^{(3)}_{3} & q^{(4)}_{1} & q^{(4)}_{2} & q^{(4)}_{3} & o^{(1)}_{1} & \cdots & o^{(1)}_{12} & o^{(2)}_{1} & \cdots & o^{(2)}_{12} \\
\hline
 D_{11} & 0 & 0 & 1 & 1 & 0 & 0 & 0 & 0 & 0 & 0 & 0 & 0 & 0 & 0 & 0 & 0 & 0 & 0 & 1 & \cdots & 1 & 1 & \cdots & 1 \\
 D_{23} & 0 & 0 & 0 & 1 & 0 & 0 & 0 & 0 & 0 & 0 & 0 & 0 & 0 & 0 & 0 & 0 & 0 & 0 & 0 & \cdots & 1 & 1 & \cdots & 1 \\
 D_{44} & 0 & 0 & 1 & 1 & 0 & 0 & 0 & 0 & 0 & 0 & 0 & 0 & 0 & 0 & 0 & 0 & 0 & 0 & 1 & \cdots & 1 & 1 & \cdots & 1 \\
 D_{56} & 0 & 0 & 0 & 1 & 0 & 0 & 0 & 0 & 0 & 0 & 0 & 0 & 0 & 0 & 0 & 0 & 0 & 0 & 1 & \cdots & 0 & 0 & \cdots & 0 \\
 D_{77} & 0 & 0 & 1 & 1 & 0 & 0 & 0 & 0 & 0 & 0 & 0 & 0 & 0 & 0 & 0 & 0 & 0 & 0 & 1 & \cdots & 1 & 1 & \cdots & 1 \\
 D_{89} & 0 & 0 & 0 & 1 & 0 & 0 & 0 & 0 & 0 & 0 & 0 & 0 & 0 & 0 & 0 & 0 & 0 & 0 & 1 & \cdots & 0 & 1 & \cdots & 0 \\
 X_{13} & 1 & 0 & 0 & 0 & 0 & 0 & 1 & 1 & 0 & 0 & 0 & 0 & 1 & 0 & 0 & 0 & 0 & 0 & 0 & \cdots & 0 & 1 & \cdots & 0 \\
 X_{25} & 1 & 1 & 0 & 0 & 0 & 0 & 1 & 0 & 1 & 1 & 0 & 1 & 0 & 1 & 0 & 0 & 1 & 0 & 0 & \cdots & 1 & 1 & \cdots & 1 \\
 X_{34} & 0 & 1 & 0 & 0 & 0 & 0 & 0 & 0 & 0 & 1 & 0 & 1 & 0 & 0 & 0 & 0 & 1 & 0 & 1 & \cdots & 0 & 0 & \cdots & 0 \\
 X_{46} & 1 & 0 & 0 & 0 & 0 & 0 & 1 & 0 & 1 & 0 & 0 & 0 & 0 & 1 & 0 & 0 & 0 & 0 & 0 & \cdots & 0 & 0 & \cdots & 0 \\
 X_{58} & 1 & 1 & 0 & 0 & 0 & 0 & 0 & 1 & 1 & 1 & 1 & 0 & 0 & 0 & 1 & 1 & 0 & 0 & 0 & \cdots & 0 & 0 & \cdots & 1 \\
 X_{67} & 0 & 1 & 0 & 0 & 0 & 0 & 0 & 0 & 0 & 1 & 1 & 0 & 0 & 0 & 0 & 1 & 0 & 0 & 0 & \cdots & 0 & 1 & \cdots & 1 \\
 X_{79} & 1 & 0 & 0 & 0 & 0 & 0 & 0 & 1 & 1 & 0 & 0 & 0 & 0 & 0 & 1 & 0 & 0 & 0 & 0 & \cdots & 0 & 0 & \cdots & 0 \\
 X_{82} & 1 & 1 & 0 & 0 & 0 & 0 & 1 & 1 & 0 & 0 & 1 & 1 & 1 & 0 & 0 & 0 & 0 & 1 & 1 & \cdots & 0 & 1 & \cdots & 0 \\
 X_{91} & 0 & 1 & 0 & 0 & 0 & 0 & 0 & 0 & 0 & 0 & 1 & 1 & 0 & 0 & 0 & 0 & 0 & 1 & 0 & \cdots & 1 & 0 & \cdots & 1 \\
 Y_{18} & 0 & 0 & 0 & 0 & 0 & 1 & 0 & 0 & 0 & 1 & 0 & 0 & 0 & 0 & 0 & 1 & 1 & 0 & 0 & \cdots & 0 & 0 & \cdots & 0 \\
 Y_{39} & 0 & 0 & 0 & 0 & 1 & 1 & 0 & 0 & 1 & 1 & 0 & 0 & 0 & 1 & 1 & 1 & 1 & 0 & 1 & \cdots & 0 & 0 & \cdots & 0 \\
 Y_{42} & 0 & 0 & 0 & 0 & 0 & 1 & 0 & 0 & 0 & 0 & 1 & 0 & 0 & 0 & 0 & 1 & 0 & 1 & 0 & \cdots & 0 & 0 & \cdots & 0 \\
 Y_{63} & 0 & 0 & 0 & 0 & 1 & 1 & 0 & 1 & 0 & 0 & 1 & 0 & 1 & 0 & 1 & 1 & 0 & 1 & 0 & \cdots & 1 & 1 & \cdots & 1 \\
 Y_{75} & 0 & 0 & 0 & 0 & 0 & 1 & 0 & 0 & 0 & 0 & 0 & 1 & 0 & 0 & 0 & 0 & 1 & 1 & 0 & \cdots & 1 & 0 & \cdots & 0 \\
 Y_{96} & 0 & 0 & 0 & 0 & 1 & 1 & 1 & 0 & 0 & 0 & 0 & 1 & 1 & 1 & 0 & 0 & 1 & 1 & 1 & \cdots & 1 & 0 & \cdots & 0 \\
 Z_{21} & 0 & 0 & 0 & 0 & 1 & 0 & 0 & 0 & 1 & 0 & 0 & 0 & 0 & 1 & 1 & 0 & 0 & 0 & 0 & \cdots & 1 & 0 & \cdots & 1 \\
 Z_{32} & 0 & 0 & 1 & 0 & 0 & 0 & 0 & 0 & 0 & 0 & 0 & 0 & 0 & 0 & 0 & 0 & 0 & 0 & 1 & \cdots & 0 & 0 & \cdots & 0 \\
 Z_{54} & 0 & 0 & 0 & 0 & 1 & 0 & 0 & 1 & 0 & 0 & 0 & 0 & 1 & 0 & 1 & 0 & 0 & 0 & 1 & \cdots & 0 & 0 & \cdots & 0 \\
 Z_{65} & 0 & 0 & 1 & 0 & 0 & 0 & 0 & 0 & 0 & 0 & 0 & 0 & 0 & 0 & 0 & 0 & 0 & 0 & 0 & \cdots & 1 & 1 & \cdots & 1 \\
 Z_{87} & 0 & 0 & 0 & 0 & 1 & 0 & 1 & 0 & 0 & 0 & 0 & 0 & 1 & 1 & 0 & 0 & 0 & 0 & 1 & \cdots & 0 & 1 & \cdots & 0 \\
 Z_{98} & 0 & 0 & 1 & 0 & 0 & 0 & 0 & 0 & 0 & 0 & 0 & 0 & 0 & 0 & 0 & 0 & 0 & 0 & 0 & \cdots & 1 & 0 & \cdots & 1 \\
\end{array}
\right)
$}~,~
\nn\\
\eea
where $o_k^{(1)}$ and $o_l^{(2)}$ are extra GLSM fields \cite{Witten:1993yc}.
The $J$- and $E$-term charge matrix is given by, 
\beal{es04k04}
&&
Q_{JE}=
\nn\\
&&
\resizebox{0.9\textwidth}{!}{$
\left(
\begin{array}{cccccc|ccc|ccc|ccc|ccc|ccc|ccc}
p_{1} & p_{2} & p_{3} & p_{4} & p_{5} & p_{6} & q^{(1)}_{1} & q^{(1)}_{2} & q^{(1)}_{3} & q^{(2)}_{1} & q^{(2)}_{2} & q^{(2)}_{3} & q^{(3)}_{1} & q^{(3)}_{2} & q^{(3)}_{3} & q^{(4)}_{1} & q^{(4)}_{2} & q^{(4)}_{3} & o^{(1)}_{1} & \cdots & o^{(1)}_{12} & o^{(2)}_{1} & \cdots & o^{(2)}_{12} \\
\hline
0 & 0 & -1 & -1 & -1 & -1 & -1 & 0 & 0 & 0 & -1 & 0 & 1 & 0 & 0 & 1 & 0 & 0 & 0 & \cdots & 1 & 0 & \cdots & 0 \\
0 & 0 & -1 & -1 & -1 & 0 & -1 & 0 & 0 & 0 & -1 & 0 & 1 & 0 & 0 & 0 & 0 & 0 & 0 & \cdots & 0 & 0 & \cdots & 1 \\
0 & 0 & -1 & -1 & -1 & 0 & -1 & 0 & 0 & 0 & -1 & 0 & 1 & 0 & 0 & 0 & 0 & 0 & 0 & \cdots & 0 & 0 & \cdots & 0 \\
0 & 0 & -1 & -1 & -1 & 0 & 0 & 0 & 0 & 0 & -1 & 0 & 0 & 0 & 0 & 0 & 0 & 0 & 0 & \cdots & 0 & 0 & \cdots & 0 \\
0 & 0 & -1 & -1 & 0 & -1 & -1 & 0 & 0 & 0 & 0 & 0 & 0 & 0 & 0 & 0 & 0 & 0 & 0 & \cdots & 0 & 0 & \cdots & 0 \\
0 & 0 & -1 & 0 & 0 & 0 & 0 & 0 & 0 & 0 & -1 & 0 & 0 & 0 & 0 & 0 & 0 & 0 & 0 & \cdots & 0 & 0 & \cdots & 0 \\
0 & 0 & -1 & 0 & 0 & 0 & 0 & 0 & 0 & 0 & -1 & 0 & 0 & 0 & 0 & 0 & 0 & 0 & -1 & \cdots & 0 & 0 & \cdots & 0 \\
0 & 0 & -1 & -1 & 0 & 0 & -1 & 0 & 0 & 0 & 0 & 0 & 0 & 0 & 0 & -1 & 0 & 0 & 0 & \cdots & 0 & 0 & \cdots & 0 \\
0 & 0 & -1 & -1 & 0 & 0 & -1 & 0 & 0 & 0 & 0 & 0 & 0 & 0 & 0 & -1 & 0 & 0 & 0 & \cdots & 0 & 0 & \cdots & 0 \\
0 & 0 & -1 & 0 & 0 & 0 & 0 & 0 & 0 & 0 & 0 & 0 & 0 & 0 & 0 & -1 & 0 & 0 & 0 & \cdots & 0 & 0 & \cdots & 0 \\
0 & 0 & -1 & 0 & 0 & 0 & 0 & 0 & 0 & 0 & 0 & 0 & 0 & 0 & 0 & -1 & 0 & 0 & -1 & \cdots & 0 & 0 & \cdots & 0 \\
0 & 0 & -1 & -1 & 0 & 0 & 0 & 0 & 0 & 0 & 0 & 0 & -1 & 0 & 0 & -1 & 0 & 0 & 0 & \cdots & 0 & 0 & \cdots & 0 \\
0 & 0 & 0 & -1 & -1 & -1 & -1 & -1 & 0 & 0 & 0 & 0 & 2 & 0 & 0 & 1 & 0 & 0 & 0 & \cdots & 0 & 0 & \cdots & 0 \\
0 & 0 & 0 & 0 & -1 & 0 & -1 & -1 & 1 & 0 & 0 & 0 & 2 & 0 & 0 & 0 & 0 & 0 & 0 & \cdots & 0 & 0 & \cdots & 0 \\
0 & 0 & 0 & 0 & -1 & 0 & 0 & -1 & 0 & 0 & 0 & 0 & 1 & 0 & 1 & 0 & 0 & 0 & 0 & \cdots & 0 & 0 & \cdots & 0 \\
0 & 0 & 0 & -1 & -1 & 0 & -1 & 0 & 0 & 0 & 0 & 0 & 1 & 0 & 0 & 0 & 0 & 0 & 0 & \cdots & 0 & 0 & \cdots & 0 \\
0 & 0 & 0 & 0 & -1 & 0 & -1 & 0 & 0 & 0 & 0 & 0 & 1 & 1 & 0 & 0 & 0 & 0 & 0 & \cdots & 0 & 0 & \cdots & 0 \\
0 & 0 & 0 & 0 & 0 & -1 & -1 & 0 & 0 & 0 & -1 & 1 & 1 & 0 & 0 & 1 & 0 & 0 & -1 & \cdots & 0 & 0 & \cdots & 0 \\
0 & 0 & 0 & 0 & 0 & -1 & 0 & 0 & 0 & 0 & -1 & 0 & 0 & 0 & 0 & 1 & 0 & 1 & 0 & \cdots & 0 & 0 & \cdots & 0 \\
0 & 0 & 0 & 0 & 0 & -1 & -1 & -1 & 0 & 0 & 0 & 0 & 2 & 0 & 0 & 1 & 0 & 0 & -1 & \cdots & 0 & 0 & \cdots & 0 \\
0 & 0 & 0 & -1 & 0 & -1 & -1 & -1 & 0 & 0 & 0 & 0 & 1 & 0 & 0 & 1 & 0 & 0 & 0 & \cdots & 0 & 0 & \cdots & 0 \\
0 & 0 & 0 & 0 & 0 & -1 & 0 & -1 & 0 & 0 & 0 & 0 & 1 & 0 & 0 & 1 & 0 & 0 & 0 & \cdots & 0 & 0 & \cdots & 0 \\
0 & 0 & 0 & 0 & 0 & -1 & -1 & 0 & 0 & 0 & 0 & 0 & 1 & 0 & 0 & 0 & 1 & 0 & -1 & \cdots & 0 & 0 & \cdots & 0 \\
0 & 1 & 0 & 0 & 0 & 0 & -1 & 0 & 0 & 0 & -1 & 0 & 1 & 0 & 0 & 0 & 0 & 0 & -1 & \cdots & 0 & 0 & \cdots & 0 \\
1 & 0 & 0 & 0 & 0 & 0 & -1 & -1 & 0 & 0 & 0 & 0 & 1 & 0 & 0 & 0 & 0 & 0 & 0 & \cdots & 0 & 0 & \cdots & 0 \\
0 & 0 & 0 & 0 & 0 & 0 & 0 & -1 & 0 & 0 & 0 & 0 & 1 & 0 & 0 & 0 & 0 & 0 & 0 & \cdots & 0 & 0 & \cdots & 0 \\
0 & 0 & 0 & 0 & 0 & 0 & 0 & -1 & 0 & 0 & 0 & 0 & 1 & 0 & 0 & 0 & 0 & 0 & -1 & \cdots & 0 & 0 & \cdots & 0 \\
0 & 0 & 0 & 0 & 0 & 0 & -1 & 0 & 0 & 0 & 0 & 0 & 1 & 0 & 0 & 0 & 0 & 0 & -1 & \cdots & 0 & 0 & \cdots & 0 \\
0 & 0 & 0 & 0 & 0 & 0 & -1 & 0 & 0 & 1 & 0 & 0 & 1 & 0 & 0 & -1 & 0 & 0 & -1 & \cdots & 0 & 0 & \cdots & 0 \\
0 & 0 & 0 & -1 & 0 & 0 & -1 & 0 & 0 & 0 & 0 & 0 & 0 & 0 & 0 & 0 & 0 & 0 & 0 & \cdots & 0 & 1 & \cdots & 0 \\
\end{array}
\right)
~,~
$}
\nn\\
\eea
and the corresponding $D$-term charge matrix is given by, 
\beal{es04k05}
&&
Q_{D}=
\nn\\
&&
\resizebox{0.9\textwidth}{!}{$
\left(
\begin{array}{cccccc|ccc|ccc|ccc|ccc|ccc|ccc}
p_{1} & p_{2} & p_{3} & p_{4} & p_{5} & p_{6} & q^{(1)}_{1} & q^{(1)}_{2} & q^{(1)}_{3} & q^{(2)}_{1} & q^{(2)}_{2} & q^{(2)}_{3} & q^{(3)}_{1} & q^{(3)}_{2} & q^{(3)}_{3} & q^{(4)}_{1} & q^{(4)}_{2} & q^{(4)}_{3} & o^{(1)}_{1} & \cdots & o^{(1)}_{12} & o^{(2)}_{1} & \cdots & o^{(2)}_{12} \\
\hline
 0 & 0 & 0 & 0 & 1 & 0 & 0 & 0 & 0 & 0 & 1 & 0 & -1 & 0 & 0 & -1 & 0 & 0 & 0 & \cdots & 0 & 0 & \cdots & 0 \\
 0 & 0 & 0 & -1 & -1 & 0 & -1 & 0 & 0 & 0 & 0 & 0 & 1 & 0 & 0 & 0 & 0 & 0 & 0 & \cdots & 0 & 0 & \cdots & 0 \\
 0 & 0 & 0 & 1 & 0 & 0 & 0 & 0 & 0 & 0 & 0 & 0 & 1 & 0 & 0 & 0 & 0 & 0 & 0 & \cdots & 0 & 0 & \cdots & 0 \\
 0 & 0 & 0 & 0 & 0 & 0 & 0 & 0 & 0 & 0 & 0 & 0 & 0 & 0 & 0 & 0 & 0 & 0 & 0 & \cdots & 0 & 0 & \cdots & 0 \\
 0 & 0 & 0 & 0 & 0 & 1 & 1 & 0 & 0 & 0 & 0 & 0 & -1 & 0 & 0 & -1 & 0 & 0 & 0 & \cdots & 0 & 0 & \cdots & 0 \\
 0 & 0 & 0 & 0 & 0 & 0 & 0 & 0 & 0 & 0 & 0 & 0 & 0 & 0 & 0 & 0 & 0 & 0 & 0 & \cdots & 0 & 0 & \cdots & 0 \\
 0 & 0 & 0 & 0 & 0 & -1 & 0 & -1 & 0 & 0 & 0 & 0 & 1 & 0 & 0 & 1 & 0 & 0 & 0 & \cdots & 0 & 0 & \cdots & 0 \\
 0 & 0 & 1 & 0 & 0 & 0 & 0 & 0 & 0 & 0 & 0 & 0 & 0 & 0 & 0 & 1 & 0 & 0 & 0 & \cdots & 0 & 0 & \cdots & 0 \\
\end{array}
\right)
$}~.~
\nn\\
\eea

\begin{figure}[ht!!]
\centering
\includegraphics[width=0.25\textwidth]{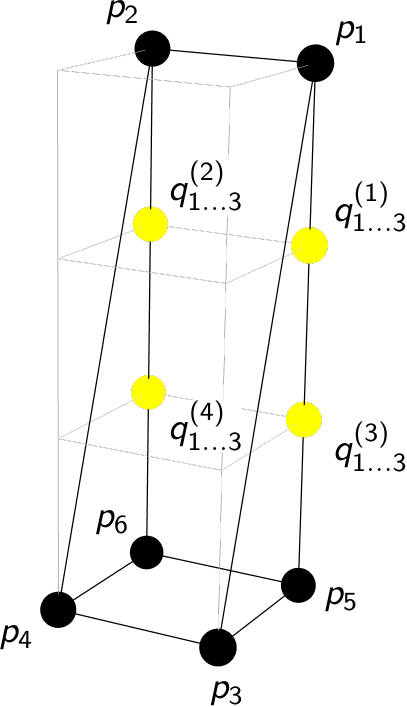}
\caption{
The toric diagram for the $D_3/\mathbb{Z}_3 \ (0,1,2,0,0)$ model.
\label{fig_425_toric}
}
\end{figure}

The corresponding toric diagram is given by,
\beal{es04k06}
&&
G_{t}=
\nn\\
&&
\resizebox{0.9\textwidth}{!}{$
\left(
\begin{array}{cccccc|ccc|ccc|ccc|ccc|ccc|ccc}
p_{1} & p_{2} & p_{3} & p_{4} & p_{5} & p_{6} & q^{(1)}_{1} & q^{(1)}_{2} & q^{(1)}_{3} & q^{(2)}_{1} & q^{(2)}_{2} & q^{(2)}_{3} & q^{(3)}_{1} & q^{(3)}_{2} & q^{(3)}_{3} & q^{(4)}_{1} & q^{(4)}_{2} & q^{(4)}_{3} & o^{(1)}_{1} & \cdots & o^{(1)}_{12} & o^{(2)}_{1} & \cdots & o^{(2)}_{12} \\
\hline
 -1 & -1 & 2 & 2 & 2 & 2 & 0 & 0 & 0 & 0 & 0 & 0 & 1 & 1 & 1 & 1 & 1 & 1 & 3 & \cdots & 3 & 2 & \cdots & 2 \\
  0 &  0 & 1 & 1 & 0 & 0 & 0 & 0 & 0 & 0 & 0 & 0 & 0 & 0 & 0 & 0 & 0 & 0 & 1 & \cdots & 1 & 1 & \cdots & 1 \\
  0 & -1 & 0 & -1 & 0 & -1 & 0 & 0 & 0 & -1 & -1 & -1 & 0 & 0 & 0 & -1 & -1 & -1 & -1 & \cdots & -1 & -1 & \cdots & -1 \\
\hline
  1 &  1 & 1 & 1 & 1 & 1 & 1 & 1 & 1 & 1 & 1 & 1 & 1 & 1 & 1 & 1 & 1 & 1 & 2 & \cdots & 2 & 2 & \cdots & 2 \\
\end{array}
\right)
$}~,~
\nn\\
\eea
and is illustrated in \fref{fig_425_toric}.

%-------------------
%-------------------

%-------------------
\begin{table}[ht!!]
\centering
\begin{tabular}{|c|c|c|c|l|}
\hline
\; & $U(1)_{f_1}$ & $U(1)_{f_2}$ & $U(1)_{f_3}$ & fugacity \\
\hline
$p_1$ & $+1$ & $0$ & $0$ &  $t_1=f_1 t$ \\
$p_2$ & $0$ & $0$ & $+1$ &  $t_2=f_3 t$\\
$p_3$ & $0$ & $0$ & $-1$ &  $t_3=f_3^{-1} t$\\
$p_4$ & $0$ & $+1$ & $0$ & $t_4=f_2 t$ \\
$p_5$ & $0$ & $-1$ & $0$ & $t_5=f_2^{-1} t$ \\
$p_6$ & $-1$ & $0$ & $0$ & $t_6=f_1^{-1} t$ \\
\hline
\end{tabular}
\caption{
Mesonic flavor symmetry of the $D_3/\mathbb{Z}_3 \ (0,1,2,0,0)$ model
and charges on the extremal GLSM fields $p_a$. 
Here, the fugacity $t$ counts the degree in extremal GLSM fields $p_a$. 
\label{tab_04k01}}
\end{table}
%-------------------

The global symmetry of the $D_3/\mathbb{Z}_3 \ (0,1,2,0,0)$ model is not enhanced and takes the following form,
\beal{es04k07}
U(1)_{f_1} \times U(1)_{f_2} \times U(1)_{f_3} \times U(1)_R
~.~
\eea
The refined Hilbert series of the mesonic moduli space for the $D_3/\mathbb{Z}_3 \ (0,1,2,0,0)$ model is given by,
\beal{es04k08}
&&
g(t_a; \mathcal{M}^{mes})=
\frac{
\left(1 - t_1 t_2 t_3 t_4 t_5 t_6\right) \left(1-t_1^3 t_2^3 t_5^3 t_6^3\right)
}{
\left(1 - t_1^3 t_2^3\right) \left(1 - t_3 t_4\right) \left(1 - t_1 t_3  t_5\right) \left(1 - t_2 t_4 t_6\right) 
}
\nn\\
&&
\hspace{1cm}
\times
\frac{1}{
\left(1-t_1 t_2 t_5 t_6\right) \left(1 - t_5^3 t_6^3\right)
}
~,~
\eea
where $t_a$ is the fugacity corresponding to the extremal GLSM field $p_a$.
We can unrefine the Hilbert series by setting $t_a=t$, which leads to the following Hilbert series,
\beal{es04k09}
g(t; \mathcal{M}^{mes})
=
\frac{1+t^6}{(1-t^2) (1-t^3)^2 (1-t^4)}
~,~
\eea
where the palindromic numerator indicates that the mesonic moduli space is Calabi-Yau.

The Hilbert series refined in terms of the mesonic flavor symmetry fugacities summarized in \tref{tab_04k01} takes the following form, 
\beal{es04k10}
&&
g(f_1,f_2,f_3,t; \mathcal{M}^{mes})=
\frac{
(1-t^6)(1-f_2^{-3}f_3^3 t^{12})
}{
(1-f_2f_3^{-1} t^2)(1-f_1f_2^{-1}f_3^{-1} t^3)(1-f_1^{-1} f_2 f_3 t^3)
}
\nn\\
&&
\hspace{1cm}
\times
\frac{1}{
(1-f_2^{-1} f_3 t^4)(1-f_1^3 f_3^3 t^6)(1-f_1^{-3}f_2^{-3} t^6)
}
~.~
\nn\\
\eea
The corresponding plethystic logarithm of the Hilbert series has the following expansion,
\beal{es04k12}
&&
\text{PL}[g(f_1,f_2,f_3,t; \mathcal{M}^{mes})]=
f_2 f_3^{-1} t^2
+ ( f_1 f_2^{-1} f_3^{-1}+f_1^{-1} f_2 f_3 ) t^3
+ f_2^{-1} f_3 t^4
\nn\\
&&
\hspace{1cm}
+ ( f_1^{-3} f_2^{-3}+f_1^{3} f_3^{3}) t^6
- t^6
- f_2^{-3} f_3^3 t^{12}
~,~
\eea
where the finite expansion indicates that the mesonic moduli space of the $D_3/\mathbb{Z}_3 \ (0,1,2,0,0)$ model 
is a complete intersection.
The first positive terms in the expansion in \eref{es04k12}
correspond to the generators of the mesonic moduli space, which are summarized with their mesonic flavor charges in \tref{tab_04k02}.
The generators form 2 defining relations for $D_3/\mathbb{Z}_3 \ (0,1,2,0,0)$, allowing us to explicitly define the mesonic moduli space as follows, 
\beal{es04k13}
\mathcal{M}^{mes} = 
\mathrm{Spec}~\mathbb{C}[A, B_1, B_2, C_1, C_2, C_3]/ \langle A C_3 - B_1 B_2 ~,~ C_1 C_2 - C_3^3 \rangle~,~
\eea
where the generators $A, B_1, B_2, C_1, C_2, C_3$ are expressed in terms of GLSM fields in \tref{tab_04k02}.

We note here that the abelian orbifold $D_3/\mathbb{Z}_3 \ (0,1,2,0,0)$ is part of a larger family of abelian orbifolds of the form $D_3/\mathbb{Z}_n \ (0,1,-1,0,0)$
as originally described in section \sref{sec049}.
\\

%-------------------
 \begin {table}[ht!!]
\centering
\begin {tabular} {|c|c|ccc|}
\hline
PL term & generator & $U(1)_{f_1}$ & $U(1)_{f_2}$ & $U(1)_{f_3}$ 
\\
\hline
\multirow{1}{*}{$+ f_2 f_3^{-1} t^{2}$}
& $A = p_3 p_4 ~$ & $0$ & $1$ & $-1$  \\
\hline
\multirow{1}{*}{$+f_1^{-1} f_2 f_3 t^{3}$}
& $B_1 = p_2 p_4 p_6 ~q_{(2)} q_{{(4)}} $ & $-1$ & $1$ & $1$  \\
\hline
\multirow{1}{*}{$+f_1 f_2^{-1} f_3^{-1} t^{3}$}
& $B_2 = p_1 p_3 p_5 ~q_{(1)} q_{{(3)}} $ & $1$ & $-1$ & $-1$  \\
\hline
\multirow{1}{*}{$+f_1^{-3} f_2^{-3}  t^{6}$}
& $C_1 = p_5^3 p_6^3 ~q_{(1)} q_{(2)} q_{{(3)}}^2 q_{{(4)}}^2 $ & $-3$ & $-3$ & $0$ \\
\hline
\multirow{1}{*}{$+f_1^{3}  f_3^{3} t^{6}$}
& $C_2 = p_1^3 p_2^3 ~q_{(1)}^2 q_{(2)}^2 q_{{(3)}} q_{{(4)}} $ & $3$ & $0$ & $3$  \\
\hline
\multirow{1}{*}{$+ f_2^{-1} f_3 t^{4}$}
& $C_3 = p_1 p_2 p_5 p_6 ~q_{(1)} q_{(2)} q_{{(3)}} q_{{(4)}} $ & $0$ & $-1$ & $1$  \\
\hline
\end{tabular}
\caption{
Generators of the $D_3/\mathbb{Z}_3 \ (0,1,2,0,0)$ model in terms of GLSM fields and their corresponding mesonic flavor symmetry charges. Here, we denote $q_{(1)}=\prod_{i=1}^{3}q_i^{(1)}$, $q_{(2)}=\prod_{i=1}^{3}q_i^{(2)}$, $q_{(3)}=\prod_{i=1}^{3}q_i^{(3)}$, $q_{(4)}=\prod_{i=1}^{3}q_i^{(4)}$, and set extra GLSM fields to 1.
 \label{tab_04k02}}
\end{table}
%-------------------

%=================================================================
\subsubsection{$D_3/ \mathbb{Z}_3 \ (0,1,2,1,2)$ \label{sec0412}}

%-------------------
\begin{figure}[H]
    \centering
    \includegraphics[width=0.5\textwidth]{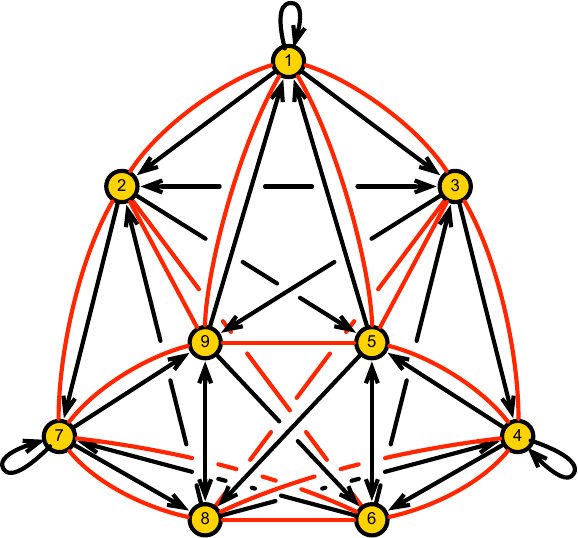}
    \caption{
    The quiver for the $D_3/ \mathbb{Z}_3 \ (0,1,2,1,2)$ model.
    \label{fig_426_quiver}
    }
\end{figure}
%-------------------

The $J$- and $E$-terms for the brane brick model corresponding to the abelian orbifold of the form 
$D_3/ \mathbb{Z}_3 \ (0,1,2,1,2)$ take the following form,
\beal{es04l01}
\begin{array}{rclccclcccc}
&& && &J& && &E& \\
&&\Lambda^{(1)}_{5 1} &:& & X_{1 3} \cdot X_{3 4} \cdot Y_{4 5} - Y_{1 2} \cdot X_{2 5}& && & D_{5 6} \cdot Z_{6 5} \cdot Z_{5 1} - Z_{5 1} \cdot D_{1 1}& \\ 
 &&\Lambda^{(1)}_{8 4} &:& & X_{4 6} \cdot X_{6 7} \cdot Y_{7 8} - Y_{4 5} \cdot X_{5 8}& && & D_{8 9} \cdot Z_{9 8} \cdot Z_{8 4} - Z_{8 4} \cdot D_{4 4}& \\ 
 &&\Lambda^{(1)}_{2 7} &:& & X_{7 9} \cdot X_{9 1} \cdot Y_{1 2} - Y_{7 8} \cdot X_{8 2}& && & D_{2 3} \cdot Z_{3 2} \cdot Z_{2 7} - Z_{2 7} \cdot D_{7 7}& \\ 
 &&\Lambda^{(2)}_{1 2} &:& & Z_{2 7} \cdot X_{7 9} \cdot X_{9 1} - X_{2 5} \cdot Z_{5 1}& && & D_{1 1} \cdot Y_{1 2} - Y_{1 2} \cdot D_{2 3} \cdot Z_{3 2}& \\ 
 &&\Lambda^{(2)}_{4 5} &:& & Z_{5 1} \cdot X_{1 3} \cdot X_{3 4} - X_{5 8} \cdot Z_{8 4}& && & D_{4 4} \cdot Y_{4 5} - Y_{4 5} \cdot D_{5 6} \cdot Z_{6 5}& \\ 
 &&\Lambda^{(2)}_{7 8} &:& & Z_{8 4} \cdot X_{4 6} \cdot X_{6 7} - X_{8 2} \cdot Z_{2 7}& && & D_{7 7} \cdot Y_{7 8} - Y_{7 8} \cdot D_{8 9} \cdot Z_{9 8}& \\ 
 &&\Lambda^{(3)}_{9 1} &:& & X_{1 3} \cdot Y_{3 9} - Y_{1 2} \cdot Z_{2 7} \cdot X_{7 9}& && & X_{9 1} \cdot D_{1 1} - Z_{9 8} \cdot D_{8 9} \cdot X_{9 1}& \\ 
 &&\Lambda^{(3)}_{3 4} &:& & X_{4 6} \cdot Y_{6 3} - Y_{4 5} \cdot Z_{5 1} \cdot X_{1 3}& && & X_{3 4} \cdot D_{4 4} - Z_{3 2} \cdot D_{2 3} \cdot X_{3 4}& \\ 
 &&\Lambda^{(3)}_{6 7} &:& & X_{7 9} \cdot Y_{9 6} - Y_{7 8} \cdot Z_{8 4} \cdot X_{4 6}& && & X_{6 7} \cdot D_{7 7} - Z_{6 5} \cdot D_{5 6} \cdot X_{6 7}& \\ 
 &&\Lambda^{(4)}_{1 3} &:& & X_{3 4} \cdot Y_{4 5} \cdot Z_{5 1} - Y_{3 9} \cdot X_{9 1}& && & D_{1 1} \cdot X_{1 3} - X_{1 3} \cdot Z_{3 2} \cdot D_{2 3}& \\ 
 &&\Lambda^{(4)}_{4 6} &:& & X_{6 7} \cdot Y_{7 8} \cdot Z_{8 4} - Y_{6 3} \cdot X_{3 4}& && & D_{4 4} \cdot X_{4 6} - X_{4 6} \cdot Z_{6 5} \cdot D_{5 6}& \\ 
 &&\Lambda^{(4)}_{7 9} &:& & X_{9 1} \cdot Y_{1 2} \cdot Z_{2 7} - Y_{9 6} \cdot X_{6 7}& && & D_{7 7} \cdot X_{7 9} - X_{7 9} \cdot Z_{9 8} \cdot D_{8 9}& \\ 
 &&\Lambda^{(5)}_{8 3} &:& & Y_{3 9} \cdot Z_{9 8} - Z_{3 2} \cdot Z_{2 7} \cdot Y_{7 8}& && & D_{8 9} \cdot X_{9 1} \cdot X_{1 3} - X_{8 2} \cdot D_{2 3}& \\ 
 &&\Lambda^{(5)}_{2 6} &:& & Y_{6 3} \cdot Z_{3 2} - Z_{6 5} \cdot Z_{5 1} \cdot Y_{1 2}& && & D_{2 3} \cdot X_{3 4} \cdot X_{4 6} - X_{2 5} \cdot D_{5 6}& \\ 
 &&\Lambda^{(5)}_{5 9} &:& & Y_{9 6} \cdot Z_{6 5} - Z_{9 8} \cdot Z_{8 4} \cdot Y_{4 5}& && & D_{5 6} \cdot X_{6 7} \cdot X_{7 9} - X_{5 8} \cdot D_{8 9}& \\ 
 &&\Lambda^{(6)}_{5 3} &:& & Z_{3 2} \cdot X_{2 5} - X_{3 4} \cdot X_{4 6} \cdot Z_{6 5}& && & D_{5 6} \cdot Y_{6 3} - Z_{5 1} \cdot Y_{1 2} \cdot D_{2 3}& \\ 
 &&\Lambda^{(6)}_{8 6} &:& & Z_{6 5} \cdot X_{5 8} - X_{6 7} \cdot X_{7 9} \cdot Z_{9 8}& && & D_{8 9} \cdot Y_{9 6} - Z_{8 4} \cdot Y_{4 5} \cdot D_{5 6}& \\ 
 &&\Lambda^{(6)}_{2 9} &:& & Z_{9 8} \cdot X_{8 2} - X_{9 1} \cdot X_{1 3} \cdot Z_{3 2}& && & D_{2 3} \cdot Y_{3 9} - Z_{2 7} \cdot Y_{7 8} \cdot D_{8 9}&
 \end{array}
 ~.~
\eea
The associated quiver is shown in \fref{fig_426_quiver}.
The $J$- and $E$-terms above are obtained from the general formula in \eref{es03c07}
with the following additional relabelling of indices, 
\beal{es04l02}
&
[1,0] \rightarrow 1~,~  [2,0] \rightarrow 2~,~  [3,0] \rightarrow 3~,~ 
&
\nn\\
&
[1,1] \rightarrow 4~,~  [2,1] \rightarrow 5~,~  [3,1] \rightarrow 6~,~ 
&
\nn\\
&
[1,2] \rightarrow 7~,~  [2,2] \rightarrow 8~,~  [3,2] \rightarrow 9~.~
&
\eea

Using the forward algorithm for brane brick models, we
obtain the $P$-matrix as follows, 
\beal{es04l03}
&&
P=
\nn\\
&&
\resizebox{0.95\textwidth}{!}{$
\left(
\begin{array}{c|cccccc|cccccc|cccccc|ccc|ccc|ccc|ccc}
 & p_{1} & p_{2} & p_{3} & p_{4} & p_{5} & p_{6} & q^{(1)}_{1} & q^{(1)}_{2} & q^{(1)}_{3} & q^{(1)}_{4} & q^{(1)}_{5} & q^{(1)}_{6} & q^{(2)}_{1} & q^{(2)}_{2} & q^{(2)}_{3} & q^{(2)}_{4} & q^{(2)}_{5} & q^{(2)}_{6} & o^{(1)}_{1} & \cdots & o^{(1)}_{3} & o^{(2)}_{1} & \cdots & o^{(2)}_{3} & o^{(3)}_{1} & \cdots & o^{(3)}_{9} & o^{(4)}_{1} & \cdots & o^{(4)}_{9} \\
\hline
 D_{11} & 0 & 1 & 0 & 1 & 0 & 0 & 0 & 0 & 0 & 0 & 0 & 0 & 0 & 0 & 0 & 0 & 0 & 0 & 1 & \cdots & 1 & 1 & \cdots & 1 & 1 & \cdots & 1 & 1 & \cdots & 1 \\
 D_{23} & 0 & 1 & 0 & 0 & 0 & 0 & 0 & 0 & 0 & 0 & 0 & 0 & 0 & 0 & 0 & 0 & 0 & 0 & 0 & \cdots & 1 & 0 & \cdots & 0 & 0 & \cdots & 1 & 1 & \cdots & 1 \\
 D_{44} & 0 & 1 & 0 & 1 & 0 & 0 & 0 & 0 & 0 & 0 & 0 & 0 & 0 & 0 & 0 & 0 & 0 & 0 & 1 & \cdots & 1 & 1 & \cdots & 1 & 1 & \cdots & 1 & 1 & \cdots & 1 \\
 D_{56} & 0 & 1 & 0 & 0 & 0 & 0 & 0 & 0 & 0 & 0 & 0 & 0 & 0 & 0 & 0 & 0 & 0 & 0 & 1 & \cdots & 1 & 0 & \cdots & 1 & 1 & \cdots & 1 & 0 & \cdots & 0 \\
 D_{77} & 0 & 1 & 0 & 1 & 0 & 0 & 0 & 0 & 0 & 0 & 0 & 0 & 0 & 0 & 0 & 0 & 0 & 0 & 1 & \cdots & 1 & 1 & \cdots & 1 & 1 & \cdots & 1 & 1 & \cdots & 1 \\
 D_{89} & 0 & 1 & 0 & 0 & 0 & 0 & 0 & 0 & 0 & 0 & 0 & 0 & 0 & 0 & 0 & 0 & 0 & 0 & 1 & \cdots & 0 & 1 & \cdots & 0 & 1 & \cdots & 0 & 1 & \cdots & 0 \\
 X_{13} & 0 & 0 & 0 & 0 & 0 & 1 & 1 & 0 & 0 & 0 & 0 & 0 & 1 & 0 & 0 & 0 & 0 & 0 & 0 & \cdots & 0 & 0 & \cdots & 0 & 0 & \cdots & 0 & 1 & \cdots & 1 \\
 X_{25} & 0 & 0 & 0 & 0 & 1 & 1 & 0 & 0 & 0 & 1 & 1 & 0 & 0 & 1 & 1 & 0 & 1 & 1 & 0 & \cdots & 1 & 0 & \cdots & 0 & 0 & \cdots & 0 & 1 & \cdots & 1 \\
 X_{34} & 0 & 0 & 0 & 0 & 1 & 0 & 0 & 0 & 0 & 1 & 0 & 0 & 0 & 1 & 0 & 0 & 1 & 1 & 1 & \cdots & 1 & 0 & \cdots & 0 & 1 & \cdots & 0 & 0 & \cdots & 0 \\
 X_{46} & 0 & 0 & 0 & 0 & 0 & 1 & 0 & 0 & 0 & 0 & 1 & 0 & 0 & 0 & 1 & 0 & 0 & 0 & 0 & \cdots & 0 & 0 & \cdots & 1 & 0 & \cdots & 0 & 0 & \cdots & 0 \\
 X_{58} & 0 & 0 & 0 & 0 & 1 & 1 & 0 & 0 & 1 & 0 & 0 & 1 & 1 & 0 & 0 & 1 & 1 & 1 & 1 & \cdots & 1 & 0 & \cdots & 1 & 0 & \cdots & 1 & 0 & \cdots & 1 \\
 X_{67} & 0 & 0 & 0 & 0 & 1 & 0 & 0 & 0 & 0 & 0 & 0 & 1 & 1 & 0 & 0 & 1 & 0 & 1 & 1 & \cdots & 0 & 0 & \cdots & 0 & 0 & \cdots & 0 & 0 & \cdots & 1 \\
 X_{79} & 0 & 0 & 0 & 0 & 0 & 1 & 0 & 0 & 1 & 0 & 0 & 0 & 0 & 0 & 0 & 0 & 1 & 0 & 0 & \cdots & 0 & 1 & \cdots & 0 & 0 & \cdots & 0 & 1 & \cdots & 0 \\
 X_{82} & 0 & 0 & 0 & 0 & 1 & 1 & 1 & 1 & 0 & 0 & 0 & 0 & 1 & 1 & 1 & 1 & 0 & 0 & 1 & \cdots & 0 & 1 & \cdots & 0 & 1 & \cdots & 0 & 1 & \cdots & 0 \\
 X_{91} & 0 & 0 & 0 & 0 & 1 & 0 & 0 & 1 & 0 & 0 & 0 & 0 & 0 & 1 & 1 & 1 & 0 & 0 & 0 & \cdots & 1 & 0 & \cdots & 0 & 0 & \cdots & 1 & 0 & \cdots & 0 \\
 Y_{12} & 1 & 0 & 0 & 0 & 0 & 0 & 1 & 0 & 0 & 0 & 0 & 0 & 1 & 0 & 0 & 0 & 0 & 0 & 1 & \cdots & 0 & 0 & \cdots & 0 & 1 & \cdots & 0 & 0 & \cdots & 0 \\
 Y_{39} & 1 & 0 & 1 & 0 & 0 & 0 & 0 & 0 & 1 & 1 & 1 & 1 & 0 & 0 & 0 & 0 & 1 & 1 & 1 & \cdots & 0 & 1 & \cdots & 1 & 1 & \cdots & 0 & 0 & \cdots & 0 \\
 Y_{45} & 1 & 0 & 0 & 0 & 0 & 0 & 0 & 0 & 0 & 0 & 1 & 0 & 0 & 0 & 1 & 0 & 0 & 0 & 0 & \cdots & 0 & 0 & \cdots & 0 & 0 & \cdots & 0 & 0 & \cdots & 0 \\
 Y_{63} & 1 & 0 & 1 & 0 & 0 & 0 & 1 & 1 & 1 & 0 & 0 & 1 & 1 & 0 & 0 & 1 & 0 & 0 & 0 & \cdots & 0 & 1 & \cdots & 0 & 0 & \cdots & 1 & 1 & \cdots & 1 \\
 Y_{78} & 1 & 0 & 0 & 0 & 0 & 0 & 0 & 0 & 1 & 0 & 0 & 0 & 0 & 0 & 0 & 0 & 1 & 0 & 0 & \cdots & 1 & 0 & \cdots & 0 & 0 & \cdots & 1 & 0 & \cdots & 0 \\
 Y_{96} & 1 & 0 & 1 & 0 & 0 & 0 & 1 & 1 & 0 & 1 & 1 & 0 & 0 & 1 & 1 & 0 & 0 & 0 & 0 & \cdots & 1 & 0 & \cdots & 1 & 1 & \cdots & 1 & 0 & \cdots & 0 \\
 Z_{27} & 0 & 0 & 1 & 0 & 0 & 0 & 0 & 0 & 0 & 1 & 1 & 1 & 0 & 0 & 0 & 0 & 0 & 1 & 0 & \cdots & 0 & 0 & \cdots & 1 & 0 & \cdots & 0 & 0 & \cdots & 1 \\
 Z_{32} & 0 & 0 & 0 & 1 & 0 & 0 & 0 & 0 & 0 & 0 & 0 & 0 & 0 & 0 & 0 & 0 & 0 & 0 & 1 & \cdots & 0 & 1 & \cdots & 1 & 1 & \cdots & 0 & 0 & \cdots & 0 \\
 Z_{51} & 0 & 0 & 1 & 0 & 0 & 0 & 0 & 1 & 1 & 0 & 0 & 1 & 0 & 0 & 0 & 1 & 0 & 0 & 0 & \cdots & 0 & 1 & \cdots & 1 & 0 & \cdots & 1 & 0 & \cdots & 0 \\
 Z_{65} & 0 & 0 & 0 & 1 & 0 & 0 & 0 & 0 & 0 & 0 & 0 & 0 & 0 & 0 & 0 & 0 & 0 & 0 & 0 & \cdots & 0 & 1 & \cdots & 0 & 0 & \cdots & 0 & 1 & \cdots & 1 \\
 Z_{84} & 0 & 0 & 1 & 0 & 0 & 0 & 1 & 1 & 0 & 1 & 0 & 0 & 0 & 1 & 0 & 0 & 0 & 0 & 0 & \cdots & 0 & 1 & \cdots & 0 & 1 & \cdots & 0 & 1 & \cdots & 0 \\
 Z_{98} & 0 & 0 & 0 & 1 & 0 & 0 & 0 & 0 & 0 & 0 & 0 & 0 & 0 & 0 & 0 & 0 & 0 & 0 & 0 & \cdots & 1 & 0 & \cdots & 1 & 0 & \cdots & 1 & 0 & \cdots & 1 \\
\end{array}
\right)
$}
~,~
\nn\\
\eea
where $o_{k}^{(1)}$, $o_{l}^{(2)}$, $o_{m}^{(3)}$ and $o_{n}^{(4)}$ correspond to extra GLSM fields \cite{Witten:1993yc}.
The $J$- and $E$-terms charge matrix is as follows,
\beal{es04l04}
&&
Q_{JE}=
\nn\\
&&
\resizebox{0.9\textwidth}{!}{$
\left(
\begin{array}{cccccc|cccccc|cccccc|ccc|ccc|ccc|ccc}
p_{1} & p_{2} & p_{3} & p_{4} & p_{5} & p_{6} & q^{(1)}_{1} & q^{(1)}_{2} & q^{(1)}_{3} & q^{(1)}_{4} & q^{(1)}_{5} & q^{(1)}_{6} & q^{(2)}_{1} & q^{(2)}_{2} & q^{(2)}_{3} & q^{(2)}_{4} & q^{(2)}_{5} & q^{(2)}_{6} & o^{(1)}_{1} & \cdots & o^{(1)}_{3} & o^{(2)}_{1} & \cdots & o^{(2)}_{3} & o^{(3)}_{1} & \cdots & o^{(3)}_{9} & o^{(4)}_{1} & \cdots & o^{(4)}_{9} \\
\hline
0 & 0 & -1 & -1 & 0 & 0 & 1 & 0 & 0 & 0 & 0 & 0 & -1 & 0 & 0 & 0 & 0 & 0 & 0 & \cdots & 0 & 1 & \cdots & 0 & 0 & \cdots & 0 & -1 & \cdots & 1 \\
0 & 0 & -1 & -1 & 0 & 0 & 1 & 0 & 0 & 0 & 0 & 0 & -1 & 0 & 0 & 0 & 0 & 0 & 0 & \cdots & 0 & 1 & \cdots & 0 & 0 & \cdots & 0 & -1 & \cdots & 0 \\
0 & 0 & -1 & -1 & 0 & 0 & 1 & 0 & 0 & 0 & 0 & 0 & -1 & 0 & 0 & 0 & 0 & 0 & 0 & \cdots & 0 & 0 & \cdots & 1 & 0 & \cdots & 0 & 0 & \cdots & 0 \\
0 & 0 & -1 & -1 & 0 & 0 & 0 & 0 & 0 & 0 & 0 & 0 & 0 & 0 & 0 & 0 & 0 & 0 & 0 & \cdots & 0 & 1 & \cdots & 0 & 0 & \cdots & 0 & -1 & \cdots & 0 \\
-1 & 0 & 0 & -1 & 0 & 0 & 2 & -1 & 0 & 0 & 0 & 0 & -1 & 0 & 0 & 0 & 0 & 0 & 0 & \cdots & 1 & 1 & \cdots & 0 & -1 & \cdots & 0 & -1 & \cdots & 0 \\
-1 & -1 & 0 & -1 & 0 & 0 & 1 & -1 & 0 & 0 & 0 & 0 & -1 & 0 & 0 & 0 & 0 & 0 & 0 & \cdots & 0 & 0 & \cdots & 0 & 0 & \cdots & 1 & 0 & \cdots & 0 \\
-1 & 0 & 0 & -1 & 0 & 0 & 1 & -1 & 0 & 0 & 0 & 0 & 0 & 0 & 0 & 0 & 0 & 0 & 0 & \cdots & 0 & 1 & \cdots & 0 & 0 & \cdots & 0 & -1 & \cdots & 0 \\
-1 & 0 & 0 & -1 & 0 & 0 & 1 & 0 & 0 & 0 & 0 & 0 & -1 & 0 & 0 & 0 & 0 & 0 & 0 & \cdots & 0 & 0 & \cdots & 0 & 0 & \cdots & 0 & -1 & \cdots & 0 \\
-1 & 0 & 0 & -1 & 0 & 0 & 1 & 0 & 0 & 0 & 0 & 0 & -1 & 0 & 0 & 0 & 0 & 0 & 0 & \cdots & 0 & 0 & \cdots & 0 & -1 & \cdots & 0 & 0 & \cdots & 0 \\
-1 & -1 & 0 & -1 & 0 & 0 & 0 & 0 & 0 & 0 & 0 & 0 & -1 & 0 & 0 & 0 & 0 & 0 & 0 & \cdots & 0 & 0 & \cdots & 0 & 0 & \cdots & 0 & 0 & \cdots & 0 \\
0 & 0 & 0 & -1 & 0 & 0 & 1 & -1 & 0 & 0 & 0 & 0 & -1 & 0 & 0 & 0 & 0 & 0 & 0 & \cdots & 0 & 0 & \cdots & 0 & 0 & \cdots & 0 & 0 & \cdots & 0 \\
0 & 0 & 0 & -1 & 0 & 0 & 0 & 0 & 0 & 0 & 0 & 0 & -1 & 0 & 0 & 0 & 0 & 0 & 0 & \cdots & 0 & 0 & \cdots & 0 & 0 & \cdots & 0 & 0 & \cdots & 0 \\
0 & 1 & -1 & 0 & 0 & 0 & 2 & 0 & 0 & 0 & 0 & 0 & -1 & 0 & 0 & 0 & 0 & 1 & 0 & \cdots & 0 & 1 & \cdots & 0 & -1 & \cdots & 0 & -1 & \cdots & 0 \\
0 & 0 & -1 & 0 & 0 & 0 & 1 & 0 & 0 & 0 & 0 & 0 & 0 & 0 & 0 & 0 & 0 & 0 & 0 & \cdots & 0 & 1 & \cdots & 0 & 0 & \cdots & 0 & -1 & \cdots & 0 \\
0 & 0 & -1 & 0 & 0 & 0 & 1 & 0 & 0 & 0 & 0 & 1 & -1 & 0 & 0 & 0 & 0 & 0 & 0 & \cdots & 0 & 0 & \cdots & 0 & 0 & \cdots & 0 & 0 & \cdots & 0 \\
0 & 1 & -1 & 0 & 0 & 0 & 1 & 0 & 0 & 0 & 1 & 0 & 0 & 0 & 0 & 0 & 0 & 0 & 0 & \cdots & 0 & 1 & \cdots & 0 & 0 & \cdots & 0 & -1 & \cdots & 0 \\
0 & 1 & -1 & 0 & 0 & 0 & 1 & 0 & 0 & 1 & 0 & 0 & 0 & 0 & 0 & 0 & 0 & 0 & 0 & \cdots & 0 & 1 & \cdots & 0 & -1 & \cdots & 0 & -1 & \cdots & 0 \\
-1 & 1 & 0 & 0 & 0 & 0 & 2 & 0 & 0 & 0 & 0 & 0 & -1 & 0 & 0 & 0 & 1 & 0 & 0 & \cdots & 0 & 0 & \cdots & 0 & -1 & \cdots & 0 & -1 & \cdots & 0 \\
-1 & 0 & 0 & 0 & 0 & 0 & 1 & 0 & 0 & 0 & 0 & 0 & 0 & 0 & 0 & 0 & 0 & 0 & 0 & \cdots & 0 & 0 & \cdots & 0 & 0 & \cdots & 0 & -1 & \cdots & 0 \\
-1 & 0 & 0 & 0 & 0 & 0 & 1 & 0 & 1 & 0 & 0 & 0 & -1 & 0 & 0 & 0 & 0 & 0 & 0 & \cdots & 0 & -1 & \cdots & 0 & 0 & \cdots & 0 & 0 & \cdots & 0 \\
0 & 1 & 0 & 0 & 1 & 0 & 2 & -1 & 0 & 0 & 0 & 0 & -1 & 0 & 0 & 0 & 0 & 0 & 0 & \cdots & 0 & 1 & \cdots & 0 & -1 & \cdots & 0 & -1 & \cdots & 0 \\
0 & 0 & 0 & 0 & 0 & 0 & 1 & -1 & 0 & 0 & 0 & 0 & 0 & 0 & 0 & 0 & 0 & 0 & 0 & \cdots & 0 & 1 & \cdots & 0 & 0 & \cdots & 0 & -1 & \cdots & 0 \\
0 & 0 & 0 & 0 & 0 & 0 & 1 & -1 & 0 & 0 & 0 & 0 & -1 & 0 & 0 & 1 & 0 & 0 & 0 & \cdots & 0 & 0 & \cdots & 0 & 0 & \cdots & 0 & 0 & \cdots & 0 \\
0 & 1 & 0 & 0 & 0 & 0 & 1 & -1 & 0 & 0 & 0 & 0 & 0 & 0 & 1 & 0 & 0 & 0 & 0 & \cdots & 0 & 1 & \cdots & 0 & 0 & \cdots & 0 & -1 & \cdots & 0 \\
0 & 1 & 0 & 0 & 0 & 0 & 1 & -1 & 0 & 0 & 0 & 0 & 0 & 1 & 0 & 0 & 0 & 0 & 0 & \cdots & 0 & 1 & \cdots & 0 & -1 & \cdots & 0 & -1 & \cdots & 0 \\
0 & 1 & 0 & 0 & 0 & 1 & 1 & 0 & 0 & 0 & 0 & 0 & -1 & 0 & 0 & 0 & 0 & 0 & 0 & \cdots & 0 & 0 & \cdots & 0 & 0 & \cdots & 0 & -1 & \cdots & 0 \\
0 & 0 & 0 & 0 & 0 & 0 & 1 & 0 & 0 & 0 & 0 & 0 & -1 & 0 & 0 & 0 & 0 & 0 & 0 & \cdots & 0 & -1 & \cdots & 0 & 0 & \cdots & 0 & 0 & \cdots & 0 \\
0 & 1 & 0 & 0 & 0 & 0 & 1 & 0 & 0 & 0 & 0 & 0 & 0 & 0 & 0 & 0 & 0 & 0 & 0 & \cdots & 0 & 0 & \cdots & 0 & 0 & \cdots & 0 & -1 & \cdots & 0 \\
0 & 1 & 0 & 0 & 0 & 0 & 1 & 0 & 0 & 0 & 0 & 0 & 0 & 0 & 0 & 0 & 0 & 0 & 0 & \cdots & 0 & 0 & \cdots & 0 & -1 & \cdots & 0 & -1 & \cdots & 0 \\
0 & 0 & 0 & 0 & 0 & 0 & 1 & 0 & 0 & 0 & 0 & 0 & -1 & 0 & 0 & 0 & 0 & 0 & 1 & \cdots & 0 & 0 & \cdots & 0 & -1 & \cdots & 0 & 0 & \cdots & 0 \\
\end{array}
\right)
$}~,~
\nn\\
\eea
and the $D$-term charge matrix takes the following form,
\beal{es04l05}
&&
Q_{D}=
\nn\\
&&
\resizebox{0.9\textwidth}{!}{$
\left(
\begin{array}{cccccc|cccccc|cccccc|ccc|ccc|ccc|ccc}
p_{1} & p_{2} & p_{3} & p_{4} & p_{5} & p_{6} & q^{(1)}_{1} & q^{(1)}_{2} & q^{(1)}_{3} & q^{(1)}_{4} & q^{(1)}_{5} & q^{(1)}_{6} & q^{(2)}_{1} & q^{(2)}_{2} & q^{(2)}_{3} & q^{(2)}_{4} & q^{(2)}_{5} & q^{(2)}_{6} & o^{(1)}_{1} & \cdots & o^{(1)}_{3} & o^{(2)}_{1} & \cdots & o^{(2)}_{3} & o^{(3)}_{1} & \cdots & o^{(3)}_{9} & o^{(4)}_{1} & \cdots & o^{(4)}_{9} \\
\hline
 0 & 0 & 0 & 0 & 0 & 0 & -1 & 1 & 0 & 0 & 0 & 0 & 0 & 0 & 0 & 0 & 0 & 0 & 0 & \cdots & 0 & 0 & \cdots & 0 & 0 & \cdots & 0 & 0 & \cdots & 0 \\
 0 & 0 & -1 & 0 & 0 & 0 & 1 & 0 & 0 & 0 & 0 & 0 & 0 & 0 & 0 & 0 & 0 & 0 & 0 & \cdots & 0 & 1 & \cdots & 0 & 0 & \cdots & 0 & -1 & \cdots & 0 \\
 0 & 1 & 0 & 0 & 0 & 0 & 1 & 0 & 0 & 0 & 0 & 0 & 0 & 0 & 0 & 0 & 0 & 0 & 0 & \cdots & 0 & 0 & \cdots & 0 & -1 & \cdots & 0 & 0 & \cdots & 0 \\
 0 & 0 & 0 & 0 & 0 & 0 & 0 & 0 & 0 & 0 & 0 & 0 & 0 & 0 & 0 & 0 & 0 & 0 & 0 & \cdots & 0 & 0 & \cdots & 0 & 0 & \cdots & 0 & 0 & \cdots & 0 \\
 0 & -1 & 0 & 0 & 0 & 0 & 0 & 0 & 0 & 0 & 0 & 0 & -1 & 0 & 0 & 0 & 0 & 0 & 0 & \cdots & 0 & -1 & \cdots & 0 & 0 & \cdots & 0 & 1 & \cdots & 0 \\
 0 & 0 & 0 & 0 & 0 & 0 & 0 & 0 & 0 & 0 & 0 & 0 & 0 & 0 & 0 & 0 & 0 & 0 & 0 & \cdots & 0 & 0 & \cdots & 0 & 0 & \cdots & 0 & 0 & \cdots & 0 \\
 -1 & 0 & 1 & 0 & 0 & 0 & 0 & 0 & 0 & 0 & 0 & 0 & 0 & 0 & 0 & 0 & 0 & 0 & 0 & \cdots & 0 & -1 & \cdots & 0 & 0 & \cdots & 0 & 0 & \cdots & 0 \\
 1 & 0 & 0 & 1 & 0 & 0 & -1 & 0 & 0 & 0 & 0 & 0 & 1 & 0 & 0 & 0 & 0 & 0 & 0 & \cdots & 0 & 0 & \cdots & 0 & 0 & \cdots & 0 & 0 & \cdots & 0 \\
\end{array}
\right)
$}
~.~
\nn\\
\eea

\begin{figure}[ht!!]
\centering
\includegraphics[width=0.3\textwidth]{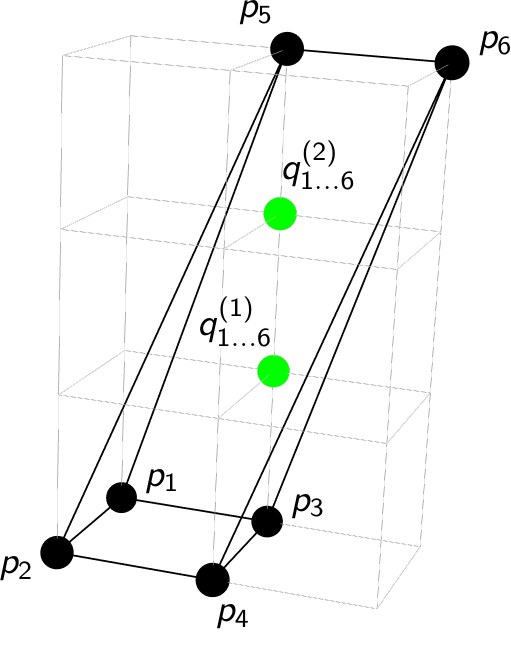}
\caption{
The toric diagram for the
$D_3/\mathbb{Z}_3 \ (0,1,2,1,2)$ model.
\label{fig_426_toric}
}
\end{figure}

The resulting toric diagram for the brane brick model is given by, 
\beal{es04l06}
&&
G_{t}=
\resizebox{0.8\textwidth}{!}{$
\left(
\begin{array}{cccccc|ccc|ccc|ccc|ccc|ccc|ccc}
p_{1} & p_{2} & p_{3} & p_{4} & p_{5} & p_{6} & q^{(1)}_{1} & \cdots & q^{(1)}_{6} & q^{(2)}_{1} & \cdots & q^{(2)}_{6} & o^{(1)}_{1} & \cdots & o^{(1)}_{3} & o^{(2)}_{1} & \cdots & o^{(2)}_{3} & o^{(3)}_{1} & \cdots & o^{(3)}_{9} & o^{(4)}_{1} & \cdots & o^{(4)}_{9} \\
\hline
1 & 0 & 1 & 0 & 1 & 1 &  1 & \cdots &  1 &  1 & \cdots &  1 &  1 & \cdots &  1 &  1 & \cdots &  1 &  1 & \cdots &  1 &  1 & \cdots &  1 \\
0 & 0 & 0 & 0 & -3 & -3 & -1 & \cdots & -1 & -2 & \cdots & -2 & -2 & \cdots & -2 & -1 & \cdots & -1 & -1 & \cdots & -1 & -2 & \cdots & -2 \\
1 & 1 & 0 & 0 & 0 & -1 &  0 & \cdots &  0 &  0 & \cdots &  0 &  1 & \cdots &  1 &  0 & \cdots &  0 &  1 & \cdots &  1 &  0 & \cdots &  0 \\
\hline
1 & 1 & 1 & 1 & 1 & 1 &  1 & \cdots &  1 &  1 & \cdots &  1 &  2 & \cdots &  2 &  2 & \cdots &  2 &  2 & \cdots &  2 &  2 & \cdots &  2 \\
\end{array}
\right)
$}
~,~
\nn\\
\eea
and is illustrated in \fref{fig_426_toric}. 

%-------------------
%-------------------

%-------------------
\begin{table}[ht!!]
\centering
\begin{tabular}{|c|c|c|c|l|}
\hline
\; & $U(1)_{f_1}$ & $U(1)_{f_2}$ & $U(1)_{f_3}$ &  fugacity \\
\hline
$p_1$ & $+1$ & $0$ & $0$ &  $t_1=f_1 t$ \\
$p_2$ & $0$ & $0$ & $+1$ &  $t_2=f_3 t$\\
$p_3$ & $0$ & $0$ & $-1$ &  $t_3=f_3^{-1} t$\\
$p_4$ & $0$ & $+1$ & $0$ & $t_4=f_2 t$ \\
$p_5$ & $0$ & $-1$ & $0$ & $t_5=f_2^{-1} t$ \\
$p_6$ & $-1$ & $0$ & $0$ & $t_6=f_1^{-1} t$ \\
\hline
\end{tabular}
\caption{
Mesonic flavor symmetry of the $D_3/\mathbb{Z}_3 \ (0,1,2,1,2)$ model 
and charges on the extremal GLSM fields $p_a$. 
Here, the fugacity $t$ counts the degree in extremal GLSM fields $p_a$. 
\label{tab_04l01}}
\end{table}
%-------------------

The global symmetry of the $D_3/\mathbb{Z}_3 \ (0,1,2,1,2)$ model
is not enhanced and takes the following form,
\beal{es04l07}
U(1)_{f_1} \times U(1)_{f_2} \times U(1)_{f_3} \times U(1)_R
~.~
\eea
Refined using GLSM fugacities $t_a$, 
the Hilbert series for the mesonic moduli space of the $D_3/\mathbb{Z}_3 \ (0,1,2,1,2)$ model
takes the following form,
\beal{es04l08}
&&
g(t_a; \mathcal{M}^{mes})=
\frac{
P(t_a; \mathcal{M}^{mes})
\left(1 - t_1 t_2 t_3 t_4 t_5 t_6\right)
}{
\left(1 - t_1^3 t_3^3\right) \left(1 - t_2 t_4\right) \left(1 - t_1^3 t_2^3 t_5^3\right) \left(1 - t_3^3 t_4^3 t_6^3\right) \left(1 - t_5^3 t_6^3\right)
}~,~
\eea
where the numerator factor $P(t_a; \mathcal{M}^{mes})$ is presented in full in appendix \sref{appCa}.
When unrefined by setting $t_a = t$, the Hilbert series takes the following form,
\beal{es04l09}
g(t)=\frac{1-t-t^2+t^3 +t^4 +t^5 -t^6 -t^7 +t^8}{(1-t) (1-t^2)^2 (1-t^9)}
~,~
\eea
where the palindromic numerator indicates that the mesonic moduli space is Calabi-Yau.

The Hilbert series refined under mesonic flavor fugacities summarized in \tref{tab_04l01}
is given as follows, 
\beal{es04l10}
&&
g(f_1,f_2,f_3,t; \mathcal{M}^{mes})
=
\frac{
(1-t^6) P(f_1,f_2,f_3,t; \mathcal{M}^{mes})
}{
(1-f_2f_3 t^2) (1-f_1^{-3} f_2^{-3} t^6) (1-f_1^3 f_3^{-3} t^6) 
}
\nn\\
&&
\hspace{1cm}
\times
\frac{1}{
(1-f_1^{-3} f_2^3 f_3^{-3} t^9) (1-f_1^3 f_2^{-3} f_3^3 t^9)
}
~,~
\eea
where the numerator factor
$P(f_1,f_2,f_3,t; \mathcal{M}^{mes})$ is presented in full in appendix \sref{appCa}.
The corresponding plethystic logarithm takes the following form, 
\beal{es04l11}
&&
\text{PL}[g(f_1,f_2,f_3,t; \mathcal{M}^{mes})]=
f_2 f_3 t^2
+ f_2^{-1} f_3^{-1} t^4
+({f_1^{2} f_2^{-1}+f_1^{-2} f_3^{-1}}) t^5
\nn\\
&&
\hspace{1cm}
+({f_1^{-3} f_2^{-3}+f_1^{3} f_3^{-3}}) t^6
+({f_1 f_2 f_3^{-3}+f_1^{-1} f_2^{-3} f_3}) t^7
+({f_1^{-1} f_2^{2} f_3^{-3}+f_1 f_2^{-3} f_3^{2}}) t^8
\nn\\
&&
\hspace{1cm}
+( {f_1^{-3} f_2^{3} f_3^{-3}+f_1^{3} f_2^{-3} f_3^{3}} ) t^9
- f_2^{-1} f_3^{-1} t^{10}
-( {f_1^{-1} f_2^{-4}+f_1 f_3^{-4}} )t^{11}
+\dots
~,~
\eea
where the infinite expansion indicates that the mesonic moduli space is not a complete intersection.
The generators of the mesonic moduli space with their corresponding mesonic flavor charges are summarized in \tref{tab_04l02}.

%-------------------
 \begin {table}[ht!!]
\centering
\begin {tabular} {|c|c|ccc|}
\hline
PL term & generator & $U(1)_{f_1}$ & $U(1)_{f_2}$ & $U(1)_{f_3}$
\\
\hline
\multirow{1}{*}{$+f_1^{-3} f_2^{-3}  t^{6}$}
& $p_5^3 p_6^3 ~q_{(1)} q_{(2)}^2 $ & $-3$ & $-3$ & $0$  \\
\hline
\multirow{1}{*}{$+f_1^{-2}  f_3^{-1} t^{5}$}
& $p_3 p_4 p_5 p_6^2 ~q_{(1)} q_{(2)} $ & $-2$ & $0$ & $-1$  \\
\hline
\multirow{1}{*}{$+f_1^{-3} f_2^{3} f_3^{-3} t^{9}$}
& $p_3^3 p_4^3 p_6^3 ~q_{(1)}^2 q_{(2)} $ & $-3$ & $3$ & $-3$  \\
\hline
\multirow{1}{*}{$+ f_2 f_3 t^{2}$}
& $p_2 p_4 ~$ & $0$ & $1$ & $1$  \\
\hline
\multirow{1}{*}{$+ f_2^{-1} f_3^{-1} t^{4}$}
& $p_1 p_3 p_5 p_6 ~q_{(1)} q_{(2)} $ & $0$ & $-1$ & $-1$  \\
\hline
\multirow{1}{*}{$+f_1^{-1} f_2^{2} f_3^{-3} t^{8}$}
& $p_1 p_3^3 p_4^2 p_6^2 ~q_{(1)}^2 q_{(2)} $ & $-1$ & $2$ & $-3$  \\
\hline
\multirow{1}{*}{$+f_1^{-1} f_2^{-3} f_3 t^{7}$}
& $p_1 p_2 p_5^3 p_6^2 ~q_{(1)} q_{(2)}^2 $ & $-1$ & $-3$ & $1$  \\
\hline
\multirow{1}{*}{$+f_1 f_2 f_3^{-3} t^{7}$}
& $p_1^2 p_3^3 p_4 p_6 ~q_{(1)}^2 q_{(2)} $ & $1$ & $1$ & $-3$  \\
\hline
\multirow{1}{*}{$+f_1^{2} f_2^{-1}  t^{5}$}
& $p_1^2 p_2 p_3 p_5 ~q_{(1)} q_{(2)} $ & $2$ & $-1$ & $0$ \\
\hline
\multirow{1}{*}{$+f_1 f_2^{-3} f_3^{2} t^{8}$}
& $p_1^2 p_2^2 p_5^3 p_6 ~q_{(1)} q_{(2)}^2 $ & $1$ & $-3$ & $2$  \\
\hline
\multirow{1}{*}{$+f_1^{3}  f_3^{-3} t^{6}$}
& $p_1^3 p_3^3 ~q_{(1)}^2 q_{(2)} $ & $3$ & $0$ & $-3$  \\
\hline
\multirow{1}{*}{$+f_1^{3} f_2^{-3} f_3^{3} t^{9}$}
& $p_1^3 p_2^3 p_5^3 ~q_{(1)} q_{(2)}^2 $ & $3$ & $-3$ & $3$  \\
\hline
\end{tabular}
\caption{Generators of the $D_3/\mathbb{Z}_3 \ (0,1,2,1,2)$ model in terms of GLSM fields and their corresponding mesonic flavor symmetry charges. Here, we denote $q_{(1)}=\prod_{i=1}^{6}q_i^{(1)}$,  $q_{(2)}=\prod_{i=1}^{6} q_i^{(2)}$,
and set extra GLSM fields to 1.
 \label{tab_04l02}}
\end{table}
%-------------------

We note here that the abelian orbifold $D_3/\mathbb{Z}_3 \ (0,1,2,1,2)$
is part of a larger family of abelian orbifolds of the form $D_3/\mathbb{Z}_n \ (0,1,-1,1,-1)$
as discussed in section \sref{sec048}.
\\

%=================================================================
\subsubsection{$D_3/ \mathbb{Z}_3 \ (1,1,1,0,0)$ \label{sec0413}}

%-------------------
\begin{figure}[H]
    \centering
    \includegraphics[width=0.5\textwidth]{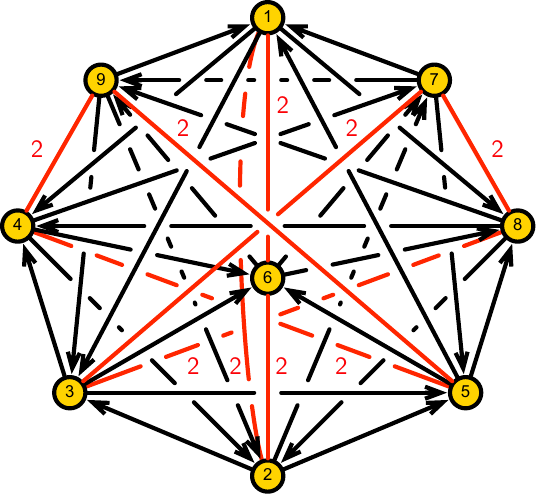}
    \caption{
    The quiver for the $D_3/ \mathbb{Z}_3 \ (1,1,1,0,0)$ model.
    \label{fig_427_quiver}
    }
\end{figure}
%-------------------

\fref{fig_427_quiver} illustrates the quiver for the $D_3/ \mathbb{Z}_2 \ (1,1,1,0,0)$ model.
The corresponding $J$- and $E$-terms take the following form, 
\beal{es04m01}
\begin{array}{rclccclcccc}
&& && &J& && &E& \\
&&\Lambda^{(1)}_{2 1} &:& & X_{1 3} \cdot X_{3 4} \cdot Y_{4 2} - Y_{1 8} \cdot X_{8 2}& && & D_{2 3} \cdot Z_{3 5} \cdot Z_{5 1} - Z_{2 7} \cdot D_{7 1}& \\ 
 &&\Lambda^{(1)}_{5 4} &:& & X_{4 6} \cdot X_{6 7} \cdot Y_{7 5} - Y_{4 2} \cdot X_{2 5}& && & D_{5 6} \cdot Z_{6 8} \cdot Z_{8 4} - Z_{5 1} \cdot D_{1 4}& \\ 
 &&\Lambda^{(1)}_{8 7} &:& & X_{7 9} \cdot X_{9 1} \cdot Y_{1 8} - Y_{7 5} \cdot X_{5 8}& && & D_{8 9} \cdot Z_{9 2} \cdot Z_{2 7} - Z_{8 4} \cdot D_{4 7}& \\ 
 &&\Lambda^{(2)}_{1 2} &:& & Z_{2 7} \cdot X_{7 9} \cdot X_{9 1} - X_{2 5} \cdot Z_{5 1}& && & D_{1 4} \cdot Y_{4 2} - Y_{1 8} \cdot D_{8 9} \cdot Z_{9 2}& \\ 
 &&\Lambda^{(2)}_{4 5} &:& & Z_{5 1} \cdot X_{1 3} \cdot X_{3 4} - X_{5 8} \cdot Z_{8 4}& && & D_{4 7} \cdot Y_{7 5} - Y_{4 2} \cdot D_{2 3} \cdot Z_{3 5}& \\ 
 &&\Lambda^{(2)}_{7 8} &:& & Z_{8 4} \cdot X_{4 6} \cdot X_{6 7} - X_{8 2} \cdot Z_{2 7}& && & D_{7 1} \cdot Y_{1 8} - Y_{7 5} \cdot D_{5 6} \cdot Z_{6 8}& \\ 
 &&\Lambda^{(3)}_{6 1} &:& & X_{1 3} \cdot Y_{3 6} - Y_{1 8} \cdot Z_{8 4} \cdot X_{4 6}& && & X_{6 7} \cdot D_{7 1} - Z_{6 8} \cdot D_{8 9} \cdot X_{9 1}& \\ 
 &&\Lambda^{(3)}_{9 4} &:& & X_{4 6} \cdot Y_{6 9} - Y_{4 2} \cdot Z_{2 7} \cdot X_{7 9}& && & X_{9 1} \cdot D_{1 4} - Z_{9 2} \cdot D_{2 3} \cdot X_{3 4}& \\ 
 &&\Lambda^{(3)}_{3 7} &:& & X_{7 9} \cdot Y_{9 3} - Y_{7 5} \cdot Z_{5 1} \cdot X_{1 3}& && & X_{3 4} \cdot D_{4 7} - Z_{3 5} \cdot D_{5 6} \cdot X_{6 7}& \\ 
 &&\Lambda^{(4)}_{7 3} &:& & X_{3 4} \cdot Y_{4 2} \cdot Z_{2 7} - Y_{3 6} \cdot X_{6 7}& && & D_{7 1} \cdot X_{1 3} - X_{7 9} \cdot Z_{9 2} \cdot D_{2 3}& \\ 
 &&\Lambda^{(4)}_{1 6} &:& & X_{6 7} \cdot Y_{7 5} \cdot Z_{5 1} - Y_{6 9} \cdot X_{9 1}& && & D_{1 4} \cdot X_{4 6} - X_{1 3} \cdot Z_{3 5} \cdot D_{5 6}& \\ 
 &&\Lambda^{(4)}_{4 9} &:& & X_{9 1} \cdot Y_{1 8} \cdot Z_{8 4} - Y_{9 3} \cdot X_{3 4}& && & D_{4 7} \cdot X_{7 9} - X_{4 6} \cdot Z_{6 8} \cdot D_{8 9}& \\ 
 &&\Lambda^{(5)}_{8 3} &:& & Y_{3 6} \cdot Z_{6 8} - Z_{3 5} \cdot Z_{5 1} \cdot Y_{1 8}& && & D_{8 9} \cdot X_{9 1} \cdot X_{1 3} - X_{8 2} \cdot D_{2 3}& \\ 
 &&\Lambda^{(5)}_{2 6} &:& & Y_{6 9} \cdot Z_{9 2} - Z_{6 8} \cdot Z_{8 4} \cdot Y_{4 2}& && & D_{2 3} \cdot X_{3 4} \cdot X_{4 6} - X_{2 5} \cdot D_{5 6}& \\ 
 &&\Lambda^{(5)}_{5 9} &:& & Y_{9 3} \cdot Z_{3 5} - Z_{9 2} \cdot Z_{2 7} \cdot Y_{7 5}& && & D_{5 6} \cdot X_{6 7} \cdot X_{7 9} - X_{5 8} \cdot D_{8 9}& \\ 
 &&\Lambda^{(6)}_{8 3} &:& & Z_{3 5} \cdot X_{5 8} - X_{3 4} \cdot X_{4 6} \cdot Z_{6 8}& && & D_{8 9} \cdot Y_{9 3} - Z_{8 4} \cdot Y_{4 2} \cdot D_{2 3}& \\ 
 &&\Lambda^{(6)}_{2 6} &:& & Z_{6 8} \cdot X_{8 2} - X_{6 7} \cdot X_{7 9} \cdot Z_{9 2}& && & D_{2 3} \cdot Y_{3 6} - Z_{2 7} \cdot Y_{7 5} \cdot D_{5 6}& \\ 
 &&\Lambda^{(6)}_{5 9} &:& & Z_{9 2} \cdot X_{2 5} - X_{9 1} \cdot X_{1 3} \cdot Z_{3 5}& && & D_{5 6} \cdot Y_{6 9} - Z_{5 1} \cdot Y_{1 8} \cdot D_{8 9}&
 \end{array}
 ~.~
 \eea
The above $J$- and $E$-terms
come from the general formula in \eqref{es03c07} 
with the following relabelling of indices, 
\beal{es04m02}
&
[1,0] \rightarrow 1~,~  [2,0] \rightarrow 2~,~  [3,0] \rightarrow 3~,~ 
&
\nn\\
&
[1,1] \rightarrow 4~,~  [2,1] \rightarrow 5~,~  [3,1] \rightarrow 6~,~ 
&
\nn\\
&
[1,2] \rightarrow 7~,~  [2,2] \rightarrow 8~,~  [3,2] \rightarrow 9~.~
&
\eea

Using the forward algorithm for brane brick models, 
we obtain the
$P$-matrix, which takes the form,
\beal{es04m03}
&&
P=
\nn\\
&&
\resizebox{0.9\textwidth}{!}{$
\left(
\begin{array}{c|cccccc|ccc|ccc|ccc|ccc|ccc|ccc}
 & p_{1} & p_{2} & p_{3} & p_{4} & p_{5} & p_{6} & q^{(1)}_{1} & q^{(1)}_{2} & q^{(1)}_{3} & q^{(2)}_{1} & q^{(2)}_{2} & q^{(2)}_{3} & o^{(1)}_{1} & \cdots & o^{(1)}_{6} & o^{(2)}_{1} & \cdots & o^{(2)}_{6} & o^{(3)}_{1} & \cdots & o^{(3)}_{6} & o^{(4)}_{1} & \cdots & o^{(4)}_{21} \\
\hline
 D_{14} & 0 & 0 & 1 & 0 & 0 & 1 & 0 & 0 & 1 & 0 & 1 & 0 & 1 & \cdots & 0 & 0 & \cdots & 0 & 1 & \cdots & 2 & 0 & \cdots & 1 \\
 D_{23} & 0 & 0 & 0 & 0 & 0 & 1 & 0 & 0 & 0 & 0 & 0 & 1 & 0 & \cdots & 0 & 0 & \cdots & 0 & 1 & \cdots & 1 & 0 & \cdots & 0 \\
 D_{47} & 0 & 0 & 1 & 0 & 0 & 1 & 1 & 0 & 0 & 1 & 0 & 0 & 0 & \cdots & 1 & 1 & \cdots & 0 & 2 & \cdots & 1 & 1 & \cdots & 1 \\
 D_{56} & 0 & 0 & 0 & 0 & 0 & 1 & 0 & 0 & 0 & 0 & 1 & 0 & 0 & \cdots & 0 & 1 & \cdots & 0 & 0 & \cdots & 1 & 1 & \cdots & 0 \\
 D_{71} & 0 & 0 & 1 & 0 & 0 & 1 & 0 & 1 & 0 & 0 & 0 & 1 & 0 & \cdots & 0 & 0 & \cdots & 1 & 1 & \cdots & 1 & 1 & \cdots & 0 \\
 D_{89} & 0 & 0 & 0 & 0 & 0 & 1 & 0 & 0 & 0 & 1 & 0 & 0 & 1 & \cdots & 0 & 0 & \cdots & 0 & 1 & \cdots & 0 & 1 & \cdots & 0 \\
 X_{13} & 1 & 0 & 0 & 0 & 0 & 0 & 0 & 0 & 1 & 0 & 0 & 0 & 1 & \cdots & 1 & 0 & \cdots & 0 & 0 & \cdots & 1 & 0 & \cdots & 1 \\
 X_{25} & 1 & 0 & 0 & 1 & 0 & 0 & 1 & 0 & 0 & 0 & 0 & 1 & 1 & \cdots & 1 & 0 & \cdots & 0 & 1 & \cdots & 0 & 0 & \cdots & 0 \\
 X_{34} & 0 & 0 & 0 & 1 & 0 & 0 & 0 & 0 & 0 & 0 & 1 & 0 & 1 & \cdots & 0 & 0 & \cdots & 0 & 0 & \cdots & 0 & 0 & \cdots & 0 \\
 X_{46} & 1 & 0 & 0 & 0 & 0 & 0 & 1 & 0 & 0 & 0 & 0 & 0 & 0 & \cdots & 1 & 1 & \cdots & 0 & 0 & \cdots & 0 & 1 & \cdots & 0 \\
 X_{58} & 1 & 0 & 0 & 1 & 0 & 0 & 0 & 1 & 0 & 0 & 1 & 0 & 1 & \cdots & 1 & 1 & \cdots & 0 & 0 & \cdots & 1 & 1 & \cdots & 1 \\
 X_{67} & 0 & 0 & 0 & 1 & 0 & 0 & 0 & 0 & 0 & 1 & 0 & 0 & 1 & \cdots & 1 & 0 & \cdots & 0 & 1 & \cdots & 0 & 0 & \cdots & 1 \\
 X_{79} & 1 & 0 & 0 & 0 & 0 & 0 & 0 & 1 & 0 & 0 & 0 & 0 & 1 & \cdots & 0 & 0 & \cdots & 0 & 0 & \cdots & 0 & 1 & \cdots & 0 \\
 X_{82} & 1 & 0 & 0 & 1 & 0 & 0 & 0 & 0 & 1 & 1 & 0 & 0 & 2 & \cdots & 2 & 0 & \cdots & 1 & 0 & \cdots & 0 & 1 & \cdots & 1 \\
 X_{91} & 0 & 0 & 0 & 1 & 0 & 0 & 0 & 0 & 0 & 0 & 0 & 1 & 0 & \cdots & 1 & 0 & \cdots & 1 & 0 & \cdots & 0 & 0 & \cdots & 0 \\
 Y_{18} & 0 & 0 & 0 & 0 & 1 & 0 & 0 & 0 & 0 & 0 & 1 & 0 & 0 & \cdots & 0 & 1 & \cdots & 0 & 0 & \cdots & 1 & 0 & \cdots & 1 \\
 Y_{36} & 0 & 1 & 0 & 0 & 1 & 0 & 1 & 0 & 0 & 0 & 1 & 0 & 0 & \cdots & 0 & 2 & \cdots & 1 & 0 & \cdots & 0 & 1 & \cdots & 0 \\
 Y_{42} & 0 & 0 & 0 & 0 & 1 & 0 & 0 & 0 & 0 & 1 & 0 & 0 & 0 & \cdots & 1 & 1 & \cdots & 1 & 0 & \cdots & 0 & 1 & \cdots & 1 \\
 Y_{69} & 0 & 1 & 0 & 0 & 1 & 0 & 0 & 1 & 0 & 1 & 0 & 0 & 1 & \cdots & 0 & 1 & \cdots & 1 & 1 & \cdots & 0 & 1 & \cdots & 1 \\
 Y_{75} & 0 & 0 & 0 & 0 & 1 & 0 & 0 & 0 & 0 & 0 & 0 & 1 & 0 & \cdots & 0 & 0 & \cdots & 1 & 0 & \cdots & 0 & 0 & \cdots & 0 \\
 Y_{93} & 0 & 1 & 0 & 0 & 1 & 0 & 0 & 0 & 1 & 0 & 0 & 1 & 0 & \cdots & 1 & 1 & \cdots & 2 & 0 & \cdots & 1 & 0 & \cdots & 1 \\
 Z_{27} & 0 & 1 & 0 & 0 & 0 & 0 & 1 & 0 & 0 & 0 & 0 & 0 & 0 & \cdots & 0 & 1 & \cdots & 0 & 1 & \cdots & 0 & 0 & \cdots & 0 \\
 Z_{35} & 0 & 0 & 1 & 0 & 0 & 0 & 1 & 0 & 0 & 0 & 0 & 0 & 0 & \cdots & 0 & 0 & \cdots & 0 & 1 & \cdots & 0 & 0 & \cdots & 0 \\
 Z_{51} & 0 & 1 & 0 & 0 & 0 & 0 & 0 & 1 & 0 & 0 & 0 & 0 & 0 & \cdots & 0 & 1 & \cdots & 1 & 0 & \cdots & 0 & 1 & \cdots & 0 \\
 Z_{68} & 0 & 0 & 1 & 0 & 0 & 0 & 0 & 1 & 0 & 0 & 0 & 0 & 0 & \cdots & 0 & 0 & \cdots & 0 & 1 & \cdots & 1 & 0 & \cdots & 1 \\
 Z_{84} & 0 & 1 & 0 & 0 & 0 & 0 & 0 & 0 & 1 & 0 & 0 & 0 & 1 & \cdots & 0 & 0 & \cdots & 1 & 0 & \cdots & 0 & 0 & \cdots & 0 \\
 Z_{92} & 0 & 0 & 1 & 0 & 0 & 0 & 0 & 0 & 1 & 0 & 0 & 0 & 0 & \cdots & 1 & 0 & \cdots & 1 & 0 & \cdots & 1 & 0 & \cdots & 1 \\
\end{array}
\right)
$}
~,~
\nn\\
\eea
where $o_k^{(1)}$, $o_l^{(2)}$, $o_m^{(3)}$ and $o_n^{(4)}$ are extra GLSM fields \cite{Witten:1993yc}.
The $J$- and $E$-term charge matrix is as follows,
\beal{es04m04}
&&
Q_{JE}=
\nn\\
&&
\resizebox{0.85\textwidth}{!}{$
\left(
\begin{array}{cccccc|ccc|ccc|ccc|ccc|ccc|ccc}
p_{1} & p_{2} & p_{3} & p_{4} & p_{5} & p_{6} & q^{(1)}_{1} & q^{(1)}_{2} & q^{(1)}_{3} & q^{(2)}_{1} & q^{(2)}_{2} & q^{(2)}_{3} & o^{(1)}_{1} & \cdots & o^{(1)}_{6} & o^{(2)}_{1} & \cdots & o^{(2)}_{6} & o^{(3)}_{1} & \cdots & o^{(3)}_{6} & o^{(4)}_{1} & \cdots & o^{(4)}_{21} \\
\hline
0 & 0 & 0 & 1 & 0 & 0 & 1 & -1 & 0 & 0 & -1 & 0 & 0 & \cdots & 0 & 0 & \cdots & 0 & 0 & \cdots & 1 & 1 & \cdots & 0 \\
0 & 0 & 0 & 1 & 0 & 1 & 1 & -1 & 0 & -1 & -1 & 0 & 0 & \cdots & 0 & 0 & \cdots & 0 & 0 & \cdots & 0 & 1 & \cdots & 1 \\
1 & 0 & 0 & 0 & 0 & 0 & 1 & -1 & 0 & 1 & 0 & 0 & 0 & \cdots & 0 & 0 & \cdots & 0 & 0 & \cdots & 0 & 0 & \cdots & 0 \\
1 & 0 & 0 & 0 & 0 & 1 & 1 & -1 & 0 & 0 & 0 & 0 & 0 & \cdots & 0 & 0 & \cdots & 0 & 0 & \cdots & 0 & 0 & \cdots & 0 \\
0 & 0 & 1 & 0 & 0 & 0 & 0 & -1 & 0 & 0 & 0 & 0 & 0 & \cdots & 0 & 0 & \cdots & 0 & 0 & \cdots & 0 & 1 & \cdots & 0 \\
3 & 0 & 0 & 1 & 0 & 2 & 2 & 0 & 0 & 2 & 0 & 0 & -1 & \cdots & 0 & 0 & \cdots & 1 & 0 & \cdots & 0 & -1 & \cdots & 0 \\
2 & 0 & 0 & 0 & 0 & 1 & 1 & 0 & 0 & 2 & 0 & 0 & -1 & \cdots & 0 & 0 & \cdots & 0 & 0 & \cdots & 0 & 0 & \cdots & 0 \\
1 & 0 & 0 & 0 & 0 & 1 & 1 & 0 & 0 & 1 & 0 & 0 & 0 & \cdots & 0 & 0 & \cdots & 0 & 0 & \cdots & 0 & 0 & \cdots & 0 \\
2 & 0 & 0 & 0 & 0 & 1 & 1 & 0 & 0 & 2 & 0 & 0 & -1 & \cdots & 0 & 0 & \cdots & 0 & 0 & \cdots & 0 & -1 & \cdots & 0 \\
0 & 0 & 0 & 1 & 0 & 1 & 1 & 0 & 0 & 0 & -1 & 0 & 0 & \cdots & 0 & 0 & \cdots & 0 & 0 & \cdots & 0 & 0 & \cdots & 0 \\
1 & 0 & 0 & 0 & 0 & 1 & 1 & 0 & 0 & 1 & 0 & 0 & 0 & \cdots & 0 & 0 & \cdots & 0 & 0 & \cdots & 0 & -1 & \cdots & 0 \\
0 & 0 & 0 & 0 & 0 & 1 & 1 & 0 & 0 & 0 & 0 & 0 & 0 & \cdots & 1 & 0 & \cdots & 0 & 0 & \cdots & 0 & 0 & \cdots & 0 \\
1 & 0 & 0 & 1 & 0 & 1 & 1 & 0 & 1 & 1 & 0 & 0 & -1 & \cdots & 0 & 0 & \cdots & 0 & 0 & \cdots & 0 & 0 & \cdots & 0 \\
1 & 0 & 0 & 0 & 0 & 0 & 0 & 0 & 0 & 1 & 0 & 0 & -1 & \cdots & 0 & 0 & \cdots & 0 & 0 & \cdots & 0 & 0 & \cdots & 0 \\
0 & 0 & 0 & 1 & 0 & 0 & 1 & -1 & 0 & 0 & -1 & 0 & 0 & \cdots & 0 & 0 & \cdots & 0 & 0 & \cdots & 0 & 1 & \cdots & 0 \\
1 & 0 & 0 & 1 & 0 & 1 & 0 & -1 & 0 & -1 & -1 & 0 & 0 & \cdots & 0 & 0 & \cdots & 0 & 0 & \cdots & 0 & 0 & \cdots & 0 \\
0 & 0 & 0 & 1 & 0 & 0 & 0 & -1 & 0 & -1 & -1 & 0 & 0 & \cdots & 0 & 0 & \cdots & 0 & 0 & \cdots & 0 & 1 & \cdots & 0 \\
1 & 0 & 0 & 0 & 0 & 0 & 0 & -1 & 0 & 0 & 0 & 0 & 0 & \cdots & 0 & 0 & \cdots & 0 & 0 & \cdots & 0 & 0 & \cdots & 0 \\
0 & 0 & 0 & 0 & 0 & 0 & 0 & -1 & 0 & 0 & -1 & 0 & 0 & \cdots & 0 & 0 & \cdots & 0 & 0 & \cdots & 0 & 1 & \cdots & 0 \\
0 & 0 & 0 & 0 & 0 & -1 & 0 & -1 & 0 & 0 & 0 & 0 & 0 & \cdots & 0 & 0 & \cdots & 0 & 0 & \cdots & 0 & 1 & \cdots & 0 \\
0 & 0 & 0 & 0 & 0 & 0 & -1 & -1 & 0 & -1 & -1 & 0 & 0 & \cdots & 0 & 0 & \cdots & 0 & 0 & \cdots & 0 & 1 & \cdots & 0 \\
-1 & 0 & 0 & 0 & 0 & 0 & 0 & -1 & 0 & -1 & -1 & 0 & 0 & \cdots & 0 & 0 & \cdots & 0 & 0 & \cdots & 0 & 1 & \cdots & 0 \\
0 & 0 & 0 & 0 & 0 & -1 & -1 & -1 & 0 & -1 & 0 & 0 & 0 & \cdots & 0 & 0 & \cdots & 0 & 1 & \cdots & 0 & 1 & \cdots & 0 \\
2 & 0 & 0 & 2 & 0 & 1 & 1 & 0 & 0 & 1 & -1 & 0 & -1 & \cdots & 0 & 0 & \cdots & 0 & 0 & \cdots & 0 & 0 & \cdots & 0 \\
2 & 0 & 0 & 1 & 0 & 1 & 0 & 0 & 0 & 1 & -1 & 0 & -1 & \cdots & 0 & 0 & \cdots & 0 & 0 & \cdots & 0 & 0 & \cdots & 0 \\
2 & 0 & 0 & 1 & 0 & 1 & 1 & 0 & 0 & 1 & 0 & 0 & -1 & \cdots & 0 & 0 & \cdots & 0 & 0 & \cdots & 0 & -1 & \cdots & 0 \\
1 & 0 & 0 & 1 & 1 & 1 & 1 & 0 & 0 & 0 & -1 & 0 & 0 & \cdots & 0 & 0 & \cdots & 0 & 0 & \cdots & 0 & 0 & \cdots & 0 \\
1 & 0 & 0 & 1 & 0 & 0 & 1 & 0 & 0 & 1 & 0 & 0 & -1 & \cdots & 0 & 0 & \cdots & 0 & 0 & \cdots & 0 & 0 & \cdots & 0 \\
1 & 0 & 0 & 0 & 0 & 0 & 1 & 0 & 0 & 1 & 0 & 1 & 0 & \cdots & 0 & 0 & \cdots & 0 & 0 & \cdots & 0 & 0 & \cdots & 0 \\
1 & 0 & 0 & 1 & 0 & 1 & 0 & 0 & 0 & 0 & -1 & 0 & 0 & \cdots & 0 & 1 & \cdots & 0 & 0 & \cdots & 0 & -1 & \cdots & 0 \\
0 & 0 & 0 & 1 & 0 & 0 & 0 & 0 & 0 & 0 & -1 & 0 & 0 & \cdots & 0 & 0 & \cdots & 0 & 0 & \cdots & 0 & 0 & \cdots & 0 \\
1 & 0 & 0 & 0 & 0 & 0 & 0 & 0 & 0 & 1 & 0 & 0 & 0 & \cdots & 0 & 0 & \cdots & 0 & 0 & \cdots & 0 & -1 & \cdots & 0 \\
2 & 1 & 0 & 1 & 0 & 1 & 0 & 0 & 0 & 1 & 0 & 0 & -1 & \cdots & 0 & 0 & \cdots & 0 & 0 & \cdots & 0 & -1 & \cdots & 0 \\
1 & 0 & 0 & 1 & 0 & 0 & 0 & 0 & 0 & 0 & 0 & 0 & -1 & \cdots & 0 & 0 & \cdots & 0 & 0 & \cdots & 0 & 0 & \cdots & 0 \\
1 & 0 & 0 & 0 & 0 & 0 & 0 & 0 & 0 & 1 & 0 & 0 & -1 & \cdots & 0 & 0 & \cdots & 0 & 0 & \cdots & 0 & 0 & \cdots & 0 \\
0 & 0 & 0 & -1 & 0 & 0 & 0 & 0 & 0 & 1 & 0 & 0 & 0 & \cdots & 0 & 0 & \cdots & 0 & 0 & \cdots & 0 & 0 & \cdots & 0 \\
1 & 0 & 0 & 0 & 0 & 0 & 0 & 0 & 0 & 1 & 0 & 0 & -1 & \cdots & 0 & 0 & \cdots & 0 & 0 & \cdots & 0 & -1 & \cdots & 0 \\
1 & 0 & 0 & 0 & 0 & 0 & -1 & 0 & 0 & 0 & 0 & 0 & -1 & \cdots & 0 & 0 & \cdots & 0 & 0 & \cdots & 0 & 0 & \cdots & 0 \\
0 & 0 & 0 & -1 & 0 & 0 & 0 & 0 & 0 & 1 & 0 & 0 & 0 & \cdots & 0 & 0 & \cdots & 0 & 0 & \cdots & 0 & -1 & \cdots & 0 \\
\end{array}
\right)
$}
~,~
\nn\\
\eea
and the corresponding $D$-term charge matrix takes the following form,
\beal{es04m05}
&&
Q_{D}=
\nn\\
&&
\resizebox{0.85\textwidth}{!}{$
\left(
\begin{array}{cccccc|ccc|ccc|ccc|ccc|ccc|ccc}
p_{1} & p_{2} & p_{3} & p_{4} & p_{5} & p_{6} & q^{(1)}_{1} & q^{(1)}_{2} & q^{(1)}_{3} & q^{(2)}_{1} & q^{(2)}_{2} & q^{(2)}_{3} & o^{(1)}_{1} & \cdots & o^{(1)}_{6} & o^{(2)}_{1} & \cdots & o^{(2)}_{6} & o^{(3)}_{1} & \cdots & o^{(3)}_{6} & o^{(4)}_{1} & \cdots & o^{(4)}_{21} \\
\hline
 -1 & 0 & 0 & 1 & 0 & 0 & 0 & 0 & 0 & -1 & -1 & 0 & 0 & \cdots & 0 & 0 & \cdots & 0 & 0 & \cdots & 0 & 1 & \cdots & 0 \\
 0 & 0 & 0 & 0 & 0 & -1 & -1 & 0 & 0 & 0 & 0 & 0 & 0 & \cdots & 0 & 0 & \cdots & 0 & 0 & \cdots & 0 & 0 & \cdots & 0 \\
 0 & 0 & 0 & -1 & 0 & 0 & -1 & 0 & 0 & 0 & 0 & 0 & 0 & \cdots & 0 & 0 & \cdots & 0 & 0 & \cdots & 0 & 0 & \cdots & 0 \\
 -1 & 0 & 0 & 0 & 0 & 0 & 0 & 0 & 0 & -1 & 0 & 0 & 1 & \cdots & 0 & 0 & \cdots & 0 & 0 & \cdots & 0 & 0 & \cdots & 0 \\
 0 & 0 & 0 & 0 & 0 & 0 & 0 & 0 & 0 & 0 & 0 & 0 & 0 & \cdots & 0 & 0 & \cdots & 0 & 0 & \cdots & 0 & -1 & \cdots & 0 \\
 0 & 0 & 0 & 0 & 0 & 0 & 0 & -1 & 0 & -1 & 0 & 0 & 0 & \cdots & 0 & 0 & \cdots & 0 & 0 & \cdots & 0 & 1 & \cdots & 0 \\
 0 & 0 & 0 & 0 & 0 & 0 & 1 & 0 & 0 & 1 & 0 & 0 & 0 & \cdots & 0 & 0 & \cdots & 0 & 0 & \cdots & 0 & 0 & \cdots & 0 \\
 1 & 0 & 0 & 0 & 0 & 0 & 0 & 1 & 0 & 1 & 1 & 0 & -1 & \cdots & 0 & 0 & \cdots & 0 & 0 & \cdots & 0 & -1 & \cdots & 0 \\
\end{array}
\right)
$}
~.~
\nn\\
\eea

\begin{figure}[ht!!]
\centering
\includegraphics[width=0.25\textwidth]{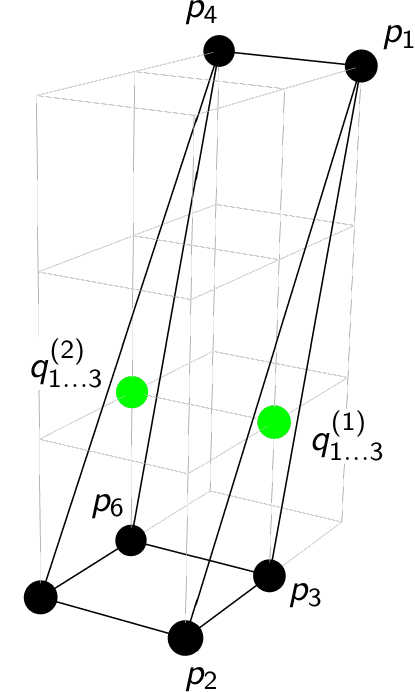}
\caption{
The toric diagram for the $D_3/\mathbb{Z}_3 \ (1,1,1,0,0)$ model.
\label{fig_427_toric}
}
\end{figure}

As a result, the toric diagram of the $D_3/ \mathbb{Z}_3 \ (1,1,1,0,0)$ model
is given by,
\beal{es04m06}
&&
G_{t}=
\nn\\
&&
\resizebox{0.85\textwidth}{!}{$
\left(
\begin{array}{cccccc|ccc|ccc|ccc|ccc|ccc|ccc}
p_{1} & p_{2} & p_{3} & p_{4} & p_{5} & p_{6} & q^{(1)}_{1} & q^{(1)}_{2} & q^{(1)}_{3} & q^{(2)}_{1} & q^{(2)}_{2} & q^{(2)}_{3} & o^{(1)}_{1} & \cdots & o^{(1)}_{6} & o^{(2)}_{1} & \cdots & o^{(2)}_{6} & o^{(3)}_{1} & \cdots & o^{(3)}_{6} & o^{(4)}_{1} & \cdots & o^{(4)}_{21} \\
\hline
0 & 0 & 0 & 1 & 1 & 1 & 0 & 0 & 0 & 1 & 1 & 1 & 1 & \cdots & 1 & 1 & \cdots & 1 & 1 & \cdots & 1 & 1 & \cdots & 1 \\
3 & 0 & 0 & 3 & 0 & 0 & 1 & 1 & 1 & 1 & 1 & 1 & 4 & \cdots & 4 & 1 & \cdots & 1 & 1 & \cdots & 1 & 2 & \cdots & 2 \\
0 & 2 & 1 & 0 & 2 & 1 & 1 & 1 & 1 & 1 & 1 & 1 & 1 & \cdots & 1 & 3 & \cdots & 3 & 2 & \cdots & 2 & 2 & \cdots & 2 \\
\hline
1 & 1 & 1 & 1 & 1 & 1 & 1 & 1 & 1 & 1 & 1 & 1 & 2 & \cdots & 2 & 2 & \cdots & 2 & 2 & \cdots & 2 & 2 & \cdots & 2 \\
\end{array}
\right)
$}
~,~
\nn\\
\eea
where \fref{fig_427_toric} illustrates the toric diagram. 

%-------------------
%-------------------

%-------------------
\begin{table}[htt!!!]
\centering
\begin{tabular}{|c|c|c|c|l|}
\hline
\; & $U(1)_{f_1}$ & $U(1)_{f_2}$ & $U(1)_{f_3}$ &  fugacity \\
\hline
$p_1$ & $+1$ & $0$ & $0$ &  $t_1=f_1 t$ \\
$p_2$ & $-1$ & $0$ & $0$ &  $t_2=f_1^{-1} t$\\
$p_3$ & $0$ & $+1$ & $0$ &  $t_3=f_2 t$\\
$p_4$ & $0$ & $-1$ & $0$ & $t_4=f_2^{-1} t$ \\
$p_5$ & $0$ & $0$ & $+1$ & $t_5=f_3 t$ \\
$p_6$ & $0$ & $0$ & $-1$ & $t_6=f_3^{-1} t$ \\
\hline
\end{tabular}
\caption{
Mesonic flavor symmetry of the $D_3/\mathbb{Z}_3 \ (1,1,1,0,0)$ model 
and charges on the extremal GLSM fields $p_a$. 
Here, the fugacity $t$ counts the degree in extremal GLSM fields $p_a$. 
\label{tab_04m01}}
\end{table}
%-------------------

The global symmetry of the $D_3/\mathbb{Z}_3 \ (1,1,1,0,0)$ model 
is not enhanced and takes the following form, 
\beal{es04m07}
U(1)_{f_1} \times U(1)_{f_2} \times U(1)_{f_3} \times U(1)_R
~.~
\eea
The refined Hilbert series of the mesonic moduli space 
takes the following form, 
\beal{es04m08}
&&
g(t_a; \mathcal{M}^{mes})=
(1+ t_1^2 t_2 t_4^2 t_5+ t_1 t_2^2 t_4 t_5^2+ t_1^2 t_3 t_4^2 t_6+ t_1 t_2 t_3 t_4 t_5 t_6+ t_2^2 t_3 t_5^2 t_6
\nn\\
&&
\hspace{1cm}
+ t_1 t_3^2 t_4 t_6^2+ t_2 t_3^2 t_5 t_6^2+ t_1^2 t_2^2 t_3^2 t_4^2 t_5^2 t_6^2)
\times
\frac{ 
\left(1 - t_1 t_2 t_3 t_4 t_5 t_6\right)
}{
\left(1 - t_1 t_2 t_3\right)
\left(1 - t_1^3 t_4^3\right) 
\left(1 - t_2^3 t_5^3\right) 
}
\nn\\
&&
\hspace{1cm}
\times
\frac{1}{
\left(1 - t_4 t_5 t_6\right) \left(1 - t_3^3 t_6^3\right)
}
~,~
\eea
where $t_a$ is the fugacity corresponding to the extremal GLSM field $p_a$.
When unrefined by setting $t_a = t$, 
the Hilbert series takes the following form,
\beal{es04m09}
g(t; \mathcal{M}^{mes})
=
\frac{1+7t^6 +t^{12}}{(1-t^3)^2 (1-t^6)^2}
~,~
\eea
where the palindromic numerator indicates that the mesonic moduli space is Calabi-Yau.

The Hilbert series refined in terms of the mesonic flavor fugacities summarized in \tref{tab_04m01} takes the following form,
\beal{es04m10}
&&
g(f_1,f_2,f_3, t; \mathcal{M}^{mes})
=
(1+ f_1f_2 f_3^{-2}t^6 +f_1^2 f_2^{-1}f_3^{-1}t^6 + f_1^{-1} f_2^2 f_3^{-1} t^6 + t^6 
\nn\\
&&
\hspace{1cm}
+ f_1f_2^{-2} f_3 t^6 + f_1^{-2} f_2 f_3 t^6 + f_1^{-1} f_2^{-1} f_3^2 t^6 + t^{12})
\times
\frac{1-t^6}{(1-f_2^{-1}t^3) (1-f_2 t^3)}
\nn\\
&&
\hspace{1cm}
\times
\frac{1}{(1-f_1^3 f_2^{-3} t^6) (1-f_2^3 f_3^{-3} t^6) (1-f_1^{-3} f_3^3 t^6)}
~,~
\eea
where the corresponding plethystic logarithm is given by, 
\beal{es04m12}
&&
\text{PL}[g(f_1,f_2,f_3,t) ]=
(f_2 + f_2^{-1}) t^3
+ (f_1^{3} f_2^{-3}+f_2^{3} f_3^{-3}+f_1 f_2 f_3^{-2}
+f_1^{2} f_2^{-1} f_3^{-1}
\nn\\
&&
\hspace{1cm}
+f_1^{-1} f_2^{2} f_3^{-1}+f_1 f_2^{-2} f_3+f_1^{-2} f_2 f_3+f_1^{-1} f_2^{-1} f_3^{2}+f_1^{-3} f_3^{3}) t^{6}
- (3 + 2 f_1 f_2 f_3^{-2} 
\nn\\
&&
\hspace{1cm}
+ 2 f_1^{2} f_2^{-1} f_3^{-1} + 2 f_1^{-1} f_2^{2} f_3^{-1} + 2 f_1 f_2^{-2} f_3 + 2 f_1^{-2} f_2 f_3 + 2 f_1^{-1} f_2^{-1} f_3^{2}
\nn\\
&&
\hspace{1cm}
+ f_1^{3} f_2^{-3}+f_1^{-3} f_2^{3}+f_1^{2} f_2^{2} f_3^{-4}+f_1^{3} f_3^{-3}+f_2^{3} f_3^{-3}+f_1^{4} f_2^{-2} f_3^{-2}+f_1^{-2} f_2^{4} f_3^{-2}
\nn\\
&&
\hspace{1cm}
+f_1^{2} f_2^{-4} f_3^{2}+f_1^{-4} f_2^{2} f_3^{2}+f_1^{-3} f_3^{3}+f_2^{-3} f_3^{3}+f_1^{-2} f_2^{-2} f_3^{4} ) t^{12}
+\dots
~.~
\eea
The above infinite expansion of the plethystic logarithm indicates that the mesonic moduli space is not a complete intersection. 
The generators of the mesonic moduli space correspond to the first positive terms of the expansion
and are summarized with their mesonic flavor charges in \tref{tab_04m02}.
\\

%-------------------
 \begin {table}[htt!!]
\centering
\begin {tabular} {|c|c|ccc|}
\hline
PL term & generator & $U(1)_{f_1}$ & $U(1)_{f_2}$ & $U(1)_{f_3}$ 
\\
\hline
\multirow{1}{*}{$+ f_2^{-1}  t^3$}
& $p_4 p_5 p_6 ~q_{(2)} $ & $0$ & $-1$ & $0$ \\
\hline
\multirow{1}{*}{$+ f_2^{3} f_3^{-3} t^{6}$}
& $p_3^3 p_6^3 ~q_{(1)} q_{(2)} $ & $0$ & $3$ & $-3$  \\
\hline
\multirow{1}{*}{$+f_1^{-1} f_2^{2} f_3^{-1} t^{6}$}
& $p_2 p_3^2 p_5 p_6^2 ~q_{(1)} q_{(2)} $ & $-1$ & $2$ & $-1$  \\
\hline
\multirow{1}{*}{$+f_1^{-2} f_2 f_3 t^{6}$}
& $p_2^2 p_3 p_5^2 p_6 ~q_{(1)} q_{(2)} $ & $-2$ & $1$ & $1$  \\
\hline
\multirow{1}{*}{$+f_1^{-3}  f_3^{3} t^{6}$}
& $p_2^3 p_5^3 ~q_{(1)} q_{(2)} {o}_{{(1)}} $ & $-3$ & $0$ & $3$  \\
\hline
\multirow{1}{*}{$+f_1 f_2 f_3^{-2} t^{6}$}
& $p_1 p_3^2 p_4 p_6^2 ~q_{(1)} q_{(2)} $ & $1$ & $1$ & $-2$  \\
\hline
\multirow{1}{*}{$+ f_2  t^3$}
& $p_1 p_2 p_3 ~q_{(1)} $ & $0$ & $1$ & $0$  \\
\hline
\multirow{1}{*}{$+f_1^{-1} f_2^{-1} f_3^{2} t^{6}$}
& $p_1 p_2^2 p_4 p_5^2 ~q_{(1)} q_{(2)} $ & $-1$ & $-1$ & $2$  \\
\hline
\multirow{1}{*}{$+f_1^{2} f_2^{-1} f_3^{-1} t^{6}$}
& $p_1^2 p_3 p_4^2 p_6 ~q_{(1)} q_{(2)} $ & $2$ & $-1$ & $-1$  \\
\hline
\multirow{1}{*}{$+f_1 f_2^{-2} f_3 t^{6}$}
& $p_1^2 p_2 p_4^2 p_5 ~q_{(1)} q_{(2)} $ & $1$ & $-2$ & $1$  \\
\hline
\multirow{1}{*}{$+f_1^{3} f_2^{-3}  t^{6}$}
& $p_1^3 p_4^3 ~q_{(1)} q_{(2)} $ & $3$ & $-3$ & $0$  \\
\hline
\end{tabular}
\caption{
Generators of the $D_3/\mathbb{Z}_3 \ (1,1,1,0,0)$ model in terms of GLSM fields and their corresponding mesonic flavor symmetry charges. Here, we denote $q_{(1)}=\prod_{i=1}^{3}q_i^{(1)}$,  $q_{(2)}=\prod_{i=1}^{3} q_i^{(2)}$, and set the extra GLSM fields to 1.
 \label{tab_04m02}}
\end{table}
%-------------------

%=================================================================
\subsubsection{$D_3/ \mathbb{Z}_3 \ (1,1,1,1,2)$ \label{sec0414}}

%-------------------
\begin{figure}[H]
    \centering
    \includegraphics[width=0.4\textwidth]{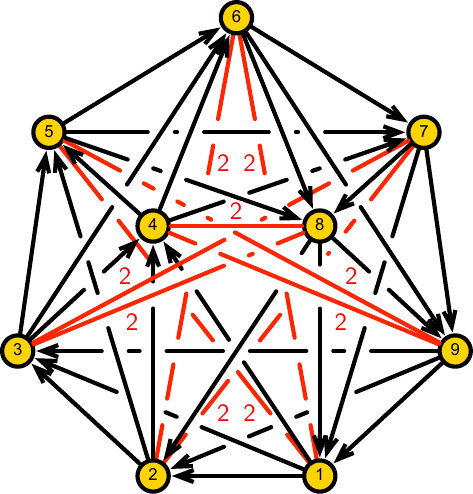}
    \caption{
    The quiver for the $D_3/ \mathbb{Z}_3 \ (1,1,1,1,2)$ model.
    \label{fig_428_quiver}
    }
\end{figure}
%-------------------

The $J$- and $E$-terms for the brane brick model corresponding to the abelian orbifold of the form
$D_3/ \mathbb{Z}_3 \ (1,1,1,1,2)$ are as follows, 
\beal{es04n01}
\begin{array}{rclccclcccc}
&& && &J& && &E& \\
&&\Lambda^{(1)}_{5 1} &:& & X_{1 3} \cdot X_{3 4} \cdot Y_{4 5} - Y_{1 2} \cdot X_{2 5}& && & D_{5 6} \cdot Z_{6 8} \cdot Z_{8 1} - Z_{5 7} \cdot D_{7 1}& \\ 
 &&\Lambda^{(1)}_{8 4} &:& & X_{4 6} \cdot X_{6 7} \cdot Y_{7 8} - Y_{4 5} \cdot X_{5 8}& && & D_{8 9} \cdot Z_{9 2} \cdot Z_{2 4} - Z_{8 1} \cdot D_{1 4}& \\ 
 &&\Lambda^{(1)}_{2 7} &:& & X_{7 9} \cdot X_{9 1} \cdot Y_{1 2} - Y_{7 8} \cdot X_{8 2}& && & D_{2 3} \cdot Z_{3 5} \cdot Z_{5 7} - Z_{2 4} \cdot D_{4 7}& \\ 
 &&\Lambda^{(2)}_{7 2} &:& & Z_{2 4} \cdot X_{4 6} \cdot X_{6 7} - X_{2 5} \cdot Z_{5 7}& && & D_{7 1} \cdot Y_{1 2} - Y_{7 8} \cdot D_{8 9} \cdot Z_{9 2}& \\ 
 &&\Lambda^{(2)}_{1 5} &:& & Z_{5 7} \cdot X_{7 9} \cdot X_{9 1} - X_{5 8} \cdot Z_{8 1}& && & D_{1 4} \cdot Y_{4 5} - Y_{1 2} \cdot D_{2 3} \cdot Z_{3 5}& \\ 
 &&\Lambda^{(2)}_{4 8} &:& & Z_{8 1} \cdot X_{1 3} \cdot X_{3 4} - X_{8 2} \cdot Z_{2 4}& && & D_{4 7} \cdot Y_{7 8} - Y_{4 5} \cdot D_{5 6} \cdot Z_{6 8}& \\ 
 &&\Lambda^{(3)}_{6 1} &:& & X_{1 3} \cdot Y_{3 6} - Y_{1 2} \cdot Z_{2 4} \cdot X_{4 6}& && & X_{6 7} \cdot D_{7 1} - Z_{6 8} \cdot D_{8 9} \cdot X_{9 1}& \\ 
 &&\Lambda^{(3)}_{9 4} &:& & X_{4 6} \cdot Y_{6 9} - Y_{4 5} \cdot Z_{5 7} \cdot X_{7 9}& && & X_{9 1} \cdot D_{1 4} - Z_{9 2} \cdot D_{2 3} \cdot X_{3 4}& \\ 
 &&\Lambda^{(3)}_{3 7} &:& & X_{7 9} \cdot Y_{9 3} - Y_{7 8} \cdot Z_{8 1} \cdot X_{1 3}& && & X_{3 4} \cdot D_{4 7} - Z_{3 5} \cdot D_{5 6} \cdot X_{6 7}& \\ 
 &&\Lambda^{(4)}_{7 3} &:& & X_{3 4} \cdot Y_{4 5} \cdot Z_{5 7} - Y_{3 6} \cdot X_{6 7}& && & D_{7 1} \cdot X_{1 3} - X_{7 9} \cdot Z_{9 2} \cdot D_{2 3}& \\ 
 &&\Lambda^{(4)}_{1 6} &:& & X_{6 7} \cdot Y_{7 8} \cdot Z_{8 1} - Y_{6 9} \cdot X_{9 1}& && & D_{1 4} \cdot X_{4 6} - X_{1 3} \cdot Z_{3 5} \cdot D_{5 6}& \\ 
 &&\Lambda^{(4)}_{4 9} &:& & X_{9 1} \cdot Y_{1 2} \cdot Z_{2 4} - Y_{9 3} \cdot X_{3 4}& && & D_{4 7} \cdot X_{7 9} - X_{4 6} \cdot Z_{6 8} \cdot D_{8 9}& \\ 
 &&\Lambda^{(5)}_{8 3} &:& & Y_{3 6} \cdot Z_{6 8} - Z_{3 5} \cdot Z_{5 7} \cdot Y_{7 8}& && & D_{8 9} \cdot X_{9 1} \cdot X_{1 3} - X_{8 2} \cdot D_{2 3}& \\ 
 &&\Lambda^{(5)}_{2 6} &:& & Y_{6 9} \cdot Z_{9 2} - Z_{6 8} \cdot Z_{8 1} \cdot Y_{1 2}& && & D_{2 3} \cdot X_{3 4} \cdot X_{4 6} - X_{2 5} \cdot D_{5 6}& \\ 
 &&\Lambda^{(5)}_{5 9} &:& & Y_{9 3} \cdot Z_{3 5} - Z_{9 2} \cdot Z_{2 4} \cdot Y_{4 5}& && & D_{5 6} \cdot X_{6 7} \cdot X_{7 9} - X_{5 8} \cdot D_{8 9}& \\ 
 &&\Lambda^{(6)}_{8 3} &:& & Z_{3 5} \cdot X_{5 8} - X_{3 4} \cdot X_{4 6} \cdot Z_{6 8}& && & D_{8 9} \cdot Y_{9 3} - Z_{8 1} \cdot Y_{1 2} \cdot D_{2 3}& \\ 
 &&\Lambda^{(6)}_{2 6} &:& & Z_{6 8} \cdot X_{8 2} - X_{6 7} \cdot X_{7 9} \cdot Z_{9 2}& && & D_{2 3} \cdot Y_{3 6} - Z_{2 4} \cdot Y_{4 5} \cdot D_{5 6}& \\ 
 &&\Lambda^{(6)}_{5 9} &:& & Z_{9 2} \cdot X_{2 5} - X_{9 1} \cdot X_{1 3} \cdot Z_{3 5}& && & D_{5 6} \cdot Y_{6 9} - Z_{5 7} \cdot Y_{7 8} \cdot D_{8 9}&
 \end{array}
 ~,~
 \eea
where the corresponding quiver diagram is shown in \fref{fig_428_quiver}.
The $J$- and $E$-terms
come from the general formula in \eqref{es03c07} 
with the following relabelling of indices, 
\beal{es04n02}
&
[1,0] \rightarrow 1~,~  [2,0] \rightarrow 2~,~  [3,0] \rightarrow 3~,~ 
&
\nn\\
&
[1,1] \rightarrow 4~,~  [2,1] \rightarrow 5~,~  [3,1] \rightarrow 6~,~ 
&
\nn\\
&
[1,2] \rightarrow 7~,~  [2,2] \rightarrow 8~,~  [3,2] \rightarrow 9~.~
&
\eea

Using the forward algorithm for brane brick models, 
we obtain the
$P$-matrix, which takes the form,
\beal{es04n03}
&&
P=
\nn\\
&&
\resizebox{0.95\textwidth}{!}{$
\left(
\begin{array}{c|cccccc|ccccccccc|ccc|ccc|ccc|ccc|ccc|ccc|ccc}
 & p_{1} & p_{2} & p_{3} & p_{4} & p_{5} & p_{6} & s_{1} & s_{2} & s_{3} & s_{4} & s_{5} & s_{6} & s_{7} & s_{8} & s_{9} & o^{(1)}_{1} & \cdots & o^{(1)}_{3} & o^{(2)}_{1} & \cdots & o^{(2)}_{3} & o^{(3)}_{1} & \cdots & o^{(3)}_{3} & o^{(4)}_{1} & \cdots & o^{(4)}_{9} & o^{(5)}_{1} & \cdots & o^{(5)}_{3} & o^{(6)}_{1} & \cdots & o^{(6)}_{3} & o^{(7)}_{1} & \cdots & o^{(7)}_{3} \\
\hline
 D_{14} & 0 & 1 & 0 & 1 & 0 & 0 & 0 & 0 & 0 & 1 & 1 & 0 & 0 & 0 & 1 & 1 & \cdots & 0 & 0 & \cdots & 0 & 1 & \cdots & 1 & 1 & \cdots & 1 & 0 & \cdots & 0 & 2 & \cdots & 1 & 1 & \cdots & 2 \\
 D_{23} & 0 & 1 & 0 & 0 & 0 & 0 & 0 & 0 & 0 & 0 & 1 & 0 & 0 & 0 & 0 & 0 & \cdots & 1 & 0 & \cdots & 1 & 1 & \cdots & 1 & 0 & \cdots & 0 & 0 & \cdots & 0 & 1 & \cdots & 1 & 1 & \cdots & 1 \\
 D_{47} & 0 & 1 & 0 & 1 & 0 & 0 & 0 & 1 & 1 & 0 & 0 & 1 & 0 & 0 & 0 & 0 & \cdots & 1 & 1 & \cdots & 0 & 2 & \cdots & 1 & 1 & \cdots & 0 & 0 & \cdots & 1 & 1 & \cdots & 2 & 1 & \cdots & 1 \\
 D_{56} & 0 & 1 & 0 & 0 & 0 & 0 & 0 & 0 & 1 & 0 & 0 & 0 & 0 & 0 & 0 & 0 & \cdots & 0 & 0 & \cdots & 0 & 0 & \cdots & 1 & 0 & \cdots & 1 & 0 & \cdots & 0 & 0 & \cdots & 1 & 0 & \cdots & 1 \\
 D_{71} & 0 & 1 & 0 & 1 & 0 & 0 & 1 & 0 & 0 & 0 & 0 & 0 & 1 & 1 & 0 & 0 & \cdots & 0 & 0 & \cdots & 1 & 1 & \cdots & 2 & 0 & \cdots & 1 & 1 & \cdots & 0 & 1 & \cdots & 1 & 2 & \cdots & 1 \\
 D_{89} & 0 & 1 & 0 & 0 & 0 & 0 & 1 & 0 & 0 & 0 & 0 & 0 & 0 & 0 & 0 & 1 & \cdots & 0 & 1 & \cdots & 0 & 1 & \cdots & 0 & 1 & \cdots & 0 & 0 & \cdots & 0 & 1 & \cdots & 0 & 1 & \cdots & 0 \\
 X_{13} & 0 & 0 & 1 & 0 & 0 & 0 & 0 & 0 & 0 & 0 & 1 & 0 & 0 & 0 & 1 & 1 & \cdots & 1 & 0 & \cdots & 0 & 0 & \cdots & 0 & 0 & \cdots & 0 & 0 & \cdots & 1 & 1 & \cdots & 1 & 0 & \cdots & 1 \\
 X_{25} & 1 & 0 & 1 & 0 & 0 & 0 & 0 & 1 & 0 & 1 & 1 & 0 & 0 & 0 & 0 & 1 & \cdots & 2 & 0 & \cdots & 1 & 1 & \cdots & 0 & 1 & \cdots & 0 & 2 & \cdots & 1 & 2 & \cdots & 1 & 1 & \cdots & 0 \\
 X_{34} & 1 & 0 & 0 & 0 & 0 & 0 & 0 & 0 & 0 & 1 & 0 & 0 & 0 & 0 & 0 & 1 & \cdots & 0 & 0 & \cdots & 0 & 0 & \cdots & 0 & 1 & \cdots & 1 & 1 & \cdots & 0 & 1 & \cdots & 0 & 0 & \cdots & 0 \\
 X_{46} & 0 & 0 & 1 & 0 & 0 & 0 & 0 & 1 & 1 & 0 & 0 & 0 & 0 & 0 & 0 & 0 & \cdots & 1 & 0 & \cdots & 0 & 0 & \cdots & 0 & 0 & \cdots & 0 & 1 & \cdots & 1 & 0 & \cdots & 1 & 0 & \cdots & 0 \\
 X_{58} & 1 & 0 & 1 & 0 & 0 & 0 & 0 & 0 & 1 & 0 & 0 & 1 & 1 & 0 & 0 & 1 & \cdots & 1 & 0 & \cdots & 0 & 0 & \cdots & 1 & 0 & \cdots & 1 & 1 & \cdots & 1 & 1 & \cdots & 2 & 0 & \cdots & 1 \\
 X_{67} & 1 & 0 & 0 & 0 & 0 & 0 & 0 & 0 & 0 & 0 & 0 & 1 & 0 & 0 & 0 & 1 & \cdots & 1 & 1 & \cdots & 0 & 1 & \cdots & 0 & 1 & \cdots & 0 & 0 & \cdots & 1 & 1 & \cdots & 1 & 0 & \cdots & 0 \\
 X_{79} & 0 & 0 & 1 & 0 & 0 & 0 & 1 & 0 & 0 & 0 & 0 & 0 & 1 & 0 & 0 & 1 & \cdots & 0 & 0 & \cdots & 0 & 0 & \cdots & 0 & 0 & \cdots & 0 & 1 & \cdots & 0 & 1 & \cdots & 0 & 1 & \cdots & 0 \\
 X_{82} & 1 & 0 & 1 & 0 & 0 & 0 & 1 & 0 & 0 & 0 & 0 & 0 & 0 & 1 & 1 & 2 & \cdots & 1 & 1 & \cdots & 0 & 0 & \cdots & 0 & 1 & \cdots & 1 & 1 & \cdots & 2 & 1 & \cdots & 1 & 0 & \cdots & 0 \\
 X_{91} & 1 & 0 & 0 & 0 & 0 & 0 & 0 & 0 & 0 & 0 & 0 & 0 & 0 & 1 & 0 & 0 & \cdots & 1 & 0 & \cdots & 1 & 0 & \cdots & 1 & 0 & \cdots & 1 & 1 & \cdots & 1 & 0 & \cdots & 1 & 0 & \cdots & 0 \\
 Y_{12} & 0 & 0 & 0 & 0 & 1 & 0 & 0 & 0 & 0 & 0 & 0 & 0 & 0 & 0 & 1 & 1 & \cdots & 0 & 1 & \cdots & 0 & 0 & \cdots & 0 & 1 & \cdots & 1 & 0 & \cdots & 1 & 0 & \cdots & 0 & 0 & \cdots & 1 \\
 Y_{36} & 0 & 0 & 0 & 0 & 1 & 1 & 0 & 1 & 1 & 1 & 0 & 0 & 0 & 0 & 0 & 0 & \cdots & 0 & 1 & \cdots & 1 & 0 & \cdots & 0 & 1 & \cdots & 1 & 2 & \cdots & 1 & 0 & \cdots & 0 & 1 & \cdots & 1 \\
 Y_{45} & 0 & 0 & 0 & 0 & 1 & 0 & 0 & 1 & 0 & 0 & 0 & 0 & 0 & 0 & 0 & 0 & \cdots & 1 & 1 & \cdots & 1 & 1 & \cdots & 0 & 1 & \cdots & 0 & 1 & \cdots & 1 & 0 & \cdots & 0 & 1 & \cdots & 0 \\
 Y_{69} & 0 & 0 & 0 & 0 & 1 & 1 & 1 & 0 & 0 & 0 & 0 & 1 & 1 & 0 & 0 & 1 & \cdots & 0 & 2 & \cdots & 1 & 1 & \cdots & 0 & 1 & \cdots & 0 & 1 & \cdots & 1 & 1 & \cdots & 0 & 2 & \cdots & 1 \\
 Y_{78} & 0 & 0 & 0 & 0 & 1 & 0 & 0 & 0 & 0 & 0 & 0 & 0 & 1 & 0 & 0 & 0 & \cdots & 0 & 0 & \cdots & 1 & 0 & \cdots & 1 & 0 & \cdots & 1 & 1 & \cdots & 0 & 0 & \cdots & 0 & 1 & \cdots & 1 \\
 Y_{93} & 0 & 0 & 0 & 0 & 1 & 1 & 0 & 0 & 0 & 0 & 1 & 0 & 0 & 1 & 1 & 0 & \cdots & 1 & 1 & \cdots & 2 & 0 & \cdots & 1 & 0 & \cdots & 1 & 1 & \cdots & 2 & 0 & \cdots & 1 & 1 & \cdots & 2 \\
 Z_{24} & 0 & 0 & 0 & 0 & 0 & 1 & 0 & 0 & 0 & 1 & 1 & 0 & 0 & 0 & 0 & 0 & \cdots & 0 & 0 & \cdots & 1 & 0 & \cdots & 0 & 0 & \cdots & 0 & 1 & \cdots & 0 & 1 & \cdots & 0 & 1 & \cdots & 1 \\
 Z_{35} & 0 & 0 & 0 & 1 & 0 & 0 & 0 & 1 & 0 & 1 & 0 & 0 & 0 & 0 & 0 & 0 & \cdots & 0 & 0 & \cdots & 0 & 1 & \cdots & 0 & 1 & \cdots & 0 & 1 & \cdots & 0 & 1 & \cdots & 0 & 1 & \cdots & 0 \\
 Z_{57} & 0 & 0 & 0 & 0 & 0 & 1 & 0 & 0 & 1 & 0 & 0 & 1 & 0 & 0 & 0 & 0 & \cdots & 0 & 1 & \cdots & 0 & 0 & \cdots & 0 & 0 & \cdots & 0 & 0 & \cdots & 1 & 0 & \cdots & 1 & 0 & \cdots & 1 \\
 Z_{68} & 0 & 0 & 0 & 1 & 0 & 0 & 0 & 0 & 0 & 0 & 0 & 1 & 1 & 0 & 0 & 0 & \cdots & 0 & 0 & \cdots & 0 & 1 & \cdots & 1 & 0 & \cdots & 0 & 0 & \cdots & 0 & 1 & \cdots & 1 & 1 & \cdots & 1 \\
 Z_{81} & 0 & 0 & 0 & 0 & 0 & 1 & 1 & 0 & 0 & 0 & 0 & 0 & 0 & 1 & 0 & 0 & \cdots & 0 & 1 & \cdots & 1 & 0 & \cdots & 0 & 0 & \cdots & 0 & 1 & \cdots & 1 & 0 & \cdots & 0 & 1 & \cdots & 0 \\
 Z_{92} & 0 & 0 & 0 & 1 & 0 & 0 & 0 & 0 & 0 & 0 & 0 & 0 & 0 & 1 & 1 & 0 & \cdots & 0 & 0 & \cdots & 0 & 0 & \cdots & 1 & 0 & \cdots & 1 & 0 & \cdots & 1 & 0 & \cdots & 1 & 0 & \cdots & 1 \\
\end{array}
\right)
$}
~,~
\nn\\
\eea
where we note that $o_k^{(1)}, \dots, o_{m}^{(7)}$ are extra GLSM fields \cite{Witten:1993yc}.
The $U(1)$ charges on the GLSM fields
corresponding to the $J$- and $E$-terms
are given by the following charge matrix, 
\beal{es04n04}
&&
Q_{JE}=
\nn\\
&&
\resizebox{0.9\textwidth}{!}{$
\left(
\begin{array}{cccccc|ccccccccc|ccc|ccc|ccc|ccc|ccc|ccc|ccc}
p_{1} & p_{2} & p_{3} & p_{4} & p_{5} & p_{6} & s_{1} & s_{2} & s_{3} & s_{4} & s_{5} & s_{6} & s_{7} & s_{8} & s_{9} & o^{(1)}_{1} & \cdots & o^{(1)}_{3} & o^{(2)}_{1} & \cdots & o^{(2)}_{3} & o^{(3)}_{1} & \cdots & o^{(3)}_{3} & o^{(4)}_{1} & \cdots & o^{(4)}_{9} & o^{(5)}_{1} & \cdots & o^{(5)}_{3} & o^{(6)}_{1} & \cdots & o^{(6)}_{3} & o^{(7)}_{1} & \cdots & o^{(7)}_{3} \\
\hline
4 & 1 & 0 & 0 & 0 & 0 & 3 & 2 & 0 & -1 & 0 & -1 & 0 & 0 & 0 & -1 & \cdots & 0 & 0 & \cdots & 0 & 0 & \cdots & 0 & 0 & \cdots & 0 & 0 & \cdots & 0 & 0 & \cdots & 0 & 0 & \cdots & 1 \\
2 & 0 & 0 & 0 & 0 & 0 & 2 & 1 & 1 & 0 & 0 & -1 & 0 & 0 & 0 & 0 & \cdots & 0 & 0 & \cdots & 0 & 0 & \cdots & 1 & 0 & \cdots & 0 & 0 & \cdots & 0 & 0 & \cdots & 0 & 0 & \cdots & 0 \\
1 & 0 & -1 & 0 & 0 & 0 & 2 & 1 & 0 & 0 & 0 & -1 & 0 & 0 & 0 & 0 & \cdots & 0 & 0 & \cdots & 0 & 0 & \cdots & 0 & 0 & \cdots & 0 & 0 & \cdots & 0 & 0 & \cdots & 1 & 0 & \cdots & 0 \\
1 & 0 & 0 & 1 & 0 & 0 & 1 & 0 & 1 & 0 & 0 & -1 & 0 & 0 & 0 & 0 & \cdots & 0 & 0 & \cdots & 0 & 0 & \cdots & 0 & 0 & \cdots & 0 & 0 & \cdots & 0 & 0 & \cdots & 0 & 0 & \cdots & 0 \\
2 & 2 & 0 & 0 & 0 & 0 & 2 & 1 & -1 & 0 & 0 & 0 & 0 & 0 & 0 & -1 & \cdots & 0 & 0 & \cdots & 0 & 0 & \cdots & 0 & 0 & \cdots & 0 & 0 & \cdots & 1 & 0 & \cdots & 0 & 0 & \cdots & 0 \\
2 & 1 & 1 & 0 & 0 & 0 & 2 & 1 & 0 & 0 & 0 & 0 & 0 & 0 & 0 & -1 & \cdots & 0 & 0 & \cdots & 0 & 0 & \cdots & 0 & 0 & \cdots & 1 & 0 & \cdots & 0 & 0 & \cdots & 0 & 0 & \cdots & 0 \\
2 & 1 & 1 & 0 & 0 & 0 & 1 & 0 & -1 & 0 & 0 & 0 & 0 & 0 & 0 & -1 & \cdots & 0 & 0 & \cdots & 0 & 0 & \cdots & 0 & 0 & \cdots & 0 & 0 & \cdots & 0 & 0 & \cdots & 0 & 0 & \cdots & 0 \\
2 & 1 & 0 & 0 & 0 & 0 & 2 & 1 & 0 & 0 & 0 & 0 & 0 & 0 & 1 & -1 & \cdots & 0 & 0 & \cdots & 0 & 0 & \cdots & 0 & 0 & \cdots & 0 & 0 & \cdots & 0 & 0 & \cdots & 0 & 0 & \cdots & 0 \\
1 & 1 & 0 & 0 & 0 & 0 & 1 & 1 & 0 & 0 & 0 & 0 & 0 & 1 & 0 & 0 & \cdots & 0 & 0 & \cdots & 0 & 0 & \cdots & 0 & 0 & \cdots & 0 & 0 & \cdots & 0 & 0 & \cdots & 0 & 0 & \cdots & 0 \\
1 & 0 & 1 & 0 & 0 & 0 & 1 & 0 & 0 & 0 & 0 & 0 & 0 & 0 & 0 & -1 & \cdots & 0 & 0 & \cdots & 0 & 0 & \cdots & 0 & 0 & \cdots & 0 & 0 & \cdots & 0 & 0 & \cdots & 0 & 0 & \cdots & 0 \\
2 & 1 & 0 & 0 & 0 & 0 & 1 & 1 & 0 & -1 & 0 & -1 & 0 & 0 & 0 & -1 & \cdots & 0 & 0 & \cdots & 0 & 0 & \cdots & 0 & 0 & \cdots & 0 & 0 & \cdots & 0 & 0 & \cdots & 0 & 0 & \cdots & 0 \\
2 & 0 & 0 & 0 & 0 & 0 & 0 & 0 & 1 & -1 & 0 & -1 & 0 & 0 & 0 & 0 & \cdots & 0 & 0 & \cdots & 0 & 0 & \cdots & 0 & 0 & \cdots & 0 & 0 & \cdots & 0 & 0 & \cdots & 0 & 1 & \cdots & 0 \\
1 & 0 & 0 & 0 & 0 & 0 & 1 & 0 & 1 & 0 & 0 & -1 & 0 & 0 & 0 & 0 & \cdots & 0 & 0 & \cdots & 0 & 0 & \cdots & 0 & 0 & \cdots & 0 & 0 & \cdots & 0 & 0 & \cdots & 0 & 0 & \cdots & 0 \\
1 & 0 & 0 & 0 & 0 & 0 & 0 & 0 & 1 & 0 & 0 & -1 & 1 & 0 & 0 & 0 & \cdots & 0 & 0 & \cdots & 0 & 0 & \cdots & 0 & 0 & \cdots & 0 & 0 & \cdots & 0 & 0 & \cdots & 0 & 0 & \cdots & 0 \\
0 & -1 & -1 & 0 & 0 & 0 & 0 & 0 & 1 & -1 & 0 & -1 & 0 & 0 & 0 & 0 & \cdots & 0 & 0 & \cdots & 0 & 0 & \cdots & 0 & 0 & \cdots & 0 & 0 & \cdots & 0 & 1 & \cdots & 0 & 0 & \cdots & 0 \\
0 & -1 & 0 & 0 & 0 & 0 & 0 & -1 & 1 & 0 & 0 & -1 & 0 & 0 & 0 & 0 & \cdots & 0 & 0 & \cdots & 0 & 1 & \cdots & 0 & 0 & \cdots & 0 & 0 & \cdots & 0 & 0 & \cdots & 0 & 0 & \cdots & 0 \\
2 & 1 & 0 & 0 & 0 & 0 & 1 & 1 & 0 & -1 & 0 & 0 & 0 & 0 & 0 & 0 & \cdots & 0 & 0 & \cdots & 1 & 0 & \cdots & 0 & 0 & \cdots & 0 & 0 & \cdots & 0 & 0 & \cdots & 0 & 0 & \cdots & 0 \\
2 & 1 & 1 & 0 & 0 & 0 & 1 & 1 & -1 & -1 & 0 & 0 & 0 & 0 & 0 & -1 & \cdots & 0 & 0 & \cdots & 0 & 0 & \cdots & 0 & 0 & \cdots & 0 & 0 & \cdots & 0 & 0 & \cdots & 0 & 0 & \cdots & 0 \\
1 & 1 & 0 & 0 & 0 & 0 & 0 & 0 & 0 & -1 & 0 & 0 & 0 & 0 & 0 & 0 & \cdots & 0 & 0 & \cdots & 0 & 0 & \cdots & 0 & 0 & \cdots & 0 & 1 & \cdots & 0 & 0 & \cdots & 0 & 0 & \cdots & 0 \\
1 & 0 & 0 & 0 & 0 & 0 & 1 & 1 & 0 & -1 & 0 & 0 & 0 & 0 & 0 & 0 & \cdots & 0 & 0 & \cdots & 0 & 0 & \cdots & 0 & 0 & \cdots & 0 & 0 & \cdots & 0 & 0 & \cdots & 0 & 0 & \cdots & 0 \\
1 & 0 & 0 & 0 & 0 & 0 & 1 & 0 & 0 & 0 & 0 & 0 & 0 & 0 & 0 & 0 & \cdots & 0 & 0 & \cdots & 0 & 0 & \cdots & 0 & 0 & \cdots & 0 & 0 & \cdots & 0 & 0 & \cdots & 0 & 0 & \cdots & 0 \\
2 & 1 & 1 & 0 & 1 & 0 & 1 & 0 & 0 & 0 & 0 & 0 & 0 & 0 & 0 & -1 & \cdots & 0 & 0 & \cdots & 0 & 0 & \cdots & 0 & 0 & \cdots & 0 & 0 & \cdots & 0 & 0 & \cdots & 0 & 0 & \cdots & 0 \\
1 & 1 & 1 & 0 & 0 & 0 & 0 & 0 & -1 & 0 & 0 & 0 & 0 & 0 & 0 & -1 & \cdots & 0 & 1 & \cdots & 0 & 0 & \cdots & 0 & 0 & \cdots & 0 & 0 & \cdots & 0 & 0 & \cdots & 0 & 0 & \cdots & 0 \\
1 & 1 & 0 & 0 & 0 & 1 & 0 & 1 & -1 & -1 & 0 & 0 & 0 & 0 & 0 & 0 & \cdots & 0 & 0 & \cdots & 0 & 0 & \cdots & 0 & 0 & \cdots & 0 & 0 & \cdots & 0 & 0 & \cdots & 0 & 0 & \cdots & 0 \\
1 & 0 & 0 & 0 & 0 & 0 & 1 & 0 & 0 & 0 & 0 & 0 & 0 & 0 & 0 & -1 & \cdots & 0 & 0 & \cdots & 0 & 0 & \cdots & 0 & 0 & \cdots & 0 & 0 & \cdots & 0 & 0 & \cdots & 0 & 0 & \cdots & 0 \\
0 & 0 & -1 & 0 & 0 & 0 & 1 & 0 & 0 & 0 & 0 & 0 & 0 & 0 & 0 & 0 & \cdots & 1 & 0 & \cdots & 0 & 0 & \cdots & 0 & 0 & \cdots & 0 & 0 & \cdots & 0 & 0 & \cdots & 0 & 0 & \cdots & 0 \\
1 & 0 & -1 & 0 & 0 & 0 & 1 & 1 & 0 & -1 & 1 & 0 & 0 & 0 & 0 & 0 & \cdots & 0 & 0 & \cdots & 0 & 0 & \cdots & 0 & 0 & \cdots & 0 & 0 & \cdots & 0 & 0 & \cdots & 0 & 0 & \cdots & 0 \\
1 & 0 & 1 & 0 & 0 & 0 & 0 & 0 & 0 & 0 & 0 & 0 & 0 & 0 & 0 & -1 & \cdots & 0 & 0 & \cdots & 0 & 0 & \cdots & 0 & 0 & \cdots & 0 & 0 & \cdots & 0 & 0 & \cdots & 0 & 0 & \cdots & 0 \\
0 & 0 & 1 & 0 & 0 & 0 & 0 & 0 & -1 & 0 & 0 & 0 & 0 & 0 & 0 & -1 & \cdots & 0 & 0 & \cdots & 0 & 0 & \cdots & 0 & 0 & \cdots & 0 & 0 & \cdots & 0 & 0 & \cdots & 0 & 0 & \cdots & 0 \\
0 & 0 & 1 & 0 & 0 & 0 & 0 & -1 & 0 & 0 & 0 & 0 & 0 & 0 & 0 & -1 & \cdots & 0 & 0 & \cdots & 0 & 0 & \cdots & 0 & 1 & \cdots & 0 & 0 & \cdots & 0 & 0 & \cdots & 0 & 0 & \cdots & 0 \\
\end{array}
\right)
$}
~.~
\nn\\
\eea
The corresponding $D$-term charge matrix takes the following form, 
\beal{es04n05}
&&
Q_{D}=
\nn\\
&&
\resizebox{0.9\textwidth}{!}{$
\left(
\begin{array}{cccccc|ccccccccc|ccc|ccc|ccc|ccc|ccc|ccc|ccc}
p_{1} & p_{2} & p_{3} & p_{4} & p_{5} & p_{6} & s_{1} & s_{2} & s_{3} & s_{4} & s_{5} & s_{6} & s_{7} & s_{8} & s_{9} & o^{(1)}_{1} & \cdots & o^{(1)}_{3} & o^{(2)}_{1} & \cdots & o^{(2)}_{3} & o^{(3)}_{1} & \cdots & o^{(3)}_{3} & o^{(4)}_{1} & \cdots & o^{(4)}_{9} & o^{(5)}_{1} & \cdots & o^{(5)}_{3} & o^{(6)}_{1} & \cdots & o^{(6)}_{3} & o^{(7)}_{1} & \cdots & o^{(7)}_{3} \\
\hline
 1 & 0 & 0 & 0 & 0 & 0 & 1 & 0 & 0 & 0 & 0 & 0 & 0 & 0 & 0 & 0 & \cdots & 0 & 0 & \cdots & 0 & 0 & \cdots & 0 & 0 & \cdots & 0 & 0 & \cdots & 0 & 0 & \cdots & 0 & 0 & \cdots & 0 \\
 -1 & -1 & -1 & 0 & 0 & 0 & -1 & 0 & 0 & -1 & 0 & 0 & 0 & 0 & 0 & 1 & \cdots & 0 & 0 & \cdots & 0 & 0 & \cdots & 0 & 0 & \cdots & 0 & 0 & \cdots & 0 & 0 & \cdots & 0 & 0 & \cdots & 0 \\
 -1 & 0 & 1 & 0 & 0 & 0 & -1 & -1 & 0 & 0 & 0 & 0 & 0 & 0 & 0 & 0 & \cdots & 0 & 0 & \cdots & 0 & 0 & \cdots & 0 & 0 & \cdots & 0 & 0 & \cdots & 0 & 0 & \cdots & 0 & 0 & \cdots & 0 \\
 0 & 0 & 0 & 0 & 0 & 0 & 0 & -1 & 0 & 1 & 0 & 0 & 0 & 0 & 0 & 0 & \cdots & 0 & 0 & \cdots & 0 & 0 & \cdots & 0 & 0 & \cdots & 0 & 0 & \cdots & 0 & 0 & \cdots & 0 & 0 & \cdots & 0 \\
 0 & 0 & 0 & 0 & 0 & 0 & 0 & 1 & -1 & 0 & 0 & 0 & 0 & 0 & 0 & 0 & \cdots & 0 & 0 & \cdots & 0 & 0 & \cdots & 0 & 0 & \cdots & 0 & 0 & \cdots & 0 & 0 & \cdots & 0 & 0 & \cdots & 0 \\
 0 & 0 & 0 & 0 & 0 & 0 & 0 & 0 & 1 & 0 & 0 & -1 & 0 & 0 & 0 & 0 & \cdots & 0 & 0 & \cdots & 0 & 0 & \cdots & 0 & 0 & \cdots & 0 & 0 & \cdots & 0 & 0 & \cdots & 0 & 0 & \cdots & 0 \\
 1 & 0 & 0 & 0 & 0 & 0 & 0 & 0 & 1 & 0 & 0 & 0 & 0 & 0 & 0 & 0 & \cdots & -1 & 0 & \cdots & 0 & 0 & \cdots & 0 & 0 & \cdots & 0 & 0 & \cdots & 0 & 0 & \cdots & 0 & 0 & \cdots & 0 \\
 -1 & 0 & 0 & 0 & 0 & 0 & -1 & 0 & -1 & 0 & 0 & 1 & 0 & 0 & 0 & 0 & \cdots & 1 & 0 & \cdots & 0 & 0 & \cdots & 0 & 0 & \cdots & 0 & 0 & \cdots & 0 & 0 & \cdots & 0 & 0 & \cdots & 0 \\
\end{array}
\right)
$}
~.~
\nn\\
\eea

\begin{figure}[ht!!]
\centering
\includegraphics[width=0.35\textwidth]{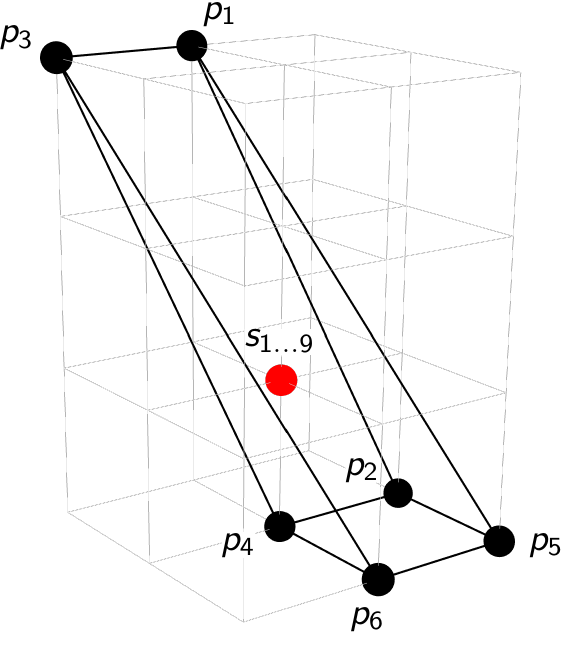}
\caption{
The toric diagram for the 
$D_3/\mathbb{Z}_3 \ (1,1,1,1,2)$ model.
\label{fig_428_toric}
}
\end{figure}

The resulting toric diagram for the
$D_3/\mathbb{Z}_3 \ (1,1,1,1,2)$ model
is given by, 
\beal{es04n06}
&&
G_{t}=
\resizebox{0.8\textwidth}{!}{$
\left(
\begin{array}{cccccc|ccc|ccc|ccc|ccc|ccc|ccc|ccc|ccc}
p_{1} & p_{2} & p_{3} & p_{4} & p_{5} & p_{6} & s_{1} & \cdots & s_{9} & o^{(1)}_{1} & \cdots & o^{(1)}_{3} & o^{(2)}_{1} & \cdots & o^{(2)}_{3} & o^{(3)}_{1} & \cdots & o^{(3)}_{3} & o^{(4)}_{1} & \cdots & o^{(4)}_{9} & o^{(5)}_{1} & \cdots & o^{(5)}_{3} & o^{(6)}_{1} & \cdots & o^{(6)}_{3} & o^{(7)}_{1} & \cdots & o^{(7)}_{3} \\
\hline
-1 & 0 & -1 & 0 & 1 & 1 & 0 & \cdots & 0 & -1 & \cdots & -1 & 1 & \cdots & 1 & 0 & \cdots & 0 & 0 & \cdots & 0 & 0 & \cdots & 0 & -1 & \cdots & -1 & 1 & \cdots & 1 \\
 0 & 1 & -1 & 0 & 1 & 0 & 0 & \cdots & 0 &  0 & \cdots &  0 & 1 & \cdots & 1 & 1 & \cdots & 1 & 1 & \cdots & 1 & 0 & \cdots & 0 &  0 & \cdots &  0 & 1 & \cdots & 1 \\
 3 & 0 &  3 & 0 & 0 & 0 & 1 & \cdots & 1 &  4 & \cdots &  4 & 1 & \cdots & 1 & 1 & \cdots & 1 & 2 & \cdots & 2 & 4 & \cdots & 4 &  4 & \cdots &  4 & 1 & \cdots & 1 \\
\hline
 1 & 1 &  1 & 1 & 1 & 1 & 1 & \cdots & 1 &  2 & \cdots &  2 & 2 & \cdots & 2 & 2 & \cdots & 2 & 2 & \cdots & 2 & 3 & \cdots & 3 &  3 & \cdots &  3 & 3 & \cdots & 3 \\
\end{array}
\right)
$}
~,~
\nn\\
\eea
where \fref{fig_428_toric} illustrates the toric diagram.

%-------------------
%-------------------

%-------------------
\begin{table}[htt!!]
\centering
\begin{tabular}{|c|c|c|c|l|}
\hline
\; & $U(1)_{f_1}$ & $U(1)_{f_2}$ & $U(1)_{f_3}$ &  fugacity \\
\hline
$p_1$ & $+1$ & $0$ & $0$ &  $t_1=f_1 t$ \\
$p_2$ & $0$ & $0$ & $+1$ &  $t_2=f_3 t$\\
$p_3$ & $0$ & $0$ & $-1$ &  $t_3=f_3^{-1} t$\\
$p_4$ & $0$ & $+1$ & $0$ & $t_4=f_2 t$ \\
$p_5$ & $0$ & $-1$ & $0$ & $t_5=f_2^{-1} t$ \\
$p_6$ & $-1$ & $0$ & $0$ & $t_6=f_1^{-1} t$ \\
\hline
\end{tabular}
\caption{
Mesonic flavor symmetry of the $D_3/\mathbb{Z}_3 \ (1,1,1,1,2)$ model 
and charges on the extremal GLSM fields $p_a$. 
Here, the fugacity $t$ counts the degree in extremal GLSM fields $p_a$. 
\label{tab_04n01}}
\end{table}
%-------------------

The global symmetry of the brane brick model is not enhanced and accordingly has the following form,
\beal{es04h07}
U(1)_{f_1} \times U(1)_{f_2} \times U(1)_{f_3} \times U(1)_R
~.~
\eea
The refined Hilbert series of the mesonic moduli space takes the following form,
\beal{es04h08}
&&
g(t_a; \mathcal{M}^{mes})
=
\frac{
P(t_a ; \mathcal{M}^{mes}) 
\left(1 - t_1 t_2 t_3 t_4 t_5 t_6\right)}{\left(1 - t_1^3 t_3^3\right) \left(1 - t_2^3 t_4^3\right) \left(1 - t_1^3 t_2^3 t_5^3\right) \left(1 - t_3^3 t_4^3 t_6^3\right) \left(1 - t_5^3 t_6^3\right)}
~,~
\eea
where $t_a$ is the fugacity corresponding to the extremal GLSM field $p_a$,
and the numerator factor
$P(t_a ; \mathcal{M}^{mes})$ is given in full in appendix \sref{appCa}.
By setting $t_a = t$, 
we can unrefine the Hilbert series such that it takes the following form,
\beal{es04n09}
&&
g(t; \mathcal{M}^{mes})
=
(1-2t +t^2 +t^3 -2t^4 +4t^5 +3t^6 -9t^7 +3t^8 +4t^9 -2 t^{10} 
\nn\\
&&
\hspace{1cm}
+t^{11} + t^{12} -2t^{13} +t^{14}) 
\times
\frac{
(1+t+t^2)
}{
(1-t) (1-t^6)^2 (1-t^9)
}
~,~
\eea
where the palindromic numerator indicates that the mesonic moduli space is Calabi-Yau. 

The Hilbert series refined in terms of
mesonic flavor fugacities summarized in \tref{tab_04n01}
is given by, 
\beal{es04n10}
&&
g(f_1,f_2,f_3,t; \mathcal{M}^{mes})
=
\frac{
P(f_1,f_2,f_3,t;\mathcal{M}^{mes})
(1-t^6)
}{
(1-f_1^{-3} f_2^{-3} t^6) (1-f_1^3 f_3^{-3} t^6) (1- f_2^3 f_3^3 t^6) 
}
\nn\\
&&
\hspace{1cm}
\times
\frac{1}{
(1-f_1^{-3} f_2^3 f_3^{-3} t^9) (1-f_1^3 f_2^{-3} f_3^3 t^9)
}
~,~
\eea
where the complete expression for the numerator factor $P(f_1,f_2,f_3,t;D_3/\mathbb{Z}_3 \ (1,1,1,1,2))$ is presented in appendix \sref{appCa}.
The corresponding plethystic logarithm is given by,
\beal{es04n12}
&&
\text{PL}[g(f_1,f_2,f_3,t; \mathcal{M}^{mes})]=
({f_1^{-1} f_2^{2}+f_2 f_3^{-2}+f_1^{-2} f_3^{-1}}) t^5
+(1 +f_1^{-3} f_2^{-3}+f_1^{3} f_3^{-3}
\nn\\
&&
\hspace{1cm}
+f_1 f_2^{-1} f_3^{-2}+f_1^{-1} f_2^{-2} f_3^{-1}+f_1^{2} f_2 f_3^{-1}+f_1^{-2} f_2^{-1} f_3+f_1 f_2^{2} f_3+f_1^{-1} f_2 f_3^{2}+f_2^{3} f_3^{3}) t^{6}
\nn\\
&&
\hspace{1cm}
+ (f_1 f_2^{-2}+f_1^{3} f_2^{-1} f_3^{-1}+f_1^{2} f_3+f_1^{-1} f_2^{-3} f_3+f_2^{-1} f_3^{2}+f_1 f_2 f_3^{3}) t^{7}
+( f_1^{3} f_2^{-2} f_3
\nn\\
&&
\hspace{1cm}
+f_1 f_2^{-3} f_3^{2}+f_1^{2} f_2^{-1} f_3^{3} ) t^8
+( f_1^{-3} f_2^{3} f_3^{-3}+f_1^{3} f_2^{-3} f_3^{3} ) t^9
- (2 f_1^{-1} f_2^2 + 2 f_2 f_3^{-2} 
\nn\\
&&
\hspace{1cm}
+ 2 f_1^{-2} f_3^{-1} + f_1^{-4} f_2^{-1}+f_1 f_3^{-4}+f_1^{-1} f_2^{-1} f_3^{-3}+f_1^{2} f_2^{2} f_3^{-3}+f_1^{-3} f_2^{-2} f_3^{-2}+f_1 f_2^{3} f_3^{-1}
\nn\\
&&
\hspace{1cm}
+f_1^{-3} f_2 f_3+f_2^{4} f_3+f_1^{-2} f_2^{3} f_3^{2}) t^{11}
+ \dots ~,~
\eea
where we note that the infinite expansion of the plethystic logarithm indicates a non-complete intersection as the mesonic moduli space for the 
$D_3/\mathbb{Z}_3 \ (1,1,1,1,2)$ model.
The generators of the mesonic moduli space correspond to the first positive terms of the expansion
and are summarized with their mesonic flavor charges in \tref{tab_04n02}.
\\

%-------------------
 \begin {table}[ht!!]
\centering
\begin {tabular} {|c|c|ccc|}
\hline
PL term & generator & $U(1)_{f_1}$ & $U(1)_{f_2}$ & $U(1)_{f_3}$ 
\\
\hline
\multirow{1}{*}{$+f_1^{-3} f_2^{-3}  t^{6}$}
& $p_5^3 p_6^3 ~s $ & $-3$ & $-3$ & $0$  \\
\hline
\multirow{1}{*}{$+f_1^{-2}  f_3^{-1} t^{5}$}
& $p_3 p_4 p_5 p_6^2 ~s $ & $-2$ & $0$ & $-1$  \\
\hline
\multirow{1}{*}{$+f_1^{-3} f_2^{3} f_3^{-3} t^{9}$}
& $p_3^3 p_4^3 p_6^3 ~s^2 $ & $-3$ & $3$ & $-3$  \\
\hline
\multirow{1}{*}{$+f_1^{-2} f_2^{-1} f_3 t^{6}$}
& $p_2 p_4 p_5^2 p_6^2 ~s $ & $-2$ & $-1$ & $1$  \\
\hline
\multirow{1}{*}{$+f_1^{-1} f_2^{2}  t^{5}$}
& $p_2 p_3 p_4^2 p_6 ~s$ & $-1$ & $2$ & $0$  \\
\hline
\multirow{1}{*}{$+f_1^{-1} f_2 f_3^{2} t^{6}$}
& $p_2^2 p_4^2 p_5 p_6 ~s $ & $-1$ & $1$ & $2$  \\
\hline
\multirow{1}{*}{$+ f_2^{3} f_3^{3} t^{6}$}
& $p_2^3 p_4^3 ~s  $ & $0$ & $3$ & $3$  \\
\hline
\multirow{1}{*}{$+f_1^{-1} f_2^{-2} f_3^{-1} t^{6}$}
& $p_1 p_3 p_5^2 p_6^2 ~s $ & $-1$ & $-2$ & $-1$  \\
\hline
\multirow{1}{*}{$+ f_2 f_3^{-2} t^{5}$}
& $p_1 p_3^2 p_4 p_6 ~s $ & $0$ & $1$ & $-2$  \\
\hline
\multirow{1}{*}{$+f_1^{-1} f_2^{-3} f_3 t^{7}$}
& $p_1 p_2 p_5^3 p_6^2 ~s $ & $-1$ & $-3$ & $1$  \\
\hline
\multirow{1}{*}{$+   t^{6}$}
& $p_1 p_2 p_3 p_4 p_5 p_6 ~s $ & $0$ & $0$ & $0$  \\
\hline
\multirow{1}{*}{$+ f_2^{-1} f_3^{2} t^{7}$}
& $p_1 p_2^2 p_4 p_5^2 p_6 ~s$ & $0$ & $-1$ & $2$  \\
\hline
\multirow{1}{*}{$+f_1 f_2^{2} f_3 t^{6}$}
& $p_1 p_2^2 p_3 p_4^2 ~s$ & $1$ & $2$ & $1$  \\
\hline
\multirow{1}{*}{$+f_1 f_2 f_3^{3} t^{7}$}
& $p_1 p_2^3 p_4^2 p_5 ~s $ & $1$ & $1$ & $3$  \\
\hline
\multirow{1}{*}{$+f_1 f_2^{-1} f_3^{-2} t^{6}$}
& $p_1^2 p_3^2 p_5 p_6 ~s$ & $1$ & $-1$ & $-2$  \\
\hline
\multirow{1}{*}{$+f_1 f_2^{-2}  t^{7}$}
& $p_1^2 p_2 p_3 p_5^2 p_6 ~s$ & $1$ & $-2$ & $0$  \\
\hline
\multirow{1}{*}{$+f_1^{2} f_2 f_3^{-1} t^{6}$}
& $p_1^2 p_2 p_3^2 p_4 ~s $ & $2$ & $1$ & $-1$  \\
\hline
\multirow{1}{*}{$+f_1 f_2^{-3} f_3^{2} t^{8}$}
& $p_1^2 p_2^2 p_5^3 p_6 ~s $ & $1$ & $-3$ & $2$  \\
\hline
\multirow{1}{*}{$+f_1^{2}  f_3 t^{7}$}
& $p_1^2 p_2^2 p_3 p_4 p_5 ~s $ & $2$ & $0$ & $1$  \\
\hline
\multirow{1}{*}{$+f_1^{2} f_2^{-1} f_3^{3} t^{8}$}
& $p_1^2 p_2^3 p_4 p_5^2 ~s $ & $2$ & $-1$ & $3$  \\
\hline
\multirow{1}{*}{$+f_1^{3}  f_3^{-3} t^{6}$}
& $p_1^3 p_3^3 ~s $ & $3$ & $0$ & $-3$  \\
\hline
\multirow{1}{*}{$+f_1^{3} f_2^{-1} f_3^{-1} t^{7}$}
& $p_1^3 p_2 p_3^2 p_5 ~s $ & $3$ & $-1$ & $-1$  \\
\hline
\multirow{1}{*}{$+f_1^{3} f_2^{-2} f_3 t^{8}$}
& $p_1^3 p_2^2 p_3 p_5^2 ~s $ & $3$ & $-2$ & $1$  \\
\hline
\multirow{1}{*}{$+f_1^{3} f_2^{-3} f_3^{3} t^{9}$}
& $p_1^3 p_2^3 p_5^3 ~s $ & $3$ & $-3$ & $3$  \\
\hline
\end{tabular}
\caption{
Generators of the $D_3/\mathbb{Z}_3 \ (1,1,1,1,2)$ model in terms of GLSM fields and their corresponding mesonic flavor symmetry charges. Here, we denote $s=\prod_{i=1}^{9}s_i$,
and set the extra GLSM fields to 1. \label{tab_04n02}
}
%-------------------

\end{table}
%-------------------

\newpage

%=================================================================
\subsubsection{$D_3/ \mathbb{Z}_2 \times \mathbb{Z}_2 \  \left(\begin{array}{@{}c@{,\;}c@{,\;}c@{,\;}c@{,\;}c@{}} 0 & 1 & 1 & 0 & 0 \\ 1 & 0 & 1 & 1 & 1 \end{array}\right)$ \label{sec0415}}

%-------------------
\begin{figure}[H]
    \centering
   \includegraphics[width=0.55\textwidth]{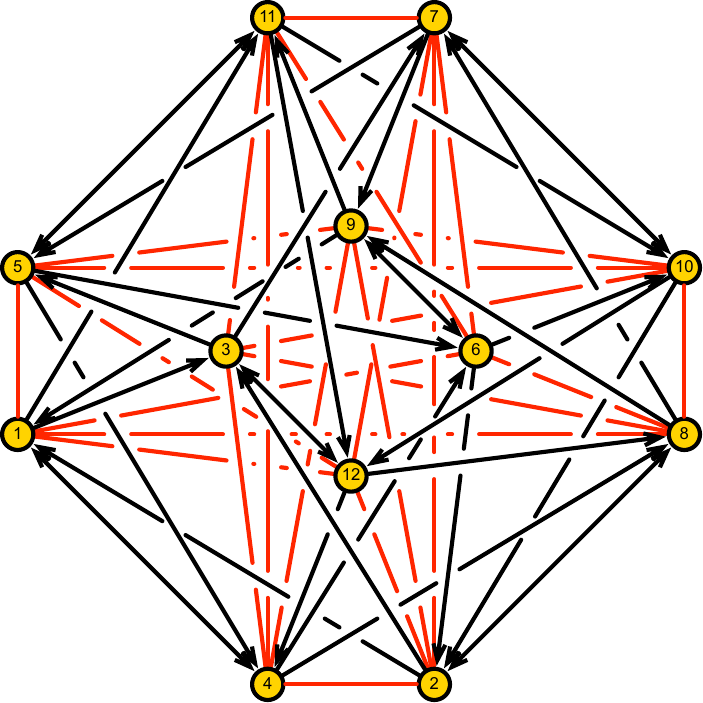}
    \caption{
    The quiver for the $D_3/ \mathbb{Z}_2 \times \mathbb{Z}_2 \ \left(\begin{array}{@{}c@{,\;}c@{,\;}c@{,\;}c@{,\;}c@{}} 0 & 1 & 1 & 0 & 0 \\ 1 & 0 & 1 & 1 & 1 \end{array}\right)$ model.
    \label{fig_429_quiver}
    }
\end{figure}
%-------------------

The $J$- and $E$-terms for the brane brick model corresponding to the abelian orbifold of the form
$D_3/ \mathbb{Z}_2 \times \mathbb{Z}_2 \ \left(\begin{array}{@{}c@{,\;}c@{,\;}c@{,\;}c@{,\;}c@{}} 0 & 1 & 1 & 0 & 0 \\ 1 & 0 & 1 & 1 & 1 \end{array}\right)$ are as follows, 
\beal{es04o01}
\begin{array}{rclccclcccc}
&& && &J& && &E& \\
&&\Lambda^{(1)}_{5, 1} &:& & X_{1, 3} \cdot X_{3, 7} \cdot Y_{7, 5} - Y_{1, 11} \cdot X_{11, 5}& && & D_{5, 6} \cdot Z_{6, 2} \cdot Z_{2, 1} - Z_{5, 4} \cdot D_{4, 1}& \\ 
 &&\Lambda^{(1)}_{2, 4} &:& & X_{4, 6} \cdot X_{6, 10} \cdot Y_{10, 2} - Y_{4, 8} \cdot X_{8, 2}& && & D_{2, 3} \cdot Z_{3, 5} \cdot Z_{5, 4} - Z_{2, 1} \cdot D_{1, 4}& \\ 
 &&\Lambda^{(1)}_{11, 7} &:& & X_{7, 9} \cdot X_{9, 1} \cdot Y_{1, 11} - Y_{7, 5} \cdot X_{5, 11}& && & D_{11, 12} \cdot Z_{12, 8} \cdot Z_{8, 7} - Z_{11, 10} \cdot D_{10, 7}& \\ 
 &&\Lambda^{(1)}_{8, 10} &:& & X_{10, 12} \cdot X_{12, 4} \cdot Y_{4, 8} - Y_{10, 2} \cdot X_{2, 8}& && & D_{8, 9} \cdot Z_{9, 11} \cdot Z_{11, 10} - Z_{8, 7} \cdot D_{7, 10}& \\ 
 &&\Lambda^{(2)}_{7, 2} &:& & Z_{2, 1} \cdot X_{1, 3} \cdot X_{3, 7} - X_{2, 8} \cdot Z_{8, 7}& && & D_{7, 10} \cdot Y_{10, 2} - Y_{7, 5} \cdot D_{5, 6} \cdot Z_{6, 2}& \\ 
 &&\Lambda^{(2)}_{10, 5} &:& & Z_{5, 4} \cdot X_{4, 6} \cdot X_{6, 10} - X_{5, 11} \cdot Z_{11, 10}& && & D_{10, 7} \cdot Y_{7, 5} - Y_{10, 2} \cdot D_{2, 3} \cdot Z_{3, 5}& \\ 
 &&\Lambda^{(2)}_{1, 8} &:& & Z_{8, 7} \cdot X_{7, 9} \cdot X_{9, 1} - X_{8, 2} \cdot Z_{2, 1}& && & D_{1, 4} \cdot Y_{4, 8} - Y_{1, 11} \cdot D_{11, 12} \cdot Z_{12, 8}& \\ 
 &&\Lambda^{(2)}_{4, 11} &:& & Z_{11, 10} \cdot X_{10, 12} \cdot X_{12, 4} - X_{11, 5} \cdot Z_{5, 4}& && & D_{4, 1} \cdot Y_{1, 11} - Y_{4, 8} \cdot D_{8, 9} \cdot Z_{9, 11}& \\ 
 &&\Lambda^{(3)}_{12, 1} &:& & X_{1, 3} \cdot Y_{3, 12} - Y_{1, 11} \cdot Z_{11, 10} \cdot X_{10, 12}& && & X_{12, 4} \cdot D_{4, 1} - Z_{12, 8} \cdot D_{8, 9} \cdot X_{9, 1}& \\ 
 &&\Lambda^{(3)}_{9, 4} &:& & X_{4, 6} \cdot Y_{6, 9} - Y_{4, 8} \cdot Z_{8, 7} \cdot X_{7, 9}& && & X_{9, 1} \cdot D_{1, 4} - Z_{9, 11} \cdot D_{11, 12} \cdot X_{12, 4}& \\ 
 &&\Lambda^{(3)}_{6, 7} &:& & X_{7, 9} \cdot Y_{9, 6} - Y_{7, 5} \cdot Z_{5, 4} \cdot X_{4, 6}& && & X_{6, 10} \cdot D_{10, 7} - Z_{6, 2} \cdot D_{2, 3} \cdot X_{3, 7}& \\ 
 &&\Lambda^{(3)}_{3, 10} &:& & X_{10, 12} \cdot Y_{12, 3} - Y_{10, 2} \cdot Z_{2, 1} \cdot X_{1, 3}& && & X_{3, 7} \cdot D_{7, 10} - Z_{3, 5} \cdot D_{5, 6} \cdot X_{6, 10}& \\ 
 &&\Lambda^{(4)}_{4, 3} &:& & X_{3, 7} \cdot Y_{7, 5} \cdot Z_{5, 4} - Y_{3, 12} \cdot X_{12, 4}& && & D_{4, 1} \cdot X_{1, 3} - X_{4, 6} \cdot Z_{6, 2} \cdot D_{2, 3}& \\ 
 &&\Lambda^{(4)}_{1, 6} &:& & X_{6, 10} \cdot Y_{10, 2} \cdot Z_{2, 1} - Y_{6, 9} \cdot X_{9, 1}& && & D_{1, 4} \cdot X_{4, 6} - X_{1, 3} \cdot Z_{3, 5} \cdot D_{5, 6}& \\ 
 &&\Lambda^{(4)}_{10, 9} &:& & X_{9, 1} \cdot Y_{1, 11} \cdot Z_{11, 10} - Y_{9, 6} \cdot X_{6, 10}& && & D_{10, 7} \cdot X_{7, 9} - X_{10, 12} \cdot Z_{12, 8} \cdot D_{8, 9}& \\ 
 &&\Lambda^{(4)}_{7, 12} &:& & X_{12, 4} \cdot Y_{4, 8} \cdot Z_{8, 7} - Y_{12, 3} \cdot X_{3, 7}& && & D_{7, 10} \cdot X_{10, 12} - X_{7, 9} \cdot Z_{9, 11} \cdot D_{11, 12}& \\ 
 &&\Lambda^{(5)}_{8, 3} &:& & Y_{3, 12} \cdot Z_{12, 8} - Z_{3, 5} \cdot Z_{5, 4} \cdot Y_{4, 8}& && & D_{8, 9} \cdot X_{9, 1} \cdot X_{1, 3} - X_{8, 2} \cdot D_{2, 3}& \\ 
 &&\Lambda^{(5)}_{11, 6} &:& & Y_{6, 9} \cdot Z_{9, 11} - Z_{6, 2} \cdot Z_{2, 1} \cdot Y_{1, 11}& && & D_{11, 12} \cdot X_{12, 4} \cdot X_{4, 6} - X_{11, 5} \cdot D_{5, 6}& \\ 
 &&\Lambda^{(5)}_{2, 9} &:& & Y_{9, 6} \cdot Z_{6, 2} - Z_{9, 11} \cdot Z_{11, 10} \cdot Y_{10, 2}& && & D_{2, 3} \cdot X_{3, 7} \cdot X_{7, 9} - X_{2, 8} \cdot D_{8, 9}& \\ 
 &&\Lambda^{(5)}_{5, 12} &:& & Y_{12, 3} \cdot Z_{3, 5} - Z_{12, 8} \cdot Z_{8, 7} \cdot Y_{7, 5}& && & D_{5, 6} \cdot X_{6, 10} \cdot X_{10, 12} - X_{5, 11} \cdot D_{11, 12}& \\ 
 &&\Lambda^{(6)}_{11, 3} &:& & Z_{3, 5} \cdot X_{5, 11} - X_{3, 7} \cdot X_{7, 9} \cdot Z_{9, 11}& && & D_{11, 12} \cdot Y_{12, 3} - Z_{11, 10} \cdot Y_{10, 2} \cdot D_{2, 3}& \\ 
 &&\Lambda^{(6)}_{8, 6} &:& & Z_{6, 2} \cdot X_{2, 8} - X_{6, 10} \cdot X_{10, 12} \cdot Z_{12, 8}& && & D_{8, 9} \cdot Y_{9, 6} - Z_{8, 7} \cdot Y_{7, 5} \cdot D_{5, 6}& \\ 
 &&\Lambda^{(6)}_{5, 9} &:& & Z_{9, 11} \cdot X_{11, 5} - X_{9, 1} \cdot X_{1, 3} \cdot Z_{3, 5}& && & D_{5, 6} \cdot Y_{6, 9} - Z_{5, 4} \cdot Y_{4, 8} \cdot D_{8, 9}& \\ 
 &&\Lambda^{(6)}_{2, 12} &:& & Z_{12, 8} \cdot X_{8, 2} - X_{12, 4} \cdot X_{4, 6} \cdot Z_{6, 2}& && & D_{2, 3} \cdot Y_{3, 12} - Z_{2, 1} \cdot Y_{1, 11} \cdot D_{11, 12}&
 \end{array}
 ~.~
 \nn\\
 \eea
The corresponding quiver diagram is shown in \fref{fig_429_quiver}.
The $J$- and $E$-terms
come from the general formula in \eqref{es03c07} 
with the following relabelling of indices, 
\beal{es04o02}
&
\left[1,\ba{c}0 \\ 0\ea\right] \rightarrow 1~,~  \left[2,\ba{c}0 \\ 0\ea\right] \rightarrow 2~,~  \left[3,\ba{c}0 \\ 0\ea\right] \rightarrow 3~,~  
\left[1,\ba{c}0 \\ 1\ea\right] \rightarrow 4~,~  
&
\nn\\
&
\left[2,\ba{c}0 \\ 1\ea\right] \rightarrow 5~,~  \left[3,\ba{c}0 \\ 1\ea\right] \rightarrow 6~,~ 
\left[1,\ba{c}1 \\ 0\ea\right] \rightarrow 7~,~  \left[2,\ba{c}1 \\ 0\ea\right] \rightarrow 8~,~ 
&
\nn\\
&
 \left[3,\ba{c}1 \\ 0\ea\right] \rightarrow 9~,~  
\left[1,\ba{c}1 \\ 1\ea\right] \rightarrow 10~,~  \left[2,\ba{c}1 \\ 1\ea\right] \rightarrow 11~,~  \left[3,\ba{c}1 \\ 1\ea\right] \rightarrow 12~.~
&
\eea

Using the forward algorithm for brane brick models, 
we obtain the
$P$-matrix, which takes the form,
\beal{es04o03}
&&
P=
\resizebox{0.85\textwidth}{!}{$
\left(
\begin{array}{c|cccccc|cc|cc|cccc|cccc|ccc}
 & p_1 & p_2 & p_3 & p_4 & p_5 & p_6 & q^{(1)}_1 & q^{(1)}_2 & q^{(2)}_1 & q^{(2)}_2 & q^{(3)}_1 & q^{(3)}_2 & q^{(3)}_3 & q^{(3)}_4 & q^{(4)}_1 & q^{(4)}_2 & q^{(4)}_3 & q^{(4)}_4 & o^{(1)}_{1} & \cdots & o^{(20)}_{16}
\\
\hline
D_{1,4} & 0 & 1 & 0 & 0 & 1 & 0 & 0 & 0 & 0 & 1 & 1 & 0 & 1 & 0 & 0 & 1 & 0 & 0 & 1
   & \cdots & 0 \\
 D_{2,3} & 0 & 0 & 0 & 0 & 1 & 0 & 0 & 0 & 1 & 0 & 0 & 0 & 1 & 0 & 0 & 0 & 0 & 0 & 0
   & \cdots & 2 \\
 D_{4,1} & 0 & 1 & 0 & 0 & 1 & 0 & 0 & 0 & 1 & 0 & 0 & 1 & 0 & 1 & 1 & 0 & 0 & 0 & 0
   & \cdots & 2 \\
 D_{5,6} & 0 & 0 & 0 & 0 & 1 & 0 & 0 & 0 & 0 & 1 & 0 & 0 & 0 & 1 & 0 & 0 & 0 & 0 & 0
   & \cdots & 0 \\
 D_{7,10} & 0 & 1 & 0 & 0 & 1 & 0 & 0 & 0 & 0 & 1 & 1 & 0 & 0 & 1 & 1 & 0 & 0 & 0 &
   0 & \cdots & 1 \\
 D_{8,9} & 0 & 0 & 0 & 0 & 1 & 0 & 0 & 0 & 1 & 0 & 0 & 0 & 0 & 1 & 0 & 0 & 0 & 0 & 1
   & \cdots & 0 \\
 D_{10,7} & 0 & 1 & 0 & 0 & 1 & 0 & 0 & 0 & 1 & 0 & 0 & 1 & 1 & 0 & 0 & 1 & 0 & 0 &
   1 & \cdots & 1 \\
 D_{11,12} & 0 & 0 & 0 & 0 & 1 & 0 & 0 & 0 & 0 & 1 & 0 & 0 & 1 & 0 & 0 & 0 & 0 & 0 &
   0 & \cdots & 0 \\
 X_{1,3} & 0 & 0 & 1 & 0 & 0 & 0 & 0 & 1 & 0 & 0 & 0 & 0 & 1 & 0 & 0 & 0 & 0 & 0 & 0
   & \cdots & 1 \\
 X_{2,8} & 0 & 0 & 1 & 0 & 0 & 1 & 1 & 0 & 0 & 0 & 0 & 0 & 1 & 0 & 0 & 1 & 1 & 0 & 1
   & \cdots & 3 \\
 X_{3,7} & 0 & 0 & 0 & 0 & 0 & 1 & 0 & 0 & 0 & 0 & 0 & 0 & 0 & 0 & 0 & 1 & 1 & 0 & 2
   & \cdots & 0 \\
 X_{4,6} & 0 & 0 & 1 & 0 & 0 & 0 & 0 & 1 & 0 & 0 & 0 & 0 & 0 & 1 & 0 & 0 & 0 & 0 & 0
   & \cdots & 1 \\
 X_{5,11} & 0 & 0 & 1 & 0 & 0 & 1 & 1 & 0 & 0 & 0 & 0 & 0 & 0 & 1 & 1 & 0 & 1 & 0 &
   1 & \cdots & 1 \\
 X_{6,10} & 0 & 0 & 0 & 0 & 0 & 1 & 0 & 0 & 0 & 0 & 0 & 0 & 0 & 0 & 1 & 0 & 1 & 0 &
   1 & \cdots & 1 \\
 X_{7,9} & 0 & 0 & 1 & 0 & 0 & 0 & 1 & 0 & 0 & 0 & 0 & 0 & 0 & 1 & 0 & 0 & 0 & 0 & 0
   & \cdots & 1 \\
 X_{8,2} & 0 & 0 & 1 & 0 & 0 & 1 & 0 & 1 & 0 & 0 & 0 & 0 & 0 & 1 & 1 & 0 & 0 & 1 & 1
   & \cdots & 0 \\
 X_{9,1} & 0 & 0 & 0 & 0 & 0 & 1 & 0 & 0 & 0 & 0 & 0 & 0 & 0 & 0 & 1 & 0 & 0 & 1 & 0
   & \cdots & 1 \\
 X_{10,12} & 0 & 0 & 1 & 0 & 0 & 0 & 1 & 0 & 0 & 0 & 0 & 0 & 1 & 0 & 0 & 0 & 0 & 0 &
   0 & \cdots & 0 \\
 X_{11,5} & 0 & 0 & 1 & 0 & 0 & 1 & 0 & 1 & 0 & 0 & 0 & 0 & 1 & 0 & 0 & 1 & 0 & 1 &
   1 & \cdots & 2 \\
 X_{12,4} & 0 & 0 & 0 & 0 & 0 & 1 & 0 & 0 & 0 & 0 & 0 & 0 & 0 & 0 & 0 & 1 & 0 & 1 &
   1 & \cdots & 1 \\
 Y_{1,11} & 0 & 0 & 0 & 1 & 0 & 0 & 0 & 0 & 0 & 0 & 1 & 0 & 0 & 0 & 0 & 0 & 1 & 0 &
   1 & \cdots & 0 \\
 Y_{3,12} & 1 & 0 & 0 & 1 & 0 & 0 & 1 & 0 & 0 & 1 & 1 & 0 & 0 & 0 & 0 & 0 & 1 & 0 &
   1 & \cdots & 0 \\
 Y_{4,8} & 0 & 0 & 0 & 1 & 0 & 0 & 0 & 0 & 0 & 0 & 0 & 1 & 0 & 0 & 0 & 0 & 1 & 0 & 0
   & \cdots & 2 \\
 Y_{6,9} & 1 & 0 & 0 & 1 & 0 & 0 & 1 & 0 & 1 & 0 & 0 & 1 & 0 & 0 & 0 & 0 & 1 & 0 & 1
   & \cdots & 2 \\
 Y_{7,5} & 0 & 0 & 0 & 1 & 0 & 0 & 0 & 0 & 0 & 0 & 1 & 0 & 0 & 0 & 0 & 0 & 0 & 1 & 0
   & \cdots & 1 \\
 Y_{9,6} & 1 & 0 & 0 & 1 & 0 & 0 & 0 & 1 & 0 & 1 & 1 & 0 & 0 & 0 & 0 & 0 & 0 & 1 & 0
   & \cdots & 1 \\
 Y_{10,2} & 0 & 0 & 0 & 1 & 0 & 0 & 0 & 0 & 0 & 0 & 0 & 1 & 0 & 0 & 0 & 0 & 0 & 1 &
   0 & \cdots & 0 \\
 Y_{12,3} & 1 & 0 & 0 & 1 & 0 & 0 & 0 & 1 & 1 & 0 & 0 & 1 & 0 & 0 & 0 & 0 & 0 & 1 &
   0 & \cdots & 3 \\
 Z_{2,1} & 1 & 0 & 0 & 0 & 0 & 0 & 1 & 0 & 1 & 0 & 0 & 0 & 0 & 0 & 0 & 0 & 0 & 0 & 0
   & \cdots & 2 \\
 Z_{3,5} & 0 & 1 & 0 & 0 & 0 & 0 & 0 & 0 & 0 & 0 & 1 & 0 & 0 & 0 & 0 & 1 & 0 & 0 & 1
   & \cdots & 0 \\
 Z_{5,4} & 1 & 0 & 0 & 0 & 0 & 0 & 1 & 0 & 0 & 1 & 0 & 0 & 0 & 0 & 0 & 0 & 0 & 0 & 0
   & \cdots & 0 \\
 Z_{6,2} & 0 & 1 & 0 & 0 & 0 & 0 & 0 & 0 & 0 & 0 & 0 & 1 & 0 & 0 & 1 & 0 & 0 & 0 & 0
   & \cdots & 0 \\
 Z_{8,7} & 1 & 0 & 0 & 0 & 0 & 0 & 0 & 1 & 1 & 0 & 0 & 0 & 0 & 0 & 0 & 0 & 0 & 0 & 1
   & \cdots & 0 \\
 Z_{9,11} & 0 & 1 & 0 & 0 & 0 & 0 & 0 & 0 & 0 & 0 & 1 & 0 & 0 & 0 & 1 & 0 & 0 & 0 &
   0 & \cdots & 0 \\
 Z_{11,10} & 1 & 0 & 0 & 0 & 0 & 0 & 0 & 1 & 0 & 1 & 0 & 0 & 0 & 0 & 0 & 0 & 0 & 0 &
   0 & \cdots & 1 \\
 Z_{12,8} & 0 & 1 & 0 & 0 & 0 & 0 & 0 & 0 & 0 & 0 & 0 & 1 & 0 & 0 & 0 & 1 & 0 & 0 &
   0 & \cdots & 2 \\
\end{array}
\right)
$}
~,~
\nn\\
\eea
where we note that $o_{k}^{(1)}, \dots, o_{m}^{(20)}$
are extra GLSM fields \cite{Witten:1993yc}.
The $J$- and $E$-term charge matrix $Q_{JE}$ can be obtained from the kernel of
the $P$-matrix above. 
Here, the $Q_{JE}$-matrix is a $411 \times 426$ dimensional matrix, which we choose not to present here.
Additionally, we have the corresponding $D$-term charge matrix, which takes the following form, 
\beal{es04o05}
&&
Q_{D}=
\resizebox{0.8\textwidth}{!}{$
\left(
\begin{array}{cccccc|cc|cc|cccc|cccc|ccc}
p_1 & p_2 & p_3 & p_4 & p_5 & p_6 & q^{(1)}_1 & q^{(1)}_2 & q^{(2)}_1 & q^{(2)}_2 & q^{(3)}_1 & q^{(3)}_2 & q^{(3)}_3 & q^{(3)}_4 & q^{(4)}_1 & q^{(4)}_2 & q^{(4)}_3 & q^{(4)}_4 & o^{(1)}_{1} & \cdots & o^{(20)}_{16}
\\
\hline
 0 & 0 & 0 & 0 & 0 & 1 & 0 & 0 & 0 & 0 & 0 & 0 & -1 & 0 & 0 & 0 & 0 & 0 & -1 & \cdots & 0 \\
 0 & 0 & 0 & 0 & 0 & 0 & 0 & 0 & 0 & 0 & 0 & 0 & 0 & 0 & 0 & 0 & 0 & 0 & 0 & \cdots & 0 \\
 0 & 0 & 0 & 0 & 0 & 0 & 0 & 0 & 1 & 0 & 0 & 0 & 1 & 0 & 0 & 0 & 0 & 0 & 0 & \cdots & 0 \\
 0 & 0 & 0 & 0 & 0 & 0 & 1 & 0 & 0 & 0 & 0 & 0 & 0 & 0 & 0 & 0 & -1 & 0 & 1 & \cdots & 0 \\
 0 & 0 & 0 & 0 & -1 & 0 & -1 & 0 & 0 & 0 & 0 & 0 & 0 & 0 & 0 & 0 & 0 & 0 & 0 & \cdots & 0 \\
 0 & 0 & 0 & 0 & 0 & 0 & -1 & 0 & -1 & 0 & 0 & 0 & 0 & 0 & 0 & 0 & 0 & 0 & -1 & \cdots & 0 \\
 0 & 0 & 0 & 0 & 0 & 0 & 0 & 0 & 0 & 0 & 0 & 0 & 0 & 0 & 0 & 0 & 0 & 0 & 0 & \cdots & 0 \\
 0 & 0 & 0 & 0 & 0 & 0 & 0 & 0 & 0 & 0 & 0 & 0 & 0 & 0 & 0 & 1 & 1 & 0 & -1 & \cdots & 0 \\
 0 & 0 & 0 & 0 & 0 & -1 & 1 & 0 & 0 & 0 & 0 & 0 & 0 & 0 & 0 & 0 & 0 & 0 & 1 & \cdots & 0 \\
 1 & 0 & 0 & 0 & 1 & 0 & 0 & 0 & 0 & 0 & 0 & 0 & 0 & 0 & 0 & 0 & 0 & 0 & 1 & \cdots & 0 \\
 -1 & 0 & 0 & 0 & 0 & 0 & 0 & 0 & 1 & 0 & 0 & 0 & 0 & 0 & 0 & 0 & 0 & 0 & 0 & \cdots & 0 \\
\end{array}
\right)
$}
~.~
\nn\\
\eea

\begin{figure}[ht!!!]
\centering
\includegraphics[width=0.4\textwidth]{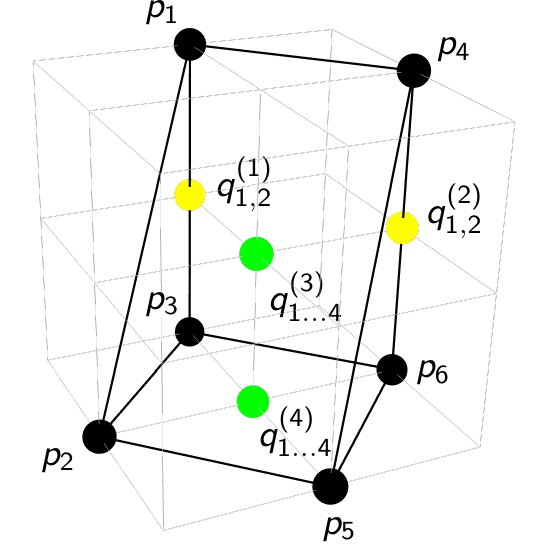}
\caption{
The toric diagram for the 
$D_3/ \mathbb{Z}_2 \times \mathbb{Z}_2 \ \left(\begin{array}{@{}c@{,\;}c@{,\;}c@{,\;}c@{,\;}c@{}} 0 & 1 & 1 & 0 & 0 \\ 1 & 0 & 1 & 1 & 1 \end{array}\right)$ model.
\label{fig_429_toric}
}
\end{figure}

The resulting toric diagram is given by,
\beal{es04o03b}
&&
G_t=
\resizebox{0.8\textwidth}{!}{$
\left(
\begin{array}{cccccc|cc|cc|cccc|cccc|ccc} 
p_1 & p_2 & p_3 & p_4 & p_5 & p_6 & 
q^{(1)}_1 & q^{(1)}_2 & 
q^{(2)}_1 & q^{(2)}_2 & 
q^{(3)}_1 & q^{(3)}_2 & q^{(3)}_3 & q^{(3)}_4 & 
q^{(4)}_1 & q^{(4)}_2 & q^{(4)}_3 & q^{(4)}_4 & 
o^{(1)}_1 & \cdots & o^{(20)}_{16} \\ \hline 
 0 & -1 & 0 & 1 & 0 & 1 & 0 & 0 & 0 & 0 & 1 & 1 & 1 & 1 & 0 & 0 & 0 & 0 & 1 &
   \cdots & 1 \\
 1 & -1 & -1 & 1 & -1 & -1 & 0 & 0 & -1 & -1 & 0 & 0 & 0 & 0 & 0 & 0 & 0 & 0 & -1 &
   \cdots & -1 \\
 1 & 0 & 1 & 0 & -1 & 0 & 1 & 1 & 0 & 0 & 0 & 0 & 0 & 0 & 0 & 0 & 0 & 0 & 0 &
   \cdots & 1 \\
 1 & 1 & 1 & 1 & 1 & 1 & 1 & 1 & 1 & 1 & 1 & 1 & 1 & 1 & 1 & 1 & 1 & 1 & 2 &
   \cdots & 4 \\
  \end{array}
\right)
$}
~,~
\nn\\
\eea
where the toric diagram is illustrated in \fref{fig_429_toric}.

%-------------------
%-------------------

%-------------------
\begin{table}[H]
\centering
\begin{tabular}{|c|c|c|c|l|}
\hline
\; & $U(1)_{f_1}$ & $U(1)_{f_2}$ & $U(1)_{f_3}$ &  fugacity \\
\hline
$p_1$ & $+1$ & $0$ & $0$ &  $t_1=f_1 t$ \\
$p_2$ & $-1$ & $0$ & $0$ &  $t_2=f_1^{-1} t$\\
$p_3$ & $0$ & $+1$ & $0$ &  $t_3=f_2 t$\\
$p_4$ & $0$ & $-1$ & $0$ & $t_4=f_2^{-1} t$ \\
$p_5$ & $0$ & $0$ & $+1$ & $t_5=f_3 t$ \\
$p_6$ & $0$ & $0$ & $-1$ & $t_6=f_3^{-1} t$ \\
\hline
\end{tabular}
\caption{
Mesonic flavor symmetry of the $D_3/ \mathbb{Z}_2 \times \mathbb{Z}_2 \ \left(\begin{array}{@{}c@{,\;}c@{,\;}c@{,\;}c@{,\;}c@{}} 0 & 1 & 1 & 0 & 0 \\ 1 & 0 & 1 & 1 & 1 \end{array}\right)$ model and charges on the extremal GLSM fields $p_a$. 
Here, the fugacity $t$ counts the degree in extremal GLSM fields $p_a$. 
\label{tab_04o01}}
\end{table}
%-------------------

The global symmetry of the brane brick model is not enhanced and takes the following form,
\beal{es04o04}
U(1)_{f_1} \times U(1)_{f_2} \times U(1)_{f_3} \times U(1)_R
~.~
\eea
The fully refined Hilbert series in terms of fugacities $t_a$ corresponding to extremal GLSM fields $p_a$
takes the following form,
\beal{es04o05b}
&&
g(t_a; \mathcal{M}^{mes})=
(1+ t_1 t_2^2 t_3 t_5+ t_1^2 t_2 t_3^2 t_4 t_6+ 2 t_1 t_2 t_3 t_4 t_5 t_6+ t_2 t_4 t_5^2 t_6
\nn\\
&&
\hspace{1cm}
+ t_1 t_3 t_4^2 t_5 t_6^2+ t_1^2 t_2^2 t_3^2 t_4^2 t_5^2 t_6^2)
\times
\frac{
\left(1 - t_1 t_2 t_3 t_4 t_5 t_6\right)
}{
\left(1 - t_1^2 t_2^2 t_3^2\right)
\left(1 - t_1^2 t_4^2\right) \left(1 - t_2^2 t_5^2\right) 
}
\nn\\
&&
\hspace{1cm}
\times
\frac{1}{
\left(1 - t_3^2 t_6^2\right)
\left(1 - t_4^2 t_5^2 t_6^2\right)
}
~.~
\eea
By setting $t_a = t$, 
we can unrefine the Hilbert series to take the following form,
\beal{es04o06}
g(t; \mathcal{M}^{mes})
=
\frac{1 - t + t^4 + t^5 - t^8 + t^9}{(1-t) (1-t^4)^2 (1-t^6)}
~,~
\eea
where we can see from the palindromic numerator that the mesonic moduli space is Calabi-Yau. 

The Hilbert series refined in terms of mesonic flavor fugacities summarized in \tref{tab_04o01}
is given by, 
\beal{es04o07}
&&
g(f_1,f_2,f_3,t; \mathcal{M}^{mes})=
(1+ f_1^{-1} f_2^{-1} f_3 t^5 + f_1^{-1} f_2 f_3 t^5 + 2t^6 + f_1 f_2^{-1} f_3^{-1} t^7
\nn\\
&&
\hspace{1cm}
+ f_1f_2f_3^{-1} t^7 + t^{12})
\times
\frac{(1-t^6) 
}{
(1-f_1^2 f_2^{-2}t^4)
(1-f_2^2 f_3^{-2}t^4)
(1-f_1^{-2} f_3^2t^4)
}
\nn\\
&&
\hspace{1cm}
\times
\frac{1}{
(1-f_2^{-2}t^6)
(1-f_2^2t^6)
}
~.~
\eea
The corresponding plethystic logarithm has the following expansion, 
\beal{es04o12}
&&
\text{PL}[g(f_1,f_2,f_3,t; \mathcal{M}^{mes})]=
(f_1^{2} f_2^{-2}+f_2^{2} f_3^{-2}+f_1^{-2} f_3^{2}) t^{4}
+ ( f_1^{-1} f_2^{-1} f_3+f_1^{-1} f_2 f_3 ) t^5
\nn\\
&&
\hspace{1cm}
+ ( 1+ f_2^2 + f_2^{-2} ) t^6 + ( f_1 f_2^{-1} f_3^{-1}+f_1 f_2 f_3^{-1} ) t^7
- ( f_1^{-2} f_3^{2}+f_1^{-2} f_2^{-2} f_3^{2}
\nn\\
&&
\hspace{1cm}
+f_1^{-2} f_2^{2} f_3^{2} ) t^{10}
- ( 2 f_1^{-1} f_2^{-1} f_3 + 2 f_1^{-1} f_2 f_3 )t^{11}
- ( 4 + f_2^2 + f_2^{-2} )t^{12} 
\nn\\
&&
\hspace{1cm}
- (2 f_1 f_2^{-1} f_3^{-1} + 2 f_1 f_2 f_3^{-1}) t^{13}
- ( f_1^{2} f_3^{-2}+f_1^{2} f_2^{-2} f_3^{-2}+f_1^{2} f_2^{2} f_3^{-2} ) t^{14}
\nn\\
&&
\hspace{1cm}
+\dots
~,~
\eea
where we note from the infinite expansion that the mesonic moduli space is not a complete intersection. 
The generators of the mesonic moduli space with their corresponding mesonic flavor charges are summarized in \tref{tab_04o02}.
\\

%-------------------
 \begin {table}[ht!!]
\centering
\begin {tabular} {|c|c|ccc|}
\hline
PL term & generator & $U(1)_{f_1}$ & $U(1)_{f_2}$ & $U(1)_{f_3}$ 
\\
\hline
\multirow{1}{*}{$+ f_2^{-2}  t^{6}$}
& $p_4^2 p_5^2 p_6^2 ~q_{(2)} q_{{(3)}} q_{{(4)}}^2 $ & $0$ & $-2$ & $0$  \\
\hline
\multirow{1}{*}{$+ f_2^{2} f_3^{-2} t^{4}$}
& $p_3^2 p_6^2 ~q_{(1)} q_{{(3)}} q_{{(4)}} $ & $0$ & $2$ & $-2$  \\
\hline
\multirow{1}{*}{$+f_1^{-1} f_2^{-1} f_3 t^{5}$}
& $p_2 p_4 p_5^2 p_6 ~q_{(2)} q_{{(3)}} q_{{(4)}} $ & $-1$ & $-1$ & $1$  \\
\hline
\multirow{1}{*}{$+f_1^{-2}  f_3^{2} t^{4}$}
& $p_2^2 p_5^2 ~q_{(2)} q_{{(3)}} $ & $-2$ & $0$ & $2$  \\
\hline
\multirow{1}{*}{$+f_1 f_2^{-1} f_3^{-1} t^{7}$}
& $p_1 p_3 p_4^2 p_5 p_6^2 ~q_{(1)} q_{(2)} q_{{(3)}} q_{{(4)}}^2 $ & $1$ & $-1$ & $-1$  \\
\hline
\multirow{1}{*}{$+   t^{6}$}
& $p_1 p_2 p_3 p_4 p_5 p_6 ~q_{(1)} q_{(2)} q_{{(3)}} q_{{(4)}} $ & $0$ & $0$ & $0$  \\
\hline
\multirow{1}{*}{$+f_1^{-1} f_2 f_3 t^{5}$}
& $p_1 p_2^2 p_3 p_5 ~q_{(1)} q_{(2)} q_{{(3)}} $ & $-1$ & $1$ & $1$  \\
\hline
\multirow{1}{*}{$+f_1^{2} f_2^{-2}  t^{4}$}
& $p_1^2 p_4^2 ~q_{(1)} q_{(2)} q_{{(4)}} $ & $2$ & $-2$ & $0$  \\
\hline
\multirow{1}{*}{$+f_1 f_2 f_3^{-1} t^{7}$}
& $p_1^2 p_2 p_3^2 p_4 p_6~ q_{(1)}^2 q_{(2)} q_{{(3)}} q_{{(4)}} $ & $1$ & $1$ & $-1$  \\
\hline
\multirow{1}{*}{$+ f_2^{2}  t^{6}$}
& $p_1^2 p_2^2 p_3^2 ~q_{(1)}^2 q_{(2)} q_{{(3)}} $ & $0$ & $2$ & $0$  \\
\hline
\end{tabular}
\caption{
Generators of the $D_3/ \mathbb{Z}_2 \times \mathbb{Z}_2 \ \left(\begin{array}{@{}c@{,\;}c@{,\;}c@{,\;}c@{,\;}c@{}} 0 & 1 & 1 & 0 & 0 \\ 1 & 0 & 1 & 1 & 1 \end{array}\right)$ model in terms of GLSM fields and their corresponding global symmetry charges. Here, we denote $q_{(1)}=q_1^{(1)}q_2^{(1)}$, $q_{(2)}=\prod_{i=1}^{4} q_i^{(2)}$, $q_{(3)}=\prod_{i=1}^{4} q_i^{(3)}$, $q_{(4)}=q_1^{(4)} q_2^{(4)}$, and set the extra GLSM fields to 1.
\label{tab_04o02}}
\end{table}
%-------------------

%=================================================================
\section{Discussion and Conclusions \label{sec06}}

In this work, 
we have presented a systematic procedure for constructing brane brick models corresponding to abelian orbifolds of toric Calabi-Yau 4-folds.
By doing so, we have extended the orbifold construction 
beyond the well-studied case of abelian orbifolds of $\mathbb{C}^4$ \cite{Franco:2015tna, Franco:2015tya}. 
Starting from a parent brane brick model corresponding to a toric Calabi-Yau 4-fold $\mathcal{M}$, 
where $\mathcal{M}$ is the mesonic moduli space when the brane brick model is abelian with $U(1)$ gauge groups,
we have shown that orbifolding by a finite abelian subgroup $\Gamma \subset SU(4)$ 
acts on the generators $z_1, \dots, z_d$ of $\mathcal{M}$. 
This induces a well-defined orbifold action on the chiral and Fermi fields 
as well as on the $J$- and $E$-terms of the parent $2d$ $(0,2)$ supersymmetric gauge theory.
The key observation here is that the orbifold action of $\Gamma$ 
satisfies the requirement that all closed paths associated to the $J$- and $E$-terms, as well as the chiral cycles formed by their products,
remain closed in the periodic quiver. 
This in turn reproduces the Calabi-Yau condition on the orbifold action of $\Gamma$.
This consistency requirement, together with the constraint that the action preserves the binomial relations defining $\mathcal{M}$,
yields an explicit formula for the $J$- and $E$-terms of the orbifolded brane brick model, 
which can be expressed entirely in terms of the data of the parent brane brick model and the orbifold action.
 
We have applied this procedure to the brane brick models corresponding to the cones over $Q^{1,1,1}$ and $D_3$, 
and have presented the resulting general families of $2d$ $(0,2)$ quiver gauge theories.
We have then verified the construction through a series of explicit examples
of brane brick models corresponding to abelian orbifolds of $Q^{1,1,1}$ and $D_3$,
with various inequivalent orbifold actions for each case.
For every example, we have calculated the $P$-matrix, 
the total charge matrix $Q_t = (Q_{JE}, Q_D)$, 
the $G_t$-matrix encoding the toric diagram of the abelian orbifold, 
the global symmetry of the mesonic moduli space of the orbifolded brane brick model, 
and the refined Hilbert series with its plethystic logarithm.
The plethystic logarithm has allowed us to identify explicitly the generators of the mesonic moduli space of the orbifolded brane brick model,
and their charges under the mesonic flavor symmetry.
The consistency of these results across all examples provides a non-trivial check of the general formulas for the $J$- and $E$-terms 
presented in sections \sref{sec032} and \sref{sec033} for the abelian orbifolds of $Q^{1,1,1}$ and $D_3$, respectively.

Beyond the construction of individual orbifold theories, we have enumerated
the inequivalent abelian orbifolds of $Q^{1,1,1}$ and $D_3$ at each order
$n=|\Gamma|$. 
We have presented explicit expressions for the generating
functions of the sequences that count these orbifolds for any order $n$.
Remarkably, this counting of inequivalent sublattices of $\mathbb{Z}^3$ matches the number of distinct and consistent orbifold actions.
 
We can conclude that the general orbifold construction presented here for brane brick models corresponding to abelian orbifolds of toric Calabi-Yau 4-folds 
significantly enlarges the set of brane brick models accessible by direct methods. 
Whereas previously the only systematic family of orbifolded brane brick models available as starting points for further constructions 
has been that of abelian orbifolds of $\mathbb{C}^4$,
our general procedure here produces families of orbifold theories built on any toric Calabi-Yau 4-fold 
whose parent brane brick model is known. 
Combined with partial resolution and the Higgs mechanism,
these new brane brick models corresponding to abelian orbifolds provide a much broader pool of 
parent theories from which brane brick models for further toric Calabi-Yau 4-folds can be reached. 
In particular, abelian orbifolds of $Q^{1,1,1}$ and $D_3$ have toric diagrams that contain a much wider variety of $\mathbb{Z}^3$ lattice polytopes 
than those of abelian orbifolds of $\mathbb{C}^4$ at comparable orbifold order.
This opens the path towards brane brick models for toric Calabi-Yau 4-folds that have previously been out of reach.

As part of future research, 
we plan to apply our method of orbifolding brane brick models 
to other parent models corresponding to a variety of toric Calabi-Yau 4-folds
and systematically study the resulting $2d$ $(0,2)$ quiver gauge theories. 
Exploring the relationship between orbifolded brane brick models under
birational transformations \cite{2012arXiv1212.1785A, 2013arXiv1309.2573G, 2013arXiv1303.3288C, 2015arXiv150105335K, 2021RSPSA.47710584C, Franco:2023flw, Ghim:2024asj, Ghim:2025zhs,Kho:2025fmp, Franco:2023tyf, Carcamo:2025shw, Kho:2025jxk, Kho:2026zwc}
as well as in the context of triality \cite{Gadde:2013lxa, Franco:2017cjj, Franco:2016qxh, Franco:2016tcm, Kho:2026geo}
are part of the key objectives for future work. 
\\
\\

%======================================================================
\section*{Acknowledgements}

We would like to thank
Jiakang Bao, Sebastian Franco, Dongwook Ghim, Amihay Hanany, Minsung Kho, Kimyeong Lee, Norton Lee, Sangmin Lee, Seong-Jin Lee, Benjamin Suzzoni and Masahito Yamazaki
for discussions on related topics. 
R.-K. S. would like to thank
the IBS Center for Geometry and Physics in Pohang,
the Beijing Institute of Mathematical Sciences and Applications
and the Yau Mathematical Sciences Center at Tsinghua University in Beijing, 
the Simons Center for Geometry and Physics at Stony Brook University,
the Korea Institute for Advanced Study in Seoul, 
the Institute of Mathematics Academia Sinica in Taipei, 
as well as the Kavli Institute for Theoretical Physics at the University of California in Santa Barbara
for their kind and generous hospitality during various stages of this work. 
R.-K. S. is supported by an Outstanding Young Scientist Grant (RS-2025-00516583) of the National Research Foundation of Korea (NRF). He is also partly supported by the BK21 Program (“Next Generation Education Program for Mathematical Sciences”, 4299990414089) funded by the Ministry of Education in Korea and the National Research Foundation of Korea (NRF).
\\

%======================================================================
\appendix

%======================================================================
\section{Numerators for Refined Hilbert Series \label{appCa}}

%----------------------------------------------------------------------
\subsection{$Q^{1,1,1}/\mathbb{Z}_2 \ (1,1,1,1,1,1,1,1)$}

\begin{quote}\raggedright \label{esAa01}
$
P(t_a; \mathcal{M}^{mes})=
1+ t_1 t_2 t_3^2 t_5^2+ t_1^2 t_3 t_4 t_5^2+ t_1 t_2 t_3 t_4 t_5^2+ t_2^2 t_3 t_4 t_5^2+ t_1 t_2 t_4^2 t_5^2-t_1^2 t_2^2 t_3^3 t_4 t_5^4-t_1^3 t_2 t_3^2 t_4^2 t_5^4-t_1^2 t_2^2 t_3^2 t_4^2 t_5^4-t_1 t_2^3 t_3^2 t_4^2 t_5^4-t_1^2 t_2^2 t_3 t_4^3 t_5^4-t_1^3 t_2^3 t_3^3 t_4^3 t_5^6+ t_1^2 t_3^2 t_5 t_6+ t_1 t_2 t_3^2 t_5 t_6+ t_2^2 t_3^2 t_5 t_6+ t_1^2 t_3 t_4 t_5 t_6+ t_2^2 t_3 t_4 t_5 t_6+ t_1^2 t_4^2 t_5 t_6+ t_1 t_2 t_4^2 t_5 t_6+ t_2^2 t_4^2 t_5 t_6-t_1^2 t_2^2 t_3^4 t_5^3 t_6-2 t_1^2 t_2^2 t_3^3 t_4 t_5^3 t_6-t_1^4 t_3^2 t_4^2 t_5^3 t_6-2 t_1^3 t_2 t_3^2 t_4^2 t_5^3 t_6-4 t_1^2 t_2^2 t_3^2 t_4^2 t_5^3 t_6-2 t_1 t_2^3 t_3^2 t_4^2 t_5^3 t_6-t_2^4 t_3^2 t_4^2 t_5^3 t_6-2 t_1^2 t_2^2 t_3 t_4^3 t_5^3 t_6-t_1^2 t_2^2 t_4^4 t_5^3 t_6+ t_1^4 t_2^2 t_3^4 t_4^2 t_5^5 t_6+ t_1^3 t_2^3 t_3^4 t_4^2 t_5^5 t_6+ t_1^2 t_2^4 t_3^4 t_4^2 t_5^5 t_6+ t_1^4 t_2^2 t_3^3 t_4^3 t_5^5 t_6+ t_1^2 t_2^4 t_3^3 t_4^3 t_5^5 t_6+ t_1^4 t_2^2 t_3^2 t_4^4 t_5^5 t_6+ t_1^3 t_2^3 t_3^2 t_4^4 t_5^5 t_6+ t_1^2 t_2^4 t_3^2 t_4^4 t_5^5 t_6+ t_1 t_2 t_3^2 t_6^2+ t_1^2 t_3 t_4 t_6^2+ t_1 t_2 t_3 t_4 t_6^2+ t_2^2 t_3 t_4 t_6^2+ t_1 t_2 t_4^2 t_6^2-t_1^3 t_2 t_3^4 t_5^2 t_6^2-t_1^2 t_2^2 t_3^4 t_5^2 t_6^2-t_1 t_2^3 t_3^4 t_5^2 t_6^2-t_1^4 t_3^3 t_4 t_5^2 t_6^2-2 t_1^3 t_2 t_3^3 t_4 t_5^2 t_6^2-4 t_1^2 t_2^2 t_3^3 t_4 t_5^2 t_6^2-2 t_1 t_2^3 t_3^3 t_4 t_5^2 t_6^2-t_2^4 t_3^3 t_4 t_5^2 t_6^2-t_1^4 t_3^2 t_4^2 t_5^2 t_6^2-4 t_1^3 t_2 t_3^2 t_4^2 t_5^2 t_6^2-3 t_1^2 t_2^2 t_3^2 t_4^2 t_5^2 t_6^2-4 t_1 t_2^3 t_3^2 t_4^2 t_5^2 t_6^2-t_2^4 t_3^2 t_4^2 t_5^2 t_6^2-t_1^4 t_3 t_4^3 t_5^2 t_6^2-2 t_1^3 t_2 t_3 t_4^3 t_5^2 t_6^2-4 t_1^2 t_2^2 t_3 t_4^3 t_5^2 t_6^2-2 t_1 t_2^3 t_3 t_4^3 t_5^2 t_6^2-t_2^4 t_3 t_4^3 t_5^2 t_6^2-t_1^3 t_2 t_4^4 t_5^2 t_6^2-t_1^2 t_2^2 t_4^4 t_5^2 t_6^2-t_1 t_2^3 t_4^4 t_5^2 t_6^2+ t_1^4 t_2^2 t_3^5 t_4 t_5^4 t_6^2+ t_1^3 t_2^3 t_3^5 t_4 t_5^4 t_6^2+ t_1^2 t_2^4 t_3^5 t_4 t_5^4 t_6^2+ t_1^5 t_2 t_3^4 t_4^2 t_5^4 t_6^2+ 2 t_1^4 t_2^2 t_3^4 t_4^2 t_5^4 t_6^2+ 4 t_1^3 t_2^3 t_3^4 t_4^2 t_5^4 t_6^2+ 2 t_1^2 t_2^4 t_3^4 t_4^2 t_5^4 t_6^2+ t_1 t_2^5 t_3^4 t_4^2 t_5^4 t_6^2+ t_1^5 t_2 t_3^3 t_4^3 t_5^4 t_6^2+ 4 t_1^4 t_2^2 t_3^3 t_4^3 t_5^4 t_6^2+ 3 t_1^3 t_2^3 t_3^3 t_4^3 t_5^4 t_6^2+ 4 t_1^2 t_2^4 t_3^3 t_4^3 t_5^4 t_6^2+ t_1 t_2^5 t_3^3 t_4^3 t_5^4 t_6^2+ t_1^5 t_2 t_3^2 t_4^4 t_5^4 t_6^2+ 2 t_1^4 t_2^2 t_3^2 t_4^4 t_5^4 t_6^2+ 4 t_1^3 t_2^3 t_3^2 t_4^4 t_5^4 t_6^2+ 2 t_1^2 t_2^4 t_3^2 t_4^4 t_5^4 t_6^2+ t_1 t_2^5 t_3^2 t_4^4 t_5^4 t_6^2+ t_1^4 t_2^2 t_3 t_4^5 t_5^4 t_6^2+ t_1^3 t_2^3 t_3 t_4^5 t_5^4 t_6^2+ t_1^2 t_2^4 t_3 t_4^5 t_5^4 t_6^2-t_1^4 t_2^4 t_3^5 t_4^3 t_5^6 t_6^2-t_1^5 t_2^3 t_3^4 t_4^4 t_5^6 t_6^2-t_1^4 t_2^4 t_3^4 t_4^4 t_5^6 t_6^2-t_1^3 t_2^5 t_3^4 t_4^4 t_5^6 t_6^2-t_1^4 t_2^4 t_3^3 t_4^5 t_5^6 t_6^2-t_1^2 t_2^2 t_3^4 t_5 t_6^3-2 t_1^2 t_2^2 t_3^3 t_4 t_5 t_6^3-t_1^4 t_3^2 t_4^2 t_5 t_6^3-2 t_1^3 t_2 t_3^2 t_4^2 t_5 t_6^3-4 t_1^2 t_2^2 t_3^2 t_4^2 t_5 t_6^3-2 t_1 t_2^3 t_3^2 t_4^2 t_5 t_6^3-t_2^4 t_3^2 t_4^2 t_5 t_6^3-2 t_1^2 t_2^2 t_3 t_4^3 t_5 t_6^3-t_1^2 t_2^2 t_4^4 t_5 t_6^3-t_1^3 t_2^3 t_3^6 t_5^3 t_6^3+ t_1^4 t_2^2 t_3^5 t_4 t_5^3 t_6^3+ t_1^2 t_2^4 t_3^5 t_4 t_5^3 t_6^3+ t_1^5 t_2 t_3^4 t_4^2 t_5^3 t_6^3+ 4 t_1^4 t_2^2 t_3^4 t_4^2 t_5^3 t_6^3+ 3 t_1^3 t_2^3 t_3^4 t_4^2 t_5^3 t_6^3+ 4 t_1^2 t_2^4 t_3^4 t_4^2 t_5^3 t_6^3+ t_1 t_2^5 t_3^4 t_4^2 t_5^3 t_6^3-t_1^6 t_3^3 t_4^3 t_5^3 t_6^3+ 3 t_1^4 t_2^2 t_3^3 t_4^3 t_5^3 t_6^3+ 3 t_1^2 t_2^4 t_3^3 t_4^3 t_5^3 t_6^3-t_2^6 t_3^3 t_4^3 t_5^3 t_6^3+ t_1^5 t_2 t_3^2 t_4^4 t_5^3 t_6^3+ 4 t_1^4 t_2^2 t_3^2 t_4^4 t_5^3 t_6^3+ 3 t_1^3 t_2^3 t_3^2 t_4^4 t_5^3 t_6^3+ 4 t_1^2 t_2^4 t_3^2 t_4^4 t_5^3 t_6^3+ t_1 t_2^5 t_3^2 t_4^4 t_5^3 t_6^3+ t_1^4 t_2^2 t_3 t_4^5 t_5^3 t_6^3+ t_1^2 t_2^4 t_3 t_4^5 t_5^3 t_6^3-t_1^3 t_2^3 t_4^6 t_5^3 t_6^3-t_1^4 t_2^4 t_3^6 t_4^2 t_5^5 t_6^3-2 t_1^4 t_2^4 t_3^5 t_4^3 t_5^5 t_6^3-t_1^6 t_2^2 t_3^4 t_4^4 t_5^5 t_6^3-2 t_1^5 t_2^3 t_3^4 t_4^4 t_5^5 t_6^3-4 t_1^4 t_2^4 t_3^4 t_4^4 t_5^5 t_6^3-2 t_1^3 t_2^5 t_3^4 t_4^4 t_5^5 t_6^3-t_1^2 t_2^6 t_3^4 t_4^4 t_5^5 t_6^3-2 t_1^4 t_2^4 t_3^3 t_4^5 t_5^5 t_6^3-t_1^4 t_2^4 t_3^2 t_4^6 t_5^5 t_6^3-t_1^2 t_2^2 t_3^3 t_4 t_6^4-t_1^3 t_2 t_3^2 t_4^2 t_6^4-t_1^2 t_2^2 t_3^2 t_4^2 t_6^4-t_1 t_2^3 t_3^2 t_4^2 t_6^4-t_1^2 t_2^2 t_3 t_4^3 t_6^4+ t_1^4 t_2^2 t_3^5 t_4 t_5^2 t_6^4+ t_1^3 t_2^3 t_3^5 t_4 t_5^2 t_6^4+ t_1^2 t_2^4 t_3^5 t_4 t_5^2 t_6^4+ t_1^5 t_2 t_3^4 t_4^2 t_5^2 t_6^4+ 2 t_1^4 t_2^2 t_3^4 t_4^2 t_5^2 t_6^4+ 4 t_1^3 t_2^3 t_3^4 t_4^2 t_5^2 t_6^4+ 2 t_1^2 t_2^4 t_3^4 t_4^2 t_5^2 t_6^4+ t_1 t_2^5 t_3^4 t_4^2 t_5^2 t_6^4+ t_1^5 t_2 t_3^3 t_4^3 t_5^2 t_6^4+ 4 t_1^4 t_2^2 t_3^3 t_4^3 t_5^2 t_6^4+ 3 t_1^3 t_2^3 t_3^3 t_4^3 t_5^2 t_6^4+ 4 t_1^2 t_2^4 t_3^3 t_4^3 t_5^2 t_6^4+ t_1 t_2^5 t_3^3 t_4^3 t_5^2 t_6^4+ t_1^5 t_2 t_3^2 t_4^4 t_5^2 t_6^4+ 2 t_1^4 t_2^2 t_3^2 t_4^4 t_5^2 t_6^4+ 4 t_1^3 t_2^3 t_3^2 t_4^4 t_5^2 t_6^4+ 2 t_1^2 t_2^4 t_3^2 t_4^4 t_5^2 t_6^4+ t_1 t_2^5 t_3^2 t_4^4 t_5^2 t_6^4+ t_1^4 t_2^2 t_3 t_4^5 t_5^2 t_6^4+ t_1^3 t_2^3 t_3 t_4^5 t_5^2 t_6^4+ t_1^2 t_2^4 t_3 t_4^5 t_5^2 t_6^4-t_1^5 t_2^3 t_3^6 t_4^2 t_5^4 t_6^4-t_1^4 t_2^4 t_3^6 t_4^2 t_5^4 t_6^4-t_1^3 t_2^5 t_3^6 t_4^2 t_5^4 t_6^4-t_1^6 t_2^2 t_3^5 t_4^3 t_5^4 t_6^4-2 t_1^5 t_2^3 t_3^5 t_4^3 t_5^4 t_6^4-4 t_1^4 t_2^4 t_3^5 t_4^3 t_5^4 t_6^4-2 t_1^3 t_2^5 t_3^5 t_4^3 t_5^4 t_6^4-t_1^2 t_2^6 t_3^5 t_4^3 t_5^4 t_6^4-t_1^6 t_2^2 t_3^4 t_4^4 t_5^4 t_6^4-4 t_1^5 t_2^3 t_3^4 t_4^4 t_5^4 t_6^4-3 t_1^4 t_2^4 t_3^4 t_4^4 t_5^4 t_6^4-4 t_1^3 t_2^5 t_3^4 t_4^4 t_5^4 t_6^4-t_1^2 t_2^6 t_3^4 t_4^4 t_5^4 t_6^4-t_1^6 t_2^2 t_3^3 t_4^5 t_5^4 t_6^4-2 t_1^5 t_2^3 t_3^3 t_4^5 t_5^4 t_6^4-4 t_1^4 t_2^4 t_3^3 t_4^5 t_5^4 t_6^4-2 t_1^3 t_2^5 t_3^3 t_4^5 t_5^4 t_6^4-t_1^2 t_2^6 t_3^3 t_4^5 t_5^4 t_6^4-t_1^5 t_2^3 t_3^2 t_4^6 t_5^4 t_6^4-t_1^4 t_2^4 t_3^2 t_4^6 t_5^4 t_6^4-t_1^3 t_2^5 t_3^2 t_4^6 t_5^4 t_6^4+ t_1^5 t_2^5 t_3^6 t_4^4 t_5^6 t_6^4+ t_1^6 t_2^4 t_3^5 t_4^5 t_5^6 t_6^4+ t_1^5 t_2^5 t_3^5 t_4^5 t_5^6 t_6^4+ t_1^4 t_2^6 t_3^5 t_4^5 t_5^6 t_6^4+ t_1^5 t_2^5 t_3^4 t_4^6 t_5^6 t_6^4+ t_1^4 t_2^2 t_3^4 t_4^2 t_5 t_6^5+ t_1^3 t_2^3 t_3^4 t_4^2 t_5 t_6^5+ t_1^2 t_2^4 t_3^4 t_4^2 t_5 t_6^5+ t_1^4 t_2^2 t_3^3 t_4^3 t_5 t_6^5+ t_1^2 t_2^4 t_3^3 t_4^3 t_5 t_6^5+ t_1^4 t_2^2 t_3^2 t_4^4 t_5 t_6^5+ t_1^3 t_2^3 t_3^2 t_4^4 t_5 t_6^5+ t_1^2 t_2^4 t_3^2 t_4^4 t_5 t_6^5-t_1^4 t_2^4 t_3^6 t_4^2 t_5^3 t_6^5-2 t_1^4 t_2^4 t_3^5 t_4^3 t_5^3 t_6^5-t_1^6 t_2^2 t_3^4 t_4^4 t_5^3 t_6^5-2 t_1^5 t_2^3 t_3^4 t_4^4 t_5^3 t_6^5-4 t_1^4 t_2^4 t_3^4 t_4^4 t_5^3 t_6^5-2 t_1^3 t_2^5 t_3^4 t_4^4 t_5^3 t_6^5-t_1^2 t_2^6 t_3^4 t_4^4 t_5^3 t_6^5-2 t_1^4 t_2^4 t_3^3 t_4^5 t_5^3 t_6^5-t_1^4 t_2^4 t_3^2 t_4^6 t_5^3 t_6^5+ t_1^6 t_2^4 t_3^6 t_4^4 t_5^5 t_6^5+ t_1^5 t_2^5 t_3^6 t_4^4 t_5^5 t_6^5+ t_1^4 t_2^6 t_3^6 t_4^4 t_5^5 t_6^5+ t_1^6 t_2^4 t_3^5 t_4^5 t_5^5 t_6^5+ t_1^4 t_2^6 t_3^5 t_4^5 t_5^5 t_6^5+ t_1^6 t_2^4 t_3^4 t_4^6 t_5^5 t_6^5+ t_1^5 t_2^5 t_3^4 t_4^6 t_5^5 t_6^5+ t_1^4 t_2^6 t_3^4 t_4^6 t_5^5 t_6^5-t_1^3 t_2^3 t_3^3 t_4^3 t_6^6-t_1^4 t_2^4 t_3^5 t_4^3 t_5^2 t_6^6-t_1^5 t_2^3 t_3^4 t_4^4 t_5^2 t_6^6-t_1^4 t_2^4 t_3^4 t_4^4 t_5^2 t_6^6-t_1^3 t_2^5 t_3^4 t_4^4 t_5^2 t_6^6-t_1^4 t_2^4 t_3^3 t_4^5 t_5^2 t_6^6+ t_1^5 t_2^5 t_3^6 t_4^4 t_5^4 t_6^6+ t_1^6 t_2^4 t_3^5 t_4^5 t_5^4 t_6^6+ t_1^5 t_2^5 t_3^5 t_4^5 t_5^4 t_6^6+ t_1^4 t_2^6 t_3^5 t_4^5 t_5^4 t_6^6+ t_1^5 t_2^5 t_3^4 t_4^6 t_5^4 t_6^6+ t_1^6 t_2^6 t_3^6 t_4^6 t_5^6 t_6^6
$
\end{quote}

%----------------------------------------------------------------------
\subsection{$Q^{1,1,1}/\mathbb{Z}_2 \ (0,0,1,1,1,1,0,0)$}

\begin{quote}\raggedright \label{esAb01}
$
P(t_a ; \mathcal{M}^{mes})=
1+ t_1^2 t_2 t_3 t_4^2-t_1^3 t_2^2 t_3 t_4^2 t_5-t_1^3 t_2 t_3^2 t_4^2 t_5-t_1^2 t_2^2 t_3 t_4^3 t_6-t_1^2 t_2 t_3^2 t_4^3 t_6-t_1 t_2 t_3 t_4 t_5 t_6+ t_1^3 t_2^3 t_3 t_4^3 t_5 t_6+ t_1^3 t_2^2 t_3^2 t_4^3 t_5 t_6+ t_1^3 t_2 t_3^3 t_4^3 t_5 t_6+ t_2 t_3 t_5^2 t_6^2-t_1^2 t_2^4 t_4^2 t_5^2 t_6^2-2 t_1^2 t_2^3 t_3 t_4^2 t_5^2 t_6^2-2 t_1^2 t_2^2 t_3^2 t_4^2 t_5^2 t_6^2-2 t_1^2 t_2 t_3^3 t_4^2 t_5^2 t_6^2-t_1^2 t_3^4 t_4^2 t_5^2 t_6^2+ t_1^4 t_2^4 t_3^2 t_4^4 t_5^2 t_6^2+ t_1^4 t_2^3 t_3^3 t_4^4 t_5^2 t_6^2+ t_1^4 t_2^2 t_3^4 t_4^4 t_5^2 t_6^2-t_1 t_2^2 t_3 t_5^3 t_6^2-t_1 t_2 t_3^2 t_5^3 t_6^2+ 2 t_1^3 t_2^4 t_3 t_4^2 t_5^3 t_6^2+ 2 t_1^3 t_2^3 t_3^2 t_4^2 t_5^3 t_6^2+ 2 t_1^3 t_2^2 t_3^3 t_4^2 t_5^3 t_6^2+ 2 t_1^3 t_2 t_3^4 t_4^2 t_5^3 t_6^2-t_1^5 t_2^4 t_3^3 t_4^4 t_5^3 t_6^2-t_1^5 t_2^3 t_3^4 t_4^4 t_5^3 t_6^2-t_2^2 t_3 t_4 t_5^2 t_6^3-t_2 t_3^2 t_4 t_5^2 t_6^3+ 2 t_1^2 t_2^4 t_3 t_4^3 t_5^2 t_6^3+ 2 t_1^2 t_2^3 t_3^2 t_4^3 t_5^2 t_6^3+ 2 t_1^2 t_2^2 t_3^3 t_4^3 t_5^2 t_6^3+ 2 t_1^2 t_2 t_3^4 t_4^3 t_5^2 t_6^3-t_1^4 t_2^4 t_3^3 t_4^5 t_5^2 t_6^3-t_1^4 t_2^3 t_3^4 t_4^5 t_5^2 t_6^3+ t_1 t_2^3 t_3 t_4 t_5^3 t_6^3+ t_1 t_2^2 t_3^2 t_4 t_5^3 t_6^3+ t_1 t_2 t_3^3 t_4 t_5^3 t_6^3-t_1^3 t_2^5 t_3 t_4^3 t_5^3 t_6^3-2 t_1^3 t_2^4 t_3^2 t_4^3 t_5^3 t_6^3-2 t_1^3 t_2^3 t_3^3 t_4^3 t_5^3 t_6^3-2 t_1^3 t_2^2 t_3^4 t_4^3 t_5^3 t_6^3-t_1^3 t_2 t_3^5 t_4^3 t_5^3 t_6^3+ t_1^5 t_2^4 t_3^4 t_4^5 t_5^3 t_6^3+ t_1^2 t_2^4 t_3^2 t_4^2 t_5^4 t_6^4+ t_1^2 t_2^3 t_3^3 t_4^2 t_5^4 t_6^4+ t_1^2 t_2^2 t_3^4 t_4^2 t_5^4 t_6^4-t_1^4 t_2^4 t_3^4 t_4^4 t_5^4 t_6^4-t_1^3 t_2^4 t_3^3 t_4^2 t_5^5 t_6^4-t_1^3 t_2^3 t_3^4 t_4^2 t_5^5 t_6^4-t_1^2 t_2^4 t_3^3 t_4^3 t_5^4 t_6^5-t_1^2 t_2^3 t_3^4 t_4^3 t_5^4 t_6^5+ t_1^3 t_2^4 t_3^4 t_4^3 t_5^5 t_6^5+ t_1^5 t_2^5 t_3^5 t_4^5 t_5^5 t_6^5
$
\end{quote}

%----------------------------------------------------------------------
\subsection{$Q^{1,1,1}/\mathbb{Z}_3 \ (0,0,1,1,2,2,0,0)$}

\begin{quote}\raggedright \label{esAc01}
$
P(t_a ; \mathcal{M}^{mes} )=
1+ t_1^3 t_2^2 t_3 t_5^3+ t_1^3 t_2 t_3^2 t_5^3-t_1^4 t_2^3 t_3 t_4 t_5^3-t_1^4 t_2^2 t_3^2 t_4 t_5^3-t_1^4 t_2 t_3^3 t_4 t_5^3-t_1 t_2 t_3 t_4 t_5 t_6-t_1^3 t_2^3 t_3 t_5^4 t_6-t_1^3 t_2^2 t_3^2 t_5^4 t_6-t_1^3 t_2 t_3^3 t_5^4 t_6+ t_1^4 t_2^4 t_3 t_4 t_5^4 t_6+ t_1^4 t_2^3 t_3^2 t_4 t_5^4 t_6+ t_1^4 t_2^2 t_3^3 t_4 t_5^4 t_6+ t_1^4 t_2 t_3^4 t_4 t_5^4 t_6+ t_2^2 t_3 t_4^3 t_6^3+ t_2 t_3^2 t_4^3 t_6^3-t_1 t_2^3 t_3 t_4^4 t_6^3-t_1 t_2^2 t_3^2 t_4^4 t_6^3-t_1 t_2 t_3^3 t_4^4 t_6^3-t_1^3 t_2^6 t_4^3 t_5^3 t_6^3-2 t_1^3 t_2^5 t_3 t_4^3 t_5^3 t_6^3-2 t_1^3 t_2^4 t_3^2 t_4^3 t_5^3 t_6^3-2 t_1^3 t_2^3 t_3^3 t_4^3 t_5^3 t_6^3-2 t_1^3 t_2^2 t_3^4 t_4^3 t_5^3 t_6^3-2 t_1^3 t_2 t_3^5 t_4^3 t_5^3 t_6^3-t_1^3 t_3^6 t_4^3 t_5^3 t_6^3+ 2 t_1^4 t_2^6 t_3 t_4^4 t_5^3 t_6^3+ 2 t_1^4 t_2^5 t_3^2 t_4^4 t_5^3 t_6^3+ 2 t_1^4 t_2^4 t_3^3 t_4^4 t_5^3 t_6^3+ 2 t_1^4 t_2^3 t_3^4 t_4^4 t_5^3 t_6^3+ 2 t_1^4 t_2^2 t_3^5 t_4^4 t_5^3 t_6^3+ 2 t_1^4 t_2 t_3^6 t_4^4 t_5^3 t_6^3+ t_1^6 t_2^6 t_3^3 t_4^3 t_5^6 t_6^3+ t_1^6 t_2^5 t_3^4 t_4^3 t_5^6 t_6^3+ t_1^6 t_2^4 t_3^5 t_4^3 t_5^6 t_6^3+ t_1^6 t_2^3 t_3^6 t_4^3 t_5^6 t_6^3-t_1^7 t_2^6 t_3^4 t_4^4 t_5^6 t_6^3-t_1^7 t_2^5 t_3^5 t_4^4 t_5^6 t_6^3-t_1^7 t_2^4 t_3^6 t_4^4 t_5^6 t_6^3-t_2^3 t_3 t_4^3 t_5 t_6^4-t_2^2 t_3^2 t_4^3 t_5 t_6^4-t_2 t_3^3 t_4^3 t_5 t_6^4+ t_1 t_2^4 t_3 t_4^4 t_5 t_6^4+ t_1 t_2^3 t_3^2 t_4^4 t_5 t_6^4+ t_1 t_2^2 t_3^3 t_4^4 t_5 t_6^4+ t_1 t_2 t_3^4 t_4^4 t_5 t_6^4+ 2 t_1^3 t_2^6 t_3 t_4^3 t_5^4 t_6^4+ 2 t_1^3 t_2^5 t_3^2 t_4^3 t_5^4 t_6^4+ 2 t_1^3 t_2^4 t_3^3 t_4^3 t_5^4 t_6^4+ 2 t_1^3 t_2^3 t_3^4 t_4^3 t_5^4 t_6^4+ 2 t_1^3 t_2^2 t_3^5 t_4^3 t_5^4 t_6^4+ 2 t_1^3 t_2 t_3^6 t_4^3 t_5^4 t_6^4-t_1^4 t_2^7 t_3 t_4^4 t_5^4 t_6^4-2 t_1^4 t_2^6 t_3^2 t_4^4 t_5^4 t_6^4-2 t_1^4 t_2^5 t_3^3 t_4^4 t_5^4 t_6^4-2 t_1^4 t_2^4 t_3^4 t_4^4 t_5^4 t_6^4-2 t_1^4 t_2^3 t_3^5 t_4^4 t_5^4 t_6^4-2 t_1^4 t_2^2 t_3^6 t_4^4 t_5^4 t_6^4-t_1^4 t_2 t_3^7 t_4^4 t_5^4 t_6^4-t_1^6 t_2^6 t_3^4 t_4^3 t_5^7 t_6^4-t_1^6 t_2^5 t_3^5 t_4^3 t_5^7 t_6^4-t_1^6 t_2^4 t_3^6 t_4^3 t_5^7 t_6^4+ t_1^7 t_2^6 t_3^5 t_4^4 t_5^7 t_6^4+ t_1^7 t_2^5 t_3^6 t_4^4 t_5^7 t_6^4+ t_1^3 t_2^6 t_3^3 t_4^6 t_5^3 t_6^6+ t_1^3 t_2^5 t_3^4 t_4^6 t_5^3 t_6^6+ t_1^3 t_2^4 t_3^5 t_4^6 t_5^3 t_6^6+ t_1^3 t_2^3 t_3^6 t_4^6 t_5^3 t_6^6-t_1^4 t_2^6 t_3^4 t_4^7 t_5^3 t_6^6-t_1^4 t_2^5 t_3^5 t_4^7 t_5^3 t_6^6-t_1^4 t_2^4 t_3^6 t_4^7 t_5^3 t_6^6-t_1^6 t_2^6 t_3^6 t_4^6 t_5^6 t_6^6-t_1^3 t_2^6 t_3^4 t_4^6 t_5^4 t_6^7-t_1^3 t_2^5 t_3^5 t_4^6 t_5^4 t_6^7-t_1^3 t_2^4 t_3^6 t_4^6 t_5^4 t_6^7+ t_1^4 t_2^6 t_3^5 t_4^7 t_5^4 t_6^7+ t_1^4 t_2^5 t_3^6 t_4^7 t_5^4 t_6^7+ t_1^7 t_2^7 t_3^7 t_4^7 t_5^7 t_6^7
$
\end{quote}

%----------------------------------------------------------------------
\subsection{$Q^{1,1,1}/\mathbb{Z}_3 \ (0,1,1,2,1,2,2,0)$}

\begin{quote}\raggedright \label{esAd01}
$
P(t_a ; \mathcal{M}^{mes})=
1+ t_1^2 t_2^2 t_3 t_4+ t_1^4 t_2^4 t_3^2 t_4^2+ t_1^2 t_2 t_3^2 t_5-t_1^5 t_2^4 t_3^2 t_4^3 t_5+ t_1^4 t_2^2 t_3^4 t_5^2-t_1^7 t_2^5 t_3^4 t_4^3 t_5^2-t_1^5 t_2^2 t_3^4 t_4 t_5^3-t_1^7 t_2^4 t_3^5 t_4^2 t_5^3-t_1^6 t_2^3 t_3^3 t_4^3 t_5^3+ t_1 t_2^2 t_4^2 t_6-t_1^4 t_2^5 t_3^3 t_4^2 t_6+ t_1 t_3^2 t_5^2 t_6-t_1^4 t_2^3 t_3^5 t_5^2 t_6-t_1^4 t_2^3 t_3^2 t_4^3 t_5^2 t_6+ t_1^7 t_2^6 t_3^5 t_4^3 t_5^2 t_6-t_1^4 t_2^2 t_3^3 t_4^2 t_5^3 t_6+ t_1^7 t_2^5 t_3^6 t_4^2 t_5^3 t_6+ t_1^2 t_2^4 t_4^4 t_6^2-t_1^5 t_2^7 t_3^3 t_4^4 t_6^2+ t_2 t_4^2 t_5 t_6^2-t_1^3 t_2^4 t_3^3 t_4^2 t_5 t_6^2-t_1^3 t_2^4 t_4^5 t_5 t_6^2+ t_1^6 t_2^7 t_3^3 t_4^5 t_5 t_6^2+ t_3 t_4 t_5^2 t_6^2-t_1^3 t_2^3 t_3^4 t_4 t_5^2 t_6^2-t_1^3 t_2^3 t_3 t_4^4 t_5^2 t_6^2+ t_1^6 t_2^6 t_3^4 t_4^4 t_5^2 t_6^2-t_1^5 t_2^4 t_3^3 t_4^4 t_5^3 t_6^2+ t_1^8 t_2^7 t_3^6 t_4^4 t_5^3 t_6^2+ t_1^2 t_3^4 t_5^4 t_6^2-t_1^5 t_2^3 t_3^7 t_5^4 t_6^2-t_1^3 t_2 t_3^3 t_4^2 t_5^4 t_6^2+ t_1^6 t_2^4 t_3^6 t_4^2 t_5^4 t_6^2-t_1^5 t_2^3 t_3^4 t_4^3 t_5^4 t_6^2+ t_1^8 t_2^6 t_3^7 t_4^3 t_5^4 t_6^2+ t_1^6 t_2^4 t_3^3 t_4^5 t_5^4 t_6^2-t_1^9 t_2^7 t_3^6 t_4^5 t_5^4 t_6^2-t_1^3 t_3^4 t_4 t_5^5 t_6^2+ t_1^6 t_2^3 t_3^7 t_4 t_5^5 t_6^2+ t_1^6 t_2^3 t_3^4 t_4^4 t_5^5 t_6^2-t_1^9 t_2^6 t_3^7 t_4^4 t_5^5 t_6^2-t_1^3 t_2^6 t_3^3 t_4^3 t_6^3-t_1^2 t_2^5 t_3 t_4^4 t_6^3-t_1^4 t_2^7 t_3^2 t_4^5 t_6^3-t_1^2 t_2^4 t_3^2 t_4^3 t_5 t_6^3+ t_1^5 t_2^7 t_3^2 t_4^6 t_5 t_6^3-t_1^4 t_2^5 t_3^4 t_4^3 t_5^2 t_6^3+ t_1^7 t_2^8 t_3^4 t_4^6 t_5^2 t_6^3-t_1^3 t_2^3 t_3^6 t_5^3 t_6^3-t_1^2 t_2^2 t_3^4 t_4 t_5^3 t_6^3-t_1^4 t_2^4 t_3^5 t_4^2 t_5^3 t_6^3-3 t_1^3 t_2^3 t_3^3 t_4^3 t_5^3 t_6^3+ 2 t_1^6 t_2^6 t_3^6 t_4^3 t_5^3 t_6^3-t_1^2 t_2^2 t_3 t_4^4 t_5^3 t_6^3+ t_1^5 t_2^5 t_3^4 t_4^4 t_5^3 t_6^3-t_1^4 t_2^4 t_3^2 t_4^5 t_5^3 t_6^3+ t_1^7 t_2^7 t_3^5 t_4^5 t_5^3 t_6^3-t_1^3 t_2^3 t_4^6 t_5^3 t_6^3+ 2 t_1^6 t_2^6 t_3^3 t_4^6 t_5^3 t_6^3-t_1^2 t_2 t_3^5 t_5^4 t_6^3-t_1^2 t_2 t_3^2 t_4^3 t_5^4 t_6^3+ t_1^5 t_2^4 t_3^5 t_4^3 t_5^4 t_6^3+ t_1^5 t_2^4 t_3^2 t_4^6 t_5^4 t_6^3-t_1^4 t_2^2 t_3^7 t_5^5 t_6^3-t_1^4 t_2^2 t_3^4 t_4^3 t_5^5 t_6^3+ t_1^7 t_2^5 t_3^7 t_4^3 t_5^5 t_6^3+ t_1^7 t_2^5 t_3^4 t_4^6 t_5^5 t_6^3+ t_1^5 t_2^2 t_3^7 t_4 t_5^6 t_6^3+ t_1^7 t_2^4 t_3^8 t_4^2 t_5^6 t_6^3-t_1^3 t_3^3 t_4^3 t_5^6 t_6^3+ 2 t_1^6 t_2^3 t_3^6 t_4^3 t_5^6 t_6^3+ t_1^5 t_2^2 t_3^4 t_4^4 t_5^6 t_6^3+ t_1^7 t_2^4 t_3^5 t_4^5 t_5^6 t_6^3+ 2 t_1^6 t_2^3 t_3^3 t_4^6 t_5^6 t_6^3-t_1^9 t_2^6 t_3^6 t_4^6 t_5^6 t_6^3-t_1 t_2^3 t_3^2 t_4^3 t_5^2 t_6^4+ t_1^4 t_2^6 t_3^5 t_4^3 t_5^2 t_6^4+ t_2^2 t_4^4 t_5^2 t_6^4-t_1^3 t_2^5 t_3^3 t_4^4 t_5^2 t_6^4+ t_1^4 t_2^6 t_3^2 t_4^6 t_5^2 t_6^4-t_1^7 t_2^9 t_3^5 t_4^6 t_5^2 t_6^4-t_1^3 t_2^5 t_4^7 t_5^2 t_6^4+ t_1^6 t_2^8 t_3^3 t_4^7 t_5^2 t_6^4-t_1 t_2^2 t_3^3 t_4^2 t_5^3 t_6^4+ t_1^4 t_2^5 t_3^6 t_4^2 t_5^3 t_6^4-t_1 t_2^2 t_4^5 t_5^3 t_6^4+ t_1^4 t_2^5 t_3^3 t_4^5 t_5^3 t_6^4+ t_3^2 t_4^2 t_5^4 t_6^4-t_1^3 t_2^3 t_3^5 t_4^2 t_5^4 t_6^4-t_1^3 t_2^3 t_3^2 t_4^5 t_5^4 t_6^4+ t_1^6 t_2^6 t_3^5 t_4^5 t_5^4 t_6^4-t_1 t_3^2 t_4^3 t_5^5 t_6^4+ t_1^4 t_2^3 t_3^5 t_4^3 t_5^5 t_6^4-t_1^3 t_2^2 t_3^3 t_4^4 t_5^5 t_6^4+ t_1^6 t_2^5 t_3^6 t_4^4 t_5^5 t_6^4+ t_1^4 t_2^3 t_3^2 t_4^6 t_5^5 t_6^4-t_1^7 t_2^6 t_3^5 t_4^6 t_5^5 t_6^4+ t_1^6 t_2^5 t_3^3 t_4^7 t_5^5 t_6^4-t_1^9 t_2^8 t_3^6 t_4^7 t_5^5 t_6^4+ t_1^4 t_2^2 t_3^6 t_4^2 t_5^6 t_6^4-t_1^7 t_2^5 t_3^9 t_4^2 t_5^6 t_6^4+ t_1^4 t_2^2 t_3^3 t_4^5 t_5^6 t_6^4-t_1^7 t_2^5 t_3^6 t_4^5 t_5^6 t_6^4-t_1^3 t_3^5 t_4^2 t_5^7 t_6^4+ t_1^6 t_2^3 t_3^8 t_4^2 t_5^7 t_6^4+ t_1^6 t_2^3 t_3^5 t_4^5 t_5^7 t_6^4-t_1^9 t_2^6 t_3^8 t_4^5 t_5^7 t_6^4-t_2^3 t_3 t_4^4 t_5^2 t_6^5+ t_1^3 t_2^6 t_3^4 t_4^4 t_5^2 t_6^5+ t_1^3 t_2^6 t_3 t_4^7 t_5^2 t_6^5-t_1^6 t_2^9 t_3^4 t_4^7 t_5^2 t_6^5-t_1^2 t_2^4 t_3^3 t_4^4 t_5^3 t_6^5+ t_1^5 t_2^7 t_3^6 t_4^4 t_5^3 t_6^5-t_1^2 t_2^4 t_4^7 t_5^3 t_6^5+ t_1^5 t_2^7 t_3^3 t_4^7 t_5^3 t_6^5-t_2 t_3^3 t_4^2 t_5^4 t_6^5+ t_1^3 t_2^4 t_3^6 t_4^2 t_5^4 t_6^5-t_1^2 t_2^3 t_3^4 t_4^3 t_5^4 t_6^5+ t_1^5 t_2^6 t_3^7 t_4^3 t_5^4 t_6^5+ t_1^3 t_2^4 t_3^3 t_4^5 t_5^4 t_6^5-t_1^6 t_2^7 t_3^6 t_4^5 t_5^4 t_6^5+ t_1^5 t_2^6 t_3^4 t_4^6 t_5^4 t_6^5-t_1^8 t_2^9 t_3^7 t_4^6 t_5^4 t_6^5+ t_1^3 t_2^3 t_3^4 t_4^4 t_5^5 t_6^5-t_1^6 t_2^6 t_3^7 t_4^4 t_5^5 t_6^5-t_1^6 t_2^6 t_3^4 t_4^7 t_5^5 t_6^5+ t_1^9 t_2^9 t_3^7 t_4^7 t_5^5 t_6^5+ t_1^5 t_2^4 t_3^6 t_4^4 t_5^6 t_6^5-t_1^8 t_2^7 t_3^9 t_4^4 t_5^6 t_6^5+ t_1^5 t_2^4 t_3^3 t_4^7 t_5^6 t_6^5-t_1^8 t_2^7 t_3^6 t_4^7 t_5^6 t_6^5+ t_1^3 t_2 t_3^6 t_4^2 t_5^7 t_6^5-t_1^6 t_2^4 t_3^9 t_4^2 t_5^7 t_6^5-t_1^2 t_3^4 t_4^3 t_5^7 t_6^5+ t_1^5 t_2^3 t_3^7 t_4^3 t_5^7 t_6^5-t_1^6 t_2^4 t_3^6 t_4^5 t_5^7 t_6^5+ t_1^9 t_2^7 t_3^9 t_4^5 t_5^7 t_6^5+ t_1^5 t_2^3 t_3^4 t_4^6 t_5^7 t_6^5-t_1^8 t_2^6 t_3^7 t_4^6 t_5^7 t_6^5-t_2^3 t_3^3 t_4^3 t_5^3 t_6^6+ 2 t_1^3 t_2^6 t_3^6 t_4^3 t_5^3 t_6^6+ t_1^2 t_2^5 t_3^4 t_4^4 t_5^3 t_6^6+ t_1^4 t_2^7 t_3^5 t_4^5 t_5^3 t_6^6+ 2 t_1^3 t_2^6 t_3^3 t_4^6 t_5^3 t_6^6-t_1^6 t_2^9 t_3^6 t_4^6 t_5^3 t_6^6+ t_1^2 t_2^5 t_3 t_4^7 t_5^3 t_6^6+ t_1^4 t_2^7 t_3^2 t_4^8 t_5^3 t_6^6+ t_1^2 t_2^4 t_3^5 t_4^3 t_5^4 t_6^6+ t_1^2 t_2^4 t_3^2 t_4^6 t_5^4 t_6^6-t_1^5 t_2^7 t_3^5 t_4^6 t_5^4 t_6^6-t_1^5 t_2^7 t_3^2 t_4^9 t_5^4 t_6^6+ t_1^4 t_2^5 t_3^7 t_4^3 t_5^5 t_6^6+ t_1^4 t_2^5 t_3^4 t_4^6 t_5^5 t_6^6-t_1^7 t_2^8 t_3^7 t_4^6 t_5^5 t_6^6-t_1^7 t_2^8 t_3^4 t_4^9 t_5^5 t_6^6+ 2 t_1^3 t_2^3 t_3^6 t_4^3 t_5^6 t_6^6-t_1^6 t_2^6 t_3^9 t_4^3 t_5^6 t_6^6+ t_1^2 t_2^2 t_3^4 t_4^4 t_5^6 t_6^6-t_1^5 t_2^5 t_3^7 t_4^4 t_5^6 t_6^6+ t_1^4 t_2^4 t_3^5 t_4^5 t_5^6 t_6^6-t_1^7 t_2^7 t_3^8 t_4^5 t_5^6 t_6^6+ 2 t_1^3 t_2^3 t_3^3 t_4^6 t_5^6 t_6^6-3 t_1^6 t_2^6 t_3^6 t_4^6 t_5^6 t_6^6-t_1^5 t_2^5 t_3^4 t_4^7 t_5^6 t_6^6-t_1^7 t_2^7 t_3^5 t_4^8 t_5^6 t_6^6-t_1^6 t_2^6 t_3^3 t_4^9 t_5^6 t_6^6+ t_1^2 t_2 t_3^5 t_4^3 t_5^7 t_6^6-t_1^5 t_2^4 t_3^5 t_4^6 t_5^7 t_6^6+ t_1^4 t_2^2 t_3^7 t_4^3 t_5^8 t_6^6-t_1^7 t_2^5 t_3^7 t_4^6 t_5^8 t_6^6-t_1^5 t_2^2 t_3^7 t_4^4 t_5^9 t_6^6-t_1^7 t_2^4 t_3^8 t_4^5 t_5^9 t_6^6-t_1^6 t_2^3 t_3^6 t_4^6 t_5^9 t_6^6-t_2^3 t_3^2 t_4^5 t_5^4 t_6^7+ t_1^3 t_2^6 t_3^5 t_4^5 t_5^4 t_6^7+ t_1^3 t_2^6 t_3^2 t_4^8 t_5^4 t_6^7-t_1^6 t_2^9 t_3^5 t_4^8 t_5^4 t_6^7-t_2^2 t_3^3 t_4^4 t_5^5 t_6^7+ t_1^3 t_2^5 t_3^6 t_4^4 t_5^5 t_6^7+ t_1 t_2^3 t_3^2 t_4^6 t_5^5 t_6^7-t_1^4 t_2^6 t_3^5 t_4^6 t_5^5 t_6^7+ t_1^3 t_2^5 t_3^3 t_4^7 t_5^5 t_6^7-t_1^6 t_2^8 t_3^6 t_4^7 t_5^5 t_6^7-t_1^4 t_2^6 t_3^2 t_4^9 t_5^5 t_6^7+ t_1^7 t_2^9 t_3^5 t_4^9 t_5^5 t_6^7+ t_1 t_2^2 t_3^3 t_4^5 t_5^6 t_6^7-t_1^4 t_2^5 t_3^6 t_4^5 t_5^6 t_6^7+ t_1^3 t_2^3 t_3^5 t_4^5 t_5^7 t_6^7-t_1^6 t_2^6 t_3^8 t_4^5 t_5^7 t_6^7-t_1^6 t_2^6 t_3^5 t_4^8 t_5^7 t_6^7+ t_1^9 t_2^9 t_3^8 t_4^8 t_5^7 t_6^7+ t_1^3 t_2^2 t_3^6 t_4^4 t_5^8 t_6^7-t_1^6 t_2^5 t_3^9 t_4^4 t_5^8 t_6^7-t_1^6 t_2^5 t_3^6 t_4^7 t_5^8 t_6^7+ t_1^9 t_2^8 t_3^9 t_4^7 t_5^8 t_6^7-t_1^4 t_2^2 t_3^6 t_4^5 t_5^9 t_6^7+ t_1^7 t_2^5 t_3^9 t_4^5 t_5^9 t_6^7+ t_1^2 t_2^4 t_3^3 t_4^7 t_5^6 t_6^8-t_1^5 t_2^7 t_3^6 t_4^7 t_5^6 t_6^8+ t_1^2 t_2^3 t_3^4 t_4^6 t_5^7 t_6^8-t_1^5 t_2^6 t_3^7 t_4^6 t_5^7 t_6^8-t_1^5 t_2^6 t_3^4 t_4^9 t_5^7 t_6^8+ t_1^8 t_2^9 t_3^7 t_4^9 t_5^7 t_6^8-t_1^5 t_2^4 t_3^6 t_4^7 t_5^9 t_6^8+ t_1^8 t_2^7 t_3^9 t_4^7 t_5^9 t_6^8-t_1^3 t_2^6 t_3^6 t_4^6 t_5^6 t_6^9-t_1^2 t_2^5 t_3^4 t_4^7 t_5^6 t_6^9-t_1^4 t_2^7 t_3^5 t_4^8 t_5^6 t_6^9-t_1^2 t_2^4 t_3^5 t_4^6 t_5^7 t_6^9+ t_1^5 t_2^7 t_3^5 t_4^9 t_5^7 t_6^9-t_1^4 t_2^5 t_3^7 t_4^6 t_5^8 t_6^9+ t_1^7 t_2^8 t_3^7 t_4^9 t_5^8 t_6^9+ t_1^5 t_2^5 t_3^7 t_4^7 t_5^9 t_6^9+ t_1^7 t_2^7 t_3^8 t_4^8 t_5^9 t_6^9+ t_1^9 t_2^9 t_3^9 t_4^9 t_5^9 t_6^9
$
\end{quote}

\begin{quote}\raggedright \label{esAd02}
$
P(f_1,f_2,f_3,t;\mathcal{M}^{mes})=
1+ \frac{t^6}{f_1^2 f_2^2}+ f_1^2 f_2^2 t^6+ \frac{t^6}{f_1^2 f_3^2}+ \frac{f_2^2 t^6}{f_3^2}+ f_1^2 f_3^2 t^6+ \frac{f_3^2 t^6}{f_2^2}+ \frac{t^{12}}{f_1^4 f_2^4}+ f_1^4 f_2^4 t^{12}+ \frac{t^{12}}{f_1^4 f_3^4}+ \frac{f_2^4 t^{12}}{f_3^4}+ f_1^4 f_3^4 t^{12}+ \frac{f_3^4 t^{12}}{f_2^4}-\frac{t^{15}}{f_1^3 f_2 f_3^5}-\frac{f_1 f_2^3 t^{15}}{f_3^5}-\frac{t^{15}}{f_1 f_2 f_3^3}-\frac{f_2 t^{15}}{f_1^5 f_3^3}-\frac{f_1 f_2 t^{15}}{f_3^3}-\frac{f_2^5 t^{15}}{f_1 f_3^3}-\frac{t^{15}}{f_1^3 f_2^5 f_3}-\frac{t^{15}}{f_1 f_2^3 f_3}-\frac{f_2 t^{15}}{f_1^3 f_3}-\frac{f_1^3 f_2 t^{15}}{f_3}-\frac{f_2^3 t^{15}}{f_1 f_3}-\frac{f_1^5 f_2^3 t^{15}}{f_3}-\frac{f_3 t^{15}}{f_1^5 f_2^3}-\frac{f_1 f_3 t^{15}}{f_2^3}-\frac{f_3 t^{15}}{f_1^3 f_2}-\frac{f_1^3 f_3 t^{15}}{f_2}-f_1 f_2^3 f_3 t^{15}-f_1^3 f_2^5 f_3 t^{15}-\frac{f_1 f_3^3 t^{15}}{f_2^5}-\frac{f_3^3 t^{15}}{f_1 f_2}-\frac{f_1^5 f_3^3 t^{15}}{f_2}-f_1 f_2 f_3^3 t^{15}-\frac{f_3^5 t^{15}}{f_1 f_2^3}-f_1^3 f_2 f_3^5 t^{15}-3 t^{18}-\frac{t^{18}}{f_1^6}-f_1^6 t^{18}-\frac{t^{18}}{f_2^6}-f_2^6 t^{18}-\frac{t^{18}}{f_3^6}-f_3^6 t^{18}-\frac{f_2 t^{21}}{f_1^3 f_3^7}-\frac{f_2^3 t^{21}}{f_1 f_3^7}-\frac{t^{21}}{f_1^7 f_2 f_3^3}-\frac{t^{21}}{f_1 f_2 f_3^3}-\frac{f_1 f_2 t^{21}}{f_3^3}-\frac{f_1 f_2^7 t^{21}}{f_3^3}-\frac{t^{21}}{f_1^7 f_2^3 f_3}-\frac{t^{21}}{f_1 f_2^3 f_3}-\frac{f_2 t^{21}}{f_1^3 f_3}-\frac{f_1^3 f_2 t^{21}}{f_3}-\frac{f_2^3 t^{21}}{f_1 f_3}-\frac{f_1^3 f_2^7 t^{21}}{f_3}-\frac{f_3 t^{21}}{f_1^3 f_2^7}-\frac{f_1 f_3 t^{21}}{f_2^3}-\frac{f_3 t^{21}}{f_1^3 f_2}-\frac{f_1^3 f_3 t^{21}}{f_2}-f_1 f_2^3 f_3 t^{21}-f_1^7 f_2^3 f_3 t^{21}-\frac{f_3^3 t^{21}}{f_1 f_2^7}-\frac{f_3^3 t^{21}}{f_1 f_2}-f_1 f_2 f_3^3 t^{21}-f_1^7 f_2 f_3^3 t^{21}-\frac{f_1 f_3^7 t^{21}}{f_2^3}-\frac{f_1^3 f_3^7 t^{21}}{f_2}+ \frac{t^{24}}{f_1^4 f_2^4}+ \frac{f_1^2 t^{24}}{f_2^4}+ \frac{t^{24}}{f_1^2 f_2^2}+ \frac{f_1^4 t^{24}}{f_2^2}+ \frac{f_2^2 t^{24}}{f_1^4}+ f_1^2 f_2^2 t^{24}+ \frac{f_2^4 t^{24}}{f_1^2}+ f_1^4 f_2^4 t^{24}+ \frac{f_2^2 t^{24}}{f_1^4 f_3^6}+ \frac{f_2^4 t^{24}}{f_1^2 f_3^6}+ \frac{t^{24}}{f_1^4 f_3^4}+ \frac{f_1^2 t^{24}}{f_3^4}+ \frac{t^{24}}{f_2^2 f_3^4}+ \frac{t^{24}}{f_1^6 f_2^2 f_3^4}+ \frac{f_2^4 t^{24}}{f_3^4}+ \frac{f_1^2 f_2^6 t^{24}}{f_3^4}+ \frac{t^{24}}{f_1^2 f_3^2}+ \frac{f_1^4 t^{24}}{f_3^2}+ \frac{t^{24}}{f_2^4 f_3^2}+ \frac{t^{24}}{f_1^6 f_2^4 f_3^2}+ \frac{f_2^2 t^{24}}{f_3^2}+ \frac{f_1^4 f_2^6 t^{24}}{f_3^2}+ \frac{f_3^2 t^{24}}{f_1^4}+ f_1^2 f_3^2 t^{24}+ \frac{f_3^2 t^{24}}{f_1^4 f_2^6}+ \frac{f_3^2 t^{24}}{f_2^2}+ f_2^4 f_3^2 t^{24}+ f_1^6 f_2^4 f_3^2 t^{24}+ \frac{f_3^4 t^{24}}{f_1^2}+ f_1^4 f_3^4 t^{24}+ \frac{f_3^4 t^{24}}{f_1^2 f_2^6}+ \frac{f_3^4 t^{24}}{f_2^4}+ f_2^2 f_3^4 t^{24}+ f_1^6 f_2^2 f_3^4 t^{24}+ \frac{f_1^2 f_3^6 t^{24}}{f_2^4}+ \frac{f_1^4 f_3^6 t^{24}}{f_2^2}+ \frac{2 t^{27}}{f_1^3 f_2^3 f_3^3}+ \frac{2 f_1^3 t^{27}}{f_2^3 f_3^3}+ \frac{2 f_2^3 t^{27}}{f_1^3 f_3^3}+ \frac{2 f_1^3 f_2^3 t^{27}}{f_3^3}+ \frac{2 f_3^3 t^{27}}{f_1^3 f_2^3}+ \frac{2 f_1^3 f_3^3 t^{27}}{f_2^3}+ \frac{2 f_2^3 f_3^3 t^{27}}{f_1^3}+ 2 f_1^3 f_2^3 f_3^3 t^{27}+ \frac{t^{30}}{f_1^4 f_2^4}+ \frac{f_1^2 t^{30}}{f_2^4}+ \frac{t^{30}}{f_1^2 f_2^2}+ \frac{f_1^4 t^{30}}{f_2^2}+ \frac{f_2^2 t^{30}}{f_1^4}+ f_1^2 f_2^2 t^{30}+ \frac{f_2^4 t^{30}}{f_1^2}+ f_1^4 f_2^4 t^{30}+ \frac{f_2^2 t^{30}}{f_1^4 f_3^6}+ \frac{f_2^4 t^{30}}{f_1^2 f_3^6}+ \frac{t^{30}}{f_1^4 f_3^4}+ \frac{f_1^2 t^{30}}{f_3^4}+ \frac{t^{30}}{f_2^2 f_3^4}+ \frac{t^{30}}{f_1^6 f_2^2 f_3^4}+ \frac{f_2^4 t^{30}}{f_3^4}+ \frac{f_1^2 f_2^6 t^{30}}{f_3^4}+ \frac{t^{30}}{f_1^2 f_3^2}+ \frac{f_1^4 t^{30}}{f_3^2}+ \frac{t^{30}}{f_2^4 f_3^2}+ \frac{t^{30}}{f_1^6 f_2^4 f_3^2}+ \frac{f_2^2 t^{30}}{f_3^2}+ \frac{f_1^4 f_2^6 t^{30}}{f_3^2}+ \frac{f_3^2 t^{30}}{f_1^4}+ f_1^2 f_3^2 t^{30}+ \frac{f_3^2 t^{30}}{f_1^4 f_2^6}+ \frac{f_3^2 t^{30}}{f_2^2}+ f_2^4 f_3^2 t^{30}+ f_1^6 f_2^4 f_3^2 t^{30}+ \frac{f_3^4 t^{30}}{f_1^2}+ f_1^4 f_3^4 t^{30}+ \frac{f_3^4 t^{30}}{f_1^2 f_2^6}+ \frac{f_3^4 t^{30}}{f_2^4}+ f_2^2 f_3^4 t^{30}+ f_1^6 f_2^2 f_3^4 t^{30}+ \frac{f_1^2 f_3^6 t^{30}}{f_2^4}+ \frac{f_1^4 f_3^6 t^{30}}{f_2^2}-\frac{f_2 t^{33}}{f_1^3 f_3^7}-\frac{f_2^3 t^{33}}{f_1 f_3^7}-\frac{t^{33}}{f_1^7 f_2 f_3^3}-\frac{t^{33}}{f_1 f_2 f_3^3}-\frac{f_1 f_2 t^{33}}{f_3^3}-\frac{f_1 f_2^7 t^{33}}{f_3^3}-\frac{t^{33}}{f_1^7 f_2^3 f_3}-\frac{t^{33}}{f_1 f_2^3 f_3}-\frac{f_2 t^{33}}{f_1^3 f_3}-\frac{f_1^3 f_2 t^{33}}{f_3}-\frac{f_2^3 t^{33}}{f_1 f_3}-\frac{f_1^3 f_2^7 t^{33}}{f_3}-\frac{f_3 t^{33}}{f_1^3 f_2^7}-\frac{f_1 f_3 t^{33}}{f_2^3}-\frac{f_3 t^{33}}{f_1^3 f_2}-\frac{f_1^3 f_3 t^{33}}{f_2}-f_1 f_2^3 f_3 t^{33}-f_1^7 f_2^3 f_3 t^{33}-\frac{f_3^3 t^{33}}{f_1 f_2^7}-\frac{f_3^3 t^{33}}{f_1 f_2}-f_1 f_2 f_3^3 t^{33}-f_1^7 f_2 f_3^3 t^{33}-\frac{f_1 f_3^7 t^{33}}{f_2^3}-\frac{f_1^3 f_3^7 t^{33}}{f_2}-3 t^{36}-\frac{t^{36}}{f_1^6}-f_1^6 t^{36}-\frac{t^{36}}{f_2^6}-f_2^6 t^{36}-\frac{t^{36}}{f_3^6}-f_3^6 t^{36}-\frac{t^{39}}{f_1^3 f_2 f_3^5}-\frac{f_1 f_2^3 t^{39}}{f_3^5}-\frac{t^{39}}{f_1 f_2 f_3^3}-\frac{f_2 t^{39}}{f_1^5 f_3^3}-\frac{f_1 f_2 t^{39}}{f_3^3}-\frac{f_2^5 t^{39}}{f_1 f_3^3}-\frac{t^{39}}{f_1^3 f_2^5 f_3}-\frac{t^{39}}{f_1 f_2^3 f_3}-\frac{f_2 t^{39}}{f_1^3 f_3}-\frac{f_1^3 f_2 t^{39}}{f_3}-\frac{f_2^3 t^{39}}{f_1 f_3}-\frac{f_1^5 f_2^3 t^{39}}{f_3}-\frac{f_3 t^{39}}{f_1^5 f_2^3}-\frac{f_1 f_3 t^{39}}{f_2^3}-\frac{f_3 t^{39}}{f_1^3 f_2}-\frac{f_1^3 f_3 t^{39}}{f_2}-f_1 f_2^3 f_3 t^{39}-f_1^3 f_2^5 f_3 t^{39}-\frac{f_1 f_3^3 t^{39}}{f_2^5}-\frac{f_3^3 t^{39}}{f_1 f_2}-\frac{f_1^5 f_3^3 t^{39}}{f_2}-f_1 f_2 f_3^3 t^{39}-\frac{f_3^5 t^{39}}{f_1 f_2^3}-f_1^3 f_2 f_3^5 t^{39}+ \frac{t^{42}}{f_1^4 f_2^4}+ f_1^4 f_2^4 t^{42}+ \frac{t^{42}}{f_1^4 f_3^4}+ \frac{f_2^4 t^{42}}{f_3^4}+ f_1^4 f_3^4 t^{42}+ \frac{f_3^4 t^{42}}{f_2^4}+ \frac{t^{48}}{f_1^2 f_2^2}+ f_1^2 f_2^2 t^{48}+ \frac{t^{48}}{f_1^2 f_3^2}+ \frac{f_2^2 t^{48}}{f_3^2}+ f_1^2 f_3^2 t^{48}+ \frac{f_3^2 t^{48}}{f_2^2}+ t^{54}
$
\end{quote}

%----------------------------------------------------------------------
\subsection{$Q^{1,1,1}/ \mathbb{Z}_2 \times \mathbb{Z}_2 \ \left(\begin{array}{@{}c@{,\;}c@{,\;}c@{,\;}c@{,\;}c@{,\;}c@{,\;}c@{,\;}c@{}} 1 & 1 & 0 & 0 & 0 & 0 & 1 & 1 \\ 0 & 0 & 1 & 1 & 1 & 1 & 0 & 0 \end{array}\right)$}

\begin{quote}\raggedright \label{esAe01}
$
P(t_a ; \mathcal{M}^{mes})=
1+ t_1^2 t_2^2 t_3 t_4+ t_1^2 t_3 t_4 t_5^2-t_1^4 t_2^2 t_3^3 t_4 t_5^2-t_1^4 t_2^2 t_3^2 t_4^2 t_5^2-t_1^4 t_2^2 t_3 t_4^3 t_5^2+ t_1 t_2 t_3^2 t_5 t_6+ t_1 t_2 t_4^2 t_5 t_6-2 t_1^3 t_2^3 t_3^2 t_4^2 t_5 t_6-2 t_1^3 t_2 t_3^2 t_4^2 t_5^3 t_6+ t_1^5 t_2^3 t_3^4 t_4^2 t_5^3 t_6+ t_1^5 t_2^3 t_3^2 t_4^4 t_5^3 t_6+ t_2^2 t_3 t_4 t_6^2-t_1^2 t_2^4 t_3^3 t_4 t_6^2-t_1^2 t_2^4 t_3^2 t_4^2 t_6^2-t_1^2 t_2^4 t_3 t_4^3 t_6^2-t_1^2 t_2^2 t_3^4 t_5^2 t_6^2+ t_3 t_4 t_5^2 t_6^2-4 t_1^2 t_2^2 t_3^3 t_4 t_5^2 t_6^2+ t_1^4 t_2^4 t_3^5 t_4 t_5^2 t_6^2-3 t_1^2 t_2^2 t_3^2 t_4^2 t_5^2 t_6^2+ 2 t_1^4 t_2^4 t_3^4 t_4^2 t_5^2 t_6^2-4 t_1^2 t_2^2 t_3 t_4^3 t_5^2 t_6^2+ 4 t_1^4 t_2^4 t_3^3 t_4^3 t_5^2 t_6^2-t_1^2 t_2^2 t_4^4 t_5^2 t_6^2+ 2 t_1^4 t_2^4 t_3^2 t_4^4 t_5^2 t_6^2+ t_1^4 t_2^4 t_3 t_4^5 t_5^2 t_6^2-t_1^2 t_3^3 t_4 t_5^4 t_6^2+ t_1^4 t_2^2 t_3^5 t_4 t_5^4 t_6^2-t_1^2 t_3^2 t_4^2 t_5^4 t_6^2+ 2 t_1^4 t_2^2 t_3^4 t_4^2 t_5^4 t_6^2-t_1^2 t_3 t_4^3 t_5^4 t_6^2+ 4 t_1^4 t_2^2 t_3^3 t_4^3 t_5^4 t_6^2-t_1^6 t_2^4 t_3^5 t_4^3 t_5^4 t_6^2+ 2 t_1^4 t_2^2 t_3^2 t_4^4 t_5^4 t_6^2-t_1^6 t_2^4 t_3^4 t_4^4 t_5^4 t_6^2+ t_1^4 t_2^2 t_3 t_4^5 t_5^4 t_6^2-t_1^6 t_2^4 t_3^3 t_4^5 t_5^4 t_6^2-2 t_1 t_2^3 t_3^2 t_4^2 t_5 t_6^3+ t_1^3 t_2^5 t_3^4 t_4^2 t_5 t_6^3+ t_1^3 t_2^5 t_3^2 t_4^4 t_5 t_6^3-t_1^3 t_2^3 t_3^6 t_5^3 t_6^3-2 t_1 t_2 t_3^2 t_4^2 t_5^3 t_6^3+ 3 t_1^3 t_2^3 t_3^4 t_4^2 t_5^3 t_6^3+ 3 t_1^3 t_2^3 t_3^2 t_4^4 t_5^3 t_6^3-2 t_1^5 t_2^5 t_3^4 t_4^4 t_5^3 t_6^3-t_1^3 t_2^3 t_4^6 t_5^3 t_6^3+ t_1^3 t_2 t_3^4 t_4^2 t_5^5 t_6^3+ t_1^3 t_2 t_3^2 t_4^4 t_5^5 t_6^3-2 t_1^5 t_2^3 t_3^4 t_4^4 t_5^5 t_6^3-t_2^2 t_3^3 t_4 t_5^2 t_6^4+ t_1^2 t_2^4 t_3^5 t_4 t_5^2 t_6^4-t_2^2 t_3^2 t_4^2 t_5^2 t_6^4+ 2 t_1^2 t_2^4 t_3^4 t_4^2 t_5^2 t_6^4-t_2^2 t_3 t_4^3 t_5^2 t_6^4+ 4 t_1^2 t_2^4 t_3^3 t_4^3 t_5^2 t_6^4-t_1^4 t_2^6 t_3^5 t_4^3 t_5^2 t_6^4+ 2 t_1^2 t_2^4 t_3^2 t_4^4 t_5^2 t_6^4-t_1^4 t_2^6 t_3^4 t_4^4 t_5^2 t_6^4+ t_1^2 t_2^4 t_3 t_4^5 t_5^2 t_6^4-t_1^4 t_2^6 t_3^3 t_4^5 t_5^2 t_6^4+ t_1^2 t_2^2 t_3^5 t_4 t_5^4 t_6^4+ 2 t_1^2 t_2^2 t_3^4 t_4^2 t_5^4 t_6^4-t_1^4 t_2^4 t_3^6 t_4^2 t_5^4 t_6^4+ 4 t_1^2 t_2^2 t_3^3 t_4^3 t_5^4 t_6^4-4 t_1^4 t_2^4 t_3^5 t_4^3 t_5^4 t_6^4+ 2 t_1^2 t_2^2 t_3^2 t_4^4 t_5^4 t_6^4-3 t_1^4 t_2^4 t_3^4 t_4^4 t_5^4 t_6^4+ t_1^2 t_2^2 t_3 t_4^5 t_5^4 t_6^4-4 t_1^4 t_2^4 t_3^3 t_4^5 t_5^4 t_6^4+ t_1^6 t_2^6 t_3^5 t_4^5 t_5^4 t_6^4-t_1^4 t_2^4 t_3^2 t_4^6 t_5^4 t_6^4-t_1^4 t_2^2 t_3^5 t_4^3 t_5^6 t_6^4-t_1^4 t_2^2 t_3^4 t_4^4 t_5^6 t_6^4-t_1^4 t_2^2 t_3^3 t_4^5 t_5^6 t_6^4+ t_1^6 t_2^4 t_3^5 t_4^5 t_5^6 t_6^4+ t_1 t_2^3 t_3^4 t_4^2 t_5^3 t_6^5+ t_1 t_2^3 t_3^2 t_4^4 t_5^3 t_6^5-2 t_1^3 t_2^5 t_3^4 t_4^4 t_5^3 t_6^5-2 t_1^3 t_2^3 t_3^4 t_4^4 t_5^5 t_6^5+ t_1^5 t_2^5 t_3^6 t_4^4 t_5^5 t_6^5+ t_1^5 t_2^5 t_3^4 t_4^6 t_5^5 t_6^5-t_1^2 t_2^4 t_3^5 t_4^3 t_5^4 t_6^6-t_1^2 t_2^4 t_3^4 t_4^4 t_5^4 t_6^6-t_1^2 t_2^4 t_3^3 t_4^5 t_5^4 t_6^6+ t_1^4 t_2^6 t_3^5 t_4^5 t_5^4 t_6^6+ t_1^4 t_2^4 t_3^5 t_4^5 t_5^6 t_6^6+ t_1^6 t_2^6 t_3^6 t_4^6 t_5^6 t_6^6
$
\end{quote}

%----------------------------------------------------------------------
\subsection{$D_3/\mathbb{Z}_3 \ (0,1,2,1,2)$}

\begin{quote}\raggedright \label{esAf01}
$
P(t_a ; \mathcal{M}^{mes})=
1+ t_1^2 t_2 t_3 t_5+ t_1^4 t_2^2 t_3^2 t_5^2+ t_1^2 t_3^3 t_4 t_6+ t_1 t_3 t_5 t_6+ t_1 t_2 t_3 t_4 t_5 t_6+ t_1^3 t_2 t_3^2 t_5^2 t_6+ t_1^3 t_2^2 t_3^2 t_4 t_5^2 t_6+ t_1^2 t_2^2 t_5^3 t_6+ t_1 t_3^3 t_4^2 t_6^2+ t_3 t_4 t_5 t_6^2+ t_1^3 t_2 t_3^4 t_4^2 t_5 t_6^2+ t_1^2 t_3^2 t_5^2 t_6^2+ t_1^2 t_2 t_3^2 t_4 t_5^2 t_6^2+ t_1^2 t_2^2 t_3^2 t_4^2 t_5^2 t_6^2+ t_1 t_2 t_5^3 t_6^2+ t_1^4 t_2^2 t_3^3 t_4 t_5^3 t_6^2+ t_1^3 t_2^2 t_3 t_5^4 t_6^2+ t_1^2 t_3^4 t_4^2 t_5 t_6^3+ t_1 t_3^2 t_4 t_5^2 t_6^3+ t_1 t_2 t_3^2 t_4^2 t_5^2 t_6^3+ t_1^3 t_2 t_3^3 t_4 t_5^3 t_6^3+ t_1^3 t_2^2 t_3^3 t_4^2 t_5^3 t_6^3+ t_1^2 t_2^2 t_3 t_4 t_5^4 t_6^3+ t_3^2 t_4^2 t_5^2 t_6^4+ t_1^2 t_2 t_3^3 t_4^2 t_5^3 t_6^4+ t_1^4 t_2^2 t_3^4 t_4^2 t_5^4 t_6^4
$
\end{quote}

\begin{quote}\raggedright \label{esAf02}
$
P(f_1,f_2,f_3,t; \mathcal{M}^{mes})=
1 + \frac{t^4}{f_2 f_3} + \frac{f_1^2 t^5}{f_2} + \frac{t^5}{f_1^2 f_3} + t^6 + \frac{f_1 f_2 t^7}{f_3^3} + \frac{f_3 t^7}{f_1 f_2^3} + \frac{f_2^2 t^8}{f_1 f_3^3} + \frac{t^8}{f_2^2 f_3^2} + \frac{f_1 f_3^2 t^8}{f_2^3} + \frac{t^9}{f_1^2 f_2 f_3^2} + \frac{f_1^2 t^9}{f_2^2 f_3} + \frac{f_1^4 t^{10}}{f_2^2} + \frac{t^{10}}{f_1^4 f_3^2} + \frac{t^{10}}{f_2 f_3} + \frac{f_1^2 t^{11}}{f_2} + \frac{t^{11}}{f_1^2 f_3} + t^{12} + \frac{f_2 t^{12}}{f_1 f_3^4} + \frac{f_1 f_3 t^{12}}{f_2^4} + \frac{f_1 f_2 t^{13}}{f_3^3} + \frac{f_3 t^{13}}{f_1 f_2^3} + \frac{t^{14}}{f_2^2 f_3^2} + \frac{t^{15}}{f_1^2 f_2 f_3^2} + \frac{f_1^2 t^{15}}{f_2^2 f_3} + \frac{t^{16}}{f_2 f_3} + \frac{t^{20}}{f_2^2 f_3^2}
$
\end{quote}

%----------------------------------------------------------------------
\subsection{$D_3/\mathbb{Z}_3 \ (1,1,1,1,2)$}

\begin{quote}\raggedright \label{esCa07}
$
P(t_a; \mathcal{M}^{mes})=
1+ t_1^2 t_2 t_3^2 t_4+ t_1 t_2^2 t_3 t_4^2+ t_1^3 t_2 t_3^2 t_5+ t_1^2 t_2^2 t_3 t_4 t_5+ t_1 t_2^3 t_4^2 t_5+ t_1^3 t_2^2 t_3 t_5^2+ t_1^2 t_2^3 t_4 t_5^2+ t_1^4 t_2^4 t_3^2 t_4^2 t_5^2+ t_1 t_3^2 t_4 t_6+ t_2 t_3 t_4^2 t_6+ t_1^2 t_2^2 t_3^3 t_4^3 t_6+ t_1^2 t_3^2 t_5 t_6+ 2 t_1 t_2 t_3 t_4 t_5 t_6+ t_2^2 t_4^2 t_5 t_6+ t_1^3 t_2^2 t_3^3 t_4^2 t_5 t_6+ t_1^2 t_2^3 t_3^2 t_4^3 t_5 t_6+ t_1^2 t_2 t_3 t_5^2 t_6+ t_1 t_2^2 t_4 t_5^2 t_6+ t_1^4 t_2^2 t_3^3 t_4 t_5^2 t_6+ 2 t_1^3 t_2^3 t_3^2 t_4^2 t_5^2 t_6+ t_1^2 t_2^4 t_3 t_4^3 t_5^2 t_6+ t_1^2 t_2^2 t_5^3 t_6+ t_1^4 t_2^3 t_3^2 t_4 t_5^3 t_6+ t_1^3 t_2^4 t_3 t_4^2 t_5^3 t_6+ t_1^2 t_3^4 t_4^2 t_6^2+ t_1 t_2 t_3^3 t_4^3 t_6^2+ t_2^2 t_3^2 t_4^4 t_6^2+ t_3 t_4 t_5 t_6^2+ 2 t_1^2 t_2 t_3^3 t_4^2 t_5 t_6^2+ 2 t_1 t_2^2 t_3^2 t_4^3 t_5 t_6^2+ t_1^3 t_2^3 t_3^4 t_4^4 t_5 t_6^2+ t_1 t_3 t_5^2 t_6^2+ t_2 t_4 t_5^2 t_6^2+ t_1^3 t_2 t_3^3 t_4 t_5^2 t_6^2+ 3 t_1^2 t_2^2 t_3^2 t_4^2 t_5^2 t_6^2+ t_1 t_2^3 t_3 t_4^3 t_5^2 t_6^2+ t_1^4 t_2^3 t_3^4 t_4^3 t_5^2 t_6^2+ t_1^3 t_2^4 t_3^3 t_4^4 t_5^2 t_6^2+ t_1 t_2 t_5^3 t_6^2+ 2 t_1^3 t_2^2 t_3^2 t_4 t_5^3 t_6^2+ 2 t_1^2 t_2^3 t_3 t_4^2 t_5^3 t_6^2+ t_1^4 t_2^4 t_3^3 t_4^3 t_5^3 t_6^2+ t_1^4 t_2^2 t_3^2 t_5^4 t_6^2+ t_1^3 t_2^3 t_3 t_4 t_5^4 t_6^2+ t_1^2 t_2^4 t_4^2 t_5^4 t_6^2+ t_1 t_3^3 t_4^2 t_5 t_6^3+ t_2 t_3^2 t_4^3 t_5 t_6^3+ t_1^2 t_2^2 t_3^4 t_4^4 t_5 t_6^3+ t_1^2 t_3^3 t_4 t_5^2 t_6^3+ 2 t_1 t_2 t_3^2 t_4^2 t_5^2 t_6^3+ t_2^2 t_3 t_4^3 t_5^2 t_6^3+ t_1^3 t_2^2 t_3^4 t_4^3 t_5^2 t_6^3+ t_1^2 t_2^3 t_3^3 t_4^4 t_5^2 t_6^3+ t_1^2 t_2 t_3^2 t_4 t_5^3 t_6^3+ t_1 t_2^2 t_3 t_4^2 t_5^3 t_6^3+ t_1^4 t_2^2 t_3^4 t_4^2 t_5^3 t_6^3+ 2 t_1^3 t_2^3 t_3^3 t_4^3 t_5^3 t_6^3+ t_1^2 t_2^4 t_3^2 t_4^4 t_5^3 t_6^3+ t_1^2 t_2^2 t_3 t_4 t_5^4 t_6^3+ t_1^4 t_2^3 t_3^3 t_4^2 t_5^4 t_6^3+ t_1^3 t_2^4 t_3^2 t_4^3 t_5^4 t_6^3+ t_3^2 t_4^2 t_5^2 t_6^4+ t_1^2 t_2 t_3^4 t_4^3 t_5^2 t_6^4+ t_1 t_2^2 t_3^3 t_4^4 t_5^2 t_6^4+ t_1^3 t_2 t_3^4 t_4^2 t_5^3 t_6^4+ t_1^2 t_2^2 t_3^3 t_4^3 t_5^3 t_6^4+ t_1 t_2^3 t_3^2 t_4^4 t_5^3 t_6^4+ t_1^3 t_2^2 t_3^3 t_4^2 t_5^4 t_6^4+ t_1^2 t_2^3 t_3^2 t_4^3 t_5^4 t_6^4+ t_1^4 t_2^4 t_3^4 t_4^4 t_5^4 t_6^4
$
\end{quote}

\begin{quote}\raggedright \label{esCb03}
$
P(f_1,f_2,f_3,t; \mathcal{M}^{mes})=
1+ \frac{f_2^2 t^5}{f_1}+ \frac{f_2 t^5}{f_3^2}+ \frac{t^5}{f_1^2 f_3}+ 2 t^6+ \frac{f_1 t^6}{f_2 f_3^2}+ \frac{t^6}{f_1 f_2^2 f_3}+ \frac{f_1^2 f_2 t^6}{f_3}+ \frac{f_3 t^6}{f_1^2 f_2}+ f_1 f_2^2 f_3 t^6+ \frac{f_2 f_3^2 t^6}{f_1}+ \frac{f_1 t^7}{f_2^2}+ \frac{f_1^3 t^7}{f_2 f_3}+ f_1^2 f_3 t^7+ \frac{f_3 t^7}{f_1 f_2^3}+ \frac{f_3^2 t^7}{f_2}+ f_1 f_2 f_3^3 t^7+ \frac{f_1^3 f_3 t^8}{f_2^2}+ \frac{f_1 f_3^2 t^8}{f_2^3}+ \frac{f_1^2 f_3^3 t^8}{f_2}+ \frac{f_2^4 t^{10}}{f_1^2}+ \frac{f_2^2 t^{10}}{f_3^4}+ \frac{f_2 t^{10}}{f_1^2 f_3^3}+ \frac{t^{10}}{f_1^4 f_3^2}+ \frac{f_2^3 t^{10}}{f_1 f_3^2}+ \frac{f_2^2 t^{10}}{f_1^3 f_3}+ \frac{2 f_2^2 t^{11}}{f_1}+ \frac{t^{11}}{f_1 f_2 f_3^3}+ \frac{2 f_2 t^{11}}{f_3^2}+ \frac{2 t^{11}}{f_1^2 f_3}+ \frac{f_1 f_2^3 t^{11}}{f_3}+ \frac{f_2 f_3 t^{11}}{f_1^3}+ 3 t^{12}+ \frac{f_1 t^{12}}{f_2 f_3^2}+ \frac{t^{12}}{f_1 f_2^2 f_3}+ \frac{f_1^2 f_2 t^{12}}{f_3}+ \frac{f_3 t^{12}}{f_1^2 f_2}+ f_1 f_2^2 f_3 t^{12}+ \frac{f_2 f_3^2 t^{12}}{f_1}+ \frac{2 f_1 t^{13}}{f_2^2}+ \frac{f_1^3 t^{13}}{f_2 f_3}+ 2 f_1^2 f_3 t^{13}+ \frac{f_3 t^{13}}{f_1 f_2^3}+ \frac{2 f_3^2 t^{13}}{f_2}+ f_1 f_2 f_3^3 t^{13}+ \frac{f_1^2 t^{14}}{f_2^4}+ \frac{f_1^3 f_3 t^{14}}{f_2^2}+ f_1^4 f_3^2 t^{14}+ \frac{f_1 f_3^2 t^{14}}{f_2^3}+ \frac{f_1^2 f_3^3 t^{14}}{f_2}+ \frac{f_3^4 t^{14}}{f_2^2}+ \frac{f_2 t^{16}}{f_1^2 f_3^3}+ \frac{f_2^3 t^{16}}{f_1 f_3^2}+ \frac{f_2^2 t^{16}}{f_1^3 f_3}+ \frac{f_2^2 t^{17}}{f_1}+ \frac{t^{17}}{f_1 f_2 f_3^3}+ \frac{f_2 t^{17}}{f_3^2}+ \frac{t^{17}}{f_1^2 f_3}+ \frac{f_1 f_2^3 t^{17}}{f_3}+ \frac{f_2 f_3 t^{17}}{f_1^3}+ 2 t^{18}+ \frac{f_1 t^{18}}{f_2 f_3^2}+ \frac{t^{18}}{f_1 f_2^2 f_3}+ \frac{f_1^2 f_2 t^{18}}{f_3}+ \frac{f_3 t^{18}}{f_1^2 f_2}+ f_1 f_2^2 f_3 t^{18}+ \frac{f_2 f_3^2 t^{18}}{f_1}+ \frac{f_1 t^{19}}{f_2^2}+ f_1^2 f_3 t^{19}+ \frac{f_3^2 t^{19}}{f_2}+ t^{24}
$
\end{quote}

%======================================================================
\bibliographystyle{JHEP}
\bibliography{mybib}
%======================================================================

\end{document}